# Theorem Proving and Algebra

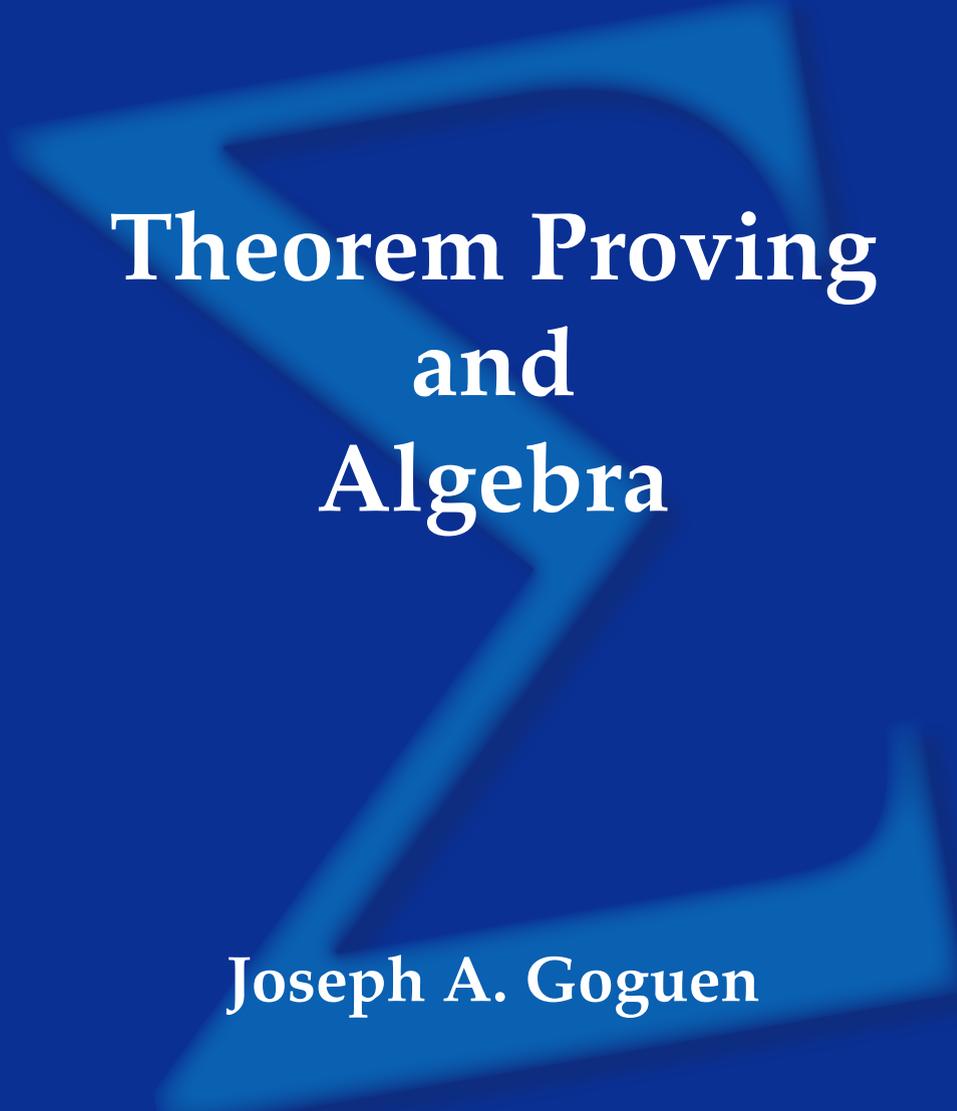

Joseph A. Goguen

# Theorem Proving
# and
# Algebra

# Theorem Proving
# and
# Algebra

Joseph A. Goguen

University of California, San Diego

Joseph A. Goguen (1941–2006)
Department of Computer Science and Engineering
University of California, San Diego
9500 Gilman Drive, La Jolla CA 92093-0114 USA

April 2006

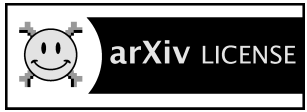

Edited by:
   Kokichi Futatsugi
   Narciso Martí-Oliet
   José Meseguer

Text reviewed and commented by:
   Kokichi Futatsugi
   Daniel Mircea Gaina
   Narciso Martí-Oliet
   José Meseguer
   Masaki Nakamura
   Miguel Palomino

Book designed and typeset by:
   Alberto Verdejo

# Contents

**6 Initial Algebras, Standard Models and Induction     159**
  6.1  Quotient and Initiality . . . . . . . . . . . . . . . . .     159
      6.1.1  Congruence, Quotient and Image . . . . . . . .     159
      6.1.2  Initiality and Freedom . . . . . . . . . . . . .     165
      6.1.3  Substitution Modulo Equations . . . . . . . . .     166
  6.2  Abstract Data Types . . . . . . . . . . . . . . . . . . .     167
      6.2.1  Motivation . . . . . . . . . . . . . . . . . . . .     167
      6.2.2  Formal Definition . . . . . . . . . . . . . . . .     168
  6.3  Standard Models are Initial Models . . . . . . . . . . .     171
  6.4  Initial Truth and Subalgebras . . . . . . . . . . . . . . .     173
  6.5  Induction . . . . . . . . . . . . . . . . . . . . . . . . .     174
      6.5.1  Simple Inductive Proofs with OBJ . . . . . . . .     176
      6.5.2  Many-Sorted Induction . . . . . . . . . . . . .     179
  6.6  A Closer Look at State, Encapsulation and Implementation . . . . . . . . . . . . . . . . . . . . . . . . . . . .     181
  6.7  Literature . . . . . . . . . . . . . . . . . . . . . . . . .     182

**7 Deduction and Rewriting Modulo Equations     183**
  7.1  Motivation . . . . . . . . . . . . . . . . . . . . . . . .     183
  7.2  Deduction Modulo Equations . . . . . . . . . . . . . .     185
      7.2.1  Some Implementation Issues . . . . . . . . . .     187
      7.2.2  Deduction Modulo Equations in OBJ3 . . . . .     189
  7.3  Term Rewriting Modulo Equations . . . . . . . . . . . .     192
      7.3.1  Some Inductive Proofs Modulo Equations . . . .     197
      7.3.2  The Propositional Calculus . . . . . . . . . . .     200
      7.3.3  Weak Rewriting Modulo Equations . . . . . . .     207
  7.4  Verification of Hardware Circuits . . . . . . . . . . . .     207
      7.4.1  Unconditional Combinatorial Circuits . . . . . .     208
      7.4.2  Conditional Triangular Systems . . . . . . . . .     214
      7.4.3  Proof Techniques . . . . . . . . . . . . . . . .     218
      7.4.4  Conditional Combinatorial Circuits . . . . . . .     220
      7.4.5  Bidirectional Circuits . . . . . . . . . . . . . .     224
  7.5  Proving Termination Modulo Equations . . . . . . . . .     227
  7.6  Proving Church-Rosser Modulo Equations . . . . . . . .     232
      7.6.1  Adding Constants . . . . . . . . . . . . . . . .     235
  7.7  Conditional Term Rewriting Modulo Equations . . . . .     235
      7.7.1  Adding Constants . . . . . . . . . . . . . . . .     240
      7.7.2  Proving Termination . . . . . . . . . . . . . .     241
      7.7.3  Proving Church-Rosser . . . . . . . . . . . . .     243
  7.8  Literature . . . . . . . . . . . . . . . . . . . . . . . . .     244

**8 First-Order Logic and Proof Planning     247**
  8.1  First-Order Signatures, Models and Morphisms . . . . .     247
  8.2  Horn Clause Logic . . . . . . . . . . . . . . . . . . . .     249
      8.2.1  (⋆) Initial Horn Models . . . . . . . . . . . . . .     250
  8.3  First-Order Logic . . . . . . . . . . . . . . . . . . . . .     253

**x** Contents

# List of Figures

# Foreword by Editors

Two of us, Futatsugi and Meseguer, had the privilege of working closely with Joseph Goguen, were influenced by his very creative and fundamental ideas, and, on the occasion of the Festschrift organized in his honor for his 65th birthday in San Diego, California, we wrote:

> *Joseph Goguen is one of the most prominent computer scientists worldwide. His numerous research contributions span many topics and have changed the way we think about many concepts. Our views about data types, programming languages, software specification and verification, computational behavior, logics in computer science, semiotics, interface design, multimedia, and consciousness, to mention just some of the areas, have all been enriched in fundamental ways by his ideas.*

Sadly, Joseph Goguen's life was cut short due to a fatal illness some days after the Festschrift Symposium in his honor, in which he could still be present. He was at that time working on *Theorem Proving and Algebra* (TPA), a long-term project still unfinished, yet quite advanced. The TPA book provides the definitive mathematical foundation for algebraic theorem proving. Professor Goguen also exposes formal methods with the OBJ language system. This is quite unique and an important feature of the book.

We are convinced that Joseph Goguen's ideas in the TPA book have a fundamental and lasting value and should be made available to the research community. Furthermore, as we explain below, they have influenced subsequent work in several algebraic languages originating in the OBJ language on which two of us, Futatsugi and Meseguer, worked closely with Joseph Goguen, namely, CafeOBJ and Maude. However, the TPA book should be *his* book, with no efforts to complete parts of the manuscript that were unfinished or in any way modify its contents.

Our approach to this task has been the usual one in editing any part of the *nachlass* of a scholar: to make only small corrections of typos or small mistakes that, clearly, the author would have himself wished to be done; and to add a few explanatory *editorial notes* —clearly marked as such, and different from the text itself— to help the reader better understand some specific points in the text: again, making the best guess possible about what the author himself might have wished to add as explanations, given the unfinished nature of the text. Both the small corrections and the editorial notes are based on careful revisions of the original text by the editors with the additional help of Daniel Mircea Gaina, Masaki Nakamura, and Miguel Palomino.

## Impact on CafeOBJ and Maude

CafeOBJ (https://cafeobj.org/) and Maude (https://maude.cs.uiuc.edu/) are two sibling languages of OBJ which draw significant inspiration from Joseph Goguen's work presented in the TPA book. In what follows we explain several ways in which the ideas in the TPA book have influenced further developments in both CafeOBJ and Maude.

**CafeOBJ** inherits from OBJ distinctive features such as user-defined mix-fix syntax, subtyping by ordered sorts, module system with parameterized module expressions, conditional rewriting with associative/commutative matching, loose and tight (or initial) semantics for modules, and theorem proving with proof scores.

CafeOBJ adds to OBJ new features such as behavioural (or observational) abstraction with hidden algebra, rewriting logic à la Maude for specifying transition systems, and their combinations with order-sorted algebra. This multiparadigm approach has a mathematical semantics based on multiple institutions. Some theorem-proving capabilities are also added, including behavioral rewriting, observational coinduction, and built-in search predicates.

Transition systems can be specified with observational abstraction or with rewriting rules in CafeOBJ. The observational style is more abstract/algebraic and there is no need to determine state configurations that are instead needed to define state transitions via rewriting rules. The built-in search predicates facilitate verification of rewriting-based transition systems. Both styles have their own merits and it is worthwhile to support both in CafeOBJ.

In the TPA book's introduction (1 Introduction) Professor Goguen states:

> *We do not pursue the lofty goal of mechanizing proofs like those of which mathematicians are justly so proud; instead, we seek to take steps towards providing mechanical assistance for proofs that are useful for computer scientists in developing software and hardware. This more modest goal has the advantage of both being achievable and having practical benefits.*

He continues (1.7 Logical Systems and Proof Scores):

> *The first step of our approach is to construct proof scores, which are instructions such that when executed (or "played"), if everything evaluates as expected, then the desired theorem is proved. A proof score is executed by applying proof measures, which progressively transform formulae in a language of goals into expressions which can be directly executed. We will see that equational logic is adequate for implementing versions of first and second-order logic in this way, as well as many other logical systems.*

and (1.8 Semantics and Soundness):

> *This text justifies proof measures for a logical system by demonstrating their soundness with respect to the notions of model and satisfaction for that system. In this sense, it places Semantics First! In fact, users are primarily concerned with truth; they want to know whether certain properties are true of certain models, which may be realized in software and/or hardware. From this point of view, proof is a necessary nuisance that we tolerate only because we have no other way to effectively demonstrate truth. Moreover, it is usually easier and more intuitive to justify proof measures on semantic rather than syntactic grounds.*

Inspired by the above-stated OBJ's proof score approach and its potential for realizing well structured and reusable proof documents, theorem proving with proof scores has been pursued extensively in CafeOBJ. Many case studies have been done in a variety of application areas and the proof scores have been found to be usable for practical theorem proving.

Constructor-based algebra and its proof calculus, which are not included in the TPA book, were formalized as a theoretical foundation for proof scores in CafeOBJ. The Constructor-based Inductive Theorem Prover (CITP) was first implemented in Maude Metalevel and its variant is now incorporated into CafeOBJ as a Proof Tree Calculus (PTcalc) subsystem. The "Semantics First!" principle played an important role in the design and implementation of PTcalc. As a result, PTcalc encourages model-based analyses and proofs of properties to be verified, which is an important merit of algebraic theorem proving.

The harmony of (1) model satisfaction semantics, (2) equational deduction, and (3) rewriting execution constitutes the core of algebraic theorem proving and the TPA book provides the most reliable and comprehensive account of it. A proof score applies proof measures by executing equations as rewriting rules to prove the model satisfaction. The harmony of semantics, deduction, and execution makes it possible to formalize effective and transparent proof measures. Major proof measures in CafeOBJ include (i) case split with exhaustive equations and (ii) well-founded induction via term refinement.

CafeOBJ's module system is basically the same as OBJ's one, except for succinct notation for inline view definition and gradually developed reliable and efficient implementation. The module system is an important feature of algebraic language systems and its power has been well appreciated. CafeOBJ's module system works significantly not only for constructing specifications and proof scores but also for preparing libraries of generic data structures and proof measures. The PTcalc subsystem of CafeOBJ, where proof nodes are modules, depends on the module system in a significant way.

**Maude** is a language based on rewriting logic, a simple computational logic to specify and program concurrent systems as initial models of rewrite theories. In Maude they are specified as system modules of the form: mod FOO is $(\Sigma, E, R)$ endm, where the rewrite theory $(\Sigma, E, R)$ specifies a concurrent system whose concurrent states belong to the algebraic data type (initial algebra) $T_{\Sigma/E}$ —with $\Sigma$ a typed signature of function symbols and $E$ a set of equations— and whose local concurrent transitions are specified by the rewrite rules $R$. When $R = \emptyset$, a rewrite theory $(\Sigma, E, R)$ becomes an equational theory $(\Sigma, E)$. In Maude this gives rise to a sublanguage of functional modules of the form fmod BAR is $(\Sigma, E)$ endfm with initial algebra semantics, which is a superset of OBJ, because Maude is based on the more expressive membership equational logic, which contains OBJ's order-sorted equational logic as a special case.

Maude naturally extends OBJ, because membership equational logic is itself a sublogic (the case $R = \emptyset$) of rewriting logic. The practical meaning of this extension is that equational logic is well-suited to specify deterministic systems, whereas rewriting logic does naturally specify non-deterministic and concurrent systems. Furthermore, Maude has the following additional features: (i) reachability analysis; (ii) LTL model checking; (iii) a strategy language to guide the execution of rewrite theories; (iv) concurrent object-oriented system specification, including external objects that allow Maude objects to interact with any other entities and support distributed implementations; (v) reflection, thanks to the existence of a universal theory that can simulate deduction in all other theories (including itself) and represent functional and system modules as *data* for meta-programming purposes; and (vi) symbolic computation features such as semantic unification, variants, symbolic reachability analysis, and SMT solving.

The ideas in Joseph Goguen's TPA book have stimulated further developments in the formal verification of Maude modules, including the following: (1) the use of reflection and of symbolic methods to automate constructor-based inductive theorem-proving verification of func-tional modules; (2) tools to check the confluence, sufficient completeness, and termination of functional modules; and (3) theorem proving of properties of system modules (rewrite theories).

All the above-described advances in CafeOBJ and Maude illustrate some of the ways in which Joseph Goguen's TPA book has stimulated further developments in algebraic verification. But they do not exhaust at all the possible ways in which this fundamental book could stimulate other readers: we are convinced that it will continue to stimulate us and others. We have undertaken the task of making it available to the research community, as it was Joseph Goguen's desire, precisely for this purpose.

December 2020

K. Futatsugi
N. Martí-Oliet
J. Meseguer

# 1 Introduction

This book can be seen either as a text on theorem proving that uses techniques from general algebra, or else as a text on general algebra illustrated and made concrete by practical exercises in theorem proving. This introductory chapter provides background and motivation, though some points may only become fully clear in light of subsequent chapters, in part because of terminology not yet defined. Section 1.13.1 is a synopsis. The book considers several different logical systems, including first-order logic, Horn clause logic, equational logic, and first-order logic with equality. Similarly, several different proof paradigms are considered. However, we do emphasize equational logic, and for simplicity we use only the OBJ3 software system, though it is used in a rather flexible manner.

We do not pursue the lofty goal of mechanizing proofs like those of which mathematicians are justly so proud; instead, we seek to take steps towards providing mechanical assistance for proofs that are useful for computer scientists in developing software and hardware. This more modest goal has the advantage of both being achievable and having practical benefits.

## 1.1 What is Theorem Proving?

You can think of theorem proving as game playing where there are very precise rules, initial positions and goals; you win if you reach the goal from the initial position by correctly following the rules. Different logical systems have different rules and different notions of position, while different problems in the same system have different goals and/or different initial positions. Playing is called *inference* or *deduction*, a move is a *step* of inference or deduction, and positions are (usually) sets of formulae; the formulae in an initial position may be called *axioms, assumptions* or *hypotheses*. Thus a *logical system* consists of a language whose sentences (which are formulae) are used to state goals and axioms, plus some rules of inference for deriving new sentences from old ones. There may also be notions of model and of satisfaction, and these will play a key role in this text. Models relate to rules of inference much as a chessboard with its pieces relates to the rules of chess; in this setting, satisfaction means that a sentence accurately describes a given position.

The most classical example is Euclidean plane geometry. More recently, first-order predicate calculus (usually called "First-Order Logic" and often abbreviated "FOL") has been the most important game in town, but there are many, many other logical systems, a few of which are seen later in this book.

## 1.2 Applications of Theorem Proving

Mechanical theorem proving has many practical applications:

- The verification of digital hardware circuits, especially VLSI, where the economic cost of design errors can be enormous.
- The design and verification of so-called *critical systems*, such as nuclear power plants, heart pacemakers, and aircraft guidance systems, where failure can endanger human life or property on a large scale.
- Tools to make programming more reliable and robust, for example, to help with debugging, modifying, and optimizing programs, based on their semantics.
- The technology of theorem proving has been used in a number of modern programming languages that are based upon logic, including OBJ (which is described in Section 1.11 and used throughout this text) and Prolog.
- The technology of theorem proving has also been used in systems for robot vision, motion planning, drug discovery, DNA sequencing, and many similar applications.

## 1.3 Logic for Foundations *versus* Logic for Applications

Logic has been mainly concerned with the foundations of mathematics since the rude shock of the paradoxes discovered around the turn of the twentieth century by Russell and others. Such foundational work tends to simplify notation, axioms, and inference rules to the bare minimum, in order to facilitate the study of meta-mathematical issues such as consistency and completeness. But logic is used in applications for completely different reasons. In particular, computer scientists and engineers build hardware and software systems that are actually used in the real worlds of science, commerce, and technology, for which very different approaches to logic are more appropriate. In particular, the logical systems used for applications are often far more complex than those used in foundations; there may be many more symbols, axioms and rules; and some data types may be "built in," such as natural numbers or lists. The ability to add new definitions and notations and then use them is also important, and some applications even require the use of more than one logical system.

## 1.4 Why Equational Logic and Algebra?

This text gives a privileged role to general algebra (also called universal algebra) and to its logic, which is equational logic.[1] One reason for this is that equational logic is the logic of substituting equals for equals, which is a basic common denominator among many different logics. Also, equational logic is attractive as the foundation for a theorem prover because of the simplicity, familiarity, and elegance of equational reasoning, and because there is a great deal of relevant theory, including the extensive literature on abstract data types. Moreover, equational reasoning can be implemented efficiently by term rewriting, which can then serve as a workhorse for a general purpose theorem prover. In addition, many other interesting and important logics can be embedded within or built on top of equational logic, as we will see.

We will also see that equational logic is ideal as a meta-logic for describing other logical systems, because the syntax of a logic is a free algebra, while the rules of deduction can be implemented by (conditional) rewrite rules. Thus, we can use equational logic at the meta level (for describing logical systems and justifying proof scores), as well as at the object level (for proving theorems).

Any computable function over any computable data structure can be defined in equational logic [11], and *order-sorted* equational logic, which adds subsorts [82], extends this to encompass the *partial* computable functions. Thus, equational logic is sufficiently powerful to describe any standard model of interest. Although not every property that one might want to prove about some real system can be expressed using just equational logic, much more can be expressed than might at first be thought. In particular, we will see that many typical results about higher-order functional programs and most of the usual digital hardware verification examples fall within this setting, and it seems reasonable to use the simplest possible logic for any given application. However, we do not restrict ourselves to the most traditional kind of equational logic, but rather extend it in various ways, as discussed further in the next section; also, Chapter 8 considers first-order logic.

---

[1]The terms "algebra" and "equational logic" in their narrow senses refer to models and deductions, respectively, but in their broad sense, both terms refer to both models and deductions.

## 1.5 What Kind of Algebra?

This text does not view algebra as having a single all encompassing logic, but rather as having a family of related logics, ranging from the classical unsorted case toward first-order logic with equality, and even second-order logic. The following are brief character sketches for certain versions that are developed in detail later on:

- **Many Sorts.** Computing applications typically involve more than one *sort* of data, and it can be awkward, or even impossible, to treat these applications adequately with unsorted algebra. Still, it is not unusual to see papers that treat only the unsorted case, perhaps with a remark that "everything generalizes easily." Although this is true in essence, Section 4.3 shows that significant difficulties can arise if the generalization is not done carefully.

- **Conditional.** Many applications involve equations that are only true under certain conditions; examples include defining the transitive closure of a relation (see Appendix C) as well as many abstract data types (see Chapter 6), and the rules of inference for first-order logic (see Chapter 8). See Chapter 3 for details.

- **Overloaded.** Many computing applications involve *overloaded* operation symbols, where arguments may have different sort patterns. Examples include overloading in ordinary programming languages (such as Ada), polymorphism in functional programming languages (such as ML) and the $\lambda$-calculus, and overwriting in object-oriented languages (such as Smalltalk and Eiffel). See Chapter 2 for more detail.

- **Ordered Sorts.** This rather substantial extension of many-sorted algebra involves having a partial ordering on the set of sorts, called the *subsort* relation, which is interpreted semantically as a *subset* relation on the sets that interpret the sorts. This has many interesting applications, including exception handling and partially defined functions.

- **Second Order.** Another substantial extension allows quantification over functions as well as over elements. This has significant applications to digital hardware verification. Surprisingly, much of general algebra extends without difficulty, as shown in Chapter 9.

- **Additional Connectives.** The basic formulae of equational logic are universally quantified equations, but we can build more complex formulae from these, using conjunction, implication, disjunction, negation, and existential quantification. Satisfaction of such formulae can be defined in terms of the satisfaction of their constituents. Chapter 8 gives the details.

- **Hidden Sorts.** Computing applications typically involve states, and it can be awkward to treat these applications in a purely functional style. Hidden sorted algebra substantially extends ordinary algebra by distinguishing sorts used for data from sorts used for states, calling them respectively visible and hidden sorts, and it changes the notion of satisfaction to *behavioral* (also called *observational*) satisfaction, so that equations need only *appear* to be satisfied under all the relevant experiments. Hidden algebra is powerful enough to give a semantics for the object paradigm, including inheritance and concurrency. See Chapter 13 for details.

Putting all this together gives possibilities that are far from classical general algebra. The major difference from the usual first and second-order logic is that the only relation symbol used is equality. However, other relations can be represented by Boolean-valued functions.

## 1.6 Term Rewriting

A significant part of this text is devoted to explaining and using term rewriting. The general idea can be expressed as follows: *Terms* (or *expressions*) over a fixed syntax $\Sigma$ form an algebra. A *rewrite rule* is a rule for rewriting some terms into others. Each rewrite rule has a *leftside* ($L$), which is an expression defining a pattern that may or may not match a given expression. A *match* of an expression $E$ to $L$ consists of a subexpression $E'$ of $E$ and an assignment of values to the variables in $L$ such that substituting those values into $L$ yields $E'$.

A rewrite rule also has a *rightside* ($R$), which is a second expression containing only variables that already occur in $L$. If there is a match of $L$ to a subexpression of a given expression, then the matched subexpression is replaced by the corresponding substitution instance of $R$. This process is called *term rewriting*, *term reduction*, or *subterm replacement*, and is the basis for the OBJ system (see Section 1.11) used in this text.

## 1.7 Logical Systems and Proof Scores

A very basic question in theorem proving is what logical system to use. The dominant modern logical system is first-order predicate logic, but advances in computer science have spawned a huge array of new logics, e.g., for database systems, knowledge representation, and the semantic web; these include variants of propositional logic, modal logic, intuitionistic logic, higher-order logic, and equational logic, among many others.

The view of this text is that the choice of logical system should be left to the user, and that a mechanical theorem prover should be a basic engine for applying rewrite rules, so that a wide variety of logical systems can be implemented by supplying appropriate definitions to the rewrite engine. This view is influenced by the theory of institutions [67], and also resembles that of the Edinburgh Logical Framework [98, 1], Paulson's Isabelle system [148], and the use of Maude as a meta-tool for theorem proving [29], in avoiding commitment to any particular logical system. However, it differs in using equational logic and term rewriting as a basis.

The first step of our approach is to construct *proof scores*, which are instructions such that when executed (or "played"), if everything evaluates as expected, then the desired theorem is proved. A proof score is executed by applying *proof measures*, which progressively transform formulae in a language of goals into expressions which can be directly executed. We will see that equational logic is adequate for implementing versions of first and second-order logic in this way, as well as many other logical systems.

This approach can be further mechanized by implementing the meta level of goals and proof measures in OBJ itself, and providing a translator to an object level of computation, also in OBJ. If each proof measure is sound, and the computer implementations are correct, then each resulting proof score is guaranteed to be sound, in the sense that if it executes as desired, then its goal has been proved. But the converse, that a proof score will prove its goal (if it is true), does not hold in general. In addition, the advanced module facilities of OBJ can be used to express the structure of proofs. The Kumo [74, 83, 73] and 2OBJ [170, 84] systems provide even more direct support for such an approach. Equational logic is demonstrably adequate for our purpose, because the syntax, rules of deduction, and rules of translation of a logical system must be computable, and therefore (see Section 1.4) can be expressed with equations.

## 1.8 Semantics and Soundness

This text justifies proof measures for a logical system by demonstrating their *soundness* with respect to the notions of *model* and *satisfaction* for that system. In this sense, it places

### Semantics First!

In fact, users are primarily concerned with *truth*; they want to know whether certain properties are *true* of certain models, which may be realized in software and/or hardware. From this point of view, *proof is a necessary nuisance* that we tolerate only because we have no other way to effectively demonstrate truth. Moreover, it is usually easier and more intuitive to justify proof measures on semantic rather than syntactic grounds.[2]

The above slogan has many further implications. For example, it suggests that to define some "expressive" (i.e., complex) syntax, write some code that is driven by it, and then call the result a "theorem prover" if it usually prints TRUE when you want it to, is unwise. Similarly, it may not be a good idea to "give a semantics" for some system after it has already been implemented (or designed), because such a semantics may well be too complex to be of much use.[3] Finally, it is dangerous to try to combine several logics, unless precise and reasonably simple notions of model and satisfaction are known for the combination. Unfortunately, many theorem-proving projects have failed to observe these basic rules of logical hygiene. Nevertheless, as emphasized by the metaphor in Section 1.1, theorem proving *is* syntactic manipulation, and hence syntax is fundamental and unavoidable for the enterprise of this book. We can state our view in a balanced way as follows:

> Semantics is fundamental at the meta level (of correctness for proof rules), while syntax is fundamental at the object level (of actual proofs).

## 1.9  Loose *versus* Standard Semantics

An important distinction concerns the intended semantics of a logical system: is it meant to capture formulae that are true of *all* models of some set of axioms, or just formulae that are true of a fixed *standard model* of those axioms? Let us call the first case *loose semantics*, and the second case *standard* (or *tight*) *semantics*. The usual first-order logic has loose semantics, and captures properties that are true of all models of some given axioms. This fits many applications. For example, a logic intended to capture group theory must be loose, since it must apply to all groups. On the other hand, a logic for arithmetic should capture properties of a single standard model, consisting of the numbers with their usual operations.

A *completeness theorem* for a logical system says that all the formulae that are true of all the intended models of a given set of formulae (the axioms) can be proved from those formulae. There are completeness theorems for some well-known loose logics, including first-order logic and equational logic. However, completeness cannot in general be expected for logics under standard semantics, because the class of formulae true of a fixed model is not in general recursively enumerable; Gödel's famous incompleteness theorem shows that this holds even for the natural numbers. On the other hand, the familiar and powerful techniques for induction are not (usually) sound for loose semantics, but only for standard semantics. It seems that completeness has been overemphasized in the theorem-proving literature, because many computer science applications actually concern properties of a standard model of some set of formulae, rather than properties of all its models.

---

[2]These points should not be seen as rejecting constructivist approaches like that of Martin-Löf and others who identify the syntax and semantics of logical systems. In fact, there is much to recommend such approaches, especially for the foundations of mathematics. But note that what we are calling soundness problems reappear in this context as *consistency* problems.

[3]Unless perhaps you are prepared to go through several iterations, modifying both the implementation and the logic until they are consistent and elegant.

This text treats theorem proving for *both* standard and loose semantics, as well as combinations of the two. Given a set $A$ of formulae and a formula $e$, we will write $A \vDash e$ to indicate that $e$ is true in all models of $A$, and $A \vDash\!\!\!\equiv e$ to indicate that $e$ is true in the standard model of $A$. The already well-developed theory of abstract data types is helpful in studying the relation $\vDash\!\!\!\equiv$. In particular, we will see that formalizing the notion of "standard model" with *initiality* leads to simple proofs at both the object and the meta levels, i.e., of properties of standard models and of theorems about such proofs. We also consider loose extensions of standard models, standard extensions of models of a loose theory, and so on recursively.

## 1.10 Human Interface Design

Logical systems are so very precise and detailed that human beings often find it difficult and/or unpleasant to use them. Usually mathematics is conducted in a quite informal way, with only infrequent reference to any underlying logical system, much as the inhabitants of a house usually ignore its foundations [51]. Indeed, unlike a house, it is not clear that mathematics really *needs* foundations, though they may help you sleep better at night.

Computers can greatly lighten the burden of rigorously following the rules of a logical system, and fully *automatic theorem proving* attempts to entirely eliminate the pain of applying rules, although of course users must still state their axioms and goals precisely. In fact, fully automatic theorem proving has not been very successful, and no new theorems of real interest to mathematics have been proved in this way.[E1] One difficulty is that users often have to trick some built-in heuristics into doing what they want. However, fully automatic theorem proving remains an important area for research. An approach at the other extreme is *proof checking*, where rules are explicitly invoked one at a time by the user, and then actually applied by the machine. This can be quite tedious, but it can detect many errors, and even correct certain errors.

This text avoids both these extremes, taking the view that humans should do the *interesting* parts of theorem proving, such as inventing proof strategies and inductive hypotheses, while machines should do the tedious parts, mechanically applying sets of *rewrite rules* (see Section 1.6 above) that lead closer to subgoals. An important advantage of this approach is that partially successful proofs may return useful information about what to try next; for example, the output may suggest a new lemma that would further advance the proof. The reader who does the exercises will see many examples of this phenomenon. A variant of this approach provides a *tactic language* in which possibly quite complex combinations of proof measures can be expressed, and a *tactic interpreter* to apply these compound tactics; many theorem-proving systems take this approach, including HOL [93], Isabelle [148], and Kumo [74, 83, 73].

Much research has been done on the use of graphics in theorem proving. But we have found that for even modest size proofs, graphical representations of proof trees are not only unhelpful, but are actually obstructive and confusing [64, 65]. Instead, we recommend *structuring* proofs, by using modules and other features of OBJ, as is much illustrated in the following; we will see that this also supports proof reuse.

## 1.11 OBJ

OBJ [47, 90] integrates specification, prototyping, and verification into a single system, with a single underlying logic, which is (first-order conditional order-sorted) equational logic. OBJ3, which is the implementation of OBJ used in this text, allows a module $P$ to be either:

1. an **object**, whose intended interpretation is a *standard model*[4] of $P$; or

2. a **theory**, whose intended interpretation is the *variety* of *all* models of $P$.

In OBJ3, objects are executable, while theories describe properties; both have sets of equations as their bodies, but the former have standard (initial algebra) semantics and are executed as rewrite rules, while the latter have loose semantics, which can be "executed" in a loose sense by applying rules of inference to derive new equations. Although theories have been studied more extensively in the theorem-proving literature, they often play a lesser role in practice, because most real applications require particular data structures and operations upon them.

OBJ also has[5] generic modules, module inheritance, and module expressions which describe interconnections of modules and actually cre-ate the described subsystem when evaluated. The OBJ module system is a practical realization of ideas originally developed in the Clear language [22, 23]; these ideas have directly influenced the module systems of the ML, Ada, C++, and Module-2 languages. OBJ's user-definable mixfix syntax allows users to tailor their notation to their application, and LaTeX [119] symbols can be used to produce pretty output. Rewriting modulo associativity and/or commutativity is also supported, and can eliminate a great deal of tedium, and the subsorts provided by order-sorted algebra (see Chapter 10) support error messages and exception handling in a smooth and convenient way.

---

[4]More precisely, we mean an *initial* model of $P$, in a sense made precise in Theorem 3.2.1 of Chapter 3.

[5]Although much of the terminology in this paragraph may be unfamiliar now, it is all defined later on.

OBJ can be used directly as a theorem prover for equational logic only because its semantics *is* the semantics of equational logic.[E2] Every OBJ computation is a proof of some theorem. It is not true that any other functional programming language would do just as well. Although most functional languages have an operational semantics that is based on higher-order rewriting, they do not have a declarative, logical semantics for all of their features. It is also important that the OBJ module facility has a rigorous semantics, as explained in Chapter 11.

OBJ began as an algebraic specification language at UCLA about 1976, and has been further developed at SRI International, Oxford [47, 90], UCSD, and several other sites [26, 33, 166] as a declarative specification and rapid prototyping language; Appendix A gives more detail on OBJ, for which see also [90, 77]. The systematic use of OBJ as a theorem prover stems from [59]. The latest members of the OBJ family are CafeOBJ [43], Maude [30], and BOBJ [76, 75]. These systems go beyond OBJ3 in significant ways which are not needed for this text (rewriting logic for Maude, hidden algebra for BOBJ, and both of these for CafeOBJ); they could be used instead of OBJ3, though some syntactic changes would be needed. This book provides everything about the syntax and semantics of OBJ3 needed for theorem proving, including practical details on getting started (for which see Appendix A).

## 1.12   Some History and Related Work

There is such a large literature on theorem proving that an adequate survey would take several volumes. Consequently, we limit the following discussion to systems that have particularly influenced the approach in this book, or that seem related in some other significant way.

Much of the early work on mechanical theorem proving was done in the context of so-called "Artificial Intelligence" ("AI"), and much of it was not very rigorous. One inspiration was to give computers some ability to reason, and then see how far this could be extended and applied; there were even dreams of replacing mathematicians by programs. The collection *Theorem Proving: After 25 Years* [14] summarizes the state of theorem proving as of about 1984. The papers by Loveland and by Wang in this collection contain many interesting historical details. For a long time, the dominant approach for first-order logic was *resolution*, introduced by Alan Robinson in 1965 [157]. This technique for loose semantics is most suitable for fully automatic theorem proving, because its representations make it hard for users to understand or use to guide proofs. Chang and Lee [27] give a readable exposition, but Leitsch [122] is more precise and up to date. This tradition is well represented by the OTTER [132] and Gandalf [174] systems.

A wide range of interesting results were first verified with the Boyer-Moore Nqthm prover [19], using clever heuristics for induction. Its basic logic is untyped first-order universally quantified with function symbols and with equality as its only predicate; users can define new data structures by induction and recursion. Users influence its behavior by requesting intermediate results to be proved in a certain order, since it recalls what it has already proved; users can also set certain parameters. However, this can be a very awkward way to control the prover. Its successor system, ACL2, is interactive instead of automatic [112].

Another tradition arises from Milner's LCF system [95, 147], in which a higher-order strongly typed functional language (namely ML) is used for writing tactics that guide the application of elementary steps of deduction for achieving goals. Soundness is guaranteed by having a type "`thm`" that can only be inhabited by formulae that have actually been proved. One problem with this approach is that, because a proof is described by a single functional expression, for difficult problems this expression can be hard to understand and to edit. Gordon's HOL system [93], which has been successful for hardware verification, is an important development in this tradition; HOL is now commonly run on Isabelle [148]. Work by Stickel on the Prolog Technology Theorem Prover [171] should also be mentioned, as should burgeoning work based on type theory, e.g., the ELF [98, 1] and IPE [156] systems from Edinburgh, and Coq from INRIA; another important development in this area is NuPRL [35]. There is also much work using term rewriting techniques generalizing Knuth-Bendix completion, some of which is discussed in Chapter 12, although we have found that inductive proofs often work better in practice.

Every approach mentioned above has some drawbacks, and so does the one in this book. In fact, every approach must be unsatisfying in some ways, because general theorem proving is recursively unsolvable. Even though there is a completeness theorem for first-order logic with loose semantics, the problem is still only *semi-decidable*; there is no way to know whether an attempt to prove a given formula will ever halt, although if there is a proof, it will eventually be found (unless the available memory is exceeded). Theorem proving for standard semantics is not in general even semi-decidable, so that any automatic prover for this domain will necessarily fail to find proofs of some true formulae, even if given arbitrarily much time. However, such results no more prevent machines from proving theorems than they do humans.

The view of theorem proving as very strict "game playing" (see Section 1.1) comes from the formalist view of the foundations of mathematics advocated by Hilbert and by Whitehead and Russell, among others. In this view, mathematics is a purely formal activity, rigidly governed by sets of rules. This view is opposed by several other schools of the philosophy of mathematics, including idealists (e.g., Platonists) and intuitionists, both of whom argue that mathematics has some inherent meaning. The mechanization of theorem proving is necessarily consistent with a formalist view, because computer manipulations are necessarily formal; but that does not mean that one has to believe the formalist position to engage in mechanical theorem proving, and in fact the author of this book does not accept the formalist position, but rather subscribes to a view like that of Wittgenstein, that mathematical proofs are a kind of (socially situated) language game [51]. In any case, formal semantics cannot capture the sort of meaning that Platonists and intuitionists talk about, because meaning in their sense is inherently non-formal. Despite all this, we will see that formal semantics can be very useful.

Some discussion of the history and literature of algebra is given in Sections 2.8 and 3.8.

## 1.13 Using this Text

This text was developed for advanced undergraduates (that is, third or fourth year), but it can also be used at the graduate level by including more of the difficult material. It can be used in courses on general algebra, on the practice of mechanical theorem proving, and on the mathematical foundations of theorem proving. The second choice (which was taken at Oxford) would give precedence to the OBJ exercises, while the first and third choices would give precedence to the mathematics; alternatively, all three goals could be pursued at once. This text could perhaps be used as the basis for a course on discrete structures, but for this purpose it should be supplemented. In any case, the exercises that use OBJ should be done if at all possible, because they give a much more concrete feeling for the more theoretical material; OBJ is available by ftp (see Appendix A for details).

The choice of what to include (or to develop, in the case of new results) has always preferred material that is directly useful in computer science, especially theorem proving, although this does sometimes require including other material that is not itself directly useful. Propositional and predicate logic, and basic set theory, are necessary for understanding this text; some prior exposure to algebra would be helpful, as would some experience with computing. Appendix C reviews some basic mathematical concepts, such as transitive closure, and many others are reviewed within the body of the text.

Results, definitions, and examples are numbered on the same counter, which is reset for each section;[E3] exercises are on a separate counter, which is also reset. Material marked "(⋆)" can be skipped without loss of continuity, and probably should be skipped on a first reading. The more difficult proofs have been relegated to Appendix B. More advanced topics that could be skipped in an introductory class include initial Horn models, order-sorted algebra and rewriting, generic modules, hidden algebra, and the object paradigm.

### 1.13.1  Synopsis

The following topics are covered: many-sorted signature, algebra and homomorphism; term algebra and substitution; equation and satisfaction; conditional equations; equational deduction and its completeness; deduction for conditional equations; the theorem of constants; interpretation and equivalence of theories; term rewriting, termination, confluence and normal form; abstract rewrite systems; standard models, abstract data types, initiality, and induction; rewriting and deduction modulo equations; first-order logic, models, and proof planning; second-order algebra; order-sorted algebra and rewriting; modules; unification and completion; and hidden algebra. In parallel with these are a gradual introduction to OBJ3, applications to group theory, various abstract data types (such as number systems, lists, and stacks), propositional calculus, hardware verification, the $\lambda$-calculus, correctness of functional programs, and other topics. Some social aspects of formal methods are discussed in Appendix D.

### 1.13.2  Novel Features

Novel features of this book include the following: the use of arrows rather than set-theoretic functions; commutative diagrams; overloaded many-sorted algebra; an emphasis on signatures for logics and term rewriting; an algebraic treatment of first-order logic; second-order general algebra; applications to VLSI (especially CMOS) transistor circuits; use of an executable specification language for proofs; the notion of proof score; systematic use of the theorem of constants; algebraic treatments of termination proofs and rewriting modulo equations. More advanced novel topics include: an algebraic treatment of parsing; an adjunction between term rewriting systems and abstract rewriting systems; an algebraic treatment of Horn clause logic and its initial models; and results on hierarchical term rewriting systems.

## 1.14 Acknowledgements

I thank the Computer Science Lab at SRI, where this research began, the Programming Research Group at Oxford University, where this text began, and the Department of Computer Science and Engineering at UCSD, where it was finished. In particular, I thank Frances Page, Joan Arnold, and Sarah Farrer for help with the diagrams and corrections. Special thanks to Dr. José Meseguer, in collaboration with whom many of the ideas behind this text were developed, and to Mr. Timothy Winkler, who implemented most of OBJ3, as well as Drs. Kokichi Futatsugi, Jean-Pierre Jouannaud, Claude Kirchner, Hélène Kirchner, David Plaisted, Joseph Tardo, Patrick Lincoln, and Aristide Megrelis, all of whom helped get OBJ3 where it is today. In addition, I thank Prof. Rod Burstall and Drs. James Thatcher, Eric Wagner, and Jesse Wright, who all helped get the theory to the point where a text like this became possible. I thank Monica Marcus, Răzvan Diaconescu, Grigore Roşu, Yoshihito Toyama, Virgil-Emil Cazanescu, and José Barros for help with Chapter 5, and Răzvan Diaconescu and Grigore Roşu for help with Chapter 8. Paulo Borba, Jason Brown, Steven Eker, Healfdene Goguen, Ranko Lazic, Alexander Leitsch, Dorel Lucanu, Sula Ma, Monica Marcus, Chiyo Matsumiya, Oege de Moor, and Adolfo Socorro have all helped spot bugs and typos. Finally, I thank the students in my classes for their patience and comments. It has been a great pleasure for me to work with all these people.

---

**A Note for Lecturers:** It is not necessary to spend time on the material in this chapter, because most of it arises naturally as the actual content of the course unfolds. Instead, it can just be assigned as reading; it should probably be assigned twice, once at the beginning and once at the end of the course.

---

# 2 Signature and Algebra

In order to prove theorems, we need formulae to express assumptions and goals, and we need precise notions of "model" and of "satisfaction of a formula by a model" to ensure correctness. This chapter develops many-sorted general (or universal) algebras as models, while the next chapter considers equations and their satisfaction by algebras.

## 2.1 Sets and Arrows

We first briefly summarize some notation that will be used throughout this book, assuming that basic set theory is already familiar. $\omega$ denotes the set $\{0, 1, 2, \ldots\}$ of all natural numbers, and $\#S$ denotes the cardinality of a finite set $S$. We let $S^*$ denote the set of all lists (or strings) of elements from $S$, including the empty list which we denote $[\,]$. We write the elements of a list in sequence without any punctuation. For example, if $S = \{a, b, c, d\}$ then some elements of $S^*$ are $a, ac, acb, d$ and $[\,]$. Notice that in this notation, $S \subseteq S^*$. Given $w \in S^*$, we let $\#w$ denote the length of $w$; in particular, $\#[\,] = 0$, and for $s \in S$ as a one element list, $\#s = 1$. Let $S^+ = S^* - \{[\,]\}$.

This book makes systematic use of **arrows** (also called **maps**), which are functions with given source and target sets (elsewhere these may be called domain and codomain sets). $f : A \to B$ designates an arrow from $A$ to $B$, i.e., a function defined on **source** $A$ with image contained in **target** $B$. For example, the successor map $s : \omega \to \omega$ is defined by $s(n) = n + 1$. We let $[A \to B]$ denote the set of all arrows from $A$ to $B$. In Appendix C, this approach is compared with the usual set-theoretic approach to functions.

Given arrows $f : A \to B$ and $g : B \to C$, then $f;g$ denotes their **composition**, which is an arrow $A \to C$ (this notation follows the conventions of computing science, rather than of mathematics). An arrow $f : A \to B$ is **injective** iff $f(a) = f(a')$ implies $a = a'$, is **surjective** iff for each $b \in B$ there is some $a \in A$ such that $f(a) = b$, and is **bijective** iff it is both injective and surjective. We let $1_A$ denote the **identity** arrow at $A$, defined by $1_A(a) = a$ for all $a \in A$. Notice that $1_A; f = f$ and $f; 1_B = f$ for any $f : A \to B$.

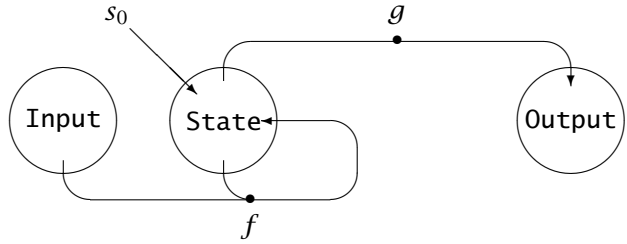

Figure 2.1: Signature for Automata

Some further set-theoretic topics are reviewed in Appendix C, which should be consulted by readers for whom the above concepts are not yet entirely comfortable.

## 2.2 Sorted Sets

Any theorem-proving problem needs a vocabulary in which to express its goals and assumptions. This vocabulary will include a *sort set S*, to classify (or "sort") the entities (or "data items") that are involved. Names for the operations involved are also needed, and for each given operation name, the sorts of arguments that it takes, and the sort of value that it returns, should be indicated. A vocabulary with such information is called a *signature*.

For example, let us consider automata. These have three sorts of entity, namely input, state, and output, so that $S = \{\text{Input}, \text{State}, \text{Output}\}$. Also, there are transition and output operations, say $f$ and $g$, plus an initial state $s_0$. It is convenient to present this structure graphically with the "ADJ diagram" in Figure 2.1, which clearly shows that $f$ takes an input and a state as its arguments and returns a state, while $g$ takes a single state as input, returning an output, and $s_0$ is a constant of sort State.

We wish to formalize the signature concept in such a way that operation symbols can be *overloaded*, that is, so that an operation symbol can have more than one type; a rather sophisticated word sometimes used to describe this phenomenon is "polymorphism." For example, $S$ might contain Nat, Bool and List, and we might want "+" to denote operations for adding natural numbers, for taking exclusive-or of booleans, and for concatenating lists.

Our first step towards formalizing all this is to capture the notion of providing a set of elements for each sort $s \in S$. For example, an automaton $A$ will have three such sets, for its elements of sorts Input, State, and Output; these three sets are denoted $A_{\text{Input}}$, $A_{\text{State}}$, and $A_{\text{Output}}$, respectively, and should be thought of as a "family" of sets of elements. Our mathematical formalization of this concept is as follows:

**Definition 2.2.1** Given a set $S$, whose elements are called **sorts** (or **indices**), an $S$-**sorted** (or $S$-**indexed**) **set** $A$ is a set-valued map with source $S$, whose value at $s \in S$ is denoted $A_s$; we will use the notation $\{A_s \mid s \in S\}$ for this. Also, we let $|A| = \bigcup_{s \in S} A_s$ and we let $a \in A$ mean that $a \in |A|$. Finally, we may sometimes write $\{a\}_s$ for the **singleton** $S$-sorted set $A$ with $A_s = \{a\}$ and with $A_{s'} = \emptyset$ for $s \neq s'$; we may also extend this notation, to write $\{a, a'\}_s$ for $\{a\}_s \cup \{a'\}_s$, etc. □

It is significant that the sets $A_s$ need not be disjoint, because it is this that supports overloading. For example, in many important examples of automata, the sets $A_{\text{Input}}$, $A_{\text{State}}$, and $A_{\text{Output}}$ are all the same, e.g., the natural numbers, or perhaps the Booleans. These sorted sets will also be used just a little later as the basis for our notion of (overloaded) signature. (Our use of the notation $\{A_s \mid s \in S\}$ should not be confused with the set of all the sets $A_s$; an $S$-indexed set $A$ is not a set of sets, but rather a map from $S$ to sets. Notations like $\langle A_s \mid s \in S \rangle$ and $\{A_s\}_{s \in S}$ might make this distinction clearer, but for this book a notation stressing the analogy of indexed sets with ordinary sets is more desirable.)

A different approach to sorting elements provides an arrow $\tau : |A| \to S$, where $\tau(a)$ gives the sort of $a \in A$. Unfortunately, this approach does not permit overloading, and therefore does not allow many important kinds of syntactic ambiguity to be studied, because they cannot even exist without overloading. Applications that require overloading include the refinement of data representations, and parsing in modern programming languages, such as Ada; in addition, most object-oriented languages allow a form of overloading that is resolved at run-time by so-called dynamic binding.

In general, concepts extend component-wise from ordinary sets to $S$-sorted sets. For example, $A \subseteq B$ means that $A_s \subseteq B_s$ for each $s \in S$, the **empty** $S$-sorted set $\emptyset$ has $\emptyset_s = \emptyset$ for each $s \in S$, and $A \cup B$ is defined by $(A \cup B)_s = A_s \cup B_s$ for each $s \in S$. Because of these component-wise definitions, many laws about sets also extend from simple sets to $S$-sorted sets. For example, we can show that $\emptyset \cup A = A$ for any $S$-sorted set $A$, by checking that it is true for each component, as follows: $(\emptyset \cup A)_s = \emptyset \cup A_s = A_s$.

**Exercise 2.2.1** Define intersection for $S$-sorted sets, and show that $A \cap B = B \cap A$, and that $A \cap (B \cup C) = (A \cap B) \cup (A \cap C)$, where $A, B, C$ are all $S$-sorted sets. □

**Definition 2.2.2** An $S$-**sorted** (or $S$-**indexed**) **arrow** $f : A \to B$ between $S$-sorted sets $A$ and $B$ is an $S$-sorted family $\{f_s \mid s \in S\}$ of arrows $f_s : A_s \to B_s$. Given $S$-sorted arrows $f : A \to B$ and $g : B \to C$, their **composition** is the $S$-sorted family $\{f_s; g_s \mid s \in S\}$ of arrows. Each $S$-sorted set $A$ has an **identity** arrow, $1_A = \{1_{A_s} \mid s \in S\}$. An $S$-sorted arrow $h : M \to M'$ is **injective** iff each component $h_s : M_s \to M'_s$ is injective, is **surjective** iff each $h_s : M_s \to M'_s$ is surjective, and is **bijective** iff it is injective and surjective. □

**Exercise 2.2.2** If $f : A \to B$ is an $S$-sorted arrow, show that $1_A; f = f$ and that $f; 1_B = f$. □

## 2.3   Signature

We are now ready for the following basic concept:

**Definition 2.3.1** Given a **sort set** $S$, then an $S$-sorted **signature** $\Sigma$ is an indexed family $\{\Sigma_{w,s} \mid w \in S^*, s \in S\}$ of sets, whose elements are called **operation symbols**, or possibly **function symbols**. A symbol $\sigma \in \Sigma_{w,s}$ is said to have **arity** $w$, **sort** $s$, and **rank** (or "type") $\langle w, s \rangle$, also written $w \to s$; in particular, any $\sigma \in \Sigma_{[],s}$ is called a **constant symbol**. (Operation and constant symbols will later be interpreted as actual operations and constants.)

A symbol $\sigma \in |\Sigma|$ is **overloaded** iff $\sigma \in \Sigma_{w,s} \cap \Sigma_{w',s'}$ for some $\langle w, s \rangle \neq \langle w', s' \rangle$. $\Sigma$ is a **ground signature** iff $\Sigma_{[],s} \cap \Sigma_{[],s'} = \emptyset$ whenever $s \neq s'$, and $\Sigma_{w,s} = \emptyset$ unless $w = []$, i.e., iff it consists only of non-overloaded constant symbols. □

Now some examples of signatures, beginning with a more formal treatment of automata:

**Example 2.3.2** (*Automata*) Because an **automaton** consists of an **input** set $X$, a **state** set $W$, an **output** set $Y$, an **initial state** $s_0 \in W$, a **transition function** $f : X \times W \to W$, and an **output function** $g : W \to Y$, we have $S = \{\text{Input}, \text{State}, \text{Output}\}$, with $\Sigma_{[],\text{State}} = \{s_0\}, \Sigma_{\text{Input State}, \text{State}} = \{f\}, \Sigma_{\text{State},\text{Output}} = \{g\}$, and $\Sigma_{w,s} = \emptyset$ for all other ranks $\langle w, s \rangle$, as shown in Figure 2.1. □

**Example 2.3.3** (*Peano Natural Numbers*) There is just one sort of interest, say $S = \{\text{Nat}\}$. To describe all natural numbers, it suffices to have symbols for the constant zero and for the successor operation, say 0 and $s$, respectively. Then we can describe the number $n$ as $n$ applications of $s$ to 0; thus, 0 is represented by 0, 1 by $s(0)$, 2 by $s(s(0))$, etc. This is sometimes called *Peano notation*; we could also speak of "cave man numbers," since this is counting in base 1. The signature has $\Sigma_{[],\text{Nat}} = \{0\}, \Sigma_{\text{Nat},\text{Nat}} = \{s\}$ and $\Sigma_{w,s} = \emptyset$ for all other ranks $\langle w, s \rangle$; or in the singleton notation of Definition 2.2.1, $\Sigma = \{0\}_{[],\text{Nat}} \cup \{s\}_{\text{Nat},\text{Nat}}$. □

Notice that the natural interpretation for Example 2.3.3 is a certain particular *standard model*, whereas any model provides a suitable interpretation for Example 2.3.2. These two kinds of semantics are called

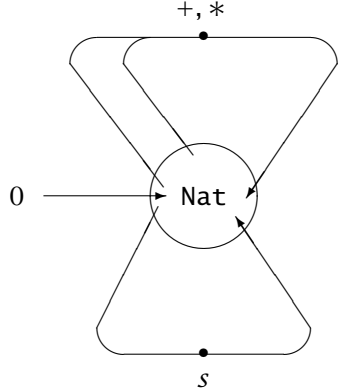

Figure 2.2: Signature for Numerical Expressions

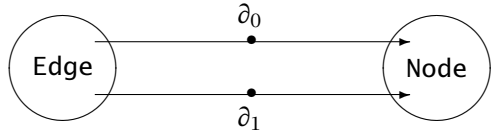

Figure 2.3: Signature for Graphs

*standard* (or *tight*) and *loose*, respectively. Later, we will make this distinction precise.

**Example 2.3.4** (*Numerical Expressions*) Again, there is just one sort of interest, say $S = \{\text{Nat}\}$, and assuming that we are only interested in the operation symbols shown in Figure 2.2, then $\Sigma_{[],\text{Nat}} = \{0\}, \Sigma_{\text{Nat},\text{Nat}} = \{s\}, \Sigma_{\text{Nat}\,\text{Nat},\text{Nat}} = \{+, *\}$, and $\Sigma_{w,s} = \emptyset$ for all other ranks $\langle w, s \rangle$. The intended semantics of this example is standard rather than loose, because there is just one intended model for these expressions. □

**Example 2.3.5** (*Graphs*) A (directed, unordered) **graph** $G$ consists of a set $E$ of **edges**, a set $N$ of **nodes**, and two arrows, $\partial_0, \partial_1 : E \to N$, which give the **source** and **target** node of each edge, respectively. Thus $S = \{\text{Edge}, \text{Node}\}$ and $\Sigma = \{\partial_0, \partial_1\}_{\text{Edge},\text{Node}}$ in the notation of Definition 2.2.1. This signature is shown in Figure 2.3. The intended semantics is loose, because there are many possible graphs. □

**Notation 2.3.6** Because a ground signature $X$ has all its sets $X_{w,s}$ empty unless $w = []$, we can identify such a signature with the $S$-indexed set $X$ with $X_s = X_{[],s}$, and we shall often do so in the following; note that in this case, $X_{s_1}$ is disjoint from $X_{s_2}$ whenever $s_1 \neq s_2$, by the definition of ground signature (Definition 2.3.1).

By our conventions about sorted sets, $|\Sigma| = \bigcup_{w,s} \Sigma_{w,s}$ and $\Sigma' \subseteq \Sigma$ means that $\Sigma'_{w,s} \subseteq \Sigma_{w,s}$ for each $w \in S^*$ and $s \in S$. Similarly, the **union** of two signatures is defined by

$$(\Sigma \cup \Sigma')_{w,s} = \Sigma_{w,s} \cup \Sigma'_{w,s}.$$

A common special case is union with a ground signature $X$. We will use the notation

$$\Sigma(X) = \Sigma \cup X$$

for this, but always assuming that $|X|$ and $|\Sigma|$ are disjoint, and $X$ is a disjoint family. When $X$ is an $S$-indexed set, the above equation may be rewritten as

$$\Sigma(X)_{[],s} = \Sigma_{[],s} \cup X_s$$

$$\Sigma(X)_{w,s} = \Sigma_{w,s} \text{ when } w \neq [] \,. \qquad \square$$

## 2.4 Signatures in OBJ

OBJ modules that are intended to be interpreted loosely begin with the keyword theory (which may be abbreviated th) and close with the keyword endth. Between these two keywords come (optional) declarations for sorts and operations, plus (as discussed in detail later on) variables, equations, and imported modules. For example, the following[1] specifies the theory of automata of Example 2.3.2:

```
th AUTOM is
  sorts Input State Output .
  op s₀ : -> State .
  op f : Input State -> State .
  op g : State -> Output .
endth
```

Notice that each of the four internal lines begins with a keyword which tells what kind of declaration it is, and terminates with a period. Any number of sorts can be declared following sorts, and operations are declared with both their arity, between the : and the ->, and their sort, following the ->. Because a constant like $s_0$ has empty arity, nothing appears between the : and the -> in its declaration. It is conventional (but not necessary) in OBJ for sort identifiers to begin with an uppercase letter, and for module names to be all uppercase.

Graphs as defined in Example 2.3.5 may be specified as follows:

```
th GRAPH is
  sorts Edge Node .
  op ∂₀ : Edge -> Node .
  op ∂₁ : Edge -> Node .
endth
```

---

[1] Actually, OBJ3 can only read ASCII characters, and the $s_0$ that you see was produced by a LATEX macro whose name consists of all ASCII characters. (ASCII provides a certain fixed set of characters with a certain fixed binary encoding.)

Also, the Peano natural numbers of Example 2.3.3 may be specified by the following:

```
obj NATP is
  sort Nat .
  op 0 : -> Nat .
  op s_ : Nat -> Nat .
endo
```

Here the keyword pair `obj ...endo` indicates that standard semantics is intended. Also, notice the use of `sort` instead of `sorts` in this example; actually, `sort` and `sorts` are synonyms in OBJ3, so that the choice of which to use is just a matter of style.

This example also uses "mixfix" syntax for the successor operation symbol: in the expression before the colon, the underbar character is a place holder, showing where the operation's arguments should go; there must be the same number of underbars as there are sorts in the arity; the other symbols before the colon go between or around the arguments. Thus, the notation `s_` defines *prefix* syntax, while `_+_` defines *infix* syntax, as in `0 + 0`; similarly, `_!` is *postfix*, `{_}` is *outfix*, and `if_then_else_fi` is general *mixfix*. When there are no underbars, a default prefix-with-parentheses syntax is assumed, as with `f` and `g` in AUTOM above. Notice that the formal definition of signature does not specify "fixity", but only arity and rank; this issue is discussed further in Section 3.7 below.

Here is an OBJ specification for the expressions over the natural numbers introduced in Example 2.3.4:

```
obj NATEXP is
  sort Nat .
  op 0 : -> Nat .
  op s_ : Nat -> Nat .
  op _+_ : Nat Nat -> Nat .
  op _*_ : Nat Nat -> Nat .
endo
```

One way to characterize the intuitive difference between loose and standard semantics is to consider what *entities* are of major interest in each case. For loose semantics (OBJ theories), the entities of greatest interest are the *models*; for example, for the theory GRAPH, we are interested in graphs, which are algebras. In such cases, we may say that the theory *denotes* the class of all graphs. On the other hand, for standard semantics (OBJ objects), we are interested in the *elements* of the standard model; for example, for the specification NATP, we are interested in the natural numbers. Of course, the algebra of all natural numbers is also of great interest, because it contains all the natural numbers, as well as certain operations upon them. In such cases, we may say that the OBJ specification *denotes* the algebra of natural numbers. (We will later see that this is only defined up to isomorphism.)

**Notation 2.4.1** If FOO is the name of an OBJ module, then we let $\Sigma^{\text{FOO}}$ denote its signature. For example, $|\Sigma^{\text{NATP}}| = \{0, s\}$. □

## 2.5 Algebras

Signatures specify the *syntax* of theorem-proving problems, but for many problems we are really interested in *semantics*, that is, in particular entities of the given sorts, and particular functions that interpret the given function symbols. This is formalized by the following basic concept:

**Definition 2.5.1** A $\Sigma$-**algebra** $M$ consists of an $S$-sorted set also denoted $M$, i.e., a set $M_s$ for each $s \in S$, plus

(0) an element $M_\sigma$ in $M_s$ for each $\sigma \in \Sigma_{[],s}$, interpreting the constant symbol $\sigma$ as an actual element, and

(1) a function $M_\sigma : M_{s_1} \times \cdots \times M_{s_n} \to M_s$ for each $\sigma \in \Sigma_{w,s}$ where $w = s_1 \ldots s_n$ (for $n > 0$), interpreting each operation symbol as an actual operation.

Together, these provide an *interpretation* of $\Sigma$ in $M$. Often we will write just $\sigma$ for $M_\sigma$. Also, we may write $M_w$ instead of $M_{s_1} \times \cdots \times M_{s_n}$. For example, using this notation we can write

$$M_\sigma : M_w \to M_s$$

for $\sigma \in \Sigma_{w,s}$. When a symbol $\sigma$ is overloaded, the notation $M_\sigma$ is ambiguous, and we may instead write $M_\sigma^{w,s}$ to explicitly indicate the rank that is intended for a particular interpretation of $\sigma$. Finally, we may sometimes write $\sigma_M$ instead of $M_\sigma$, especially in examples.

The set $M_s$ is called the **carrier** of $M$ of sort $s$. □

**Example 2.5.2** For example, we could have $S = \{\text{Int}, \text{Bool}\}$ with the symbols $0, 1$ in both $\Sigma_{[],\text{Int}}$ and $\Sigma_{[],\text{Bool}}$. Then we might have a $\Sigma$-algebra $M$ with $M_0^{[],\text{Int}} = 0$ and $M_0^{[],\text{Bool}} = 0$. □

**Example 2.5.3** (*Automata*) An **automaton** is a $\Sigma$-algebra where $\Sigma = \Sigma^{\text{AUTOM}}$ is the signature of Example 2.3.2, i.e., it consists of an input set $X$, a state set $W$, an output set $Y$, an initial state $s_0 \in W$, a transition function $f : X \times W \to W$, and an output function $g : W \to Y$.

Here is a simple $\Sigma^{\text{AUTOM}}$-algebra $A$: let $A_{\text{Input}} = A_{\text{State}} = A_{\text{Output}} = \omega$, the natural numbers; let $(s_0)_A = 0$, let $f_A(m,n) = m + n$, and let $g_A(n) = n + 1$. Then $A$ is an automaton whose state records the sum of the inputs received, and whose output is one more than the sum of the inputs received. AUTOM denotes the class of all automata. □

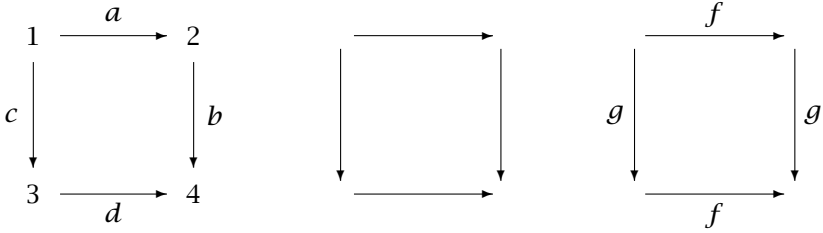

Figure 2.4: Views of a Graph

**Example 2.5.4** (*Expressions*) The standard semantics for Example 2.3.4 is a one-sorted algebra $E$ whose carrier consists of all well-formed expressions in $0, s, +, *$, such as $s(+(s(0), s(s(0))))$. Here the value of $+_E(e_1, e_2)$ is just the expression $+(e_1, e_2)$; similarly, the value of $s_E(e)$ is $s(e)$, of $*_E(e_1, e_2)$ is $*(e_1, e_2)$, and the interpretation $0_E$ of the symbol 0 is 0. Of course, there is another $\Sigma^{\mathsf{NATEXP}}$-algebra which has carrier $\omega$, and which interprets $0, s, +, *$ as the expected operations on natural numbers. However, the intended denotation is the algebra of all well-formed expressions — or more precisely, any isomorphic algebra, as will be discussed later on.  □

**Example 2.5.5** (*Graphs*) If we let $\Sigma$ be the signature $\Sigma^{\mathsf{GRAPH}}$ of Example 2.3.5, then a $\Sigma$-algebra $G$ consists of a set $E$ of **edges**, a set $N$ of **nodes**, and two arrows, $\partial_0, \partial_1 : E \to N$, which give the **source** and **target** node of each edge, respectively, that is, $G$ is a (directed, unordered) **graph**, which we may write as $(E, N, \partial_0, \partial_1)$.

A typical graph is shown to the left in Figure 2.4; here $E = \{a, b, c, d\}$, $N = \{1, 2, 3, 4\}$, $\partial_0(a) = \partial_0(c) = 1$, $\partial_1(a) = \partial_0(b) = 2$, $\partial_1(c) = \partial_0(d) = 3$, and $\partial_1(b) = \partial_1(d) = 4$. It is usual to draw such a graph as shown in the center of Figure 2.4, omitting the names of nodes and edges, so that labels can be attached instead, as shown in the rightmost diagram of Figure 2.4, and as explained further in Example 2.5.6 below.  □

**Example 2.5.6** (*Labelled Graphs*) To the signature of Example 2.3.5, let us add a single new sort Nlabel, and a single new operation symbol $l \in \Sigma_{\mathsf{Node}, \mathsf{Nlabel}}$. An algebra over this signature is a **node labelled graph**, and may be written $(E, N, L, \partial_0, \partial_1, l)$. The most typical interpretations are strict in $L$, but loose in everything else. An algebra with the underlying graph shown in the left of Figure 2.4 and with node labels from $\omega$ is shown at the left of Figure 2.4. We can also label edges, by adding another sort Elabel and another operation symbol, $l' \in \Sigma_{\mathsf{Edge}, \mathsf{Elabel}}$. Thus, in the rightmost diagram of Figure 2.4, $l'(a) = l'(d) = f$ and $l'(c) = l'(b) = g$.  □

**Exercise 2.5.1** Write an OBJ specification for the node labelled graphs of Example 2.5.6.  □

**Example 2.5.7** (*Overloading*) Now let's consider an example with overloading, given by the following OBJ code:

```
th OL is
  sorts Nat Bool .
  ops 0 1 : -> Nat .
  ops 0 1 : -> Bool .
  op s_ : Nat -> Nat .
  op n_ : Bool -> Bool .
  op _+_ : Nat Nat -> Nat .
  op _+_ : Bool Bool -> Bool .
endth
```

Here the keyword ops indicates that a number of operations with the same rank will be defined together.[2] Writing this signature out the hard way, we have $S = \{\text{Nat}, \text{Bool}\}$, $\Sigma_{[],\text{Bool}} = \Sigma_{[],\text{Nat}} = \{0,1\}$, $\Sigma_{\text{Bool},\text{Bool}} = \{n\}$, $\Sigma_{\text{Nat},\text{Nat}} = \{s\}$, $\Sigma_{\text{Bool Bool},\text{Bool}} = \Sigma_{\text{Nat Nat},\text{Nat}} = \{+\}$, and $\Sigma_{w,s} = \emptyset$ for all other ranks $\langle w, s \rangle$. Then 0 and 1 are overloaded, and so is +.

One algebra for this signature, usually denoted $T_\Sigma$, has the natural number terms in its carrier of sort Nat, and the Boolean terms in its carrier of sort Bool. Many of these terms are *ambiguous* in the sense that there is no unique $s \in S$ such that they lie in $T_{\Sigma,s}$. For example, the terms $0 + 1$ and $1 + (1 + 0)$ are ambiguous, as of course are 0 and 1; but $s(0)$ and $1 + (n(0) + 0)$ are unambiguous. (Proposition 3.7.2 in Section 3.7 will give a necessary and sufficient condition for non-ambiguity.) We will also see later how to disambiguate terms. □

## 2.6 Term Algebras

The terms over a given signature $\Sigma$ form a $\Sigma$-algebra which will be especially useful and important to us in the following. Indeed, it is a kind of "universal" $\Sigma$-algebra, which can serve as a standard model for specifications that do not have any equations.

**Definition 2.6.1** Given an $S$-sorted signature $\Sigma$, then the $S$-sorted set $T_\Sigma$ of all (**ground**) $\Sigma$-**terms** is the smallest set of lists over the set $|\Sigma| \cup \{(,)\}$ (where ( and ) are special symbols disjoint from $\Sigma$) such that

(0) $\Sigma_{[],s} \subseteq T_{\Sigma,s}$ for all $s \in S$, and

(1) given $\sigma \in \Sigma_{s_1...s_n,s}$ and $t_i \in T_{\Sigma,s_i}$ for $i = 1,...,n$ then $\sigma(t_1...t_n) \in T_{\Sigma,s}$. □

---

[2]When the operations are not constants, parentheses are needed to separate the different operation forms.

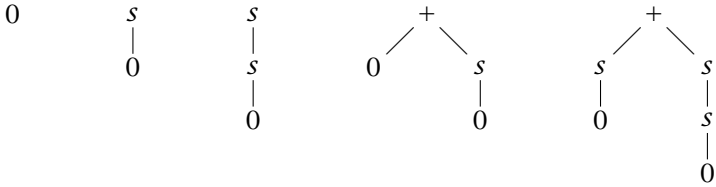

Figure 2.5: Trees for Some Terms

Notice that this representation of terms does not use mixfix syntax, but rather uses a default prefix-with-parentheses syntax; nevertheless, we will use mixfix notation in examples, and will later give some theory to support its use. Also, we will usually omit the underbars on the parentheses. For example, using the signature $\Sigma^{\text{NATEXP}}$, we can form terms like $0$, $s(0)$, $s(s(0))$, $0 + s(0)$, and $s(0) + s(s(0))$. It is common to picture such terms as node labelled trees, as shown in Figure 2.5. This correspondence is made precise in Example 3.6.3 below.

Notice also that the carriers of $T_\Sigma$ need not be disjoint when $\Sigma$ is overloaded. For example, if $\Sigma$ is the signature of Example 2.5.7, then $(T_\Sigma)_{\text{Int}}$ and $(T_\Sigma)_{\text{Bool}}$ both contain 0 and 1.

We can use an operation symbol $\sigma$ in $\Sigma$ as a "constructor," that is, as a template into whose argument slots terms of appropriate sorts can be placed, yielding new terms. For example, if $t_1$ and $t_2$ are two $\Sigma$-terms of sort $s$, and if $+$ is in $\Sigma_{ss,s}$, then $+(t_1, t_2)$ is another $\Sigma$-term, constructed by placing $t_1$ and $t_2$ into the form $+(\_,\_)$. Similarly, we can think of a constant symbol $\sigma \in \Sigma_{[],s}$ as constructing the constant term $\sigma$ itself. In this way, $T_\Sigma$ becomes a $\Sigma$-algebra. More precisely now,

**Definition 2.6.2** We can view $T_\Sigma$ as a $\Sigma$-algebra as follows:

(0) interpret $\sigma \in \Sigma_{[],s}$ in $T_\Sigma$ as the singleton list $\sigma$, and

(1) interpret $\sigma \in \Sigma_{s_1...s_n,s}$ in $T_\Sigma$ as the operation which sends $t_1, \ldots, t_n$ to the list $\sigma\underline{(}t_1 \ldots t_n\underline{)}$, where $t_i \in T_{\Sigma,s_i}$ for $i = 1, \ldots, n$.

Thus, $(T_\Sigma)_\sigma(t_1, \ldots, t_n) = \sigma\underline{(}t_1 \ldots t_n\underline{)}$, and from here on we usually use the first notation. $T_\Sigma$ is called the **term algebra**, or sometimes the **word algebra**, over $\Sigma$. □

**Example 2.6.3** Let us consider the term algebra for the signature $\Sigma^{\text{AUTOM}}$ of Example 2.3.2: since there are *no* terms of sort Input, the only term of sort State is $s_0$, and hence the only term of sort Output is $g(s_0)$; that is, $T_{\Sigma^{\text{AUTOM}},\text{Input}} = \emptyset$, while $T_{\Sigma^{\text{AUTOM}},\text{State}} = \{s_0\}$, and $T_{\Sigma^{\text{AUTOM}},\text{Output}} = \{g(s_0)\}$. The moral of this example is that the term algebras of signatures that are intended to be interpreted loosely are not necessarily very interesting. In fact, Example 2.5.4 is a more typical term algebra. □

## 2.7 Unsorted Signature and Algebra

Most expositions of logic treat the *unsorted* case, rather than the *many-sorted* case. Indeed, there has been a belief that many-sorted logic is just a special case of unsorted logic. However, this fails for a variety of reasons, including that many-sorted theorem proving can be much more efficient than unsorted theorem proving, and that different rules of deduction must be used in certain cases.

This subsection describes unsorted algebra, and informally shows that it is equivalent to one-sorted algebra, i.e., to the special case where $\#S = 1$. Rules of deduction and some other logics are considered later.

**Definition 2.7.1** An **unsorted signature** $\Sigma$ is a family $\{\Sigma_n \mid n \in \omega\}$ of sets, whose elements are called **operation symbols**; $\sigma \in \Sigma_0$ is called a **constant symbol**, and $\sigma \in \Sigma_n$ is said to have **arity** $n$. □

Notice that overloading is impossible for unsorted signatures.

**Definition 2.7.2** Given an unsorted signature $\Sigma$, then a $\Sigma$-**algebra** $M$ is a set, also denoted $M$ and called the **carrier**, together with

(0) an element $M_\sigma \in M$ for each $\sigma \in \Sigma_0$, and

(1) an arrow $M_\sigma : M^n \to M$ for each $\sigma \in \Sigma_n$ with $n > 0$. □

Note that when $n = 0$, then $M^0$ is (by convention defined to be) a one-point set, and so $M_\sigma : M^0 \to M$ determines an element of $M$ in its image, which can be considered its value.

We now consider the relationship with the one-sorted case. Let us assume that we are given a sort set $S = \{s\}$. Then an $S$-sorted set is a family $\{M_s \mid s \in S\}$ consisting of a single set $M_s$ and an $S$-sorted arrow $h : M_s \to M'_s$ is a family $\{h_s \mid s \in S\}$ consisting of a single arrow $h_s : M_s \to M'_s$. So one-sorted sets and arrows are essentially the same thing as ordinary sets and arrows.

Again assuming that $S = \{s\}$, an $S$-sorted signature is a family $\{\Sigma_{\langle w,s \rangle} \mid \langle w,s \rangle \in S^* \times S\}$, which can be identified with the unsorted signature $\Sigma' = \{\Sigma'_n \mid n \in \omega\}$ where $\Sigma'_n = \Sigma_{s^n,s}$. Hence, a $\Sigma$-algebra is essentially the same thing as a $\Sigma'$-algebra. In the following, we will usually identify the unsorted and one-sorted concepts.

## 2.8 Literature

The use of many-sorted structures is important in computing science because it can model the way that programming and specification languages keep track of the types of entities. Also, syntax in general forms a many-sorted algebra; this observation will later be helpful in formalizing various logical systems. However, most of the mathematics literature and much of the computing literature treat only unsorted algebra, which is inadequate for the applications on which this book focusses.

The development and use of algebras in the technical sense was a major advance in mathematics. Alfred North Whitehead [182] prefigured this revolution in the late nineteenth century.

Around 1931, Emmy Nöther laid the foundations for what is now called "modern" or "abstract algebra" by systematizing and widely applying the concepts of algebra and particularly homomorphism; for this reason she has been called "the mother of modern algebra" [20, 127]. In 1935, Garrett Birkhoff [12] gave the now standard definitions for the unsorted case, and proved the important completeness and variety theorems. Perhaps the classic mathematical reference for general (unsorted) algebra is Cohn [32]; this book also discusses some category theory and some applications to theoretical computing science.

Many-sorted algebra seems to have been first studied by Higgins [101] in 1963; Benabou [8] gave an elegant category-theoretic development around 1968. The use of sorted sets for many-sorted algebra seems notationally simpler than alternative approaches, such as [101], [8] and [13]; it was introduced by the author in lectures at the University of Chicago in 1968, and first appeared in print in [52]. The definition of signature with overloading (Definition 2.3.1) was first developed in these early lectures, but the idea only reveals its full potential in order-sorted algebra, which adds subsorts, as discussed in Chapter 10. Ideas in the papers [24] and [137] also contributed to the treatment of many-sorted general algebra that is given in this text.

Our systematic use of arrows is influenced by category theory, for which see, e.g., [126, 63].

"ADJ diagrams" were introduced in [87] as a way to visualize the many-sorted signatures used in the theory of abstract data types by the "ADJ group," which was originally defined to be the set {Goguen, Thatcher, Wagner, Wright}. (See [58] for some historical remarks on ADJ.) The name "ADJ diagram" is due to Cliff Jones.

OBJ began as an algebraic specification language at UCLA about 1976 [53, 55, 85], and was further developed at SRI International [47, 90] and several other sites [26, 166, 33] as a declarative specification and rapid prototyping language; Appendix A gives more detail on OBJ3, following [90, 77]. The use of OBJ as a theorem prover stems from [59], as further developed in [62].

> **A Note for Lecturers:** When lecturing on the material in this chapter, it may help to begin with examples (automata, natural number expressions, graphs, etc.), first giving their ADJ diagrams, then an intuitive explanation, then their OBJ3 syntax, then some models, and then some computations in those models. This is because some students without a sufficient mathematics background can find the formalities of $S$-sorted sets, arrows, and so on, rather difficult. After these topics have been treated, then the formal definitions of signature and algebra can be introduced. It helps to motivate the material by reminding students frequently that signatures provide the syntax for a domain within which we wish to prove theorems, and that algebras provide the semantics (models).

# 3 Homomorphism, Equation and Satisfaction

Homomorphisms can express many important relationships between algebras, including *isomorphism*, in which two structures differ only in how they represent their elements, as well as the subalgebra and quotient algebra relationships. In addition, we will use homomorphisms to characterize standard models and to define substitutions, two basic concepts that will play an important role throughout this book.

## 3.1 Homomorphism and Isomorphism

Homomorphisms formalize the idea of interpreting one $\Sigma$-algebra into another, by mapping elements to elements in such a way that all sorts, operations, and constants are preserved. This concept may already be familiar from linear transformations, which map vectors to vectors in such a way as to preserve the (constant) vector $\underline{0}$, as well as the operations of vector addition and scalar multiplication. The following equations express this,

$$T(\underline{0}) = \underline{0}$$
$$T(\underline{a} + \underline{b}) = T(\underline{a}) + T(\underline{b})$$
$$T(r \bullet \underline{a}) = r \bullet T(\underline{a})$$

where $\underline{a}, \underline{b}$ are vectors, $r$ is a scalar, and $\bullet$ is scalar multiplication. The general notion is:

**Definition 3.1.1** Given an $S$-sorted signature $\Sigma$ and $\Sigma$-algebras $M, M'$, a $\Sigma$-**homomorphism** $h : M \to M'$ is an $S$-sorted arrow $h : M \to M'$ such that the following **homomorphism condition** holds:

(0) $h_s(M_c) = M'_c$ for each constant symbol $c \in \Sigma_{[],s}$, and

(1) $h_s(M_\sigma(m_1,\ldots,m_n)) = M'_\sigma(h_{s_1}(m_1),\ldots,h_{s_n}(m_n))$ whenever $n > 0$, $\sigma \in \Sigma_{s_1\ldots s_n,s}$ and $m_i \in M_{s_i}$ for $i = 1,\ldots,n$.

The **composition** of Σ-homomorphisms $g : M \to M'$ and $h : M' \to M''$ is their composition as $S$-sorted arrows, denoted $g;h : M \to M''$.

If $h : M' \to M$ is an inclusion[1] and a homomorphism, then $M'$ is said to be a **sub-Σ-algebra** of $M$; in this case, $h$ may be called an **inclusion homomorphism**. □

**Exercise 3.1.1** Show that a sub-Σ-algebra $M'$ of $M$ is a subset ($S$-indexed, of course) that is closed under all the operations in Σ, i.e., that satisfies: (1) $M'_s \subseteq M_s$ for all $s \in S$; and (2) for every $\sigma \in \Sigma_{s_1...s_n,s}$, $M_\sigma(a_1,\ldots,a_n) \in M'_s$ whenever $a_i \in M'_{s_i}$. □

Note that to cover signatures with overloading, we should really have written the two conditions of Definition 3.1.1 as:

(0) $h_s(M_c^{[],s}) = M_c'^{[],s}$ for each constant symbol $c \in \Sigma_{[],s}$, and

(1) $h_s(M_\sigma^{w,s}(m_1,\ldots,m_n)) = M_\sigma'^{w,s}(h_{s_1}(m_1),\ldots,h_{s_n}(m_n))$ whenever $w = s_1 \ldots s_n$, $n > 0$, $\sigma \in \Sigma_{w,s}$ and $m_i \in M_{s_i}$ for $i = 1,\ldots,n$.

**Exercise 3.1.2** Show that a composition of two Σ-homomorphisms is a Σ-homomorphism, and that the identity $1_M$ on a Σ-algebra $M$ is a Σ-homomorphism. □

It may be interesting to see explicitly what are the homomorphisms of graphs and automata as defined in the previous chapter:

**Example 3.1.2** Given two graphs (in the sense of Example 2.5.5), say $G = (E, N, \partial_0, \partial_1)$ and $G' = (E', N', \partial_0', \partial_1')$, then a homomorphism $h : G \to G'$ consists of two arrows, $h_E : E \to E'$ and $h_N : N \to N'$ that satisfy the homomorphism condition for each $\sigma \in \Sigma$. In this case, there are just two $\sigma$ in Σ, and the corresponding equations are

$$h_N(\partial_0(e)) = \partial_0'(h_E(e))$$
$$h_N(\partial_1(e)) = \partial_1'(h_E(e))$$

for $e \in E$. These equations say that graph homomorphisms preserve source and target. □

**Example 3.1.3** If $A = (X, W, Y, f, g, s_0)$ and $A' = (X', W', Y', f', g', s_0')$ are two automata (in the sense of Example 2.5.3), then a homomorphism $h : A \to A'$ consists of three arrows, which we may denote $h_{\text{Input}} : X \to X'$, $h_{\text{State}} : W \to W'$, and $h_{\text{Output}} : Y \to Y'$, satisfying the following three equations

$$h_{\text{State}}(f(x,s)) = f'(h_{\text{Input}}(x), h_{\text{State}}(s))$$
$$h_{\text{Output}}(g(s)) = g'(h_{\text{State}}(s))$$
$$h_{\text{State}}(s_0) = s_0'$$

which just say that automaton homomorphisms preserve the operations of automata. □

---

[1] For those not already familiar with it, this notion is defined in Appendix C.

Thus, Σ-homomorphisms are arrows that preserve the structure of Σ-algebras. One of the most important kinds of homomorphism is the *isomorphism*, which provides a translation between the data representations of two Σ-algebras that are "abstractly the same." Before giving a formal definition, we illustrate the concept with the following:

**Example 3.1.4** Let us consider two different ways of representing the natural numbers, each a variant of the Peano representation of Example 2.3.3. For the first, we have a one-sorted algebra $P$ whose carrier consists of the lists $0, s\, 0, s\, s\, 0, \ldots$, in which $0$ denotes the list $0$, and the operation $s$ maps an expression $e$ to the expression $s\, e$. For the second, we have an algebra $P'$ whose carrier consists of the lists $0, 0', 0'', \ldots$. We can now define $h : P \to P'$ recursively by the equations

$$h(0) = 0$$
$$h(s\, e) = h(e)'.$$

Intuitively, $P$ and $P'$ provide two different representations of the same thing, and $h$ describes a translation between these representations.  □

**Definition 3.1.5** A Σ-homomorphism $h : M \to M'$ is a Σ-**isomorphism** iff there is another Σ-homomorphism $g : M' \to M$ such that $h; g = 1_M$ and $g; h = 1_{M'}$ (i.e., such that for each $s \in S$, $g_s(h_s(m)) = m$ for all $m \in M_s$ and $h_s(g_s(m')) = m'$ for all $m' \in M'_s$). In this case, $g$ is called the **inverse** of $h$, and is denoted $h^{-1}$; also, we write $M \cong_\Sigma M'$ if there exists a Σ-isomorphism between $M$ and $M'$, and we may omit the subscript Σ if it is clear from context.  □

**Exercise 3.1.3** Prove that $h$ as defined in Example 3.1.4 above really is an isomorphism.  □

**Example 3.1.6** We now consider the binary representation of the natural numbers, forming a one-sorted algebra $B$. Its carrier consists of the symbol $0$ plus all (finite) lists of 0's and 1's not beginning with 0; 0 denotes the list 0 (i.e., $B_0 = 0$); and the operation $s$ is binary addition of 1. Then there is an arrow $h : P \to B$ with $P$ as defined in Example 3.1.4, such that

$$h(0) = 0$$
$$h(s\, e) = 1 + h(e).$$  □

**Exercise 3.1.4** Prove that $h$ as defined in Example 3.1.6 is an isomorphism.  □

**Exercise 3.1.5** Prove that a Σ-homomorphism $h$ is an isomorphism iff each $h_s$ is bijective.  □

The following summarizes some of the most useful properties of isomorphisms:

**Proposition 3.1.7** If $f : M \to M'$ and $g : M' \to M''$ are $\Sigma$-isomorphisms, then

(a) $(f^{-1})^{-1} = f$.
(b) $(f;g)^{-1} = g^{-1};f^{-1}$.
(c) $1_M^{-1} = 1_M$.
(d) $\cong_\Sigma$ is an equivalence relation[2] on the class of all $\Sigma$-algebras.  □

**Exercise 3.1.6** Prove the assertions in Proposition 3.1.7.  □

In Chapter 6, we will see that if there is an injective $\Sigma$-homomorphism $h : M \to M'$, then $M$ is isomorphic to a subalgebra of $M'$, and if there is a surjective $\Sigma$-homomorphism $h : M \to M'$, then $M'$ is isomorphic to a quotient algebra of $M$; the converses also hold. These results are Corollaries 6.1.8 and 6.1.9, respectively, and their converses are given in Exercise 6.1.2.

**Definition 3.1.8** An $S$-sorted arrow $f : M \to M'$ is a **left inverse** iff there is another $S$-sorted arrow $g : M' \to M$ such that $f;g = 1_M$. In this case, we also say that $g$ is a **right inverse** of $f$; we may also say that $f$ **has** a right inverse and that $g$ **has** a left inverse.  □

**Exercise 3.1.7** Show that if an $S$-sorted arrow has a right inverse then it is injective, and if it has a left inverse then it is surjective.  □

**Exercise 3.1.8** Show that if an $S$-sorted arrow is injective and has a left inverse, then it is bijective. Similarly, show that if an arrow is surjective and has a right inverse, then it is bijective.  □

These results imply the following, which will be very useful in certain proofs later on:

**Proposition 3.1.9** An injective $\Sigma$-homomorphism with a left inverse is an isomorphism, and so is a surjective $\Sigma$-homomorphism with a right inverse.  □

### 3.1.1 Unsorted Homomorphisms

We now consider homomorphisms for unsorted algebra in the sense of Section 2.7.

**Definition 3.1.10** Given unsorted $\Sigma$-algebras $M$ and $M'$, then a $\Sigma$-**homomorphism** $h : M \to M'$ is an arrow $M \to M'$, also denoted $h$, such that

(0) $h(M_\sigma) = M'_\sigma$ whenever $\sigma \in \Sigma_0$, and

---
[2]This concept is defined in Appendix C.

(1) $h(M_\sigma(m_1,\ldots,m_n)) = M'_\sigma(h(m_1),\ldots,h(m_n))$ whenever $\sigma \in \Sigma_n$ and $m_i \in M$ for $i = 1,\ldots,n$ with $n > 0$. □

Consistently with the results of Section 2.7, such a $\Sigma$-homomorphism is essentially the same thing as a one-sorted $\Sigma'$-homomorphism, where $\Sigma_n = \Sigma'_{s^n,s}$ with $S = \{s\}$.

## 3.2  Initiality of the Term Algebra

For many signatures $\Sigma$, the term algebra $T_\Sigma$ of Section 2.6 has a very special (and important) property: There is a unique way to interpret each of its elements in any $\Sigma$-algebra $M$. For example, if we let $\Sigma = \Sigma^{\text{NATEXP}}$ and let $M = \omega$, then $t = s(s(s(s(0))) * (s(0) + s(0)))$ should be interpreted as 7. And if we let $M = \{\textit{true}, \textit{false}\}$, with + interpreted as "or", with * interpreted as "and", $s$ as "not", and 0 as *false*, then $t$ should be interpreted as *false*. This section formalizes this property and explores some of its consequences. The key property of $T_\Sigma$ is stated below; its proof is given in Appendix B. We later construct a $\Sigma$-algebra that has this property for $\Sigma$ that may be overloaded (Theorem 3.2.10).

**Theorem 3.2.1** (*Initiality*) Given a signature $\Sigma$ without overloading and any $\Sigma$-algebra $M$, there is a unique $\Sigma$-homomorphism $T_\Sigma \to M$. □

The property that there is a unique $\Sigma$-homomorphism to any other $\Sigma$-algebra is called **initiality**, and any such algebra is called an **initial algebra**. We may think of the operation symbols in a signature $\Sigma$ as elementary operations or *commands* (or *microinstructions*), and then think of $T_\Sigma$ as the collection of all expressions (or simple *programs*) formed from $\Sigma$, and finally think of a $\Sigma$-algebra $M$ as a *machine* (or *microprocessor*) that can execute the commands in $\Sigma$. For example, a constant symbol $f$ in $\Sigma$ can be thought of as an instruction to load the value of $f$ in $M$. Then Theorem 3.2.1 tells us that each such simple program has one and only one value when *executed* on $M$. Thus initiality expresses a very basic intuition about computation on a machine.

We will later see that any two initial algebras are isomorphic, so that initiality defines a "standard model" for a signature that is unique up to the renaming of its elements.

Many interesting arrows arise as unique $\Sigma$-homomorphisms from some $\Sigma$-algebra; indeed, defining arrows by initiality is essentially the same as defining functions by induction. Let us consider some examples.

**Example 3.2.2** (*Evaluating Terms over the Naturals*) If $\Sigma$ is the signature $\Sigma^{\text{NATEXP}}$ of Example 2.3.4, then we can give $\omega$ the structure of a $\Sigma$-algebra in which the operation symbol 0 is interpreted as the number 0, the operation symbol $s$ is interpreted as the successor operation, the operation symbol + is interpreted as the addition function, and ∗ as multiplication. Then the unique $\Sigma$-homomorphism from $T_\Sigma$ to $\omega$ computes the values of the arithmetic expressions in $T_\Sigma$ in exactly the expected way. □

**Exercise 3.2.1** Compute the values of the terms in Figure 2.5 in the algebra of Example 3.2.2. □

**Example 3.2.3** If $\Sigma = \Sigma^{\mathsf{NATEXP}}$ then $T_\Sigma$ is *not* isomorphic to $\omega$ but for $\Sigma' = \Sigma^{\mathsf{NATP}}$, then $T_{\Sigma'}$ *is* isomorphic to $\omega$. Indeed, $T_{\Sigma'}$ is the natural numbers in Peano notation. □

**Example 3.2.4** (*Depth of a Term*) Given an arbitrary signature $\Sigma$, we can make the natural numbers into a $\Sigma$-algebra $\Omega$ by letting $\Omega_s = \omega$ for each $s \in S$, and by interpreting

(0) each $\sigma \in \Sigma_{[],s}$ as $0 \in \Omega_s$, and

(1) each $\sigma \in \Sigma_{s_1...s_n,s}$ for $n > 0$ as the arrow which sends the $n$ natural numbers $i_1, \ldots, i_n$ to the number $1 + max\{i_1, \ldots, i_n\}$.

Then the unique $\Sigma$-homomorphism $d : T_\Sigma \to \Omega$ computes the **depth** of $\Sigma$-terms, that is, the maximum amount of nesting in terms. □

**Exercise 3.2.2** Compute the depth of the terms shown in Figure 2.5 (page 25) using the algebra of Example 3.2.4. □

**Example 3.2.5** (*Size of a Term*) Let $\Sigma$ be the signature $\Sigma^{\mathsf{NATEXP}}$ of Example 2.3.4, let $\omega$ be the carrier of an algebra $A$ in which $0 \in \Sigma$ is interpreted as 1, $s_A(n) = n + 1$, $+_A(m, n) = *_A(m, n) = m + n + 1$. Then the unique $\Sigma$-homomorphism $h : T_\Sigma \to A$ computes the **size** of a term, that is, the number of operation symbols that occur in it. □

**Exercise 3.2.3** Compute the size of the terms shown in Figure 2.5 (page 25) using the algebra of Example 3.2.5. □

**Exercise 3.2.4** Use initiality to define an arrow from $\Sigma$-terms which gives the number of *interior* (i.e., non-leaf) nodes in the corresponding tree. □

This way of defining functions is a special case of a much more general method called *initial algebra semantics* [52, 88]. This method regards terms in $T_\Sigma$ as objects to which some meaning is to be assigned, constructs a $\Sigma$-algebra $M$ of suitable meanings, and then lets the unique $\Sigma$-homomorphism that automatically exists do the work. For example, $T_\Sigma$ might contain the various syntactic elements of a programming language in its various sorts, such as expressions, procedures, and of course programs, with $M$ containing suitable denotations for these, e.g., in the style of denotational semantics [92, 161]; many examples of this approach are given in [77].

**Example 3.2.6** (*Final Algebra*) There is a trivial but interesting algebra denoted $F_\Sigma$ that can be constructed for any signature $\Sigma$: let $(F_\Sigma)_s = \{s\}$ for each sort $s \in S$; and given $\sigma \in \Sigma_{w,s}$, let $F_\sigma(s_1,\ldots,s_n) = s$ when $w = s_1\ldots s_n$. Then the unique homomorphism $T_\Sigma \to F_\Sigma$ gives the *sort* of a $\Sigma$-term.  □

The following generalizes this to any signature $\Sigma$ and any $\Sigma$-algebra; the unique $\Sigma$-homomorphism again gives the sorts of elements.

**Proposition 3.2.7** Given any signature $\Sigma$ and any $\Sigma$-algebra $M$, there is one and only one $\Sigma$-homomorphism $h : M \to F_\Sigma$.

**Proof:** Given $m \in M_s$, we have to define $h(m) = s$ because $h(m)$ must be in $(F_\Sigma)_s = \{s\}$. It is straightforward to check that this gives a $\Sigma$-homomorphism.  □

This property is called **finality**; it is dual to initiality.

**Exercise 3.2.5** Let $\Sigma$ be the signature $\Sigma^{\mathsf{NATP}}$ of Example 2.3.3, and let $D$ be the $\Sigma$-algebra with carrier $\{0,1\}$, with $0_D = 0$, with $s_D(0) = 1$ and with $s_D(1) = 0$. Give a direct proof that there is one and only one $\Sigma$-homomorphism $T_\Sigma \to D$.  □

**Example 3.2.8** When there is overloading, terms do not always have a unique sort or parse. For example, if $\Sigma$ is the signature of Example 2.5.7, then 0 and 1 are *ambiguous* in the sense that there is no unique $s \in S$ such that they lie in $(T_\Sigma)_s$; the terms $0+1$, $1+(1+0)$ and many others are also ambiguous, although for example, the terms $s(0)$ and $1+(n(0)+0)$ are unambiguous. Proposition 3.7.2 in Section 3.7 below gives a necessary and sufficient condition on $\Sigma$ such that no $\Sigma$-terms are ambiguous.

For $\Sigma$ as in Example 2.5.7, $T_\Sigma$ is initial even though it has ambiguous terms. However, there are overloaded signatures such that $T_\Sigma$ is *not* initial. For example, let $S = \{A, B, C\}$, let $\Sigma_{[],A} = \Sigma_{[],B} = \{0\}$, and let $\Sigma_{A,C} = \Sigma_{B,C} = \{f\}$. Now define a $\Sigma$-algebra $M$ as follows: $M_A = \{0\}$; $M_B = \{0\}$; $M_C = \{0,1\}$; $M_0^{[],A} = 0$; $M_0^{[],B} = 0$; $M_f^{A,C}(0) = 0$; $M_f^{B,C}(0) = 1$. Then there is no $\Sigma$-homomorphism $h : T_\Sigma \to M$, because the term $f(0)$ has two distinct parses of the same sort, which are $T_f^{A,C}(T_0^{[],A})$ and $T_f^{B,C}(T_0^{[],B})$. Therefore $f(0)$ must be mapped to two different elements of $M$,

$$h(T_f^{A,C}(T_0^{[],A})) = M_f^{A,C}(M_0^{[],A}) = 0 \;,$$
$$h(T_f^{B,C}(T_0^{[],B})) = M_f^{B,C}(M_0^{[],B}) = 1 \;,$$

which is impossible.  □

Because initiality is so important for this book, the above example means that $T_\Sigma$ is not adequate for our purposes. However, there is a closely related $\Sigma$-algebra that *is* initial for any signature; its terms are annotated with their sorts.

**Definition 3.2.9** Given any $S$-sorted signature $\Sigma$, the $S$-sorted set $\overline{T}_\Sigma$ of all **sorted (ground)** $\Sigma$-**terms** is the smallest set of lists over the set $S \cup |\Sigma| \cup \{\cdot, \underline{(}, \underline{)}\}$ (where $\cdot$, $\underline{(}$ and $\underline{)}$ are special symbols disjoint from $\Sigma$) such that

  (0) if $\sigma \in \Sigma_{[],s}$ then $\sigma \cdot s \in \overline{T}_{\Sigma,s}$ for all $s \in S$, and
  
  (1) if $\sigma \in \Sigma_{s_1 \ldots s_n, s}$ and $t_i \in \overline{T}_{\Sigma, s_i}$ for $i = 1, \ldots, n$ then $\sigma \cdot s\underline{(}t_1 \ldots t_n\underline{)} \in \overline{T}_{\Sigma, s}$.

The $S$-sorted set $\overline{T}_\Sigma$ can be given the structure of a $\Sigma$-algebra in the same way that $T_\Sigma$ was:

  (0) interpret $\sigma \in \Sigma_{[],s}$ in $\overline{T}_\Sigma$ as the singleton list $\sigma \cdot s$, and
  
  (1) interpret $\sigma \in \Sigma_{s_1 \ldots s_n, s}$ in $\overline{T}_\Sigma$ as the function that sends $t_1, \ldots, t_n$ to the list $\sigma \cdot s\underline{(}t_1 \ldots t_n\underline{)}$, where $t_i \in \overline{T}_{\Sigma, s_i}$ for $i = 1, \ldots, n$ with $n > 0$.

As before, we will usually write $\sigma \cdot s(t_1, \ldots, t_n)$ instead of $\sigma \cdot s\underline{(}t_1 \ldots t_n\underline{)}$. Call $\overline{T}_\Sigma$ the **(sort) annotated term algebra** over $\Sigma$. □

The following result is proved in Appendix B:

**Theorem 3.2.10** (*Initiality*) Given any signature $\Sigma$ and any $\Sigma$-algebra $M$, there is a unique $\Sigma$-homomorphism $\overline{T}_\Sigma \to M$. □

It follows that $T_\Sigma$ is "almost an initial $\Sigma$-algebra," because its terms differ from those of $\overline{T}_\Sigma$ only in the sort annotations; for many signatures, including most of those that come up in practice, $T_\Sigma$ actually is initial. This motivates the following

**Convention 3.2.11** We will usually write terms without sort annotation, and will usually annotate operations only in so far as necessary to determine a unique fully annotated term with the given partial annotation. Moreover, we will usually write $T_\Sigma$ when we really mean $\overline{T}_\Sigma$. □

Proposition 3.2.12 below characterizes when $T_\Sigma$ is initial. It uses the following:

**Exercise 3.2.6** Show that the arrow $h : \overline{T}_\Sigma \to T_\Sigma$ that strips sort annotations off operation symbols is a $\Sigma$-homomorphism. It may be defined as follows:

  (0) $h_s(\sigma \cdot s) = \sigma$ for $\sigma \in \Sigma_{[],s}$, and
  
  (1) $h_s(\sigma \cdot s\underline{(}t_1 \ldots t_n\underline{)}) = \sigma\underline{(}h_{s_1} t_1 \ldots h_{s_n} t_n\underline{)}$ for $\sigma \in \Sigma_{s_1 \ldots s_n, s}$ and $t_i \in \overline{T}_{\Sigma, s_i}$ for $i = 1, \ldots, n$ with $n > 0$.

Also, show that this homomorphism is surjective. □

**Proposition 3.2.12** The term algebra $T_\Sigma$ is initial iff any two distinct sorted terms of the same sort remain distinct after the sorts are stripped off operation symbols.

**Proof:** That the unique $\Sigma$-homomorphism $h : \overline{T}_\Sigma \to T_\Sigma$ strips off sorts is due to its homomorphic property, and since it is surjective, it is an isomorphism iff it is injective. □

Recall that a sufficient condition for $T_\Sigma$ to be initial is that $\Sigma$ has no overloading. The above says that it is initial iff whatever overloading may be present does not produce ambiguity. From this, it follows by induction that it suffices for there to be no overloaded constants.

### 3.2.1 Initiality and Structural Induction

Structural induction [21] is an important proof technique that is closely related to initiality. To prove that a certain property $P$ is true of all $\Sigma$-terms, we (0) prove that $P$ is true of all the constants in $\Sigma$, and then (1) prove that if $P$ is true of $t_1, \ldots, t_n$ (of appropriate sorts), then $P$ is true of $\sigma(t_1, \ldots, t_n)$, for every nonconstant symbol $\sigma$ in $\Sigma$. It then follows that $P$ is true of all (ground) $\Sigma$-terms, because they all can be built from the symbols in $\Sigma$, working upward from the constants. Somewhat more formally, let $P \subseteq T_\Sigma$ be the ($S$-sorted) subset of all $\Sigma$-terms for which the desired property holds. Then the two properties to be shown imply that $P$ is a $\Sigma$-algebra. But it can be shown (see the next paragraph) that $T_\Sigma$ does not have any *proper* $\Sigma$-subalgebras, from which it follows that $P = T_\Sigma$.

More formally the following steps are required for proving that some $S$-indexed family $P$ of predicates holds for all $\Sigma$-terms by **structural induction**:

(0) prove that $P_s$ holds for every $\sigma \in \Sigma_{[],s}$, for each $s \in S$; and

(1) prove that $P_s$ holds for $\sigma(t_1, \ldots, t_n)$ for all $\sigma \in \Sigma_{s_1 \ldots s_n, s}$ where $P_{s_i}$ is assumed to hold for each $t_i$, a $\Sigma$-term of sort $s_i$, for $i = 1, \ldots, n$.

This proof method is carefully stated and validated in Chapter 6 (Theorem 6.4.4), but the idea is as follows: let $i$ denote the inclusion of $P$ into $T_\Sigma$ and let $h : T_\Sigma \to P$ be the unique $\Sigma$-homomorphism given by initiality. Then $h; i : T_\Sigma \to T_\Sigma$ is also a $\Sigma$-homomorphism. But initiality implies that there is only one $\Sigma$-homomorphism $T_\Sigma \to T_\Sigma$, which is necessarily the identity $1_{T_\Sigma}$, because that *is* a $\Sigma$-homomorphism. Therefore $h; i = 1_{T_\Sigma}$, and so by Exercise 3.1.8, $i$ is a $\Sigma$-isomorphism, because it is injective and a right inverse.

## 3.3 Equation and Satisfaction

This section defines the basic concepts of equation, and of satisfaction of an equation by an algebra. This will give us a semantic notion of truth for equational logic, and hence a standard by which to judge the soundness of rules of deduction for that system. A number of examples are also given.

To discuss equations, we need terms with variables. It can seem quite difficult to say exactly what a variable actually *is* in some branches of mathematics. But in general algebra, this is not so hard: A $\Sigma$-term with variables in $X$ is just an element of $T_{\Sigma(X)}$ where $X$ is a ground signature (see Notation 2.3.6) disjoint from $\Sigma$; that is, a variable is just a new constant symbol.

**Definition 3.3.1** A $\Sigma$-**equation** consists of a ground signature[3] $X$ of **variable symbols** (disjoint from $\Sigma$) plus two $\Sigma(X)$-terms of the same sort $s \in S$; we may write such an equation abstractly in the form

$$(\forall X)\, t = t'$$

and concretely in the form

$$(\forall x, y, z)\, t = t'$$

when (for example) $|X| = \{x, y, z\}$ and the sorts of $x, y, z$ can be inferred from their uses in $t$ and in $t'$. A **specification** is a pair $(\Sigma, A)$, consisting of a signature $\Sigma$ and a set $A$ of $\Sigma$-equations. A $\Sigma$-specification is a specification whose signature is $\Sigma$. □

**Example 3.3.2** (*Semigroups*) This specification has just one sort, say Elt, and just one operation, say _*_ : Elt Elt -> Elt, which must obey the associative law,

$$(x * y) * z = x * (y * z)$$

where $x, y, z$ are variables of sort Elt. If we let $X$ be the ground signature with $X_{\text{Elt}} = \{x, y, z\}$, then this can be written more accurately as

$$(\forall X)\, (x * y) * z = x * (y * z)$$

and slightly less formally as

$$(\forall x, y, z)\, (x * y) * z = x * (y * z)\,.$$

In OBJ, we would write

```
th SEMIGROUP is sort Elt .
  op _*_ : Elt Elt -> Elt .
  vars X Y Z : Elt .
  eq (X * Y)* Z = X *(Y * Z).
endth
```

This follows the convention that variable names begin with an uppercase letter; like the convention for sort names, it is not enforced by the OBJ3 system. However, the systematic use of these conventions does help users to distinguish sort and variable names from keywords and operation symbols, and thus helps make specifications more readable. (For a diagram of the signature of this specification, delete the edges labelled by e and by -1 from Figure 3.1.) □

---

[3] Recall that this means that all of the variable symbols in $X$ are distinct.

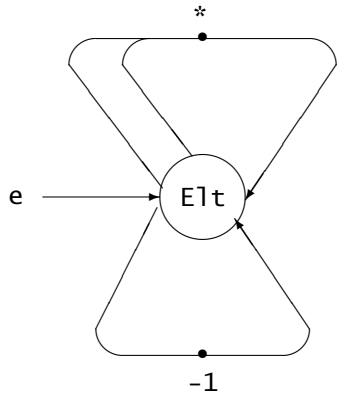

Figure 3.1: Signature for Groups

**Example 3.3.3** (*Monoids*) Similarly, monoids are specified as follows:

```
th MONOID is sort Elt .
  op e : -> Elt .
  op _*_ : Elt Elt -> Elt .
  vars X Y Z : Elt .
  eq X * e = X .
  eq e * X = X .
  eq (X * Y)* Z = X *(Y * Z).
endth
```

This theory denotes the class of all monoids. (We no longer give a set-theoretic version.) Our convention names both objects and sorts with the *singular* version of the structure involved, rather than the plural; thus, we write MONOID and Elt rather than MONOIDS and Elts. □

**Exercise 3.3.1** Write out a formal set-theoretic definition of the above OBJ specification of monoids, in the style of Example 2.5.3. □

**Example 3.3.4** (*Groups*) It is little more work to specify groups than to specify monoids:

```
th GROUP is sort Elt .
  op e : -> Elt .
  op _⁻¹ : Elt -> Elt .
  op _*_ : Elt Elt -> Elt .
  vars X Y Z : Elt .
  eq X * e = X .
  eq X *(X ⁻¹) = e .
  eq (X * Y)* Z = X *(Y * Z).
endth
```

Notice that only half of the usual pairs of equations for the identity and inverse laws are given. We will later prove that the other halves follow from these laws (and *vice versa*). The signature for this specification is shown in Figure 3.1.

If $h : G' \to G$ is an inclusion homomorphism, then $G'$ is said to be a **subgroup** of $G$. □

**Example 3.3.5** (*Integers*) We can specify the integers as follows:

```
obj INT is sort Int .
  op 0  : -> Int .
  op s_ : Int -> Int .
  op p_ : Int -> Int .
  var I : Int .
  eq s p I = I .
  eq p s I = I .
endo
```

Here s_ is the successor operation, and p_ is the predecessor operation.[4] This specification defines the algebra of integers, with the given operations; we may also say that it denotes the class of all standard models of the integers, as initial algebras with these operations.

The following specification also defines addition and negation on the integers:

```
obj INT is sort Int .
  op 0   : -> Int .
  op s_  : Int -> Int .
  op p_  : Int -> Int .
  op -_  : Int -> Int .
  op _+_ : Int Int -> Int .
  vars I J : Int .
  eq s p I = I .
  eq p s I = I .
  eq - 0 = 0 .
  eq - s I = p - I .
  eq - p I = s - I .
  eq I + 0 = I .
  eq I + s J = s(I + J).
  eq I + p J = p(I + J).
endo
```

Here -_ is the negation operation and _+_ is addition. It is interesting to notice that the integers with 0 as identity, -_ as inverse and _+_ as "multiplication" form a group; however, we do not yet have the tools needed to prove this. □

Our commitment to semantics requires that we not only formalize what equations *are*, but also what they *mean*. This is done by the concept of satisfaction, which will use the following:

---

[4]As will be discussed in more detail later, neither of these operations is a "constructor" in the usual sense, because there are non-trivial relations between them.

**Notation 3.3.6** Recall that a Σ-algebra $M$ provides an interpretation for each operation symbol in Σ, and in particular, for each constant symbol in Σ. If $X$ is a ground signature (e.g., a set of variables), then an interpretation for $X$ is just a (many-sorted) arrow $a : X \to M$. Thus a Σ-algebra $M$ and an arrow $a : X \to M$ give an interpretation in $M$ for all of $\Sigma(X)$, allowing $M$ to be seen as a $\Sigma(X)$-algebra. Theorem 3.2.1 now gives a unique $\Sigma(X)$-homomorphism from the initial $\Sigma(X)$-algebra $T_{\Sigma(X)}$ to $M$ as a $\Sigma(X)$-algebra, using $a$.

In such a situation, we call $a : X \to M$ an **interpretation** or an **assignment** of the variable symbols in $X$, and we let $\overline{a} : T_{\Sigma(X)} \to M$ denote the unique extension of $a$ to a $\Sigma(X)$-homomorphism from the term algebra $T_{\Sigma(X)}$. □

**Definition 3.3.7** A Σ-algebra $M$ **satisfies**[E4] a Σ-equation $(\forall X)\, t = t'$ iff for any assignment $a : X \to M$ we have $\overline{a}(t) = \overline{a}(t')$ in $M$. In this case we write

$$M \vDash_\Sigma (\forall X)\, t = t' \,.$$

We call ⊨ a **"2-bar turnstile,"** and we generally omit the subscript Σ when it is clear from context. A Σ-algebra $M$ **satisfies** a *set* $A$ of Σ-equations iff it satisfies each $e \in A$, and in this case we write

$$M \vDash_\Sigma A \,.$$

We may also say that $M$ is a $P$-algebra, and write

$$M \vDash P$$

where $P$ is a specification $(\Sigma, A)$. The class of all algebras that satisfy $P$ is called the **variety** defined by $P$, and we may also say that the **denotation** of $P$ is this variety.

Finally, for $A$ a set of Σ-equations, we let

$$A \vDash_\Sigma (\forall X)\, t = t'$$

mean that $M \vDash_\Sigma A$ implies $M \vDash_\Sigma (\forall X)\, t = t'$. □

**Example 3.3.8** If Σ is the signature $\Sigma^{\mathtt{MONOID}}$ of Example 3.3.3, then a Σ-algebra is a monoid iff it satisfies the equations in Example 3.3.3, i.e., iff it satisfies the specification `MONOID` of monoids. The denotation of the theory `MONOID` is the variety of all monoids. For example, given a set $S$, recall that $S^*$ denotes the set of all lists of elements from $S$, including the empty list $[\,]$. Then $S^*$ is a monoid with $e = [\,]$ and with $*$ interpreted as concatenation of lists: for any choice of $x, y, z \in S^*$, it is true that $(x * y) * z = x * (y * z)$, because each term yields the concatenation of the three lists. □

**Example 3.3.9** If Σ is the signature $\Sigma^{\mathtt{GROUP}}$ of Example 3.3.4, then a Σ-algebra is a group iff it satisfies the equations in Example 3.3.4, i.e., iff it satisfies the specification GROUP. For example, $S^*$ satisfies the first and third axioms of GROUP, because it is a monoid, but there is *no* way to define an operation $i : S^* \to S^*$ such that $x * i(x) = [\,]$ for all $x \in S^*$, because a concatenation of two lists always yields a list that is at least as long as its arguments. Indeed, the *only* concatenation that yields the value $[\,]$ is $[\,] * [\,]$. This is an example of non-satisfaction. □

**Exercise 3.3.2** Let $S$ be a set.

1. Show that the bijections $f : S \to S$ form a group under composition (of functions), with identity $1_S$.

2. Let $G$ be a group of bijections on a set $S$, and let $F \subseteq S$, called a "figure." Show that $\{f \in G \mid f(F) = F\}$ is a subgroup of $G$, called the *group of symmetries* of $F$. □

**Exercise 3.3.3** An **endomorphism** of a $\Sigma$-algebra $M$ is a $\Sigma$-homomorphism $h : M \to M$, and an **automorphism** is an endomorphism that is bijective (i.e., an isomorphism).

1. Show that the set of all endomorphisms of a given $\Sigma$-algebra $M$ has the structure of a monoid under composition.

2. Show that the set of all automorphisms of a given $\Sigma$-algebra $M$ has the structure of a group under composition. □

**Example 3.3.10** A rather cute specification that is not very well known defines pairs of natural numbers using just one constant, two unary operations, and a single equation:

```
obj 2NAT is sort 2Nat .
  op 0 : -> 2Nat .
  ops (s1_) (s2_) : 2Nat -> 2Nat .
  var P : 2Nat .
  eq s1 s2 P = s2 s1 P .
endo
```

We can show that pairs of natural numbers are an initial algebra for 2NAT as follows: Let $P_{\text{2Nat}} = \{\langle m, n\rangle \mid m, n \in \omega\}$, let $0_P$ be $\langle 0, 0\rangle$, let $s_1$ send $\langle m, n\rangle$ to $\langle sm, n\rangle$, and let $s_2$ send $\langle m, n\rangle$ to $\langle m, sn\rangle$. Now if $M$ is any 2NAT-algebra, then define $h : P \to M$ to send $\langle m, n\rangle$ to (the value that is denoted by the term) $s_1^m s_2^n 0$ in $M$. Then $h$ is a $\Sigma^{\text{2NAT}}$-homomorphism because $h(0_P) = h(\langle 0, 0\rangle) = 0_M$, and $h(s_1\langle m, n\rangle) = s_1 h(\langle m, n\rangle)$ because each equals $s_1^{m+1} s_2^n 0$ in $M$, and similarly for preservation of $s_2$. We leave it as an exercise to show that if $g : P \to M$ is also a $\Sigma^{\text{2NAT}}$-homomorphism, then necessarily $g = h$.

Another 2NAT-algebra has as carrier the set consisting of all terms of the form $s_1^m s_2^n 0$, with $s_1(s_1^m s_2^n 0) = s_1^{m+1} s_2^n 0$ and $s_2(s_1^m s_2^n 0) = s_1^m s_2^{n+1} 0$. This algebra can be shown to be initial in a very similar way, and therefore it is isomorphic to $P$. □

The fact that variables are just unconstrained constants suggests that equations with variables can be regarded as ground equations in which the variables are treated as new constants. The following is a formal statement of this basic intuition:

**Theorem 3.3.11** (*Theorem of Constants*) Given a signature $\Sigma$, a ground signature $X$ disjoint from $\Sigma$, a set $A$ of $\Sigma$-equations, and $t, t' \in T_{\Sigma(X)}$, then

$$A \vDash_\Sigma (\forall X)\, t = t' \quad \text{iff} \quad A \vDash_{\Sigma \cup X} (\forall \emptyset)\, t = t'\,.$$

**Proof:** Each condition is equivalent to the condition that $\overline{a}(t) = \overline{a}(t')$ for every $\Sigma(X)$-algebra $M$ satisfying $A$ and every assignment $a : X \to M$.[E5]  $\square$

It is very pleasing that this proof is so simple. This is because it is based on the *semantics* of satisfaction, rather than some particular rules of deduction, and because it exploits the *initiality* of the term algebra. The intuition behind this result is again that variables "are" constants about which we do not know anything.

Although Example 3.3.4 gives perhaps the most common way to specify groups, there are many other equivalent ways. Therefore it is interesting to examine what "equivalent" means in this context. Of course, we first define equivalence semantically, and after that give syntactic measures for proving equivalence. This section considers only the special case where the two specifications have the same signature; Section 4.10 extends this to allow different signatures.

**Definition 3.3.12** $\Sigma$-specifications $P$ and $P'$ are **equivalent** iff for each $\Sigma$-algebra $M$,

$$M \vDash P \quad \text{iff} \quad M \vDash P'\,.$$

We can now define a **theory** to be an equivalence class of specifications.  $\square$

It is usual to identify a specification $P$ with the theory that it represents, that is, with its equivalence class; thus, we may say that GROUP *is* the theory of groups, rather than saying that GROUP *represents* (or *presents*) the theory of groups.

**Example 3.3.13** (*Left Groups*) Example 3.3.4 specified the theory GROUP of groups with right identity and inverse equations; here is a specification with left-handed versions of these equations:

```
th GROUPL is sort Elt .
  op e : -> Elt .
  op _⁻¹ : Elt -> Elt .
  op _*_ : Elt Elt -> Elt .
  vars X Y Z : Elt .
  eq e * X = X .
```

```
    eq (X ⁻¹) * X = e .
    eq X *(Y * Z) = (X * Y)* Z .
  endth
```

In Chapter 4 we will prove that GROUPL is equivalent to GROUP. □

## 3.4 Conditional Equations and Satisfaction

There are many cases where some equation (or other formula) holds only in certain conditions. The extension of equations and their satisfaction to the conditional case is straightforward.

**Definition 3.4.1** A **conditional $\Sigma$-equation** consists of a ground signature $X$ disjoint from $\Sigma$, a finite set $C$ of pairs of $\Sigma(X)$-terms, and a pair $t, t'$ of $\Sigma(X)$-terms; we will use the notation

$$(\forall X)\ t = t'\ \text{if}\ C\ .$$

Given a $\Sigma$-algebra $M$, define[E6]

$$M \vDash_\Sigma (\forall X)\ t = t'\ \text{if}\ C$$

to mean that, given any interpretation $a : X \to M$, if $\overline{a}(u) = \overline{a}(v)$ for each $\langle u, v \rangle \in C$, then $\overline{a}(t) = \overline{a}(t')$. If $A$ is a set of conditional $\Sigma$-equations, then we say that $M$ satisfies $A$ iff $M$ satisfies each equation in $A$, and if $e$ is a conditional equation, then $A \vDash_\Sigma e$ iff $M \vDash_\Sigma e$ whenever $M \vDash_\Sigma A$. □

Conditional equations make sense even when $C$ is not finite; but without that restriction, neither equational deduction nor term rewriting with such equations would be possible in finite space, and in particular, we could not write down (finite) proof scores that use equations with infinite conditions.

**Fact 3.4.2** Given any $\Sigma$-equation $e = (\forall X)\ t = t'$, let $e' = (\forall X)\ t = t'\ \text{if}\ \emptyset$. Then for each $\Sigma$-algebra $M$, $M \vDash_\Sigma e$ iff $M \vDash_\Sigma e'$. □

Consequently, we can regard any ordinary equation as a conditional equation with the empty condition, and *vice versa*; we will feel free to do this hereafter.

The following result gives us a technique for proving conditional equations with equational deduction, and hence with reduction; the application of this result for deduction is given in Theorem 4.8.4.

**Proposition 3.4.3** Given a conditional $\Sigma$-equation $(\forall X)\ t = t'\ \text{if}\ C$ and a set $A$ of $\Sigma$-equations, then

$$A \vDash_\Sigma (\forall X)\ t = t'\ \text{if}\ C\ \text{iff}\ (A \cup C') \vDash_{\Sigma(X)} (\forall \emptyset)\ t = t'\ ,$$

where $C'$ is defined to be $\{(\forall \emptyset)\ u = v \mid \langle u, v \rangle \in C\}$.

**Proof:** Each condition is equivalent to the following:

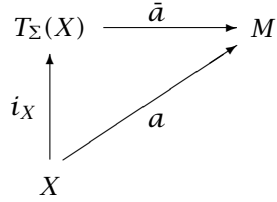

Figure 3.2: Free Algebra Property

for each $\Sigma$-algebra $M$ and interpretation $a : X \to M$, if $M \models A$ and if $\overline{a}(u) = \overline{a}(v)$ for each $\langle u, v \rangle \in C$, then $\overline{a}(t) = \overline{a}(t')$,

where $\overline{a} : T_{\Sigma(X)} \to M$ is the unique homomorphism. □

Once again, initiality enables us to give a very simple proof.

## 3.5 Substitution

The substitution of terms into other terms will play a basic role in later chapters, especially those on deduction and term rewriting. To help define this concept, we first consider the so-called *free algebras*, using a technique already employed in setting up Definition 3.3.7. Given a signature $\Sigma$ and a ground signature $X$ disjoint from $\Sigma$, we can form the $\Sigma(X)$-algebra $T_{\Sigma(X)}$ and then view it as a $\Sigma$-algebra by just "forgetting" about the constants in $X$; this works because $T_{\Sigma(X)}$ already has all the operations it needs to be a $\Sigma$-algebra, and it does no harm that it also has some others. Let us denote this $\Sigma$-algebra by $T_\Sigma(X)$. It is called the **free $\Sigma$-algebra generated by** (or **over**) $X$, and it has the following characteristic property, called **free generation by** (or **over**) $X$ (see also Figure 3.2):

**Proposition 3.5.1** Given a signature $\Sigma$, a ground signature $X$ disjoint from $\Sigma$, a $\Sigma$-algebra $M$, and a map $a : X \to M$, there is a unique $\Sigma$-homomorphism $\overline{a} : T_\Sigma(X) \to M$ which extends $a$, in the sense that $\overline{a}_s(x) = a_s(x)$ for each $s \in S$ and $x \in X_s$. (This property is illustrated in Figure 3.2, where $i_X$ is the $S$-sorted inclusion.) We may call $a$ an **assignment** from $X$ to $M$.

**Proof:** Let $j$ be the interpretation for $\Sigma$ in $M$. Then combining $j$ with $a$ gives an interpretation for $\Sigma(X)$ in $M$, and hence makes $M$ into a $\Sigma(X)$-algebra. Therefore, by initiality of $T_{\Sigma(X)}$ there is a unique $\Sigma(X)$-homomorphism from $T_{\Sigma(X)}$ to $M$. But this is exactly the same thing as a $\Sigma$-homomorphism from $T_\Sigma(X)$ to $M$ that extends $a$. □

We have already noted that a $\Sigma$-term with variables in an $S$-sorted ground signature $Y$ is just an element of $T_\Sigma(Y)$. Then an assignment $a : X \to T_\Sigma(Y)$ assigns $\Sigma$-terms with variables from $Y$ to variables from $X$ in a way that respects the sorts involved, and the $\Sigma$-homomorphism $\overline{a} : T_\Sigma(X) \to T_\Sigma(Y)$ given by Proposition 3.5.1 *substitutes* a term $a(x)$ for each variable $x \in X$ into each term $t$ in $T_\Sigma(X)$, yielding a term $\overline{a}(t)$ in $T_\Sigma(Y)$. Hence we have the following:

**Definition 3.5.2** A **substitution** of $\Sigma$-terms with variables in $Y$ for variables in $X$ is an arrow $a : X \to T_\Sigma(Y)$; we may also use the notation $a : X \to Y$. The **application** of $a$ to $t \in T_\Sigma(X)$ is $\overline{a}(t)$. Given substitutions $a : X \to T_\Sigma(Y)$ and $b : Y \to T_\Sigma(Z)$, their **composition** $a; b$ (as substitutions) is the $S$-sorted arrow $a; \overline{b} : X \to T_\Sigma(Z)$. □

**Notation 3.5.3** The following notation makes substitutions look less abstract: Given $t \in T_\Sigma(X)$ and given $a : X \to T_\Sigma(Y)$ such that $|X| = \{x_1, \ldots, x_n\}$ and $a(x_i) = t_i$ for $i = 1, \ldots, n$, then we may write $\overline{a}(t)$ in the form

$$t(x_1 \leftarrow t_1, x_2 \leftarrow t_2, \ldots, x_n \leftarrow t_n),$$

and whenever $t_i$ is the variable $x_i$, we can omit the pair $x_i \leftarrow t_i$. □

**Exercise 3.5.1** Let $\Sigma$ be the signature of Example 2.3.4, let $X = \{x, y, z\}_{[],\text{Nat}}$, let $t = x + s(s(y) + z)$, let $Y = \{u, v\}_{[],\text{Nat}}$, and define $a : X \to T_\Sigma(Y)$ by $a(x) = u + s(v)$, $a(y) = 0$, and $a(z) = v + s(0)$. Now compute $\overline{a}(t)$. □

**Exercise 3.5.2** If $i_X : X \to T_\Sigma(X)$ is the inclusion, show that $\overline{i_X}(t) = t$ for each $t \in T_\Sigma(X)$. □

**Exercise 3.5.3** Given a substitution $a : X \to T_\Sigma(Y)$, show that $i_X; a = a$ and $a; i_Y = a$. □

**Notation 3.5.4** Because $i_X$ serves as an identity for the composition of substitutions, we may write $1_X$ for $i_X$ in the following. □

It is natural to expect that term substitution is associative, in the sense that given substitutions $a : W \to T_\Sigma(X)$, $b : X \to T_\Sigma(Y)$ and $c : Y \to T_\Sigma(Z)$, we have $(a; b); c = a; (b; c)$. We will see in the next section that there is a simple and beautiful proof of this using the free property of term algebras.

**Exercise 3.5.4** Is substitution commutative? I.e., given $a, b : X \to T_\Sigma(X)$, does $a; b = b; a$? Give a proof or a counterexample. □

## 3.6 Pasting and Chasing

There is a very nice way to graphically represent systems of equations, such as those that arise from the homomorphism condition; this is the method of *commutative diagrams*. We will later see that this method not only allows us to represent systems of equations graphically, but also to reason about them graphically; such reasoning with diagrams is often called *diagram chasing* because of the characteristic way in which one follows arrows around the diagram. In order to explain this in a precise way, we first need some further concepts from graph theory.

**Definition 3.6.1** A **path** $p$ in a graph $G$ is a list $e_1, \ldots, e_m$ of edges of $G$ such that $\partial_1(e_i) = \partial_0(e_{i+1})$ for $i = 1, \ldots, m-1$; the **source** of $p$ is $\partial_0(p) = \partial_0(e_1)$ and the **target** of $p$ is $\partial_1(p) = \partial_1(e_m)$; we will write $p : n \to n'$ when $\partial_0(p) = n$ and $\partial_1(p) = n'$. If $m = 0$ then $p$ would be the empty list [ ], and $\partial_0([\,])$ and $\partial_1([\,])$ would not be defined; so instead, we make the source and target of [ ] explicit, writing $[\,]_n$. Given $p : n \to n'$ and $q : n' \to n''$, we define their **composition** $p; q : n \to n''$ to be their concatenation as lists. Note that $[\,]_n$ is an **identity** for this composition, in the sense that $[\,]_n; p = p$ and $p; [\,]_m = p$, for any path $p$ with source $n$ and target $m$; in practice, we may omit the subscripts on [ ].

A graph $G$ is a **tree** iff it has a node $r$, called its **root**, such that for each node $n$ of $G$, there is a *unique* path from $r$ to $n$ in $G$. □

One motivation for being so formal about all of this is that mechanical theorem proving is necessarily formal to this extent.

**Exercise 3.6.1** Show that the root of a tree is necessarily unique. □

**Exercise 3.6.2** Given a graph $G$ with at most one edge between each pair of nodes, show that a path $p = e_1 \ldots e_k$ in $G$ is uniquely determined by the sequence $n_0 n_1 \ldots n_{k-1} n_k$ of the nodes that it passes through, where $n_0 = \partial_0(e_1)$, $n_1 = \partial_1(e_1) = \partial_0(e_2)$, ..., $n_{k-1} = \partial_1(e_{k-1}) = \partial_0(e_k)$ and $n_k = \partial_1(e_k)$. □

**Definition 3.6.2** A **diagram (of (sorted) sets)** is a graph whose nodes are labelled by (sorted) sets, and whose edges are labelled by (sorted) arrows, in such a way that

- if the arrow $f : A \to A'$ labels the edge $e : n \to n'$, then the label of $n$ is $A$ and the label of $n'$ is $A'$.

A diagram **commutes** iff

- whenever $p, q$ are two paths, at least one of which has length at least 2, each with (say) source $n$ and target $n'$, such that the labels along the edges of $p$ are $f_1, \ldots, f_m$ and along $q$ are $g_1, \ldots, g_k$, then $f_1; \ldots; f_m = g_1; \ldots; g_k$.

That is, the arrows obtained by composition along any two (non-trivial) paths from one node to another are equal. □

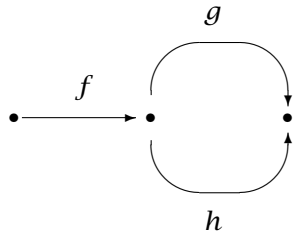

Figure 3.3: Length One Paths

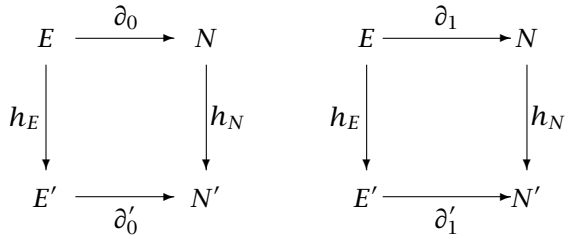

Figure 3.4: Commutative Diagrams for Graph Homomorphism

Thus, a commutative diagram is a geometrical presentation of a system of (non-trivial) equations among arrows. The reason for excluding paths of length 1 in the above definition is that we want diagrams of the form in Figure 3.3 to say that $f;g = f;h$, without also saying that $g = h$.

For example, we can express the two equations which say that an $S$-sorted arrow $h : G \to G'$ is a graph homomorphism by the two commutative diagrams shown in Figure 3.4, in which $h_E$ and $h_N$ denote the Edge and Node components of $h$, respectively. Similarly, the three diagrams in Figure 3.5 express the conditions for a sorted arrow to be an automaton homomorphism (here $h, i, j$ denote the Input, State and Output components of the homomorphism, respectively).

We now give a more complex example (it can be skipped at first reading):

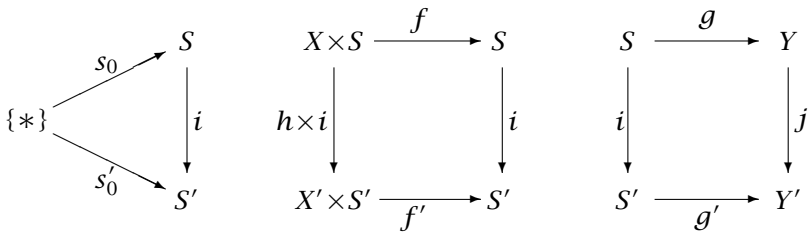

Figure 3.5: Commutative Diagrams for Automaton Homomorphism

**Example 3.6.3 ($\star$)** (*Tree of a Term*) Given a signature $\Sigma$, we will construct a "$\Sigma$-algebra of labelled graphs," some of whose elements will be the trees that represent $\Sigma$-terms. Let $S$ be the sort set of $\Sigma$, and recall that $|\Sigma| = \bigcup_{w,s} \Sigma_{w,s}$. Now define $\mathcal{G}_\Sigma$ to be the $\Sigma$-algebra where, for each $s \in S$, $\mathcal{G}_{\Sigma,s}$ is the set of all node labelled graphs $G = (E, N, L, \partial_0, \partial_1, l)$ having $E, N \subseteq \omega^*$ and $L = |\Sigma|$, with each $\sigma \in \Sigma_{[],s}$ interpreted as the graph $G_\sigma = (\emptyset, \{[\,]\}, L, \partial_0, \partial_1, l_\sigma)$ where $l_\sigma([\,]) = \sigma$, and each $\sigma \in \Sigma_{s_1...s_m,s}$ interpreted as the arrow which sends graphs $G_1, \ldots, G_m$ to the graph $G = (E, N, L, \partial_0, \partial_1, l)$ in which

- $N = \{[\,]\} \cup \bigcup_{i=1}^m i \cdot N_i$, where $\cdot$ denotes concatenation for lists of naturals, where $N_i$ is the node set of the graph $G_i$
- $E = N - \{[\,]\}$
- $\partial_0(i_1 \ldots i_n) = i_1 \ldots i_{n-1}$
- $\partial_1(i_1 \ldots i_n) = i_1 \ldots i_n$
- $l([\,]) = \sigma$ and
- $l(i_1 \ldots i_n) = l_{i_1}(i_2 \ldots i_n)$, where $n > 0$ and $l_k$ is the label function of $G_k$ for $k = 1, \ldots, m$.

Then the unique $\Sigma$-homomorphism $h : T_\Sigma \to \mathcal{G}_\Sigma$ sends each $\Sigma$-term to its $|\Sigma|$-labelled tree representation. □

**Exercise 3.6.3 ($\star$)** Show that the trees shown in Figure 2.5 actually do arise in the manner of Example 3.6.3 from the terms shown after Definition 2.6.1. □

Commutative diagrams are a well established proof technique in modern algebra, and are increasingly used in computing science as well. In this technique, a diagram represents a system of simultaneous equations (among compositions of arrows[5]), and geometrical operations on diagrams correspond to algebraic operations on systems of equations. One such operation is called "pasting," because geometrically it amounts to pasting commutative diagrams together, whereas algebraically it amounts to combining systems of equations.

For example, we can prove that the composition of two graph homomorphisms is a graph homomorphism by diagram pasting, rather than by calculation (as in Exercise 3.1.2): assume that we are given homomorphisms $h : G \to G'$ and $h' : G' \to G''$. Then Figure 3.4 shows the diagrams for $h$; those for $h'$ are similar, but with an additional $'$ symbol everywhere. For the operation symbol $\partial_0$, Figure 3.6 shows the two diagrams, let us call them $P_1$ and $P_2$, that we wish to paste together, with their common subdiagram $P_0$ and their union $P$. The fact that $P$ commutes then gives us that the rightmost diagram commutes, which is what we really want. (The case of $\partial_1$ is similar.)

---

[5]Later we will see how equations among other kinds of entities fall into the same framework.

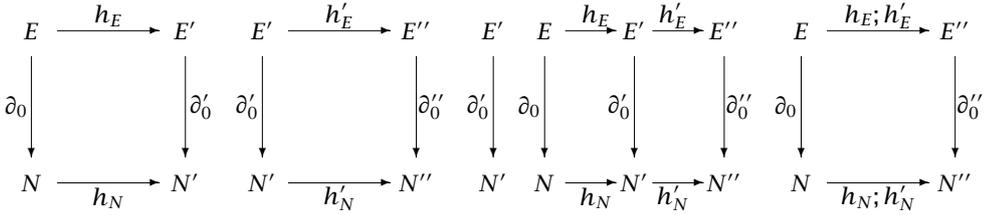

Figure 3.6: Commutative Diagrams for Graph Homomorphism Proof

It is easy to give a formal justification for this assertion using equational reasoning. The leftmost two squares represent the two equations

$$\partial_0; h_N = h_E; \partial_0'$$
$$\partial_0'; h_N' = h_E'; \partial_0''$$

and so we can prove commutativity of the square in which we are interested as follows:

$$\partial_0; (h_N; h_N') = (\partial_0; h_N); h_N' =$$
$$(h_E; \partial_0'); h_N' = h_E; (\partial_0'; h_N') =$$
$$h_E; (h_E'; \partial_0'') = (h_E; h_E'); \partial_0''.$$

Geometrically, this argument simply says that the functions along each outside path are equal to the function along the path through the central edge, namely $h_E; \partial_0'; h_N'$.

This argument is typical of those used to justify diagram pasting. It is also a typical (though rather simple) *diagram chase*. Usually such arguments are done geometrically, preferably on a black (or white) board, and are omitted in written documents.

**Example 3.6.4** Although pasting commutative diagrams works just as well with triangles, pentagons and other polygons as it does with squares, it is worth remarking that there are cases where the union of a collection of commutative diagrams is not commutative. Hence some caution must be observed. Consider, for example, the diagram in Figure 3.7, in which $\mathbb{N}$ is the natural numbers, $\mathbb{Z}$ is the integers, each edge labelled 1 is the identity on $\mathbb{N}$, each diagonal is the inclusion map $\mathbb{N} \to \mathbb{Z}$, and the four outer maps $(a, b, c, d)$ are arbitrary except that they restrict to the identity on $\mathbb{N}$. For example, we might choose the following, for $i \in \mathbb{Z}$,

$$a(i) = \begin{cases} i & \text{if } i \in \mathbb{N} \\ -1 & \text{if } i \notin \mathbb{N} \end{cases}$$

$$b(i) = \begin{cases} i & \text{if } i \in \mathbb{N} \\ -2 & \text{if } i \notin \mathbb{N} \end{cases}$$

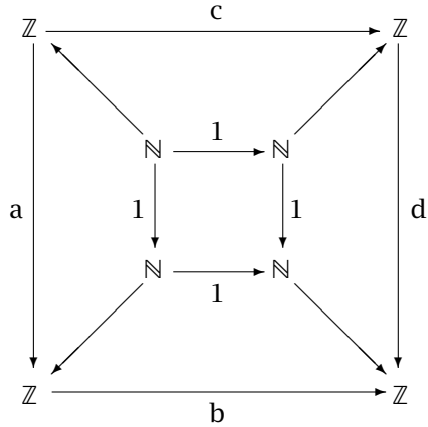

Figure 3.7: A Non-Commutative Diagram

$$c(i) = \begin{cases} i & \text{if } i \in \mathbb{N} \\ -3 & \text{if } i \notin \mathbb{N} \end{cases}$$

$$d(i) = \begin{cases} i & \text{if } i \in \mathbb{N} \\ -4 & \text{if } i \notin \mathbb{N} \end{cases}$$

Then for all $i \in \mathbb{Z}$, $(a;b)(i) = b(i)$ and $(c;d)(i) = d(i)$; so in particular, $(a;b)(-2) = -2$ while $(c;d)(-2) = -4$. So $a;b \neq c;d$. □

Because "diagram chasing" refers to arguments made using diagrams, diagram pasting may be considered a particular kind of diagram chasing. Another common form involves using initiality (or freeness) to argue that because there are two arrows between two nodes (with some property), they must be equal. Both are illustrated in the following elegant proof of the associativity of substitution:

**Proposition 3.6.5** (*Associativity of Substitution*) Given substitutions $a : W \to T_\Sigma(X)$, $b : X \to T_\Sigma(Y)$, $c : Y \to T_\Sigma(Z)$, then

$$(a;b);c = a;(b;c).$$

**Proof:** The assertion to be proved translates to $(a;\overline{b});\overline{c} = a;\overline{(b;\overline{c})}$, where ";" indicates composition of ordinary (many-sorted) arrows. By the usual associative law for such arrows, it suffices to show that

$$\overline{b};\overline{c} = \overline{(b;\overline{c})}.$$

By the uniqueness condition of Proposition 3.5.1, the above equation will follow from showing that $\overline{b};\overline{c}$ is a Σ-homomorphism extending $b;\overline{c}$.

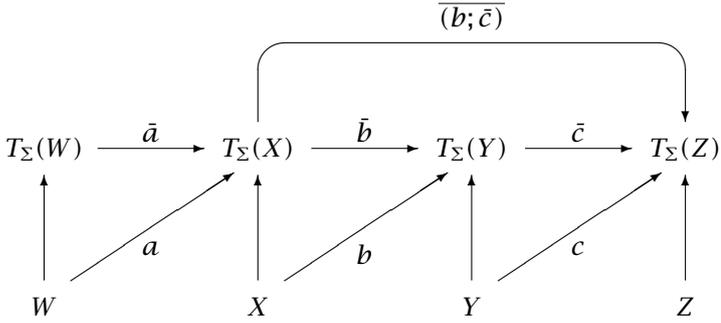

Figure 3.8: Associativity of Substitution Proof Diagram

If we let $i : X \to T_\Sigma(X)$ denote the injection, then what we have to show is that

$$i; (\overline{b}; \overline{c}) = b; \overline{c} \;.$$

But this follows from $i; \overline{b} = b$, which is just commutativity of the middle bottom triangle. □

This proof looks much simpler and more elegant if done by chasing the right hand two thirds of the diagram in Figure 3.8 on a white or blackboard. By contrast, to prove the result by direct manipulation of the set-theoretic representation of terms would require several pages of very tedious calculation. It is worth drawing out the following key element of this in the proof, because it is needed later on:

**Corollary 3.6.6** Given substitutions $a : W \to T_\Sigma(X)$, $b : X \to T_\Sigma(Y)$, then $\overline{a}; \overline{b} = \overline{(a; \overline{b})}$. □

## 3.7 (⋆) Parse

Because a signature $\Sigma$ can be overloaded, $\Sigma$-terms can also be overloaded, and it is useful to characterize when this can happen.

**Definition 3.7.1** An $S$-sorted signature $\Sigma$ is **regular** iff $\sigma \in \Sigma_{w,s} \cap \Sigma_{w,s'}$ implies $s = s'$. A $\Sigma$-term $t$ is **overloaded** iff there are distinct sorts $s, s' \in S$ such that $t \in T_{\Sigma,s} \cap T_{\Sigma,s'}$. □

Notice that regularity implies in particular that all constant symbols are distinct.

**Proposition 3.7.2** A signature $\Sigma$ is regular iff there are no overloaded $\Sigma$-terms.

**Proof:** Assume that $\Sigma$ is regular. By induction on the depth of terms, we will show that $\Sigma$-terms $t, t'$ of depths $\leq d$ with distinct sorts $s, s'$ must be different. For $d = 0$, suppose that $t = \sigma$ and $t' = \sigma'$; then $w = w' = [\,]$ and the result follows directly from regularity. For depth $d > 0$, suppose that $t = \sigma(t_1, \ldots, t_m)$ and $t' = \sigma'(t'_1, \ldots, t'_n)$. Because $\Sigma$-terms are lists, and lists have unique factorizations, $t = t'$ implies $\sigma = \sigma'$, $m = n$ and $t_i = t'_i$ for $i = 1, \ldots, n$. Now the induction hypothesis implies that $t_i$ and $t'_i$ have the same sort for $i = 1, \ldots, n$. Therefore $w = w'$, and hence regularity gives $s = s'$.

Conversely,[E7] suppose that $t = \sigma(t_1, \ldots, t_n) \in T_{\Sigma,s} \cap T_{\Sigma,s'}$ is a *minimal overloading*, in the sense that $s \neq s'$ and none of the $t_i$ are overloaded. Then necessarily $\sigma \in \Sigma_{w,s} \cap \Sigma_{w,s'}$ where $w = s_1 \ldots s_n$ and $s_i$ is the sort of $t_i$ (for $i = 1, \ldots, n$). Thus $\Sigma$ is not regular. □

This result does not consider ambiguities due to mixfix syntax, because it only uses the prefix-with-parenthesis syntax. These more elaborate kinds of ambiguities are considered below.

**Exercise 3.7.1** Give a simple non-regular signature $\Sigma$ and a simple overloaded $\Sigma$-term. □

We now generalize the definitions of signature and term to the case of mixfix syntax:

**Definition 3.7.3** Let $\mathcal{A}$ be some fixed set of characters that does not include the underbar character "_" or the three special symbols ·, (, and ). Then a **form** is a list in $(\mathcal{A} \cup \{\_\})^*$, and the **arity** of a form is the number of _'s that occur in it. A (many-sorted) **mixfix signature** is an indexed family $\{\Sigma_{w,s} \mid w \in S^*, s \in S\}$ for some set $S$ of **sorts**, where each $\Sigma_{w,s}$ is a set of forms of arity $\#w$. □

**Example 3.7.4** Let $\mathcal{A} = \{a, b, +\}$. Then the following are all forms:

$$a,\ a\_,\ \_a,\ a\_b,\ \_+\_,\ \_\_\ .$$

The first has arity 0, the next three have arity 1, and the last two have arity 2. The first defines syntax for a constant, while the second through last (respectively) define syntax for prefix, postfix, outfix, infix, and juxtaposition operations. □

We now give a recursive construction for mixfix $\Sigma$-terms:

**Definition 3.7.5** If $\Sigma$ is an $S$-sorted mixfix signature disjoint from $S$, then the $S$-sorted set[E8] $M_\Sigma$ of all **mixfix (ground) $\Sigma$-terms** is the smallest set of lists over the set $\mathcal{A} \cup \{\cdot, (,)\}$ such that

(0) if $f \in \Sigma_{[\,],s}$ then $f \cdot s \in M_{\Sigma,s}$ for all $s \in S$, and

(1) if $f \in \Sigma_{s_1 \ldots s_n, s}$ for $n > 0$ and $t_i \in M_{\Sigma, s_i}$ for $i = 1, \ldots, n$ then

$$(k_1 t_1 k_2 \ldots k_n t_n k_{n+1}) \cdot s \in M_{\Sigma,s}\ ,$$

where $f = k_1 \_ k_2 \_ \ldots k_n \_ k_{n+1}$ (note that some of the $k_i$ may be the empty list).

As with terms in $\overline{T}_\Sigma$, we will usually omit the sort annotation unless it is necessary. □

Every mixfix signature $\Sigma$ is also an ordinary signature. However, the $\Sigma$-terms will look rather different in the two cases. For example, if $\_+\_ \in \Sigma_{ss,s}$ and $a, b \in \Sigma_{[],s}$, then $a + b$ is in $M_\Sigma$ but not in $T_\Sigma$, whereas $\_+\_(a, b)$ is in $T_\Sigma$ but not in $M_\Sigma$. If extra clarity is needed, we will let $\underline{\Sigma}$ denote the ordinary signature corresponding to a mixfix signature $\Sigma$.

**Definition 3.7.6** Given a mixfix signature $\Sigma$ disjoint from $S$, we can give $M_\Sigma$ the structure of a $\underline{\Sigma}$-algebra in the following way:

(0) interpret $f \in \Sigma_{[],s}$ in $M_\Sigma$ as the singleton list $f \cdot s$, and

(1) interpret $f \in \Sigma_{s_1...s_n,s}$ with $n > 0$ in $M_\Sigma$ as the function sending $t_1, \ldots, t_n$ to the list $(k_1 t_1 k_2 \ldots k_n t_n k_{n+1}) \cdot s$, where $t_i \in M_{\Sigma,s_i}$ for $i = 1, \ldots, n$ and $f = k_1\_k_2\_\ldots k_n\_k_{n+1}$.

Thus, we have that $(M_\Sigma)_f(t_1, \ldots, t_n) = (k_1 t_1 k_2 \ldots k_n t_n k_{n+1}) \cdot s$, although it will usually be written $k_1 t_1 k_2 \ldots k_n t_n k_{n+1}$. □

It follows that there is a unique $\Sigma$-homomorphism[6] $T_\Sigma \to M_\Sigma$; the following uses this fact in comparing the two kinds of $\Sigma$-terms:

**Definition 3.7.7** A mixfix signature $\Sigma$ is **sort ambiguous** iff the carriers of $M_\Sigma$ are non-disjoint, and is **mixfix ambiguous** iff the unique homomorphism $h : T_{\underline{\Sigma}} \to M_\Sigma$ is non-injective. Given $m \in M_\Sigma$, if $h(t) = m$ then $t$ is said to be a **parse** of $m$. □

**Exercise 3.7.2** . Suppose that a mixfix signature $\Sigma$ has a single sort $A$, and also has

$$\Sigma_{[],A} = \{a\}, \text{ and}$$
$$\Sigma_{A,A} = \{a\_, a\_a, \_a\},$$

with all other $\Sigma_{w,s} = \emptyset$. Then

1. Show that $\Sigma$ is mixfix ambiguous; in particular, show that $aaa$ has five distinct parses, and write each one out.

2. How many parses does $aaaa$ have?

3. Noting that $\Sigma$ is not sort ambiguous, construct an ordinary many-sorted signature that is sort ambiguous. □

**Exercise 3.7.3** Show that a term over a mixfix signature is mixfix ambiguous iff it has at least two distinct parses. □

---

[6]As usual, $T_\Sigma$ here really means $\overline{T}_\Sigma$.

## 3.8 Literature

Relatively few books develop many-sorted general algebra in any depth. Bergstra *et al.* [9], Ehrig and Mahr [45] and van Horenbeck [181] each develop a certain amount for the algebraic specification theory which is their main concern. I am not aware of any mathematics text that develops many-sorted general algebra in any detail. The notation and approach of this chapter continues that of the previous chapter, following ideas from [52] as further developed in [137, 78] and other publications.

Initial algebra semantics (as discussed in Section 3.2) originated in [52], and was further developed in [88]. It can be seen as an algebraic formulation of the so-called attribute semantics of Knuth [116].

It is common to treat both variables and substitutions either intuitively, or else with extreme logical formalism; the approach given here tries to find a middle ground. Our Theorem of Constants (Theorem 3.3.11) is analogous to a well-known result in first-order logic. This result is not usually treated in the computing or the general algebra literature, although it is not difficult, and it plays an important role in justifying proofs by term rewriting.

It is known that conditional equations have more expressive power than unconditional equations, in the sense that there are algebras that are initial models of a specification having conditional equations that are not initial models of any specification having only unconditional equations [176].

The proof of associativity for substitution (Theorem 3.6.5) follows [89], and is the same as the more abstract proof which a category theorist would call "associativity of composition in a Kleisli category" (the necessary concepts are beyond the scope of this book, but may be found, for example, in [126]). Structural induction was introduced to computer science by Burstall [21] in 1969.

The formulation of theory equivalence at the end of Section 3.3 is a more concrete and many-sorted version of Lawvere's category-theoretic formulation for theories [121]; see [130] and [5] for further information on Lawvere theories, and see Section 4.10 for some techniques for proving equivalence.

The emphasis on models and satisfaction in this and the previous chapter (as well as in subsequent chapters) was influenced by the theory of institutions [67], which axiomatizes the notion of "logical system" using satisfaction. But as Wittgenstein is said to have remarked,

> *Is a proof not also part of an institution?*

and as Thomas Jefferson said on July 12, 1816,

> *Laws and institutions must go hand in hand with the progress of the human mind.*

And indeed, laws and proofs play a major role in the rest of this book, as they must in any study of theorem proving.

---

**A Note for Lecturers:** Emphasizing the "microprocessor" interpretation of initiality in the discussion after Theorem 3.2.1 can considerably sharpen students' intuitions, and some concrete examples with drawings can strengthen this process.

The proof of Proposition 3.6.5 should be done as a live diagram chase on the board; this is a lot of fun, and it is also the best way to bring out the essential simplicity of this proof.

---

# 4 Equational Deduction

This chapter considers how to correctly deduce new equations from old ones. We give several finite sets of rules for equational deduction that are both *sound*, i.e., truth preserving, and *complete* for loose semantics, in the sense that every equation that is true in all models of a given set of equations can be deduced from that set using these rules. Such results are important because they say that we can find out what is true by using formal, finitary manipulations of finite syntactic objects, whereas the semantic definition of truth (by satisfaction, Definition 3.3.7 of the previous chapter) in general requires examining infinite sets of infinite objects (since algebras in general have infinite carriers); obviously, such an examination cannot be done on any real computer, which has only a finite amount of memory. Nonetheless, satisfaction remains fundamental, because it provides the standard of correctness for deduction. Equational deduction also has many important applications in computer science and elsewhere (see Section 4.12).

## 4.1 Rules of Deduction

Equational deduction is reasoning with just the properties of equality. Basic properties of equality include the following:

(1) Anything is equal to itself; this is the *reflexivity* of equality.

(2) If $t$ equals $t'$, then $t'$ equals $t$; this is the *symmetry* of equality.

(3) If $t$ equals $t'$ and $t'$ equals $t''$, then $t$ equals $t''$; this is the *transitivity* of equality.

(4) If $t_1$ equals $t'_1$ and $t_2$ equals $t'_2$ ... and $t_n$ equals $t'_n$, and if $t$ has variables $x_1, \ldots, x_n$, then the result of substituting $t_i$ for $x_i$ in $t$ equals the result of substituting $t'_i$ for $x_i$ in $t$; this is called the *congruence* property of equality, and may be paraphrased as saying that substituting equal expressions into the same expression yields equal expressions.

(5) If $t$ equals $t'$ where $t$ and $t'$ involve variables $x_1, \ldots, x_n$, and if $t_1, \ldots, t_n$ are terms, then the result of substituting $t_i$ for $x_i$ in $t$ equals the result of substituting $t_i$ for $x_i$ in $t'$; this is called the *substitutivity* (or *instantiation*) property of equality, and may be paraphrased as saying that any substitution instance of an equation is an equation.

In these properties, the various $t$'s are terms, which may involve variables; furthermore, both (4) and (5) involve substituting terms for variables. We will see that it is necessary to keep careful track of variables during equational deduction, or else soundness can be lost. This motivates the following:

**Notation 4.1.1** We will use the notation of Definition 3.3.1 for equations that appear in deduction, writing $(\forall X)\ t = t'$, where all variables in $t$ and $t'$ are taken from $X$. Also, we will write $\theta : X \to T_\Sigma(Y)$ for a substitution of terms from $T_\Sigma(Y)$ for variables in $X$, as in Definition 3.5.2. Finally, we adopt the convention that if $t \in T_\Sigma(X)$ is a $\Sigma$-term with variables in $X$, then the result of substituting $\theta(x)$ for each $x$ in $X$ into $t$ may be written $\theta(t)$, rather than $\overline{\theta}(t)$ as in Definition 3.5.2.  □

The simple example below reviews notation and concepts, in preparation for more complex material to come.

**Example 4.1.2** Suppose there is just one sort, say Elt, and that $X$ has three variables of that sort, say $x, y, z$. Let $Y = \{x, w\}$ and define $\theta : X \to T_\Sigma(Y)$ by $\theta(x) = x^{-1-1}$, $\theta(y) = w^{-1}$, $\theta(z) = x * x^{-1}$. Now if $t = (x * y)^{-1}$, then $\theta(t) = (x^{-1-1} * w^{-1})^{-1}$.  □

We can now give the following formal versions of the above properties of equality:

**Definition 4.1.3** Given a signature $\Sigma$ and a set $A$ of $\Sigma$-equations, called the **axioms** or **assumptions**, the following **rules of deduction** define the $\Sigma$-equations that are **deducible** (or **provable** or **inferable**) (**from** $A$):

(0) *Assumption.* Each equation in $A$ is deducible.

(1) *Reflexivity.* Each equation of the form
$$(\forall X)\ t = t$$
is deducible.

(2) *Symmetry.* If
$$(\forall X)\ t = t'$$
is deducible, then so is
$$(\forall X)\ t' = t.$$

(3) *Transitivity.* If the equations
$$(\forall X)\ t = t',\ (\forall X)\ t' = t''$$
are deducible, then so is
$$(\forall X)\ t = t''.$$

(4) *Congruence.* If $\theta, \theta' : Y \to T_\Sigma(X)$ are substitutions such that for each $y \in Y$, the equation
$$(\forall X)\ \theta(y) = \theta'(y)$$
is deducible, then given any $t \in T_\Sigma(Y)$, the equation
$$(\forall X)\ \theta(t) = \theta'(t)$$
is also deducible.

(5) *Instantiation.* If
$$(\forall Y)\ t = t'$$
is in $A$, and if $\theta : Y \to T_\Sigma(X)$ is a substitution, then the equation
$$(\forall X)\ \theta(t) = \theta(t')$$
is deducible.  □

The next section will give a formal definition of equational deduction using these rules, and will define a relation $A \vdash e$ indicating that $e$ can be deduced from $A$. But first, we illustrate what it is that we wish to formalize:

**Example 4.1.4** (*Left Groups*) Suppose we want to prove that the right inverse law

$$(\forall x)\ x * x^{-1} = e$$

holds in the specification GROUPL of Example 3.3.4, which is reproduced below, except that the variables A, B, and C are used instead of X, Y, and Z, and a **precedence declaration** has been added for the inverse operation -1.

Precedence provides a way to declare that some operation symbols are "stronger" or "more binding" than others. For example, the usual conventions for mathematical notation assume that $x * x^{-1}$ means $x * (x^{-1})$ rather than $(x * x)^{-1}$, because $^{-1}$ binds more tightly than $*$. In OBJ3, precedence is defined by giving a natural number p as an "attribute" of an operation, in the form [prec p] following the operation's sort. Lower precedence means tighter binding, and a binary infix operation symbol like * has a default precedence of 41. (We do not give a formal treatment of precedence in this text, although it is not difficult to do so; see [90] for further discussion.)

```
th GROUPL is sort Elt .
  op _*_ : Elt Elt -> Elt .
  op e : -> Elt .
  op _-1 : Elt -> Elt [prec 2] .
  var A B C : Elt .
  eq e * A = A .
  eq A -1 * A = e .
  eq A *(B * C) = (A * B)* C .
endth
```

Let G.1, G.2 and G.3 denote the three equations in the specification GROUPL above, in their order of appearance.

As an illustration, let us apply rule (4) to the equation

$$(\forall x)\, e = (x^{-1-1} * x^{-1})\,,$$

with $X = \{x\}$ and $Y = \{z, x\}$ and $t = z * (x * x^{-1})$. Then we want $\theta(z) = e$ and $\theta'(z) = x^{-1-1} * x^{-1}$, while $\theta(x) = \theta'(x) = x$, so that the result is the equation

$$(\forall x)\, e * (x * x^{-1}) = (x^{-1-1} * x^{-1}) * (x * x^{-1})\,.$$

The following is a deduction for the right identity law, using the rules of Definition 4.1.3. In this display, some of the deduced equations are named by bracketed numbers given to their left. Also, each step of deduction is indicated by a long horizontal line, to the right of which the rule used is indicated, together with the names of any equations used (in addition to the one above the line) after the word "on".

[1] $\dfrac{\overline{(\forall x)\, e * (x * x^{-1}) = (x * x^{-1})}}{(\forall x)\, (x * x^{-1}) = e * (x * x^{-1})}$ (5) on G.1
(2)

[2] $\dfrac{\overline{(\forall x)\, (x^{-1-1} * x^{-1}) = e}}{(\forall x)\, e = (x^{-1-1} * x^{-1})}$ (5) on G.2
(2)

[3] $\dfrac{\overline{(\forall x)\, e * (x * x^{-1}) = (x^{-1-1} * x^{-1}) * (x * x^{-1})}}{(\forall x)\, (x * x^{-1}) = (x^{-1-1} * x^{-1}) * (x * x^{-1})}$ (4) on [2] with $t = z * (x * x^{-1})$
(3) on [1]

[4] $\dfrac{\overline{(\forall x)\, (x^{-1-1} * x^{-1}) * (x * x^{-1}) = ((x^{-1-1} * x^{-1}) * x) * x^{-1}}}{(\forall x)\, (x * x^{-1}) = ((x^{-1-1} * x^{-1}) * x) * x^{-1}}$ (5) on G.3
(3) on [3]

[5] $\dfrac{\overline{(\forall x)\, x^{-1-1} * (x^{-1} * x) = (x^{-1-1} * x^{-1}) * x}}{\dfrac{(\forall x)\, (x^{-1-1} * x^{-1}) * x = x^{-1-1} * (x^{-1} * x)}{\dfrac{(\forall x)\, ((x^{-1-1} * x^{-1}) * x) * x^{-1} = (x^{-1-1} * (x^{-1} * x)) * x^{-1}}{(\forall x)\, (x * x^{-1}) = (x^{-1-1} * (x^{-1} * x)) * x^{-1}}}}$ (5) on G.3
(2)
(4) with $t = z * x^{-1}$
(3) on [4]

[6] $\dfrac{\overline{(\forall x)\, (x^{-1} * x) = e}}{\dfrac{(\forall x)\, (x^{-1-1} * (x^{-1} * x)) * x^{-1} = (x^{-1-1} * e) * x^{-1}}{(\forall x)\, (x * x^{-1}) = (x^{-1-1} * e) * x^{-1}}}$ (5) on G.2
(4) with $t = (x^{-1-1} * z) * x^{-1}$
(3) on [5]

[7] $\dfrac{\overline{(\forall x)\, x^{-1-1} * (e * x^{-1}) = (x^{-1-1} * e) * x^{-1}}}{\dfrac{(\forall x)\, (x^{-1-1} * e) * x^{-1} = x^{-1-1} * (e * x^{-1})}{(\forall x)\, (x * x^{-1}) = x^{-1-1} * (e * x^{-1})}}$ (5) on G.3
(2)
(3) on [6]

[8] $\dfrac{\overline{(\forall x)\, e * x^{-1} = x^{-1}}}{\dfrac{(\forall x)\, x^{-1-1} * (e * x^{-1}) = x^{-1-1} * x^{-1}}{(\forall x)\, (x * x^{-1}) = x^{-1-1} * x^{-1}}}$ (5) on G.1
(4) with $t = x^{-1-1} * z$
(3) on [7]

[9] $\dfrac{\overline{(\forall x)\, (x^{-1-1} * x^{-1}) = e}}{(\forall x)\, (x * x^{-1}) = e}$ (5) on G.2
(3) on [8]

   This rather tedious proof is illustrated in the "proof tree" shown in Figure 4.1, in which each arrow indicates the application of the rule of deduction with which it is labelled (the last step is omitted). Section 4.5 will give a more powerful rule that will allow us to give a much easier proof of this result.  □

## 4.2  Equational Proof

The rules of equational deduction are used to *prove* or *deduce* a new equation $e$ from a given set $A$ of equations, by repeatedly applying the rules (0–5) to previously deduced equations. We let $A \vdash e$ mean that $e$ is deducible from $A$. A *proof* for the assertion $A \vdash e$ is a sequence of rule applications that really proves $e$ from $A$. A fully annotated proof provides all of the information that is used in each step of deduction, including the name of the rule involved, and any substitutions that are used. We formalize this as follows:

**Definition 4.2.1** Given a signature $\Sigma$ and a set $A$ of $\Sigma$-equations, a **(bare) proof** from $A$ is a sequence $e_1, \ldots, e_n$ of $\Sigma$-equations where each $e_i$ is deducible from $A \cup \{e_1, \ldots, e_{i-1}\}$ by a single application of a single rule; then we say that $e_1, \ldots, e_n$ is a **(bare) proof of $e_n$ from** $A$. A **(fully) annotated proof** is a sequence $a_1, \ldots, a_n$, where each $a_i$ has one of the following forms:

(0) $\langle e_i, (0) \rangle$ where $e_i \in A$.

(1) $\langle e_i, (1) \rangle$ where $e_i$ is of the form
$$(\forall X)\ t = t.$$

(2) $\langle e_i, e_j, (2) \rangle$ where $j < i$ and $e_i$ is of the form
$$(\forall X)\ t = t'$$
and $e_j$ is of the form
$$(\forall X)\ t' = t.$$

(3) $\langle e_i, e_j, e_k, (3) \rangle$ where $j, k < i$, and $e_i$ is of the form
$$(\forall X)\ t = t''$$
and $e_j, e_k$ are of the forms
$$(\forall X)\ t = t',\ (\forall X)\ t' = t''$$
respectively.

(4) $\langle e_i, \theta, \theta', \varphi, t, (4) \rangle$ where $t \in T_\Sigma(Y)$, where $\varphi : Y \to \omega$, and where $\theta, \theta' : Y \to T_\Sigma(X)$ are substitutions such that for each $y \in Y$, each equation
$$(\forall X)\ \theta(y) = \theta'(y)$$
is some $e_{\varphi(y)}$ where $\varphi(y) < i$, and $e_i$ is of the form
$$(\forall X)\ \theta(t) = \theta'(t).$$

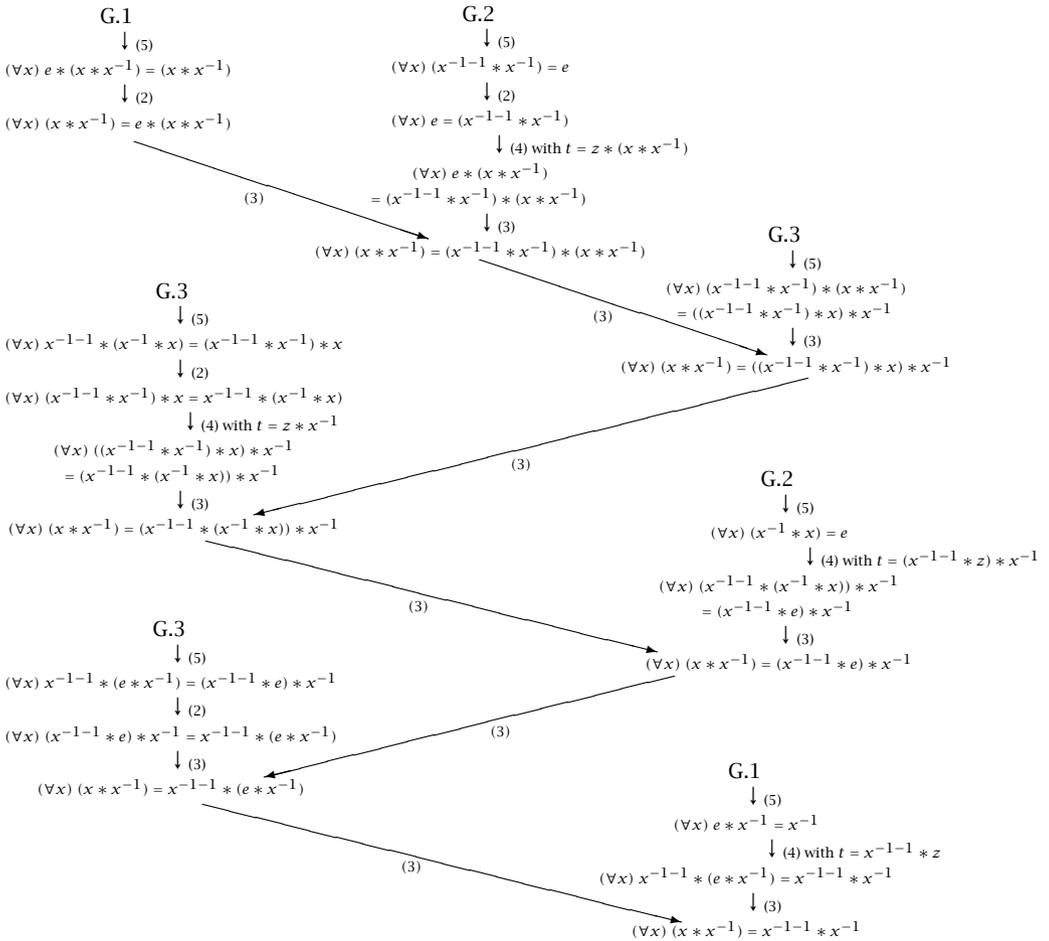

Figure 4.1: A Proof Tree

(5) $\langle e_i, \theta, e, t, t', (5) \rangle$ where $t, t' \in T_\Sigma(Y)$, and where $\theta : Y \to T_\Sigma(X)$ is a substitution such that $e \in A$ has the form
$$(\forall Y)\ t = t'$$
and $e_i$ is of the form
$$(\forall X)\ \theta(t) = \theta(t').$$

Let $A \vdash_\Sigma e$, or usually $A \vdash e$ when $\Sigma$ is clear from context, indicate that there is a proof of $e$ from $A$. We will also use notations like "$A \vdash^{(0-4)} e$" or "$A \vdash^{(1-3,5)} e$" to indicate that $e$ is deducible using only the rules (0-4), or (1-3) plus (5), respectively; by default, $A \vdash e$ will mean $A \vdash^{(0-5)} e$. Also, we let $\overline{A}$ denote the set of all equations deducible from $A$ using rules (0-5), and call it the **deductive closure** or **theory** of $A$.    □

Notice that if $a_1, \ldots, a_n$ is a fully annotated proof, then the sequence $e_1, \ldots, e_n$ of the first components of the $a_i$ is a bare proof.

We now illustrate these concepts by proving that any equation that can be deduced using (0) can also be deduced using (5). First, suppose that $(\forall X)\ t = t'$ is in $A$, let $Y = X$, and define $\theta : X \to T_\Sigma(X)$ by $\theta(x) = x$ for all $x \in X$. Then (by Exercise 3.5.2) $\theta(t) = t$ and $\theta(t') = t'$, so (5) tells us that $(\forall X)\ t = t'$ is deducible, as desired. The following is a formal statement of what we have just shown:

**Fact 4.2.2** Given a set $A$ of $\Sigma$-equations and a $\Sigma$-equation $e$, then $e \in A$ implies $A \vdash^{(5)} e$.    □

From this we get the following:

**Fact 4.2.3** Given a set $A$ of $\Sigma$-equations, then for any $\Sigma$-equation $e$,
$$A \vdash^{(0-5)} e \ \text{iff} \ A \vdash^{(1-5)} e\ .$$

**Proof:** Let $P$ be a proof of $e$ from $A$ using (0-5). Then for each use of the rule (0) in $P$, substitute the corresponding use of (5) according to Fact 4.2.2, resulting in a proof $P'$ of $e$ from $A$ that does not use rule (0).    □

Thus we can get by with a set of five rules, instead of six. We will see later on that there are many other rule sets for equational deduction, and that the number of rules can be further reduced. A more abstract formulation of deduction is given in (the optional) Section 4.11.

## 4.3  Soundness and Counterexamples

This section shows that equational deduction is *sound*, in the sense that if we can deduce $e$ from $A$, then $e$ is true in all models of $A$. To this end, the following lemmas show the soundness of each rule separately:

**Lemma 4.3.1** Given any $\Sigma$-algebra $M$ such that $M \models A$, and given $e \in A$, then $M \models e$.

**Proof:** This is immediate from the definition of satisfaction. □

**Lemma 4.3.2** Given any $\Sigma$-algebra $M$ and given $t \in T_\Sigma(X)$, then $M \models (\forall X)\ t = t$.

**Proof:** Let $a : X \to M$. Then certainly $\bar{a}(t) = \bar{a}(t)$. □

**Lemma 4.3.3** Given any $\Sigma$-algebra $M$ and given $t, t' \in T_\Sigma(X)$, then $M \models (\forall X)\ t = t'$ implies $M \models (\forall X)\ t' = t$.

**Proof:** Let $a : X \to M$. Then $\bar{a}(t) = \bar{a}(t')$ implies $\bar{a}(t') = \bar{a}(t)$. □

**Lemma 4.3.4** Given any $\Sigma$-algebra $M$ and given $t, t', t'' \in T_\Sigma(X)$, then $M \models (\forall X)\ t = t'$ and $M \models (\forall X)\ t' = t''$ imply $M \models (\forall X)\ t = t''$.

**Proof:** Let $a : X \to M$. Then $\bar{a}(t) = \bar{a}(t')$ and $\bar{a}(t') = \bar{a}(t'')$ imply $\bar{a}(t) = \bar{a}(t'')$. □

**Lemma 4.3.5** Given any $\Sigma$-algebra $M$, given $t \in T_\Sigma(Y)$, and given $\theta, \theta' : Y \to T_\Sigma(X)$ such that $M \models (\forall X)\ \theta(y) = \theta'(y)$ for each $y \in Y$, then $M \models (\forall X)\ \theta(t) = \theta'(t)$.

**Proof:** Let $a : X \to M$. Then $\bar{a}(\theta(y)) = \bar{a}(\theta'(y))$, i.e., $\theta;\bar{a}(y) = \theta';\bar{a}(y)$ for each $y \in Y$. But now the freeness of $T_\Sigma(Y)$ implies that $\theta;\bar{a}(t) = \theta';\bar{a}(t)$, i.e., that $\bar{a}(\theta(t)) = \bar{a}(\theta'(t))$. □

**Lemma 4.3.6** Given any $\Sigma$-algebra $M$, given $t, t' \in T_\Sigma(Y)$ such that $M \models (\forall Y)\ t = t'$, and given $\theta : Y \to T_\Sigma(X)$, then $M \models (\forall X)\ \theta(t) = \theta(t')$.

**Proof:** Let $a : X \to M$. Then $\theta;\bar{a} : Y \to M$, and so $M \models (\forall Y)\ t = t'$ implies $(\theta;\bar{a})(t) = (\theta;\bar{a})(t')$, i.e., $\bar{a}(\theta(t)) = \bar{a}(\theta'(t'))$, i.e., $M \models (\forall X)\ \theta(t) = \theta(t')$. □

**Exercise 4.3.1** The notation used in the proof above conceals a use of Corollary 3.6.6. Identify the gap and show how to fill it using this result. □

We can now use induction on proof length to show soundness of equational deduction:

**Proposition 4.3.7** (*Soundness*) Given a set $A$ of $\Sigma$-equations, a $\Sigma$-equation $e$, and a $\Sigma$-algebra $M$, then $M \models A$ and $A \vdash^{(0-5)} e$ imply $M \models e$.

**Proof:** Let $M$ be a $\Sigma$-algebra such that $M \models A$.

If $e$ has a proof of length 1 from $A$, then $e$ is derived using exactly one instance of exactly one of the rules (0–5); then Lemmas 4.3.1–4.3.6 show that $M \models e$ for each of these six cases, thus concluding the base of the induction.

For the inductive step, assume that if $e$ has a proof of length $n$ then $M \models e$, and let $e'$ have a proof $e_1, \ldots, e_{n+1}$ of length $n+1$. The inductive hypothesis gives us that $M \models e_i$, for $i = 1, \ldots, n$, and from this we can conclude that $M \models e_{n+1}$ by applying one of Lemmas 4.3.1–4.3.6. □

### 4.3.1 What is a Counterexample?

Suppose we are given a set $A$ of axioms and an equation $e$, all over the same signature $\Sigma$, and we want to prove that $e$ *cannot* be deduced from $A$. (For example, we may have put some effort into proving $A \vdash e$ without success, and now suspect that it is not possible.) The impossibility of giving a proof can be demonstrated by giving a *counterexample*, which is a $\Sigma$-algebra $M$ that satisfies $A$ but does not satisfy $e$. The proof that counterexamples work only depends on the *soundness* of deduction, because if $A \vdash e$ then for any $\Sigma$-algebra $M$, if $M \models A$ then $M \models e$. Therefore if $M \models A$ but $M \models e$ is false, we cannot have $A \vdash e$.

In order to show that $M \models e$ is false, we need only give a single assignment where $e$ fails: if $e$ is $(\forall X)\ t = t'$, then we need only exhibit $\theta : X \to M$ such that $\theta(t) \neq \theta(t')$. Thus, the way to show that an equation cannot be proved is to give an algebra $M$, an assignment $\theta$ into that algebra, and a proof that the assignment has different values on the two terms of the equation. We will use this in the next subsection and elsewhere.

### 4.3.2 The Need for Quantifiers

The most common formulations of equation and equational deduction do not involve explicit universal quantifiers for variables. However, we will show that explicit quantifiers[E9] are necessary for an adequate treatment of satisfaction. Our demonstration will use the following specification:

```
th FOO is sorts B A .
  ops T F : -> B .
  ops (_∨_) (_&_) : B B -> B .
  op ¬_ : B -> B [prec 2] .
  op foo : A -> B .
  var B : B . var A : A .
  eq B ∨ ¬ B = T .
  eq B & ¬ B = F .
  eq B ∨ B = B .
  eq B & B = B .
  eq ¬ F =  T .
  eq ¬ T =  F .
  eq ¬ foo(A) = foo(A) .
endth
```

The OBJ3 keyword `ops` allows two or more operation symbols having the same rank to be declared together; for non-constant operation symbols, parentheses must be used to separate the different operation forms. The notation T, F, ∨, &, ¬, and the first four equations should be familiar from Boolean algebra, and we can think of foo as a kind of "test" on elements of sort A. This example therefore resembles specifications found in many applications, except perhaps for the last equation.

Now consider the $\Sigma^{\text{FOO}}$-algebra $I$ with $I_A = \emptyset$ and $I_B = \{T, F\}$, where T, F are distinct, and where &, $\vee$, $\neg$ are interpreted as expected for the booleans (F $\vee$ F = F, etc.), and where foo is the empty function. (This is actually the *initial* $\Sigma^{\text{FOO}}$-algebra.) It is easy to check that $I$ satisfies the equation $(\forall x)$ F = T where $x$ is of sort A, and that $I$ does not satisfy the equation $(\forall \emptyset)$ F = T. Since these two equations have different meanings, they cannot be identified, and therefore the quantifier really is necessary.

Example 4.3.8 below will show that with unsorted equational deduction, the unquantified equation F = T can be proved from the equations in FOO, from which, given the above discussion of $I$, it follows that unsorted equational deduction is in general *not sound*. This refutes the apparently common misconception that unsorted and many-sorted equational deduction are equivalent; see [78] for a detailed discussion of this issue.

To make our discussion precise, we need an explicit formulation of unsorted equational deduction. Recall that a $\Sigma$-equation consists of a ground signature $X$ disjoint from $\Sigma$, plus two terms $t, t' \in T_{\Sigma(X),s}$ for some sort $s$; that is, a $\Sigma$-equation is a triple $\langle X, t, t' \rangle$, by convention written $(\forall X)\ t = t'$. By contrast, equations in **unsorted equational logic** do not have explicit quantifiers; they are just pairs $\langle t, t' \rangle$, conventionally written in the form $t = t'$. The **unsorted rules of deduction** are exactly the same as the many-sorted rules (1–5) of Definition 4.1.3 except that all quantifiers (e.g., $(\forall X)$ and $(\forall Y)$) are omitted.

**Example 4.3.8** (*An Unsound Deduction*) We will show that unsorted equational deduction can prove an equation that is untrue in some models of the specification FOO above. We apply the unsorted versions of the rules (1–5), letting F.1, ...,F.7 denote the equations in FOO in the order of their appearance, and letting $x$ be a new variable symbol:

$$[0] \quad \cfrac{\overline{\neg\mathsf{foo}(x) = \mathsf{foo}(x)}\ \ (5)\text{ on F.7}}{\mathsf{foo}(x) = \neg\mathsf{foo}(x)}\ (2)$$

$$[1] \quad \cfrac{\cfrac{\cfrac{\cfrac{\cfrac{\overline{\mathsf{foo}(x) \vee \neg\mathsf{foo}(x) = \mathsf{T}}\ \ (5)\text{ on F.1}}{\mathsf{foo}(x) \vee \mathsf{foo}(x) = \mathsf{foo}(x) \vee \neg\mathsf{foo}(x)}\ \ (4)\text{ on }[0]\text{ with }t = \mathsf{foo}(x) \vee z}{\mathsf{foo}(x) \vee \mathsf{foo}(x) = \mathsf{T}}\ (3)}{\mathsf{foo}(x) \vee \mathsf{foo}(x) = \mathsf{foo}(x)}\ (5)\text{ on F.3}}{\mathsf{foo}(x) = \mathsf{foo}(x) \vee \mathsf{foo}(x)}\ (2)}{\mathsf{foo}(x) = \mathsf{T}}\ (3)$$

$$\cfrac{\cfrac{\overline{\mathsf{foo}(x)\,\&\,\mathsf{foo}(x) = \mathsf{foo}(x)}\ \ (5)\text{ on F.4}}{\mathsf{foo}(x) = \mathsf{foo}(x)\,\&\,\mathsf{foo}(x)}\ \ (2)}{\mathsf{foo}(x)\,\&\,\mathsf{foo}(x) = \mathsf{foo}(x)\,\&\,\neg\,\mathsf{foo}(x)}\ \ (4)\text{ on [0] with } t = \mathsf{foo}(x)\,\&\,z$$

[2] $\quad \mathsf{foo}(x) = \mathsf{foo}(x)\,\&\,\neg\,\mathsf{foo}(x) \quad$ (3)

[3] $\quad \cfrac{\cfrac{\cfrac{\overline{\mathsf{foo}(x)\,\&\,\neg\,\mathsf{foo}(x) = \mathsf{F}}\ \ (5)\text{ on F.2}}{\mathsf{foo}(x) = \mathsf{F}}\ \ (3)\text{ on [2]}}{\mathsf{F} = \mathsf{foo}(x)}\ \ (2)}{\mathsf{F} = \mathsf{T}}\ \ (3)\text{ on [1]}$

The algebra $I$ is a counterexample to the equation $\mathsf{F} = \mathsf{T}$ that was proved above. But since the proof really *does* use the unsorted rules of deduction correctly, we must conclude that these rules are not sound for this many-sorted algebra. It should however be noted that the unsorted rules of deduction are sound and complete for the classical case (studied by Birkhoff and others) where only unsorted (i.e., one-sorted) algebras are used as models. □

We will see later that by adding quantifiers to the proof, we get a proof of $(\forall x)\ \mathsf{F} = \mathsf{T}$, and we will also see that this does *not* mean that $\mathsf{F} = \mathsf{T}$ is satisfied by all models of FOO. The counterexample is only possible because $I_A = \emptyset$, and indeed, it can be shown that $\mathsf{F} = \mathsf{T}$ does hold in every model of FOO that has all of its carriers non-empty. Moreover, it can be shown that unsorted equational deduction is sound if restricted to models that have all their carriers non-empty. Hence, it might seem that the way out is just to restrict signatures so that no carrier can possibly be empty; for example, this approach is advocated in [109]. But such a restriction would exclude many important examples, such as the theory of partially ordered sets. Another possible way out (and this is the approach of classical logic) is simply to require that all models have all their carriers non-empty. However, we do not want to abandon the possibility of empty carriers, because then not all specifications will have initial models, as demonstrated by the above example, and many others. It therefore follows that we cannot use the unsorted rules of deduction with their unsorted notation for equations, and instead must use a version of many-sorted equational deduction in which equations have explicit quantifiers.

## 4.4 Completeness

The main result about equational deduction is that it is *complete* for loose semantics. The following extensions of the notation for satisfaction enable us to state this in a simple way:

**Definition 4.4.1** Let $A$ and $A'$ be sets of $\Sigma$-equations, and let $e$ be a $\Sigma$-equation. Then we write $A \vDash_\Sigma e$ iff for all $\Sigma$-algebras $M$, $M \vDash_\Sigma A$ implies $M \vDash_\Sigma e$, that is, iff every $\Sigma$-algebra that satisfies $A$ also satisfies $e$. Also, we write $A \vDash_\Sigma A'$ iff $A \vDash_\Sigma e'$ for all $e' \in A'$. Similarly, we write $A \vdash_\Sigma A'$ iff $A \vdash_\Sigma e'$ for all $e' \in A'$. □

Now the main result:

**Theorem 4.4.2** (*Completeness*) Given a signature $\Sigma$ and a set $A$ of $\Sigma$-equations, then for any $\Sigma$-equation $e$,

$$A \vdash e \quad \text{iff} \quad A \vDash e \,.$$
□

One direction of this equivalence is the soundness of the rules, which has already been proved (Proposition 4.3.7); it says that anything that can be proved by equational deduction really is true of all models.

The other direction, which is completeness in the narrow sense, is much more difficult, and is proved in Appendix B (actually, the more general case of conditional order-sorted equations is proved there). Theorem 4.4.2 is very comforting, because it says every equation $e$ that is true in all models of $A$ can be deduced using our rules. For example, we can conclude that every equation that is true of all groups can be proved from the group axioms.

We will soon see that there are other rule sets that can make proofs much easier than they are with (1-5). In fact, the particular rules (1-5) were chosen because each rule is relatively simple and intuitive, and because this formulation facilitates proving the completeness theorem.

The following slightly more general formulation of completeness follows from Theorem 4.4.2 and Definition 4.4.1:

**Corollary 4.4.3** Let $A$ and $A'$ be sets of $\Sigma$-equations. Then $A \vdash A'$ iff $A \vDash A'$. □

Before leaving this section, we show transitivity for the extended notion of satisfaction given in Definition 4.4.1:

**Fact 4.4.4** Let $A, A', A''$ be sets of $\Sigma$-equations. Then

$$A \vDash A' \text{ and } A' \vDash A'' \text{ imply } A \vDash A''.$$

**Proof:** We are assuming that $M \vDash A$ implies $M \vDash A'$ and that $M \vDash A'$ implies $M \vDash A''$. Therefore, by transitivity of implication, $M \vDash A$ implies $M \vDash A''$. □

## 4.5 Subterm Replacement

A specialized rule of inference using subterm replacement is the basis for *term rewriting*, a powerful technique for mechanical inference that is discussed in the next chapter. We will develop this rule gradually, starting with a special case of rule (4) in which only one variable is substituted for.

Suppose (using the notation of Definition 2.2.1) that $X = Y \cup \{z\}_s$ where $z \notin Y$ and that $\theta, \theta' : X \to T_\Sigma(Y)$ are substitutions such that $\theta(y) = \theta'(y) = y$ for all $y \in Y$ and such that the equation $(\forall Y)\, \theta(z) = \theta'(z)$ is deducible. Since $(\forall Y)\, y = y$ is deducible for all $y \in Y$, the rule (4) implies that, for any $t_0 \in T_\Sigma(Y \cup \{z\}_s)$,

$$(\forall Y)\, t_0(z \leftarrow t_1) = t_0(z \leftarrow t_2)$$

is also deducible, where $t_1 = \theta(z)$ and $t_2 = \theta'(z)$, noting that $t_1, t_2$ have the same sort $s$ as $z$. Therefore the following rule is sound, because we have shown that it is a special case of (4):

(4$_1$) *One Variable Congruence.* Given $t_0 \in T_\Sigma(Y \cup \{z\})$ where $z \notin Y$, if
$$(\forall Y)\, t_1 = t_2$$
is of sort $s$ and is deducible, then
$$(\forall Y)\, t_0(z \leftarrow t_1) = t_0(z \leftarrow t_2)$$
is also deducible.

**Example 4.5.1** Let us use the specification F00 of Example 4.3.8. Consider the equation $(\forall x)\, \mathsf{foo}(x) \vee \neg\mathsf{foo}(x) = \mathsf{T}$, which is shown deducible in Example 4.3.8. Now let $t = \mathsf{foo}(x) \vee z$. Then rule (4$_1$) gives us that

$$(\forall x)\, \mathsf{foo}(x) \vee (\mathsf{foo}(x) \vee \neg\mathsf{foo}(x)) = \mathsf{foo}(x) \vee \mathsf{T}$$

is also deducible.  □

We can get the effect of (4) by repeated applications of (4$_1$) (i.e., the formal proof is by induction on the number of variables in $X$, using the transitivity of equality). Notice that in (4$_1$), $t_1$ (respectively, $t_2$) is substituted for *all* occurrences of $z$ in $t_0$; there may be many such occurrences, or none. We will see later that OBJ3 implements the case where there is *exactly one* occurrence of $z$.

**Proposition 4.5.2** Given a set $A$ of $\Sigma$-equations, then for any $\Sigma$-equation $e$,
$$A \vdash^{(3,4)} e \quad \text{iff} \quad A \vdash^{(3,4_1)} e\,.  \qquad \square$$

That is, (4) and (4$_1$) are interchangeable so long as (3) is present. This gives the following:

**Corollary 4.5.3** Given a set $A$ of $\Sigma$-equations, then for any $\Sigma$-equation $e$,
$$A \vdash^{(1-5)} e \quad \text{iff} \quad A \vdash^{(1-3,4_1,5)} e\,.  \qquad \square$$

And of course, both rule sets are complete, by Theorem 4.4.2. Our next step is to combine (4$_1$) and (5) into the following rule:

(6) *Forward Subterm Replacement.* Given $t_0 \in T_\Sigma(X \cup \{z\}_s)$ with $z \notin X$, and given a substitution $\theta : Y \to T_\Sigma(X)$, if
$$(\forall Y)\, t_1 = t_2$$
is of sort $s$ and is in $A$, then
$$(\forall X)\, t_0(z \leftarrow \theta(t_1)) = t_0(z \leftarrow \theta(t_2))$$
is also deducible.

**Exercise 4.5.1** Show that rule (0) is a special case of rule (6). □

**Exercise 4.5.2** Show that if $A \neq \emptyset$ then rule (1) is a special case of rule (6). □

**Exercise 4.5.3** Show that (5) is a special case of (6), but ($4_1$) is not. □

The following symmetrical variant of (6) is just as useful:

(-6) *Reverse Subterm Replacement.* Given $t_0 \in T_\Sigma(X \cup \{z\}_s)$ with $z \notin X$, and given a substitution $\theta : Y \to T_\Sigma(X)$, if
$$(\forall Y)\ t_2 = t_1$$
is of sort $s$ and is in $A$, then
$$(\forall X)\ t_0(z \leftarrow \theta(t_1)) = t_0(z \leftarrow \theta(t_2))$$
is also deducible.

The soundness of (-6) follows from that of (6) by first using (2) on the equation $(\forall X)\ t_2 = t_1$ and then applying (6). For clarity and emphasis, we may write (+6) instead of (6). We now combine (+6) and (-6) into a single rule, as follows:

(±6) *Bidirectional Subterm Replacement.* Given $t_0 \in T_\Sigma(X \cup \{z\}_s)$ with $z \notin X$, and given a substitution $\theta : Y \to T_\Sigma(X)$, if either
$$(\forall Y)\ t_1 = t_2 \text{ or } (\forall Y)\ t_2 = t_1$$
is of sort $s$ and is in $A$, then
$$(\forall X)\ t_0(z \leftarrow \theta(t_1)) = t_0(z \leftarrow \theta(t_2))$$
is also deducible.

This rule is sound because it is the disjunction of two sound rules. It includes (5) and basic cases[1] of (2) and (4). In fact, the following can be shown (see Appendix B):

**Theorem 4.5.4** For any set $A$ of $\Sigma$-equations and any (unconditional) $\Sigma$-equation $e$,
$$A \vdash e \text{ iff } A \vdash^{(1,3,\pm 6)} e\ .$$
□

This result says that $\vdash$ is the reflexive and transitive closure[2] of $\vdash^{(\pm 6)}$; consequently, we might write (±6*) instead of (1, 3, ±6). It is equivalent to take the reflexive, symmetric and transitive closure of (6), which justifies writing ($6^=$). Based on this, we could get a single rule of deduction based on (6) that is complete all by itself. However, this rule would be rather complex, and we do not give it here.

Theorem 4.5.4 has the important consequence that the reflexive, transitive closure of (±6) is complete, by Theorem 4.4.2.

---
[1] These are the cases where the deduced equation in the premise is actually in $A$.
[2] Appendix C contains a brief review of this concept.

**Exercise 4.5.4** Given a set $A$ of $\Sigma$-equations, show that for any $\Sigma$-equation $e$,
$$A \vdash e \text{ iff } A \vdash^{(1,2,3,6)} e \ .$$

**Hint:** Show $A \vdash^{(\pm 6)} e$ iff $A \vdash^{(2,6)} e$. □

**Example 4.5.5** (*Groups*) Now let us use this to prove the right inverse law
$$(\forall x)\, x * x^{-1} = e$$
for the specification GROUPL. By the Theorem of Constants (Theorem 3.3.11), it suffices to introduce a new constant $a$ and then prove the equation
$$(\forall \emptyset)\, a * a^{-1} = e.$$

Let GL.1, GL.2, GL.3 denote the three equations in GROUPL. Then:

| | | |
|---|---|---|
| [1] | $a * a^{-1} = e * (a * a^{-1})$ | (-6) on GL.1 |
| [2] | $= (a^{-1 -1} * a^{-1}) * (a * a^{-1})$ | (-6) on GL.2 with $A = a^{-1}$ |
| [3] | $= ((a^{-1 -1} * a^{-1}) * a) * a^{-1}$ | (6) on GL.3 |
| [4] | $= (a^{-1 -1} * (a^{-1} * a)) * a^{-1}$ | (-6) on GL.3 |
| [5] | $= (a^{-1 -1} * e) * a^{-1}$ | (6) on GL.2 |
| [6] | $= a^{-1 -1} * (e * a^{-1})$ | (-6) on GL.3 |
| [7] | $= (a^{-1 -1} * a^{-1})$ | (6) on GL.1 |
| [8] | $= e$ | (6) on GL.2 |

This proof is *much* simpler than that given in Section 4.1. Also, notice that each step builds on the one before it, which makes the proof much easier to understand. □

The rules (+6), (-6) and (±6) can all be specialized to the case where $t_0$ has exactly one occurrence of $z$, which corresponds to what OBJ3 implements. In particular, the specialized form of (±6) is the following rule:

(±$6_1$) *Bidirectional One Occurrence Subterm Replacement.* Given $t_0 \in T_\Sigma(X \cup \{z\}_s)$ with exactly one occurrence of $z$ where $z \notin X$, and given a substitution $\theta: Y \to T_\Sigma(X)$, if either
$$(\forall Y)\, t_1 = t_2 \text{ or } (\forall Y)\, t_2 = t_1$$
is of sort $s$ and is in $A$, then
$$(\forall X)\, t_0(z \leftarrow \theta(t_1)) = t_0(z \leftarrow \theta(t_2))$$
is also deducible.

It is a bit tricky to formalize the concept that $t_0 \in T_\Sigma(X \cup \{z\}_s)$ has *exactly one* occurrence of $z$. One way is to use the initial algebra approach of Section 3.2. Letting $\Sigma' = \Sigma(X \cup \{z\}_s)$, we define a $\Sigma'$-homomorphism $\#_z: T_{\Sigma'} \to \omega$ which counts the number of occurrences of $z$ in terms, by giving $\omega$ a $\Sigma'$-structure, using the convention that $\omega$ denotes an $S$-sorted set of copies of the natural numbers, where $S$ is the sort set of $\Sigma$: if $\sigma \in \Sigma$ has $n > 0$ arguments, define $\sigma$ on $\omega$ by $\sigma(i_1,\ldots,i_n) = i_1 + \cdots + i_n$; if $\sigma$ is a constant in $\Sigma$, define $\sigma$ on $\omega$ to be 0; and finally, define $x \in X$ to be 0, and $z$ to be 1, in $\omega$.

Now we can state the main result of this section:

**Theorem 4.5.6** Given a set $A$ of $\Sigma$-equations, then for any $\Sigma$-equation $e$,

$$A \vdash^{(1-5)} e \text{ iff } A \vdash^{(1,3,\pm 6_1)} e .$$

**Proof:** We have already shown the soundness of rules (1), (3) and ($\pm 6_1$). Therefore $A \vdash^{(1,3,\pm 6_1)} e$ implies $A \models e$, and then the completeness of $\vdash$ gives us that $A \vdash e$. For the converse, by Theorem 4.5.4 it suffices to show that we can derive the rule ($\pm 6$) from rules (1), (3), and ($\pm 6_1$). This can be done by using induction and rule (3) for the two cases (6) and (-6) separately. □

Because of Theorem 4.4.2, this result implies that any equation that is true of all groups can be proved using just ($\pm 6_1$) plus transitivity and reflexivity.

**Exercise 4.5.5** Show that if we weaken (6) to "at most one occurrence," then (1) is a special case of this weaker rule.[E10]  □

**Corollary 4.5.7** Given a set $A$ of $\Sigma$-equations, then for any $\Sigma$-equation $e$,

$$A \vdash e \text{ iff } A \vdash^{(1,2,3,6_1)} e .$$

**Proof:** This follows from Theorem 4.5.6, since (2) and ($6_1$) are equivalent to ($\pm 6_1$). □

Completeness of $\vdash_A^{(1,2,3,6_1)}$ means that every equation valid for $A$ can be proved from $A$ without ever having to apply a rule backwards. This perhaps surprising result motivates and justifies term rewriting, a computational method based on $\vdash_A^{(1,3,6_1)}$ which is the topic of the next chapter.

### 4.5.1 (⋆) An Alternative Congruence Rule

This section shows that an apparently weaker congruence rule is in fact equivalent to the original formulation (4) on page 58. The new rule is:

(4′) *Congruence.* Given $\sigma \in \Sigma_{s_1\ldots s_n,s}$ and given deducible equations
$(\forall X)\ t_i = t_i'$ of sort $s_i$ for $i = 1,\ldots,n$, then
$(\forall X)\ \sigma(t_1,\ldots,t_n) = \sigma(t_1',\ldots,t_n')$
is also deducible.

**Proposition 4.5.8** Given a set $A$ of $\Sigma$-equations, then for any $\Sigma$-equation $e$,

$$A \vdash^{(1-5)} e \text{ iff } A \vdash^{(1-3,4',5)} e \,.$$
$$A \vdash^{(4)} e \text{ iff } A \vdash^{(4')} e \,.$$

**Proof:** Since the first assertion follows from the second by induction on the length of proofs, it suffices to prove the second assertion.

If $A \vdash^{(4')} e$, then $A \vdash^{(4)} e$, because $e$ necessarily has the form $(\forall X)\, \sigma(t_1,\ldots,t_n) = \sigma(t'_1,\ldots,t'_n)$, and in rule (4) we can take $t = \sigma(y_1,\ldots,y_n)$ with $\theta(y_i) = t_i$ and $\theta'(y_i) = t'_i$ for $i = 1,\ldots,n$.

For the converse, assume $A \vdash^{(4)} e$. We use structural induction (see Section 3.2.1) on the form of $t$ to show that $A \vdash^{(4')} e$. For the base, if $t \in Y$, then $(\forall X)\, \theta(t) = \theta'(t)$ is deducible by hypothesis. Now suppose that $t = \sigma(t_1,\ldots,t_n)$ and that $(\forall X)\, \theta(t_i) = \theta'(t_i)$ is deducible for $i = 1,\ldots,n$. Then (4') gives us that $(\forall X)\, \sigma(\theta(t_1),\ldots,\theta(t_n)) = \sigma(\theta'(t_1),\ldots,\theta'(t_n))$ is deducible, i.e., that $(\forall X)\, \theta(t) = \theta'(t)$ is deducible. □

Notice that this result implies that $\vdash^{(1-3,4',5)}$ is complete.

### 4.5.2 Discussion

We have given many rules of deduction for many-sorted equational logic, and shown that various subsets are complete for loose semantics. The first variant, consisting of the rules (1–5) in Definition 4.1.3, is not very convenient for calculation, but each rule is relatively intuitive, and this system is convenient for proving the completeness theorem. The final variant, consisting of rules (1), (3) and ($\pm 6_1$), is much more convenient for calculation, although the rule ($\pm 6_1$) may seem somewhat complex at first. Some intermediate variants helped to bridge the gap between these two.

## 4.6 Deduction using OBJ

OBJ3 not only supports writing theories (such as that of groups), but also deducing new equations from theories, by applying subterm replacement. This section introduces some features of OBJ3 that are useful for such proofs, through an example proving the right inverse law for the following left-handed theory GROUPL for groups (it is the same as the theory GROUPL in Example 4.1.4 on page 59):

```
th GROUPL is sort Elt .
  op _*_ : Elt Elt -> Elt .
  op e : -> Elt .
  op _-1 : Elt -> Elt [prec 2] .
  var A B C : Elt .
```

```
    eq e * A = A .
    eq A -1 * A = e .
    eq A *(B * C) = (A * B)* C .
  endth
```

OBJ3 automatically assigns numbers to equations. If there are no other equations in the current environment, then the above equations will be numbered, starting from 1, in the order that they occur, and can be referred to as "GROUPL.1", "GROUPL.2" and "GROUPL.3", or more compactly, as ".1", ".2" and ".3", provided that GROUPL is the module currently in focus.

You can also give your own name to an equation, by placing that name in square brackets in front of the equation. For example, if you had written

```
    [e] eq e * A = A .
    [i] eq A -1 * A = e .
    [a] eq A *(B * C) = (A * B)* C .
```

in GROUPL, then you could refer to these equations with the names "GROUPL.e", "GROUPL.i" and "GROUPL.a", or more compactly, with ".e", ".i" and ".a". When there are multiple modules around, this can be much more convenient, because introducing new modules can cause the numbers of old equations to change, whereas the user-assigned names will not change.

As in Example 4.5.5, the proof given below exploits the Theorem of Constants to get rid of a quantifier, and instead reason with a new constant. Assuming that GROUPL has just been read into OBJ3, the command "open ." permits us to begin working within the module GROUPL, and "op a : -> Elt" temporarily adds a new constant symbol "a" of sort Elt, so that we can form terms that involve this symbol (it represents the universally quantified variable). The command "start a * a -1 ." declares an initial term to which subterm replacement can be applied, yielding a series of equal new terms.

The lines beginning with "***>" are comments, in this case used to say what term we *expect* the command above it to produce. Each apply command applies the rule $(6_1)$ or $(-6_1)$ to the term produced by the command above it. In each case, an equation in GROUPL is mentioned, either in the form .n or in the form -.n, depending on whether $(6_1)$ or $(-6_1)$ is to be used. A substitution $\theta$ is indicated in the form "with A = a -1", and "at term" indicates that $t_0 = z$ in rule $(\pm 6_1)$. Other subterms (i.e., other, non-trivial choices of $t_0$) can be selected using so-called "occurrence notation." For example, the left subterm of $(a * b) * (c * d)$ is selected with (1), and the right subterm with (2); moreover, $b$ is selected with (1 2), $c$ with (2 1), and $d$ with (2 2). Finally,

"close"[3] exits this special mode of OBJ3, forgetting any operations and equations that may have been added. Further details about apply appear in Section 7.2.2.

```
open .
op a : -> Elt .
start a * a -1 .
apply -.1 at term .
  ***> should be: e * (a * a -1)
apply -.2 with A = (a -1) at (1) .
  ***> should be: (a -1 -1 * a -1) * (a * a -1)
apply .3 at term .
  ***> should be: ((a -1 -1 * a -1)* a)* a -1
apply -.3 at (1) .
  ***> should be: (a -1 -1 * (a -1 * a)) * a -1
apply .2 at (1 2) .
  ***> should be: (a -1 -1 * e) * a -1
apply -.3 at term .
  ***> should be: a -1 -1 * (e * a -1)
apply .1 at (2) .
  ***> should be: a -1 -1 * a -1
apply .2 at term .
  ***> should be: e
close
```

**Exercise 4.6.1** Try this yourself.                                    □

In conjunction with Theorem 4.5.6, the completeness theorem implies that *every* equation that is true in the theory of groups can be proved using OBJ3 in the style of Example 4.5.5; therefore, we know that the proofs requested in Exercises 4.6.2–4.6.4 are possible (provided the equations are true).

The following additional features of OBJ3 are also useful in doing such proofs: To add an equation that you have just proved, while still inside an open...close environment, you can just type (for example)

```
[ri] eq A * A -1 = e .
```

and then use this equation in proving another. You can also add more variables, e.g.,

```
vars C D : Elt .
```

And of course you can add more operations. At any time, you can see what the current environment contains just by typing

```
show .
```

---

[3]Note that open *must* be followed by a period, while close *must not* be followed by a period!

You may be surprised to see that a version of the Booleans has been automatically included; it will not be included if before the specification you type

```
set include BOOL off .
```

The command "`show rules .`" causes all the equations currently known to OBJ to be printed together with their numbers and names (if any). In examples that involve importing other modules, it can be hard to predict what ordering OBJ will give to the rules. Therefore, it is important to check what order they actually have, if you want to apply them by their numbers. (A potentially confusing point is that in the terminology of OBJ, "equations" are also called "rules," because they are applied as rewrite rules in OBJ computations; thus, one must be careful to distinguish whether a given instance of the word "rule" means "rule of deduction" in talk about the *theory of* OBJ, or "rewrite rule" in talk about *computations in* OBJ.)

**Exercise 4.6.2** Prove the right identity law for the specification GROUPL. □

**Exercise 4.6.3** Prove the left identity law for the specification GROUP. □

**Exercise 4.6.4** Prove the left inverse law for the specification GROUP. □

**Exercise 4.6.5** Use OBJ3 to prove that T = F for the specification FOO of Example 4.3.8. What *equation* have you really proved? Does it follow that T = F in every FOO-algebra? Why? □

## 4.7 Two More Rules of Deduction

This section gives two more rules of deduction for equational logic; they throw an interesting light on cases like that of Example 4.3.8.

(7) *Abstraction.* If
$$(\forall X)\ t_1 = t_2$$
is deducible from $A$, and if $Y$ is a ground signature disjoint from $X$, then
$$(\forall X \cup Y)\ t_1 = t_2$$
is also deducible from $A$.

(This rule also applies when $X = \emptyset$, where there are originally no variables and some are added.)

For the next rule, we need a preliminary concept: let us say that a sort $s \in S$ is **void** in a signature $\Sigma$ iff $(T_\Sigma)_s = \emptyset$.

(8) *Concretion.* If
$$(\forall X \cup Y)\ t_1 = t_2$$
is deducible from $A$, if no sort of a variable in $Y$ is void in $\Sigma$ and if $t_1, t_2 \in T_\Sigma(X)$, then
$$(\forall X)\ t_1 = t_2$$
is also deducible from $A$.

**Exercise 4.7.1** Show the soundness of rule (7). □

**Fact 4.7.1** Rule (8) is sound.

**Proof:** We have to show that if $M \models (\forall X \cup Y)\ t_1 = t_2$, then $M \models (\forall X)\ t_1 = t_2$; i.e., that if $\overline{a}(t_1) = \overline{a}(t_2)$ for all assignments $a : X \cup Y \to M$, then $\overline{b}(t_1) = \overline{b}(t_2)$ for all $b : X \to M$. This will follow if we can extend any $b : X \to M$ to some $a : X \cup Y \to M$. But we can always pick an arbitrary element $m_y \in M_s$ for each $y \in Y_s$, and then set $b(y) = m_y$, unless there are some $s \in S$ and $y \in Y_s$ such that $M_s = \emptyset$. However, the non-voidness of each $s \in S$ such that there is some $y \in Y_s$ guarantees that this cannot happen. □

From the proof, we see that it is unsound to remove a quantifier over a void sort, because there really can exist models where the carrier of that sort is void. For example, in Example 4.3.8, we are unable to apply the concretion rule to remove the variable in the equation $(\forall x)\ F = T$, because the sort A is void. We have seen that the resulting equation $F = T$ is not satisfied by the model $I$, although the quantified version is satisfied.

By contrast, the abstraction rule (7) is sound if some, or even all, sorts in $Y$ are void.

## 4.8 Conditional Deduction and its Completeness

We can deduce unconditional equations from conditional equations using rules of deduction very similar to those in Definition 4.1.3, except that rule (5) must be modified to account for conditional equations in $A$. (Recall that conditional equations and their satisfaction have already been defined in Section 3.4.) We will see later that the resulting rule set is complete. Here is the modified rule:

(5C) *Conditional Instantiation.* If
$$(\forall Y)\ t = t'\ \text{if}\ C$$
is in $A$, and if $\theta : Y \to T_\Sigma(X)$ is a substitution such that $(\forall X)\ \theta(u) = \theta(v)$ is deducible for each pair $\langle u, v \rangle \in C$, then
$$(\forall X)\ \theta(t) = \theta(t')$$
is deducible.

We will write concrete instances of conditional rules in forms like

$$(\forall x, y, z, w) \; x + z = y + w \; \texttt{if} \; x = y, \; z = w \, ,$$

separating pairs in the condition by commas, and using the equality sign.

We now show that the rule (5C) is sound:

**Lemma 4.8.1** Given a $\Sigma$-algebra $M$ satisfying $A$ and a substitution $\theta : Y \to T_\Sigma(X)$, if $M \models (\forall X) \; \overline{\theta}(u) = \overline{\theta}(v)$ for all $\langle u, v \rangle \in C$, then also $M \models (\forall X) \; \overline{\theta}(t) = \overline{\theta}(t')$.

**Proof:** Let $a : X \to M$ and assume that $M \models (\forall X) \; \overline{\theta}(u) = \overline{\theta}(v)$ for each $\langle u, v \rangle \in C$. Then $\overline{(\theta; a)}(u) = \overline{(\theta; a)}(v)$, and so by the definition of conditional satisfaction, $\overline{(\theta; a)}(t) = \overline{(\theta; a)}(t')$, i.e., $\overline{a}(\overline{\theta}(t)) = \overline{a}(\overline{\theta}(t'))$, by Corollary 3.6.6. □

The proof above shows that the rule (5C) is sound even if $C$ is infinite; but of course, in that case we could never write a finite proof score using the rule. Here are some examples of the use of rule (5C):

**Example 4.8.2** In the context of a specification for the natural numbers with a Boolean-valued inequality function $>$, consider the conditional equation

$$(\forall x, y, z) \; x = y \; \texttt{if} \; z * x = z * y, \; z > 0 = \textit{true} \, ,$$

and suppose that at some point we have deduced that $5 * a = 5 * b$. Then we can use the above conditional equation and rule (5C) to deduce that $a = b$, since $5 > 0 = \textit{true}$.

In the context of the same specification, now consider the equation

$$(\forall x, y, z) \; x > z = \textit{true} \; \texttt{if} \; x > y = \textit{true}, \; y > z = \textit{true} \, .$$

Then if at some point we have deduced that $a + b > a + c = \textit{true}$ and $a + c > d = \textit{true}$, then we can use rule (5C) and the above to deduce that $a + c > d = \textit{true}$. □

Let us write $A \vdash^C_\Sigma e$ if $e$ is deducible from $A$ using the rules (1, 2, 3, 4, 5C); also as usual, let us omit the subscript $\Sigma$ and the superscript $C$ if they are clear from context. As with Proposition 4.3.7 in Section 4.3, it now follows by induction on the length of derivations that $\vdash^C$ is sound. As with deduction for unconditional equations, we have a completeness theorem:

**Theorem 4.8.3** (*Completeness*) Given a set $A$ of (possibly conditional) $\Sigma$-equations, then for any unconditional $\Sigma$-equation $e$,

$$A \vdash^C e \; \text{iff} \; A \models e.$$
□

The proof is given in Appendix B. (Actually, Appendix B proves the more general Theorem 10.3.2 for the case where the equations in $A$ may be order-sorted.)

The following result gives the most generally useful approach to proving a conditional equation from a given set of (possibly conditional) equations:

**Theorem 4.8.4** Given a set $A$ of (possibly conditional) $\Sigma$-equations and a conditional $\Sigma$-equation $(\forall X)\ t = t'\ \text{if}\ C$, let $A' = A \cup \{(\forall \varnothing)\ u = v \mid \langle u, v \rangle \in C\}$. Then

$$A \vDash_{\Sigma} (\forall X)\ t = t'\ \text{if}\ C \quad \text{iff} \quad A' \vdash^{C}_{\Sigma(X)} (\forall \varnothing)\ t = t'\ .$$

**Proof:** By letting $C' = \{(\forall \varnothing)\ u = v \mid \langle u, v \rangle \in C\}$ and using Proposition 3.4.3, $A \vDash_{\Sigma} (\forall X)\ t = t'\ \text{if}\ C$ is equivalent to

$$A \cup C' \vDash_{\Sigma(X)} (\forall \varnothing)\ t = t'\ ,$$

which by Theorem 4.8.3 is in turn equivalent to

$$A \cup C' \vdash^{C}_{\Sigma(X)} (\forall \varnothing)\ t = t'\ ,$$

as desired. □

This result, although not very difficult, gives an important completeness theorem for conditional equations, since it says that any conditional equation $e$ that is satisfied by all models of $A$ can be proved by equational deduction from $A$ plus the conditions of $e$. In practice, it is often possible to do the proof using just (conditional) rewriting.

**Exercise 4.8.1** Show that the result of modifying Theorem 4.8.4 by defining $A'$ to be $A \cup \{(\forall X)\ u = v \mid \langle u, v \rangle \in C\}$, and then asserting

$$A \vDash_{\Sigma} (\forall X)\ t = t'\ \text{if}\ C \quad \text{iff} \quad A' \vdash^{C}_{\Sigma} (\forall X)\ t = t'$$

is false. □

### 4.8.1 Deduction with Conditional Equations in OBJ3

Conditional equations in OBJ are a special case of conditional equations as defined above: there is only one pair $\langle u, v \rangle$ in the set $C$ of conditions, and it must have sort `Bool` with $v = \text{true}$. As a result, conditional equations in OBJ3 can have the simplified syntactic form

```
eq t = t' if u .
```

where $u$ is a term of sort `Bool`. However, this is not really much of a restriction, because equalities (as well as inequalities) can (usually) be expressed as Boolean terms, and $u$ can also be a conjunction of conditions.

For example, the transitive law for a relation > viewed as a Boolean-valued function takes the following form:

```
eq X > Z = true if X > Y and Y > Z .
```

The OBJ3 commands for explicitly applying conditional equations have exactly the same syntax as for unconditional equations. However, the rewrite will not actually be done unless is there not only a match of the leftside, but OBJ3 is also able to reduce the condition to `true`. There are two modes within which this reduction might be accomplished:

1. If `reduce conditions` is `off` (which is the default) then the focus for application shifts to the condition, and the user can explicitly invoke `apply` to try to prove that the condition equals `true`.

2. If `reduce conditions` is set `on` (which must be done explicitly, using the `set` command) then OBJ3 will just compute the normal form of the condition, and apply the equation iff that form is `true`.

If a conditional equation is applied when the focus is on the condition of a previously applied equation, then the focus shifts to the condition of the latest equation; these foci can be nested arbitrarily deeply, and a given focus is abandoned in favour of the previous one iff the proof that it is `true` has been completed. OBJ3 does all this automatically when `reduce conditions` is `on`, as can be seen by setting `trace on`.

**Exercise 4.8.2** Recall that a function $f : A \to B$ is injective iff it satisfies the conditional equation

$$(\forall x, y) \; x = y \text{ if } f(x) = f(y) ,$$

and is a right inverse iff there is a function $g : B \to A$ such that $g; f = 1_B$, i.e., such that

$$(\forall x) \; f(g(x)) = x$$

is satisfied. Now do the following:

(a) Write an OBJ theory which expresses the two assumptions above.

(b) Write an OBJ proof score for showing that

$$(\forall y) \; g(f(y)) = y$$

holds under these assumptions.

(c) Explain why this proof score proves the equation in (b).

Note that this proves Exercise 3.1.8 in Section 3.1 for the unsorted case.

□

**Exercise 4.8.3** Given the following code

```
obj INT is sort Int .
  ops (inc_)(dec_): Int -> Int .
  op 0 : -> Int .
  vars X Y : Int .
  eq inc dec X = X .
  eq dec inc X = X .
  op _+_ : Int Int -> Int .
  eq 0 + Y = Y .
  eq (inc X)+ Y = inc(X + Y).
  eq (dec X)+ Y = dec(X + Y).
endo
```

give an OBJ proof score for the conditional equation

$$(\forall x, y) \; x = y \; \text{if inc } x = \text{inc } y$$

and justify it. □

## 4.9 Conditional Subterm Replacement

This section develops subterm replacement for the case of conditional equations. The basic rule, generalizing rule (6) is as follows:

(6C) *Forward Conditional Subterm Replacement.* Given $t_0 \in T_\Sigma(X \cup \{z\}_s)$ with $z \notin X$, if
$$(\forall Y) \; t_1 = t_2 \; \text{if } C$$
is of sort $s$ and is in $A$, and if $\theta : Y \to T_\Sigma(X)$ is a substitution such that $(\forall X) \; \theta(u) = \theta(v)$ is deducible for each pair $\langle u, v \rangle \in C$, then
$$(\forall X) \; t_0(z \leftarrow \theta(t_1)) = t_0(z \leftarrow \theta(t_2))$$
is also deducible.

**Exercise 4.9.1** Show that rule (6C) is sound. □

**Exercise 4.9.2** Show that rule (0) is a special case of (6C). □

**Exercise 4.9.3** Show that if $A \neq \emptyset$ then rule (1) is a special case of (6C). □

**Exercise 4.9.4** Show that (5C) is a special case of (6C). □

**Exercise 4.9.5** Show that $(4_1)$ is not a special case of (6C). □

As with unconditional rewriting, there is also a very useful symmetrical variant of (6C):

(-6C) *Reverse Conditional Subterm Replacement.* Given $t_0 \in T_\Sigma(X \cup \{z\}_s)$ with $z \notin X$, if
$$(\forall Y) \; t_2 = t_1 \; \text{if } C$$

is of sort $s$ and is in $A$, and if $\theta : Y \to T_\Sigma(X)$ is a substitution such that $(\forall X)\, \theta(u) = \theta(v)$ is deducible for each pair $\langle u, v \rangle \in C$, then
$$(\forall X)\, t_0(z \leftarrow \theta(t_1)) = t_0(z \leftarrow \theta(t_2))$$
is of sort $s$ and also deducible.

The soundness of (-6C) follows from that of (6C) by first using (2) on the equation $(\forall X)\, t_2 = t_1$ and then applying (6C). For clarity and emphasis, we may write (+6C) instead of (6C). We now combine (+6C) and (-6C) into a single rule, as follows:

($\pm$6C) *Bidirectional Conditional Subterm Replacement.* Given $t_0 \in T_\Sigma(X \cup \{z\}_s)$ with $z \notin X$, if either
$$(\forall Y)\, t_1 = t_2 \text{ if } C \quad \text{or} \quad (\forall Y)\, t_2 = t_1 \text{ if } C$$
is of sort $s$ and is in $A$, and if $\theta : Y \to T_\Sigma(X)$ is a substitution such that $(\forall X)\, \theta(u) = \theta(v)$ is deducible for each pair $\langle u, v \rangle \in C$, then
$$(\forall X)\, t_0(z \leftarrow \theta(t_1)) = t_0(z \leftarrow \theta(t_2))$$
is also deducible.

This rule is sound because it is the disjunction of two sound rules. It includes (5C) and basic cases (where the given equation is in $A$) of (2) and (4). In fact, the following can be shown:

**Theorem 4.9.1** (*Completeness of Subterm Replacement*) For any set $A$ of (possibly conditional) $\Sigma$-equations and any unconditional $\Sigma$-equation $e$,
$$A \vdash^C e \quad \text{iff} \quad A \vdash^{(1,3,\pm 6C)} e \,. \qquad \square$$

This result follows from the more general Theorem 10.3.3 proved in Appendix B, for the case where the equations in $A$ may be order-sorted. Note also that Theorem 4.5.4 is a special case of Theorem 4.9.1, and hence also of Theorem 10.3.3.

The rules (+6C), (-6C) and ($\pm$6C) can each be specialized to the case where $t_0$ has exactly one occurrence of $z$. In particular, the specialized form of ($\pm$6C) is the following rule:

($\pm 6_1 C$) *Bidirectional Conditional One Occurrence Subterm Replacement.* Given $t_0 \in T_\Sigma(X \cup \{z\}_s)$ with exactly one occurrence of $z$ where $z \notin X$, and given a substitution $\theta : Y \to T_\Sigma(X)$, if either
$$(\forall Y)\, t_1 = t_2 \text{ if } C \quad \text{or} \quad (\forall Y)\, t_2 = t_1 \text{ if } C$$
is of sort $s$ and is in $A$, then
$$(\forall X)\, t_0(z \leftarrow \theta(t_1)) = t_0(z \leftarrow \theta(t_2))$$
is deducible if for each pair $\langle u, v \rangle \in C$
$$(\forall X)\, \theta(u) = \theta(v)$$
is also deducible.

We can now obtain the following:

**Corollary 4.9.2** Given a set $A$ of (possibly conditional) $\Sigma$-equations, then for any unconditional $\Sigma$-equation $e$,

$$A \vdash^{(1-5C)} e \text{ iff } A \vdash^{(1,3,\pm 6_1 C)} e \text{ iff } A \vdash^{(1,2,3,6_1 C)} e \ .$$

**Proof:** We have already shown the soundness of rules (1), (3) and ($\pm 6_1 C$). Therefore $A \vdash^{(1,3,\pm 6_1 C)} e$ implies $A \vDash e$, and then the completeness of $\vdash^C$ gives us that $A \vdash^C e$. For the converse, by Theorem 4.9.1 it suffices to show that we can derive the rule ($\pm 6C$) from rules (1), (3), and ($\pm 6_1 C$). This can be done by using induction and rule (3) for the two cases (+6C) and (-6C) separately. The second "iff" follows as in Corollary 4.5.7.  □

Results in this section provide a foundation for conditional term rewriting, discussed in Section 5.8 of the next chapter. In particular, the above result, extending Corollary 4.5.7, gives completeness of the transitive, reflexive, symmetric closure of forward conditional subterm replacement, which becomes conditional term rewriting when symmetry is dropped.

## 4.10 (⋆) Specification Equivalence

This section discusses the equivalence of specifications that may involve different signatures. For example, the following gives a rather different specification of groups:

**Example 4.10.1** If we define $a/b = a * b^{-1}$ in the theory GROUPL of groups in Example 4.1.4, and then try to find enough properties of this operation to define groups, we might get the following:

```
th GROUPD- is sort Elt .
  op _/_ : Elt Elt -> Elt .
  var A B C : Elt .
  eq A /(B / B) = A .
  eq (A / A)/(B / C) = C / B .
  eq (A / C)/(B / C) = A / B .
endth
```

Even though these equations are enough, this specification is *not* equivalent to GROUPL, because the empty set is a model of GROUPD-, whereas it is not a model of GROUPL. However, if we add an identity to the specification, we do get an equivalent theory:

```
th GROUPD is sort Elt .
  op _/_ : Elt Elt -> Elt .
  op e : -> Elt .
  var A B C : Elt .
  eq A /(B / B) = A .
  eq (A / A)/(B / C) = C / B .
```

```
      eq (A / C)/(B / C) = A / B .
      eq (A / A) = e .
    endth
```

The last axiom says that e is an identity. □

But how can we *prove* that GROUPD and GROUPL are equivalent? We will certainly need a more general definition of equivalence than the one given in Section 3.3, because the signatures of the two specifications are different.

Before we can give this definition, we need some more notation. Each $\Sigma$-algebra has an interpretation of each operation symbol $\sigma \in \Sigma$ as an actual operation; we show how this extends to an interpretation for $\Sigma$-terms with variables. Given $w = s_1 \ldots s_n \in S^*$, we let $^wX$ denote an $S$-sorted ground signature disjoint from $\Sigma$ such that $\#(^wX_s) = \#\{i \mid s_i = s\}$. One way to construct such a signature is to let $|^wX| = \{x_1, \ldots, x_n\}$ where $n = \#w$, and then let $^wX_s = \{x_i \mid s_i = s\}$. For example, if $S = \{a, b, c\}$ and $w = abbac$, then $^wX$ has $^wX_a = \{x_1, x_4\}, ^wX_b = \{x_2, x_3\}$ and $^wX_c = \{x_5\}$. We shall use this construction in the following:

**Definition 4.10.2** Given a signature $\Sigma$, the signature of all **derived** $\Sigma$-operations is the $S$-sorted signature $Der(\Sigma)$ with

$$Der(\Sigma)_{w,s} = T_\Sigma(^wX)_s \text{ for all } w \in S^* \text{ and } s \in S .$$

Any $t \in Der(\Sigma)_{w,s}$ defines an actual operation $M_t : M^w \to M_s$ on any $\Sigma$-algebra $M$ as follows: given $a \in M^w$, there is a naturally corresponding $S$-indexed map $a : {^wX} \to M$, which lets us view $M$ as a $\Sigma(^wX)$-algebra; hence there is a unique $\Sigma(^wX)$-homomorphism $\bar{a} : T_{\Sigma(^wX)} \to M$ which lets us define $M_t(a)$ to be $\bar{a}(t)$. This is called the **derived operation** defined by $t$. In this way, we can view any $\Sigma$-algebra $M$ as a $Der(\Sigma)$-algebra, also denoted $M$. □

**Definition 4.10.3** Given signatures $\Sigma$ and $\Sigma'$ with sort sets $S$ and $S'$ respectively, then a **signature morphism** (or **map**) $\varphi : \Sigma \to \Sigma'$ consists of a map $f : S \to S'$ and an $S^* \times S$-indexed map $g$ with components $g_{w,s} : \Sigma_{w,s} \to \Sigma'_{f(w),f(s)}$ where $f$ is extended to lists by $f([]) = []$ and $f(s_1 \ldots s_n) = f(s_1) \ldots f(s_n)$. Given $s \in S, w \in S^*$ we may write $\varphi(s)$ and $\varphi(w)$ instead of $f(s)$ and $f(w)$, respectively; and given $\sigma \in \Sigma_{w,s}$, we may write $\varphi(\sigma)$ instead of $g(\sigma)$.

Given a signature morphism $\varphi : \Sigma \to \Sigma'$ and a $\Sigma'$-algebra $M$, we get a $\Sigma$-algebra, called the **reduct** of $M$ under $\varphi$ and denoted $\varphi M$, as follows:

- Given $s \in S$, let $(\varphi M)_s = M_{\varphi(s)}$;
- Given $\sigma \in \Sigma_{w,s}$, let $(\varphi M)_\sigma = M_{\varphi(\sigma)} : M^{\varphi(w)} \to M_{\varphi(s)}$.

In particular, given a signature morphism $\varphi : \Sigma \to Der(\Sigma')$ and a $\Sigma'$-algebra $M$, we can view $M$ as a $Der(\Sigma')$-algebra by Definition 4.10.2, and then get a $\Sigma$-algebra denoted $\varphi M$ from the construction above.

We will call a signature morphism $\varphi : \Sigma \to Der(\Sigma')$ a **derivor** from $\Sigma$ to $\Sigma'$.  □

It follows that any derivor $\varphi : \Sigma \to Der(\Sigma')$ induces a unique $\Sigma$-homomorphism $T_\Sigma \to \varphi T_{\Sigma'}$, because $\varphi T_{\Sigma'}$ is a $\Sigma$-algebra by the above. Let us denote this homomorphism $\overline{\varphi}$.

**Definition 4.10.4** An **interpretation** of specifications $\varphi : (\Sigma, A) \to (\Sigma', A')$ is a derivor $\varphi : \Sigma \to Der(\Sigma')$ such that for every $\Sigma'$-algebra $M'$,

$M' \vDash A'$ implies $\varphi M' \vDash A$.

Specifications $(\Sigma, A)$ and $(\Sigma', A')$ are **equivalent** iff there exist interpretations $\varphi : (\Sigma, A) \to (\Sigma', A')$ and $\psi : (\Sigma', A') \to (\Sigma, A)$ such that

$\varphi(\psi M) = M$
$\psi(\varphi M') = M'$

for all $(\Sigma, A)$-algebras $M$ and $(\Sigma', A')$-algebras $M'$.  □

If specifications $(\Sigma, A)$ and $(\Sigma', A')$ are equivalent, then any $(\Sigma, A)$-algebra can be seen as a $(\Sigma', A')$-algebra, and *vice versa*. We will see that the specifications GROUPL and GROUPD are equivalent in this sense. Also, Exercises 4.6.1–4.6.4 show that the specifications GROUP and GROUPL are equivalent in the sense of the less general definition of equivalence given in Section 3.3, which suffices because the signatures are the same.

It is worth remarking that we can get an even more general definition of equivalence by weakening the conditions in the above definition to the following:

$\varphi(\psi M) \vDash e$  iff  $M \vDash e$
$\psi(\varphi M') \vDash e'$  iff  $M' \vDash e'$

for all $\Sigma$-equations $e$ and $\Sigma'$-equations $e'$. Another generalization (which however involves some category theory) is to require the two categories of models to be isomorphic.

**Exercise 4.10.1** Use OBJ3 to prove that the specifications GROUPL and GROUPD (of Example 4.10.1) are equivalent in the sense of Definition 4.10.4.  □

**Exercise 4.10.2** Show that if two specifications are equivalent in the sense of Section 3.3, then they are equivalent in the sense of Definition 4.10.4.  □

**Exercise 4.10.3** Show that replacing the three equations in GROUPD– by the single equation

```
eq A /((((A / A)/ B)/ C)/(((A / A)/ A)/ C)) = B .
```

yields an equivalent specification.  □

**Definition 4.10.5** A derivor $\varphi : \Sigma \to Der(\Sigma')$ induces a signature morphism $\varphi^* : Der(\Sigma) \to Der(\Sigma')$ as follows: suppose that $\varphi = \langle f, g \rangle$, and note that $g_{w,s} : \Sigma_{w,s} \to T_{\Sigma'}(f^{(w)}X)_{f(s)}$ for each pair $\langle w, s \rangle$. Define $\varphi^*_w : T_\Sigma({}^wX) \to \varphi T_{\Sigma'}(f^{(w)}X)$ for each $w \in S^*$ to send $x_i$ (of sort $w_i$) in ${}^wX_s$ to $x_i$ (of sort $f(w_i)$) in $f^{(w)}X_{f(s)}$, noting that $T_\Sigma({}^wX)$ is the free $\Sigma$-algebra generated by ${}^wX$ and that $\varphi T_{\Sigma'}(f^{(w)}X)$ is also a $\Sigma$-algebra. Then, noting that $\varphi^*_w$ is an $S$-sorted map, the collection $\varphi^*_{w,s}$ forms a signature morphism $\varphi^* : Der(\Sigma) \to Der(\Sigma')$.

Now given interpretations $\varphi : (\Sigma, A) \to (\Sigma', A')$ and $\psi : (\Sigma', A') \to (\Sigma'', A'')$, we can define their **composition** as interpretations to be the signature morphism $\varphi; \psi^* : \Sigma \to Der(\Sigma'')$. □

**Exercise 4.10.4** Show that the composition of two interpretations $\varphi : (\Sigma, A) \to (\Sigma', A')$ and $\psi : (\Sigma', A') \to (\Sigma'', A'')$ is also an interpretation. □

**Exercise 4.10.5** Let $i_\Sigma : \Sigma \to Der(\Sigma)$ send $\sigma \in \Sigma_{w,s}$ to the term $\sigma(x_1, \ldots, x_n) \in T_\Sigma({}^wX)_s$ where $w$ has length $n$. Then show that $i_\Sigma{}^* = 1_{Der(\Sigma)}$ and that $i_\Sigma$ serves as an identity for the composition of interpretations. □

In Section 5.9 we will need a more syntactical formulation of interpretation. We first give some auxiliary notions:

**Definition 4.10.6** Let $X$ be an $S$-sorted variable set and let $f : S \to S'$ be a map. Then the $S'$-sorted variable set denoted $fX$ is defined for $s' \in S'$ by

$$(fX)_{s'} = \bigcup \{X_s \mid f(s) = s'\} .$$

Note that $fX$ is again a variable set because the variable symbols in $X$ are all distinct.

Now let $t$ be a $\Sigma$-term with variables in $X$ and let $\varphi = (f, g) : \Sigma \to \Sigma'$ be a signature morphism. Then we extend $\varphi$ to a function $\overline{\varphi} : T_\Sigma(X) \to \varphi T_{\Sigma'}(fX)$ as follows:

$\overline{\varphi}(x) = x$ for $x \in X$
$\overline{\varphi}(\sigma) = g_{[],s}(\sigma)$ for $\sigma \in \Sigma_{[],s}$
$\overline{\varphi}(\sigma(t_1, \ldots, t_n)) = g_{w,s}(\sigma)(\overline{\varphi}t_1, \ldots, \overline{\varphi}t_n)$ for $\sigma \in \Sigma_{w,s}$
  where $w = s_1 \ldots s_n$ .

Finally, let $e$ be a $\Sigma$-equation $(\forall X)\ t = t'$ and let $\varphi = (f, g) : \Sigma \to \Sigma'$ be a signature morphism. Then the $\Sigma'$-equation denoted $\varphi e$ is defined[E11] to be

$$(\forall\ fX)\ \overline{\varphi}t = \overline{\varphi}t' .$$

In this context, we may also write $\varphi X$ instead of $fX$. Note that $\overline{\varphi} : T_\Sigma(X) \to \varphi T_{\Sigma'}(\varphi X)$ is a $\Sigma$-homomorphism, and that it could also have been defined using the freeness of $T_\Sigma(X)$. □

We will need the following result:

**Theorem 4.10.7** (*Satisfaction Condition*) Given a signature morphism $\varphi : \Sigma \to \Sigma'$ and a $\Sigma'$-algebra $M'$, then for any $\Sigma$-equation $e$,

$$M' \vDash_{\Sigma'} \varphi e \text{ iff } \varphi M' \vDash_{\Sigma} e \:.\qquad\square$$

A proof may be found in [66]; it is not trivial. A general discussion of the importance of this kind of result for abstract model theory is given in [67]. We can now state the result that we have been working towards:

**Theorem 4.10.8** A derivor $\varphi : \Sigma \to Der(\Sigma')$ is an interpretation of specifications $\varphi : (\Sigma, A) \to (\Sigma', A')$ iff for each $\Sigma$-equation $e$,

$$A \vDash_{\Sigma} e \text{ implies } A' \vDash_{\Sigma'} \varphi e \:.$$

**Proof:** Call the conditions of Definition 4.10.4 and of this theorem (A) and (B), respectively. To show (A) implies (B), we assume (A) and $A \vDash e$, and then show that $A' \vDash \varphi e$, i.e., that

$$M' \vDash A' \text{ implies } M' \vDash \varphi e \:.$$

So assuming $M' \vDash A'$, (A) gives us $\varphi M' \vDash A$, and then $A \vDash e$ gives us $\varphi M' \vDash e$. Now we apply the satisfaction condition to obtain $M' \vDash \varphi e$, as desired.

For the converse, we assume (B) and $M' \vDash A'$, and wish to show that $\varphi M' \vDash A$. If we let $e \in A$, then $A \vDash e$, so that (B) reduces to $M' \vDash A'$ implies $M' \vDash \varphi e$. Our assumption then gives $M' \vDash \varphi e$, and the satisfaction condition gives $\varphi M' \vDash e$. Therefore $\varphi M' \vDash A$.  $\square$

The following is now a direct consequence of the completeness of equational deduction:

**Corollary 4.10.9** A derivor $\varphi : \Sigma \to Der(\Sigma')$ is an interpretation of specifications $\varphi : (\Sigma, A) \to (\Sigma', A')$ iff for each equation $e \in A$,

$$A \vdash e \text{ implies } A' \vdash \varphi e \:.$$

Furthermore, a pair of derivors $\varphi : \Sigma \to Der(\Sigma')$ and $\psi : \Sigma' \to Der(\Sigma)$ is an equivalence of the specifications $(\Sigma, A)$ and $(\Sigma', A')$ iff

$$A \vdash e \text{ implies } A' \vdash \varphi e$$
$$A' \vdash e' \text{ implies } A \vdash \psi e'$$

for each $e \in A$ and $e' \in A'$, and in addition

$$\varphi(\psi M) = M$$
$$\psi(\varphi M') = M'$$

for each $(\Sigma, A)$-algebra $M$ and $(\Sigma', A')$-algebra $M'$.  $\square$

**Exercise 4.10.6** Given a specification $(\Sigma, A)$, show that its closure $(\Sigma, \overline{A})$ is given by
$$\overline{A} = \{e \mid M \models A \text{ implies } M \models e \text{ for all } \Sigma\text{-algebras } M\} \;;$$
that is, $\overline{A}$ equals the set of all equations that are true of all models of $A$. Now show that if $\varphi$ and $\psi$ are an equivalence of specifications $(\Sigma, A)$ and $(\Sigma', A')$, then
$$\overline{\varphi A} = \overline{A'}$$
$$\overline{\psi A'} = \overline{A} \;.$$
□

## 4.11 (⋆) A More Abstract Formulation of Deduction

This section gives a more abstract formulation of deduction; it is rather specialized, and can safely be skipped on a first reading or in an introductory course.

**Definition 4.11.1** Let *Sen* be a set whose elements are called **sentences**. Then an **inference rule over** *Sen* is a pair $\langle H, c \rangle$ where $H \subseteq Sen$ is finite and $c \in Sen$; we call $H$ the **hypothesis** of the rule, and $c$ its **conclusion**. A subset $C \subseteq Sen$ is **closed** under $\langle H, c \rangle$ iff $H \subseteq C$ implies $c \in C$. Given a set $R$ of inference rules, a set $C$ of sentences is **closed under** $R$ iff $C$ is closed under each rule in $R$. Two sets $R, R'$ of inference rules are called **equivalent** iff they have the same closed sets of sentences.

Given a set $R$ of inference rules, a **proof** for a sentence $e$ is a finite sequence of sentences, $e_1 e_2 \ldots e_n$ with $e_n = e$ such that for $i = 1, \ldots, n$ there exists a rule $\langle H, e_i \rangle$ with $H \subseteq \{e_1, \ldots, e_{i-1}\}$; in this case, we say that $e$ can be **proved using** $R$. Let $Th(R)$ denote the set of all sentences that can be proved using $R$.   □

In the case of equational logic, *Sen* would be the set of all equations over a given signature, and a rule (e.g., congruence) is represented by the set of its instances for all ground equations. A rule with no hypothesis, such as reflexivity, has $H = \emptyset$.

**Fact 4.11.2** If $C_i$ is closed under a set $R$ of inference rules for $i \in I$, then $\bigcap_{i \in I} C_i$ is also closed under $R$.   □

**Proposition 4.11.3** Given any set $R$ of inference rules, $Th(R)$ is the least set of sentences that is closed under $R$.

**Proof:** We first show that $Th(R)$ is closed under $R$. Let $\langle H, c \rangle$ be in $R$ with $H = \{e_1, \ldots, e_n\}$ such that $H \subseteq Th(R)$. Then for $i = 1, \ldots, n$, there exists a proof $p_i$ of $e_i$. It now follows that the sequence $p_1 \ldots p_n c$ is a proof of $c$, so that $c \in Th(R)$.

Now assuming that $T \subseteq Sen$ is closed under $R$, we will show that $Th(R) \subseteq T$. Let $e \in Th(R)$ and let $e_1 \ldots e_n$ be a proof of $e = e_n$. We have to show $e \in T$. For this purpose, we show by induction that $e_i \in T$ for $i = 1, \ldots, n$. Suppose that $e_j \in T$ for each $j < i$. Because $e_1 \ldots e_i$ is a proof, there exists a rule $\langle H, e_i \rangle$ such that $H \subseteq \{e_1, \ldots, e_{i-1}\}$. Since $H \subseteq T$ and $T$ is closed under $R$, we have that $e_i \in T$. □

This result can be used to prove a property of $Th(R)$ by showing that the set of sentences having that property is closed under $R$. This proof technique may be called "structural induction over proof rules" (see also the discussion in Section 3.2.1).

## 4.12 Literature

It is typical of logical systems that there are many different variants of their rules of deduction, each variant more suitable for some purposes than for others. Although the intuitions behind equational logic are very familiar (basically, substituting equals for equals in equals), the details can be surprisingly subtle, and there are many errors in the published literature. In fact, it is also typical that reasoning about deducibility (such reasoning is called "proof theory") can be very subtle, and the reasons for emphasizing semantics in this text include avoiding such reasoning as much as possible, as well as checking its soundness. (The proof of Theorem 4.5.4 in Appendix B is a good example of proof-theoretic reasoning.)

In 1935 Birkhoff [12] first proved a completeness theorem for equational logic, in the unsorted case. Example 4.3.8, showing that the unsorted rules can be unsound for many-sorted algebras that may have empty carriers, is from [78] and [137], which first gave rules of deduction that are sound and complete for the general case. The rules in Section 4.7 are also from [78]. The discussion of completeness (especially Theorem 4.4.2) follows [136], although the proof in Appendix B is from [82] for the order-sorted case. The discussion in Section 4.3 follows that in [80]. A more detailed historical discussion of various versions of completeness for equational logic is given in [136], along with further discussion of the non-equivalence of one-sorted and many-sorted equational logic that was demonstrated in Section 4.3.2.

Subterm replacement is the logical basis for term rewriting, which is the topic of Chapter 5. Some historical remarks on term rewriting are also given there. The extension of OBJ3 to permit applying rules one at a time was done at Oxford by Timothy Winkler mostly during September 1989.

The discussion of deduction with conditional equations in Sections 4.8 and 4.9 parallels that in the preceding sections for unconditional equations. The conditional Completeness Theorem (4.8.3) is a special case of the corresponding theorem for the order-sorted case, first proved in (the first version of) [82]. Theorems 4.9.1 and 4.8.4 seem not to appear in the literature, though they are certainly important, and will probably not surprise experts in the field.

Lawvere's categorical formulation of algebraic theories for the unsorted case [121] embodies the same notion of specification equivalence as that discussed in Section 4.10; however, the formulation that we give is new. Readers with a categorical background may be interested to note that *Der* is a left adjoint to the forgetful functor from Lawvere theories to signatures; many properties in Section 4.10 follow from this property. The composition operation for interpretations is the Kleisli category composition. The one-equation specification of groups in Section 4.10 is due to Higman and Neumann [102]. The satisfaction condition (Theorem 4.10.7) plays a central role in the theory of institutions, which is an abstract theory of the relationship between syntax and semantics that has been used for a number of computing science applications, including specification [67].

Equational deduction has many applications in computer science and other areas. For example, it is applied to modelling and verifying several different kinds of hardware circuits in Chapter 5, and it has even been used directly as a model of computation [70, 135].

I thank Prof. Virgil Emil Cazanescu for the formulations in Section 4.11, and for several other valuable suggestions, including finding several bugs in this chapter and suggesting fixes for them.

---

**A Note to Lecturers:** When lecturing on Section 4.5, it would make sense to treat all the preceding rules and results as leading up to rule ($\pm 6_1$) and Theorem 4.5.6. In this case, the variant rules (such as ($4_1$)) and the results about them can be regarded as parts of the proof of Theorem 4.5.6, and assigned as reading rather than covered in lectures.

Much of the material on conditional equations can be covered quickly by relying on the analogy with the unconditional case. However, rule (5C) and the two completeness theorems (4.8.3 and 4.8.4) should be covered explicitly.

The material in Section 4.10 is difficult and should not be attempted in an introductory course. Section 4.11 is rather specialized and somewhat abstract, and should also be omitted in an introductory course. Section 4.5.1 is also rather specialized.

---

# 5 Rewriting

This chapter studies *term rewriting*, the restricted form of equational deduction in which equations are applied in the forward direction only, starting from a given term and "chaining" with the transitivity of equality, that is, repeatedly applying the rule ($+6_1$), or ($+6_1C$) for the conditional case, in the style of Example 4.5.5. Corollary 4.5.7 (or Corollary 4.9.2) shows that term rewriting with symmetry is complete for equational logic. Without symmetry, completeness is lost, but we will see that certain natural assumptions restore it, giving a decision procedure for equality of ground terms under loose semantics, and a computational semantics for sets of equations. Term rewriting has important applications in algebraic specification, computer algebra, the λ-calculus, implementation of declarative languages, and much else. OBJ implements term rewriting with a command which, given a term, searches for a match of a rule to a subterm of that term, applies the rule, and iterates this process until there are no more matches; the final term (if one exists) is considered the *result* of the computation. This process is sound, even when not complete.

This chapter mainly presents and proves results that are useful for theorem proving; many are new, especially those on termination and conditional term rewriting. Abstract rewrite systems are also discussed, with optional sections (5.8.3 and 5.9) on Noetherian orderings and on the relationship between term rewriting and abstract rewriting systems.

## 5.1 Term Rewriting

Syntactically, rewrite rules are a special kind of equation; their definition (5.1.3 below) uses:

**Definition 5.1.1** Given $t \in T_\Sigma(X)$, the set of **variables in** $t$, denoted $var(t)$, is the least[1] subsignature $Y \subseteq X$ such that $t \in T_\Sigma(Y)$. □

---

[1] We can define $var(t)$ formally using initial algebra semantics (Section 3.2). Let $V$ be the $S$-sorted set with each $V_s$ the set of all finite subsets of the elements of $X$, given $\Sigma(X)$-structure by $V_x = \{x\} \in V_s$ for $x \in X_s$, $V_\sigma = \emptyset \in V_s$ for $\sigma \in \Sigma_{[],s}$, and

**Notation 5.1.2** From now on, we will usually just say "$\Sigma$-term" for what we were previously careful to call a "$\Sigma$-term with variables," i.e., for $t \in T_\Sigma(X)$ for some ground signature $X$ without overloading. A $\Sigma$-term without variables will be called a **ground term**. □

Notice that $t$ is a ground term iff $var(t) = \emptyset$. Of course, every $\Sigma$-term $t$ is a ground term over any signature that contains $\Sigma(var(t))$. The usual literature on term rewriting is not very careful with bookkeeping for the variables involved, but we have seen in Section 4.3.2 that this can be very important for theorem proving.

**Definition 5.1.3** A $\Sigma$-**rewrite rule** is a $\Sigma$-equation $(\forall X)\ t_1 = t_2$ with $var(t_2) \subseteq var(t_1) = X$. It follows that the notation $t_1 \to t_2$ is unambiguous, because $X$ is determined by $t_1$. A $\Sigma$-**term rewriting system** (abbreviated $\Sigma$-**TRS**) is a set of $\Sigma$-rewrite rules;[2] we may denote such a system $(\Sigma, A)$, and we may omit $\Sigma$ here and elsewhere if it is clear from context. □

Most equations that users write in OBJ are rewrite rules with variables exactly those that occur in its leftside. For example, all the equations in all of our group specifications are rewrite rules in this way. Notice that if some equation is not a rewrite rule, then its converse (with its left and right sides reversed) may be a rewrite rule.[3]

The rule ($+6_1$) of Chapter 4 replaces exactly one subterm, moving in the forward direction only. We now further restrict it to equations that are rewrite rules, to get the following:

(RW) *Rewriting.* Given $t_0 \in T_\Sigma(\{z\}_s \cup Y)$ with exactly one occurrence of $z$, and given a substitution $\theta : X \to T_\Sigma(Y)$, if $t_1 \to t_2$ is a $\Sigma$-rewrite rule of sort $s$ in $A$ with $var(t_1) = X$, then
$(\forall Y)\ t_0(z \leftarrow \theta(t_1)) = t_0(z \leftarrow \theta(t_2))$
is deducible.

As usual, it is assumed that $\{z\}_s$ and $Y$ are disjoint. This rule is sound because it is a restriction of a rule that we have already proved to be sound. The successive application of this rule to a term gives a method of reasoning that is formalized in the following:

**Definition 5.1.4** Given a $\Sigma$-TRS $A$, the **one-step rewriting** relation is defined for $\Sigma$-terms $t, t'$ by $t \overset{1}{\Rightarrow}_A t'$ iff there exist a rule $t_1 \to t_2$ of sort $s$ in $A$, a

---

$V_\sigma(v_1, \ldots, v_n) = \bigcup_{i=1}^{n} v_i$ for $\sigma \in \Sigma_{w,s}$ where $w = s_1 \ldots s_n$ and $v_i \in V_{s_i}$. Then *var* is the restriction to $T_\Sigma(X)$ of the unique $\Sigma(X)$-homomorphism $T_{\Sigma(X)} \to V$.

[2] It is common to assume that the leftside of a rewrite rule is not just a single variable, because rules of this kind, which we call **lapse rules**, are not very useful and moreover have some bad properties. (The term "lapse" is a joke based on the facts that a rule with its rightside a variable is called a **collapse rule**, and that "co" indicates duality.) However, few results actually need the no lapse assumption, and we will invoke it only where necessary. A TRS that has no lapse rules will be called **lapse free**.

[3] However, this is not necessarily the case; for example, the variables that occur in the two sides could even be disjoint.

term $t_0 \in T_\Sigma(\{z\}_s \cup Y)$ with exactly one occurrence of the variable $z$, and a substitution $\theta : X \to T_\Sigma(Y)$ where $X = var(t_1)$, such that

$$t = t_0(z \leftarrow \theta(t_1)) \text{ and } t' = t_0(z \leftarrow \theta(t_2)) \,.$$

In this case, the pair $(t_0, \theta)$ is called a **match to** (a subterm of) $t$ **by** (the term $t_1$ of) the rule $t_1 \to t_2$. The **term rewriting** relation is the transitive, reflexive closure[4] of the one-step rewriting relation, for which we write $t \stackrel{*}{\Rightarrow}_A t'$ and say that $t$ **rewrites to** $t'$ (**under** $A$). We may also write $t \stackrel{+}{\Rightarrow}_A t'$ if $t$ rewrites to $t'$ in one or more steps (i.e., $\stackrel{+}{\Rightarrow}_A$ is the transitive closure of $\stackrel{1}{\Rightarrow}_A$), and $\stackrel{*}{\Leftrightarrow}_A$ for the transitive, reflexive, symmetric cloure of $\Rightarrow_A$. We may omit the subscript $A$ if it is clear from context, and we may also write $\Rightarrow$ instead of $\stackrel{1}{\Rightarrow}$. The subterm $\theta(t_1)$ of $t$ is sometimes called the **redex** (for **red**ucible **ex**pression) of the rewrite $t \stackrel{1}{\Rightarrow} t'$, and the subterm $\theta(t_2)$ of $t'$ is sometimes called the **contractum** of the rewrite, while $t$ is called the **source** and $t'$ the **target** or the **result** of the rewrite.

□

For reasons that will soon become clear, it is worth emphasizing that the above definitions apply only to ground terms, i.e., to $T_\Sigma$. We consider rewriting terms with variables later.

**Example 5.1.5** Consider the following specification for the natural numbers with addition:

```
obj NATP+ is sort Nat .
  op 0 : -> Nat .
  op s_ : Nat -> Nat [prec 2] .
  op _+_ : Nat Nat -> Nat .
  var N M : Nat .
  eq N + 0 = N .
  eq N + s M = s(N + M) .
endo
```

We can tell OBJ to regard this specification as a TRS and apply its equations as rewrite rules to a term $t$, just by giving the command

```
reduce t .
```

where the final period is required, and must be separated from the term by a space unless the last character of $t$ is a parenthesis. Note that, by default, reductions are executed in the context of the most recently preceding module. Also, "`reduce`" can be abbreviated "`red`". Here are two examples:

```
red s s 0 + s s 0 .
red s(s s 0 + s s s 0)+ 0 .
```

---
[4]This concept is reviewed in Appendix C.

The results are as you would expect, namely s s s s 0 and s s s s s s 0, respectively, i.e., $2 + 2 = 4$ and $(1 + (2 + 3)) + 0 = 6$. The OBJ output from the first reduction looks as follows,

```
==========================================
reduce in NATP+ : s (s 0) + s (s 0)
rewrites: 3
result Nat: s (s (s (s 0)))
==========================================
```

and the steps of this reduction are

```
s s 0 + s s 0   ⇒
s(s s 0 + s 0) ⇒
s s(s s 0 + 0) ⇒
s s(s s 0)   =
s s s s 0 .
```

If the trace mode is on, then OBJ will display each rewriting step of each reduction it executes. The commands

```
set trace on .
set trace off .
```

respectively turn trace mode on and off.   □

**Exercise 5.1.1** Show the rewriting steps for the second reduction in Example 5.1.5.   □

Examples like the above suggest how term rewriting can be considered a model of computation: given a $\Sigma$-term $t_1$ in $T_\Sigma(X)$, each rewrite in a sequence

$$t_1 \overset{1}{\Rightarrow} t_2 \overset{1}{\Rightarrow} t_3 \overset{1}{\Rightarrow} \ldots$$

is considered a step of computation, and a term that cannot be rewritten any further is considered a result. Note that given $t$ with $var(t) = X$, all these computations occur in the $\Sigma$-algebra $T_\Sigma(X)$ or $T_\Sigma(Y)$ for any $X \subseteq Y$. The following formalizes this notion of a computation result:

**Definition 5.1.6** Given a $\Sigma$-TRS $A$, a $\Sigma$-term $t$ is **irreducible**, also called a **normal** or **reduced form** (**under** $A$), iff there is no match to $t$ by any rule in $A$. If $t \overset{*}{\Rightarrow} t'$ and $t'$ is a normal form, then we say that $t'$ is a **normal** (or **reduced**) **form of** $t$ (**under** $A$).   □

Here is another example of computation with rewrite rules:

**Example 5.1.7** The following simple version of the Booleans is sufficient for computing the values of ground terms that involve only true, false, and, and not:

```
obj ANDNOT is sort Bool .
  ops true false : -> Bool .
  op _and_ : Bool Bool -> Bool .
  op not_ : Bool -> Bool [prec 2].
  var X : Bool .
  eq true and X = X .
  eq false and X = false .
  eq not true = false .
  eq not false = true .
endo
```

The following are some sample reductions using this code:

```
red not true and not false .
red not (true and not false) .
red (not not true and true) and not false .
```

We can also run reductions that involve variables, e.g.,

```
red X and not X .
red not not X .
```

(We don't need to declare X here because it is already declared in ANDNOT, but any other variables would need to be declared.) The results of these two reductions show that this specification is not powerful enough to prove every true Boolean equation with variables.    □

**Exercise 5.1.2** Show the rewrites and the results for each reduction in Example 5.1.7.    □

Although this book contains many "natural" examples of TRS's, artificial examples like the two below can be illuminating, for example, as counterexamples to conjectured general results, or to illustrate certain concepts:

**Example 5.1.8** Consider a TRS $A$ having just one sort, with $a, b, c, d$ constants of that sort, and with the following rewrite rules: $a \to b$, $a \to c$, $b \to a$, and $b \to d$.    □

**Exercise 5.1.3** Show that $a \stackrel{+}{\Rightarrow} c$, $a \stackrel{+}{\Rightarrow} d$, and also that $a \stackrel{+}{\Rightarrow} a$, for the TRS of Example 5.1.8.    □

The rest of this section explores the crucial relationship between term rewriting and equational deduction. We first extend rewriting to terms with variables $X$, which we define as ground term rewriting in $T_{\Sigma(X)}$ and denote by $\Rightarrow_{A,X}$ when rules in $A$ are used.

**Proposition 5.1.9** Given $t, t' \in T_\Sigma(Y)$, $Y \subseteq X$, and a $\Sigma$-TRS $A$, then $t \Rightarrow_{A,X} t'$ iff $t \Rightarrow_{A,Y} t'$, and in both cases $var(t') \subseteq var(t)$.

**Proof:** The converse implication is easy, so we assume $t \Rightarrow_{A,X} t'$ with $t = t_0(z \leftarrow \theta(t_1))$, $t' = t_0(z \leftarrow \theta(t_2))$ for $t_1 = t_2 \in A$ with $\theta : var(t_1) \to T_\Sigma(X)$ and $t_0 \in T_\Sigma(X \cup \{z\})$. Since $t, t' \in T_\Sigma(Y)$, we must have $t_0 \in T_\Sigma(Y \cup \{z\})$ as well as $\theta(t_1), \theta(t_2) \in T_\Sigma(Y)$, so that $\theta : var(t_1) \to T_\Sigma(Y)$. Therefore $t \Rightarrow_{A,Y} t'$, and $var(t') \subseteq var(t)$ since $var(t_2) \subseteq var(t_1)$.  □

**Corollary 5.1.10** Given $t, t' \in T_\Sigma(Y)$, $Y \subseteq X$, and $\Sigma$-TRS $A$, then $t \stackrel{*}{\Rightarrow}_{A,X} t'$ iff $t \stackrel{*}{\Rightarrow}_{A,Y} t'$, and in both cases $var(t') \subseteq var(t)$.

**Proof:** By induction using Proposition 5.1.9.  □

Thus both $\Rightarrow_{A,X}$ and $\stackrel{*}{\Rightarrow}_{A,X}$ restrict and extend well over $X$, so we can drop the variable set subscript and write just $t \stackrel{*}{\Rightarrow}_A t'$, with the understanding that any $X$ such that $var(t) \subseteq X$ may be used.

**Exercise 5.1.4** Show for any finite $X$, there are $A, t, t'$ such that $t \Rightarrow_A t'$ but $t \Rightarrow_{A,X} t'$ fails.  □

The following gives soundness for term rewriting:

**Proposition 5.1.11** Given $\Sigma$-TRS $A$ and $t_1, t_2 \in T_\Sigma(X)$, then $t_1 \stackrel{*}{\Rightarrow}_A t_2$ implies $A \vdash (\forall X)\, t_1 = t_2$.

**Proof:** Each single step of rewriting is an application of the rule (RW), so soundness follows from the fact that (RW) is a special case of the rule ($+6_1$) of Chapter 4, which we have already shown sound. Induction then extends this from $\stackrel{1}{\Rightarrow}_A$ to $\stackrel{*}{\Rightarrow}_A$.  □

**Definition 5.1.12** Given $\Sigma$-TRS $A$ and $t_1, t_2 \in T_\Sigma(X)$, write $t_1 \downarrow_{A,X} t_2$ iff there is some $\Sigma$-term $t$ such that $t_1 \stackrel{*}{\Rightarrow}_A t$ and $t_2 \stackrel{*}{\Rightarrow}_A t$; in this case, we say that $t_1$ and $t_2$ are **convergent**, or **converge** (**to** $t$). We may refer to a configuration $t_1 \stackrel{*}{\Rightarrow}_A t, t_2 \stackrel{*}{\Rightarrow}_A t$ as a **join** or a **V**.  □

**Proposition 5.1.13** Given $\Sigma$-TRS $A$ and $t_1, t_2 \in T_\Sigma(Y)$ with $Y \subseteq X$, then $t_1 \downarrow_{A,X} t_2$ iff $t_1 \downarrow_{A,Y} t_2$, so the variable set subscript can be dropped. Moreover, $t_1 \downarrow_A t_2$ implies $A \vdash (\forall X)\, t_1 = t_2$.  □

**Exercise 5.1.5** Prove Proposition 5.1.13.  □

**Proposition 5.1.14** Given $\Sigma$-TRS $A$ and $t, t' \in T_\Sigma(X)$, then $t \stackrel{*}{\Leftrightarrow}_A t'$ iff there are $t_1, \ldots, t_n \in T_\Sigma(X)$ such that $t \downarrow_A t_1$ and $t_i \downarrow_A t_{i+1}$ for $i = 1, \ldots, n-1$ and $t_n \downarrow_A t'$.

**Proof:** If $R$ denotes the transitive closure of $\downarrow_A$ then we wish to show that $t \stackrel{*}{\Leftrightarrow}_A t'$ iff $tRt'$. Since $\Rightarrow_A \subseteq \downarrow_A \subseteq \stackrel{*}{\Leftrightarrow}_A$, it follows that $\stackrel{*}{\Rightarrow}_A \subseteq R \subseteq \stackrel{*}{\Leftrightarrow}_A$. But $R$ is reflexive and symmetric because $\downarrow_A$ is. Therefore $R = \stackrel{*}{\Leftrightarrow}_A$.  □

Corollary 5.1.10 and Proposition 5.1.13 show that $\overset{*}{\Rightarrow}_{A,X}$ and $\downarrow_{A,X}$ restrict and extend reasonably over variables, whereas $\overset{*}{\Leftrightarrow}_{A,X}$ does not, because in Example 5.1.15 below, $\mathsf{T} \overset{*}{\Leftrightarrow}_{\mathsf{FOO},\{x\}} \mathsf{F}$ holds but $\mathsf{T} \overset{*}{\Leftrightarrow}_{\mathsf{FOO},\emptyset} \mathsf{F}$ fails. Nevertheless, it makes sense to let $t \overset{*}{\Leftrightarrow}_A t'$ mean there exists an $X$ such that $t \overset{*}{\Leftrightarrow}_{A,X} t'$, and we shall do so.

**Example 5.1.15** Using the specification of Example 4.3.8 and letting $A = \mathsf{FOO}$, then over the signature $\Sigma(\{x\})$ where $x$ has sort $A$, we have

$$\mathsf{T} \downarrow \mathsf{foo}(x) \vee \neg\mathsf{foo}(x) \downarrow \mathsf{foo}(x) \downarrow \mathsf{foo}(x) \& \neg\mathsf{foo}(x) \downarrow \mathsf{F},$$

which implies the valid equation $(\forall x)\ \mathsf{T} = \mathsf{F}$, but not the invalid equation $(\forall \emptyset)\ \mathsf{T} = \mathsf{F}$. □

In fact, $\overset{*}{\Leftrightarrow}_A$ is complete, even though $\overset{*}{\Leftrightarrow}_{A,X}$ may not be when $X$ is finite:

**Theorem 5.1.16** Given $\Sigma$-TRS $A$ and $t, t' \in T_\Sigma(X)$, then $t \overset{*}{\Leftrightarrow}_A t'$ if and only if $A \vdash (\forall X)\ t = t'$.

**Proof:** By Proposition 5.1.11, we need only prove the converse direction, so assume we have a proof for $A \vdash (\forall X)\ t = t'$. Any such proof necessarily starts with (1) and then chains forward using $(3, 6_1)$ until an application of (2) occurs. Unless this chain gives $t'$ or is a dead end, its final term must also be the final term of another chain, in which case we have a join. Similarly, the whole proof is a set of joins, which can only be put together without dead ends if they form a sequence as in Proposition 5.1.14, which then gives $t \overset{*}{\Leftrightarrow}_A t'$. □

One might think that Proposition 5.1.14 could give a decision procedure for $\overset{*}{\Leftrightarrow}_A$ and hence for equality under $A$, since only term rewriting is involved, but this is not the case, because it can be difficult to find the appropriate $t_i$. In fact, the problem is unsolvable, for reasons discussed in Section 5.10.

## 5.2 Canonical Form

When we compute, we usually hope to get a unique well-defined answer in the end. The following gives one necessary condition for this to occur for every term using a given set of rules; we also give a version that applies to all terms of a particular sort.

**Definition 5.2.1** A $\Sigma$-TRS $A$ is **terminating** (also called **Noetherian**) iff there is no infinite sequence $t_1, t_2, t_3, \ldots$ of $\Sigma$-terms such that

$$t_1 \overset{1}{\Rightarrow} t_2 \overset{1}{\Rightarrow} t_3 \overset{1}{\Rightarrow} \ldots .$$

Similarly, $A$ is **ground terminating** iff there is no such infinite sequence of ground terms, and $A$ is **terminating** (or **ground terminating**) **for sort** $s$ iff there is no such sequence (of ground terms), all of sort $s$. □

For example, if we add a commutative law for addition to the specification NATP+ of Example 5.1.5, then there are computations like the following that do not terminate:

$$0 + s\,0 \;\overset{1}{\Rightarrow}\; s\,0 + 0 \;\overset{1}{\Rightarrow}\; 0 + s\,0 \;\overset{1}{\Rightarrow}\; \ldots \;.$$

If $A$ is terminating, then every $\Sigma$-term has a normal form, but some $\Sigma$-terms may have more than one normal form. However, the following condition will guarantee the *uniqueness* of normal forms for terminating TRS's; again we also give a version for terms of a particular sort.

**Definition 5.2.2** A $\Sigma$-TRS $A$ is **Church-Rosser**[5] (also called **confluent**) iff for every $\Sigma$-term $t$, whenever $t \overset{*}{\Rightarrow} t_1$ and $t \overset{*}{\Rightarrow} t_2$, there is some $\Sigma$-term $t_3$ such that $t_1 \overset{*}{\Rightarrow} t_3$ and $t_2 \overset{*}{\Rightarrow} t_3$. Similarly, $A$ is **Church-Rosser for sort** $s$ iff this condition holds for all $t$ of sort $s$. A $\Sigma$-TRS $A$ is **canonical** (or sometimes **convergent** or **complete**) iff it is terminating and Church-Rosser, and is **canonical for sort** $s$ iff it is terminating and Church-Rosser for sort $s$. In these cases, normal forms may also be called **canonical forms**.

A $\Sigma$-TRS $A$ is **ground Church-Rosser** (also called **ground confluent**) iff for every ground $\Sigma$-term $t$, whenever $t \overset{*}{\Rightarrow} t_1$ and $t \overset{*}{\Rightarrow} t_2$, there is a ground $\Sigma$-term $t_3$ such that $t_1 \overset{*}{\Rightarrow} t_3$ and $t_2 \overset{*}{\Rightarrow} t_3$. Similarly, $A$ is **ground canonical** iff it is ground terminating and ground Church-Rosser, and is **ground Church-Rosser for sort** $s$ iff the condition holds for all $t$ of sort $s$. In these cases, normal forms may also be called **ground canonical forms**.

Similarly, a TRS $A$ is **locally Church-Rosser** (or **locally confluent**) iff for every $\Sigma$-term $t$, whenever $t \overset{1}{\Rightarrow} t_1$ and $t \overset{1}{\Rightarrow} t_2$, there is a $\Sigma$-term $t_3$ such that $t_1 \overset{*}{\Rightarrow} t_3$ and $t_2 \overset{*}{\Rightarrow} t_3$. Also, a TRS $A$ is **ground locally Church-Rosser** iff the above condition holds for all ground terms $t$, and is **locally Church-Rosser for sort** $s$ or (**ground locally Church-Rosser for sort** $s$) iff it holds for all (ground) terms $t$ of sort $s$. □

The generalizations to a particular sort are needed because rewriting with a many-sorted TRS may well have different properties for different sorts. Figures 5.1(a) and 5.1(b) illustrate the Church-Rosser and local Church-Rosser properties.

The Church-Rosser property is shown graphically in Figure 5.1(a). For example, our discussion of FOO from Example 4.3.8 shows that the

---

[5]But the French school does not use the terms "Church-Rosser" and "confluent" synonymously, e.g., [109].

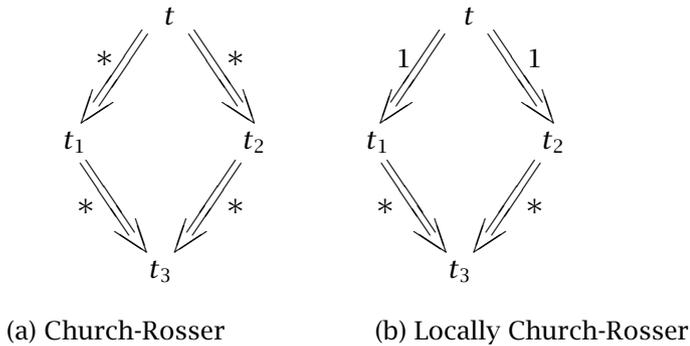

(a) Church-Rosser    (b) Locally Church-Rosser

Figure 5.1: Church-Rosser Properties

resulting TRS is *not* Church-Rosser, because if we let $t = \text{foo}(x)$ & $\neg\text{foo}(x)$, then $t \stackrel{*}{\Rightarrow} F$ and $t \stackrel{*}{\Rightarrow} \text{foo}(x)$, each of which is irreducible; however, this system is terminating. The TRS of Example 5.1.8 is not terminating; it is also not Church-Rosser, because both $c$ and $d$ are normal forms of $a$. The following result is immediate from Definitions 5.2.1 and 5.2.2:

**Fact 5.2.3** If $(\Sigma, A)$ is Church-Rosser, then it is ground Church-Rosser; if it is terminating then it is ground terminating; and if it is canonical then it is ground canonical. □

However, a ground Church-Rosser TRS is not necessarily Church-Rosser, and a ground canonical TRS is not necessarily canonical, as the following shows:

**Example 5.2.4** Consider the following variant of the theory of monoids, in which the direction of the associative law has been reversed:

```
th RMON is
  sort Elt .
  op e : -> Elt .
  op _*_ : Elt Elt -> Elt .
  vars X Y Z : Elt .
  eq X * e = X .
  eq (X * Y)* Z = X *(Y * Z).
endth
```

Viewed as a TRS, this has only one reduced ground term, namely e, so it is certainly ground Church-Rosser, ground locally Church-Rosser, ground terminating, and ground canonical. However, it is not Church-Rosser; for example, the term (X * e) * X rewrites to both X * X and X * (e * X), each of which is reduced. □

**Exercise 5.2.1** Show that the following specification of the Peano natural numbers with addition gives a ground canonical TRS that is not Church-Rosser, and hence not canonical:

```
obj RNATP+ is sort Nat .
  op 0 : -> Nat .
  op s_ : Nat -> Nat [prec 2] .
  op _+_ : Nat Nat -> Nat .
  vars X Y Z : Nat .
  eq 0 + X = X .
  eq (s X) + Y = s(X + Y) .
  eq X + (Y + Z) = (X + Y) + Z .
endo
```
□

**Exercise 5.2.2** Show that if a TRS $A$ has a lapse rule of sort $s$ with its rightside a ground term, then $A$ is Church-Rosser for sort $s$. □

**Exercise 5.2.3** Show that the TRS of Example 5.1.8 is locally Church-Rosser. □

We call the next result a "Theorem" and give its proof in detail, even though it is trivial, because it is such a fundamental result about term rewriting:

**Theorem 5.2.5** If a $\Sigma$-TRS $A$ is canonical, then every $\Sigma$-term $t$ has a unique normal form, denoted $[\![t]\!]_A$, or just $[\![t]\!]$ if $A$ is clear from context, and called the canonical form of $t$.

**Proof:** Each $\Sigma$-term $t$ has at least one normal form by the Noetherian property. Suppose that $t_1$ and $t_2$ are two normal forms for $t$. Then by the Church-Rosser property, because $t \stackrel{*}{\Rightarrow} t_1$ and $t \stackrel{*}{\Rightarrow} t_2$, there is a term $t_3$ such that $t_1 \stackrel{*}{\Rightarrow} t_3$ and $t_2 \stackrel{*}{\Rightarrow} t_3$. But because $t_1$ and $t_2$ are both normal forms, we get that $t_1 = t_3$ and $t_2 = t_3$, and hence that $t_1 = t_2$. □

For example, it will follow from later results that the TRS of Example 5.1.5 is canonical, and that its ground normal forms all have the form $s\,s\,\ldots\,0$, with zero or more $s$'s. The results below bring out the very important consequence of Theorem 5.2.5 that every canonical TRS has a natural procedure for deciding the equality of ground terms; the proposition below is proved in Section 5.7 as a consequence of the more abstract Proposition 5.7.6 there.

**Proposition 5.2.6** If $A$ is a Church-Rosser TRS, then $A \models (\forall X)\, t = t'$ iff $t \downarrow t'$. □

**Corollary 5.2.7** If $A$ is a canonical TRS, then

$$A \models (\forall X)\, t = t' \text{ iff } [\![t]\!]_A = [\![t']\!]_A,$$

where the last equality is syntactical identity.

**Proof:** By Proposition 5.2.6, because $t \downarrow t'$ iff $[\![t]\!]_A = [\![t']\!]_A$ when $A$ is canonical. □

This says that when $A$ is canonical, we can decide whether or not an equation $(\forall X)\ t = t'$ is satisfied by all models of $A$ just by comparing the canonical forms of its two sides. Equivalently, it says that we can decide whether or not terms $t, t'$ can be proved equal using the equations in $A$, just by checking whether or not their canonical forms are identical. OBJ provides a function that does exactly this: $t\ ==\ t'$ computes the normal forms of $t$ and $t'$, and then returns `true` if these terms are identical, and `false` otherwise.

For a simple illustration using the TRS of Example 5.1.5, if we show that the terms $s\,s\,0 + s\,s\,s\,0$ and $s(s\,s\,0 + s\,s\,0)$ each reduce to the same thing (namely $s\,s\,s\,s\,s\,0$), then they are provably equal, i.e.,

$$(\forall \varnothing)\ s\,s\,0 + s\,s\,s\,0 = s(s\,s\,0 + s\,s\,0),$$

holds for all models; this is very conveniently done by executing

```
red s s 0 + s s s 0 == s(s s 0 + s s 0) .
```

However, this method *cannot* be used to prove either

$$(\forall x)\ x + s\,s\,s\,0 = s\,s\,s\,0 + x,$$

or the more general equation

$$(\forall x, y)\ x + y = y + x,$$

and, in fact, both of these are false in some models of NATP+.

The situation is as follows: Term rewriting over a canonical $\Sigma$-TRS gives a decision procedure for equality of $\Sigma$-terms with respect to *loose* semantics. But often we are really interested in the *initial* semantics of a specification, e.g., NATP+. In such a case, we can decide the equality of any two *ground* $\Sigma$-terms, i.e., of any two elements of the initial $\Sigma$-algebra, but we cannot decide the equality of two $\Sigma$-terms whose variables are restricted to range over ground terms only, i.e., over the initial algebra. For example, the commutative law is true for every pair $x, y$ of ground terms over NATP+, and hence it is true for the initial algebra, but this cannot be proved just by reduction, it requires *induction*. In fact, the commutative law is *not true* for every choice of elements from every model of NATP+. Thus it is important to remember that this kind of decision procedure only decides equality for loose semantics.

### Exercise 5.2.4

1. Use the Completeness Theorem to prove that there are models of NATP+ where the commutative law fails, without explicitly giving such a model.

2. Now give such a model. □

Clearly it is useful to know if a specification is canonical as a TRS. Sections 5.5 and 5.6 will give several useful tools for proving termination and confluence, respectively. However, it is important (and perhaps surprising) to note that for proving equality, it is *not necessary* for a TRS $A$ to be Church-Rosser or even terminating: if $t$ and $t'$ have equal normal forms under $A$ then the equation $(\forall X)\ t = t'$ is provable from $A$, whether or not $A$ is canonical; canonicity is only needed to guarantee that if $[\![t]\!]_A \neq [\![t']\!]_A$ then the equation $(\forall X)\ t = t'$ is *not* provable from $A$. This explains why OBJ does not require checking confluence or termination before it accepts code, and why we may use the notation $[\![t]\!]_A$ to denote an arbitrary normal form of $t$ even when $A$ is not canonical. We have found that in practice, it is more irritating than useful to prove canonicity, although it would be desirable to have algorithms that could help with this on an optional basis. Experience also shows that OBJ specifications written for many application areas, including programming, are essentially always canonical.

OBJ also provides a function =/= that is the negation of ==. However, it can be dangerous to use when $A$ is not Church-Rosser for the sort of the terms involved, because even if the terms are provably equal, OBJ could compute different normal forms for them. But the result will always be correct if $A$ is canonical for the common sort of $t, t'$ and the subset of rules actually used in computing $t$ =/= $t'$ (however, complications can arise for conditional rules, as discussed in Section 5.8).

**Example 5.2.8** (*Groups*) This OBJ code gives a canonical[6] TRS for the theory of groups:

```
th GROUPC is sort Elt .
  op _*_ : Elt Elt -> Elt .
  op e : -> Elt .
  op _-1 : Elt -> Elt [prec 2] .
  var A B C : Elt .
  eq e * A = A .
  eq A -1 * A = e .
  eq A * e = A .
  eq e -1 = e .
  eq (A * B)* C = A *(B * C) .
  eq A -1 -1 = A .
  eq A * A -1 = e .
  eq A *(A -1 * B) = B .
  eq A -1 *(A * B) = B .
  eq (A * B)-1 = B -1 * A -1 .
endth
```

For example, suppose we want to know whether or not the equation

$$(\forall w, x, y, z)\ ((w * x) * (y * z))^{-1} = ((z^{-1} * y^{-1}) * x^{-1}) * w^{-1}$$

---

[6]Termination is shown in Exercise 5.5.4, while the Church-Rosser property is shown in Exercise 5.6.7.

is true in all groups. In OBJ, we can open the module GROUPC, introduce the variables, and then reduce the left and right sides to see if they have the same normal form:

```
open .
  vars W X Y Z : Elt .
  red ((W * X)*(Y * Z))-1 .
  red ((Z -1 * Y -1)* X -1)* W -1 .
close
```

We could also use the OBJ built-in operation == for this:

```
open .
  vars W X Y Z : Elt .
  red ((W * X)*(Y * Z))-1 == ((Z -1 * Y -1)* X -1)* W -1 .
close
```

This tells whether or not the equation is true of all groups, but it does not tell us what the normal forms are, and that additional information is often useful when trying to build a proof.

Alternatively, using the Theorem of Constants, we can add new constants $a, b, c, d$, and then reduce the left and right sides to see if they have the same normal form. This is equivalent because variables are really just new constants. Here is how that looks in OBJ:

```
open .
  ops a b c d : -> Elt .
  red ((a * b)*(c * d))-1 .
  red ((d -1 * c -1)* b -1)* a -1 .
close
```

It is not necessary to use open and close for this example; we could instead define a new module which enriches GROUPC, and then do the reduction in that context, as follows:

```
th GROUPC+ is
  inc GROUPC .
  ops a b c d : -> Elt .
endth
red ((a * b)*(c * d))-1 == ((d -1 * c -1)* b -1)* a -1 .
```

This uses a feature of OBJ that we have not yet discussed: the previously defined theory GROUPC is "imported" into the current theory by the declaration "including GROUPC", here abbreviated "inc GROUPC"; the effect is exactly the same as if the code in GROUPC were copied into GROUPC+.

New material introduced within an open...close pair is forgotten after the close. If you want it to be added to the module in focus and retained as part of it for future use, you should instead use the pair

openr...close. If you introduced the module GROUPC+ after having done the above proof within an openr...close pair, then you would get parsing errors, because now there would be *two* copies each of a,b,c,d (you can see the problem by typing "show ."). In order to get around this, you could re-enter the theory GROUPC, which has the effect of restoring it to its original state; OBJ will then warn you that GROUPC is being "redefined," but this should not worry you because that is exactly what you wanted to do. Another approach is to type

```
select GROUPC .
red ((a * b)*(c * d))-1 == ((d -1 * c -1)* b -1)* a -1 .
```

This returns focus to the original GROUPC module, which will have retained (one copy of each of) the constants a,b,c,d, provided you previously used openr...close; otherwise you will get a parse error. Thus we see that there is considerable flexibility in how OBJ can be used in proofs of this kind. □

**Exercise 5.2.5** Experiment with the new features of OBJ introduced above, including "select", "openr...close" and "include". □

**Exercise 5.2.6** Assuming canonicity of the TRS of Example 5.2.8, check whether or not the following equations are true of all groups:

1. $(\forall x, y, z)\ (x * y)^{-1} * (x * z) = y^{-1} * z$.
2. $(\forall x, y)\ (x^{-1} * y^{-1})^{-1} = y * x$.
3. $(\forall x, y, z)\ ((x^{-1} * e) * (y^{-1} * z)^{-1})^{-1} = ((y^{-1} * e) * (z * x))^{-1}$.
□

**Exercise 5.2.7** Given the following theory of monoids,

```
th MONOID is
  sort Elt .
  op e : -> Elt .
  op _*_ : Elt Elt -> Elt .
  vars X Y Z : Elt .
  eq X * e = X .
  eq e * X = X .
  eq (X * Y)* Z = X *(Y * Z).
endth
```

check whether or not the following equations are true of all monoids, assuming canonicity of the rules in MONOID:

1. $(\forall x, y)\ (x * e) * y = y * x$.
2. $(\forall x, y, z, w)\ x * (y * (z * w)) = ((x * y) * z) * w$.
3. $(\forall x, y, z)\ ((x * e) * (y * z) = (x * y) * (z * e)$.

Exercise 5.6.7 shows that MONOID is a Church-Rosser TRS. □

The following fundamental result tells us that for canonical specifications, the ground canonical forms give an initial algebra. This is another justification for the use of irreducible terms as the results of computations. The proof makes good use of the initiality of $T_\Sigma$.

**Theorem 5.2.9** If a specification $P = (\Sigma, A)$ is a ground canonical TRS, then the canonical forms of ground terms under $A$ form a $P$-algebra, called the **canonical term algebra** of $P$, denoted $N_P$ or $N_{\Sigma,A}$ or just $N_A$, in the following way:

(0) interpret $\sigma \in \Sigma_{[],s}$ as $[\![\sigma]\!]$ in $N_{P,s}$; and

(1) interpret $\sigma \in \Sigma_{s_1...s_n,s}$ with $n > 0$ as the function that sends $(t_1,\ldots,t_n)$ with $t_i \in N_{P,s_i}$ to $[\![\sigma(t_1,\ldots,t_n)]\!]$ in $N_{P,s}$.

Furthermore, if $M$ is any $P$-algebra, there is one and only one $\Sigma$-homomorphism $N_P \to M$.

**Proof:** Since $N_P$ is a $\Sigma$-algebra by definition, we check that it satisfies $A$. Given $(\forall X)\ t = t'$ in $A$ and $a : X \to N_P$ we also get $b : X \to T_\Sigma$ since[7] $N_P \subseteq T_\Sigma$. Moreover, $\overline{a}(t) = [\![\overline{b}(t)]\!]$ for any $t \in T_\Sigma(X)$ because $[\![\overline{b}(\_)]\!]$ is a $\Sigma$-homomorphism since $[\![\_]\!]$ is, and there is a unique $\Sigma$-homomorphism $T_\Sigma(X) \to N_P$ that extends $a$. Applying the given rule to $t$ with the substitution $b$ gives $\overline{b}(t) \overset{1}{\Rightarrow}_A \overline{b}(t')$, and so these two terms have the same canonical form, i.e., $[\![\overline{b}(t)]\!] = [\![\overline{b}(t')]\!]$. Therefore $\overline{a}(t) = \overline{a}(t')$ for every $a$, and so we are done.

Now let $h : T_\Sigma \to M$ be the unique $\Sigma$-homomorphism. Noting that $N_P \subseteq T_\Sigma$, let us define $g : N_P \to M$ to be the restriction of $h$ to $N_P$. We now check that $g$ is a $\Sigma$-homomorphism, using structural induction on $\Sigma$:

(0) Given $\sigma \in \Sigma_{[],s}$, we get $g(\sigma_{N_P}) = h([\![\sigma]\!])$ by definition. Now Proposition 5.1.16 gives us that $h(\sigma) = h([\![\sigma]\!])$ because $\sigma \overset{*}{\Rightarrow}_A [\![\sigma]\!]$. Therefore $g(\sigma_{N_P}) = \sigma_M$, as desired, because $h(\sigma) = \sigma_M$ since $h$ is a $\Sigma$-homomorphism.

(1) Given $\sigma \in \Sigma_{s_1...s_n,s}$ with $n > 0$, by definition we get

$$g(\sigma_{N_P}(t_1,\ldots,t_n)) = h([\![\sigma(t_1,\ldots,t_n)]\!]).$$

Then Proposition 5.1.16 gives us that

$$h(\sigma(t_1,\ldots,t_n)) = h([\![\sigma(t_1,\ldots,t_n)]\!]),$$

---

[7]The assignments $a, b$ are different functions (in the sense of Appendix C) because they have different targets, and even though $a, b$ have the same values, the functions $\overline{a}, \overline{b}$ have quite different values.

and the fact that $h$ is a $\Sigma$-homomorphism gives us

$$g(\sigma_{N_P}(t_1,\ldots,t_n)) = \sigma_M(h(t_1),\ldots,h(t_n)) = \sigma_M(g(t_1),\ldots,g(t_n)),$$

as desired.

To show uniqueness, suppose $g' : N_P \to M$ is another $\Sigma$-homomorphism. Let $r : T_\Sigma \to N_P$ be the map that sends $t$ to $[\![t]\!]$, and note that it is a $\Sigma$-homomorphism by the definition of $[\![\_]\!]$. Next, note that if $i : N_P \to T_\Sigma$ denotes the inclusion $\Sigma$-homomorphism, then $i;r = 1_{N_P}$. Finally, note that $r;g = r;g' = h$, by the uniqueness of $h$. It now follows that $i;r;g = i;r;g'$, which implies $g = g'$. □

**Exercise 5.2.8** Draw a commutative diagram that brings out the simple equational character of the above uniqueness argument. □

**Exercise 5.2.9** Show that any terminating TRS is lapse free. □

## 5.3 Adding Constants

This section discusses the preservation of some basic properties of a TRS when new constants are added to its signature. This is important because it allows us to conclude that a TRS is terminating from a proof that it is ground terminating, and it can also justify using the Theorem of Constants in theorem proving. Although we may speak of adding "variables" to a signature, of course they are really constants. Recalling that $(\Sigma, A)$ indicates that $A$ is a $\Sigma$-TRS, if $X$ is a suitable variable set, it is convenient to let $A(X)$ denote the TRS $(\Sigma(X), A)$. We begin with a simple but important result:

**Proposition 5.3.1** If a TRS $A$ is terminating, then so is $A(X)$, for any signature of constants $X$ for $\Sigma$. On the other hand, if A is Church-Rosser or locally Church-Rosser, then so is $A(X)$. □

The intuition is that the variables in $X$ are really just constants; however the formal justifications in Propositions 5.3.4 and 5.3.1 use abstract rewrite systems, which have not yet been introduced at this point in the chaper.

Although proofs of the Church-Rosser property generally apply to the non-ground case, so that one does not have to worry about added constants causing trouble, this is not so for termination, which is usually easier to prove for the ground case. By definition (or Fact 5.2.3), any terminating TRS is also ground terminating, but the converse does not hold, as shown by the following simple but important TRS (and also by Example 5.2.4):

**Example 5.3.2** Let $\Sigma$ have just one sort, one unary function symbol $f$, and one rule,

$$f(Z) \to f(f(Z)) \,,$$

where $Z$ is a variable. Because there are no ground terms, there are no ground rewrite sequences at all, and so this TRS is necessarily ground terminating. However, the term $f(X)$ is the start of an infinite rewrite sequence, so it is not terminating; similarly, if we add a constant to the signature, the resulting TRS also fails to terminate. □

This motivates the following:

**Definition 5.3.3** A signature $\Sigma$ is **non-void** iff $(T_\Sigma)_s \neq \emptyset$ for each sort $s$. □

That is, $\Sigma$ is non-void iff each of its sorts is non-void in the sense of Section 4.7. A simple sufficient condition is that each sort has a constant; also, a signature cannot be non-void if it has no constants at all. The next result follows from a more abstract version, Proposition 5.8.10, on page 131:

**Proposition 5.3.4** If $\Sigma$ is non-void, then a TRS $(\Sigma, A)$ is ground terminating iff $(\Sigma(X), A)$ is ground terminating, where $X$ is any signature of constants for $\Sigma$. Also $(\Sigma(X), A)$ is ground terminating if $(\Sigma, A)$ is ground terminating.[E12] If $\Sigma$ is non-void, then a TRS is ground terminating iff it is terminating. □

Therefore when $\Sigma$ is non-void, we know that $A(X)$ is terminating if we know that $A$ is ground terminating; also, when $\Sigma(X)$ is non-void, it is sufficient that $A(X)$ is ground terminating.

Although every Church-Rosser TRS is ground Church-Rosser, the converse is false, and the same holds for the local Church-Rosser property. Moreover, these converse implications fail even when the signature is non-void, as shown by the following:

**Example 5.3.5** Let $\Sigma$ have just one sort, with one constant $a$ plus three unary function symbols, $f, g, h$, and with the following rules:

$$\begin{array}{ll} f(X) \to g(X) & g(a) \to a \\ f(X) \to h(X) & h(a) \to a \end{array}$$

Then every ground term has the same normal form, namely $a$, so this TRS is certainly ground Church-Rosser and also ground locally Church-Rosser; furthermore, it is terminating and ground terminating. However, it is neither Church-Rosser nor locally Church-Rosser. □

(Example 5.3.5 is also a counterexample to the conjecture that every ground locally Church-Rosser TRS is locally Church-Rosser; it also shows that even assuming termination does not help.) Nevertheless, we have the following:

**Proposition 5.3.6** Given a sort set $S$, let $X_S^\omega$ be the signature of constants with $(X_S^\omega)_s = \{x_s^i \mid i \in \omega\}$ for each $s \in S$ (that is, $X_S^\omega$ has a countable number of distinct variable symbols for each sort in $S$, different from those in $\Sigma$). Then a TRS $(\Sigma, A)$ is Church-Rosser iff $(\Sigma(X_S^\omega), A)$ is ground Church-Rosser. Also, $(\Sigma, A)$ is locally Church-Rosser iff $(\Sigma(X_S^\omega), A)$ is ground locally Church-Rosser. □

The proof is given in Section 5.7, as an application of ideas developed there.

The following example and exercise show that the above result is optimal with respect to the number of additional constants.

**Example 5.3.7** Let $\Sigma$ have one sort and binary function symbols, $f, g, h$ with the rules:

$$f(X, Y) \to g(X, Y) \qquad g(X, X) \to X$$
$$f(X, Y) \to h(X, Y) \qquad h(X, X) \to X$$

This TRS is neither Church-Rosser nor locally Church-Rosser, but it is ground Church-Rosser and ground locally Church-Rosser. If we add one constant, the resulting TRS remains ground Church-Rosser and ground locally Church-Rosser, but if we add two constants, that TRS is neither ground Church-Rosser nor ground locally Church-Rosser. So it is not sufficient for $A(\{a\})$ to have the ground property in order for $A$ to have the non-ground property. □

**Exercise 5.3.1** Generalize the above example to show that adding two constants will not suffice. Then show that there is no natural number $n$ such that $n$ additional constants will suffice. □

**Exercise 5.3.2** Apply the results in this section to discuss the situation resulting from adding constants to the theory F00 of Example 4.3.8, and then using rewriting of ground terms to prove equations with variables. □

## 5.4 Evaluation Strategies

In general, a large tree will have many different sites where rewrite rules might apply, and the choice of which rules to try at which sites can strongly affect both efficiency and termination. Most modern functional programming languages have a uniform lazy (i.e., top-down, or outermost, or call-by-name) semantics. But because raw lazy evaluation is slow, lazy evaluation enthusiasts have built clever compilers that figure out when an "eager" (i.e., bottom-up, or innermost, or call-by-value) evaluation can be used with exactly the same result; this is called "strictness analysis" (for example, see [141, 110]).

OBJ is much more flexible, because each operator can be given its own evaluation strategy. Syntactically, a **local strategy**, also called an **E-strategy** (E is for "evaluation"), is a sequence of integers in parentheses, given as an operator attribute following the keyword `strategy`, or just `strat` for short. For example, OBJ's built-in conditional operator has the following declaration

```
op if_then_else_fi : Bool Int Int -> Int [strat (1 0)] .
```

which says its local strategy is to evaluate its first argument until it is reduced, and then apply rules at the top (indicated by "0"). Similarly,

```
op _+_ : Int Int -> Int [strat (1 2 0)] .
```

indicates that `_+_` on `Int` has strategy (1 2 0), which evaluates both arguments before attempting to add them.

Moreover, the flexibility of local evaluation strategies requires minimum effort, because OBJ determines a default strategy for each operator if none is explicitly given. This default strategy is computed very quickly, because only a very simple form of strictness analysis is done, and it is surprisingly effective, though of course it does not fit all possible needs. In OBJ3, the default local strategy for a given operator is determined from its equations by requiring that all argument places that contain a non-variable term in some rule are evaluated before equations are applied at the top. If an operator with a user-supplied local strategy has a tail recursive rule (in the weak sense that the top operator occurs in its rightside), then it may apply an optimization that repeatedly applies that rule, and thus violates the strategy. In those rare cases where it is desirable to prevent this optimization from being applied, you can just give an explicit local strategy that does not have an initial 0.

There are actually two ways to get lazy evaluation. The simplest is to omit a given argument number from the strategy; then that argument is not evaluated unless some rewrite exposes it from underneath the given operator. For example, taking this approach to "lazy cons" gives

```
op cons : Sexp Sexp -> Sexp [strat (0)] .
```

The second approach involves giving a negative number -j in a strategy, which indicates that the $j^{th}$ argument is to be evaluated "on demand," where a "demand" is an attempt to match a pattern to the term that occurs in the $j^{th}$ argument position. Under this approach, lazy cons has the declaration

```
op cons : Sexp Sexp -> Sexp [strat (-1 -2)] .
```

Then a `reduce` command at the top level of OBJ is interpreted as a top-level demand that may force the evaluation of certain arguments. This second approach cannot be applied to operators with an associative or commutative attribute.

A local strategy is called **non-lazy** if it requires that all arguments of its operator are reduced in some order, and either the operator has no rules, or the strategy ends with a final "0". In general, for the results of a reduction command to actually be fully reduced, it is necessary that all local strategies be non-lazy. All of the default local strategies computed by the system are non-lazy.

### 5.4.1 Memoization

Giving an operator the memo attribute causes the results of evaluating a term headed by this operator to be saved, and then used if this term needs to be reduced later in the same context [139]. In OBJ3, users can give any operator the memo attribute, and memoization is implemented efficiently with hash tables. More precisely, given a memoized operator symbol f and given a term f(t1, ...,tn) to be reduced (possibly as part of some larger term), a table entry for f(t1, ...,tn) giving its fully reduced value is added to the memo table, and entries giving this fully reduced value are also added for each term f(r1, ...,rn) that, according to the evaluation strategy for f, could arise while reducing f(t1, ...,tn) just before a rule for f is applied at the top. This is necessary because at that moment the function symbol f could disappear. In some cases, memoizing these intermediate reductions is more valuable than memoizing just the original expression.

For example, if f has the strategy (2 3 0 1 0), let r be the reduced form of the term f(t1,t2,t3,t4), and let $ri$ be the reduced form of $ti$ for $i = 1, 2, 3$. Then the memo table will contain the following pairs:

```
(f(t1,t2,t3,t4),r)
(f(t1,r2,r3,t4),r)
(f(r1,r2,r3,t4),r)
```

Memoization gives the effect of structure sharing for common subterms, and this can greatly reduce term storage requirements in some problems. Whether or not the memo tables are re-initialized before each reduction can be controlled with the top level commands

```
set clear memo on .
set clear memo off .
```

The default is that the tables are *not* reinitialized. However, they can be reinitialized at any time with the command

```
do clear memo .
```

Of course, none of this has any effect on the *result* of a reduction, but only on its speed. A possible exception to this is the case where the definitions of operators appearing in the memo table have been altered. (When rules are added to an open module, previous computations may become obsolete. Therefore, you may need to explicitly give the command "do clear memo .") Memoization is an area where term-rewriting-based systems seem to have an advantage over unification-based systems like Prolog.

## 5.5 Proving Termination

It is known that (ground) termination is undecidable; that is, there is no algorithm which, given a TRS, can decide whether or not it is terminating. Nonetheless one can often prove termination by assigning a "weight" $\rho(t)$ to each term $t$, i.e., by giving a function[8] $\rho : T_\Sigma \to \omega$, such that $\rho(t) > \rho(t')$ whenever $t \stackrel{1}{\Rightarrow} t'$. Because there are no infinite strictly decreasing sequences of natural numbers, it follows that if such a function exists, then the TRS is terminating. The converse also holds under a rather mild assumption.

**Proposition 5.5.1** A $\Sigma$-TRS $A$ is ground terminating if there is a function $\rho : T_\Sigma \to \omega$ such that for all ground $\Sigma$-terms $t, t'$, if $t \stackrel{1}{\Rightarrow}_A t'$ then $\rho(t) > \rho(t')$. Moreover, the converse holds when $A$ is **globally finite**, in the sense that for each term, there are only a finite number of rewrite sequences that begin with it.

**Proof:** If $t_1 \stackrel{1}{\Rightarrow} t_2 \stackrel{1}{\Rightarrow} t_3 \stackrel{1}{\Rightarrow} \cdots \stackrel{1}{\Rightarrow} t_n$, then $\rho(t_1) > \rho(t_2) > \rho(t_3) > \cdots > \rho(t_n)$, that is, a sequence of $n - 1$ proper rewrites gives a strictly decreasing sequence of natural numbers, of length $n$. Hence, because there are no infinite strictly decreasing natural number sequences, there cannot be infinite proper rewrite sequences; therefore $A$ is terminating.

For the converse, assume $A$ is terminating and let $\rho(t)$ be the maximum of the lengths of all rewrite sequences that reduce $t$ to a normal form; there are only a finite number of these, because of global finiteness. Then $t \stackrel{1}{\Rightarrow}_A t'$ implies $\rho(t) \geq 1 + \rho(t')$.  □

The converse direction of this result is of mainly theoretical interest, since it can be difficult to prove global finiteness without knowing termination. But note that a terminating TRS is globally finite if it has a finite rule set.

**Example 5.5.2** Here is a simple TRS[9] showing that the converse of Proposition 5.5.1 does not hold without the additional assumption of global finiteness. The signature $\Sigma$ has just one sort, say $s$, with $\Sigma_{[],s} = \omega$ and $\Sigma_{w,s} = \emptyset$ for $w \neq []$; therefore $T_\Sigma = \omega$. The rule set $A$ is

$$0 \to n \quad \text{for each } n > 0$$
$$n \to n - 1 \quad \text{for each } n > 1 \ .$$

---

[8]In the many-sorted case, the target is the $S$-sorted set with each component $\omega$. We will see later that it is sometimes convenient to replace $\omega$ by certain other ordered sets.

[9]I thank Prof. Yoshihito Toyama for providing this example. Note that $A$ is infinite here.

Then there is a rewrite sequence $0 \Rightarrow n \Rightarrow n - 1 \Rightarrow \cdots \Rightarrow 1$ of length $n$ for every $n > 0$. Now suppose there is a function $\rho : T_\Sigma \to \omega$ such that $t \overset{1}{\Rightarrow} t'$ implies $\rho(t) > \rho(t')$, and let $\rho(0) = K$. But because there is a rewrite sequence of length $K + 1$ beginning $0 \Rightarrow K + 1$, we get that $K = \rho(0) > \rho(K + 1) > \rho(K) > \rho(K - 1) > \cdots > \rho(1)$, which is impossible. □

There are two difficulties with using Proposition 5.5.1: (1) it can be hard to find an appropriate function $\rho$; and (2) it can be hard to prove the required inequalities. We discuss the first difficulty a little later; regarding the second, it is natural to reduce it by using the structure of terms, as in the following definition and result:

**Definition 5.5.3** Given a poset $P$ and $\rho : T_\Sigma \to P$, a $\Sigma$-rewrite rule $r : t \to t'$ of sort $s$ is **strict $\rho$-monotone** iff $\rho(\theta(t)) > \rho(\theta(t'))$ for each applicable ground substitution $\theta$. An operation symbol $\sigma \in \Sigma$ is **strict $\rho$-monotone** iff $\rho(t) > \rho(t')$ implies $\rho(t_0(z \leftarrow t)) > \rho(t_0(z \leftarrow t'))$ for each $t, t' \in T_\Sigma$ and any $t_0 \in T_\Sigma(\{z\}_s)$ of the form $\sigma(t_1, \ldots, t_n)$ where each $t_i$ except one is ground, and that one is just $z$. $\Sigma$-**substitution is strict $\rho$-monotone** if $\rho(t) > \rho(t')$ implies $\rho(t_0(z \leftarrow t)) > \rho(t_0(z \leftarrow t'))$ for any $t, t' \in T_\Sigma$ and $t_0 \in T_\Sigma(\{z\}_s)$ having a single occurrence of $z$. In any of the above, we speak of **weak $\rho$-monotonicity** if $>$ is replaced by $\geq$. □

The first definition says that every application of the rule is weight decreasing; the second says that any weight decreasing substitution decreases the weight of the result; and the third says the same for a single operation symbol.[E13]

**Proposition 5.5.4** Given a $\Sigma$-TRS $A$, if there is a function $\rho : T_\Sigma \to \omega$ such that each rule in $A$ is strict $\rho$-monotone, and $\Sigma$-substitution is strict $\rho$-monotone, then $A$ is ground terminating.

**Proof:** If $t_1 \overset{1}{\Rightarrow} t_2$ then the two assumptions imply that $\rho(t_1) > \rho(t_2)$, because $t_1 = t_0(z \leftarrow \theta(t))$ and $t_2 = t_0(z \leftarrow \theta(t'))$. Therefore $A$ is ground terminating by Proposition 5.5.1. □

The third condition in Definition 5.5.3 is actually a special case of the second that is sufficient to imply it; this fact can greatly simplify many termination proofs; the proof of the following result is given in Appendix B:

**Proposition 5.5.5** Given a $\Sigma$-TRS $A$ and a function $\rho : T_\Sigma \to \omega$, then $\Sigma$-substitution is strict $\rho$-monotone if every operation symbol in $\Sigma$ is strict $\rho$-monotone; the same holds for weak $\rho$-monotonicity. □

We can now directly combine Propositions 5.5.4 and 5.5.5 to get the following useful result:

**Proposition 5.5.6** Given a $\Sigma$-TRS $A$, if (1) each rule in $A$ is strict $\rho$-monotone, and (2′) each $\sigma \in \Sigma$ is strict $\rho$-monotone, then $A$ is ground terminating. □

Note that Section 5.3 gives easy-to-check conditions for a ground terminating TRS to be terminating, so it is not a problem that the above results, and others that follow, only show ground termination.

It is very natural to reduce the tedium of defining an appropriate function $\rho$ by using initial algebra semantics, that is, by giving $\omega$ a $\Sigma$-algebra structure and letting $\rho$ be the unique $\Sigma$-homomorphism. Then to prove that $A$ is terminating, we can prove that each rule is weight decreasing, and that each operation $\omega_\sigma$ on $\omega$ is strict $\rho$-monotone in each argument on $\omega$. The two hypotheses can be stated using variables that range over terms of the appropriate sorts, and the resulting inequalities can be proved by rewriting, as illustrated by examples in this and Section 5.8.2. A subtle point is that if we know that $\rho(t) \geq 1$ for all $t$, then we can assume that all variables are $> 1$ when proving the inequalities over $\omega$ that are induced by the two hypotheses, as illustrated in the following examples.

**Example 5.5.7** Consider the TRS for Boolean conjunction that corresponds to the following OBJ specification:

```
obj AND is sort Bool .
  ops tt ff : -> Bool .
  op _&_ : Bool Bool -> Bool .
  var X : Bool .
  eq X & tt = X .
  eq tt & X = X .
  eq X & ff = ff .
  eq ff & X = ff .
endo
```

Let $\Sigma$ denote its signature, and give $\omega$ the structure of a $\Sigma$-algebra by defining $\omega_{tt} = \omega_{ff} = 1$ and $\omega_\&(m, n) = m + n$. Then by Proposition 5.5.6, it suffices to prove the following, where $\rho$ is the unique $\Sigma$-homomorphism $T_\Sigma \to \omega$,

$\rho(x \,\&\, tt) > \rho(x)$
$\rho(tt \,\&\, x) > \rho(x)$
$\rho(x \,\&\, ff) > \rho(ff)$
$\rho(ff \,\&\, x) > \rho(ff)$

for all $x \in T_\Sigma$; and

$\rho(t) > \rho(t')$ implies $\rho(t_1 \,\&\, t) > \rho(t_1 \,\&\, t')$
$\rho(t) > \rho(t')$ implies $\rho(t \,\&\, t_2) > \rho(t' \,\&\, t_2)$

for all ground $\Sigma$-terms $t, t', t_1, t_2$. (The first four inequalities come from (1) and the next two from (2′).) Each of these six assertions can be proved straightforwardly, using $\rho(t) \geq 1$ for all $t$. For example, the first and third amount to

$$\rho(x) + 1 > \rho(x)$$
$$\rho(x) + 1 > 1$$

while the fifth amounts to

$$i > j \text{ implies } k + i > k + j \,.$$

All the other cases are similar, and so this TRS is ground terminating. Therefore it is terminating by Proposition 5.3.4, and since Exercise 5.6.7 shows it is Church-Rosser, it is also canonical. □

**Exercise 5.5.1** Fill in the missing cases and details in Example 5.5.7. □

**Exercise 5.5.2** Show that the TRS of Example 5.1.5 is terminating by applying Proposition 5.5.6 with $\rho$ the unique homomorphism into $\omega$ satisfying the following:

$$\rho(0) = 2$$
$$\rho(s\,t) = 1 + \rho(t)$$
$$\rho(t + t') = 1 + \rho(t)\rho(t') \,.$$
□

**Exercise 5.5.3** Show that the TRS of Example 5.1.7 is terminating. □

**Exercise 5.5.4** Show that the group TRS of Example 5.2.8 is terminating by applying Proposition 5.5.6 with $\rho$ the unique homomorphism into $\omega$ satisfying the following:

$$\rho(e) = 2$$
$$\rho(t^{-1}) = 2^{\rho(t)}$$
$$\rho(t * t') = \rho(t)^2 \rho(t') \,.$$

You may also enjoy mechanizing the proofs using OBJ, in the manner illustrated in Exercise 5.5.5 below. □

Termination proofs in the literature tend to use polynomials, but the initial algebra point of view makes it evident that any monotone function at all can be used, e.g., exponentials, as in the function for group inverse in the above exercise. Moreover, Proposition 5.5.6 generalizes to partially ordered sets other than $\omega$, provided they are **Noetherian**, in the sense that they have no infinite sequence of strictly decreasing elements; this is discussed in detail in Section 5.8.3 below. OBJ can often be used to do termination proofs based on Proposition 5.5.6, because as we have seen, these boil down to proving a set of inequalities, which are often rather fun in themselves. We illustrate this in the following:

**Exercise 5.5.5** The purpose of this exercise is to show termination for the following specification of the function $half(n)$, which computes the largest natural number $k$ such that $2k \leq n$:

```
obj NATPH is sort Nat .
  op 0 : -> Nat .
  op s_ : Nat -> Nat [prec 2] .
  op _+_ : Nat Nat -> Nat .
  op half : Nat -> Nat .
  var N M : Nat .
  eq N + 0 = N .
  eq N + s M = s(N + M) .
  eq half(0) = 0 .
  eq half(s 0) = 0 .
  eq half(s s M) = s half(M) .
endo
```

The first step is to define an appropriate function $\rho : T_\Sigma \to \omega$, where $\Sigma$ is the signature of NATPH.

1. Give a fourth equation which, when added to the three below, uniquely defines a function $\rho$, which should furthermore satisfy the properties in items 2–5 below:

$$\begin{aligned} \rho(0) &= 2 \\ \rho(s(t)) &= 2 + \rho(t) \\ \rho(t + t') &= 2 + \rho(t)\rho(t') . \end{aligned}$$

Explain why this uniquely defines $\rho$.

The following should first be proved by hand, and then proved using OBJ proof scores based on the object NATP+*> given below:

2. $\rho(t) > \rho(t')$ implies $\rho(s(t)) > \rho(s(t'))$ for every $t, t' \in T_\Sigma$.

3. $\rho(t) > \rho(t')$ implies $\rho(t + t'') > \rho(t' + t'')$ for every $t, t', t'' \in T_\Sigma$.

4. $\rho(t) > \rho(t')$ implies $\rho(half(t)) > \rho(half(t'))$.

5. $\rho(\theta(t)) > \rho(\theta(t'))$ for each rule $t \to t'$ in NATPH and each substitution $\theta$.

**Hints:** You may assume that adding the equations

```
  eq L + M > L + N = M > N .
  cq L * M > L * N = M > N if L > 0 .
  cq s M    > N    = M > N if M > N .
  cq L + M > L    = true   if M > 0 .
  cq M     > 0    = true   if M > s 0 .
```

has already been justified by earlier OBJ proofs; you may also need some other similar lemmas. (We will later see how to prove such results by induction.) You may need to use the fact that $\rho(t) > 1$ for any term $t$.

```
obj NATP+*> is sort Nat .
  ops 0 1 2 : -> Nat .
  op s_ : Nat -> Nat [prec 1] .
  eq 1 = s 0 .
  eq 2 = s 1 .
  vars L M N : Nat .
  op _+_ : Nat Nat -> Nat [assoc comm prec 3] .
  eq M + 0 = M .
  eq M + s N = s(M + N).
  op _*_ : Nat Nat -> Nat [assoc comm prec 2] .
  eq M * 0 = 0 .
  eq M * s N = M * N + M .
  eq L * (M + N) = L * M + L * N .
  op _>_ : Nat Nat -> Bool .
  eq M > M = false .
  eq s M > 0 = true .
  eq s M > M = true .
  eq 0 > M = false .
  eq s M > s N = M > N .
endo
```

6. Explain why the above results prove that NAPTH is terminating. □

**Exercise 5.5.6** Give mechanical proofs for the other termination examples in this section. □

## 5.6 Proving Church-Rosser

Like termination, the Church-Rosser property is undecidable for term rewriting systems. But once again, there are useful techniques for many special cases. We start with the following important result, the clever proof of which is due to Barendregt [3]; this proof is easier to understand when done dynamically on a whiteboard or applet, than statically on paper.

**Proposition 5.6.1** (*Newman Lemma*) If a TRS $A$ is terminating, then it is Church-Rosser if and only if it is locally Church-Rosser.

**Proof:** The "only if" direction is trivial. For the converse, let us call a term $t$ **ambiguous** iff it has (at least) two distinct normal forms. If we can show that when $t$ is ambiguous, there is some ambiguous $t'$ such that $t \stackrel{1}{\Rightarrow} t'$, it then follows that if there are any ambiguous terms, then the system is non-terminating. Hence by contradiction, the system cannot have any ambiguous terms. The Church-Rosser property follows from this.

We now prove the auxiliary claim. Assume that $t$ is ambiguous, and let $t_1$ and $t_2$ be two distinct normal forms for $t$, where $t \stackrel{1}{\Rightarrow} t'_1 \stackrel{*}{\Rightarrow} t_1$ and

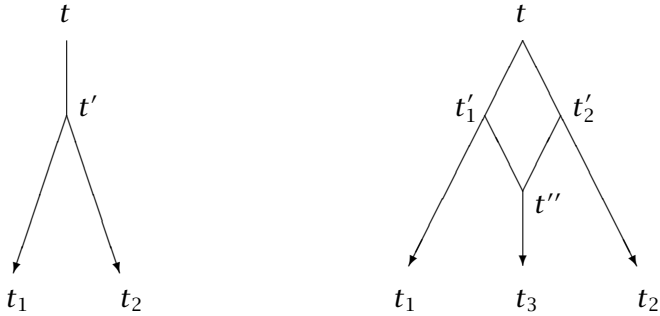

Figure 5.2: Barendregt's Proof of the Newman Lemma

$t \overset{1}{\Rightarrow} t'_2 \overset{*}{\Rightarrow} t_2$. If $t'_1 = t'_2$, let $t' = t'_1$. If $t'_1 \neq t'_2$, apply local confluence to get $t''$ such that $t \overset{1}{\Rightarrow} t'_1 \overset{*}{\Rightarrow} t''$ and $t \overset{1}{\Rightarrow} t'_2 \overset{*}{\Rightarrow} t''$, and let $t_3$ be a normal form for $t''$. Then $t_3$ is also a normal form for $t$, so that $t_3 \neq t_1$ or $t_3 \neq t_2$. If $t_3 \neq t_1$ we let $t' = t'_1$ and if $t_3 \neq t_2$ we let $t' = t'_2$. See Figure 5.2. □

Notice that although the TRS of Example 5.1.8 is locally Church-Rosser by Exercise 5.2.3, it is neither terminating nor Church-Rosser, nor does it have unique normal forms for all its terms, by Exercise 5.1.3. This shows that confluence and local confluence are not equivalent concepts.

Theorem 5.6.9 below gives a method for showing the local Church-Rosser property, and then Corollary 5.6.10 applies the Newman lemma to get a method for showing the Church-Rosser property, and hence the canonicity, of a terminating TRS. Chapter 12 will show how to make this method into a general algorithm.

**Definition 5.6.2** A TRS is **left linear** iff no rule has more than one instance of the same variable in its leftside, and is **right linear** iff no rule has more than one instance of the same variable in its rightside.

Two rules with leftsides $t, t'$ **overlap** iff there are substitutions $\theta, \theta'$ such that $\theta(t_0) = \theta'(t')$ with $t_0$ a subterm of $t$ not just a variable, i.e., with $t = t_1(z \leftarrow t_0)$ where $t_1$ has just one occurrence of the new variable $z$ and $t_0$ is not a variable. If the two rules are actually the same, it is additionally required that $t_1 \neq z$, and the rule is called **self-overlapping**. A TRS is **overlapping** iff it has two rules (possibly the same) that overlap, and then the term $\theta(t_0) = \theta'(t')$ is called an **overlap** of $t, t'$; otherwise the TRS is **non-overlapping**.

A TRS is **orthogonal** iff it is left linear and non-overlapping. □

**Example 5.6.3** The idempotent rule, of the form $B + B = B$, is not left linear, but the associative and commutative rules are left linear.

We now show that the associative and commutative rules overlap. Let their leftsides be $t = (A + B) + C$ and $t' = A + B$, respectively. Let $t_0$ be the subterm $(A + B)$ of $t$; then $t_0$ is non-trivial because $t_1 = z + C$. Define substitutions $\theta, \theta'$ as follows:

$$\begin{aligned} \theta(A) &= a & \theta'(A) &= a \\ \theta(B) &= b & \theta'(B) &= b \\ \theta(C) &= C. \end{aligned}$$

Then $\theta(t_0) = \theta'(t') = a + b$.

The associative rule is self-overlapping. As before, let its leftside be $t = (A + B) + C$. Let $t_0$ be the subterm $(A + B)$ of $t$; then $t_0$ is non-trivial because $t_1 = z + C$. Now define the substitutions $\theta$ and $\theta'$ by:

$$\begin{aligned} \theta(A) &= a + b & \theta'(A) &= a \\ \theta(B) &= c & \theta'(B) &= b \\ \theta(C) &= C & \theta'(C) &= c. \end{aligned}$$

Then $\theta(t_0) = \theta'(t') = (a + b) + c$. □

**Exercise 5.6.1** Prove that the commutative rule is not self-overlapping.  □

The rather complex proof of the following theorem is given in Appendix B:

**Theorem 5.6.4** (*Orthogonality*)[E14] A TRS is Church-Rosser if it is lapse free and orthogonal. □

**Exercise 5.6.2** Show that a lapse rule overlaps with any non-lapse rule. Give a TRS showing that the lapse free hypothesis is needed in Theorem 5.6.4.
□

**Example 5.6.5** The TRS of the object NATP+ of Example 5.1.5 is lapse free and orthogonal, and therefore Church-Rosser. To prove this, we check for overlap of each rule with itself and each other rule; this gives 4 cases, and the reader can verify that each one fails because of incompatible function symbols. Combining this with Exercise 5.5.2, it follows that NATP+ is canonical. □

**Exercise 5.6.3**   1. Show that the TRS AND of Example 5.5.7 is not orthogonal.

   2. Show that the TRS's MONOID of Exercise 5.2.7 and GROUPC of Example 5.2.8 are not orthogonal. □

Chapter 12 shows that proofs of canonicity by orthogonality can be fully mechanized, by using an algorithm that checks if a given pair of rules is overlapping, and noting that it is trivial to check left linearity.

**Example 5.6.6** (*Combinatory logic*) The motivation for this classical logic is similar to that for the lambda calculus, namely to axiomatize a theory of functions, in this case certain collection of higher-order functions called **combinators**. Here we give it as an equational theory.

The basic operation is to *apply* one combinator to another. The traditional notation for this (which might seem a bit confusing at first) is simple juxtaposition, i.e., the syntactic form denoted __ in OBJ. For example, A B means apply A to B; this might be more explicitly written something like App(A,B), or A . B. This calculus has just one sort, which is denoted T in the OBJ code below; it is the type of functions. Thus, particular combinators will be constants of sort T, even though they represent functions.

The attribute gather (E e) of the operation __ makes it parse left associatively ([90] gives a detailed explanation of how these "gathering patterns" work in OBJ3). For example, A B C would be more explicitly written as App(App(A,B),C), or (A . B) . C. Finally, the let construction used below is just a convenient shorthand for first declaring a constant and then letting it equal a given term; OBJ3 computes the sort for the constant by parsing the term.

```
obj COMBL is sort T .
  op __ : T T -> T [gather (E e)].
  ops S K I : -> T .
  vars L M N : T .
  eq K M N = M .
  eq I M = M .
  eq S M N L = (M L)(N L).
endo
open .
  ops m n p : -> T .
  red S K K m == I m .
  red S K S m == I m .
  red S I I I m == I m .
  red K m n == S(S(K S)(S(K K)K))(K(S K K)) m n .
  red S m n p ==
    S(S(K S)(S(K(S(K S)))(S(K(S(K K)))S))))(K(K(S K K))) m n p .
  red S(K K) m n p == S(S(K S)(S(K K)(S(K S)K)))(K K) m n p .
  let X = S I .
  red X X X X m == X(X(X X)) m .
close
```

The last reduction takes 27 rewrites, which is more than one would like to do by hand.  □

**Exercise 5.6.4** The following refer to Example 5.6.6, and if possible should be done with OBJ.

1. Define $B = S(KS)K$. Then show that $Bxyz = x(yz)$, and hence that $Bxy$ is the composition of functions $x$ and $y$.

2. Define $C$ by $Cxyz = zxy$ and prove that $SI(KK) = CIK$.

3. Define $\omega = SII$ and show that $\omega x = xx$, so that $\omega\omega \stackrel{*}{\Rightarrow} \omega\omega$, which implies that the TRS COMBL is non-terminating.  □

**Exercise 5.6.5** Show that the TRS COMBL of Example 5.6.6 is orthogonal and so Church-Rosser. □

**Exercise 5.6.6** Show that the following TRS's are orthogonal, and since already known to be terminating, therefore canonical:

1. NATPH of Exercise 5.5.5; and
2. ANDNOT of Example 5.1.7. □

Unfortunately, many important Church-Rosser TRS's are not orthogonal, and therefore cannot be checked using Theorem 5.6.4. The following material, which builds on concepts in Definition 5.6.2, and which is further developed in Chapter 12, is much more powerful. The next result is proved in Chapter 12:

**Proposition 12.0.1** If terms $t, t'$ overlap at a subterm $t_0$ of $t$, then there is a **most general overlap** $p$, in the sense that any other overlap of $t, t'$ at $t_0$ is a substitution instance of $p$. □

Note that if the leftsides $t, t'$ of two rules in a TRS have the overlap $\theta(t_0) = \theta'(t')$, then the term $\theta(t)$ can be rewritten in two ways (one for each rule).

**Definition 5.6.7** A most general overlap (in the sense of Proposition 12.0.1) is called a **superposition** of $t$ and $t'$, and the pair of rightsides resulting from applying the two rules to the term $\theta(t)$ is called a **critical pair**. If the two terms of a critical pair can be rewritten to a common term using rules in $A$, then that critical pair is said to **converge** or to **be convergent**. □

Theorem 5.6.9 below is our main result, while the following illustrates the definition above:

**Example 5.6.8** The fourth and sixth rules of Example 5.2.8 overlap. Their leftsides are $t =$ e -1 and $t' =$ A -1 -1, while $t_0 =$ A -1, $\theta(A) =$ e, $\theta' = \emptyset$ (the empty substitution), the superposition is e -1, and the critical pair is e, e -1, each term of which rewrites to e, so that the two different rewrites of $t$ also both yield e. □

**Theorem 5.6.9** (*Critical Pair Theorem*) A TRS is locally Church-Rosser if and only if all its critical pairs are convergent.

**Sketch of Proof:** The converse is easy. Suppose that all critical pairs converge, and consider a term with two distinct rewrites. Then their redexes are either disjoint or else one of them is a subterm of the other, since if two subterms of a given term are not disjoint, one must be contained in the other. If the redexes are disjoint, then the result of applying both rewrites is the same in either order. If the redexes are not disjoint, then either the rules overlap (in the sense of Definition 5.6.2), or else the subredex results from substituting for a variable in the leftside of the rule producing the larger redex. In the first case, the result terms of the two rewrites rewrite to a common term by hypothesis, since the overlap is a substitution instance of the overlap of some critical pair by Proposition 12.0.1. In the second case, the result of applying both rules is the same in either order, though the subredex may have to be rewritten multiple (or zero) times if the variable involved is non-linear. □

The full proof is in Appendix B. This and the Newman Lemma (Proposition 5.6.1) give:

**Corollary 5.6.10** A terminating TRS is Church-Rosser if and only if all its critical pairs are convergent, in which case it is also canonical. □

Chapter 12 introduces unification, an algorithm that can be used to compute all critical pairs of a TRS, and hence to decide the Church-Rosser property for any terminating TRS.

**Exercise 5.6.7** Use Corollary 5.6.10 to show the Church-Rosser property, and hence the canonicity, of the following TRS's:

1. GROUPC of Example 5.2.8;
2. AND of Example 5.5.7; and
3. MONOID of Exercise 5.2.7. □

## 5.7 Abstract Rewrite Systems

Many important results about term rewriting are actually special cases of much more general results about a binary relation on a set. Although this abstraction of term rewriting to the one-step rewrite relation ignores the structure of terms, it still includes a great deal. The classical approach takes an unsorted view of the elements to be rewritten, but here we generalize to sorted sets of elements, enabling applications to many-sorted term rewriting and equational deduction that appear to be new.

**Definition 5.7.1** An **abstract rewrite system** (abbreviated **ARS**) consists of a (sorted) set $T$ and a (similarly sorted) binary relation $\rightarrow$ on $T$, i.e., $\rightarrow \subseteq T \times T$. We may denote such a system as a pair $(T, \rightarrow)$, or possibly as a triple $(S, T, \rightarrow)$, if the sort set $S$ needs to be emphasized.

An ARS $(T, \rightarrow)$ is **terminating** if and only if there is no infinite sequence $a_1, a_2, \ldots$ of elements of $T$ such that $a_i \rightarrow a_{i+1}$ for $i = 1, 2, \ldots$ (note that $a_i \in T_s$ and $a_i \to_s a_{i+1}$ for the same $s \in S$). Also, $t \in T$ is called **reduced**, or a **reduced form**, or a **normal form**, iff there is no $t' \in T$ such that $t \to t'$.

Let $\stackrel{*}{\to}$ denote the reflexive, transitive closure of $\to$. Then $t$ is called a **normal form of** $t_0 \in T$ iff $t_0 \stackrel{*}{\to} t$ and $t$ is a normal form. Let $t_1 \downarrow t_2$ mean there is some $t \in T$ such that $t_1 \stackrel{*}{\to} t$ and $t_2 \stackrel{*}{\to} t$; we say $t_1, t_2$ **are convergent**, or **converge to** $t$. An ARS is **Church-Rosser** (also called **confluent**) iff for every $t \in T$, whenever $t \stackrel{*}{\to} t_1$ and $t \stackrel{*}{\to} t_2$ then $t_1 \downarrow t_2$. An ARS is **canonical** iff it is terminating and Church-Rosser; in this case, normal forms are called **canonical forms**. Let $\stackrel{*}{\leftrightarrow}$ denote the reflexive, symmetric, transitive closure of $\to$, and let $\stackrel{0,1}{\to}$ denote the reflexive closure of $\to$.

An ARS is **locally Church-Rosser** (or **locally confluent**) iff for every $t \in T$, whenever $t \to t_1$ and $t \to t_2$ then $t_1 \downarrow t_2$. An ARS is **globally finite** iff for every $t \in T$, there are only a finite number of distinct maximal rewrite sequences (finite or infinite) that begin with $t$. □

We can relativize all these concepts to a single sort $s$ just as in Definitions 5.2.1 and 5.2.2, to take account of the fact that rewriting over different sorts may have different properties. Three of the more useful TRS results that generalize to ARS's are as follows:

**Theorem 5.7.2** Given a canonical ARS, every $t \in T$ has a unique normal form, denoted $[\![t]\!]$ and called the canonical form of $t$. □

**Proposition 5.7.3** Given an ARS $(T, \to)$ and $t, t' \in T$, then $t \stackrel{*}{\leftrightarrow} t'$ iff there are $t_1, \ldots, t_n \in T$ such that $t \downarrow t_1$ and $t_i \downarrow t_{i+1}$ for $i = 1, \ldots, n-1$ and $t_n \downarrow t'$. □

**Proposition 5.7.4** (*Newman Lemma*) A terminating ARS is Church-Rosser iff it is locally Church-Rosser. Hence any ARS that is terminating and locally Church-Rosser is canonical. □

These results are proved essentially the same way as the corresponding TRS results (the second generalizes Proposition 5.1.14). Alternatively, they can be proved directly from the TRS results, by using the connection between ARS's and TRS's that we now describe; a more detailed discussion of this connection appears in Section 5.9.

Given a TRS $\mathcal{T} = (\Sigma, A)$ where $\Sigma$ has sort set $S$, we get an ARS $(S, T, \to)$ by letting $T_s = (T_\Sigma)_s$ and for $t_1, t_2 \in T_s$, defining $t_1 \to_s t_2$ iff $t_1 \stackrel{1}{\Rightarrow}_A t_2$; denote this ARS by $R(\mathcal{T})$. It is suitable for dealing with ground properties of its TRS.

**Exercise 5.7.1** Prove that a TRS $\mathcal{T}$ is ground terminating iff the ARS $R(\mathcal{T})$ is terminating. Prove that a TRS $\mathcal{T}$ is ground Church-Rosser iff the ARS $R(\mathcal{T})$ is Church-Rosser. Also prove corresponding results for the local Church-Rosser and canonicity properties. □

Next, an ARS $\mathcal{A} = (T, \rightarrow)$ gives rise to a TRS $F(\mathcal{A}) = (\Sigma^T, A^\rightarrow)$ as follows: define $\Sigma^T$ by letting $\Sigma^T_{[],s} = T_s$ and $\Sigma_{w,s} = \emptyset$ for all other pairs $w, s$, with the rules in $A^\rightarrow$ the equations $(\forall \emptyset)\ t_1 = t_2$ such that $t_1 \rightarrow_s t_2$ in $\mathcal{A}$ for some sort $s$.

**Exercise 5.7.2** Prove that an ARS $\mathcal{A}$ is terminating iff the TRS $F(\mathcal{A})$ is ground terminating. Do the same for the Church-Rosser, local Church-Rosser, and canonicity properties. □

It is usually easier to prove results about ARS's than about TRS's, but like most bridges, this one can be used in either direction, as illustrated in the following:

**Exercise 5.7.3** Prove Theorem 5.7.2 and Proposition 5.7.4 by reducing them to the corresponding results for TRS's. Also do the reverse for the ground case. □

The following gives another tool for showing the Church-Rosser property:

**Proposition 5.7.5** (*Hindley-Rosen Lemma*) For each $i \in I$, let $(T, \rightarrow_i)$ be a Church-Rosser ARS, and assume that for all $i, j \in I$ the relations $\rightarrow_i$ and $\rightarrow_j$ **commute** in the sense that for all $a, b, c \in T$, if $a \xrightarrow{*}_i b$ and $a \xrightarrow{*}_j c$ then there is some $d \in T$ such that $b \xrightarrow{*}_j d$ and $c \xrightarrow{*}_i d$. Now define $\rightarrow$ on $T$ by $a \rightarrow b$ iff there is some $i \in I$ such that $a \rightarrow_i b$. Then $(T, \rightarrow)$ is Church-Rosser.

**Proof:** We begin by showing that it suffices to prove this result for the case where $I$ has just two indices. First, notice that since any particular rewrites $a \xrightarrow{*} b$ and $a \xrightarrow{*} c$ can only involve a finite set of relations $\rightarrow_i$, it suffices to consider finite sets $I$ of indices. Now assuming that Hindley-Rosen holds for index sets of cardinality 2, we show that it holds for any finite cardinality $k$, by induction on $k$. For $k = 1$, there is nothing to prove. Now assume Hindley-Rosen for some $k > 2$, and suppose we are given relations $\rightarrow_i$ for $i = 1, \ldots, k+1$ which are Church-Rosser and commute. Then $\rightarrow_0 = \bigcup_{i=1}^{k} \rightarrow_i$ is Church-Rosser by the induction hypothesis. Therefore if we show that $\rightarrow_0$ and $\rightarrow_{k+1}$ commute, we are done by Hindley-Rosen for $k = 2$.

To show that $\rightarrow_0$ and $\rightarrow_{k+1}$ commute, let $a \xrightarrow{*}_0 b$ and $a \xrightarrow{*}_{k+1} c$. The proof is by induction on the length $n$ of the rewrite sequence for $a \xrightarrow{*}_0 b$. If $n = 1$, we apply Hindley-Rosen for $k = 2$. Now assuming the hypothesis for some $n > 1$, we prove it for $a \xrightarrow{*}_0 b$ of length $n + 1$. Let the first rewrite in the sequence for $a \xrightarrow{*}_0 b$ be $a \xrightarrow{*}_i b_1$. Then by Hindley-Rosen for $k = 2$, there is some $d_1$ such that $c \xrightarrow{*}_i d_1$ and $b_1 \xrightarrow{*}_{k+1} d_1$. We now conclude the proof by applying the induction hypothesis to $b_1 \xrightarrow{*}_0 b$ and $b_1 \xrightarrow{*}_{k+1} d_1$, noting that the former has length $n$, to get $d$ such that $b \twoheadrightarrow_{k+1}^* d$ and $d_1 \twoheadrightarrow_0^* d$, and hence also $c \twoheadrightarrow_0^* d$ (see Figure 5.3, in which every arrow has an omitted $*$, and each downward arrow is $\twoheadrightarrow_{k+1}^*$).

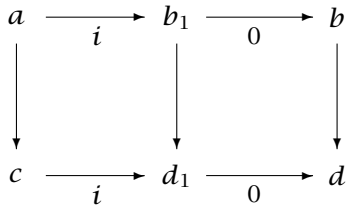

Figure 5.3: Hindley-Rosen Proof Reduction

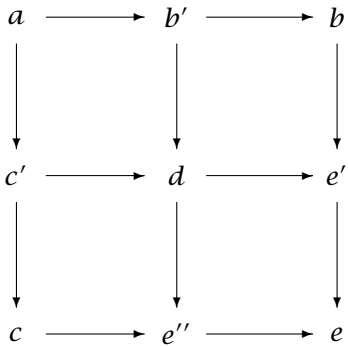

Figure 5.4: Hindley-Rosen Proof for $k = 2$

We now prove Hindley-Rosen for $k = 2$. Let $\to_1$ and $\to_2$ be commuting Church-Rosser relations, let $\to \,=\, \to_1 \cup \to_2$, and let $\to_0$ be arbitrarily many applications of $\to_1$ followed by arbitrarily many applications of $\to_2$, i.e., $\to_0 \,=\, \twoheadrightarrow_1^* \circ \twoheadrightarrow_2^*$. Then the argument suggested by Figure 5.4 shows that $\to_0$ is Church-Rosser, where all leftside horizontal arrows are $\twoheadrightarrow_1^*$, all rightside horizontal arrows are $\twoheadrightarrow_2^*$, all top downward arrows are $\twoheadrightarrow_1^*$, and all bottom downward arrows are $\twoheadrightarrow_2^*$. Next, one can check that $\to \,\subseteq\, \to_0 \,\subseteq\, \twoheadrightarrow^*$, which implies that $\twoheadrightarrow^* \,=\, \twoheadrightarrow_0^*$. Therefore $\twoheadrightarrow^*$ is also Church-Rosser, and hence so is $\to$ (by Exercise 5.7.6). □

**Exercise 5.7.4** Show that there is no analog of the Hindley-Rosen Lemma for termination: give terminating ARS's $(T, \to_1)$ and $(T, \to_2)$ which commute in the sense of Proposition 5.7.5, such that $(T, \to)$ is not terminating, where $\to \,=\, \to_1 \cup \to_2$. □

**Exercise 5.7.5** Prove a more convenient version of Hindley-Rosen, which replaces commutation with the following notion of **strong commutation**:

if $a \to_i b$ and $a \to_j c$ then there is some $d \in T$ such that $b \twoheadrightarrow_j^{0,1} d$ and $c \twoheadrightarrow_i^* d$

where $\stackrel{0,1}{\to}_j$ indicates the reflexive closure of $\to_j$. □

Although the definition of strong commutation is assymetrical, in practice it is used in situations where it holds symmetrically, in both orders.

**Exercise 5.7.6** Prove that an ARS $(T, \to)$ is Church-Rosser iff $(T, \stackrel{*}{\to})$ is Church-Rosser. □

There are also ARS versions of Proposition 5.2.6 and its Corollary 5.2.7. For the first of these, recall that $\stackrel{*}{\leftrightarrow}$ denotes the reflexive, symmetric, transitive closure of the relation $\to$.

**Proposition 5.7.6** If $(A, \to)$ is a Church-Rosser ARS, then $t \stackrel{*}{\leftrightarrow} t'$ iff $t \downarrow t'$.

**Proof:** We use induction on the number $n$ of rewrites involved in $t \stackrel{*}{\leftrightarrow} t'$. If $n = 0$ then $t = t'$ so that $t \downarrow t'$ trivially. Now suppose that $t \stackrel{*}{\leftrightarrow} t'$ with $n+1$ rewrites. We consider 2 cases, (1) $t \stackrel{*}{\leftrightarrow} t'' \leftarrow t'$, and (2) $t \stackrel{*}{\leftrightarrow} t'' \to t'$, where in both cases, $t \stackrel{*}{\leftrightarrow} t''$ in $n$ rewrites, so that $t \downarrow t''$ by the induction hypothesis, say with $t_0$ such that $t \stackrel{*}{\to} t_0$ and $t'' \stackrel{*}{\to} t_0$. For case (1), since $t'' \stackrel{*}{\to} t_0$ and $t' \to t''$, we get $t \downarrow t'$. For case (2), from $t'' \stackrel{*}{\to} t_0$ and $t'' \stackrel{*}{\to} t'$, the Church-Rosser property gives $t_1$ such that $t_0 \stackrel{*}{\to} t_1$ and $t' \stackrel{*}{\to} t_1$, from which it follows that $t \stackrel{*}{\to} t_1$, so that $t \downarrow t'$. □

**Corollary 5.7.7** If $(T, \to)$ is a canonical ARS, then $t \stackrel{*}{\leftrightarrow} t'$ iff $t \downarrow t'$ iff $[\![t]\!] = [\![t']\!]$.

**Proof:** This is because $t \downarrow t'$ iff $[\![t]\!] = [\![t']\!]$ for a canonical ARS, noting that the equality used here is syntactical identity. □

Propositions 5.8.16 on page 134 and 5.8.19 on page 135 give ways to prove termination of ARS's.

So far we have related ARS's to ground term rewriting; extending the relationship to non-ground rewriting can be somewhat tricky, because we must take account not only of the sorts of terms, but also of the sets of variables that appear in terms through universal quantification. As a first example, we provide the proof promised in Section 5.3 for the following result:

**Proposition 5.3.6** Given a sort set $S$, let $X_S^\omega$ be the signature of constants with $(X_S^\omega)_s = \{x_s^i \mid i \in \omega\}$, i.e., with a countable set of distinct variable symbols for each $s \in S$. Then a TRS $(\Sigma, A)$ is Church-Rosser iff $(\Sigma(X_S^\omega), A)$ is ground Church-Rosser. Also, $(\Sigma, A)$ is locally Church-Rosser iff $(\Sigma(X_S^\omega), A)$ is ground locally Church-Rosser.

**Proof:** Given a term $t$ with $var(t) = X$, let $T = T_\Sigma(X)$ and let $G = T_\Sigma(X_S^\omega)$. That the properties for $(\Sigma, A)$ imply the corresponding ground properties for $(\Sigma(X_S^\omega), A)$ is direct. For the converse, form the ARS's $\mathcal{T} = (T, \to_T)$ and $\mathcal{G} = (G, \to_G)$ using rewriting with $A$ on $T$ and on $G$, respectively. Let $f : X \to X_S^\omega$ be an injection, which we can without loss of generality assume is an inclusion, and let $f$ also denote its free extension to terms which is again an inclusion, $T \to G$. Then $t \to_T t'$ iff $t \to_G t'$, and hence $t \xrightarrow{*}_T t'$ iff $t \xrightarrow{*}_G t'$. Now suppose that $(\Sigma(X_S^\omega), A)$ is ground Church-Rosser and let $t \xrightarrow{*}_T t_1$ and $t \xrightarrow{*}_T t_2$. Then $t \xrightarrow{*}_G t_1$ and $t \xrightarrow{*}_G t_2$, with $X = var(t)$. Therefore there exists $t_3$ such that $t_1 \xrightarrow{*}_G t_3$ and $t_2 \xrightarrow{*}_G t_3$, and hence $t_1 \xrightarrow{*}_T t_3$ and $t_2 \xrightarrow{*}_T t_3$, from which it follows that $\mathcal{T}$ is Church-Rosser, and hence that $(\Sigma, A)$ is Church-Rosser, since $t$ was arbitrary. An analogous proof works for the local Church-Rosser property. □

The above proof involves a rewrite relation explicitly indexed over the sorts in $S$, and implicitly indexed over variable sets $X$. To make the latter explicit, we could index over $I = \mathcal{P}_F(X_S^\omega) \times S$, where $\mathcal{P}_F(U)$ denotes the set of all finite subsets of $U$, and where for simplicity all variables are assumed drawn from the fixed signature $X_S^\omega$ introduced above.

For the next result, we need the following construction: Given a TRS $\mathcal{T} = (\Sigma, A)$, define the ARS $N(\mathcal{T}) = (T, \to)$ by $T_s = (T_\Sigma(X_S^\omega))_s$ and $t \to_s t'$ iff $t \xRightarrow{1}_A t'$, for $t, t' \in T_s$. We now apply this machinery to get the proofs that were promised for some results in Section 5.2:

**Proposition 5.2.6** If $\mathcal{T} = (\Sigma, A)$ is a Church-Rosser TRS, then $A \models (\forall X)\, t = t'$ iff $t \downarrow t'$.

**Proof:** First form the ARS $N(\mathcal{T})$ as described above, and notice that $A \vdash (\forall X)\, t = t'$ iff $t \xrightarrow{*}_s t'$ where $t, t'$ both have sort $s$. Now apply Proposition 5.7.6 to $N(\mathcal{T})$, and finally appeal to the Completeness Theorem. □

**Definition 5.7.8** Let $(T, \to)$ and $(T', \to')$ be ARS's. Then $(T', \to')$ is a **sub-ARS** of $(T, \to)$ iff $T' \subseteq T$ and $t_1 \to' t_2$ implies $t_1 \to t_2$. Also, an **ARS isomorphism** of $(T, \to)$ and $(T', \to')$ is a bijective function $f : T \to T'$ such that $t_1 \to t_2$ iff $f(t_1) \to' f(t_2)$. □

**Exercise 5.7.7** Show that if $(T, \to)$ is a terminating ARS and if $(T', \to')$ is a sub-ARS of $(T, \to)$, then $(T', \to')$ is also terminating. Show that by contrast, a sub-ARS of a Church-Rosser ARS need not be Church-Rosser. Also show that if two ARS's are isomorphic, then each of them is terminating, or Church-Rosser, or locally Church-Rosser iff the oher one is. □

We now prove the result stated in Section 5.3 about adding constants to a TRS:

**Proposition 5.3.4** If $\Sigma$ is non-void, then a TRS $(\Sigma, A)$ is ground terminating iff $(\Sigma(X), A)$ is ground terminating, where $X$ is any signature of constants for $\Sigma$. Also $(\Sigma(X), A)$ is ground terminating if $(\Sigma, A)$ is ground terminating,[E15] and if $\Sigma$ is non-void, then a TRS is ground terminating iff it is terminating.

**Proof:** The reader should first check that a TRS $(\Sigma, A)$ is terminating iff $N(\Sigma, A)$ is. Next given an $S$-sorted set $Y$ of constants and a signature isomorphism $f : X_S^\omega \to Y$, we can show that $N(\Sigma, A)$ and $(T_\Sigma(Y), \to_A)$ are isomorphic ARSs, from which it follows by Exercise 5.7.7 that one of them is terminating iff the other is. Finally, since $X$ is countable, $X_S^\omega$ and $X_S^\omega \cup X$ are isomorphic, from which it follows that the TRS $(\Sigma, A)$ is terminating iff $(\Sigma(X), A)$ is, since $N(\Sigma(X), A) = (T_\Sigma(X \cup X_S^\omega), \to_A)$. □

**Proposition 5.3.1** If a TRS $A$ is Church-Rosser or locally Church-Rosser, then so is $A(X)$, for any suitable countable set $X$ of constants.

**Proof:** Let $\mathcal{P}$ stand for either of the above properties. The reader should check that a TRS $(\Sigma, A)$ is $\mathcal{P}$ iff $N(\Sigma, A)$ is $\mathcal{P}$. Next given an $S$-sorted set $Y$ of constants and a signature isomorphism $f : X_S^\omega \to Y$, $N(\Sigma, A)$ and $(T_\Sigma(Y), \to_A)$ are isomorphic ARSs, from which it follows by Exercise 5.7.7 that one of them is $\mathcal{P}$ iff the other is. Finally, since $X$ is countable, $X_S^\omega$ and $X_S^\omega \cup X$ are isomorphic, from which it follows that the TRS $(\Sigma, A)$ is $\mathcal{P}$ iff $(\Sigma(X), A)$ is $\mathcal{P}$, since $N(\Sigma(X), (T_\Sigma(X \cup X_S^\omega), \to_A))$. □

## 5.8 Conditional Term Rewriting

Conditional term rewriting arises naturally from the desire to implement algebraic specifications that have conditional equations in the same way that unconditional rewriting implements unconditional equational specifications. There are many examples of such specifications, and they are very useful in practice, as well as strictly more expressive [176]. Just as unconditional rewrite rules are a special kind of unconditional equation, so conditional rewrite rules are a special kind of conditional equation:

**Definition 5.8.1** A **conditional $\Sigma$-rewrite rule** is a conditional $\Sigma$-equation

$$(\forall X)\ t_1 = t_2\ \text{if}\ C$$

such that $var(t_2) \cup var(C) \subseteq var(t_1) = X$,
where $var(C) = \bigcup_{\langle u,v \rangle \in C}(var(u) \cup var(v))$.

A **conditional $\Sigma$-term rewriting system** (abbreviated $\Sigma$-**CTRS**) is a set of (possibly) conditional $\Sigma$-rewrite rules; we denote such a system by $(\Sigma, A)$, and we may omit $\Sigma$ here and elsewhere if it is clear from context. □

Notation and terminology for conditional term rewriting follow those for the unconditional case. Instead of $(\forall Y)\ t_1 = t_2$ if $C$, we usually write $(\forall Y)\ t_1 \to t_2$ if $C$, and in concrete cases we may write $(\forall Y)\ t_1 \to t_2$ if $u = v$, $(\forall Y)\ t_1 \to t_2$ if $u = v, u' = v'$, etc. The notation $t_1 \to t_2$ if $C$ is unambiguous because $X$ is determined by $t_1$. Also, when $t \stackrel{1}{\Rightarrow} t'$ using a rule in $A$ having leftside $\ell$, with substitution $\theta$ where $t = t^0(z \leftarrow \theta(\ell))$, then the pair $(t^0, \theta)$, is called a **match to** (a subterm of) $t$ by that rule. Unconditional rules are the special case where $C = \emptyset$.

Unfortunately, there is no easy way to generalize the rule (RW) for term rewriting to the conditional case, e.g., by specializing the rule (+6C) from Section 4.9 to replace exactly one subterm using a substitution instance of a conditional rewrite rule. This is because the conditions must be checked, which may lead to further conditional term rewriting, including further condition checking, and so on recursively. We therefore need a recursive definition of the conditional term rewriting relation, and so will define (sorted) relations temporarily denoted $R_k$ and $\overline{R}_k$, with the rewriting relation the union of the $R_k$ and with the $\overline{R}_k$ used for evaluating conditions.

**Definition 5.8.2** Given a CTRS $(\Sigma, A)$ and a set $X$ of variables, let $R_0 = \overline{R}_0 = \{\langle t, t \rangle \mid t \in T_\Sigma(X)\}$, and for $k \geq 0$, let $\langle t, t' \rangle \in R_{k+1}$ iff there exist a conditional rule $(\forall Y)\ t_1 = t_2$ if $C$ of sort $s$ in $A$ and a substitution $\theta : Y \to T_\Sigma(X)$ such that $t = t^0(z \leftarrow \theta(t_1))$ and $t' = t^0(z \leftarrow \theta(t_2))$ for some $t^0 \in T_\Sigma(X \cup \{z\}_s)$, and such that for each $\langle u, v \rangle \in C$ there is some $r$ such that $\langle \theta(u), r \rangle, \langle \theta(v), r \rangle \in \overline{R}_k$. Also let $\overline{R}_{k+1} = (R_{k+1} \cup \overline{R}_k)^*$ and let $R = \bigcup_{k=0}^{\infty} R_k$. Then $R$ is the **conditional term rewriting relation**, hereafter denoted $t \stackrel{1}{\Rightarrow}_A t'$. As usual, $\stackrel{*}{\Rightarrow}_A$ denotes its transitive reflexive closure, and when $X = \emptyset$ we get the ground case. □

Note that it is possible to go into an infinite loop when evaluating the condition of an instance of a rule, in which case the corresponding head is simply not included in $R$. Note also that there exists $r$ such that $\langle \theta(u), r \rangle, \langle \theta(v), r \rangle \in \overline{R}_k$ iff $\theta(u), \theta(v)$ converge under $R_{k-1}$.[E16] It is not hard to check that $t \stackrel{1}{\Rightarrow}_A t'$ iff $\langle t, t' \rangle \in R_k$ for some $k > 0$, and that $R^* = R$. The soundness results in Proposition 5.8.8 are also straightforward. However, we do not prove these results here, because they follow from more general results in Section 7.7.

Many results developed earlier in this chapter for unconditional term rewriting extend to the conditional case. The easiest extensions use the fact that CTRS's give rise to ARS's just as in the unconditional case, so we can directly apply general ARS definitions and results to conditional term rewriting. More specifically, if $P = (\Sigma, A)$ is a CTRS, we let $R(P)$ be the ARS $(T, \to)$ with $T_s = T_\Sigma(X)_s$ and with $t \to_s t'$ iff $t \Rightarrow_A t'$ for $t, t'$ of sort $s$. This gives us the correct notions of termina-tion, normal form, Church-Rosser, local Church-Rosser, and canonicity for CTRS's. For example, $P$ is terminating iff $R(P,s)$ is terminating for each $s \in S$. As with ordinary TRS's, we let $X = \emptyset$ for the ground case, and we choose $X$ large enough for the general case, e.g., $X_S^\omega$ as defined in Proposition 5.3.6. The ARS results Theorem 5.7.2, Proposition 5.7.3, and Proposition 5.7.4 give the following:

**Theorem 5.8.3** Given a canonical CTRS, every $t \in T$ has a unique normal form, denoted $[\![t]\!]$ and called the canonical form of $t$. □

**Proposition 5.8.4** Given CTRS $(\Sigma, A)$ and $t, t' \in T_\Sigma(X)$, then $t \overset{*}{\Leftrightarrow}_A t'$ iff there are terms $t_1, \ldots, t_n \in T_\Sigma(X)$ such that $t \downarrow_A t_1$, $t_i \downarrow_A t_{i+1}$ for $i = 1, \ldots, n-1$, and $t_n \downarrow_A t'$. □

**Proposition 5.8.5** (*Newman Lemma*) A terminating CTRS is Church-Rosser if and only if it is locally Church-Rosser. □

Other results generalize, not through ARS's, but because their proofs generalize. We begin with two results from Section 5.1 that connect rewriting with deduction:

**Proposition 5.8.6** For $t, t' \in T_\Sigma(Y)$, $Y \subseteq X$, and $(\Sigma, A)$ a CTRS, $t \Rightarrow_{A,X} t'$ iff $t \Rightarrow_{A,Y} t'$, and in both cases $var(t') \subseteq var(t)$. □

**Corollary 5.8.7** For $t, t' \in T_\Sigma(Y)$, $Y \subseteq X$, and $(\Sigma, A)$ a CTRS, $t \overset{*}{\Rightarrow}_{A,X} t'$ iff $t \overset{*}{\Rightarrow}_{A,Y} t'$, and in both cases $var(t') \subseteq var(t)$; moreover, $t \downarrow_{A,X} t'$ iff $t \downarrow_{A,Y} t'$. □

As before, this shows that $\overset{*}{\Rightarrow}_{A,X}$ and $\downarrow_{A,X}$ restrict and extend well over variables, which permits dropping the variable set subscripts. On the other hand, noting that TRS's are a special case of CTRS's, Example 5.1.15 shows that $\overset{*}{\Leftrightarrow}_{A,X}$ does not restrict and extend well over variables. The next result gives soundness:

**Proposition 5.8.8** Given CTRS $(\Sigma, A)$ and $t, t' \in T_\Sigma(X)$, then $t \overset{*}{\Rightarrow}_A t'$ implies $A \vdash (\forall X)\, t = t'$. Also $t \downarrow_A t'$ and $t \overset{*}{\Leftrightarrow}_A t'$ both imply $A \vdash (\forall X)\, t = t'$. □

We cannot hope for completeness here, because it is possible, for a condition $t_1 = t_2$, that $A \vdash t_1 = t_2$ but $t_1 \downarrow t_2$ fails. The literature includes several different ways to define conditional term rewriting; the one in Definition 5.8.2 is called **join conditional rewriting**. OBJ implements a special case of this, where there is just one condition in each rule, with its leftside a `Bool`-sorted term, and its implicit rightside the constant `true`. Although conditional rewriting can be difficult, the special case implemented in OBJ is much more efficient, because the convergence of a condition can be checked just by rewriting its leftside term.

Perhaps surprisingly, OBJ's restrictions do not limit its power in practice. In particular, evaluation of an OBJ equation of the form $t = t'$ if $u == v$ agrees with Definition 5.8.2 despite the implicit `true` on the rightside of the condition, because of the operational semantics of ==. In any case, soundness implies that any rewriting computation *is* a proof, so if you get the result you want, then you have proved the result you wanted to prove, whether or not the CTRS that you used was Church-Rosser or terminating. Also, it is rare in practice that when OBJ evaluates $u == v$, a term $r$ exists such that $u \stackrel{*}{\Rightarrow}_A r$ and $v \stackrel{*}{\Rightarrow}_A r$, but OBJ does not find this $r$, because $u$ and $v$ reduce to different normal forms, or because at least one of them does not terminate. There is an important obstacle to soundness for non-canonical CTRS's: if == occurs in a negative position (such as =/=) in a condition, then failure of == to find a common reduced form may lead to its negation returning an unsound `true`. However, soundness can be guaranteed for such conditional rules if $A$ is canonical for the sorts of terms that occur in such positions, using the subset of rules that are actually applied in the particular computation. As noted before, some uses of == have to be considered carefully, because they can take one outside the mathematical semantics of OBJ.

Theorem 5.2.9 on initiality of the algebra of normal forms also generalizes; we do not prove it here, because it is a special case of Theorem 7.7.8, which is proved in Section 7.7.

**Theorem 5.8.9** If a (conditional) specification $P = (\Sigma, A)$ is a ground canonical CTRS, then the canonical forms of ground terms under $A$ form a $P$-algebra called the **canonical term algebra of** $P$ and denoted $N_P$, in the following way:

(0) interpret $\sigma \in \Sigma_{[],s}$ as $[\![\sigma]\!]$ in $N_{P,s}$; and

(1) interpret $\sigma \in \Sigma_{s_1...s_n,s}$ with $n > 0$ as the function that sends $(t_1, \ldots, t_n)$ with $t_i \in N_{P,s_i}$ to $[\![\sigma(t_1, \ldots, t_n)]\!]$ in $N_{P,s}$.

Furthermore, if $M$ is any $P$-algebra, there is one and only one $\Sigma$-homomorphism $N_P \to M$. □

## 5.8.1 Adding Constants

This subsection extends results from Section 5.3 from TRS's to CTRS's, on how properties can change when new constants are added. As before, these results are important because they help us conclude that rewriting systems terminate, are Church-Rosser, local Church-Rosser, or canonical; moreover they can justify using the Theorem of Constants in theorem proving. Proposition 5.3.1 extends to the conditional case to support this assertion; the proof appears in Appendix B:

**Proposition 5.8.10** A CTRS $(\Sigma(X), A)$ is ground terminating, where $X$ is a signature of constants if $(\Sigma, A)$ is ground terminating.[E17] Moreover, if $\Sigma$ is non-void, then $(\Sigma, A)$ is ground terminating iff $(\Sigma(X), A)$ is ground terminating.  □

As with the unconditional case, proofs of the (local) Church-Rosser property usually cover the general case, not just the ground case, so that constants can be added without worry. The following result is still of interest:

**Proposition 5.8.11** A CTRS $(\Sigma, A)$ is (locally) Church-Rosser if and only if the CTRS $(\Sigma(X_S^\omega), A)$ is (locally) ground Church-Rosser, where $X_S^\omega$ is as defined in Proposition 5.3.6.  □

The proof is omitted since it is the same as that for Proposition 5.3.6 on page 108 for the unconditional case.

### 5.8.2  Proving Termination

Proving termination of a CTRS can be much more difficult than for the unconditional case. But we can often reduce to the unconditional case, and then apply the techniques of Section 5.5. In the following result, the "unconditional version" of a conditional rule is defined to be the rule obtained by deleting its condition:

**Proposition 5.8.12** Given a CTRS $C$, let $C^U$ be the TRS whose rules are those of $C$ with their conditions (if any) removed. Then $C$ is terminating (or ground terminating) if $C^U$ is.

**Proof:** Any rewrite sequence of $C$ is also a rewrite sequence of $C^U$ and therefore finite.  □

Notice that the normal forms of $C$ may be different from those of $C^U$, because in general $C^U$ has more rewrites than $C$. The following illustrates the use of this result to prove termination of a CTRS, and because we give rather a lot of detail, it can also serve as a review of the technique of Proposition 5.5.6:

**Example 5.8.13** The function max, which gives the maximum of two natural numbers, is often defined using conditional equations as follows:

```
obj NATMAX is sort PNat .
  op 0 : -> PNat .
  op s_ : PNat -> PNat .
  op _<=_ : PNat PNat -> Bool .
  op max : PNat PNat -> PNat .
  vars N M : PNat .
  eq 0 <= N = true .
  eq s N <= 0 = false .
```

```
    eq s N <= s M = N <= M .
    cq max(N,M) = N if M <= N .
    cq max(N,M) = M if N <= M .
  endo
```

We will show that this CTRS is terminating. It suffices to prove that the corresponding unconditional TRS is ground terminating, by Propositions 5.8.10 and 5.8.12. (It is interesting to notice that this TRS is not Church-Rosser, although the original CTRS is Church-Rosser.)

We define $\rho : T_\Sigma \to \omega$ by initiality, by making $\omega$ a $\Sigma$-algebra as follows: $\omega_{\text{true}} = \omega_{\text{false}} = \omega_0 = 1$; $\omega_{\text{s}}(N) = N + 1$; $\omega_{<=}(N,M) = 1 + N + M$; and $\omega_{\max}(N,M) = 2 + N + M$. To apply Proposition 5.5.6, we must check a number of inequalities. The following arise from condition (1), and must hold for any $x, y \in T_{\Sigma,\text{PNat}}$:

$$\begin{array}{lll}
\rho(0 <= x) & > & \rho(\text{true}) \\
\rho(\text{s}x <= 0) & > & \rho(\text{false}) \\
\rho(\text{s}x <= \text{s}y) & > & \rho(x <= y) \\
\rho(\max(x,y)) & > & \rho(x) \quad \text{if } (x <= y) = \text{true} \\
\rho(\max(x,y)) & > & \rho(y) \quad \text{if } (y <= x) = \text{false} .
\end{array}$$

For condition (2'), we must check the following for any $x, y, z \in T_{\Sigma,\text{PNat}}$ under the assumption that $\rho(x) > \rho(y)$:

$$\begin{array}{lll}
\rho(\text{s}x) & > & \rho(\text{s}y) \\
\rho(x <= z) & > & \rho(y <= z) \\
\rho(z <= x) & > & \rho(z <= y) \\
\rho(\max(x,z)) & > & \rho(\max(y,z)) \\
\rho(\max(z,x)) & > & \rho(\max(z,y)) .
\end{array}$$

All of these translate to inequalities over the natural numbers that are easily checked mechanically, e.g., with appropriate reductions under the following definition, noting that we must introduce only the syntax of NATMAX, not its equations, and that the version of NAT used, in this case NATP+*>, must contain enough facts about addition and > to make the proofs work:

```
obj NATMAXPF is sort PNat .
  pr NATP+*> .
  op 0 : -> PNat .
  op s_ : PNat -> PNat .
  op _<=_ : PNat PNat -> Bool .
  op max : PNat PNat -> PNat .
  op r : PNat -> Nat .
  op r : Bool -> Nat .
  vars X Y : PNat .
  eq r(0) = 1 .
  eq r(true) = 1 .
  eq r(false) = 1 .
```

```
  eq r(s X) = s r(X).
  eq r(X <= Y) = s(r(X) + r(Y)).
  eq r(max(X,Y)) = s s(r(X) + r(Y)).
endo
```

Thus the following proves the first set of inequalities, where we have introduced constants to eliminate the universal quantifiers, and then a lemma that was not already in NATP+*>:

```
openr .
  ops x y : -> PNat .
  vars N M : Nat .
  eq s(N + M) > N = true .
  red r(0 <= x) > r(true).
  red r(s x <= 0) > r(false).
  red r(s x <= s y) > r(x <= y).
  red r(max(x,y)) > r(x).
  red r(max(x,y)) > r(y).
close
```

The last two equations are true without their conditions, although the conditions could have been added as assumptions for the proofs if they had been needed. □

**Exercise 5.8.1** Give mechanical proofs for the second set of inequalities in Example 5.8.13, similar to those given for the first set of inequalities there. □

The conclusion of Proposition 5.8.12 is that infinite rewrite sequences cannot occur. However, a related phenomenon *can* occur for terminating CTRS's, whereby a process of determining whether a conditional rewrite applies does not stop. This is illustrated in the following:

**Example 5.8.14** Let $\Sigma$ have just one sort plus four constants, $a, b, c, d$, and let $A$ contain the following two conditional rewrite rules:

$$a \to b \quad \text{if} \quad c = d$$
$$c \to d \quad \text{if} \quad a = b$$

Then given the term $a$, to check if the first rule applies, we must consult the second, which in turn requires that we consult the first rule again, etc., etc. According to the formal definition of conditional term rewriting, the result of such an infinite regress is simply that the original rule does not apply to the given term; so this does *not* lead to non-termination in the sense of Definition 5.2.1, and in fact, this CTRS is terminating, as is easily seen using Proposition 5.8.12. Intuitively, neither rule applies, because neither condition can ever be satisfied. What does occur here, is that a certain algorithm that might be used to implement conditional term rewriting fails to terminate.[10] In fact, each of $a, b, c, d$ are reduced forms under this CTRS (however, only $b$ and $d$ are reduced under its unconditional version).

It is possible to get the same phenomenon with just one rule. Let $\Sigma$ now have one sort plus two constants, $a, b$, and one unary function $s$. Then the rule

$$a \to b \quad \text{if} \quad s(a) = s(b)$$

leads to an infinite condition evaluation regress similar to that of the above two-rule example. Here too the CTRS is terminating, and $a, b$ are both reduced, for similar reasons. □

**Exercise 5.8.2** Write out the details of the infinite regress and of the termination proof for the one-rule CTRS above. □

**Example 5.8.15** The following OBJ code for the two examples above aborts, producing the error message "`Value stack overflow.`" because of infinite conditional evaluation regress:

```
obj CTRS1 is sort S .
  ops a b c d : -> S .
  cq a = b if c == d .
  cq c = d if a == b .
endo
red a .
obj CTRS2 is sort S .
  ops a b : -> S .
  op s : S -> S .
  cq a = b if s(a) == s(b) .
endo
red a .
```
□

Of course, it is interesting to know when condition evaluation terminates, as well as when rewriting terminates, but we do not address that problem here.

Proposition 5.5.1 on page 111 generalizes to abstract rewrite systems, and hence applies to the conditional case just as well as to the unconditional case, and Example 5.5.2 again shows the necessity of global finiteness for the converse.

**Proposition 5.8.16** An ARS $\mathcal{A}$ on a set $T$ is terminating if there is a function $\rho : T \to \omega$ such that for all $t, t' \in T$, if $t \to_A t'$ then $\rho(t) > \rho(t')$. Furthermore, if $\mathcal{A}$ is globally finite, then $\mathcal{A}$ is terminating iff such a function exists. □

---

[10]Although it is easy to design an algorithm that does terminate on simple examples of this kind, just by checking for loops, it is impossible to write an algorithm that works for all examples, because the problem is unsolvable.

With this we can generalize Proposition 5.5.6 to CTRS's, using the following terminology:

**Definition 5.8.17** Given a poset $P$ and $\rho : T_\Sigma \to P$, then a conditional $\Sigma$-rewrite rule $t \to t'$ if $C$ is **strict $\rho$-monotone** iff $\rho(\theta(t)) > \rho(\theta(t'))$ for each applicable ground substitution $\theta$ such that $\theta(u) \downarrow \theta(v)$ for each $\langle u, v \rangle \in C$; we speak of **weak $\rho$-monotonicity** if $>$ is replaced by $\geq$ above. See Definition 5.5.3 on page 112 for related concepts. □

**Proposition 5.8.18** Given a CTRS $(\Sigma, A)$, if there is a function $\rho : T_\Sigma \to \omega$ such that

(1) each rule in $A$ is strict $\rho$-monotone,[11] and

(2') each $\sigma \in \Sigma$ is strict $\rho$-monotone,

then $A$ is ground terminating. □

The proof is like that of Proposition 5.5.6 on page 113, and is therefore omitted. Rather than give an example using this result now, we will further generalize it to the very common case of a (C)TRS that we know terminates, to which we add some new rules, and then want to show that the resulting system also terminates. The following easy but useful ARS result is the basis for this generalization:

**Proposition 5.8.19** Let $\mathcal{A}$ be an ARS on a set $T$, let $\mathcal{B}$ be a terminating "base" ARS contained in $\mathcal{A}$, and let $\mathcal{N}$ denote the "new" rewrites of $\mathcal{A}$ on $T$, i.e., let $\to_\mathcal{N} = \to_\mathcal{A} - \to_\mathcal{B}$. Then $\mathcal{A}$ is terminating if there is a function $\rho : T \to \omega$ such that

(1) if $t \xrightarrow{1}_\mathcal{B} t'$ then $\rho(t) \geq \rho(t')$, and

(2) if $t \xrightarrow{1}_\mathcal{N} t'$ then $\rho(t) > \rho(t')$.

**Proof:** Any $\mathcal{A}$-rewrite sequence can be put in the form $t_1 \xrightarrow{*}_\mathcal{B} t_2 \xrightarrow{1}_\mathcal{N} t_3 \xrightarrow{*}_\mathcal{B} t_4 \xrightarrow{1}_\mathcal{N} \cdots$, from which it follows that $\rho(t_1) \geq \rho(t_2) > \rho(t_3) \geq \rho(t_4) > \cdots$. Hence there is some $k$ such that no $\mathcal{N}$ rewrite applies to $t_k$. Because $\mathcal{B}$ is terminating, there can only be a finite number of rewrites after $t_k$, so the sequence must be finite. □

The two levels of this result can be iterated to form a multi-level hierarchy, in which one proves the termination of each layer assuming the one below it. The following is a conditional hierarchical version of Proposition 5.5.6; of course, it also applies to unconditional TRS's. The proof is not entirely trivial.

---

[11] Recall that the inequality only needs to hold when all the conditions of the rule converge.

**Theorem 5.8.20** Let $(\Sigma, A)$ be a CTRS with $\Sigma$ non-void, let $(\Sigma, B)$ be a terminating sub-CTRS of $(\Sigma, A)$, and let $N = A - B$. If there is a function $\rho : T_\Sigma \to \omega$ such that

(1) every rule in $B$ is weak $\rho$-monotone,

(2) every rule in $N$ is strict $\rho$-monotone,

(3) every $\sigma \in \Sigma$ is strict $\rho$-monotone,

then $A$ is ground terminating.

**Proof:** We will use Proposition 5.8.19. Let $\mathcal{A}, \mathcal{B}, \mathcal{N}$ be the ARS's for $A, B, N$ respectively, on the (indexed) set $T = T_\Sigma$ and let $\Sigma'$ be the minimal signature for $B$. Notice that rules in $B$ may apply to terms with operations in $\Sigma - \Sigma'$; therefore $\mathcal{B}$ must apply such rewrites. This means we cannot assume termination for $\mathcal{B}$, and hence to apply Proposition 5.8.19, we must first establish that assumption for $\mathcal{B}$ on $T_\Sigma$. We will do this by induction on the depth of nesting of new operation symbols in a $\Sigma$-term $t$.

For the base case, a $\Sigma$-term $t$ has depth zero iff it contains no operations in $\Sigma - \Sigma'$, and then we have termination by our assumption that $B$ is ground terminating on $T_{\Sigma'}$.

Next, suppose $t$ is a $\Sigma$-term with depth $d > 0$ of the form $g(t_1, \ldots, t_n)$ with $g \in \Sigma - \Sigma'$ and with each $t_1, \ldots, t_n$ of depth less than $d$. Then by the inductive assumption, rewriting with $\mathcal{B}$ is terminating on each $t_i$, and hence is terminating on $t$, because only a lapse rule in $B$ could be applied at the top of $t$, and any such application will reduce us to the case of the previous paragraph, because the rightside of the lapse rule must be a ground term, or else $B$ would not be terminating.

Now consider the general case of a $\Sigma$-term $t$ with depth $d > 0$, which will have the form $t = t_0(z_1 \leftarrow t_1, \ldots, z_n \leftarrow t_n)$ with $t_0$ involving only operations in $\Sigma'$, with each $t_1, \ldots, t_n$ of depth $d$ or less, and with the top operation of each $t_i$ in $\Sigma - \Sigma'$. Then any rewrite of $t$ must either be inside of some $t_i$ or else inside of $t_0$. There can only be a finite number of rewrites of the first kind, by the argument of the previous paragraph, and there can only be a finite number of rewrites of the second kind, noting that our signature is non-void and applying Proposition 5.8.10 on page 131 of Section 5.8.1, which generalizes Proposition 5.3.4, about the effect on termination of adding constants for the conditional case, because the $z_i$ are just new constants. Hence rewriting with $\mathcal{B}$ terminates on any such term $t$, and we have therefore proved termination of $\mathcal{B}$.

Next, observe that our assumptions (1) and (3) above imply assumption (1) of Proposition 5.8.19, by the same reasoning that was used to prove Proposition 5.5.5 in Appendix B. Similarly our assumptions (2) and (3) imply assumption (2) of Proposition 5.8.19. □

In many cases, $\rho$ is already defined on $B$, and we only need check the conditions for the new rules and new operations. Notice that if a CTRS $A$ has a terminating sub-CTRS $B$ such that the new rules in $N = A - B$ cannot be used in evaluating the conditions of rules in $N$, then infinite condition evaluation regress cannot occur.

We first apply Theorem 5.8.20 to a case where all the new rules are unconditional; here the hierarchical specification greatly simplifies the termination proof, using the fact that termination was previously shown for the base system.

**Example 5.8.21** Suppose we are given some (C)TRS $B$ for the natural numbers that we already know is terminating, such as NATP+, and then define the Fibonacci numbers over $B$ by:

```
obj FIBO is pr NATP+ .
  op f : Nat -> Nat .
  var N : Nat .
  eq f(0) = 0 .
  eq f(s 0) = s 0 .
  eq f(s s N) = f(s N) + f(N).
endo
```

Letting $\Sigma$ be the signature for the union TRS $A$, and $\Sigma'$ the signature for $B$ (which is NATP+), we define $\rho : T_\Sigma \to \omega$ by letting each $\sigma \in \Sigma'$ have its usual meaning in $\omega$, and letting $\omega_f(N) = 2^N$. Then for $t \in T_{\Sigma'}$, the value of $\rho(t)$ is the number that it denotes, and so all the $B$-rules are weak monotone (in fact, with equality). For condition (2), strict monotonicity of the three $N$-rules for the Fibonacci function follows from the corresponding inequalities, of which the most interesting is the third,
$$4 \cdot 2^N > 3 \cdot 2^N.$$
Condition (3) of Theorem 5.8.20 is easy to check from the definitions of the functions defined on $\omega$. Hence this TRS is terminating. □

Notice that proving termination for the specification of a function like Fibonacci gives much more than just termination of the underlying algorithm, because it applies to terms with any number of occurrences of the function, in any combination with functions from the base rewriting system, to any level of nesting.

We next do an example with a conditional rule such that the method of Proposition 5.8.12 cannot be used, because the unconditional version of this CTRS fails to terminate; this example is the bubblesort algorithm.

**Example 5.8.22** Assume that the following specification for lists of natural numbers has been shown to be terminating as a TRS:

```
obj NATLIST is sorts Nat List .
  op 0 : -> Nat .
  op s : Nat -> Nat .
  op _<_ : Nat Nat -> Bool .
  op nil : -> List .
  op _._ : Nat List -> List .
  *** vars and eqs omitted ...
endo
```

Now add to this the following new operation and rule, which define the so called bubblesort algorithm for sorting lists of naturals:

```
obj BSORT is pr NATLIST .
  op sort : List -> List .
  vars N M : Nat .
  var L : List .
  cq sort(N .(M . L)) = sort(M .(N . L)) if M < N .
endo
```

The conditional rewrite rule above switches two adjacent list elements iff they are out of order. We want to show that this hierarchical CTRS is terminating. Notice that the above equation without the condition is definitely not terminating; for example, the list 1 . 2 . nil can be rewritten to 2 . 1 . nil which can be rewritten to the original list, etc., etc. Even though we only sketch the proof, the specification really needs to have an operation and equation such as

```
sorted : List -> Bool .
cq sort(L) = L if sorted(L) .
```

to get rid of the sort function symbol when the list is finally sorted. However, the essence of bubble sort is the conditional rule in the BSORT module above.

The $\Sigma$-algebra structure of $\omega$ is defined by interpreting the operations on the naturals as themselves, interpreting true, false, and nil as 0, letting $\omega_<(N,M) = N + M$, letting $\omega_{\text{sort}}(L) = L + 1$, and letting $\omega_.(N,L) = d(N,L) + d(L)$, where $d$ is the "displacement" function, i.e., the number of pairs that are out of order, defined by

$$\begin{aligned} d(\text{nil}) &= 0 \\ d(N.L) &= d(N,L) + d(L) \\ d(N,\text{nil}) &= 0 \\ d(N,M.L) &= 1 + d(N,L) \quad \text{if} \quad N > M \\ d(N,M.L) &= d(N,L) \quad \text{if} \quad N \le M . \end{aligned}$$

Proving strict monotonicity of the new rule depends on the lemma

$$d(N.(M.L)) = 1 + d(M.(N.L)) \text{ if } N < M,$$

which is not hard to prove by case analysis. The strict monotonicity of $\omega_*$ can be checked from the definition of $d$. It is easy to check the other monotonicity conditions for both rules and operations, and so we are done. By the way, we can actually write the above definition of $d$ in OBJ and then define

```
eq sorted(L) = d(L) == 0 .
```
□

**Exercise 5.8.3** Show that the equations defining $d$ in Example 5.8.22 above are terminating, when viewed as rewrite rules over NATLIST. **Hint:** Show that the unconditional version is terminating with Theorem 5.8.20 and then apply Proposition 5.8.12. □

**Exercise 5.8.4** Give OBJ proofs for the results of Example 5.8.22 and Exercise 5.8.3. □

It should not be thought that proving termination of conditional term rewriting systems is always an easy task. While the results given in this subsection seem adequate for the most common examples, there are many others for which they are not. The following two examples constitute a somewhat entertaining partial digression on non-proofs of non-termination.

**Example 5.8.23** We give two TRS's that are ground terminating separately but combine to give a TRS that is not; this is called "Toyama's example" [177]. We also prove that Theorem 5.8.20 could never be used to demonstrate the termination of this TRS.

```
obj B is sort S .
  ops 0 1 : -> S .
  op f : S S S -> S .
  var X : S .
  eq f(0,1,X) = f(X,X,X) .
endo
obj A is pr B .
  op g : S S -> S .
  vars X Y : S .
  eq g(X,Y) = X .
  eq g(X,Y) = Y .
endo
```

A term that demonstrates the non-ground termination of this TRS is

$$t = f(g(0,1), g(0,1), g(0,1)),$$

which rewrites first to $f(0, g(0,1), g(0,1))$, then to $f(0, 1, g(0,1))$, and then back to the initial term $t$. Note that the equation in B could also have been given as the conditional equation

```
cq f(X,Y,Z) = f(Z,Z,Z) if X == 0 and Y == 1 .
```

Now suppose we have $\rho : T_\Sigma \to \omega$ (where $\Sigma$ is the signature of A) that is weak $\rho$-monotone on the rule in B, and strict $\rho$-monotone on the new rules of A, such that all operations in $\Sigma$ are strict $\rho$-monotone. Let us write $[t]$ for $\rho(t)$. Then

$$\begin{array}{ll} [t] & = \\ [f(g(0,1),g(0,1),g(0,1))] & > \\ [f(0,g(0,1),g(0,1))] & > \\ [f(0,1,g(0,1))] & \geq [t] \,, \end{array}$$

which is a contradiction. Therefore Theorem 5.8.20 could never be used to prove ground termination of this TRS (which is of course consistent with the fact that this TRS is not ground terminating). □

**Example 5.8.24** Using the same technique as in Example 5.8.23, we sketch a proof that Theorem 5.8.20 cannot be used to prove ground termination of the specification for the greatest common divisor given below (which is essentially Euclid's algorithm), viewed as a hierarchical CTRS over some suitable terminating specification NAT of the natural numbers with subtraction and >, where $\rho$ is defined homomorphically. Termination of this CTRS is proved in Example 5.8.32 using much more sophisticated methods.

```
obj GCD is pr NAT .
  op gcd : Nat Nat -> Nat .
  vars M N : Nat .
  eq gcd(M,0) = M .
  eq gcd(0,N) = N .
  cq gcd(M,N) = gcd(M - N, N) if M >= N and N > 0 .
  cq gcd(M,N) = gcd(M, N - M) if N >= M and M > 0 .
endo
```

The proof will be by contradiction, so we assume that there are a $\Sigma$-algebra structure on $\omega$ and a weight function $\rho : T_\Sigma \to \omega$ satisfying all the conditions of Theorem 5.8.20, where $A$ is GCD plus NAT, $\Sigma$ is the signature of $A$, and $B$ is NAT.

We first prove a lemma, that $M(x,y) \geq x$ for all $x,y$ in $\omega$, where $M(x,y)$ denotes the function $\omega_-(x,y)$ on $\omega$. The proof is by contradiction, so we suppose that there exist $x_0, y_0$ such that $M(x_0, y_0) < x_0$. Let $M(x_0, y_0) = x_1$. Then $x_1 < x_0$, and the strict monotonicity of $M$ implies that $M(x_1, y_0) < M(x_0, y_0) = x_1$. Similarly letting $x_{n+1} = M(x_n, y_0)$, we get $x_{n+1} < x_n$ for all $n \geq 0$. But this is impossible because $\omega$ is Noetherian (i.e., has no infinite strictly decreasing sequences). By the same reasoning, the analoguous inequality holds for the function $G(x,y) = \omega_{gcd}$.

Now we are ready for the main part of the proof, in which we write $[t]$ for $\rho(t)$, as well as $M$ for $\omega_-$ and $G$ for $\omega_{gcd}$ as above. Let $x, y$ be natural number terms (i.e., ground terms in the base rewriting system NAT) with $x > y$. Then

$$\begin{array}{ll} [gcd(x,y)] & > \\ [gcd(x-y,y)] & = \\ G([x-y],[y]) & = \\ G(M([x],[y]),[y]) & \geq \\ G([x],[y]) & = [gcd(x,y)]\,, \end{array}$$

which is a contradiction (the first step results from monotonicity when applying the first conditional rule, and the next to last step uses the lemma for $M$, and then for $G$). □

The final calculation in the above example suggests that the reason this termination proof method fails for gcd is the monotonicity requirement for operations combined with homomorphicity. Because these assumptions are not needed for Proposition 5.8.19, the possibility remains of applying something like that result directly, as is done in the next subsection.

**Exercise 5.8.5** Assuming a suitable terminating specification NAT for the natural numbers with inequality >, prove termination of the following CTRS for binary search trees:

```
obj BTREE is sort BTree .
  pr NAT .
  op empty : -> BTree .
  op make : BTree Nat BTree -> BTree .
  op insert :    Nat BTree -> BTree .
  vars T1 T2 T3 : BTree .
  vars N M : Nat .
  eq insert(M,empty)          = make(empty,M,empty) .
  cq insert(M,make(T1,N,T2)) = make(insert(M,T1),N,T2)
     if N > M .
  cq insert(M,make(T1,N,T2)) = make(T1,N,insert(M,T2))
     if M > N .
endo
```
□

### 5.8.3 (⋆) Noetherian Orderings

This subsection develops the remark after Example 5.5.4 that it is useful to allow weight functions that take values in Noetherian partial orderings other than $\omega$ (see Appendix C for a review of partially ordered sets, also called **posets**), where a poset is **Noetherian** (also called **well founded**) iff it has no infinite sequence of strictly decreasing elements. The key observation is that Proposition 5.8.19 and Theorem 5.8.20 generalize to any Noetherian poset, because their proofs depend only on the Noetherian property; note also that a different Noetherian ordering could be used for each sort, since we are really dealing with a sorted set of posets. As with using $\omega$, the key intuition is that rewrites should strictly decrease weight. Some examples, including the greatest common divisor as computed by Euclid's algorithm, need rather complicated orderings. To help with this, we introduce some ways to build new orderings out of old ones, such that if the old orderings are Noetherian then so are the new ones. Unfortunately, much of the material in this subsection is rather technical.

**Definition 5.8.25** Let $P, Q$ be posets, with both their orderings denoted $\geq$. Then their (**Cartesian**) **product poset**, denoted $P \times Q$, has as its elements the pairs $(p, q)$ with $p \in P$ and $q \in Q$, and has $(p, q) \geq (p', q')$ iff $p \geq p'$ and $q \geq q'$. Their **lexicographic product**, here denoted $P \oslash Q$, again has as elements the pairs $(p, q)$ with $p \in P$ and $q \in Q$ (we may use the notation $p \oslash q$), but now ordered by $(p, q) \geq (p', q')$ iff $p > p'$ or else $p = p'$ and $q \geq q'$. To avoid confusion with pairs $(p, q) \in P \times Q$, we will hereafter use the notation $p \oslash q$ for elements of $P \oslash Q$. The **sum** of posets $P_1, P_2$, denoted $P_1 + P_2$, has as its elements pairs[12] $(i, p)$ with $p \in P_i$ for $i = 1$ or $i = 2$, ordered by $(i, p) \geq (i', p')$ iff $i = i'$ and $p \geq p'$ in $P_i$. A poset $Q$ is a **subposet** of a poset $P$ iff $Q \subseteq P$ and $q \geq q'$ in $Q$ iff $q \geq q'$ in $P$, for all $q, q' \in Q$. □

The following result is rather straightforward to prove:

**Proposition 5.8.26** If $P, Q$ are both Noetherian posets, then so are $P \times Q$, $P \oslash Q$ and $P + Q$. Moreover, the **discrete ordering** on any set $X$, defined by $x \geq y$ iff $x = y$, is also a Noetherian poset, and any subposet of a Noetherian poset is Noetherian. □

**Example 5.8.27** Motivated by the applications of term rewriting to verifying hardware circuits that are developed in Section 7.4, a system $T$ of $\Sigma(X)$-equations is said to be **triangular** iff $X$ is finite, there is a subset of $X$ called **input variables**, say $i_1, \ldots, i_n$, and there is an ordering of the non-input variables, say $p_1, \ldots, p_m$, such that the equations in $T$ have the form

$$p_k = t_k(i_1, \ldots, i_n, p_1, \ldots, p_{k-1}) \quad \text{for } k = 1, \ldots, m,$$

where each $t_k$ is a $\Sigma(X)$-term involving only input variables and those non-input variables $p_j$ with $j < k$ (in particular, $t_1$ must contain only input variables).

We first prove that any triangular system $T$ is terminating as a TRS. Let $P = \oslash_{i=1}^{m} \omega$, the $m$-fold lexicographic product of $\omega$ with itself, and define $\rho : T_\Sigma \to P$ by letting $\rho(t) = (\ell_m, \ldots, \ell_1)$ where $\ell_k$ is the number of occurrences of $p_k$ in $t$. Rewriting a $\Sigma(X)$-term $t$ with any equation in $T$ will decrease $\rho(t)$, because it will decrease the number $\ell_k$ of occurrences of the non-input variable $p_k$ in the rule's leftside by one, while possibly increasing the numbers $\ell_j$ of occurrences of variables $p_j$ with $j < k$. Therefore $T$ is terminating by Proposition 5.5.1 generalized to Noetherian posets.

---

[12]If $P_1$ and $P_2$ are disjoint, then the elements of $P_1 + P_2$ can be taken as just those in $P_1 \cup P_2$. The purpose of the construction with the pairs $(i, p)$ is just to enforce disjointness in case $P_1, P_2$ were not already disjoint.

Next, we use the Newman Lemma (Proposition 5.7.4) to show that $T$ is Church-Rosser, by proving that the local Church-Rosser property holds. For this purpose, we first note that if a $\Sigma(X)$-term $t$ can be rewritten in two distinct ways, it must have the form $t_0(z_1 \leftarrow p_i, z_2 \leftarrow p_j)$ where $z_1, z_2$ are distinct new variables, each occurring just once in $t_0$. To prove this, pick one of the rewrites and note that, since its redex is a non-input variable, $t$ must have the form $t'_0(z_1 \leftarrow p_i)$. Because there is just one rule for each non-input variable, the redex for the second rewrite is disjoint from that for the first, so that $t'_0(z_1 \leftarrow p_i)$ and hence $t$, has the form $t_0(z_1 \leftarrow p_i, z_2 \leftarrow p_j)$, for which we will use the shorter notation $t_0(p_i, p_j)$. It now follows that the two rewrites have the forms $t \Rightarrow t_0(t_i, p_j)$ and $t \Rightarrow t_0(p_i, t_j)$. Therefore each target term can be rewritten to $t_0(t_i, t_j)$ by applying the other rule once. We now conclude that any triangular system is canonical.

Finally, we show that the only variables that can occur in a normal form of a triangular system are input variables, by proving the contrapositive: If a term $t$ contains a non-input variable, then it can be rewritten using the rule with that variable as its leftside, and hence it is not reduced. □

A more complex construction of a new Noetherian poset from an old one is given by multisets. Intuitively, multisets generalize ordinary sets by allowing elements to occur multiple times. A multiset is often defined to be a function $A : D \to \omega^+$, where $D$ is the domain and $A(d)$ is the *multiplicity* of $d \in D$. It is common to use set notation for multisets, so that for example the multiset denoted by $\{1, 1, 2\}$ would have $D = \{1, 2\}$, with $A(1) = 2$ and $A(2) = 1$, indicating two instances of 1 and one of 2. Then the most natural notion of a *submultiset* of $A$ would be a subset $D'$ of $D$ and a function $A' : D' \to \omega^+$ such that $A'(d) \leq A(d)$ for all $d \in D'$; for example, $\{1, 1\} \leq \{1, 2, 1\}$.

However, this approach is inadequate for our applications, which require multisets of elements drawn from a Noetherian poset $P$, with an ordering such that, for example where $P$ is $\omega$ with the usual ordering, $\{1\} < \{2\} < \{3\}$ and $\{1, 2\} < \{2, 2\}$. Also, in term rewriting theory, the phrase "multiset ordering" usually refers to an ordering that allows even more possibilities, such as $\{1, 1, 1\} < \{2\}$ and $\{2, 1, 1, 1\} < \{2, 2\}$; but because our applications do not need this extra sophistication, we will develop only a somewhat simplified special case.

Our mathematical formulation of multisets involves a possibly surprising reversal of the approach sketched above, in that we dispense with $\omega^+$, and instead rely on abstract sets whose elements represent instances of elements of $P$. For example, $\{1, 2, 1\}$ is represented by the function $A : \{x, y, z\} \to P$ with $A(x) = 1, A(y) = 2$, and $A(z) = 1$.

**Definition 5.8.28** Given a poset $P$ with an ordering $\geq$, then a **multiset** over $P$ is a function $A : X \to P$ with **underlying set** $X$; call a multiset $A : X \to P$ **finite** iff its underlying set is finite; the **empty multiset**, denoted $\emptyset$, has the empty underlying set. Given multisets $A : X \to P$ and $B : Y \to P$, define $A \geq B$ iff there is an injective function $f : Y \to X$ such that $A(f(y)) \geq B(y)$ for all $y \in Y$. Let $\mathcal{M}(P)$ denote the class of all finite multisets,[13] where all multisets $A, B$ such that $A \geq B$ and $B \geq A$ are identified.[14] □

**Exercise 5.8.6** Prove the following, where $P$ is $\omega$ with the usual ordering,

$$\begin{array}{rcccl} \{1, 2\} & > & \{1\} & > & \emptyset \\ \{1, 2\} & > & \{1, 1\} & > & \{1\} \\ \{3, 3\} & > & \{2, 3\} & > & \{1, 2\}, \end{array}$$

and where (as in Appendix C) $A > B$ means $A \geq B$ and $A \neq B$ (which, because of the equivalence on multisets, means $A \geq B$ and not $B \geq A$). However, it is not possible to show (for example) that

$$\{2, 2\} > \{2, 1, 1, 1\},$$

which would be required by the more usual and powerful multiset ordering. □

Although this multiset ordering is weaker than the usual one, it is easier to reason about, and is sufficient for the applications in this chapter.

**Proposition 5.8.29** If $P$ is a Noetherian poset, then so is $\mathcal{M}(P)$.

**Proof:** Reflexivity is easy. For anti-symmetry, use the lemma that $A \geq B$ and $B \geq A$ iff there is some bijective $f : Y \to X$ such that $A(f(y)) = B(y)$ for all $y \in Y$. For transitivity, given $A \geq B \geq C$ with underlying sets $X, Y, Z$ and injections $f : Z \to Y$ and $g : Y \to X$, then $f;g : Z \to X$ is also injective and satisfies $A(g(f(z))) \geq B(f(z)) \geq C(z)$ for all $z \in Z$.

For the Noetherian property, suppose that $A_1 > A_2 > \cdots > A_n > \cdots$ is an infinite strictly decreasing sequence, where $A_i$ has underlying set $X_i$. This gives rise to an infinite sequence of injections $X_1 \leftarrow X_2 \leftarrow \cdots \leftarrow X_n \leftarrow \cdots$. Then because $X_1$ is finite, there must exist some $n$ such that (up to isomorphism) $X_n = X_{n+k}$ for all $k \geq 1$. Then for each $k \geq 1$ there must exist some $x \in X_n$ such that $A_{n+k}(x) > A_{n+k+1}(x)$. But for each particular $x \in X_n$, there can only be a finite number of such $k$ because $P$ is Noetherian. Now because $X_n$ is finite, there can only be a finite number of pairs $(k, x)$ such that the above inequality holds, which contradicts our initial assumption. □

---

[13] In order to avoid set-theoretic worries, it is desirable to restrict the underlying sets that are used, for example, to finite subsets of $\omega$.

[14] So technically speaking, we have an ordering on the quotient set. To obtain the multiset ordering that is more usual in term rewriting, the restriction to injective functions should be relaxed to asserting of $f : Y \to X$ that if $f(y) = f(y')$ with $y \neq y'$ then $A(f(y)) > B(y)$ and $A(f(y')) > B(y')$.

We now make one further identification, of $p \in P$ with $\{p\} \in \mathcal{M}(P)$, noting that $p \geq p'$ in $P$ iff $\{p\} \geq \{p'\}$ in $\mathcal{M}(P)$, so that the inclusion map $P \subseteq \mathcal{M}(P)$ is order preserving. Therefore defining $\mathcal{M}^{n+1}(P) = \mathcal{M}(\mathcal{M}^n(P))$ with of course $\mathcal{M}^1(P) = \mathcal{M}(P)$, we get

$$P \subseteq \mathcal{M}(P) \subseteq \mathcal{M}^2(P) \subseteq \cdots \subseteq \mathcal{M}^n(P) \subseteq \cdots,$$

and can therefore form the union of all these to get

$$\mathcal{M}^\omega(P) = \bigcup_n \mathcal{M}^n(P),$$

the union ordering on which is called the **nested multiset ordering**.[15]

**Fact 5.8.30** The nested multiset ordering $\mathcal{M}^\omega(P) = \bigcup_n \mathcal{M}^n(P)$ is Noetherian if $P$ is.

**Proof:** Each $\mathcal{M}^n(P)$ is Noetherian by induction using Proposition 5.8.29, and each element of the union lies in $\mathcal{M}^n(P)$ for some least $n$, as do all elements less than any given element. □

Similar constructions are used in Exercise 5.8.8 and Example 5.8.32 below.

**Exercise 5.8.7** Given a poset $P$ and an equivalence relation $\equiv$ on the carrier of $P$ (which is also denoted $P$), let $P/\equiv$ denote the set $P/\equiv$ ordered by $[p] \leq [q]$ for $p, q \in P$ iff $p' \leq q'$ for some $p' \equiv p$ and $q' \equiv q$. Show that if $P/\equiv$ is a poset, and if it has only a finite number of non-trivial equivalence classes, then it is Noetherian if $P$ is. Give an example showing that the hypothesis about a finite number of non-trivial equivalence classes is necessary. **Hint:** Let $P$ have $a_1 > a_2$, $a'_2 > a_3$, $a'_3 > a_4, \ldots$, and then identify $a_i$ with $a'_i$ for $i = 1, 2, 3, \ldots$. □

**Exercise 5.8.8** Given a poset $P$, let $\bot$ be a new element not already in $P$, and let $P^\bot$ denote the poset having underlying set $P \cup \{\bot\}$ with the ordering of $P$ plus $\bot < p$ for all $p \in P$. Show that $P^\bot$ is Noetherian if $P$ is.
Now given a Noetherian poset $P$ with a unique least element $\bot$, form $\oslash^2 P = P \oslash P$, and identify $p \in P$ with the element $p \oslash \bot \in \oslash^2 P$, so that there is an order-preserving inclusion $P \subseteq \oslash^2 P$. Iterate this to obtain $\oslash^n P \subseteq \oslash^{n+1} P$, and note that each $\oslash^n P$ is Noetherian for each $n$ by induction and Proposition 5.8.26. Now form $\oslash^\omega P = \bigcup_n \oslash^n P$, show that $\oslash^\omega P$ corresponds to the usual lexicographic ordering on the set of finite strings from $P$, and give an example showing that $\oslash^\omega P$ in general is *not* Noetherian. **Hint:** If $b > a$, then $b > ab > aab > aaab > \cdots$. □

---

[15]Those who know some category theory may recognize this as a colimit construction. The result proved in the next sentence is that this colimit of an increasing sequence of Noetherian posets is Noetherian. We also note that the product and sum constructions of Definition 5.8.25 are the categorical product and coproduct.

A straightforward generalization of Proposition 5.8.19 requires defining a weight function $\rho : T_\Sigma \to P$ where $P$ is a Noetherian poset, and then showing that each new rewrite is strict $\rho$-monotone and each old rewrite is weak $\rho$-monotone. A less straightforward generalization weakens the assumption that $P$ is Noetherian to assuming that each particular item and everything to which it can be rewritten lie within some Noetherian subposet of $P$.

**Proposition 5.8.31** Let $\mathcal{A}$ be an ARS on ($S$-indexed) set $T$, let $\mathcal{B}$ be a terminating "base" ARS contained in $\mathcal{A}$, let $\mathcal{N}$ denote the "new" rewrites of $\mathcal{A}$ on $T$ (i.e., $\to_\mathcal{N} = \to_\mathcal{A} - \to_\mathcal{B}$), and let $P$ be a poset.[16] Then $\mathcal{A}$ is terminating if there is a function $\rho : T \to P$ such that

(1) if $t \xrightarrow{1}_\mathcal{B} t'$ then $\rho(t) \geq \rho(t')$,

(2) if $t \xrightarrow{1}_\mathcal{N} t'$ then $\rho(t) > \rho(t')$, and

(3) $P$ is Noetherian, or if not, then for each $t \in T_{\Sigma,s}$ there is a Noetherian poset $P_s^t \subseteq P_s$ such that $t \xrightarrow{*}_\mathcal{A} t'$ implies $\rho(t') \in P_s^t$.

**Proof:** By exactly the same reasoning that was used for Proposition 5.8.19. □

**Example 5.8.32** (⋆) We show termination of the GCD CTRS of Example 5.8.24 using Proposition 5.8.31 with a rather complex ordering. To define this ordering, for a given poset $P$, let

$$\mathcal{N}(P) = (P \times P) + (\omega \times \omega) \oslash P + \mathcal{M}(P),$$

with its ordering given by Definition 5.8.25. Then $\mathcal{N}(P)$ is Noetherian if $P$ is by Propositions 5.8.26 and 5.8.29, and because $P \subseteq \mathcal{M}(P)$ is an order-preserving inclusion, so is $P \subseteq \mathcal{N}(P)$. Therefore defining $\mathcal{N}^{n+1}(P) = \mathcal{N}(\mathcal{N}^n(P))$ with $\mathcal{N}^0(P) = P$, we have that each $\mathcal{N}^n(P)$ is Noetherian if $P$ is. However, the union $\mathcal{N}^\omega(P)$ of the chain

$$P \subseteq \mathcal{N}(P) \subseteq \mathcal{N}^2(P) \subseteq \cdots \subseteq \mathcal{N}^n(P) \subseteq \cdots$$

is in general *not* Noetherian, for the reasons considered in Exercise 5.8.8.

The case in which we are most interested takes $P = \omega^\perp$. For each $\mathcal{N}^n(\omega^\perp)$, identify $p \in P$ with $p \oslash \perp \in \oslash^2 P$, and with $p \oslash \perp \oslash \perp \in \oslash^3 P$, etc., recursively as in Exercise 5.8.8, and also identify $\perp$ in $P$ with $(\perp, \perp)$

---

[16]When $T$ is $S$-indexed, a family $\{P_s \mid s \in S\}$ of posets, although in practice they are often all the same.

in $(P \times P)$ and with $\emptyset$ in $\mathcal{M}(P)$; then the result of these identifications is Noetherian for each $n$, by Exercise 5.8.7. Finally, for all $p, q$, add the inequalities

1. $(p, q)$ > $p, q$ if $p, q \neq \bot$
2. $(m, n) \oslash (p, q)$ > $p, q$ if $p, q \neq \bot$.

To simplify notation, denote the resulting poset at level $n$ by $N_n$ and the union by $N$; let $N_0 = \{\emptyset\}$. We leave the reader to check that the above new inequalities do not violate the poset axioms or the Noetherian condition for each $N_n$.

In order to apply[17] Proposition 5.8.31, let $T$ be $T_\Sigma$ where $\Sigma$ is the total signature of GCD, let $\mathcal{A}$ consist of all rewrites induced on $T$ by the rules in GCD, let $\mathcal{B}$ consist of all rewrites induced on $T$ by the rules in NAT, and let $\mathcal{N} = \mathcal{A} - \mathcal{B}$. We must define a weight function $\rho : T \to N$ such that each rewrite in $\mathcal{N}$ is strict $\rho$-monotone, such that each rewrite in $\mathcal{B}$ is weak $\rho$-monotone, and such that for each $\Sigma$-term, everything to which it can be rewritten lies within some fixed Noetherian subposet $N_t$ of $N$. We take $N_t$ be the poset that was denoted $N_n$ above, with $n = 3d$, where $d$ is the maximum depth of nesting of $gcd$'s inside of $t$. In the following, we let $T^d$ denote the set of ground $\Sigma$-terms of maximum nesting depth not greater than $d$, we let $\Sigma'$ denote the signature of NAT, and we let $g$ abbreviate $gcd$. Notice that for any $d$, rewriting on $T^d$ with $\mathcal{A}$ always remains within $T^d$, because none of the rules in GCD or NAT can increase the depth of nesting of $gcd$'s in terms.

We give a recursive definition for $\rho$, in which a subterm of $t \in T$ is called **top** if it is a maximal subterm of $t$ having $g$ as its head, and as before we write $[t]$ for $\rho(t)$:

(a) $[n] = \emptyset$ (the empty multiset)    if $n$ is a $\Sigma'$ term
(b) $[g(t, t')] = (m, n) \oslash ([t], [t'])$    where $t, t'$ reduce to Peano terms $m, n$
(c) $[t] = \{[t_1], \ldots, [t_n]\}$    if $t$ is not top and $t_1, \ldots, t_n$ are its top subterms.

By a "Peano term" we mean a term of the form $s \ldots s0$. We can show that rewriting on $T_{\Sigma'}$ always terminates with such a term using an argument[E18] like that given in Example 5.5.5, and then we can show that $\mathcal{B}$ is terminating on $T_\Sigma$ with an argument like that given in Theorem 5.8.20 on page 136, noting that $\Sigma$ is non-void and using Proposition 5.3.4. To apply (b), we need to know that the $\Sigma$-terms $t$ and $t'$ reduce to Peano terms under $\mathcal{A}$, whereas in general we don't even know whether rewriting with $\mathcal{A}$ terminates for arbitrary $\Sigma$-terms. Therefore we should demonstrate termination with a Peano term result along with conditions (1) and (2) of Proposition 5.8.31, as part of our induction on the maximum depth of nesting of *gcd*'s in Σ-terms.

---

[17]Generalizing the proof in Example 5.8.24 to poset weights shows that the generalization of Proposition 5.8.19 to poset weights (stated below as Theorem 5.8.33) cannot be made to work for this example.

Our induction hypothesis is the conjunction of four subsidiary hypotheses: ($A_d$) rewriting on $T^d$ with $\mathcal{B}$ preserves weight; ($B_d$) rewriting on $T^d$ with $\mathcal{N}$ is strict monotone; ($C_d$) the weights of terms in $T^d$ always lie in $N_{3d}$; and ($D_d$) rewriting on $T^d$ with $\mathcal{A}$ always terminates with a Peano term. Notice that ($A_d$) implies that rewriting with $\mathcal{B}$ is weak monotone.

The base case takes $d = 0$ and considers $t \in T^0 = T_{\Sigma'}$. By (a) of the definition of $\rho$, we have $[t] = \emptyset$; therefore rewrites with old rules are weight preserving, and weights remain within $N_0 = \{\emptyset\}$ because rewriting remains within $T^0$. Also, because no new rules can be applied to $t \in T_{\Sigma'}$ and no operations from $\Sigma - \Sigma'$ can be introduced by rewriting $\Sigma'$-terms, rewrites using new rules are vacuously strict monotone, because there aren't any.

The induction step assumes the four induction hypotheses ($A_d$, $B_d$, $C_d$, $D_d$) for some $d > 0$. We first prove a preliminary lemma, which says that any rewrite induced by applying a new rule at the top of a term in $T^{d+1}$ is strict monotone. For the first rule, $g(M, 0) = M$, because any $t \in T^d$ reduces to a Peano term (say) $m$ by ($D_d$), we get $[g(t, 0)] = (m, 0) \oslash ([t], \bot)$, while for the rightside, we get just $[t]$. Therefore $[g(t, 0)] > [t]$ by the inequality 2. The argument for the second rule, $g(0, N) = N$, is the same. Of the two conditional rules in GCD, we check only the first, because the second follows the same way. This rule is $g(M, N) = g(M - N, N)$ if $M \geq N$ and $N > 0$. By ($D_d$), $t, t'$ reduce to Peano terms, say $m, n$; then $[g(t, t')] = (m, n) \oslash ([t], [t'])$, while for the rightside we have $[g(t-t', t')] = (m-n, n) \oslash ([t-t'], [t'])$, and the desired inequality follows because $n > 0$, so that $(m, n) > (m - n, m)$, and hence $(m, n) \oslash ([t], [t']) > (m - n, n) \oslash ([t - t'], [t'])$, by the definitions of the product and lexicographic orderings.

The induction step for the first two inductive assertions has two cases. The first case considers $t = g(t_1, t_2)$ with $t_1, t_2 \in T^d$. Then by (b), $[t] = (n_1, n_2) \oslash ([t_1], [t_2])$, where the reduced forms of $t_1, t_2$ are respectively $n_1, n_2$, which are Peano terms by ($D_d$). For assertion ($A_{d+1}$), any application of an old rule is weight preserving because it rewrites either $t_1$ or $t_2$, which preserves the weight of $t$ by ($A_d$). For assertion ($B_{d+1}$), any application of a new rule at the top is strict monotone by our lemma, and otherwise is strict monotone by ($B_d$).

The second case of the induction step considers a Σ-term $t$ having depth $d + 1$ of the form $t_0(g_1, \ldots, g_k)$ where $k > 0$ and each $g_i$ has the form $g(t_{i,1}, t_{i,2})$ with $t_{i,j} \in T^d$ and $t_0 \in T_{\Sigma'}(\{z_1, \ldots, z_k\})$. Then $[g_i] = (n_{i,1}, n_{i,2}) \oslash ([t_{i,1}], [t_{i,2}])$ as in the first case, and so we have $[t] = \{(n_{1,1}, n_{1,2}) \oslash ([t_{1,1}], [t_{1,2}]), \ldots, (n_{k,1}, n_{k,2}) \oslash ([t_{k,1}], [t_{k,2}])\}$ by (c). For assertion ($A_{d+1}$), once again any application of an old rule preserves the weight of $t$, because it either rewrites some $t_{i,j}$, which preserves the weight of $t$ because it preserves the weight of $g_i$ by $(A_d)$, or else it rewrites within $t_0$, which also preserves the weight of $t$, because of (c). For assertion $(B_{d+1})$, any application of a new rule is strict monotone on any such $t$, either by $(B_d)$, or else by the lemma.

Finally, we consider the remaining two inductive assertions. For $(C_{d+1})$, when rewriting with $\mathcal{A}$ on $T^d$, all weights remain within $N_{3d+3} = N_{3(d+1)}$ because of the form of $[t]$ and $(C_d)$. For $(D_{d+1})$, rewriting always terminates for terms in $T^{d+1}$ by Proposition 5.8.31 with $T = T^{d+1}$ and $P = N_{3d+3}$, plus the induction hypotheses; moreover, the result must be a Peano term because of the form of $[t]$, using $(D_d)$ and (a).

It now follows that $(A_d, B_d, C_d, D_d)$ hold for every $d \geq 0$, and in particular that rewriting on any ground $\Sigma$-term $t \in T = \bigcup_d T^d$ necessarily terminates with a Peano term as its result.  □

The main result of Section 5.8.2, Theorem 5.8.20 on page 136, generalizes to Noetherian orderings, though we do not use this result in this book:

**Theorem 5.8.33** Let $(\Sigma, A)$ be a CTRS with $\Sigma$ non-void, let $(\Sigma, B)$ be a terminating sub-CTRS of $(\Sigma, A)$, let $P$ be a poset, and let $N = A - B$. If there is a function $\rho : T_{\Sigma,B} \to P$ such that

(1) each rule in $B$ is weak $\rho$-monotone,

(2) each rule in $N$ is strict $\rho$-monotone,

(3) each operation in $\Sigma$ is strict $\rho$-monotone, and

(4) $P$ is Noetherian, or at least, then for each $t \in (T_{\Sigma,B})_s$ there is some Noetherian poset $P^t_s \subseteq P_s$ such that $t \overset{*}{\Rightarrow}_A t'$ implies $\rho(t') \in P^t_s$.

then $(\Sigma, A, B)$ is ground terminating.  □

The proof depends on the straightforward generalization from $\rho : T_\Sigma \to \omega$ to $\rho : T_\Sigma \to P$ of results in Section 5.8.2 and hence is omitted here.

### 5.8.4  Proving Church-Rosser

When applying a conditional rule $t = t'$ **if** $u == v$ over some theory $A$ in OBJ, it is possible that a term $r$ exists such that $\theta(u) \overset{*}{\Rightarrow}_A r$ and $\theta(v) \overset{*}{\Rightarrow}_A r$, but rewriting does not find this $r$ because $\theta(u)$ and $\theta(v)$ reduce to different normal forms. In fact, it cannot be guaranteed that OBJ will always evaluate a condition $u == v$ to `true` when $\theta(u) \downarrow \theta(v)$ unless the set of rules that can be used for evaluating conditions is both Church-Rosser and terminating. This provides some motivation for checking the Church-Rosser property in the conditional case. But as we continue to emphasize, because of soundness, any rewriting computation is a proof, so if you do get the result that you want, then you have proved the result that you wanted to prove, whether or not the CTRS is terminating or Church-Rosser. In our experience, practical examples can usually be handled without bothering to check canonicity, though of course it is comforting.

This subsection extends the techniques presented for proving confluence in Section 5.6 to the conditional case. Many basic results extend just because they follow directly from the corresponding results about ARS's, as discussed at the beginning of Section 5.8. The situation for Proposition 5.3.6 is slightly different; the proof, which involves passing to two different ARS's, generalizes to the conditional case without any change.

**Proposition 5.8.34** Given a sort set $S$, let $X_S^\omega$ be the ground signature with $(X_S^\omega)_s = \{x_s^i \mid i \in \omega\}$ for each $s \in S$. Then a CTRS $(\Sigma, A)$ is Church-Rosser iff the CTRS $(\Sigma(X_S^\omega), A)$ is ground Church-Rosser. Similarly, $(\Sigma, A)$ is locally Church-Rosser iff the CTRS $(\Sigma(X_S^\omega), A)$ is ground locally Church-Rosser. □

**Example 5.8.35** The analog of Proposition 5.8.12 for confluence is not true, not even for orthogonal CTRS's. Let $C$ be the CTRS corresponding to the following equations (with $x$ a variable):

$$f(x) = a \quad \text{if} \quad x = f(x)$$
$$b = f(b) \ .$$

Then $b \Rightarrow f(b) \Rightarrow a$ and $f(b) \stackrel{*}{\Rightarrow} f(a)$. However, it is not true that $f(a) \downarrow a$. Hence $C$ is not Church-Rosser. However, $C^U$ is Church-Rosser. Note also that $C$ is orthogonal. (This example is due to Bergstra and Klop [10].) □

The Orthogonality Theorem (Theorem 5.6.4) can be generalized to CTRS's by generalizing the notion of non-overlapping, but we do not do so here, because orthogonality is a rather strong property, and in any case the generalization of the Newman Lemma handles most examples of practical interest; to maximize practicality, we give a hierarchical version that allows checking the key properties "incrementally," that is, one level at a time, as was previously done for termination.

**Proposition 5.8.36** Let $A$ be a terminating $\Sigma$-CTRS, let $B$ be a "base" Church-Rosser CTRS contained in $A$, and let $N = A - B$. Then $A$ is Church-Rosser (and hence canonical) if:

(1) $N$ is locally Church-Rosser, i.e., if $t \Rightarrow_N t_1$ and $t \Rightarrow_N t_2$ then there is some $t'$ such that $t_1 \stackrel{*}{\Rightarrow}_N t'$ and $t_2 \stackrel{*}{\Rightarrow}_N t'$; and

(2) if $t \Rightarrow_B t_1$ and $t \Rightarrow_N t_2$ then there is some $t'$ such that $t_1 \stackrel{*}{\Rightarrow}_N t'$ and $t_2 \stackrel{*}{\Rightarrow}_B t'$.

**Proof:** (1) and (2) imply that $A$ is locally Church-Rosser, and then the Newman Lemma gives the full Church-Rosser property.  □

Note that by the original Newman Lemma, it would be equivalent to assume that $B$ is locally Church-Rosser. Condition (2) is a local version of the Hindley-Rosen property from Proposition 5.7.5. We now give some applications of the above result.

**Example 5.8.37** We first consider the maximum function of Example 5.8.13, which has already been shown terminating. Assume that the natural number part of this CTRS, here denoted $B$, has been shown Church-Rosser. Then it remains to check conditions (1) and (2). Condition (1) can be checked completely mechanically by using the Knuth-Bendix algorithm [117] (see Chapter 12), and condition (2) can be checked by a variant of the same algorithm. But here we give a rather informal argument, which will serve to motivate the more formal developments of Chapter 12. The idea is to determine which rules could give rise to the two given rewrites, and then show the existence of a suitable $t'$ for each such case. Note that unless the two rewrites overlap, it is straightforward to see that $t'$ exists. e will abbreviate $max$ by just $m$ in the detailed arguments below.

For (1), we suppose that $t \Rightarrow_N t_1$ and $t \Rightarrow_N t_2$. The only way that two new rules can overlap is if the redex has the form $m(u,u)$ for some $\Sigma$-term $u$ with $t_1 = t_2 = u$, so that we have $t_1 \stackrel{*}{\Rightarrow}_N t'$ and $t_2 \stackrel{*}{\Rightarrow}_N t'$ with $t' = u$. For (2), we consider $t \Rightarrow_B t_1$ and $t \Rightarrow_N t_2$. But it is impossible for a new rule to overlap with a base rule in this specification, because the leftsides of the rules in $B$ and $N$ have disjoint function symbols; so there is nothing to check. Thus Proposition 5.8.36 implies that this specification is Church-Rosser, and hence canonical.

Now we consider the greatest common divisor function of Example 5.8.24, which was already shown terminating in Example 5.8.32. For (1), note that in this specification, there is no overlap between new rules, so there is nothing to check. Similarly for (2), there is no overlap between new and old rules, because the leftsides involve disjoint function symbols, and so again there is nothing to check. Therefore Proposition 5.8.36 shows that this specification is Church-Rosser, and hence canonical.  □

There is also a version of Proposition 5.8.36 that does not assume termination, but instead requires stronger confluence conditions; however we are usually less interested in the Church-Rosser property if termination does not hold, so we do not give this result here.

**Exercise 5.8.9** Given a CTRS $C$, let $C^U$ be the TRS whose rules are those of $C$ with their conditions (if any) removed. Then $C$ is Church-Rosser (or ground Church-Rosser) if $C^U$ is. □

## 5.9 (⋆) Relation between Abstract and Term Rewriting Systems

This section describes a relationship between abstract rewriting systems and term rewriting systems, including a construction of each from the other, and proves that these two constructions are the best possible with respect to each other, in a sense that is made precise by saying that they form an "adjoint pair of functors." However, the notion of adjointness is not needed to understand our statement of this result, and in fact, no category theory at all is used in this section, although we do mention categories and functors in some exercises.[18]

We can say much more about the relationship between the TRS's and ARS's than in Section 5.7 after we introduce morphisms of TRS's and ARS's. First, we add a little more information[E19] to TRS's, by including a sort $s$ from its signature. Then TRS morphisms are interpretations (in the sense of Definition 4.10.4) that preserve the designated sort and all one-step rewrites.

**Definition 5.9.1** A **TRS morphism** $(\Sigma, A) \to (\Sigma', A')$ is a signature morphism $h : \Sigma \to Der(\Sigma')$ such that whenever $t_0 \stackrel{1}{\Rightarrow}_A t_1$ then $\overline{h}(t_0) \stackrel{1}{\Rightarrow}_{A'} \overline{h}(t_1)$, where $\overline{h}$ is as defined just after Definition 4.10.3, the unique $\Sigma$-homomorphism to $T_{\Sigma'}$ obtained by looking at $T_{\Sigma'}$ first as a $Der(\Sigma')$-algebra, and then through $h$, as the reduct $\Sigma$-algebra $hT_{\Sigma'}$ (see page 84 for this).

Given TRS morphisms $h : (\Sigma, A) \to (\Sigma', A')$ and $h' : (\Sigma', A') \to (\Sigma'', A'')$, their **composition** $h; h' : (\Sigma, A) \to (\Sigma'', A'')$ is defined to be their composition $h; h'^* : \Sigma \to Der(\Sigma'')$ as derivors in the sense of Definition 4.10.5. □

**Exercise 5.9.1** Show that the composition of TRS morphisms is a TRS morphism. Use Exercise 4.10.5 to show that $i_\Sigma : \Sigma \to Der(\Sigma)$ (sending $\sigma \in \Sigma_{w,s}$ to the term $\sigma(x_1, \ldots, x_n) \in T_\Sigma(^w X)_s$) is the identity for TRS morphism composition. Show that TRS morphism composition is associative (whenever the compositions involved are defined). These results show that TRS's form a category; let us denote it $\mathbb{T}RS$. □

**Definition 5.9.2** An **ARS morphism** $(S, T, \to) \to (S', T', \to')$ is a pair $(f, g)$ where $f : S \to S'$ and $g_s : T_s \to T'_{f(s)}$ for each $s \in S$, such that $t_0 \to t_1$ implies $g(t_0) \to' g(t_1)$. Given ARS morphisms $(f, g) : (S, T, \to) \to$

---

[18]A category is just a collection of objects (such as TRS's) and maps (also often called morphisms) between them, such that certain axioms are satisfied. There are many places to learn about these concepts, including [91, 6] and [126]; [63] discusses the intuitive meanings of these and other categorical concepts.

$(S', T', \rightarrow')$ and $(f', g') : (S', T', \rightarrow') \longrightarrow (S'', T'', \rightarrow'')$, then their **composition** is the pair $(f; f', \{g_s; g'_{f(s)} \mid s \in S\}) : (S, T, \rightarrow) \longrightarrow (S'', T'', \rightarrow'')$.  □

**Exercise 5.9.2** Show that the composition of ARS morphisms is an ARS morphism. Show that the pair of identity maps $(1_S, 1_T)$ serves as an identity for $(S, T, \rightarrow)$ under ARS morphism composition. Show that ARS morphism composition is associative (when the compositions involved are defined). These results show ARS's form a category; let us denote it $\mathbb{A}RS$.  □

Now we are ready for the first of our two main constructions:

**Definition 5.9.3** Let $R$ send a TRS $(\Sigma, A)$ to the ARS $(T_\Sigma, \overset{1}{\Rightarrow}_A)$, and send a TRS morphism $h : (\Sigma, A) \rightarrow (\Sigma', A')$, i.e., $h : \Sigma \rightarrow Der(\Sigma')$, to $R(h) = (f, \overline{h}) : T_\Sigma \rightarrow T_{\Sigma'}$, where $f$ is the sort component of the signature morphism $\overline{h}$.  □

**Exercise 5.9.3** Show that $R(h)$ is an ARS morphism, and that $R : \mathbb{T}RS \rightarrow \mathbb{A}RS$ preserves composition and identities, i.e., that it is a functor.

**Hint:** To show that $R(h)$ is an ARS morphism, check that $t_0 \overset{1}{\Rightarrow}_A t_1$ implies $\overline{h}(t_0) \overset{1}{\Rightarrow}_{A'} \overline{h}(t_1)$.  □

Here is our second construction:

**Definition 5.9.4** Let $F$ send an ARS $(S, T, \rightarrow)$ to $(\Sigma^T, A^\rightarrow)$, where $\Sigma^T$ is defined by $\Sigma^T_{[],s} = T_s$ and $\Sigma^T_{w,s} = \emptyset$ for all other $w, s$, and where $A^\rightarrow$ contains a rewrite rule $(\forall \emptyset)\, t_0 = t_1$ iff $t_0 \rightarrow t_1$ in $(S, T, \rightarrow)$. Also, if $(f, g) : (S, T, \rightarrow) \longrightarrow (S', T', \rightarrow')$ is an ARS morphism, then define $F(f, g) : (\Sigma^T, A^\rightarrow) \rightarrow (\Sigma^{T'}, A^{\rightarrow'})$ to be $h : \Sigma^T \rightarrow Der(\Sigma^{T'})$ defined by $h_s(t) = g_{f(s)}(t) \in \Sigma^{T'}_{[],f(s)} = T'_{f(s)}$ for $t \in \Sigma^T_{[],s} = T_s$.  □

**Exercise 5.9.4** Show that $F(f, g)$ is a TRS morphism, and that $F : \mathbb{A}RS \rightarrow \mathbb{T}RS$ preserves composition and identities, i.e., that it is a functor.

**Hint:** To show that $F(f, g)$ is a TRS morphism, show that $t_0 \rightarrow_{A^\rightarrow} t_1$ implies $h(t_0) \rightarrow_{A^{\rightarrow'}} h(t_1)$ where $h = F(f, g)$.  □

Before stating the main result, we need the following:

**Fact 5.9.5** $R(F(\mathcal{A})) = \mathcal{A}$, for any ARS $\mathcal{A}$.

**Proof:** If $\mathcal{A} = (S, T, \rightarrow)$, then $F(\mathcal{A}) = (\Sigma^T, A^\rightarrow)$ where $\Sigma^T_{[],s} = T_s$ and $\Sigma_{w,s} = \emptyset$ for all other $w, s$, and where the rewrite rule $(\forall \emptyset)\, t_1 = t_2$ is in $A^\rightarrow$ iff $t_1 \rightarrow t_2$ in $\mathcal{A}$. Then $R(F(\mathcal{A})) = (T_{\Sigma^T}, \overset{1}{\Rightarrow}_{A^\rightarrow}) = (T, \rightarrow)$.  □

The theorem below says that the functor $F$ is left adjoint to $R$, but it is stated as a so-called "universal property" that does not use any category theory. (Figure 5.5 shows the traditional commutative diagram for this property.)

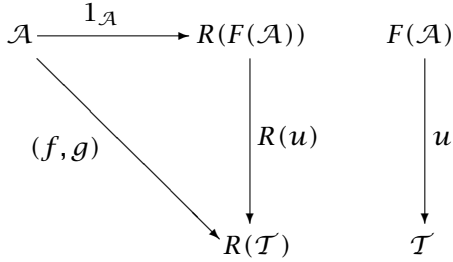

Figure 5.5: Universal Property of $F(\mathcal{A})$

**Theorem 5.9.6** For every ARS $\mathcal{A}$, TRS $\mathcal{T}$ and ARS morphism $(f, g) : \mathcal{A} \to R(\mathcal{T})$, there is a unique TRS morphism $u : F(\mathcal{A}) \to \mathcal{T}$ such that $R(u) = (f, g)$.

**Proof:** Let $\mathcal{A}$ be $(S, T, \to)$ and let $\mathcal{T}$ be $(\Sigma', A')$. Then $F(\mathcal{A}) = (\Sigma^T, A^\to)$. If we assume that $R(u) = (f, g)$ as morphisms $(S, T, \to) \longrightarrow (T_{\Sigma'}, \overset{1}{\Rightarrow}_{A'})$, then for $u : (\Sigma^T, A^\to) \to (\Sigma', A')$, which is really $u : \Sigma^T \to Der(\Sigma')$, we must have $u_{[],s}(t) = g_s(t)$ for each $t \in \Sigma^T_{[],s} = T_s$, and that all the other maps $\Sigma^T_{w,s} \to T_{\Sigma',s'}(^w X)$ are empty. Furthermore, with this definition of $u$, we have that $R(u) = (f, g)$, because $R(u) = \overline{u} : T_{\Sigma^T} = T \to T_{\Sigma'}$ where $u_s(t) = g_s(t)$ for $t \in T_s$.

Finally, to show that $u$ is a TRS morphism, we must show that $t_1 \overset{1}{\Rightarrow}_{A^\to} t_2$ implies $u(t_1) \overset{1}{\Rightarrow}_{A'} u(t_2)$. But $t_1 \overset{1}{\Rightarrow}_{A^\to} t_2$ iff $t_1 \to t_2$, and since $(f, g)$ is an ARS morphism, from this we get $g_s(t_1) \to_{A'} g_s(t_2)$, which occurs iff $u(t_1) \to_{A'} u(t_2)$. □

**Exercise 5.9.5** Substitute "$t_0 \overset{1}{\Rightarrow}_A t_1$ implies $\overline{h}(t_0) \overset{*}{\Rightarrow}_{A'} \overline{h}(t_1)$" for "$t_0 \overset{1}{\Rightarrow}_A t_1$ implies $\overline{h}(t_0) \overset{1}{\Rightarrow}_{A'} \overline{h}(t_1)$" in Definition 5.9.1, substitute "$t_0 \to t_1$ implies $g(t_0) \overset{*}{\to}' g(t_1)$" for "$t_0 \to t_1$ implies $g(t_0) \to' g(t_1)$" in Definition 5.9.2, and then show that Theorem 5.9.6 still holds. Give an interpretation for this result. Show that the local Church-Rosser property is not preserved by either $F$ or $R$ when morphisms are generalized in this way. □

## 5.10 Literature

Term rewriting captures a basic computational aspect of equational logic, and is fundamental for theorem proving. However, expositions of term rewriting typically have a combinatorial, syntactic flavor, rather than an algebraic, semantic flavor. This is due in part to the historical fact that term rewriting arose as an abstraction of the lambda calculus, especially the so called Normalization (i.e., Church-Rosser) Theorem, which was first proved by Church and Rosser, and which is the origin of the "Church-Rosser property."

There is a very large literature on term rewriting. This chapter does not faithfully represent that literature, because it emphasizes results that are of practical value for theorem proving, as opposed to results that are largely of theoretical interest, and it leans heavily towards algebra. Huet and Oppen gave a good survey that developed some of the connections with algebra [109]. Klop [114, 115], Dershowitz and Jouannaud [39], and Plaisted [151] have also written useful surveys; the latter two describe some more recent developments. A nice self-contained introductory textbook has been written by Baader and Nipkow [2]. Newman proved his lemma for the unsorted case in 1942 [143]; the elegant proof given here is due to Barendregt [3], but generalized to overloaded many-sorted rewriting. Theorem 5.2.9 is from [56]; it expresses a fundamental connection between term rewriting and algebra. The Noetherian condition is named after Emmy Noether, the great pioneer in abstract algebra mentioned in Section 2.8.

Combinatory logic was developed by Schönfinkel [160] to eliminate bound variables from predicate logic, and was later independently developed further by Haskell Curry as a foundation for mathematics that he called "Illative Combinatory Logic" [37, 38]. Combinatory logic also plays an important role in implementing functional programming languages, as described in [4], [178] and many other places. The so-called categorical combinators developed more recently by Curien and others have played a similar role [36].

Although not discussed here, the lambda calculus is a TRS closely related to combinatory logic. It was developed by Alonzo Church [28] as a calculus of functions, again as part of a foundational programme for mathematics. This TRS played a key role in formalizing the notion of computability (the so-called Church-Turing thesis), and following work of Landin [120] and Strachey [172], it became the basis for the "denotational semantics" [162] method for defining the meaning of programming languages. Lambda calculus has also been an important influence on the design of programming languages, including Lisp [131] and more recently, higher-order functional languages like ML [99], Miranda [179], and Haskell [107].

Term rewriting plays a basic role in proving properties of abstract data types, including their correctness and implementation, and also in the study of their computability properties [137]. In addition, term rewriting has played an important role in developing languages that combine the functional and logic paradigms, through an operational semantics based on so called narrowing [79, 40, 105]. As far back as 1951, Evans [46] used term rewriting to prove the decidability of the equational theory called "loops."

Most expositions of term rewriting do not make the signature explicit, so that the distinction between (for example) confluence and ground confluence can seem mysterious, and various confusions can easily arise. Similarly, it is not usual to be careful about the variables and constants involved in rewriting a given term. The results of Propo-sitions 5.3.4 and 5.3.6, and of Corollary ??,[E20] which address these issues, do not seem to be in the literature, nor do the corresponding results for the conditional case, Propositions 5.8.10 and 5.8.11. This is presumably because they cannot even be stated without the additional care for variables and constants that we have taken.

Section 5.4 on evaluation strategies has been largely taken from [90]. There is an interesting literature on proving termination of rewriting when operations have local strategies, for example, see [50], which cites many other papers.

The unsolvability of equality mentioned in connection with Proposition 5.1.14 is shown by the unsolvability of the so-called word problem for groups, posed by Max Dehn in 1911: given a group presentation, determine whether or not two terms over the generators are equal in that equational theory; this was shown unsolvable by Petr Novokov [144] and William Boone [16] in the 1950s. Unsolvability of equality also follows from the word problem for semigroups, posed by Axel Thue in 1914 and shown unsolvable by Emil Post in 1947 [153].

That orthogonality implies Church-Rosser (Proposition 5.6.4) has been proved many times, perhaps first by Rosen [158], but our version may be the first that goes beyond the unsorted case, and our proof also appears to be novel. The term "orthogonal" is due to Dershowitz. Many other results in this chapter are also new, in the same limited sense that they are proved for overloaded many-sorted rewriting. Theorem 5.6.9 and the notions of superposition and critical pair are part of a larger story about unification and the Knuth-Bendix method covered in Chapter 12; that material appears in this chapter because of its value for showing the Church-Rosser property. Hindley's original proof of the Hindley-Rosen Lemma appears in [103].

The literature includes several different notions of conditional rewriting, e.g., see the survey of Klop [115]. The join conditional rewriting approach of our Definition 5.8.2 is the most satisfactory for OBJ because it includes the computations done in the common case when == occurs in a condition. The alternative notions are either less general, or else are too general, for example, going beyond term rewriting by requiring conditions to be evaluated using the full power of equational deduction.

Although Propositions 5.5.1 and 5.8.16 are very simple, they express the fundamental relationship between termination and weight functions, and they have not been emphasized in the literature, and may even be partially new. Results 5.8.18, 5.8.19, 5.8.20, and 5.8.33 all appear to be new, and have practical value for proving termination, especially Theorems 5.8.20 and 5.8.33, which also handle conditional rules.

The results on constructing Noetherian orderings in Section 5.8.3 are standard, although the particular constructions given for the mul-tiset and lexicographic orderings may be new. The observation that colimits appear in several places is new, as is the termination criterion in Theorem 5.8.31 and its application in Example 5.8.32. Though the proof in this example is a bit elaborate for a result that is intuitively relatively obvious, it does provide a fairly thorough illustration of the machinery introduced in Section 5.8.3. The material on the Church-Rosser property in Section 5.8.4 may be new; although the special case of Proposition 5.8.36 with $B = \emptyset$ is of course familiar, the generalization to hierarchical CTRS's is very useful in practice.

The use of ARS's to study TRS's is standard in the literature, although terminology and definitions vary. Klop [114, 115] considers sets with an indexed family of relations (as in Proposition 5.7.5), calling them "abstract reduction systems," and using the name "replacement system" for the case of just one relation, which we call an abstract rewrite system; actually, our formulation is a bit more general, because it is $S$-indexed, which enables some novel applications to many-sorted term rewriting and equational deduction. The results in Section 5.9 are new, especially Theorem 5.9.6 on the adjoint relation between ARS's and TRS's. This material suggests many questions for further research, such as exploring properties of the two categories involved, and more ambitiously, reformulating term rewriting theory in a more categorical style.

José Meseguer [134] has developed rewriting logic, which gives sound and complete rules of inference for term rewriting; these rules are the same as those for equational deduction, except that the symmetry law is omitted. This logic can also be seen as a logic for the term rewriting model of computation, and as such has many interesting applications, including a comprehensive unification of different theories of concurrency, a nice operational semantics for inference systems, and a uniform meta-logic in which inference systems can be described and implemented [29].

I thank Prof. Virgil-Emil Cazanescu for his help with the proofs of Propositions 5.2.6 and 5.3.4, and Dr. Răzvan Diaconescu for help with the proof of Theorem 5.9.6. I also thank José Barros and Răzvan Diaconescu for their help with some of the examples, Kai Lin for several very useful discussions, as well as for significant help with the examples in Section 5.8.2, and Grigore Roşu for the proof of the Orthogonality Theorem (Theorem 5.6.4) in Appendix B, and for several valuable suggestions. Finally, I thank Ms. Chiyo Matsumiya and especially Prof. Yoshihito Toyama, and Dr. Monica Marcus, for their valuable comments and corrections to this chapter. The proof of Theorem 5.6.9 in Appendix B is due to Dr. Marcus.

> **A Note to Lecturers:** This chapter contains a great deal of material, some of which is rather difficult. Except in the case of an advanced course of some duration, the lecturer will have to omit a fair amount, certainly including all the starred sections. Beyond that, the material to be covered may be determined by the taste of the lecturer and the choice of material to be covered from later chapters. In particular, it is safe to omit most of the detailed material on proving termination and the Church-Rosser property, since little of that is needed for later chapters. It could also be a good idea to interleave material from this chapter with parts of Chapter 6, to create a bit more variety.

# 6 Initial Algebras, Standard Models and Induction

This chapter shows that every equational specification has an initial algebra, gives further characterizations for these structures, and justifies and illustrates the use of induction for verifying their properties. It also investigates abstract data types and standard models for equational specifications, showing that they are initial models. Congruence and quotients are important technical tools, and we prove some of their main properties.

## 6.1 Quotient and Initiality

This section discusses congruences, quotients, initial and free algebras satisfying equations, and then substitutions modulo equations. Main results include the so-called homomorphism theorem, the universal characterization of quotients, and the existence of initial algebras.

### 6.1.1 Congruence, Quotient and Image

Initial algebras for specifications with equations are constructed as quotients of term algebras, a construction that relies upon the following:

**Definition 6.1.1** A $\Sigma$-**congruence relation** on a $\Sigma$-algebra $M$ is an $S$-sorted equivalence relation $\equiv\ =\ \{\equiv_s \mid s \in S\}$ on $M$, for $S$ the sort set of $\Sigma$, where each $\equiv_s$ is an equivalence relation on $M_s$ such that whenever $\sigma \in \Sigma_{s_1 \ldots s_n, s}$ then

$$a_i \equiv_{s_i} a'_i \text{ for } i = 1, \ldots, n \text{ implies } M_\sigma(a_1, \ldots, a_n) \equiv_s M_\sigma(a'_1, \ldots, a'_n),$$

for $a_i, a'_i \in M_{s_i}$ for $i = 1, \ldots, n$. □

**Example 6.1.2** Define a signature $\Sigma$ by the OBJ fragment,

```
sorts Nat Bool .
op 0 : -> Nat .
op s : Nat -> Nat .
ops T F : -> Bool .
op odd : Nat -> Bool .
```

and let $N$ be the $\Sigma$-algebra with $N_{\mathsf{Nat}} = \omega$, with $N_{\mathsf{Bool}} = \{T, F\}$, and with the operations interpreted as expected. Then we can define a $\Sigma$-congruence $Q_8$ on $N$ as follows: $nQ_{8,\mathsf{Nat}}n'$ iff $n - n'$ is divisible by 8; and $bQ_{8,\mathsf{Bool}}b'$ iff $b = b'$. We can also define another $\Sigma$-congruence $Q_2$ on $N$ as follows: $nQ_{2,\mathsf{Nat}}n'$ iff $n - n'$ is divisible by 2; and $bQ_{2,\mathsf{Bool}}b'$ iff $b = b'$. □

**Exercise 6.1.1** In the context of Example 6.1.2, prove that $Q_8$ and $Q_2$ are $\Sigma$-congruences on $N$. Now define $Q_3$ to mean having the same remainder under division by 3, and show that $Q_3$ is *not* a $\Sigma$-congruence on $N$. □

**Proposition 6.1.3** Given a $\Sigma$-algebra $M$ and a $\Sigma$-congruence $\equiv$ on $M$, the **quotient**[1] of $M$ by $\equiv$, denoted $M/\equiv$, is a $\Sigma$-algebra, interpreting constant symbols $\sigma \in \Sigma_{[],s}$ as $[M_\sigma]$, and operations $\sigma \in \Sigma_{s_1...s_n,s}$ with $n > 0$ as maps sending $[a_1], \ldots, [a_n]$ to $[M_\sigma(a_1, \ldots, a_n)]$, for $a_i \in M_{s_i}$.

**Proof:** We have to show that $[M_\sigma(a_1, \ldots, a_n)]$ is well defined. So let us assume, for $a_i, a_i' \in M_{s_i}$, that $a_i \equiv_{s_i} a_i'$ for $i = 1, \ldots, n$, i.e., that $[a_i] = [a_i']$. Then the definition of congruence gives us that $[M_\sigma(a_1, \ldots, a_n)] = [M_\sigma(a_1', \ldots, a_n')]$. □

**Example 6.1.4** The equivalence classes of ground $\Sigma$-terms under a set $A$ of $\Sigma$-equations form a nice $\Sigma$-algebra, which is in fact a quotient of the term algebra by a congruence based on equational deduction as follows: Given $\Sigma$-terms $t, t'$ with variables in $X$, let

$$t \simeq_A^X t' \quad \text{iff} \quad A \vdash (\forall X)\ t = t' \ .$$

where $\simeq_A^X$ for sort $s$ has $t, t'$ also of sort $s$. That $\simeq_A^X$ is an equivalence relation follows directly from rules (1), (2), (3) of Definition 4.1.3 on page 58, which are the reflexivity, symmetry and transitivity of equational deduction, respectively. In the special case where $X = \emptyset$, we write just $\simeq_A$ instead of $\simeq_A^\emptyset$. Now we take the quotient $T_\Sigma/\simeq_A$ as an $S$-sorted set, and let $[t]_A$ or (usually) just $[t]$, denote the equivalence class of a term $t$ under $\simeq_A$.

For these equivalence classes to form a $\Sigma$-algebra, we need to give interpretations for the constant and operation symbols in $\Sigma$. It seems clear that we should interpret the constant symbol $\sigma \in \Sigma_{[],s}$ as $[\sigma]$, and interpret $\sigma \in \Sigma_{s_1...s_n,s}$ with $n > 0$ as the map sending $[t_1], \ldots, [t_n]$ to $[\sigma(t_1,\ldots,t_n)]$, where $t_i \in T_{\Sigma,s_i}$ for $i = 1,\ldots,n$. But it may not be clear that this definition makes sense. For, if we had picked some other $t'_1,\ldots,t'_n$ such that $[t_i] = [t'_i]$, then we would need to know that

$$[\sigma(t_1,\ldots,t_n)] = [\sigma(t'_1,\ldots,t'_n)]$$

in order to know that the proposed interpretation for $\sigma$ gives the same result, no matter which representatives we happen to have chosen for the equivalence classes. Translating back to the notation of equational deduction, the property we need is

$$A \vdash (\forall \emptyset) \; t_i = t'_i \text{ for } i = 1,\ldots,n \text{ implies}$$

$$A \vdash (\forall \emptyset) \; \sigma(t_1,\ldots,t_n) = \sigma(t'_1,\ldots,t'_n) \;.$$

---

[1]Appendix C reviews this construction for $S$-sorted sets.

But this follows directly from the rule of deduction (4) of Definition 4.1.3: let $X = \{x_1,\ldots,x_n\}$ with $x_i$ of sort $s_i$, let $Y = \emptyset$, let $\theta(x_i) = t_i$, let $\theta'(x_i) = t'_i$, and let $t = \sigma(x_1,\ldots,x_n)$; then $(\forall \emptyset) \; \sigma(t_1,\ldots,t_n) = \sigma(t'_1,\ldots,t'_n)$ is deducible, because $\theta(t) = \sigma(t_1,\ldots,t_n)$ and $\theta'(t) = \sigma(t'_1,\ldots,t'_n)$. Thus $T_\Sigma/\simeq_A$ is a $\Sigma$-algebra, in fact an initial $(\Sigma, A)$-algebra, though we do not prove it directly in this way. □

**Definition 6.1.5** Given a $\Sigma$-homomorphism $h : M \to M'$, the **kernel** of $h$ is the $S$-sorted family of equivalence relations $\equiv_h$ on $M$, defined on $M_s$ by $a \equiv_{h,s} a'$ iff $h_s(a) = h_s(a')$; the kernel of $h$ may be denoted $ker(h)$. The **image** of $h$, denoted $im(h)$ or $h(M)$, is the $\Sigma$-subalgebra of $M'$ with $h(M)_s = h_s(M_s)$ for each $s \in S$, and with operations those of $M'$ suitably restricted. □

**Proposition 6.1.6** The kernel of a $\Sigma$-homomorphism $h : M \to M'$ is a $\Sigma$-congruence, and its image is a $\Sigma$-algebra.

**Proof:** Each $\equiv_{h,s}$ is an equivalence relation, for any $S$-indexed function $h : M \to M'$. To prove the congruence property, let $\sigma \in \Sigma_{w,s}$ with $w = s_1 \ldots s_n$ and assume $a_i \equiv_{h,s} a'_i$, i.e., that $h_s(a_i) = h_s(a'_i)$ for $i = 1,\ldots,n$. Then

$$h_s(M_\sigma(a_1,\ldots,a_n)) = M'_\sigma(h_{s_1}(a_1),\ldots,h_{s_n}(a_n)) =$$
$$M'_\sigma(h_{s_1}(a'_1),\ldots,h_{s_n}(a'_n)) = h_s(M_\sigma(a'_1,\ldots,a'_n)) \;,$$

so that

$$M_\sigma(a_1,\ldots,a_n) \equiv_{h,s} M_\sigma(a'_1,\ldots,a'_n) \;,$$

as desired.

For the second assertion, we first check condition (2) of the definition of subalgebra given in Exercise 3.1.1, let $\sigma \in \Sigma_{w,s}$ with $w = s_1 \ldots s_n$, let $b_i \in h(M)_{s_i}$ for $i = 1,\ldots,n$, and let $a_i \in M_{s_i}$ such that $b_i = f_{s_i}(a_i)$ for $i = 1,\ldots,n$. Then $M'_\sigma(b_1,\ldots,b_n) \in h(M)_s$ since $M'_\sigma(b_1,\ldots,b_n) = h_s(M_\sigma(a_1,\ldots,a_n))$. □

The following is one of the most important elementary results of general algebra. Due to Emmy Noether in its original form, called the "first isomorphism theorem," it relates homomorphisms, quotients, and subalgebras in a very elegant (and useful) way.

**Theorem 6.1.7** (*Homomorphism Theorem*) For any $\Sigma$-homomorphism $h : M \to M'$, there is a $\Sigma$-isomorphism

$$M/ker(h) \cong_\Sigma im(h) .$$

**Proof:** Let $\equiv$ denote $ker(h)$, let $Q$ denote $M/\equiv$, and define $f : Q \to h(M)$ as follows: given some $\equiv$-class $c$, let $f(c)$ be $h(m)$ where $m$ is any element of $M$ such that $[m] = c$; by the definition of $\equiv$, if $m_1, m_2$ are two such elements, then $h(m_1) = h(m_2)$, so that $f$ is well-defined. Also, $f$ is surjective, since for any $h(m) \in h(M)$, we have $f([m]) = h(m)$. So it remains to show that $f$ is a $\Sigma$-homomorphism and is injective.

For the first, if $\sigma \in \Sigma_{[],s}$ then $f(Q_\sigma) = f([M_\sigma]) = h(M_\sigma) = M'_\sigma$. Also, if $\sigma \in \Sigma_{w,s}$ with $w = s_1 \ldots s_k$ then

$$f(Q_\sigma([m_1],\ldots,[m_k])) = f([M_\sigma(m_1,\ldots,m_k)]) = \\ h(M_\sigma(m_1,\ldots,m_k)) = M'_\sigma(h(m_1),\ldots,h(m_k)) = \\ M'_\sigma(f([m_1]),\ldots,f([m_k])).$$

To show that $f$ is injective, assume that $[m_1] \neq [m_2]$ but $f([m_1]) = f([m_2])$, which by definition of $f$ means that $h(m_1) = h(m_2)$, which by definition of $\equiv$ means $[m_1] = [m_2]$, contradicting our assumption. □

The following two corollaries and one exercise spell out some easy consequences of the above:

**Corollary 6.1.8** If $h : M \to M'$ is an injective $\Sigma$-homomorphism, then $M$ is isomorphic to the subalgebra $h(M)$ of $M'$. □

**Corollary 6.1.9** If $h : M \to M'$ is a surjective $\Sigma$-homomorphism, then $M'$ is isomorphic to the quotient $M/ker(h)$ of $M$. □

**Exercise 6.1.2** Show that the converses of the above two corollaries also hold, i.e., show that $M$ is isomorphic to a subalgebra of $M'$ iff there is an injective $\Sigma$-homomorphism $h : M \to M'$, and show that $M'$ is isomorphic to a quotient of $M$ iff there is a surjective $\Sigma$-homomorphism $h : M \to M'$. □

**Example 6.1.10** There is a nice example of the homomorphism theorem in automaton theory. Define a **state system** to consist of an **input set** $X$, a **state set** $Z$, and a **transition function** $t : X \times Z \to Z$; it is conventional to use a tuple notation $(X, Z, t)$ for such systems. Recall that $X^*$ is the set of all finite sequences from $X$, with the empty sequence denoted []. We can extend $t$ to a function $t : X^* \times Z \to Z$, by defining $t([\,], z) = z$ and $t(wx, z) = t(x, t(w, z))$ for $x \in X$, $w \in X^*$, $z \in Z$; this gives the state reached from $z$ after a sequence of inputs. State systems with input set $X$ are $\Sigma$-algebras with $\Sigma$ the one-sorted signature having $\Sigma_1 = X$, and $\Sigma_n = \emptyset$ for all $n \neq 1$, where the "action" of $x \in X$ on $z \in Z$ is defined to be $t(x, z)$. It is conventional to write $x \cdot z$ instead of $t(x, z)$, and also to extend this notation to write $w \cdot z$ instead of $t(w, z)$ for $w \in X^*$.

Next, define an **automaton** to be a state system plus a function $o : Z \to Y$, and define the **behavior** of an automaton $A = (X, Z, t, o)$ **at state** $z \in Z$ to be the function $b_z : X^* \to Y$ defined by $b_z(w) = o(t(w, z))$. Now let $B$ be the $\Sigma$-algebra of all possible behaviors for $A$, with carrier $[X^* \to Y]$, by defining $(x \cdot b)(w) = b(xw)$ for $x \in X$, $b \in B$, $w \in X^*$. Then the function $b$ that sends $z \in Z$ to $b_z \in B$ is a $\Sigma$-homomorphism, and thus Theorem 6.1.7 gives the $\Sigma$-isomorphism

$$A/ker(b) \;\cong_\Sigma\; im(b) \,.$$

The $\Sigma$-algebra $A/ker(b)$ is called the **minimal realization** of $A$; let us denote it $M$. The above isomorphism says that $M$ is the state system with the minimal set of states that realizes the same behaviors as $A$. We can extend $M$ to an automaton by defining $M_o([z]) = o(z)$; the reader may check that this is well defined, in that if $[z] = [z']$ then $o(z) = o(z')$.

The literature often adds an **initial state** $\sigma \in Z$ to state systems and/or automata; then the signature $\Sigma$ is extended by adding $\Sigma_0 = \{\sigma\}$, and the algebra $B$ is extended by adding $B_\sigma = b_\sigma$. Attention is often restricted to automata that are **reachable** in the sense that for every $z \in Z$, there is some $w \in X^*$ such that $w \cdot \sigma = z$, and minimal realizations for such automata are again obtained by taking the quotient by the kernel of the behavior map. The kernel of $b$ is often called the **Nerode equivalence**, after Anil Nerode, who first defined it and the minimal realization of a machine, though in a different way. □

A slightly more general notion of quotient than that developed above starts with an arbitrary relation on a $\Sigma$-algebra:

**Definition 6.1.11** Given a $\Sigma$-algebra $M$ and a subset $R_s$ of $M_s \times M_s$ for each sort $s$ of $\Sigma$, let $\equiv_R$ be the $\Sigma$-**congruence generated by** $R$ **on** $M$, which is the least $\Sigma$-congruence on $M$ that contains $R$, and let $M/R$ denote the quotient $M/\equiv_R$. □

The relation $\equiv_R$ exists because any intersection of $\Sigma$-congruences on $M$ containing $R$ is another such, necessarily the least, and the intersection is non-empty because $M \times M$ is a $\Sigma$-congruence on $M$ containing $R$. The following states a fundamental property of quotients:

**Proposition 6.1.12** Given a $\Sigma$-algebra $M$ and a relation $R$ on $M$, then the quotient map $q : M \to M/R$ satisfies the following:

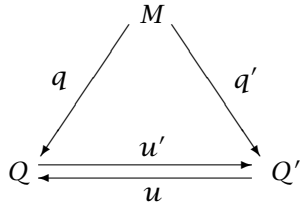

Figure 6.1: Proof for Uniqueness of Quotients

(1) $R \subseteq ker(q)$; and

(2) if $h : M \to B$ is a $\Sigma$-homomorphism such that $R \subseteq ker(h)$ then there is a *unique* $\Sigma$-homomorphism $u : M/R \to B$ such that $q;u = h$.

**Proof:** For (1), it suffices to note that $ker(q)$ is the least congruence containing $R$. For (2), Theorem 6.1.7 gives $M/ker(h) \cong im(h) \subseteq B$, so $R \subseteq ker(h)$ implies $ker(q) \subseteq ker(h)$, which implies by Lemma 6.1.13 below that $h$ factors as $q;q';i$ where $q' : Q \to M/ker(h)$ with $Q = M/ker(h)$, and where $i : ker(h) \to B$ is the inclusion. Therefore we get $u = q';i : Q \to B$ such that $q;u = h$. To show uniqueness, if $u' : A \to B$ such that $q;u' = h$, then $q;u' = q;u$ and so the surjectivity of $q$ implies $u = u'$. □

**Lemma 6.1.13** Given $\Sigma$-congruences $\equiv, \equiv'$ on a $\Sigma$-algebra $M$ with $\equiv\, \subseteq\, \equiv'$, let $Q, Q'$ and $q, q'$ be the respective quotients and quotient maps for $M$. Then $q'$ factors as $q;q''$, where $q''$ is also surjective.

**Proof:** Define $q''([m]) = [m]'$ where $[m]'$ is the $\equiv'$ congruence class of $m \in M$. This is well defined because if $[m] = [m']$ then $[m]' = [m']'$ since $\equiv\, \subseteq\, \equiv'$. Then $q'(q(m)) = [m]' = q'(m)$. □

An assertion that a unique map exists satisfying certain conditions is often called a **universal property**; the above is an example, as are initiality assertions (e.g., Theorems 3.2.1, 3.2.10 and 6.1.15). In each case, the universal property characterizes a structure uniquely up to isomorphism. Proposition 6.2.1 shows this for initial algebras, and the following proves it for quotients:

**Proposition 6.1.14** If both $q : M \to Q$ and $q' : M \to Q'$ satisfy conditions (1) and (2) of Proposition 6.1.12, then $Q$ and $Q'$ are isomorphic.

**Proof:** First notice that if $B = M$ in Proposition 6.1.12, then uniqueness implies $u = 1_M$ since $1_M$ satisfies (1) and (2) in this case. Now under our assumptions, we get $\Sigma$-homomorphisms $u, u'$ such that $q;u = q'$ and $q';u' = q$, from which it follows that $q';u';u = q'$ and $q;u;u' = q$, which by our initial remark implies that $u';u = 1_Q$ and $u;u' = 1_{Q'}$, so that $Q$ and $Q'$ are isomorphic. See Figure 6.1. □

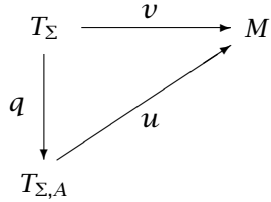

Figure 6.2: Proof for Initiality with Equations

### 6.1.2 Initiality and Freedom

The following is the main result of this section. It says that given a set $A$ of $\Sigma$-equations, there is a $\Sigma$-algebra $T_{\Sigma,A}$, also denoted $T_P$ when $P = (\Sigma, A)$, with the property that given any $\Sigma$-algebra $M$ satisfying $A$, there is a unique $\Sigma$-homomorphism $T_{\Sigma,A} \to M$; i.e., it says that every equational specification has an initial model. Note that, strictly speaking, we are dealing with sorted (or annotated) terms here, in the sense of Definition 3.2.9.

**Theorem 6.1.15** (*Initiality*) Given a set $A$ of (possibly conditional) $\Sigma$-equations, let $\equiv$ be the $\Sigma$-congruence on $T_\Sigma$ generated by the relation $R$ having the components

$$R_s = \{ \langle t, t' \rangle \mid A \vdash (\forall \emptyset)\, t = t' \text{ where } t, t' \text{ are of sort } s \}.$$

Then $T_\Sigma/R$, denoted $T_{\Sigma,A}$, is an initial $(\Sigma, A)$-algebra.

**Proof:** [E21] Given any $(\Sigma, A)$-algebra $M$, let $v : T_\Sigma \to M$ be the unique homomorphism, and note that $R \subseteq ker(v)$ because $M \models A$ implies $M \models (\forall \emptyset)\, t = t'$ for every $\langle t, t' \rangle \in R$, by the soundness of equational deduction. Now let $v = q; u$ with $u : T_{\Sigma,A} \to M$ be the factorization of $v$ given by Proposition 6.1.12; see Figure 6.2. This shows existence. For uniqueness, if also $u' : T_{\Sigma,A} \to M$, then $q; u' = v$ by the initiality of $T_\Sigma$, and $R \subseteq ker(v)$ because $M \models A$. Therefore $u = u'$ by (2) of Proposition 6.1.12. □

In fact, Example 4.3.8 shows that $R$ is a $\Sigma$-congruence, there denoted $\simeq_A$. Theorem 6.1.15 is very fundamental in algebraic specification; it is basic to the theory of abstract data types developed later in this chapter, as well as to the theory of rewriting modulo equations in Chapter 7, and several other topics. Moreover, it has a satisfying intuitive interpretation similar to that given for $T_\Sigma$ in Section 3.2: we can view $T_{\Sigma,A}$ as a kind of "universal language" of (simple expression-like) programs for the $\Sigma$-algebras that satisfy $A$, which are the "processors" that are able to correctly evaluate the programs in $T_{\Sigma,A}$. Initiality says that every such program has a unique result on each such processor. Let's consider an example, which will help to motivate our approach to substitutions modulo equations in Section 6.1.3:

**Example 6.1.16** Let $\Sigma = \Sigma^{\mathsf{GROUPL}}(\{a, b, c\})$ and let $A$ contain just the associative law. Then expressions like $a * b * c$ and $(a * a * b * b)^{-1}$ have unique interpretations in any group in which $a, b, c$ have been given interpretations. For example, if $M$ is the group of non-zero rational numbers under multiplication with $a = 3$, $b = 2$ and $c = 2$, then the first expression has value 12, while the second has value $\frac{1}{36}$. □

We now generalize freeness to the case where there are equations. Given an $S$-sorted set $X$, define $T_{\Sigma,A}(X)$ to be the algebra $T_{\Sigma \cup X}/\simeq_A$ viewed as a $\Sigma$-algebra, with the $A$-equivalence class of $t$ denoted $[t]_A$. This algebra is called the **free $(\Sigma, A)$-algebra generated by** $X$, and as was the case with Theorem 3.2.1, it has the following important universal property:

**Theorem 6.1.17** Given a set $A$ of $\Sigma$-equations, a $\Sigma$-algebra $M$ satisfying $A$, and an assignment $a : X \to M$, there is a unique $\Sigma$-homomorphism $\overline{a} : T_{\Sigma,A}(X) \to M$ that extends $a$, i.e., such that $\overline{a}(x) = a(x)$ for all $x \in X$.

**Proof:** A $(\Sigma \cup X, A)$-algebra $\overline{M}$ is exactly the same thing as a $(\Sigma, A)$-algebra $M$ and an assignment $a : X \to M$. For $T_{\Sigma,A}(X)$, the assignment is the injective morphism $i_x : X \to T_{\Sigma,A}(X)$ that sends $x$ to $[x]_A$. Theorem 3.2.1 implies that there is a unique $(\Sigma \cup X)$-homomorphism $\overline{a} : T_{\Sigma \cup X, A} \to \overline{M}$, which is exactly the same as saying that there is a unique $\Sigma$-homomorphism $\overline{a} : T_{\Sigma,A}(X) \to M$ that extends $a$. □

**Exercise 6.1.3** Show that any two $\Sigma$-algebras that satisfy the freeness universal property of Theorem 6.1.17 are $\Sigma$-isomorphic, and show that any $\Sigma$-algebra that is $\Sigma$-isomorphic to $T_{\Sigma,A}(X)$ also satisfies the same universal property. □

The following consequence of the above free algebra theorem and the homomorphism theorem will be used in Chapter 7:

**Proposition 6.1.18** Every $(\Sigma, A)$-algebra is a quotient of a free $(\Sigma, A)$-algebra.

**Proof:** Let $M$ be a $(\Sigma, A)$-algebra, and let $|M|$ denote its underlying ($S$-indexed) carrier set. We will show that $M$ is a quotient of the free algebra $P = T_{\Sigma,A}(|M|)$. Let $h$ be the unique $\Sigma$-homomorphism $\overline{a}$ from $P$ to $M$ given by Theorem 6.1.17 with $a$ the identity map on $|M|$. Notice that $h$ is surjective, because $a$ is already surjective. Then by Corollary 6.1.9, $P/\equiv_h$ is isomorphic to $M$. □

### 6.1.3 Substitution Modulo Equations

We conclude this section with substitution modulo $A$, generalizing Section 3.5; this also is needed in Chapter 7.

**Definition 6.1.19** Given a set $A$ of $\Sigma$-equations, a **substitution modulo** $A$ of $(\Sigma, A)$-terms with variables in $Y$ for variables in $X$ is an arrow $a : X \to T_{\Sigma,A}(Y)$; we may use the notation $a : X \to Y$. The **application** of $a$ to $t \in T_{\Sigma,A}(X)$ is $\overline{a}(t)$. Given substitutions $a : X \to T_{\Sigma,A}(Y)$ and $b : Y \to T_{\Sigma,A}(Z)$, then their **composition** (as substitutions), denoted $a;b$, is the $S$-sorted arrow $a; \overline{b} : X \to T_{\Sigma,A}(Z)$.  □

Again as in Section 3.5, an alternative notation makes this look more familiar: Given $t \in T_{\Sigma,A}(X)$ and $a : X \to T_{\Sigma,A}(Y)$ such that $|X| = \{x_1, \ldots, x_n\}$ and $a(x_i) = [t_i]_A$ for $i = 1, \ldots, n$, then $\overline{a}(t)$ can also be written $t(x_1 \leftarrow t_1, x_2 \leftarrow t_2, \ldots, x_n \leftarrow t_n)$, and whenever $t_i$ is just $x_i$, the pair $x_i \leftarrow t_i$ can be omitted from this notation.

**Exercise 6.1.4** The following assume a set $A$ of $\Sigma$-equations.

1. If $i_X : X \to T_{\Sigma,A}(X)$ is the inclusion, show that $\overline{i_X}([t]_A) = [t]_A$ for each $[t]_A \in T_{\Sigma,A}(X)$.

2. Given a substitution $a : X \to T_{\Sigma,A}(Y)$, show that $i_X; a = a$ and that $a; i_Y = a$; as before, this justifies writing $1_X$ for $i_X$.

3. Show that substitution modulo $A$ is associative, in the sense that given substitutions $a : W \to T_{\Sigma,A}(X)$, $b : X \to T_{\Sigma,A}(Y)$ and $c : Y \to T_{\Sigma,A}(Z)$, then $(a;b); c = a; (b;c)$.
   **Hint:** The "magical" proof for the ordinary case (Proposition 3.6.5) generalizes, using the free property of term algebras modulo $A$ (Theorem 6.1.17).

4. Show that Corollary 3.6.6, asserting $\overline{a}; \overline{b} = \overline{(a;b)}$, generalizes to substitution modulo $A$.  □

## 6.2  Abstract Data Types

This section motivates abstract data types from the viewpoint of software engineering, then gives a precise definition for this concept, and finally proves some of its most basic properties, especially that an abstract data type is uniquely determined by its specification as an initial algebra, and that abstract data types are indeed abstract. Section 6.6 discusses some limitations of this chapter's approach, placing it along a path having important further developments.

### 6.2.1  Motivation

It is well known that most of the effort in programming goes into debugging and maintenance (i.e., into improving and updating programs) [15]. Therefore anything that can be done to ease these processes has enormous economic potential. One step in this direction is to "encapsulate data representations"; this means to make the actual structure of data invisible, and to provide access to it *only* through a given set of operations which retrieve and modify the hidden data structure. Then the implementing code can be changed without having any effect on other code that uses it. On the other hand, if client code relies on properties of the representation, it may be extremely hard to track down all the consequences of modifying a given data structure (say, changing a doubly linked list to an array), because the client code may be scattered all over the program, without any clear identifying marks. The so-called Y2K problem is one relatively dramatic example of this phenomenon.

An encapsulated data structure with its accompanying operations is called an *abstract data type*. The crucial advance was to recognize that operations should be associated with data representations; this is exactly the same insight that advanced algebra from mere *sets* to *algebras*, which are sets *with* their associated operations. In software engineering this insight seems to have been due to David Parnas [146], and in algebra to Emmy Noether [20, 127]. The parallel between developments in software engineering and in abstract algebra is a major subtheme of this chapter.

A theory of abstract data types should enable us to check whether or not implementations are correct, by verifying their properties. This chapter presents some of the basics of such a theory. Abstract data types also provide the foundation for many theorem-proving problems: before we can prove something about the natural numbers, or about lists, we need a precise characterization of the structure that is involved. Even results about groups often use the natural numbers. More elaborate problems in computer science, such as proving the correctness of a compiler, usually involve more elaborate data structures, such as queues, stacks, arrays, or lists of stacks of integers. We usually want such proofs to be independent of how the underlying data types happen to be represented; for example, we are usually not interested in properties of the decimal or binary representations of natural numbers, but instead are interested in abstract properties of the natural numbers, like the commutativity of addition.

### 6.2.2   Formal Definition

We have already seen several examples where the $\Sigma$-term algebra $T_\Sigma$ serves as a standard model for a specification $P = (\Sigma, \emptyset)$ with no equations. For example, if $\Sigma = \Sigma^{\text{NATP}}$ (from Example 2.3.3) then $T_\Sigma$ is the natural numbers in Peano notation, and if $\Sigma = \Sigma^{\text{NATEXP}}$ (from Example 2.3.4) then $T_\Sigma$ consists of all expressions formed from the operation symbols 0, s, + and *.

There are also many examples that need equations, such as Example

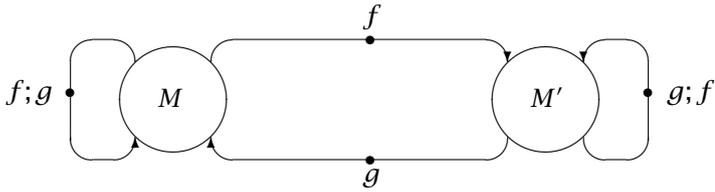

Figure 6.3: Uniqueness of Initial Algebras

5.1.5, the natural numbers with addition, for which we now repeat the OBJ code:

```
obj NATP+ is
  sort Nat .
  op 0 : -> Nat .
  op s_ : Nat -> Nat .
  op _+_ : Nat Nat -> Nat .
  var N M : Nat .
  eq N + 0 = N .
  eq N + s M = s(N + M) .
endo
```

Theorem 6.1.15 tells us that such specifications do indeed have initial models, in which the elements of the carriers are the equivalence classes of terms modulo the equations. However, Theorem 5.2.9 gives a different initial algebra for specifications that are also canonical as term rewriting systems, namely as the normal forms of terms under reduction. Moreover, a specification like NATP+ may well have still other representations that are preferred, such as natural numbers in the usual decimal positional notation. The choice of representation is just a matter of convenience, because all initial algebras are "essentially the same" in the sense that they are isomorphic algebras, as shown by the following:

**Proposition 6.2.1** Given a specification $P = (\Sigma, A)$, any two initial $P$-algebras are $\Sigma$-isomorphic; in fact, if $M$ and $M'$ are two initial $P$-algebras, then the unique $\Sigma$-homomorphisms $M \to M'$ and $M' \to M$ are both isomorphisms, and indeed, are inverse to each other.

**Proof:** The diagram in Figure 6.3 pertains to this proof. Because $M$ and $M'$ are both initial, there are $\Sigma$-homomorphisms $f : M \to M'$ and $g : M' \to M$. Thus there are $\Sigma$-homomorphisms $f;g : M \to M$ and $g;f : M' \to M'$. But because the identity on $M$ is a $\Sigma$-homomorphism and there is a unique $\Sigma$-homomorphism from $M$ to $M$ by the initiality of $M$, we necessarily have $f;g = 1_M$. Similarly, $g;f = 1_{M'}$. □

For example, if $P = (\Sigma, A)$ is NATP+, then the $\Sigma$-algebra $N_P$ of normal forms under $A$ (of Theorem 5.2.9) and the $\Sigma$-algebra $T/\simeq_A$ of equiva-lence classes of ground terms under $A$ are isomorphic, and in fact, both are isomorphic to $\omega$.

The following result shows that satisfaction of an equation by an algebra is an "abstract" property, in the sense that it is independent of how the algebra happens to be represented; more precisely, it is invariant under isomorphism. This is fortunate, because these are usually the properties in which we are most interested. This and Proposition 6.2.1 imply that exactly the same equations are true of any one initial $P$-algebra as any other.

**Proposition 6.2.2** Given isomorphic $\Sigma$-algebras $M$ and $M'$, and given a $\Sigma$-equation $e$, then

$$M \models e \text{ iff } M' \models e \,.$$

**Proof:** Let $h : M \to M'$ be an isomorphism, let $e$ be $(\forall X)\, t = t'$ and let $a : X \to M$ be an interpretation of $X$ in $M$. Then $\overline{a}(t) = \overline{a}(t')$ implies $h(\overline{a}(t)) = h(\overline{a}(t'))$. Moreover, any interpretation $b : X \to M'$ is of the form $a; h$ for some $a : X \to M$, namely $a = b; g$, where $g$ is the inverse of $h$. Hence $\overline{a}(t) = \overline{a}(t')$ for all $a : X \to M$ implies $\overline{b}(t) = \overline{b}(t')$ for all $b : X \to M'$. The converse implication follows by symmetry. □

The word "abstract" in the phrase "abstract algebra" means "uniquely defined up to isomorphism." In abstract group theory, we are not interested in properties of representations of groups, but only in those that hold up to isomorphism. Because Proposition 6.2.1 implies that all the initial models of a specification $P = (\Sigma, E)$ are abstractly the same in precisely this sense, the word "abstract" in "abstract data type" has *exactly* the same meaning. This is not a mere pun, but a significant fact about software engineering.

Another fact which strongly suggests that we are on the right track is that any computable abstract data type has an equational specification; moreover, this specification tends to be reasonably simple and intuitive in practice. The following result from [137] somewhat generalizes the original version due to Bergstra and Tucker [11] ($M$ is **reachable** iff the unique $\Sigma$-homomorphism $T_\Sigma \to M$ is surjective):

**Theorem 6.2.3** (*Adequacy of Initiality*) Given any computable reachable $\Sigma$-algebra $M$ with $\Sigma$ finite, there is a finite specification $P = (\Sigma', A')$ such that $\Sigma \subseteq \Sigma'$, such that $\Sigma'$ has the same sorts as $\Sigma$, and such that $M$ is $\Sigma$-isomorphic to $T_P$ viewed as a $\Sigma$-algebra. □

We do not here define the concept of a "computable algebra," but it corresponds to what one would intuitively expect: all carrier sets are decidable and all operations are total computable functions; see [137]. What this result tells us is that all of the data types that are of interest in computer science can be defined using initiality, although sometimes it may be necessary to add some auxiliary functions. All of this motivates the following:

**Definition 6.2.4** The **abstract data type** (abbreviated **ADT**) defined by a specification $P$ is the class of all initial $P$-algebras. □

## 6.3 Standard Models are Initial Models

We now address the basic question of what a standard model is by giving two intuitively motivated properties of a standard model, and then showing that any model satisfying these properties is in fact an initial model; because we already know that initial models are unique up to isomorphism, this settles the question.

Suppose we are given a theorem-proving problem that involves a signature $\Sigma$ and a set $A$ of equations that characterize the operations in $\Sigma$. Suppose further that $M$ is a standard model of $P = (\Sigma, A)$, and let $h$ denote the unique $\Sigma$-homomorphism $T_\Sigma \to M$. Then the two properties are as follows:

1. **No Junk**. For each $m \in M$, there is some $t \in T_\Sigma$ such that $m = h(t)$.

2. **No Confusion**. Given $t, t' \in T_\Sigma$, then $h(t) = h(t')$ iff $A \vdash (\forall \emptyset)\, t = t'$.

The intuitive justification for these principles is as follows: Because the elements of $M$ are supposed to represent the entities that exist in the "world" of the problem, it would be wrong to allow entirely new entities. Similarly, it is necessary that all entities are distinct unless it follows from the statement of the problem that they must be the same.

For example, consider the "Missionaries and Cannibals" problem, in which $n$ Missionaries and $n$ Cannibals are on one shore of a river, and all of them wish to get to the other shore, using a boat which can hold at most $k$ people. If ever there are more Missionaries than Cannibals, either on one shore or the other, or in the boat, then all the Cannibals present in that place are converted to Christianity. The problem is to get everyone to the other shore without any conversions.

Clearly, it would not be legitimate to postulate a bridge over which everyone could just walk, or a second larger boat into which everyone could fit; this would be "junk." Similarly, it would not be legitimate to postulate some number of extra Cannibals to stand guard. A different kind of illegitimate solution would simply assume that all the Missionaries are actually the same individual, with a number of different names; this would be a "confusion" of identities.

We can also give a "fair mystery story" interpretation of these two conditions: the first says that the butler didn't do it unless he was actually introduced into the story as a suspect ("no *deus ex machina*"), while the second says that all the characters are distinct unless the author has explicitly said otherwise ("no artificial aliases"). Thus, if the clues point to two different characters, the author would be cheating if he resolved the apparent conflict by saying that these two characters are really the same. Rather, he should give sufficient evidence to narrow the suspects down to just one.

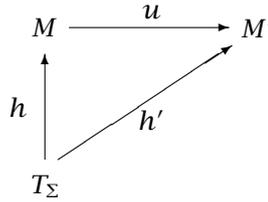

Figure 6.4: No Junk, No Confusion Proof

**Theorem 6.3.1** Given a specification $P = (\Sigma, A)$, then a $\Sigma$-algebra $M$ has no junk and no confusion relative to $P$ iff it is an initial $P$-algebra.

**Proof:** The diagram in Figure 6.4 pertains to this proof.

If $M$ is an initial $P$-algebra, then $M \cong T_\Sigma/\simeq_A$ and the no junk and no confusion conditions are obvious for this algebra.

For the converse, we first show that if a $\Sigma$-algebra $M$ has no junk and no confusion relative to $P = (\Sigma, A)$, then it satisfies $A$. Let $(\forall X)\, t = t'$ be in $A$, and let $\theta : X \to M$ be a substitution; then we must show that $\theta(t) = \theta(t')$ in $M$. Let $h : T_\Sigma \to M$ be the unique $\Sigma$-homomorphism, and let $|X| = \{x_1, \ldots, x_n\}$. By no junk, we may assume that $\theta(x_i) = h(t_i)$ for some $t_i \in T_\Sigma$. Because

$$A \vdash (\forall \emptyset)\, t(x_1 \leftarrow t_1, \ldots, x_n \leftarrow t_n) = t'(x_1 \leftarrow t_1, \ldots, x_n \leftarrow t_n)$$

follows by equational deduction, and because

$$\theta(t) = h(t(x_1 \leftarrow t_1, \ldots, x_n \leftarrow t_n))$$

and

$$\theta(t') = h(t'(x_1 \leftarrow t_1, \ldots, x_n \leftarrow t_n)) ,$$

the no confusion condition gives us that $\theta(t) = \theta(t')$, as desired.

To show that $M$ is initial, let $M'$ be another $P$-algebra, and let $h : T_\Sigma \to M$ and $h' : T_\Sigma \to M'$ be the unique homomorphisms. Now given $m \in M$, by no junk we may assume that $m = h(t)$ for $t \in T_\Sigma$ and then define $u : M \to M'$ by $u(m) = h'(t)$. To show that this is well-defined, let us also assume that $h(t) = h(t')$; then by no confusion, $A \vdash (\forall \emptyset)\, t = t'$, and so $M' \models (\forall \emptyset)\, t = t'$, which implies that $h'(t) = h'(t')$, as desired. Moreover, $h; u = h'$ by construction, and $u$ is unique because if it exists, it must satisfy the equation $h; u = h'$, which as we have just seen determines its value. □

## 6.4 Initial Truth and Subalgebras

This section defines initial satisfaction, and proves the fundamental result that an initial algebra has no proper subalgebras. The proof is, I think, suprisingly simple and beautiful, and many important results about induction will follow from it in subsequent sections of this chapter.

**Definition 6.4.1** Given a specification $P = (\Sigma, A)$ and a $\Sigma$-equation $e$, we say that $P$ **initially satisfies** $e$ iff $T_P \models e$; in this case we write $P \models_\cong e$ or $A \models_{\Sigma} e$, and we may omit the subscript $\Sigma$ when it is clear from context. □

Notice that this is a *semantic* property. Because anything that is true of all models is certainly true of initial models, we have that

$$P \models e \text{ implies } P \models_\cong e$$

(where $P \models e$ means that $A \models_\Sigma e$). However, the converse does *not* hold:

**Example 6.4.2** Let $\Sigma$ contain a constant $0$, a unary function symbol $s$, and a binary function symbol $+$, which we will write with infix notation; let $A$ contain the equations

$$(\forall n)\, 0 + n = n$$
$$(\forall m, n)\, s(m) + n = s(m + n) \,.$$

Then the commutative equation

$$(\forall m, n)\, m + n = n + m$$

holds in $T_P$ but does not hold in *every* $P$-algebra. For example, it does not hold in the $\Sigma$-algebra $M$ with carrier all strings of $a$'s and $b$'s, with $0 \in \Sigma$ denoting the empty string in $M$, with $m + n$ denoting the concatenation of the string $n$ after the string $m$, and with $s$ sending a string $m$ to the string $a + m$. For example, $a + b \neq b + a$, because $ab \neq ba$. □

Theorem 6.4.4 below is another fundamental property of initial algebras, with a very simple proof. We will soon see that this result provides the foundation for proofs by induction. But first, we need the following:

Recall that given a $\Sigma$-algebra $M$, another $\Sigma$-algebra $M'$ is a **subalgebra** (or **sub-$\Sigma$-algebra**) of $M$ iff there is an inclusion $M' \to M$ that is a $\Sigma$-homomorphism. In this case, we may write $M' \subseteq M$. A subalgebra $M'$ of $M$ is **proper** iff $M' \neq M$ (i.e., iff $M'_s \neq M_s$ for some $s \in S$).

**Example 6.4.3** If $\omega$ denotes the natural numbers, and $\mathbb{Z}$ denotes the integers (positive, negative and zero), and if $\Sigma = \Sigma^{\mathsf{NATP}}$, then $\omega$ is a sub-$\Sigma$-algebra of $\mathbb{Z}$. □

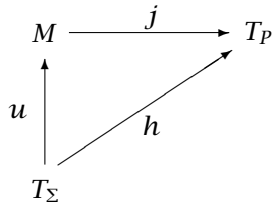

Figure 6.5: No Proper Subalgebra Proof

**Exercise 6.4.1** [E22] Given a $\Sigma$-algebra $M$ and given a subset $M'_s \subseteq M_s$ for each $s \in S$, show that $M'$ gives the carriers of a subalgebra of $M$ if and only if $M_\sigma(m_1, \ldots, m_k) \in M'_s$ for each $\sigma \in \Sigma$, where $m_i \in M'_{s_i}$ for $i = 1, \ldots, k$ and where $\sigma$ has arity $s_1 \ldots s_k$ and sort $s$. □

**Exercise 6.4.2** Show that if $M$ is a $(\Sigma, A)$-algebra and if $M' \subseteq M$ is a sub-$\Sigma$-algebra, then $M'$ is also a $(\Sigma, A)$-algebra. □

**Theorem 6.4.4** If $P = (\Sigma, A)$ is a specification, then an initial $P$-algebra has no proper sub-$\Sigma$-algebras.

**Proof:** The diagram in Figure 6.5 pertains to this proof. Let $h : T_\Sigma \to T_P$ be the unique $\Sigma$-homomorphism, which is surjective by hypothesis, let $j : M \to T_P$ be the inclusion for a sub-$\Sigma$-algebra $M$, and let $u : T_\Sigma \to M$ be the unique $\Sigma$-homomorphism. Then by initiality of $T_\Sigma$ and because $j$ is a $\Sigma$-homomorphism, $u; j = h$. Hence $j$ is also surjective, and so $T_P$ has no proper sub-$\Sigma$-algebra. Therefore, no other initial $P$-algebra can have a proper sub-$\Sigma$-algebra, because they are all isomorphic. □

## 6.5 Induction

In general, pure equational deduction is inadequate for proving properties of standard models, and many properties require the use of induction. Fortunately, there exist powerful induction principles for initial models, that let us prove that some predicate holds for all values by proving that it holds for each constructor whenever it holds for all arguments of that constructor. These generalize Peano induction from the natural numbers to arbitrary data types, and can be considered forms of "structural induction" [21], as discussed in Section 3.2.1.

The results in this section follow [137], and justify using induction in proof scores for unsorted signatures; Section 6.5.2 extends this to the many-sorted case; much more general results on induction, including its use on first-order formulae, are given in Chapter 8. The following basic definition applies to both the unsorted and many-sorted cases:

**Definition 6.5.1** Given a signature $\Sigma$ and a $\Sigma$-algebra $M$, then a subsignature $\Phi \subseteq \Sigma$ is a **signature of constructors** for $M$ iff the unique $\Phi$-homomorphism $h : T_\Phi \to M$ is surjective, and is a **signature of unique constructors** iff $h$ is a $\Phi$-isomorphism. A signature $\Phi$ of constructors is **minimal** iff no proper subsignature $\Phi' \subset \Phi$ has $T_{\Phi'} \to M$ surjective. A **signature of constructors** for a specification $P = (\Sigma, A)$ is a signature of constructors for $T_P$. □

If a $\Sigma$-algebra $M$ has a signature of constructors, then it has a minimal signature of constructors; however, it may have more than one. Every specification has a signature of constructors and at least one minimal signature of constructors. Clearly, using a minimal signature of constructors will require less effort in proofs. Although specifications need not in general have signatures of unique constructors, the specifications that arise in practice often have a unique minimal signature of unique constructors. The properties of reachability and induction correspond to the "no junk" and "no confusion" conditions that together are equivalent to initiality [24, 137].

**Example 6.5.2** If $\Sigma = \Sigma^{\text{NATP+}}$, then $\Phi = \Sigma^{\text{NATP}}$ is a minimal signature of constructors and also a signature of unique constructors for $T_{\text{NATP+}}$. □

You may wish to review Section 2.7 on unsorted algebra before reading the next result.

**Theorem 6.5.3** (*Structural Induction I*) Given an unsorted specification $P = (\Sigma, A)$ and a signature $\Phi$ of constructors for $P$, let $V$ be a subset of $T_P$. Then $V = T_P$ if

(0) $c \in \Phi_0$ implies $[c] \in V$, and

(1) $f \in \Phi_n$ for $n > 0$ and $[t_i] \in V$ for $i = 1, \ldots, n$ imply $[f(t_1, \ldots, t_n)] \in V$.

**Proof:** Because $T_P$ has no proper subalgebras by Proposition 6.4.4 and because $V \subseteq T_P$, we need only show that $V$ is closed under $\Phi$; but that is exactly what conditions (0) and (1) say. □

This very simple proof is possible because we have taken initial algebra semantics as our starting point. Note that a complete inductive proof using a signature $\Phi$ of constructors must include a proof that $\Phi$ in fact is a signature of constructors, i.e., that there is a surjective $\Phi$-homomorphism $h : T_\Phi \to T_P$. Often this surjective property will be proved using structural induction. But if $A$ is ground canonical and all its normal forms are $\Phi$-terms, then $h$ is just the isomorphism of $N_P$ with $T_P$ that is discussed in Section 5.2.9.

The usual formulation of induction follows easily from Theorem 6.5.3:

**Corollary 6.5.4** (*Structural Induction II*) Given an unsorted specification $P = (\Sigma, A)$ and a signature $\Phi$ of constructors for $P$, let $Q(x)$ be a $\Sigma(\{x\})$-sentence. Then $A \models_\Sigma (\forall x)\, Q(x)$ if

(0) $c \in \Phi_0$ implies $A \models_\Sigma Q(c)$, and

(1) $f \in \Phi_n$ for $n > 0$ and $t_i \in T_\Sigma$ for $i = 1, \ldots, n$ and $A \models_\Sigma Q(t_i)$ for $i = 1, \ldots, n$ imply $A \models_\Sigma Q(f(t_1, \ldots, t_n))$.

**Proof:** This follows from Theorem 6.5.3 by letting $V = \{h(x) \mid A \models_\Sigma Q(x)\}$, where $h: T_\Phi \to T_P$ is the surjective $\Phi$-homomorphism that exhibits $\Phi$ as a signature of constructors. □

Chapter 8 is much more precise about the notion of "sentence," but for now it suffices to think of sentences as including equations and their combinations under conjunction and implication. Corollary 6.5.4 justifies the use of simple induction in proof scores, and is used in examples below.

**Example 6.5.5** When $\Sigma = \Sigma^{\mathsf{NATP}}$, the above result states exactly the usual principle of induction for the natural numbers. □

Induction is basic to many theorem-proving systems, including the Boyer-Moore theorem prover [19], although not in the same form as above. Experience [59] shows that it is often easier to prove results by structural induction, as justified by the above results, than by so-called inductionless induction [56, 137, 140] using Knuth-Bendix completion, because structural induction arguments do not require showing the termination of new rule sets, and do not produce uncontrollable explosions of strange new rules that may gradually become less and less relevant; Garland and Guttag [49] report a similar experience.

Note that inductive proof techniques are *not valid* for loose semantics, because (in general) the results proved by induction are not true of all models, but only of the standard (initial) models. Also, it is usually much easier to directly exploit the close connection between rewrite rules, initiality, and induction than to try to remain within a "loose" semantics framework by axiomatizing a standard model with explicit reachability and induction schemata, because the first of these requires existential quantification (e.g., Skolem functions) and the second requires second-order quantification.

### 6.5.1 Simple Inductive Proofs with OBJ

The following proof scores use the inductive proof techniques introduced above. The first two examples import the following code for the natural numbers with addition:

```
obj NAT is sort Nat .
  op 0 : -> Nat .
  op s_ : Nat -> Nat [prec 1] .
  op _+_ : Nat Nat -> Nat [prec 3] .
  vars M N : Nat .
  eq M + 0 = M .
  eq M + s N = s(M + N) .
endo
```

**Example 6.5.6** (*Associativity of Addition*) The following score proves that addition of natural numbers is associative:

```
open NAT .
ops l m n : -> Nat .
***> base case, n=0: l+(m+0)=(l+m)+0
reduce l + (m + 0) == (l + m) + 0 .
***> induction step
eq l + (m + n) = (l + m) + n .
reduce l + (m + s n) == (l + m) + s n .
close
```

Therefore we have proved the equation $(\forall L, M, N)\ L + (M + N) = (L + M) + N$. □

**Example 6.5.7** (*Commutativity of Addition*) [E23] The following proof score shows that addition of natural numbers is commutative:

```
open NAT .
ops m n : -> Nat .

***> first lemma0: 0 + n = n, by induction on n
***> base for lemma0, n=0
reduce 0 + 0 == 0 .
***> induction step
eq 0 + n = n .
reduce 0 + s n == s n .
*** thus we can assert
eq 0 + N = N .

***> show lemma1: s m + n = s(m + n), again by induction on n
***> base for lemma1, n=0
reduce s m + 0 == s(m + 0) .
***> induction step
eq s m + n = s(m + n) .
reduce s m + s n == s(m + s n) .
*** thus we can assert
eq s M + N = s(M + N).

***> show m + n = n + m, again by induction on n
***> base case, n=0
```

```
        reduce m + 0 == 0 + m .
        ***> induction step
        eq m + n = n + m .
        reduce m + s n == s n + m .
        close
```

Of course, we should not assert commutativity as a rewrite rule, or we may get non-terminating behavior.  □

We will see in Chapter 7 that the above results imply that we can use associative-commutative rewriting for addition in doing more complex examples.

It is interesting to contrast the above proofs with corresponding proofs due to Paulson in Cambridge LCF [147]. The LCF proofs are much more complex, in part because LCF allows partial functions, and then must prove them total, whereas functions are automatically total (on their domain) in equational logic.

**Exercise 6.5.1** [E24] Use OBJ3 to prove that the equation $(\forall N)\ 0 + M = M$ holds for the specification NAT above.  □

**Exercise 6.5.2** Given the following code:

```
obj INT is sort Int .
  ops (inc_)(dec_): Int -> Int .
  op 0 : -> Int .
  vars X Y : Int .
  eq inc dec X = X .
  eq dec inc X = X .
  op _+_ : Int Int -> Int .
  eq 0 + Y = Y .
  eq (inc X) + Y = inc(X + Y).
  eq (dec X) + Y = dec(X + Y).
endo
```

  (a) What set of algebras does it denote? What are its signatures of contructors (if any)? What are its minimal signatures of contructors (if any)? What are its signatures of unique constructors (if any)?

  (b) Give an OBJ proof score for the equation $(\forall Y)\ Y + 0 = Y$, and justify it.  □

**Exercise 6.5.3** This question refers to the same code as Exercise 6.5.2.

  (a) Explain how to represent $(-3) + 4$, and explain how OBJ would reduce it.

  (b) Give an OBJ proof score for $(\forall X, Y)\ X + (\text{dec } Y) = \text{dec } (X + Y)$, and justify it.

  (c) Give an OBJ proof score for $(\forall X, Y)\ X + Y = Y + X$, and justify it.

  □

### 6.5.2 Many-Sorted Induction

This section extends the results of Section 6.5 to the many-sorted case; hence, $\Sigma$ is an $S$-sorted signature throughout. Again, much more general results may be found in Chapter 8.

**Definition 6.5.8** Given a $\Sigma$-algebra $M$ and $s \in S$, then a subsignature $\Delta \subseteq \Sigma$ is **inductive for** $s$ **over** $M$ iff

(0) each $\delta \in \Delta$ has sort $s$, and

(1) $\Sigma' = \{\sigma \in \Sigma \mid sort(\sigma) \neq s\} \cup \Delta$ is a signature of constructors for $M$.

A signature $\Delta$ that is inductive for $s$ over $M$ is **minimal** iff it is an inductive signature for $s$ over $M$ such that no proper subsignature is inductive for $s$ over $M$. Given a specification $P = (\Sigma, A)$, a signature $\Delta$ is **inductive for** $s$ **over** $P$ iff it is inductive for $s$ over $T_P$. □

Of course, we want to do as little work as possible in an inductive proof. A minimal inductive signature allows this.

**Exercise 6.5.4** Show that a signature $\Delta$ is a minimal inductive signature for $s$ over $M$ iff $\Delta \subseteq \Phi$ where $\Phi$ is a minimal signature of constructors for $M$. □

**Theorem 6.5.9** (*Structural Induction I'*) Given a specification $P = (\Sigma, A)$ and a signature $\Delta$ that is inductive for sort $s$ over $P$, let $V \subseteq T_{P,s}$. Then $V = T_{P,s}$ if

(0) $c \in \Delta_{[],s}$ implies $[c] \in V$, and

(1) $f \in \Delta_{s_1 \ldots s_n, s}$ for $n > 0$ and $t_i \in T_{\Sigma, s_i}$ for $i = 1, \ldots, n$ with $[t_i] \in V$ if $s_i = s$ imply $[f(t_1, \ldots, t_n)] \in V$.

**Proof:** We first define an $S$-sorted set $M$ by $M_s = V$ and $M_{s'} = T_{P,s'}$ for $s' \neq s$. Then (0) and (1) tell us that $M$ is a $\Sigma'$-algebra, where $\Sigma'$ is as in Definition 6.5.8. Now because $T_P$ has no proper sub-$\Sigma'$-algebras by Proposition 6.4.4, we conclude that $V = T_{P,s}$. □

As before, the more familiar formulation of induction follows immediately:

**Corollary 6.5.10** (*Structural Induction II'*) Given a specification $P = (\Sigma, A)$ and a signature $\Delta$ that is inductive for sort $s$ over $P$, let $Q(x)$ be a $\Sigma(\{x\})$-sentence where $x$ is a variable of sort $s$. Then $A \models_\Sigma (\forall x) Q(x)$ if

(0) $c \in \Delta_{[],s}$ implies $A \models_\Sigma Q(c)$, and

(1) $f \in \Delta_{s_1 \ldots s_n, s}$ for $n > 0$ and $t_i \in T_{\Sigma, s_i}$ for $i = 1, \ldots, n$ and $A \models_\Sigma Q(t_i)$ when $s_i = s$ imply $A \models_\Sigma Q(f(t_1, \ldots, t_n))$.

**Proof:** This follows directly from Theorem 6.5.9 by letting $V = \{x \in T_{P,s} \mid A \models_\Sigma Q(x)\}$.  □

Actually, "if" can be replaced by "iff" in both Theorem 6.5.9 and Corollary 6.5.10; however, these converse "completeness" results do not seem to be useful in practice. Also, note that we can generalize all this a bit further by considering an $S$-sorted set $V = \{V_s \mid s \in S\}$ defined by $Q = \{Q_s(x) \mid s \in S\}$.

**Exercise 6.5.5** [E25] Consider the following specification:

```
obj SET is sort Set .
  pr NAT .
  op {} : -> Set .
  op ins : Nat Set -> Set .
  op _U_ : Set Set -> Set [id: ({})] .
  vars N N' : Nat .
  vars S S' : Set .
  eq ins(N,ins(N',S)) = ins(N',ins(N,S)).
  eq ins(N,ins(N,S)) = ins(N,S).
  eq ins(N,S) U S' = ins(N,S U S').
endo
```

where NAT is the Peano natural numbers. Then do the following:

(a) Write a specification for SET that involves neither module importation (the line "`pr NAT`") nor attributes ("`[id:{}]`"). What would result from executing the following in OBJ3?

```
open SET .
op s0 : -> Set .
red ins(0,{}) .
red {} U {} .
red ins(0,ins(0,{})) .
red ins(0,ins(0,s0)) .
```

(b) Give an inductive signature for the sort Set over SET which is minimal, and one which is not. Explain why.

(c) Give a manual proof that the equation $(\forall S, S')\ S \cup S' = S' \cup S$ holds for SET.

(d) Give an OBJ proof score for this equation.  □

We can formulate induction as a general rule of inference. Let the notation $A \models_\Sigma e$ indicate that $e$ can be proved from $A$ using the new rule given below plus the usual rules for $\vdash_\Sigma$, and assume that $\Delta$ is inductive for sort $s$ over $P = (\Sigma, A)$. Then the new rule is:

(I) Given $t, t' \in T_\Sigma(\{x\})$ with $x$ of sort $s$, if

$$A \models_\Sigma (\forall \emptyset)\, t(x \leftarrow c) = t'(x \leftarrow c)$$

for each $c \in \Delta_{[],s}$, and if

$$A \models_\Sigma (\forall \emptyset)\, t(x \leftarrow t_i) = t'(x \leftarrow t_i)$$

for $i = 1, \ldots, n$ and $f \in \Delta_{s_1 \ldots s_n, s}$ imply

$$A \models_\Sigma (\forall \emptyset)\, t(x \leftarrow f(t_1, \ldots, t_n)) = t'(x \leftarrow f(t_1, \ldots, t_n)),$$

then $A \models_\Sigma (\forall x)\, t = t'$.

Corollary 6.5.10 and the soundness of $\vdash$ show that $\models_\Sigma$ is sound for initial truth, i.e., that

$$A \models_\Sigma e \text{ implies } A \models\!\!\!\models_\Sigma e\,.$$

However, $\models_\Sigma$ cannot be complete, i.e., in general the converse implication does not hold [129]. In fact, the set of equations satisfied by $T_P$ is not in general even recursively enumerable, for reasons discussed in Section 5.10.

## 6.6   A Closer Look at State, Encapsulation and Implementation

Initial semantics works very well for static data structures like integers, lists, booleans, and even vectors and matrices, that are typically passed as values in programs unless they are very large. But such an approach is more awkward for dynamic data structures that have dedicated storage and commands that change the internal representation, which is not viewed directly, but only through "attributes." For example, it is usually more appropriate to view stacks as "state machines" with an encapsulated internal state, and with "top" as an attribute. Although initial models exist for any reasonable specification of stacks, real stacks are more likely to be implemented by a model that is not initial, such as a pointer plus an array. This means that a notion of implementation is needed that differs from initial models. Moreover, in considering (for example) stacks of integers, the sorts for stacks and for integers have a different character, since the latter can still be modelled initially as data. Although an initial framework has been successfully used for some applications of this kind (e.g., see [86]), it is really better to take a viewpoint that explicitly distinguishes between "visible" sorts for data and "hidden" sorts for states, and that defines an implementation to be any model satisfying certain natural constraints; the hidden algebra developed in Chapter 13 takes such an approach.

## 6.7 Literature

The notions of quotient and image algebra, like the notions of homomorphism and isomorphism, developed as abstractions of the corresponding basic notions for groups and rings, mainly from lectures by Emmy Nöther in the late 1920s; the same holds for the Homomorphism Theorem (Theorem 6.1.7). The development of minimal automata in Example 6.1.10 using the homomorphism theorem may be original, but algebraic treatments of automata go back to Anil Nerode and others in the earliest days of theoretical computer science [142, 154]. The universal properties of quotient and free algebras (Proposition 6.1.12 and Theorem 6.1.17) are notable for the smooth way that consequences can be drawn from them, as illustrated by the proof of Theorem 6.1.15 and the results at the end of Section 6.1; several proofs in the chapter also make good use of diagrams. The concept of universal property comes from category theory, where it takes the elegant form of an adjoint pair of functors (see [128] or the end of [126]); it was also developed by Bourbaki in a more concrete form closer to the one in this text, which is also similar to that in the excellent abstract algebra text of Mac Lane and Birkhoff [128].

The importance of initiality for computing has developed gradually. The term "initial algebra semantics" and its first applications, to Knuthian attribute semantics, appear in [52], while the first applications to abstract data types are in [87]; a more complete and rigorous exposition is given in [86]. The terminology "no junk" and "no confusion" and Theorem 6.3.1 are from [24]. Many examples of initiality can be found in [89] and [137]; the latter especially develops connections with induction and computability. Results on the computational adequacy of initiality first appeared in the work of Bergstra and Tucker [11]. The principle that *standard models are initial models* extends beyond equational logic, for example, to standard models of Horn clause logic as used in (pure) Prolog; [79] discusses this and some other applications of the principle. A significant generalization of algebraic induction is given in Section 8.7.

# 7 Deduction and Rewriting Modulo Equations

This chapter generalizes term rewriting to the situation where some equations $B$ are "built in" as part of matching, rather than used as rewrite rules; the phrase "*modulo B*" refers to this. We also introduce terms, equations, and deduction modulo $B$, give a decision procedure for the propositional calculus (using rewriting modulo associativity and commutativity), generalize theory from Chapter 5, including ways to prove termination and Church-Rosser, and apply all this to several kinds of digital hardware, among other things.

## 7.1 Motivation

The reader may well have felt that the repeated use of the associative law in Examples 4.1.4 and 4.5.5 (as well as Exercises 4.6.1–4.6.4) was rather tedious, and perhaps even unnecessary. A more precise way to express this feeling is to say that because the associative law tells us that "parentheses are unnecessary," it should be unnecessary to move parentheses around in deductions that assume this law. For example, the two expressions

$$(a^{-1\,-1} * a^{-1}) * (a * a^{-1})$$

$$(a^{-1\,-1} * (a^{-1} * a)) * a^{-1}$$

are "obviously" equal to each other, because they have the same unparenthesised form, namely

$$a^{-1\,-1} * a^{-1} * a * a^{-1} \ .$$

**Example 7.1.1** Eliminating parentheses in the proof of Example 4.5.5 gives the following,

| | | |
|---|---|---|
| [1] | $a * a^{-1} = e * a * a^{-1}$ | by (−6) on GL.1 |
| [2] | $= a^{-1\,-1} * a^{-1} * a * a^{-1}$ | by (−6) on GL.2 with $A = a^{-1}$ |
| [3] | $= a^{-1\,-1} * e * a^{-1}$ | by (6) on GL.2 |
| [4] | $= a^{-1\,-1} * a^{-1}$ | by (6) on GL.1 |
| [5] | $= e$ | by (6) on GL.2 |

which is a very significant simplification. □

Because we want to be precise throughout this book, we must ask what it *means* to "build in" associativity this way. One approach is to consider expressions that differ only in their parenthesization to be *equivalent*; the equivalence classes should then form an algebra such that we can do deduction on the classes in essentially the same way that we do deduction on terms. For example, the following is an equivalence class of terms modulo associativity:

$$\{ (a * b) * (c * d),\ a * (b * (c * d)),\ a * ((b * c) * d),$$
$$((a * b) * c) * d,\ (a * (b * c)) * d \}.$$

We also want the unparenthesized expression $a * b * c * d$ to serve as surface syntax of this class, representing it to users in the familiar way.

Because there are applications other than associativity, it is worthwhile to develop the theory at a general level. For example, if we want to study commutative groups, which in addition to the usual group laws, also satisfy the commutative law,

$$(\forall X, Y)\ X * Y = Y * X,$$

then we need to avoid using this equation as a rewrite rule, because it can lead to non-terminating computations, such as

$$a * b \overset{1}{\Rightarrow} b * a \overset{1}{\Rightarrow} a * b \overset{1}{\Rightarrow} \cdots.$$

This observation means it is impossible to give a canonical term rewriting system for the theory of commutative groups. However, by regarding terms that differ only in the order of their factors as equivalent, we can build in commutativity and thus avoid non-termination. For example, $(a * b) * c$ has the following class of equivalent terms

$$\{ (a * b) * c,\ (b * a) * c,\ c * (a * b),\ c * (b * a) \},$$

and moreover, the class $\{a * b,\ b * a\}$ is a normal form under rewriting modulo commutativity.

There are also many examples, including commutative groups, where it is useful to identify terms that are the same up to the order of their factors *and* their parenthesization. Thus, there are at least three interesting cases: associativity, commutativity, and both together, which are often abbreviated A, C, and AC, respectively.

**Exercise 7.1.1** Prove that there are 12 terms in the equivalence class modulo AC of the term $(a*b)*c$, and write them all out. Also prove there are 5 terms in the equivalence class of $(a*b)*(c*d)$ modulo associativity.

□

## 7.2 Deduction Modulo Equations

The "*Semantics First*" slogan of Chapter 1 implies that a discussion of deduction should be preceded by a discussion of satisfaction, as a standard against which to test soundness and completeness. We now do this for deduction with a set $A$ of equational axioms, modulo a set $B$ of equations. The first step is to define the kind of equation involved; we begin with $B$-equivalence classes of $\Sigma$-terms, i.e., with elements of $T_{\Sigma,B}(X)$, as defined in Section 6.1.1.

**Definition 7.2.1** Given a set $B$ of (possibly conditional) $\Sigma$-equations, a (**conditional**) **$\Sigma$-equation modulo $B$**, or **$(\Sigma, B)$-equation**, is a 4-tuple $\langle X, t, t', C \rangle$ with $t, t' \in T_{\Sigma,B}(X)$ and $C$ a finite set of pairs from $T_{\Sigma,B}(X)$, usually written $(\forall X)\ t =_B t'$ `if` $C$; we may use the same notation with $t, t', C$ all $\Sigma$-terms to represent their $B$-equivalence classes, and we may drop the $B$ subscripts. We write just $(\forall X)\ t =_B t'$ when $C = \emptyset$ and call it an **unconditional equation**.

Given a $(\Sigma, B)$-algebra $M$, **$\Sigma$-satisfaction modulo $B$**, written $M \models_{\Sigma,B} (\forall X)\ t =_B t'$ `if` $C$, is defined by $\bar{a}(t) = \bar{a}(t')$ whenever $\bar{a}(u) = \bar{a}(v)$ for each $\langle u, v \rangle \in C$, for all $a : X \to M$. Theorem 6.1.17 provides the unique $\Sigma$-homomorphism $\bar{a} : T_{\Sigma,B}(X) \to M$ extending $a$. Given a set $A$ of $(\Sigma, B)$-equations, let $A \models_{\Sigma,B} e$ mean that $M \models_{\Sigma,B} A$ implies $M \models_{\Sigma,B} e$ for all $B$-models $M$. We may drop the subscripts $\Sigma$ and/or $B$ if they are clear from context.

□

We can get class deduction versions of the rules for equational inference in Chapter 4, just by replacing each occurrence of $T_\Sigma$ by $T_{\Sigma,B}$ and each occurrence of $=$ by $=_B$, assuming that all axioms in $A$ are $(\Sigma, B)$-equations. We denote the $B$-class deduction version of rule $(i)$ by $(i_B)$, and we let $A \vdash_{\Sigma,B} e$ indicate **deduction modulo $B$**, also called **class deduction**, which is deduction using the rules $(1_B), (2_B), (3_B), (4_B)$ and $(5C_B)$ to deduce $e$ from $A$ modulo $B$; as above we may drop either or both subscripts $\Sigma$ and $B$ if they are clear from context. For example, here is the class deduction version of rule $(6C)$:

($6C_B$) *Forward Conditional Subterm Replacement.* Given $t_0 \in T_{\Sigma,B}(X \cup \{z\}_s)$ with $z \notin X$, if
$(\forall Y)\ t_1 =_B t_2$ `if` $C$
is of sort $s$ and is in $A$, and if $\theta : Y \to T_{\Sigma,B}(X)$ is a substitution such that $(\forall X)\ \theta(u) =_B \theta(v)$ is deducible for each pair $\langle u, v \rangle \in C$, then

$$(\forall X)\ t_0(z \leftarrow \theta(t_1)) =_B t_0(z \leftarrow \theta(t_2))$$

is also deducible.

Note that this uses substitution modulo $B$, Definition 6.1.19 of Chapter 6.

The next result lets us carry over soundness and completeness results from Chapter 4 to class deduction. It says that deduction from $A$ modulo $B$ is equivalent to deduction from $A \cup B$ on representatives, and that satisfaction of an equation modulo $B$ is equivalent to ordinary satisfaction by a representative of the equation. To express this more precisely, given a $\Sigma$-equation $e$ of the form $(\forall X)\ t = t'$, let $[e]$ denote its modulo $B$ version, $(\forall X)\ [t] =_B [t']$, and similarly for conditional equations. Given a set $A$ of $\Sigma$-equations, let $[A]$ denote the set of modulo $B$ versions of equations in $A$; for more clarity, we may also write $[t]_B$, $[e]_B$ and $[A]_B$.

**Proposition 7.2.2** (*Bridge*) Given sets $A, B$ of $\Sigma$-equations and another $\Sigma$-equation $e$ (with $A, B$ and $e$ possibly conditional), then

$$[A] \vdash_B [e] \quad \text{iff} \quad A \cup B \vdash e\ .$$

Furthermore, given any $(\Sigma, B)$-algebra $M$ and a (possibly conditional) $\Sigma$-equation $e$, then

$$M \vDash_{\Sigma, B} [e] \quad \text{iff} \quad M \vDash_\Sigma e\ .$$

**Proof:** For the first assertion, if $e_1, \ldots, e_n$ is a proof of $e$ from $A \cup B$, then the subsequence $[e_{i_1}], \ldots, [e_{i_k}]$ formed by omitting all steps that used $B$, and then taking the $B$-classes of those equations that remain, is a proof of $[e]$ from $[A]$, and conversely, any proof $[e_1], \ldots, [e_k]$ of $[e]$ from $[A]$ can be filled out to become a proof of $e$ from $A \cup B$, by choosing representatives for each $[e_i]$ and adding intermediate steps that use $B$.

For the second assertion, let $q : T_\Sigma(X) \to T_{\Sigma,B}(X)$ be the quotient map, let $e$ be the equation $(\forall X)\ t = t'$ if $C$, and (just for now) let $\tilde{a}$ denote the extension of $a : X \to M$ to a $\Sigma$-homomorphism $T_{\Sigma,B}(X) \to M$, with $\overline{a}$ the usual $\Sigma$-homomorphism $T_\Sigma(X) \to M$. Then

$$(*) \quad q; \tilde{a} = \overline{a}\ ,$$

by the universal property of Theorem 6.1.17, because both sides are $\Sigma$-homomorphisms $T_{\Sigma,B}(X) \to M$ that agree on $X$, since $q(x) = [x]$ for $x \in X$ implies $\tilde{a}(q(x)) = a(x)$ for $x \in X$. Then $M \vDash e$ iff for all $a : X \to M$, $\overline{a}(t) = \overline{a}(t')$ whenever $\overline{a}(u) = \overline{a}(v)$ for all $\langle u, v \rangle \in C$, and composing with $q$ gives us that for all $a : X \to M$, $\tilde{a}([t]) =_B \tilde{a}([t'])$ whenever $\tilde{a}([u]) =_B \tilde{a}([v])$ for all $\langle u, v \rangle \in C$, because of $(*)$. But this says that $M \vDash_B [e]$. □

**Theorem 7.2.3** (*Completeness*) Given sets $A, B$ of $\Sigma$-equations and another $\Sigma$-equation $e$ (all possibly conditional), then the following are equivalent:

(1) $[A] \vdash_B [e]$     (2) $[A] \vDash_B [e]$     (3) $A \cup B \vdash e$     (4) $A \cup B \vDash e$

**Proof:** The first part of Proposition 7.2.2 gives the equivalence of (1) with (3), the Completeness Theorem (4.8.4) gives the equivalence of (3) with (4), and the second part of Proposition 7.2.2 gives the equivalence of (2) with (4). □

We also have the following completeness result, which (with the Theorem of Constants) justifies the calculation in Example 7.1.1, since each step there is an instance of rule ($\pm 6C_B$):

**Theorem 7.2.4** Given sets $A, B$ of (possibly conditional) $\Sigma$-equations and an unconditional $\Sigma$-equation $e$, then

$$[A] \vdash_B [e] \quad \text{iff} \quad [A] \vdash_B^{(1_B, 3_B, \pm 6C_B)} [e] \ .$$

Moreover, $[A] \vdash_B^{(1_B, 3_B, \pm 6C_B)} [e]$ iff $M \vDash_B [e]$ for all $(\Sigma, A \cup B)$-algebras $M$.

**Proof:** The two predicates in the first assertion are equivalent to $(A \cup B) \vdash e$ and $(A \cup B) \vdash^{(1,3,\pm 6C)} e$, respectively, the latter by reasoning analoguous to that in Proposition 7.2.2; hence they are equivalent by Theorem 4.9.1. The second assertion now follows by Theorem 7.2.3. □

### 7.2.1 Some Implementation Issues

When $B$ consists of just the associative law, every equivalence class in any $T_\Sigma(X)/\simeq_B$ is finite; however, there is no upper bound for the number of terms that may be in these classes. Moreover, there are specifications where the equivalence classes are actually infinite. For example, if $B$ contains an identity law ($x * 1 = x$), then equivalence classes modulo $B$ are infinite; the same holds for an idempotent law ($x * x = x$). Consequently, equivalence classes are not feasible representations for systems like OBJ, either for surface syntax seen by users, or for concrete data structures used internally for calculation. However, they are fundamental for semantics.

When $B$ consists of the associative law, $B$-equivalence classes have a simple natural representation based on omitting parentheses; for example, the equivalence class of $(a * b) * c$ can be represented to users as $a * b * c$ and internally as $*(a, b, c)$, which avoids the ambiguity of the simpler list representation $(a, b, c)$ if more than one binary operation is declared associative. Similarly, bags and sets of terms can represent terms modulo AC, and AC plus idempotency, respectively, and these can be implemented using standard concrete data structures.

However, there is no optimal linear representation for built in commutativity, since any linear representation must choose some ordering for subterms. Theorem 7.2.3 shows that equivalence classes can be represented by terms, and in fact this is what OBJ3 does, for both calculation and surface syntax, except for associativity, where omitting parentheses is the default; however, this default can be turned off with the command "`set print with parens off`".

Another difficulty involved with implementing deduction modulo $B$ with equivalence classes is that occurrences of a variable in $[t]_B$ may not be well defined, since different representatives of the class may have different numbers of occurrences. For example, if $B$ contains an idempotent law for a binary operation $*$, then the terms

$$x, \ x * x, \ (x * x) * x, \ x * (x * x), \ \ldots$$

are all $B$-equivalent, so that the class $[x]_B$ contains terms with $n$ occurrences of $x$ for every $n > 0$. Similarly, if $B$ contains a zero law $(x * 0 = 0)$, then $[x * 0]_B$ contains infinitely many terms, e.g., with $n$ occurrences of $x$ for every $n \geq 0$. In such cases subterm replacement cannot make sense at a single instance of a variable, i.e., the rule $(\pm 6_1)$ does not generalize to arbitrary equational theories $B$; this is unfortunate because this rule is the basis for term rewriting.

However, we can make single occurrence rewriting work on $B$-classes with two additional assumptions. Recall that an equation $(\forall X) \ t = t'$ is **balanced** iff $var(t) = var(t') = X$, and note that if an equation in $B$ does not have the same variables in its two sides, then deduction modulo $B$ may require finding values for the unmatched variables, which in general cannot be done automatically. Also, recall that an equation $(\forall X) \ t = t'$ is **linear** iff $t$ and $t'$ each have at most one occurrence of each variable in $X$. For example, associative, commutative and identity laws are all both linear and balanced, while an idempotent law is balanced but not linear, and a zero law is linear but not balanced.

**Fact 7.2.5** Given $B$ linear balanced and $[t] \in T_{\Sigma,B}(X)$, then every term $t'$ that is $B$-equivalent to $t$ has the same number of occurrences of any $x \in X$ as $t$ does.

**Proof:** This is because the number of occurrences of a variable symbol is the same in the result of applying a linear balanced equation to a term as it was in the original term. □

Therefore if $B$ is linear balanced, the following makes sense:

$(\pm 6_{1,B})$ *Bidirectional Single Subterm Replacement Modulo B.* Given $t_0 \in T_{\Sigma,B}(\{z\}_s \cup Y)$ with exactly one occurrence of $z$, where $z \notin Y$, and given a substitution $\theta : X \to T_{\Sigma,B}(Y)$, if
$$(\forall X) \ t_1 =_B t_2 \quad \text{or} \quad (\forall X) \ t_2 =_B t_1$$
is of sort $s$ and is in $A$, then

$(\forall Y)\ t_0(z \leftarrow \theta(t_1)) =_B t_0(z \leftarrow \theta(t_2))$
is deducible.

This suggests the following notion of class rewriting modulo $B$ based on single subterm replacement: If $B$ is linear balanced and $A$ is a set of modulo $B$ rewrite rules, define an abstract rewriting system on the ($S$-sorted) set $T_{\Sigma,B}$ of $B$-equivalence classes of ground $\Sigma$-terms, by $c \overset{1}{\Rightarrow}_{[A/B]} c'$ iff there is some $c_0 \in T_{\Sigma,B}(\{z\}_s)$ such that $c = c_0(z \leftarrow \theta(t_1))$ and $c' = c_0(z \leftarrow \theta(t_2))$, for some (modulo $B$) substitution $\theta$ and some rule $t_1 \to t_2$ of sort $s$ in $A$. Later in this chapter, we show that class rewriting can be defined without restricting $B$ to be linear or balanced. As noted above, class rewriting is impractical, because classes can be very large, even infinite; nevertheless our later general version provides a semantic standard for the correctness of efficient implementations like that in OBJ3, which rewrites representive terms rather than classes. This is discussed in detail in Section 7.3, and also appears in the next subsection.

### 7.2.2 Deduction Modulo Equations in OBJ3

OBJ3 implements deduction modulo any combination of A, C and I (where "I" stands for identity[1]), for any subset of binary operations in the signature; the equations in $B$ are declared using attributes of operations. For example, an operation modulo CI is declared by

```
op _*_ : S S -> S [comm id: e] .
```

Note that the identity constant "e" must be declared explicitly, because there could be other constants of an appropriate sort. The keyword "assoc" is used for associativity. We illustrate OBJ's rewriting modulo associativity with the following proof of the right inverse law for left groups, as in Example 7.1.1; the reader may wish to first review Section 4.6.

**Example 7.2.6** (*Right Identity for Left Groups*) We must first give a new version of the specification that treats associativity as a built in equation rather than as a rewrite rule. Then we do the proof itself, beginning with a constant for the universal quantifier. The "range" notation, as in "[2 .. 3]", is explained after the proof.

```
th GROUPLA is sort Elt .
  op _*_ : Elt Elt -> Elt [assoc] .
  op e :    -> Elt .
  op _-1 : Elt -> Elt [prec 2] .
  var A : Elt .
  [lid]  eq e * A = A .
```

---

[1] For the non-commutative case, this means both the left and right identity laws.

```
    [linv] eq A -1 * A = e .
  endth
open .                          ***> first prove the right inverse law:
  op a : -> Elt .
  start a * a -1 .
  apply -.lid at term .       ***> should be: e * a * a -1
  apply -.linv with A = (a -1) at [1] .
                              ***> should be: a -1 -1 * a -1 * a * a -1
  apply .linv at [2 .. 3] .   ***> should be: a -1 -1 * e * a -1
  apply .lid at [2 .. 3] .    ***> should be: a -1 -1 * a -1
  apply .linv at term .       ***> should be: e
  [rinv] eq A * A -1 = e .    ***> add the proven equation:
  start a * e .               ***> now prove the right identity law:
  apply -.linv with A = a at [2] .  ***> should be: a * a -1 * a
  apply .rinv at [1 .. 2] .   ***> should be: e * a
  apply .lid at term .        ***> should be: a
close
```

The keyword "`term`" indicates application of the rule at the top of the current term (i.e., with $t_0 = z$ in rule $(\pm 6)_B$), while the notation "`[2]`" indicates application at its second subterm, and "`[2 .. 3]`" indicates the subterm consisting of the second and third subterms; the selected rule is applied at most once, and fails if the selected subterm does not match. The next section shows how this example can be simplified even further by using term rewriting modulo equations, in addition to deduction modulo equations. □

We now discuss `apply` in somewhat more detail; a complete description is given in [90]. The notations "`[k]`" and "`[k .. n]`" are used for binary operations that are associative only; incidentally, "`[k .. k]`" is equivalent to "`[k]`", and "`[]`" is not allowed. Because OBJ3 represents terms modulo A, C, and AC with ordinary terms, if you know how the representing term is parenthesized, then for each case you can select subterms using the parenthetic occurrence notation of Section 4.6. Thus, instead of "`[2]`" above, we could have written "`(2)`"; however, the square bracket range notation is preferable. You can see the parenthesization with the command "`apply print at term`" provided "`print with parens`" is on. The default parenthesization takes the right most subterm as the innermost; note that applying an equation may cause re-parenthesization closer to the default form, even in subterms disjoint from the redex.

Occurrence notation must be used for selecting subterms for operations that are commutative only. Note that "`()`" is a valid occurrence, and is equivalent to "`term`"; another synonym is "`top`". The "set" notations "`{k}`" and "`{k,...,n}`" are used for AC binary operations, analogously to "`[k]`" and "`[k .. n]`" for associative operations.

Because specifications can have multiple binary operations with varying combinations of modulo attributes, a notation is needed for composing the three selection methods discussed above. For example, sup-pose *, +, # are respectively associative, AC, and without attributes, and consider the term $t = (a * b)\#(a + b + (a * c))$. Then the following

```
    [1] of {3} of (2) ,
```
selects the third subterm $a$, while
```
    {3,1} of (2)
```
will select the subterm $(a * c) + a$, also causing the representing term to be rearranged. A final "`of term`" is optional for composite selectors.

If we knew that the representation of $t$ was parenthesized as $(a * b)\#(a + (b + (a * c))))$, then we could also do the first selection above using the occurrence notation (2 3 1); however, the second selection above cannot be done with occurrence notation. Note the reversal of order between "`[1] of {3} of (2)`" and "`(2 3 1)`". Note also that the command
```
    apply print at {3,1} of (2)
```
will cause the representation of $t$ to be rearranged as $(a * b)\#(((a * c) + a) + b)$, even though no deduction is involved.

Instead of "`at`", the keyword "`within`" can be used, indicating a single application at some proper subterm of the selected term; this can be convenient when there is a unique subterm that matches. A summary of the syntax for `apply` is produced by the command "`apply ?`". To better understand this material, in addition to the two exercises below, the reader should also look at Example 7.3.6 on page 194.

**Exercise 7.2.1** (1) Do the proof in Example 7.2.6, everywhere replacing range notation with occurrence notation. (2) Do the proof of Example 7.2.6 using "`within`" wherever possible. □

**Exercise 7.2.2** The following is part of the calculus of relations (Appendix C has some basics):

```
th REL is sort Rel .
  op I : -> Rel .
  op _U_ : Rel Rel -> Rel [assoc comm].
  op _;_ : Rel Rel -> Rel [assoc].
  vars R R1 R2 : Rel .
  eq I ; R = R .   eq R ; I = R .
  eq R ;(R1 U R2) = (R ; R1) U (R ; R2).
  eq (R1 U R2); R = (R1 ; R) U (R2 ; R).
endth
```

These operations and laws constitute a so-called **semi-ring**. Now add the recursive definition

```
op _*_ : Rel Nat -> Rel .
var R : Rel .   var N : Nat .
eq R * 0 = I .
eq R * s N = (R * N);(I U R).
```

and prove that if R * N = R * s N, then also

```
(R U I);(R * N) = (R * N) .
```

(In this case, R * N is the transitive reflexive closure of R.) Use the OBJ3 `apply` feature to do all the calculations, although the induction itself remains outside the OBJ3 framework. **Hint:** Use induction to prove that

```
R ;(R * N) = (R * N); R
```

for all R and all N. □

## 7.3 Term Rewriting Modulo Equations

Term rewriting is the main computational engine for theorem proving in this book, and this section develops rewriting modulo $B$, which has significant advantages over the class rewriting sketched in Section 7.2.1: (1) it uses ordinary rules instead of modulo $B$ rules; (2) $B$ needs be neither balanced nor linear; (3) infinite classes of terms are avoided; and (4) occurrences make sense. We assume throughout that $B$ is unconditional.

**Definition 7.3.1** A **modulo term rewriting system**, abbreviated **MTRS**, consists of a signature $\Sigma$, a set $B$ of $\Sigma$-equations, and a $\Sigma$-term rewriting system $A$, written $(\Sigma, A, B)$.

Given an MTRS $(\Sigma, A, B)$, define an ARS on $T_{\Sigma,B}$ by $c \Rightarrow_{[A/B]} c'$ iff there are $t, t' \in T_\Sigma$ such that $c = [t]$, $c' = [t']$, and $t \Rightarrow_A t'$. This relation is called **one-step class rewriting**, and its transitive, reflexive closure is **class rewriting**.

Given an MTRS $(\Sigma, A, B)$, define an ARS on $T_\Sigma$ by $t \Rightarrow_{A/B} t'$ iff there are $t_1, t_2 \in T_\Sigma$ such that $t \simeq_B t_1$, $t' \simeq_B t_2$, and $t_1 \Rightarrow_A t_2$. This relation is called **one-step rewriting with $A$ modulo $B$**, and its transitive, reflexive closure is **rewriting modulo $B$**.

We extend to rewriting terms with variables by extending $\Sigma$ to $\Sigma(X)$, and in this case write $c \Rightarrow_{[A/B],X} c'$ and $t \Rightarrow_{A/B,X} t'$, which are defined on $T_{\Sigma(X),B}$ and $T_{\Sigma(X)}$ respectively. □

The proof of the following is similar to those of Proposition 5.1.9 and Corollary 5.1.10:

**Proposition 7.3.2** Given $t, t' \in T_\Sigma(Y)$, $Y \subseteq X$ and MTRS $(\Sigma, A, B)$, then $t \Rightarrow_{A/B,X} t'$ iff $t \Rightarrow_{A/B,Y} t'$, and in both cases $var(t') \subseteq var(t)$. Also $t \stackrel{*}{\Rightarrow}_{A/B,X} t'$ iff $t \stackrel{*}{\Rightarrow}_{A/B,Y} t'$, and in both cases $var(t') \subseteq var(t)$. □

Thus both $\Rightarrow_{A/B,X}$ and $\stackrel{*}{\Rightarrow}_{A/B,X}$ restrict and extend reasonably over variables, so that we can drop the subscript $X$, with the understanding that any $X$ such that $var(t) \subseteq X$ may be used. On the other hand, $\stackrel{*}{\Leftrightarrow}_{A/B,X}$ does not restrict and extend reasonably, as shown by Example 5.1.15. Thus, we define $t \stackrel{*}{\Leftrightarrow}_A t'$ to mean there exists an $X$ such that $t \stackrel{*}{\Leftrightarrow}_{A,X} t'$. Example 5.1.15 also shows bad behavior for $\simeq^X_{A/B}$ (defined by $t \simeq^X_{A/B} t'$ iff $A \cup B \models (\forall X)\ t = t'$), although the concretion rule (8) of Chapter 4 (extended to rewriting modulo) implies $\simeq^X_{A,B}$ does behave reasonably when the signature is non-void. Defining $\downarrow_{A/B,X}$ as usual from the ARS of an MTRS $(\Sigma, A, B)$, we can generalize Proposition 5.1.13 to show that $\downarrow_{A/B,X}$ also restricts and extends reasonably, again allowing the subscript $X$ to be dropped:

**Proposition 7.3.3** Given $t_1, t_2 \in T_\Sigma(Y)$, $Y \subseteq X$ and MTRS $(\Sigma, A, B)$, then we have $t_1 \downarrow_{A/B,X} t_2$ if and only if $t_1 \downarrow_{A/B,Y} t_2$, and moreover, these imply $A \cup B \vdash (\forall X)\ t_1 = t_2$. □

There are good implementations of rewriting modulo $B$ for those $B$ that are actually available through attributes in OBJ. Although rewriting modulo $B$ accurately describes what OBJ does, it is not how OBJ actually does it, because this would require much needless search for matches; Section 7.3.3 gives some details of what OBJ3 really does, which of course is equivalent to rewriting modulo $B$. The following (generalizing 5.1.11 and 5.1.16) says rewriting modulo $B$ is equivalent to class rewriting, and that both are semantically sound and complete:

**Theorem 7.3.4** (*Completeness*) Given an MTRS $(\Sigma, A, B)$ and $t, t' \in T_\Sigma(X)$, then

$$t \Rightarrow_{A/B} t' \quad \text{iff} \quad [t] \Rightarrow_{[A/B]} [t']\ ,$$

$$t \stackrel{*}{\Rightarrow}_{A/B} t' \quad \text{iff} \quad [t] \stackrel{*}{\Rightarrow}_{[A/B]} [t']\ ,$$

$$[t] \stackrel{*}{\Rightarrow}_{[A/B]} [t'] \quad \text{implies} \quad [A] \vdash (\forall X)\ [t] = [t']\ ,$$

$$t \stackrel{*}{\Leftrightarrow}_{A/B} t' \quad \text{iff} \quad A \cup B \vdash (\forall X)\ t = t'\ .$$

Thus $\stackrel{*}{\Leftrightarrow}_{A/B}$ is complete for satisfaction of $A \cup B$, and $\stackrel{*}{\Leftrightarrow}_{[A/B]}$ is complete for satisfaction of $A$ modulo $B$. Moreover, $\stackrel{*}{\Leftrightarrow}_{A/B}$ and $\simeq^X_{A \cup B}$ are equal relations on terms with variables in $X$.

**Proof:** The first assertion follows from the definitions of $\Rightarrow_{A/B}$ and $\Rightarrow_{[A/B]}$ (Definition 7.3.1), and the second follows from the first by induction. The third follows from $\Rightarrow_{[A/B]}$ being a rephrasing of the rule $(+6_B)$. The forward direction of the fourth follows from the second and third, plus (1) implies (4) of Theorem 7.2.3, while the converse follows from Theorem 5.1.16 for $A \cup B$ and the definition of $\stackrel{*}{\Rightarrow}_{A/B}$. The fifth and sixth follow from the fourth plus Theorem 7.2.3. □

**Example 7.3.5** (*Right Identity for Left Groups*) Although the proof in Example 7.2.6 using built in associativity is simpler than the original proof in Section 4.6, it can be simplified even further, by using rewriting modulo associativity, replacing the last two `apply` commands in both the first and the second subproof by "`apply reduction within term .`"

```
open GROUPLA .
op a : -> Elt .
start a * a -1 .          ***> first prove the right inverse law:
apply -.lid at term .     ***> should be: e * a * a -1
apply -.linv with A = (a -1) within term .
                          ***> should be: a -1 -1 * a -1 * a * a -1
apply reduction within term .  ***> should be: e
[rinv] eq A * A -1 = e .  ***> add the proven equation
start a * e .             ***> now prove the right identity law
apply -.linv with A = a within term .  ***> should be: a * a -1 * a
apply reduction within term .          ***> should be: a
close
```

Now the first subproof takes only three `apply` commands, and the second only two. But it is something of an accident that this works, because a different ordering of rewrites could have blocked this proof by *undoing* some prior backward applications; thus, the proof style of Example 7.2.6 represents a more safe and sure way to proceed. □

**Example 7.3.6** The definition of ring is as follows (e.g., see [128], Chapter IV), noting that the multiplication * need not be commutative:

```
th RING is sort R .
  ops 0 1 : -> R .
  op _+_ : R R -> R [assoc comm idr: 0 prec 5] .
  op _*_ : R R -> R [assoc idr: 1 prec 3] .
  op -_ : R      -> R [prec 1] .
  vars A B C : R .
  [ri] eq A + (- A) = 0 .
  [ld] eq A *(B + C) = A * B + A * C .
  [rd] eq (B + C) * A = B * A + C * A .
endth
```

We will prove that $(\forall A)\ A*0 = 0$. For this, we should turn on the `print with parens` feature so that when rules are shown, we can check how they are parenthesized.

```
open .
  ops a b c : -> R .
  show rules .
  start a * 0 .
  apply -.6 at top .
  apply -.ri with A = a * a at 1 .
  apply -.ld with A = a at [1 .. 3] .
  apply red at term .
close
```

Rule .6 is (0 + X_id) = X_id, which was generated by the "idr: 0" declaration (we learn this from the output of the `show rules` command). The result of the final reduction is 0, and so the proof is done.

□

Modulo $B$ versions of the basic term rewriting concepts are obtained from the ARS definitions applied to the class rewriting relation:

**Definition 7.3.7** Given an MTRS $\mathcal{M} = (\Sigma, A, B)$, then $A$ is **terminating modulo** $B$ iff $\Rightarrow_{[A/B]}$ is terminating, and is **Church-Rosser modulo** $B$ iff $\Rightarrow_{[A/B]}$ is Church-Rosser. Ground terminating, ground Church-Rosser, canonical, ground canonical, etc. modulo $B$ are defined similarly, and we also say that $\mathcal{M}$ is terminating, Church-Rosser, canonical, etc.   □

From this and Theorem 7.3.4 we get the following:

**Proposition 7.3.8** An MTRS $\mathcal{M} = (\Sigma, A, B)$ is terminating iff $\Rightarrow_{A/B}$ is terminating, i.e., iff there is no infinite sequence $t_1, t_2, t_3, \ldots$ of $\Sigma$-terms such that $t_1 \Rightarrow_{A/B} t_2 \Rightarrow_{A/B} t_3 \Rightarrow_{A/B} \cdots$. Similarly, $\mathcal{M}$ is ground terminating iff there is no such infinite sequence of ground $\Sigma$-terms. Also, $\mathcal{M}$ is Church-Rosser iff whenever $t \stackrel{*}{\Rightarrow}_{A/B} t_1$ and $t \stackrel{*}{\Rightarrow}_{A/B} t_1'$, there exist $t_2, t_2'$ equivalent modulo $B$ such that $t_1 \stackrel{*}{\Rightarrow}_{A/B} t_2$ and $t_1' \stackrel{*}{\Rightarrow}_{A/B} t_2'$, and is ground Church-Rosser iff this property holds for all ground terms $t$. Moreover, a $\Sigma$-term $t$ is a normal form for $\mathcal{M}$ iff there is no $t'$ such that $t \Rightarrow_{A/B} t'$, and is a normal form for a $\Sigma$-term $t'$ iff $t$ is a normal form and $t' \stackrel{*}{\Rightarrow}_{A/B} t$. Finally, $t$ is a normal form for $\Rightarrow_{A/B}$ iff $[t]_B$ is a normal form for $\Rightarrow_{[A/B]}$.   □

The following generalizes Theorem 5.2.9 to rewriting modulo $B$:

**Theorem 7.3.9** Given a ground canonical MTRS $(\Sigma, A, B)$, if $t_1, t_2$ are two normal forms of a ground term $t$ under $\Rightarrow_{A/B}$ then $t_1 \simeq_B t_2$. Moreover, the $B$-equivalence classes of ground normal forms under $\Rightarrow_{A/B}$ form an initial $(\Sigma, A \cup B)$-algebra, denoted $N_{\Sigma, A/B}$ or $N_{A/B}$, as follows, where $\llbracket t \rrbracket$ denotes any arbitrary normal form of $t$, and $\llbracket t \rrbracket_B$ denotes the $B$-equivalence class of $\llbracket t \rrbracket$:

(0) interpret $\sigma \in \Sigma_{[],s}$ as $\llbracket \sigma \rrbracket_B$ in $N_{\Sigma, A/B, s}$; and

(1) interpret $\sigma \in \Sigma_{s_1 \ldots s_n, s}$ with $n > 0$ as the function that sends $(\llbracket t_1 \rrbracket_B, \ldots, \llbracket t_n \rrbracket_B)$ with $t_i \in T_{\Sigma, s_i}$ to $\llbracket \sigma(t_1, \ldots, t_n) \rrbracket_B$ in $N_{\Sigma, A/B, s}$.

Finally, $N_{\Sigma, A/B}$ is $\Sigma$-isomorphic to $T_{\Sigma, A \cup B}$.

**Proof:** For convenience, write $N$ for $N_{\Sigma, A/B}$. The first assertion follows from the ARS result Theorem 5.7.2, using Theorem 7.3.4. Note that $\sigma_N$ is well defined by (1), because of the first assertion, plus the fact that $\simeq_B$ is a $\Sigma$-congruence relation.

Next, we check[E26] that $N$ satisfies $A \cup B$. Satisfaction of $B$ is by definition of $N$ as consisting of $B$-equivalence classes of normal forms. Now let $(\forall X)\, t = t'$ be in $A$; we need $\overline{a}(t) = \overline{a}(t')$ for all $a : X \to N$. Let $b : X \to T_{\Sigma, B}$ denote the extension of $a$ from target $N$ to $T_{\Sigma, B}$. Then $\overline{a}(t) = \llbracket \overline{b}(t) \rrbracket_B$ for any $t \in T_{\Sigma, B}$ because $\llbracket \overline{b}(\_) \rrbracket_B$ is a $\Sigma$-homomorphism since $\llbracket \_ \rrbracket_B$ is, and there is a unique $\Sigma$-homomorphism $T_{\Sigma, B}(X) \to N$ that extends $a$. Now applying the given rule to $t$ with the substitution $b$ gives $\overline{b}(t) \Rightarrow_{A/B} \overline{b}(t')$, so these two terms have the same canonical form, i.e., $[\![\overline{b}(t)]\!]_B = [\![\overline{b}(t')]\!]_B$ and thus $\overline{a}(t) = \overline{a}(t')$, as desired.

Next, let $M$ be an arbitrary $(\Sigma, A \cup B)$-algebra, and let $h : T_{\Sigma,B} \to M$ be the unique $\Sigma$-homomorphism. Noting that $N \subseteq T_{\Sigma,B}$, let $g : N \to M$ be the restriction of $h$ to $N$. We now prove that $g$ is a $\Sigma$-homomorphism by structural induction over $\Sigma$:

(0) Given $\sigma \in \Sigma_{[],s}$, we get $g(\sigma_N) = h([\![\sigma]\!]_B)$ by definition. Then Theorem 7.3.4 gives $h([\sigma]) = h([\![\sigma]\!]_B)$ because $[\sigma] \overset{*}{\Rightarrow}_{[A/B]} [\![\sigma]\!]_B$, and then $h([\sigma]) = \sigma_M$ because $h$ is a $\Sigma$-homomorphism. Therefore $g(\sigma_N) = \sigma_M$, as desired.

(1) Given $\sigma \in \Sigma_{s_1\ldots s_n,s}$ with $n > 0$, we get

$$g(\sigma_N([\![t_1]\!]_B, \ldots, [\![t_n]\!]_B)) = h([\![\sigma(t_1,\ldots,t_n)]\!]_B)$$

by definition. Then Theorem 7.3.4 gives

$$h([\sigma(t_1,\ldots,t_n)]) = h([\![\sigma(t_1,\ldots,t_n)]\!]_B),$$

so that

$$\begin{aligned}h([\sigma(t_1,\ldots,t_n)]) &= \sigma_M(h([t_1]),\ldots,h([t_n])) \\ &= \sigma_M(g([\![t_1]\!]_B),\ldots,g([\![t_n]\!]_B))\end{aligned}$$

because $h$ is a $\Sigma$-homomorphism. Therefore

$$g(\sigma_N([\![t_1]\!]_B,\ldots,[\![t_n]\!]_B)) = \sigma_M(g([\![t_1]\!]_B),\ldots,g([\![t_n]\!]_B)),$$

as desired.

For uniqueness, suppose $g' : N \to M$ is another $\Sigma$-homomorphism. Let $r : T_{\Sigma,B} \to N$ be the map sending $[t]$ to $[\![t]\!]_B$, and note that it is a $\Sigma$-homomorphism by the definition of $[\![\_]\!]_B$. Next, note that if $i : N \to T_{\Sigma,B}$ denotes the inclusion $\Sigma$-homomorphism, then $i;r = 1_N$. Finally, note that $r;g = r;g' = h$, by the uniqueness of $h$. It now follows that $i;r;g = i;r;g'$, which implies $g = g'$. The last assertion follows since both are initial $(\Sigma, A \cup B)$-algebras. □

**Theorem 7.3.10** Given a canonical MTRS $(\Sigma, A, B)$, then

$$[A] \vdash_B (\forall X)\ t =_B t' \quad \text{iff} \quad A \cup B \vdash (\forall X)\ t = t' \quad \text{iff} \quad [\![t]\!] \simeq^X_B [\![t']\!],$$

where as before, $[\![t]\!]$ denotes an arbitrary normal form of $t$ under $\Rightarrow_{A/B}$.

**Proof:** The first "iff" is Theorem 7.2.3. The "if" direction of the second "iff" is straightforward. For its "only if", let $h$ and $h'$ be the unique $\Sigma(X)$-homomorphisms from $T_{\Sigma(X)}$ to $T_{\Sigma(X),A\cup B}$ and $N_{\Sigma(X),A/B}$, respectively. Then $A \cup B \vdash (\forall X)\ t = t'$ implies $h(t) = h(t')$. But $T_{\Sigma(X),A\cup B}$ is $\Sigma(X)$-isomorphic to $N_{\Sigma(X),A/B}$ by Theorem 7.3.9. Therefore $h'(t) = h'(t')$, i.e., $t$ and $t'$ have the same normal form modulo $B$. □

An important consequence of the above theorem is that we can define a function == that works for canonical MTRS's the same way that the function == described in Section 5.2 works for ordinary canonical TRS's, namely, $t == t'$ returns true over an MTRS $(\Sigma, A, B)$ iff $t, t'$ are provably equal under $A \cup B$ as an equational theory. This function is implemented in OBJ3 by computing the normal forms of $t, t'$ and checking whether they are equal modulo $B$. Note that even when $(\Sigma, A, B)$ is not canonical, if $t == t'$ does return true then $t$ and $t'$ are equal under $(\Sigma, A, B)$, again just as for ordinary TRS's. The function =/= is also available for MTRS's in OBJ3, but just as for ordinary TRS's, it is dangerous if the system is not canonical for the sort involved. The use of == is illustrated by proofs in the following subsection.

### 7.3.1 Some Inductive Proofs Modulo Equations

This section gives more inductive proofs along the lines of those in Section 6.5.1, but using associative-commutative rewriting. Example 6.5.7 implies we can use AC rewriting for addition, and Exercises 7.3.1 and 7.3.2 below imply we can also use AC rewriting for multiplication.

**Example 7.3.11** (*Formula for* $1 + \cdots + n$) We give an inductive proof of a formula for the sum of the first $n$ positive numbers,

$$1 + 2 + \cdots + n = n(n+1)/2 \,,$$

using Exercises 7.3.1 and 7.3.2 by giving + and * the attributes assoc and comm. This saves us from having to worry about the ordering and grouping of subterms within expressions. The second module defines the function $sum(n) = 1 + \cdots + n$. (What we actually prove is that $sum(n) + sum(n) = n(n+1)$.)

```
obj NAT is sort Nat .
  op 0 : -> Nat .
  op s_ : Nat -> Nat [prec 1] .
  op _+_ : Nat Nat -> Nat [assoc comm prec 3] .
  vars M N : Nat .
  eq M + 0 = M .
  eq M + s N = s(M + N).
  op _*_ : Nat Nat -> Nat [assoc comm prec 2] .
  eq M * 0 = 0 .
  eq M * s N = M * N + M .
endo
obj SUM is protecting NAT .
  op sum : Nat -> Nat .
  var N : Nat .
  eq sum(0) = 0 .
  eq sum(s N) = s N + sum(N) .
endo
```

```
    open .
      ops m n : -> Nat .
      ***> base case
      reduce sum(0) + sum(0) == 0 * s 0 .
      ***> induction step
      eq sum(n) + sum(n) = n * s n .
      reduce sum(s n) + sum(s n) == s n * s s n .
    close
```

The line "`protecting NAT`" indicates that the natural numbers are imported in "protecting" mode, which means that there are supposed to have no junk and no confusion for the sort Nat in the models of SUM.

We can also use this example to illustrate how unsuccessful proof scores can yield hints about lemmas to prove. If we try the same proof score as above, but without the `assoc` and `comm` attributes for multiplication, then the base case works, but the induction step fails, with the two sides being `s (s (n + n + (n * n) + n))` and `s (s (n + (s n * n) + n))`. (The reduction actually evaluates to a rather non-informative `false`. However, we can get the desired information either by reducing the two sides separately, or else by replacing `_==_` by a Boolean-valued operation `_eq_` satisfying the single equation `N eq N = true`). The difference between these comes from the terms `n + (n * n)` and `(s n * n)`, the equality of which differs from the second law for `*` by commutativity. This might suggest either proving the lemma

$$s N * N = N + N * N ,$$

or else proving the commutativity of `*`. The induction step goes through either way, and we also discover that associativity of multiplication is not needed here. The two lemmas in the proof of the commutativity of addition (Example 6.5.7) were arrived at in the same way. □

**Exercise 7.3.1** Use OBJ3 to show the associativity of multiplication of natural numbers. □

**Exercise 7.3.2** Use OBJ3 to show the commutativity of multiplication of natural numbers. □

**Example 7.3.12** (*Fermat's Little Theorem for $p = 3$*) The "little Fermat theorem" says that

$$x^p \equiv x \pmod{p}$$

for any prime $p$, i.e., that the remainder of $x^p$ after division by $p$ equals the remainder of $x$ after division by $p$. The following OBJ3 proof score for the case $p = 3$ assumes we have already shown that multiplication is associative and commutative. This is a nice example of an inductive proof where there are non-trivial relations among the constructors;[2] for in this example, unlike the usual natural numbers, s s s 0 = 0.

```
obj NAT3 is sort Nat .
  op 0 : -> Nat .
  op s_ : Nat -> Nat [prec 1] .
  op _+_ : Nat Nat -> Nat [assoc comm prec 3] .
  vars L M N : Nat .
  eq M + 0 = M .
  eq M + s N = s(M + N) .
  op _*_ : Nat Nat -> Nat [assoc comm prec 2] .
  eq M * 0 = 0 .
  eq M * s N = M * N + M .
  eq L * (M + N) = L * M + L * N .
endo
open .
  var M : Nat .
  eq M + M + M = 0 .
  op x : -> Nat .
  ***> base case, x = 0
  red 0 * 0 * 0 == 0 .
  ***> induction step
  eq x * x * x = x .
  red s x * s x * s x == s x .
close
```

The first equation after the open quotients the natural numbers to the naturals modulo 3. □

**Exercise 7.3.3** Let $(A, \oplus)$ be an **Abelian semigroup**, i.e., suppose that $\oplus$ is a binary associative, commutative operation on $A$. Now use OBJ for the following:

1. Define $\bigoplus_{1 \leq i \leq n} a(i)$, where $a(i) \in A$ for $1 \leq i \leq n$ and $n > 0$.

2. If $a(i), b(i) \in A$ for $1 \leq i \leq n$, prove that

$$\bigoplus_{1 \leq i \leq n} (a(i) \oplus b(i)) = (\bigoplus_{1 \leq i \leq n} a(i)) \oplus (\bigoplus_{1 \leq i \leq n} b(i))$$

**Hint:** use the following declarations in OBJ:

```
op _+_ : A A -> A [assoc comm] .
ops a b : Nat -> A .
```

3. Give an example of this formula where $A$ is the integers and $\oplus$ is addition. Explain how to extend $\bigoplus_{1 \leq i \leq n} a(i)$ to the case where $n = 0$, and generalize this to the case of an arbitrary Abelian semigroup. □

---

[2] I thank Dr. Immanuel Kounalis for doubting that OBJ3 could handle non-trivial relations on constructors, and then presenting the challenge to prove this result.

### 7.3.2 The Propositional Calculus

A very nice application of term rewriting modulo equations is a decision procedure for the propositional calculus. One way to define the propositional calculus is with the following equational theory, which is written in OBJ3; we override the default inclusion of the Booleans in order to avoid ambiguous parsing for and, or, etc.; the imported module TRUTH provides OBJ's built in sort Bool with just the two constants true and false, and basic built in operations like ==.

```
set include BOOL off .
obj PROPC is protecting TRUTH .
  op _and_ : Bool Bool -> Bool [assoc comm prec 2] .
  op _xor_ : Bool Bool -> Bool [assoc comm prec 3] .
  vars P Q R : Bool .
  eq P and false = false .
  eq P and true = P .
  eq P and P = P .
  eq P xor false = P .
  eq P xor P = false .
  eq P and (Q xor R) = (P and Q) xor (P and R).

  op _or_ : Bool Bool -> Bool [assoc comm prec 7] .
  op not_ : Bool -> Bool [prec 1] .
  op _implies_ : Bool Bool -> Bool [prec 9] .
  op _iff_ : Bool Bool -> Bool [assoc prec 11] .
  eq P or Q = (P and Q) xor P xor Q .
  eq not P = P xor true .
  eq P implies Q = (P and Q) xor P xor true .
  eq P iff Q = P xor Q xor true .
endo
```

The main part of this specification involves only connectives and and xor; the second part defines the remaining propositional connectives in terms of these two, plus the constant true. Because it is already known that these equations (including those for AC) are one way to define the propositional calculus, we know that the above really is a theory of the propositional calculus. The following result (due to Hsiang [106]) explains why PROPC is important:

**Theorem 7.3.13** As a term rewriting system, PROPC is canonical modulo $B$, where $B$ consists of the associative and commutative laws for xor and and. □

This is proved in Exercise 12.1.3. Note also that this $B$ is linear balanced. Moreover,

**Fact 7.3.14** The initial algebra of PROPC has just two elements, namely true and false.

**Proof:** Given Theorem 7.3.13, it suffices to determine the reduced forms, by Theorem 7.3.9. The terms true and false are reduced because no rules apply to them, and a case analysis of the eight terms built from true and false using and and xor shows that they all reduce to either true or false. (We can ignore the other operations because they are defined in terms of and and xor.) □

It follows from this that PROPC really does protect its imported module TRUTH, as its specification claims (the "protecting" notion was defined[E27] in Chapter 6).

Our commitment to semantics demands that before going further, we should make it clear what we mean by saying that an equation is "true" in the propositional calculus:

**Definition 7.3.15** An equation $e$ is a **theorem** of the propositional calculus iff $T_{\text{PROPC}} \models e$, and a $T_{\Sigma\text{PROPC}}(X)$-term $t$ is a **tautology** iff $(\forall X)\ t = true$ is a theorem of the propositional calculus. Formulae $t, t'$ are **equivalent** iff $(\forall X)\ t = t'$ is a theorem of the propositional calculus. □

We can prove $T_{\text{PROPC}} \models (\forall X)\ t = t'$ directly, by checking whether $\overline{a}(t) = \overline{a}(t')$ for all $a : X \to T_{\text{PROPC}}$. Because $T_{\text{PROPC}}$ has exactly two elements, true and false, by Fact 7.3.14, there are exactly $2^N$ cases to check when $X$ has $N$ variables, one case for each possible assignment $a$. This is essentially Ludwig Wittgenstein's well-known method of *truth tables*, also called (*Boolean*) *case analysis*; it is easy to apply this method by hand when $N$ is small.

**Example 7.3.16** To illustrate the method of truth tables, let's check whether or not the equation $(\forall P, Q)\ P\ and\ (P\ or\ Q) = P\ or\ Q$ is a theorem of the propositional calculus. Since there are two variables, there are four possible assignments; these appear in the left two columns, serving as labels for the four rows of the table below:

| P | Q | P or Q | P and (P or Q) | equal? |
|---|---|--------|----------------|--------|
| true | true | true | true | yes |
| true | false | true | true | yes |
| false | true | true | false | no |
| false | false | false | false | yes |

Thus we see that the equation is false, and that $P = false$, $Q = true$ is a counterexample. Of course, such calculations do not require an elaborate LaTeX tabular format. □

Theorem 7.3.13 plus some results from Chapter 5 that are generalized to rewriting modulo $B$ later in this chapter will imply that PROPC($X$), the enrichment of PROPC by $X$, is canonical for any $X$. However, the canonical forms of this MTRS are not as well known as they should be. The following is needed to describe those forms:

**Definition 7.3.17** A formula of the propositional calculus is an **exclusive normal form** (abbreviated **ENF**) iff it has the form

$$E_1 \text{ xor } E_2 \text{ xor } \ldots \text{ xor } E_n ,$$

where each $E_i$ (called an **exjunct**) has the form

$$C_{i,1} \text{ and } C_{i,2} \text{ and } \ldots \text{ and } C_{i,k_i} ,$$

where each $C_{i,j}$ (called a **conjunct**) either has the form $P$ or else the form *not P*, where $P$ is a variable of sort Prop; by convention, we say that the empty ENF ($n = 0$ above) is *false*, and that the empty exjunct ($k_i = 0$ above) is *true*. Those $P$ that occur in an exjunct with a preceding *not* are said to occur **negatively**, and those that occur without it **positively**. Given a set $X$ of variable symbols, an exjunct $E$ is **complete** (with respect to $X$) iff each variable in $X$ appears in $E$, and an ENF is **complete** (with respect to $X$) iff each of its exjuncts is complete.   □

The result mentioned above may now be stated as follows:

**Proposition 7.3.18** Every formula of the propositional calculus is equivalent to a unique (modulo $B$) **irredundant positive ENF**, defined to be an ENF having only positive conjuncts, involving (at most) the same variables, with no repeated conjuncts and no repeated exjuncts; these irredundant positive ENFs are the canonical forms of PROPC. Moreover, every formula of the propositional calculus is equivalent to a unique (modulo $B$) complete ENF involving (exactly) the same variables that it has.

**Proof:** The first assertion follows from noticing that no rule of PROPC($X$) applies to any irredundant positive ENF, so that these forms are reduced, and noticing that any formula using only *xor* and *and* that is not an irredundant positive ENF can be rewritten using one of the rules in the first part of PROPC, and so cannot be canonical. Therefore the irredundant positive ENFs must be its canonical forms.

For the second assertion, the complete ENF of a formula can be obtained from its irredundant positive ENF as follows: for each exjunct, if some variable $x$ does not appear in it, exjoin to it the conjunct *x and not x*, and then simplify the resulting term using only the distributive and idempotent laws; the result will be a complete ENF that is equivalent to the original irredundant positive ENF, because only rules from PROPC were used. This equivalence and the first assertion imply that distinct complete ENFs are inequivalent, and that every term is equivalent to a unique (modulo $B$) complete ENF.   □

Under the correspondence of the above proof, a term $t$ has the irredundant positive ENF *true* iff its complete ENF contains all $2^N$ exjuncts. More generally, the exjuncts in the complete ENF of $t$ correspond to those rows in its truth table where it is true. As an example, we find the complete ENF of the term $x \text{ xor } y$, using a simplified notation with

+ for *xor*, juxtaposition for *and*, and overbar for negation: calculating with PROPC, modulo AC for both binary operations, we have $x = x + y\bar{y} = xy + x\bar{y}$ and $y = y + x\bar{x} = yx + y\bar{x}$, so that the complete ENF for $x + y$ is $xy + x\bar{y} + \bar{x}y$. Similarly, the complete ENF for $x\bar{y} + xz$ is $xyz + x\bar{y}z + x\bar{y}\bar{z}$.

**Corollary 7.3.19** Two propositional calculus formulae over variables $X$ are provably equal in PROPC($X$) iff they yield the same Boolean value for every assignment of Boolean values for the variables in $X$.

**Proof:** Two terms are provably equal iff they have the same complete ENF, the conjuncts of which give exactly the Boolean assignments to $X$ for which the terms are true. [E28]  □

**Proposition 7.3.20** Given a set $X$ of $N$ Boolean variables, PROPC has a free algebra on $X$ generators, and it has $2^{2^N}$ elements.

**Proof:** Recall that the free PROPC-algebra on $X$ generators is the initial algebra of PROPC($X$) viewed as a PROPC-algebra. By the proof of Proposition 7.3.18, the normal forms of PROPC($X$) are in bijective correspondence with the complete exclusive normal forms. Each complete exclusive normal form can be seen as a *set* of complete disjuncts, and then it is easy to see that there are $2^N$ different complete disjuncts, and therefore $2^{2^N}$ different sets of complete disjuncts.  □

The elements of this free algebra can be seen as all of the possible Boolean functions on $N$ variables, noting that $N$ variables can take $2^N$ configurations, each of which can have 2 values.

**Exercise 7.3.4** Let $\mathbb{B}$ be the set containing *true* and *false*, let $\Sigma$ be the signature of PROPC($X$), and let $M$ be the $\Sigma$-algebra with carrier $[[X \to \mathbb{B}] \to \mathbb{B}]$, with operations from PROPC defined "pointwise" on functions from Boolean operations on $\mathbb{B}$ (e.g., with $\text{xor}_M(f, g)(a) = f(a) \text{ xor } g(a)$ for $a : X \to \mathbb{B}$, and with $x \in X$ interpreted as $x_M(a) = a(x)$. Show that $M$ is a free PROPC-algebra on $X$ generators.  □

The PROPC MTRS has the very special property that we can decide *whether or not* non-ground equations hold in the initial algebra just by comparing the canonical forms of their left- and rightsides; this property is unfortunately as rare as it is useful. Note that canonicity only allows showing that an equation *does* hold for the initial algebra; it decides whether or not the equation holds for *all* models. The following provides a precise formulation of what it means to say that PROPC gives a *decision procedure for the propositional calculus*:

**Definition 7.3.21** A TRS $(\Sigma, A)$ is **reduction complete** iff it is canonical and for any $\Sigma$-equation $e$, say $(\forall X)\ t_1 = t_2$, we have $T_{\Sigma,A} \models e$ iff $[\![t_1]\!] = [\![t_2]\!]$. An MTRS $(\Sigma, A, B)$ is **reduction complete** iff it is canonical and for any $\Sigma$-equation $e$, say $(\forall X)\ t_1 = t_2$, we have $T_{\Sigma,A\cup B} \models e$ iff $[\![t_1]\!] \simeq_B^X [\![t_2]\!]$.  □

**Exercise 7.3.5** Show that the TRS's from Examples 5.1.7 and 5.5.7 are not reduction complete. □

**Theorem 7.3.22** The MTRS PROPC is reduction complete.

**Proof:** Let $E = A \cup B$. By the completeness theorem, it will suffice to prove that, for any $\Sigma$-equation $e$,

$$T_{\Sigma,E} \models e \text{ iff } M \models e$$

for every $(\Sigma, E)$-algebra $M$. That the second condition implies the first is immediate. For the converse, we first treat the free $(\Sigma, E)$-algebras. We will use contradiction, and so we suppose that $T_{\Sigma,E}$ satisfies $e$ but that $T_{\Sigma,E}(Z)$ does not satisfy $e$. Then there exists an $a : X \to T_{\Sigma,E}(Z)$ such that $\overline{a}(t) \neq \overline{a}(t')$. By Exercises 7.3.4 and 6.1.3, we can take $T_{\Sigma,E}(Z)$ to be $[[Z \to \mathbb{B}] \to \mathbb{B}]$ with pointwise operations, where $\mathbb{B} = \{true, false\}$; similarly, we can take $T_{\Sigma,E}$ to be $\mathbb{B}$. Let $u = \overline{a}(t)$ and let $u' = \overline{a}(t')$. Since $u \neq u'$, there exists some $b : Z \to \mathbb{B}$ such that $u(b) \neq u'(b)$. Now defining $c = a; \overline{b} : X \to \mathbb{B}$ we get $\overline{c} = \overline{a}; \overline{b} : T_{\Sigma,E}(X) \to \mathbb{B}$, by 4. of Exercise 6.1.4. Next, if we define $\hat{b} : [[Z \to \mathbb{B}] \to \mathbb{B}] \to \mathbb{B}$ by $\hat{b}(u) = u(b)$, then the reader can check that $\hat{b}$ is a $\Sigma$-homomorphism such that $\hat{b}(z) = b(z)$ where the first $z$ is the function in $[[Z \to \mathbb{B}] \to \mathbb{B}]$ defined in Exercise 7.3.4. Then $\hat{b} = \overline{b}$ since there is just one such $\Sigma$-homomorphism extending $b$. Therefore $\overline{c}(t) = \overline{b}(\overline{a}(t)) = \overline{b}(u) = \hat{b}(u) = u(b)$, and similarly $\overline{c}(t') = u'(b)$. Therefore $\overline{c}(t) \neq \overline{c}(t')$, contradicting our assumption that $\mathbb{B}$ satisfies $e$. We next show the desired implication for any $(\Sigma, E)$-algebra $M$. By Proposition 6.1.18, there is some $Z$ such that $q : T_{\Sigma,E}(Z) \to M$ is surjective. But then $T_{\Sigma,E}(Z) \models e$ implies $M \models e$, and so we are done. □

It is easy to apply this result in OBJ, because its built in operation == returns `true` iff its two arguments have normal forms that are equivalent modulo the attributes declared for the operations involved; an alternative, which is justified in Exercise 7.3.8, is just to reduce the expression `t iff t'`.

**Exercise 7.3.6** Use OBJ3 to determine whether or not the following are tautologies of the propositional calculus:

 1. *P implies (P implies P)* .
 2. *P implies (P implies not P)* .
 3. *not P implies (P implies not P)* .
 4. *(P implies Q) implies Q* .
 5. *P iff P iff P* .
 6. *P iff P iff P iff P* .

    7. *(P implies Q) implies (Q implies Q)*.

Now use truth tables to check at least three of the above. □

**Exercise 7.3.7** Use OBJ3 to determine whether or not the following are theorems of the propositional calculus:

1. $(\forall P)$ *P = not not P*.
2. $(\forall\ P, Q)$ *P or Q = not P xor not Q*.
3. $(\forall\ P)$ *P = P iff P*.
4. $(\forall\ P, Q, R)$ *P implies (Q and R) = (P implies Q) and (P implies R)*.
5. $(\forall\ P, Q)$ *not (P and Q) = not P or not Q*.
6. $(\forall\ P, Q, R)$ *P implies (Q or R) = (P implies Q) or (P implies R)*.

Also use truth tables to check at least three of them. □

**Exercise 7.3.8** Show that $(\forall X)\ t = t'$ is a theorem of the propositional calculus iff $t$ `iff` $t'$ is a tautology, iff $\text{not}(t\ \text{xor}\ t')$ is a tautology. □

**Exercise 7.3.9** Show that if $\Sigma^{\text{PROPC}}$-formulae $t, t'$ are equivalent, then $t$ is a tautology iff $t'$ is. □

**Exercise 7.3.10** Show that if $X \subseteq Y$ then $(\forall X)\ t = t'$ is a theorem of the propositional calculus iff $(\forall Y)\ t = t'$ is. □

**Definition 7.3.23** A formula of the propositional calculus is a **disjunctive normal form** (abbreviated **DNF**) iff it has the form

    $D_1$ *or* $D_2$ *or* ... *or* $D_n$,

where each $D_i$ (called a **disjunct**) has the form

    $C_{i,1}$ *and* $C_{i,2}$ *and* ... *and* $C_{i,k_i}$,

where each $C_{i,j}$ (called a **conjunct**) either has the form *P* or else the form *not P*, where *P* is a variable of sort Prop; by convention, we say that the empty DNF ($n = 0$ above) is *false* and the empty disjunct ($k_i = 0$ above) is *true*. Those *P* that occur in a disjunct with a preceding *not* occur **negatively**, and those that occur without it occur **positively**. Given a set $X$ of variable symbols, a disjunct $C$ is **complete** (with respect to $X$) iff each variable in $X$ appears in $C$, and a DNF is **complete** (with respect to $X$) iff each of its disjuncts is complete. □

It follows from the above conventions that both *true* and *false* are DNFs. It also follows that if $X$ has $N$ elements, then each complete disjunct has $N$ conjuncts. The following is well known:

**Proposition 7.3.24** Every formula of the propositional calculus is equivalent to a DNF having (at most) the same variables, and to a unique (modulo *B*) complete DNF having (exactly) the same variables.  □

A nice proof of the above uses the MTRS in Exercise 7.3.11 below to rewrite formulae to disjunctive normal form; this MTRS is shown terminating in Example 7.5.10, and Church-Rosser in Exercise 12.1.1; hence this MTRS is canonical. Exercise 12.1.2 shows that its reduced forms are DNF's.

**Exercise 7.3.11** Choose five non-trivial formulae of the propositional calculus and use the TRS below to find their DNFs; explain why the reduced forms of this TRS are necessarily correct if they are DNFs, without using the as yet unproved result that the TRS is canonical.

```
obj DNF is protecting TRUTH .
  op _and_ : Bool Bool -> Bool [assoc comm prec 2] .
  op _or_ : Bool Bool -> Bool [assoc comm prec 3] .
  op not_ : Bool -> Bool [prec 1] .
  vars P Q R : Bool .
  eq P and false = false .
  eq P and true = P .
  eq P and P = P .
  eq P or false = P .
  eq P or true = true .
  eq P or P = P .
  eq not false = true .
  eq not true = false .
  eq P or not P = true .
  eq not not P = P .
  eq not(P and Q) = not P or not Q .
  eq not(P or Q)  = not P and not Q .
  eq P and (Q or R) = (P and Q) or (P and R).
  op _xor_ : Bool Bool -> Bool [assoc comm prec 7] .
  op _implies_ : Bool Bool -> Bool [prec 9] .
  op _iff_ : Bool Bool -> Bool [assoc prec 11] .
  eq P xor Q = (P and not Q) or (not P and Q).
  eq P implies Q = not P or Q .
  eq P iff Q = (P and Q) or (not P and not Q) .
  eq P and not P = false .
endo
```

Please note that once again, BOOL should not be included. Although this TRS is similar to PROPC, it has a quite different purpose; in particular, the formula *(p and q) or (p and not q)* is reduced under DNF but not under PROPC, where it has the canonical form *p*. Similarly, *not p* is reduced under DNF but not under PROPC, where it has canonical form *p xor true*.  □

### 7.3.3  Weak Rewriting Modulo Equations

This section brings us closer to how OBJ3 actually implements term rewriting modulo equations, with the following weaker relation which overcomes the inefficiency of Definition 7.3.1, because it only requires matching on subterms of the source term:

**Definition 7.3.25**  Given an MTRS $(\Sigma, A, B)$, for $t, t' \in T_\Sigma(X)$, we say $t$ **weakly rewrites to** $t'$ **under** (or **with**) $A$ **modulo** $B$ **in one step** iff there exist a rule $t_1 \to t_2$ of sort $s$ in $A$ with variables $Y$, a term $t_0 \in T_\Sigma(\{z\}_s \cup X)$, and a substitution $\theta : Y \to T_\Sigma(X)$ such that $t = t_0(z \leftarrow t_*)$ and $t_* \simeq_B \theta(t_1)$ and $t' = t_0(z \leftarrow \theta(t_2))$. In this case we write $t \overset{1}{\Rightarrow}_{A,B} t'$. The relation **weakly rewrites under** $A$ **modulo** $B$ is the reflexive, transitive closure of $\overset{1}{\Rightarrow}_{A,B}$, denoted $\overset{*}{\Rightarrow}_{A,B}$.  □

As usual, $\Rightarrow_{A,B}$ gives an abstract rewriting system, so we automatically get the appropriate notions of termination, Church-Rosser, canonical, and local Church-Rosser for weak term rewriting modulo $B$, in both the general and the ground cases; and of course we also get the usual collection of results by specializing the general results about abstract rewrite systems.

It is clear that weak rewriting modulo $B$ implies rewriting modulo $B$, i.e., $\Rightarrow_{A,B} \subseteq \Rightarrow_{A/B}$. But the following example shows that weak rewriting modulo $B$ is strictly weaker than rewriting modulo $B$, so that its reflexive, transitive, symmetric closure cannot be complete.

**Example 7.3.26**  Let $\Sigma$ have one sort with a binary operation $+$ and constants $0, a, b$, let $A$ contain the left zero law, $0 + X = X$, and let $B$ contain the associative law. Then $(a + 0) + b \Rightarrow_{A/B} a + b$ because $(a + 0) + b \simeq_B a + (0 + b)$ and $a + (0 + b) \to_A a + b$, but $(a + 0) + b$ is a normal form for $\Rightarrow_{A,B}$. Therefore $\Rightarrow_{A,B}$ really is weaker than $\Rightarrow_{A/B}$.  □

Despite this incompleteness, many $B$ have "completion procedures," which given a set $A$ of rewrite rules, produce another set $A'$ such that rewriting with $\overset{*}{\Rightarrow}_{A/B}$ and with $\overset{*}{\Rightarrow}_{A',B}$ always yield $B$-equivalent terms. In fact this is how OBJ3 actually implements rewriting modulo some equations [90, 113]. This allows users to think of computation as being done with $\overset{*}{\Rightarrow}_{A/B}$ even though it is really done with $\overset{*}{\Rightarrow}_{A',B}$; in OBJ3, the new rules generated by completion can be seen with the "`show all rules .`" command. Completion will be discussed in Chapter 12. Hereafter, we study $\Rightarrow_{A/B}$ since it describes what OBJ3 does, though not how.

## 7.4  Verification of Hardware Circuits

Hardware verification is a natural application for equational logic, because both circuits and their behaviors are described by sets of equations in a very natural way; moreover equational logic is simple and well understood, with efficient algorithms for many relevant decision problems. We first treat so called **combinatorial** (or **combinational**) circuits, which have no loops or memory, illustrated by combinations of "logic gates" and also by simple CMOS transistor circuits. Bidirectional logic circuits are then treated in Section 7.4.5; these may have loops and memory. Chapter 9 extends equational logic with second-order universal quantification, enabling us to verify so-called sequential circuits, which have time-dependent behavior. Much more could be said about solving the equations that arise from hardware circuits, but our intention here is not to develop a complete theory, but rather to provide a collection of enticing applications for term rewriting modulo equations.

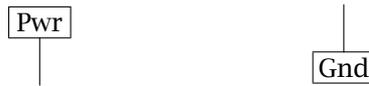

Figure 7.1: Power and Ground

Our circuit models involve two voltage levels, called "power" and "ground," diagramed as shown in Figure 7.1, and identified with `true` and `false`, respectively. Wires in circuits are assumed to have either the value power or else the value ground, and are modeled by Boolean variables. Wires that are directly connected must share the same voltage level, and are therefore represented by the same variable. These assumptions allow us to use the term rewriting decision procedure for the propositional calculus in our proofs.[3]

## 7.4.1 Unconditional Combinatorial Circuits

The simplest kind of combinatorial circuit features a direct "flow" from some given input wires, through some logic gates, to some output wires. The gates are modeled by the corresponding Boolean functions, and in this situation, all the computation can be done by the propositional calculus decision procedure, as illustrated in the following:

**Example 7.4.1** (*1-bit Full Adder*) Figure 7.2 is a circuit diagram for the usual 1-bit full adder, which produces sum and carry output bits from three inputs, two of which are bits of the numbers to be added, with the third a carry bit from a previous half adder. The "rounded boxes" are gates that compute various logical functions. A flat input side indicates conjunction, while a concave input side indicates disjunction, and a doubled concave input side indicates XOR (exclusive or). The equations in the OBJ module FADD below say the same thing as this diagram, using constants for the values of wires: i1, i2, cin are inputs, and cout, sout are the outputs. To verify that this circuit has the logical behavior of a full adder, we must prove the following first-order formula (Section 7.4.3 explains why we prove this particular formula):

$$(\forall Z)\,(T \Rightarrow e_1 \wedge e_2)\,,$$

where $T$ specifies the circuit, $Z$ consists of its variables, and $e_1, e_2$ are the two equations

```
cout = (i1 and i2) or (i1 and cin) or (i2 and cin)

sout = (i1 and i2 and cin) or (i1 and not i2 and not cin) or
       (not i1 and i2 and not cin) or (not i1 and not i2 and cin)
```

with the equations in $T$ as in the module FADD below.

---

[3]Although this approach ignores issues such as load (i.e., current flow), resistance, timing, and capacitance, it does fully capture the logical aspects of circuits. Moreover, it seems likely that many other issues can be handled by using larger sets of values on wires, for example, in the style of Winskel [183], and that these larger value sets can also be implemented with term rewriting.

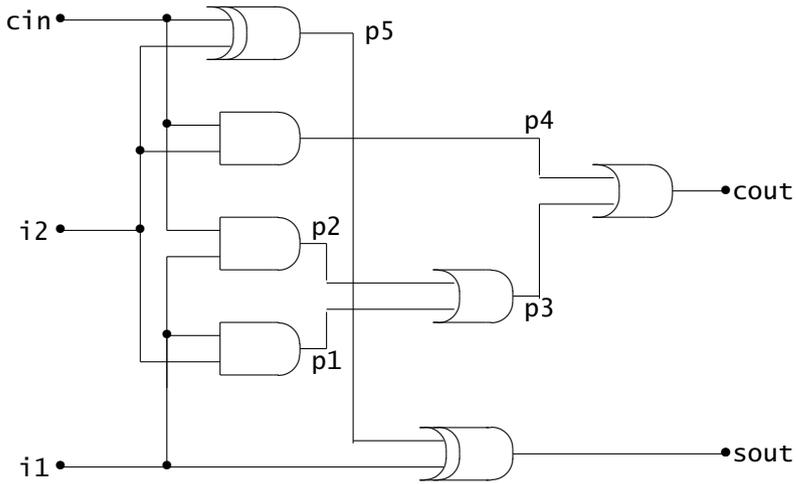

Figure 7.2: 1-bit Full Adder

The following OBJ proof score for this verification first introduces constants for the variables representing the wires, then gives the equations that describe the circuit, and finally checks whether the output variables satisfy their specifications by using reduction over the PROPC Boolean decision procedure. Proposition 7.4.3 below and familiar results on first-order logic (fully explained in Chapter 8) justify that this score actually proves the desired formula.

```
th FADD is extending PROPC .
  ops i1 i2 cin p1 p2 p3 p4 p5 cout sout : -> Bool .
  eq p1 = i1 and i2 .
```

```
      eq p2 = i1 and cin .
      eq p3 = p1 or p2 .
      eq p4 = cin and i2 .
      eq p5 = cin xor i2 .
      eq cout = p3 or p4 .
      eq sout = i1 xor p5 .
    endth
    reduce cout iff (i1 and i2) or (i1 and cin) or (i2 and cin) .
    reduce sout iff (i1 and i2 and cin) or
                    (i1 and not i2 and not cin) or
                    (not i1 and i2 and not cin) or
                    (not i1 and not i2 and cin).
```

No manual application of rules is needed, since OBJ does all the work. By contrast, [25] gives a six step, one and a half page outline of a proof for just the sout formula. □

The equations for this circuit have a very special form, which is described in the following, as a first step towards an algebraic theory of hardware circuits:

**Definition 7.4.2** A set $T$ of $\Sigma(Z)$-equations is an **unconditional triangular propositional system** iff $\Sigma$ is the signature of PROPC (the propositional calculus specification[4] given in Section 7.3.2), $Z$ is a finite set of constants of sort Bool called **variables**, there is a subset $X$ of $Z$, say $x_1,\ldots,x_n$, called the **input variables**, and there is an ordering of the rest of $Z$, say $y_1,\ldots,y_m$, called the **dependent variables**, such that $T$ consists of equations having the form $y_k = t_k$ for $k = 1,\ldots,m$, where each $t_k$ is a $\Sigma(Z)$-term involving only input variables and those non-input variables $y_j$ with $j < k$; there must be exactly one equation for each $k$. In addition, some of the non-input variables may be designated as **output variables**, with the rest being called **internal** (or "test point") **variables**. □

Hereafter we may omit "unconditional" and we may also use the phrases "combinatorial system" and "triangular propositional system" interchangeably, often omitting the word "propositional." The following display demonstrates why we chose the term "triangular":

$$y_1 = t_1(x_1,\ldots,x_n)$$
$$\ldots\ldots$$
$$y_k = t_k(x_1,\ldots,x_n,y_1,\ldots,y_{k-1})$$
$$\ldots\ldots\ldots$$
$$y_m = t_m(x_1,\ldots,x_n,y_1,\ldots,y_{k-1},\ldots,y_{m-1})$$

Note that each equation has sort Bool since that is the only sort in $\Sigma(Z)$, and that $t_1$ can only contain input variables. Also, the equations in a triangular system are usually considered (implicitly) quantified by ($\forall \emptyset$), which means that, although we call them variables, all the $x_i$ and $y_i$ are technically constants; however, they will sometimes be universally quantified in formulae that describe the intended behavior of hardware circuits. In particular, we are interested in solving the equations in the triangular system over the initial model of PROPC, so that the variables in the triangular system are constrained to be either `true` or `false`, i.e., power or ground; we will see that these solutions correspond to certain $\Sigma(X)$-models of the system. Here PROPC has initial semantics, while triangular systems over it have loose semantics.

---

[4]In practice we may let $\Sigma$ and PROPC contain some propositional functions not in the original version of PROPC that could have been defined in it, such as *p nor q = not(p or q)*.

Triangular systems are not in general term rewriting systems over $\Sigma$, because the rightsides in general contain variables that do not occur in the leftsides. However, they are MTRS's over $\Sigma(Z)$, since for this signature the "variables" are really constants. Example 5.8.27 showed that unconditional triangular systems (in the sense of that example) are canonical as TRS's, and that the only variables in their normal forms are input variables. We want similar results for triangular systems over PROPC. We will generalize techniques from Chapter 5 to rewriting modulo $B$ to show that enriching PROPC with a triangular system $T$ again yields a canonical system, modulo the same $B$ used for PROPC (Theorem 7.7.24). We use this in the following:

**Proposition 7.4.3** Given an unconditional triangular system[E29] $T$ with variables $Z$, let $B$ be the associative and commutative laws for `and` and `xor`, $P$ the equations of PROPC except $B$, and $A = T \cup P$. Then the following are equivalent, for $t, t'$ any $\Sigma(Z)$-terms:

1. PROPC $\vDash_\Sigma (\forall Z)\ (T \Rightarrow t = t')$ ;

2. $(A \cup B) \vDash_{\Sigma(Z)} (\forall \emptyset)\ t = t'$ ;

3. $(t \text{ iff } t') \stackrel{*}{\Rightarrow}_{A/B} \text{true}$ ;

4. $[\![t]\!]_A \simeq_B [\![t']\!]_A$ ;

5. $t \downarrow_{A/B} t'$ ;

6. $(t == t') \stackrel{*}{\Rightarrow}_{A/B} \text{true}$ .

**Proof:** We omit subscripts $B$ from $\simeq_B$, $=_B$, $\Rightarrow_{A/B}$, $\Rightarrow_{T/B}$ and $\Rightarrow_{P/B}$. Conditions 1. and 2. are equivalent by rules of first-order logic.[5] Conditions 2. and 4. are equivalent by Theorem 7.3.10. Conditions 3. and 4. are equivalent because $([\![t]\!]_T \text{ iff } [\![t']\!]_T) \stackrel{*}{\Rightarrow}_P \text{true}$ iff $[\![[\![t]\!]_T]\!]_P \simeq [\![[\![t']\!]_T]\!]_P$ by Exercise 7.3.8, and $[\![[\![t]\!]_T]\!]_P \simeq [\![t]\!]_A$ because both $T$ modulo $B$ and $A$ modulo $B$ are canonical, by Theorem 7.7.24. Finally, 4. and 5. are equivalent by Corollary 5.7.7, and 4. and 6. are equivalent by the definition of ==. □

---

[5]Chapter 8 gives full details, including the first-order version of the Theorem of Constants, which gives us that $P \vDash_{\Sigma(Z)} (\forall \emptyset)\ (T \Rightarrow t = t')$, and an implication elimination rule which moves $T$ over to conjoin with $P$.

This result justifies proving universally quantified implications for triangular systems as in Example 7.4.1, by reducing $t$ iff $t'$ to true, to conclude that $(\forall Z)\ (T \Rightarrow t = t')$. We call this the "method of reduction." Note that the variables in $T, t, t'$ are constants for reduction, while those in PROPC are not.

**Fact 7.4.4** The canonical forms of an unconditional triangular propositional system contain only input variables.

**Proof:** We prove the contrapositive: If a term $t$ contains an occurrence of a non-input variable $y_k$, then $t$ is not reduced, because the rewrite rule $y_k = t_k$ can be applied to it. □

Proof scores like that in Example 7.4.1 can be generated completely automatically from the circuit diagram and the sentences to be proved, because there is an exact correspondence between circuit diagrams like that in Figure 7.2 and triangular propositional systems. Although it would be too tedious to spell out this correspondence in detail here, we note that if a circuit cannot be put in triangular form, then either it is not combinatorial because it has some loops, or else it has internal variables that should instead have been declared as input variables, or vice versa.

**Exercise 7.4.1** Design a circuit with inputs $a_1, a_2, a_3$, and with one output $z$ which is true iff exactly two of the inputs are true; you may use any (2 input) logic gates you like. Prove that your design is correct using the method of reduction and OBJ3. □

**Exercise 7.4.2** Design a circuit using only NOT, NAND and OR gates having inputs $a_1, a_2, a_3, a_4$, and an output $z$ which is true iff exactly two of the inputs are true. Use as few gates as you can (there is a solution with just 19). Prove the correctness of your design using OBJ3. □

**Definition 7.4.5** A **Boolean** (or **ground**) **solution** to a triangular system is an assignment of Boolean values to all its variables such that all its equations are satisfied. A system is **consistent** iff each assignment of Boolean values to input variables extends to a Boolean solution. A system is **underdetermined** iff it is consistent and some assignment of Boolean values to input variables extends to more than one Boolean solution. Two systems are **Boolean equivalent** iff they have exactly the same Boolean solutions. A **model** of a triangular system is a $\Sigma(Z)$-model that satisfies PROPC and the equations of the system (considered quantified by $(\forall\varnothing)$), and a **Boolean** (or **protected**) **model** is a model having the set $\{true, false\}$ as its carrier. □

We are interested in Boolean solutions because these correspond to possible behaviors of the circuit. They are bijective with Boolean models, by letting $a : Z \to \{true, false\}$ correspond to the model that interprets each $z \in Z$ as $a(z)$; we might say that Boolean models satisfy the so-called Law of the Excluded Middle, in that the only values allowed are `true` and `false`, with everything else excluded. It follows that two systems are Boolean equivalent iff they have exactly the same Boolean models.

Underdetermination is similar to the situation for a system of linear equations where there are more variables than (independent) equations.[6] The next subsection will show that transistors are consistent and underdetermined; this is possible because these are conditional rather than unconditional systems. A circuit consisting of an inverter (i.e., negation) with its output connected to its input is inconsistent, because its equation

```
p = not p
```

has no solutions; but no system containing such an equation can be triangular. It is also possible for an unwise choice of input variables to produce inconsistency. For example, a system that contains the equation

```
i1 = not i2
```

is unsolvable if both `i1` and `i2` are input variables. To avoid this, one of these two variables could instead be declared internal.

**Proposition 7.4.6** Every unconditional triangular propositional system is consistent, and no unconditional triangular propositional system is underdetermined.

**Proof:** The proof is by induction. Let $a$ be a Boolean assignment to the input variables. Then by the first equation, $t_1(a(x_1), \ldots, a(x_n))$ gives a value for $y_1$, let's denote it $a(y_1)$. Similarly, $t_2(a(x_1), \ldots, a(x_n), a(y_1))$ gives a value for $y_2$, denoted $a(y_2)$. And so on, until we get a value

$$a(y_m) = t_m(a(x_1), \ldots, a(x_n), a(y_1), \ldots, a(y_{m-1}))$$

for $y_m$. This assignment $a$ on $Z$ is a solution by construction, and since its values are computed directly from the equations, it is the only possible solution extending the original assignment. □

---

[6]This is more than an analogy, because systems of linear equations over the field $\mathbb{Z}_2$ (the two-element field of the integers modulo 2), with 0 representing `false` and 1 representing `true`, describe certain kinds of circuit. Our systems are more general since their equations may be non-linear (they are multi-linear) and/or conditional.

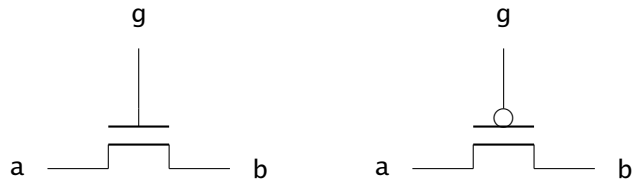

Figure 7.3: The Two Kinds of MOS Transistor: n on Left, p on Right

### 7.4.2 Conditional Triangular Systems

This section considers conditional triangular propositional systems and their solutions. A key difference between conditional and unconditional systems is that a conditional system may fail to determine the values of some of its variables under some conditions. The most basic (and most important) example of such a system is a transistor.

**Example 7.4.7** The two main types of transistor are the *n-transistor* and the *p-transistor*, diagramed as shown in Figure 7.3, and having logical behavior given by conditional equations of respective forms

$$a = b \text{ if } g = \text{true}$$
$$a = b \text{ if } g = \text{false}$$

where the variables $a, b, g$ are Boolean, with $a, g$ as inputs and $b$ as output. The system consisting of a single n-transistor is consistent and underdetermined, because if $g = \text{false}$, then according to its equation (the first above), $b$ can have any value, no matter what value $a$ has; the same holds for p-transistors with $g = \text{true}$. (These single equation systems are clearly triangular.) □

More complex systems can be built by putting several transistors together, as illustrated in examples in Section 7.4.4 and thereafter. We now generalize Definition 7.4.2 to conditional equations, and then develop some theory for such systems.

**Definition 7.4.8** A system $T$ of (possibly conditional) $\Sigma(Z)$-equations is a (**conditional**) **triangular** (**propositional**) **system** iff $\Sigma$ is the signature of PROPC (the propositional calculus specification of Section 7.3.2), $Z$ is a finite set of constants of sort Bool called **variables**, there is a subset $X$ of $Z$, say $x_1, \ldots, x_n$, called the **input variables**, and there is an ordering of the rest of $Z$, say $y_1, \ldots, y_m$, called the **dependent variables**, such that $T$ consists of equations of the form $y_k = t \text{ if } C$, where each $t$ is a $\Sigma(Z)$-term involving only input variables and variables $y_j$ with $j < k$, where each $C$ is a finite set of pairs of such terms, and where there may be any number (including zero) of equations for each $k$. In addition, some of the dependent variables may be designated as **output** **variables**, with the remaining dependent variables being called **internal** (or "test point") variables.

If the rightside of each pair in each $C$ is true, then the system has **Boolean conditions**. A triangular system $T$ has **consistent conditions** iff for each variable $y_k$ and for each assignment of Boolean values to input variables, if more than one condition of an equation with leftside $y_k$ is true, then the corresponding rightsides are not provably equal (using PROPC and all previous equations in $T$) to different Boolean values. A triangular system has **disjoint conditions** iff whenever $C, C'$ are the conditions of distinct equations with the same dependent variable as leftside, then $C \wedge C'$ is provably false for each assignment of Boolean values to input variables. A triangular system is **total** iff every dependent variable has equations, and for each $k$ and each choice of Boolean values for its input variables, the disjunction of the conditions in its equations with leftside $y_k$, say $C_k = C_{k,1} \vee C_{k,2} \vee \cdots \vee C_{k,\ell}$, is provably true, where each $C_{k,i}$ is considered the conjunction of its pairs as equations; otherwise the system is called **partial**. □

When a conditional triangular system has all $C_k = \emptyset$, it is equivalent to an unconditional system. We may assume that triangular systems have Boolean conditions, since this is convenient and entails no loss of generality. As with unconditional triangular systems, the equations are usually considered to be quantified by $(\forall \emptyset)$, with all variables considered as constants, although they may sometimes appear universally quantified, e.g., in formulae that describe intended circuit behavior. As in the unconditional case, conditional triangular systems are not in general rewriting systems over $\Sigma$, but are over $\Sigma(Z)$. However, unlike the unconditional case, conditional triangular systems are not always Church-Rosser, which motivates the next result. Since the concepts in Definition 7.4.5, including solution of a system, consistent system, and underdetermined system, carry over completely unchanged to conditional triangular systems, we do not repeat this material here.

**Proposition 7.4.9** A conditional triangular system is consistent if its conditions are consistent. Moreover, a conditional triangular system with disjoint conditions has consistent conditions, and also is Church-Rosser.

**Proof:** Let $a$ be an assignment to the input variables. Then if the condition of the first equation is true for that assignment, its rightside gives a value $t_1(a(x_1), \ldots, a(x_n))$ for $y_1$, let's denote it $a(y_1)$. Otherwise, if there is a subsequent equation with leftside $y_1$ the condition of which is true, let the value of its rightside be $a(y_1)$; if there is no such equation, pick an arbitrary value for $a(y_1)$. Similarly, we get a value $a(y_2)$ for $y_2$, and so on, until we get a value $a(y_m)$ for $y_m$. The resulting assignment $a$ is by construction a solution.

The second and third assertions follow since for each assignment, each independent variable can be rewritten in at most one way. □

The next result generalizes Proposition 7.4.3 to conditional triangular systems, and applying its equivalence of 1. and 2. to conditional rewrite rules reassures us that our join semantics for conditional term rewriting modulo equations is adequate for our hardware applications (see Section 7.7 for the technical details of this semantics).

**Proposition 7.4.10** Using the notation of Proposition 7.4.3, the following are equivalent for any conditional triangular system[E30] $T$ with variables $Z$ that is Church-Rosser as a rewrite system, for $t, t'$ any $\Sigma(Z)$-terms:

1. $t \downarrow_{A/B} t'$ ;
2. $[\![t]\!]_A \simeq_B [\![t']\!]_A$ ;
3. $(t == t') \stackrel{*}{\Rightarrow}_{A/B}$ true ;
4. $(A \cup B) \vdash_{\Sigma(Z)} (\forall \emptyset)\ t = t'$ ;
5. PROPC $\models_\Sigma (\forall Z)\ (T \Rightarrow t = t')$ ;
6. $(t \text{ iff } t') \stackrel{*}{\Rightarrow}_{A/B}$ true .

**Proof:** We omit subscripts $B$ from $\simeq_B, =_B, \Rightarrow_{A/B}, \Rightarrow_{T/B}$ and $\Rightarrow_{P/B}$. $A$ is terminating by Proposition 7.7.19 and Church-Rosser by hypothesis, so it is canonical. Therefore 1. and 2. are equivalent by Corollary 5.7.7, and 2. and 3. are equivalent by definition of ==. Also, 1. implies 4. and 4. implies 2., while 4. and 5. are equivalent by the Completeness Theorem and first-order logic (see footnote 5). Finally, 2. and 6. are equivalent, since $([\![t]\!]_T \text{ iff } [\![t']\!]_T) \stackrel{*}{\Rightarrow}_P$ true iff $[\![[\![t]\!]_T]\!]_P \simeq [\![[\![t']\!]_T]\!]_P$ by Exercise 7.3.8, and $[\![[\![t]\!]_T]\!]_P \simeq [\![t]\!]_A$ because $T$ modulo $B$ and $A$ modulo $B$ are canonical. □

This result justifies proving sentences of the form $(\forall Z)\ (T \Rightarrow t = t')$ by reducing $t$ iff $t'$ to true when $T$ has conditional equations. However, for hardware verification problems, this reduction will not in general work without using case analysis on the input variables, for reasons that are discussed in Section 7.4.3.

For underdetermined systems, it is often convenient to use parameters in solutions.

**Definition 7.4.11** A **general (Boolean) solution** of a conditional triangular system $T$ with dependent variables $y_1, \ldots, y_m$ is a family $f_k(x_1, \ldots, x_n, w_1, \ldots, w_\ell)$ of terms for[7] $k = 1, \ldots, m$ such that for every assignment of Boolean values $a_1, \ldots, a_n$ to the input variables $x_1, \ldots, x_n$ and of

---

[7]The requirement that a general solution include all dependent variables, not just the output variables, is reasonable, because a circuit designer should know what all his wires are supposed to do; indeed, the redundancy involved in checking the mutual consistency of solutions for all internal variables is desirable in itself.

Boolean values $b_1, \ldots, b_\ell$ to the **parameter variables** $w_1, \ldots, w_\ell$, the values of $f_k(a_1, \ldots, a_n, b_1, \ldots, b_\ell)$ are a Boolean solution extending the original input variable assignment. A **most general (Boolean) solution** of $T$ is a general Boolean solution such that the Boolean ground solutions of $T$ are exactly its Boolean substitution instances with their corresponding input assignments. A set $F$ of equations has the **form of an unparameterized general solution** iff its equations are $y_k = f_k(x_1, \ldots, x_n)$ for $k = 1, \ldots, m$, and has the **form of a parameterized general solution** iff its equations are $y_k = f_k(x_1, \ldots, x_n, w_1, \ldots, w_\ell)$ for $k = 1, \ldots, m$. □

To check if a family of terms for dependent variables is a general solution of a system $T$, it is by definition sufficient to substitute the terms for the corresponding variables in each equation of $T$, and check if the two sides are equal for all Boolean values of the input and parameter variables. Notice that equations having the form of general solutions are in particular unconditional triangular systems.

**Example 7.4.12** Consider the following conditional triangular system $T$,

$y_1 = x_1$ or $x_2$
$y_2 =$ not $y_1$ if not $x_1$

which is underdetermined, since $y_2$ can have any value when $x_1$ is true. The following is a proposed general solution $F$ for $T$,

$y_1 = x_1$ or $x_2$
$y_2 = ($ not $x_1$ and not $x_2)$ or $(w$ and $x_1)$

where $w$ is a parameter variable. We can check that $F$ is a most general solution of $T$ by enumerating the solutions of $T$ and then checking that they all are substitution instances of $F$. Using the format $(x_1, x_2, y_1, y_2)$, and representing true by 1 and false by 0, the solutions of $T$ are $(0, 0, 0, 1)$, $(0, 1, 1, 0)$, $(1, 0, 1, \star)$, and $(1, 1, 1, \star)$, where $\star$ can be either 0 or 1. The reader may now verify that exactly the same set of six Boolean 4-tuples arises from $F$. □

**Proposition 7.4.13** Every unconditional triangular system has a most general solution, obtained by progressively substituting its equations into later equations; this solution has no parameters, and is unique in the sense that the corresponding terms for each $y_i$ are equal under PROPC. Such systems have exactly $2^n$ Boolean solutions, where $n$ is the number of input variables.

**Proof:** The construction is like that of Proposition 7.4.9, except that we do the substitutions with the terms in the triangular system, instead of with Boolean values. First, let $f_1$ be $t_1$. Next, substitute $f_1$ for each instance of $y_1$ in $t_2$ and call the result $f_2$, noting that both $f_1$ and $f_2$ contain no non-input variables. Then, in $t_3$, substitute $f_1$ for each instance of $y_1$ and $f_2$ for each instance of $y_2$ and call the result $f_3$, noting that it too contains no non-input variables. Continuing in this way, after the appropriate substitutions $t_k$ becomes $f_k$, which by induction also contains no non-input variables. Since this construction involves only equational reasoning, the result is sound, and because each $f_k$ involves only input variables, the result is indeed a solution with no parameters.

Moreover, this process is reversible, i.e., we can also derive the original triangular system from this solution. Therefore the two sets of equations are equivalent as theories. Because equivalent theories have exactly the same models, they also have exactly the same Boolean models, and hence exactly the same Boolean solutions. For uniqueness, Corollary 7.3.19 implies that two $\Sigma(X)$-terms are equivalent if they are equal for all Boolean values of the input variables.

Finally, the form of a general solution ensures that it produces exactly one Boolean solution for each Boolean assignment to the input variables, and since there are $2^n$ such distinct assignments, that is also the number of Boolean solutions. □

### 7.4.3 Proof Techniques

Throughout this subsection, $\Sigma$ denotes the signature of PROPC, and all terms are over $\Sigma(Z)$ for some variable symbols $Z$ containing input variables $X = \{x_1, \ldots, x_n\}$ and parameter variables $W = \{w_1, \ldots, w_\ell\}$ (if any). If $T$ is a finite set of equations (perhaps conditional), let $\overline{T}$ denote the conjunction of the equations in $T$ without quantification, and with conditional equations represented as implications. Note that an assignment $a : Z \to M$ to a $\Sigma(Z)$-model $M$ is a solution of $T$ iff $\overline{a}(\overline{T})$ is true in $M$, where $\overline{a}(\overline{T})$ denotes the truth value of $\overline{T}$ in $M$ under $a$.

**Proposition 7.4.14** If $T$ is a (conditional) triangular system, then a set $F$ of equations having the form of an unparameterized general solution is a most general solution for $T$ if the formula $(\forall Z)(\overline{F} \Leftrightarrow \overline{T})$ can be proved assuming PROPC.

**Proof:** The formula says that the two sets of equations are equivalent as theories extending PROPC, which implies that they have exactly the same models, and therefore in particular, have exactly the same Boolean models, and hence exactly the same Boolean solutions. □

In examples, this formula can be proved by checking it for all possible Boolean values of the input and parameter variables, since values of the other variables are determined by these using reduction, due to the forms of $T$ and $F$. There are $2^{n+\ell}$ cases to check, which is manageable in comparison with $2^{n+\ell+m}$, which could be much larger for a complex circuit.

The formula $(\forall Z)(\overline{F} \Rightarrow \overline{T})$ says that if the $y_k$ are defined according to $F$, then they satisfy $T$. Its converse, $(\forall Z)(\overline{T} \Rightarrow \overline{F})$, says that if some assignment of values to $Z$ satisfies $T$, then it also satisfies $F$; this is the "most general" part of "most general solution." Together these give the equivalence of the two theories.[8] Note that it can be satisfied without $F$ actually being a solution, for example, if $T$ is inconsistent (i.e., has no solutions), then the converse formula is valid for any $F$ whatsoever. Nevertheless, Proposition 7.4.16 shows that under some mild assumptions, it suffices to prove just the "most general" part of the equivalence; this explains why we only proved that direction in Example 7.4.1, and why we do the same in several examples below. Recall that $\mathbb{B} = \{\texttt{true}, \texttt{false}\}$.

**Lemma 7.4.15**  If $T$ is a total triangular system with consistent conditions, then every Boolean assignment $i : X \to \mathbb{B}$ to input variables extends to a unique solution $i^* : Z \to \mathbb{B}$ for $T$.

**Proof:**  Since $T$ is total and has consistent conditions, the conditions of at least one equation with leftside $y_1$ evaluate to true, and all the rightsides of such equations evaluate to the same value. Therefore the value of $y_1$ is uniquely determined by the values of $i$. Similarly, the value of $y_2$ is uniquely determined by the values of $i$ and $y_1$, and so on by induction, so that all the dependent variables in a solution are uniquely determined.  □

**Proposition 7.4.16**  If $T$ is a total triangular system with consistent conditions and if $F$ has the form of an unparameterized general solution, then $(\forall Z)(\overline{T} \Rightarrow \overline{F})$ implies $(\forall Z)(\overline{T} \Leftrightarrow \overline{F})$, and hence implies that $F$ is a most general solution.

**Proof:**  Given a Boolean assignment $i : X \to \mathbb{B}$ for input variables, there are unique extensions $i_T^*$ and $i_F^*$ that are solutions to $T$ and $F$, by Lemma 7.4.15, noting that $F$ is also total with consistent conditions. The formula in the hypothesis implies that $i_T^* = i_F^*$ for any $i : X \to \mathbb{B}$, so we can write just $i^*$. Uniqueness implies that $a : Z \to \mathbb{B}$ is a solution to $T$ iff $a = i^*$ when the restriction of $a$ to $X$ is $i$; the same holds for $F$. Hence, $a : Z \to \mathbb{B}$ is a solution to $T$ iff it is a solution to $F$, from which $(\forall Z)(\overline{F} \Leftrightarrow \overline{T})$ follows by Corollary 7.3.19.  □

In practice, it suffices to prove $(\forall X)(\overline{T} \Rightarrow \overline{F})$, regarding the input variables as quantified and the dependent variables as constants, whose values are determined by $X$; this is considerably easier by case analysis, since there are considerably fewer cases than required by $Z$.

The situation is more complex for parameterized solutions, where for $Z' = Z \cup W$, with $W$ the parameter variables, the relevant formula is $(\forall Z)(((\exists W)\overline{F}) \Leftrightarrow \overline{T})$, for which it often suffices to prove just $(\forall Z)((\overline{T} \Rightarrow (\exists W)\overline{F}))$, noting that $(\forall Z)(((\exists W)\overline{F}) \Rightarrow \overline{T})$ is equivalent to $(\forall Z')(\overline{F} \Rightarrow \overline{T})$. We omit details, which are similar to those for Proposition 7.4.16.

---

[8] Using the inference rule $((\forall Z)P) \land ((\forall Z)Q) = (\forall Z)(P \land Q)$, which is discussed in Chapter 8.

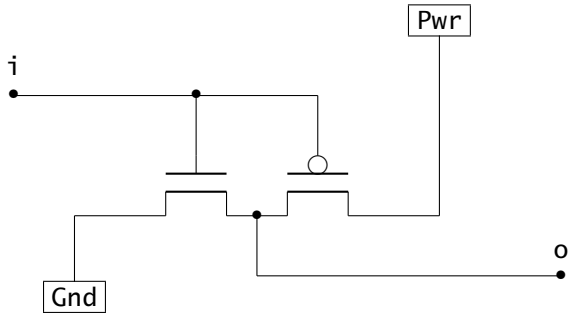

Figure 7.4: A CMOS NOT Gate

The method of reduction of Proposition 7.4.3 and its extension to conditional equations in Theorem 7.4.10, provide a way to verify universally quantified implications of the kind considered here. (But note that rewriting with conditional rules modulo equations is not treated until Section 7.7.) Although often not applicable, the method of reduction is efficient when it is applicable, and Boolean case analysis is available when it is not. Also note that while proving a conditional equation by checking that its two sides reduce to the same thing, we should assume that its condition is true when doing the reduction, as is of course familiar mathematical practice (Chapter 8 gives a formal treatment). All this together gives a powerful and flexible tool set for verifying hardware circuits.

### 7.4.4 Conditional Combinatorial Circuits

This subsection gives some examples of conditional triangular systems and their solutions, mainly so-called "CMOS" circuits, which use n- and p-transistors in balanced pairs.

**Example 7.4.17** (NOT *Gate*) We prove that the CMOS circuit shown in Figure 7.4 implements a NOT gate (also called a *negation*, or an *inverter* gate), i.e., we prove that $o = not\ i$ is a most general solution for $T$, which contains the two conditional equations of the module NOT below, describing the behavior of the two transistors:

```
th NOT is extending PROPC .
  ops i o : -> Bool .
  cq o = true  if not i .
  cq o = false if i .
endth
```

We show that negation is a most general solution by case analysis on the variable i (the validity of case analysis was shown in Section 7.4.3):

```
open NOT .  eq i = true .
  red o iff not i .
close
open NOT .  eq i = false .
  red o iff not i .
close
```

Since both reductions give true and the circuit is easily seen to be total and consistent, the proof is done by Proposition 7.4.16. Although it is now unnecessary, we can also show directly that negation is a solution, by assuming it as an equation and then checking that the two conditional equations of the circuit hold:

```
th BEH is ex PROPC .
  ops i o : -> Bool .
  eq o = not i .
endth
open .  eq not i = true .
  red o == true .
close
open .  eq i = true .
  red o == false .
close                                                                  □
```

**Example 7.4.18** (XOR *Gate*) We show that the six-transistor circuit of Figure 7.5 realizes the exclusive or (XOR) function, i.e., that $(\forall Z)\ (T \Leftrightarrow o = i1\ xor\ i2)$, where $T$ consists of the equations in the module XOR below, describing the behavior of the circuit, and where $Z$ contains the four variables i1, i2, p1, o. The variables i1 and i2 are inputs, while p1 is internal and o is the output. To model this circuit, we write one conditional equation for each transistor:

```
th XOR is extending PROPC .
  ops i1 i2 p1 o : -> Bool .
  cq p1 = false if i1 .
  cq p1 = true  if not i1 .
  cq o = i1 if not i2 .
  cq o = p1 if i2 .
  cq o = i2 if not i1 .
  cq o = i2 if p1 .
endth
```

Example 7.4.17 shows that the internal variable p1 is the negation of i1, but we nevertheless prove this again in the present context. Because this circuit is total, Proposition 7.4.16 implies that it suffices to show that xor is most general:

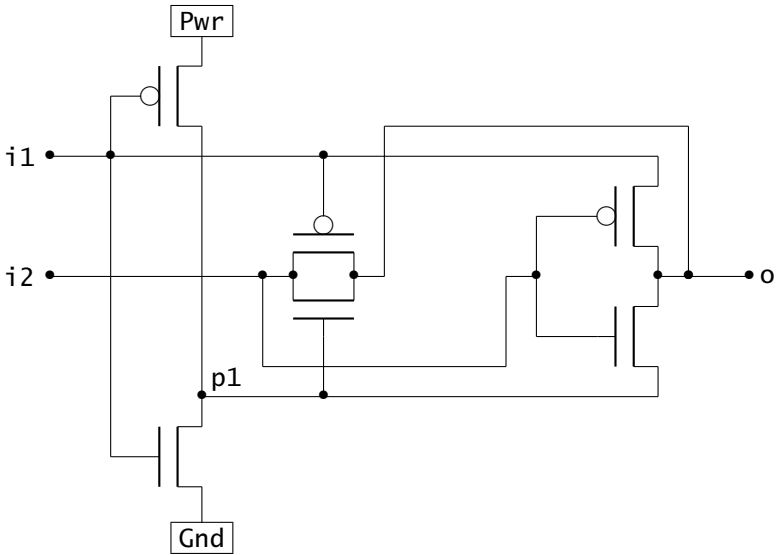

Figure 7.5: A CMOS XOR Gate

```
open XOR .  eq i1 = false .  eq i2 = true .
  red p1 iff not i1 .
  red o  iff i1 xor i2 .
close
open XOR .  eq i1 = true .  eq i2 = false .
  red p1 iff not i1 .
  red o  iff i1 xor i2 .
close
open XOR .  eq i1 = false .  eq i2 = true .
  red p1 iff not i1 .
  red o  iff i1 xor i2 .
close
open XOR .  eq i1 = false .  eq i2 = false .
  red p1 iff not i1 .
  red o  iff i1 xor i2 .
close
```

Since all reductions give `true`, the proof is done.   □

We can summarize our method for verifying a proposed unparameterized most general solution for a total consistent conditional system as follows: use case analysis on the input variables and then reduction to prove that the circuit equations imply the solution equations. This does not require human intervention: the OBJ proof score can be generated automatically from a circuit diagram and a proposed solution. Moreover, it is a kind of decision procedure under the conditions of Proposition 7.4.16, because an unparameterized system fails to be a most general solution for a circuit iff reduction fails to be true for some case of some equation (but the system could still be a solution, even though not most general). In some cases, this method can be more efficient than more traditional approaches, and it works for combinatorial circuits built from any (combinatorial) components, including e.g., both transistors and logic gates.

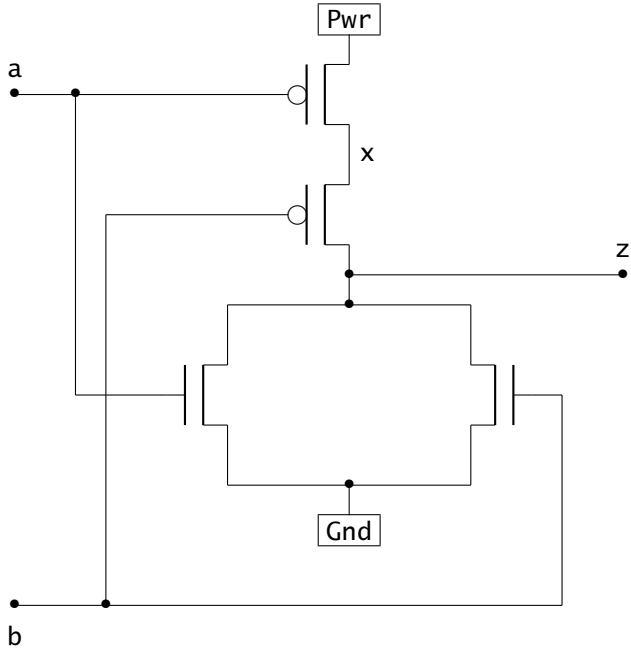

Figure 7.6: A CMOS NOR Gate

**Exercise 7.4.3** Use OBJ to prove correctness of the NOR circuit shown in Figure 7.6 (where *x nor y* = *not*(*x or y*)). □

The method for checking solutions described above extends to partial solutions, where the desired behavior of a circuit is not determined for some input conditions. This can arise in specifications where it is known that certain inputs will never occur, so it doesn't matter what the circuit does in these cases. Partial specifications are preferable to choosing arbitrary values for these inputs, since this allows engineers more freedom to produce better designs.

**Definition 7.4.19** A **partial solution** is a family of conditional equations, each of the form

$$y_k = f_k(x_1,\ldots,x_n) \text{ if } c_k(x_1,\ldots,x_n) ,$$

where $y_k$ is a non-input variable, where $x_1, \ldots, x_n$ are the input variables, and where $f_k$ and $c_k$ are propositional terms. The cases where the predicates $c_k(x_1, \ldots, x_n)$ are not `true` correspond to what are often called **don't care conditions**.  □

Although we do not give all the details here, when verifying a partial solution, it is necessary to determine when the condition of a conditional equation is `true` on Boolean models: first, substitute the expressions of the proposed solution into the condition; then compute the complete disjunctive normal form (see Proposition 7.3.24) of the result; next, consider each disjunct as a "case," in which the input variables that occur positively (i.e., those that are not negated) must be `true`, while those that occur negatively must be `false`; and finally, check that for each case the two sides evaluate to the same reduced form, using the values for the variables that belong to that case. For example, the condition of the sixth equation in Example 7.4.18 is `p1`; substituting the equation for `p1` yields `not i1`, which is already in disjunctive normal form, with only one disjunct and hence only one case, in which `i1 = false`; the last `open` above sets up this assumption before evaluating the equation. More complex conditions can easily arise, and can be handled the same way. It is also valid to use exclusive normal form in the same manner.

The problem of verifying a proposed solution is inherently simpler than the problem of finding a solution, because we can look at the latter as having the form

$(\forall X)(\exists N)$ PROPC $\models_\Sigma T$ ,

where $T$ is a propositional system and $X, N$ are its input and non-input variables, respectively. Although standard solution methods like Gaussian elimination require linearity for their completeness, essentially the same method of successive substitution and simplification can often be used to find solutions for general propositional systems; sometimes even most general solutions can be found this way. Note that the above formula only concerns Boolean solutions; a formula defining general solutions would require second-order variables.

**Exercise 7.4.4** Design and verify a circuit that implements the conditional equation $a = b$ `if` $c = d$. (Note that this is a partial specification.)  □

## 7.4.5 Bidirectional Circuits

Recall that bidirectional circuits may involve feedback loops, in the sense that they are not triangular. It may not be obvious that a function-based formalism can deal with bidirectional logic circuits, due to the input/output character of functions. But equational logic is based on the *relation* of equality, which is symmetric, and conditional equations provide additional expressive power. Section 7.4.4 showed that this framework is sufficient for MOS transistor circuits. We now generalize beyond the simple input-output flow-oriented structure of triangular systems, to circuits described by arbitrary conditional propositional systems, but with a designated subset of "external" variables, instead of designated input and output variables as in Definition 7.4.8. The transistor and the cell (Example 7.4.21 below) are examples.

**Definition 7.4.20** A **propositional system** (or **propositional theory**) over a variable set $Z$ is a finite set of (possibly conditional) $\Sigma(Z)$-equations, where $\Sigma$ is the signature of PROPC, with a designated subset of $Z$ called the **external variables**, some of which may be **input** variables, and with the remainder called the **internal variables**. □

The notions of **Boolean solution**, **consistency**, **underdetermination**, **Boolean equivalence**, and **Boolean model** for propositional systems are the same as in Definition 7.4.5, and the bijection between Boolean solutions and Boolean models also carries over, so that two propositional systems are Boolean equivalent iff they have the same Boolean models. Moreover, the notions of general solution and most general solution in Definition 7.4.11 also carry over.

The task of proving that some equations $E$ are satisfied if the equations $T$ describing a circuit are satisfied has the form $(\forall Z)(\overline{T} \Rightarrow \overline{E})$, and can therefore be proved by showing PROPC $\models_{\Sigma(Z)} (\overline{T} \Rightarrow \overline{E})$, which in turn can be proved by showing PROPC $\cup\, T \models_{\Sigma(Z)} E$, in which the variables in the equations of PROPC remain variables, while those in $T$ and $E$ become constants. The method of Proposition 7.4.16 is not available, because neither set of equations is in triangular form. Therefore to show that a proposed solution is most general, it is necessary both to prove that it is a solution by proving the formula $(\forall Z)(\overline{T} \Rightarrow \overline{E})$ as above, and to prove the converse formula, $(\forall Z)(\overline{E} \Rightarrow \overline{T})$; of course, we hope these can be done by reduction based on the forms PROPC $\cup\, T \models_{\Sigma(Z)} E$ and PROPC $\cup\, E \models_{\Sigma(Z)} T$, but in general some case analysis is required.

**Example 7.4.21** (*Cell*) Figure 7.7 is a CMOS circuit for a simple 1-bit memory cell. This circuit is underdetermined, and in fact is **bistable**, i.e., it has exactly two distinct Boolean solutions, which are its possible stably persistent memory states. The system describing this circuit is

```
p1 = true if not p2
p1 = false if p2
p2 = true if not p1
p2 = false if p1
```

where p1 and p2 are external variables (so there are no internal variables). This system is equivalent to

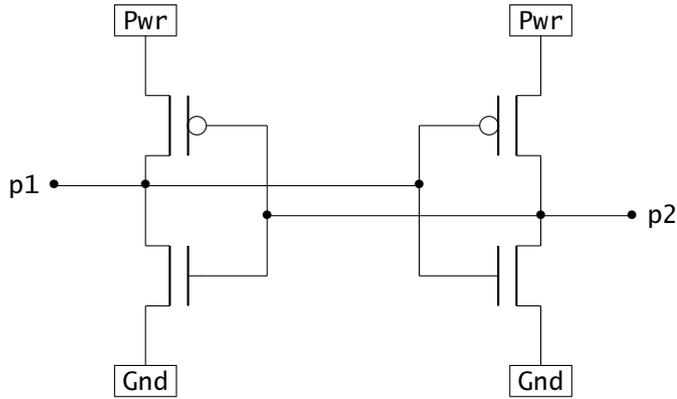

Figure 7.7: A 1-bit CMOS Cell

```
p1 = not p2
p2 = not p1
```

as well as to

```
p1 = not p2 .
```

This circuit has no designated inputs or outputs because the wires `p1` and `p2` are used bidirectionally, for both reading and writing. Because solutions are Boolean expressions in the internal variables, of which there are none, and because there are only two Boolean expressions that depend on no variables, namely `true` and `false`, it is clear that the two systems

```
p1 = true
p2 = false
```

and

```
p1 = false
p2 = true
```

are the only possible unparameterized solutions for this circuit, and it is also clear that they do indeed satisfy the equations.

We can introduce a parameter `q` for a most general parameterized solution

```
p1 = not q
p2 = q
```

and observe that it is indeed a most general solution because it has exactly the above two Boolean solutions as its Boolean instances.

We can also give a mechanical proof that the above parameterized system is indeed a solution. The OBJ3 proof score for this is a bit more subtle than previous examples because the cases when the condition of a conditional equation is true must be expressed in terms of the parameter variable q.

```
th BEH is extending PROPC .
  ops p1 p2 q : -> Bool .
  eq p1 = not q .
  eq p2 = q .
endth
*** p1 = true if not p2 .
open BEH .  eq q = false .
  reduce p1 iff true .
close
*** p1 = false if p2 .
open BEH .  eq q = true .
  reduce p1 iff false .
close
*** p2 = true if not p1 .
open BEH .  eq q = true .
  reduce p2 iff true .
close
*** p2 = false if p1 .
open BEH .  eq q = false .
  reduce p2 iff false .
close
```

Since all these reductions give true, we are done.                    □

**Exercise 7.4.5** Design and verify a CMOS circuit that can store and read two bits.                                                                          □

## 7.5  Proving Termination Modulo Equations

This section considers ways to prove termination modulo equations, generalizing results from Chapter 5 on rewriting with unconditional rules, and illustrating their use on examples from earlier in this book. We first note that any TRS result proved from an ARS result will of course generalize to MTRS's, although this doesn't get us very far with termination proofs. We first show that it is not necessary that a terminating TRS is also terminating modulo $B$.

**Example 7.5.1** Let $A$ contain the rules $a + b \to c$ and $c \to b + a$ where $a, b, c$ are constants, $+$ is binary, and there is just one sort. It is not hard to see that $A$ is terminating. Now let $B$ contain the commutative law for $+$. Then $a + b \Rightarrow^1_{A/B} a + b$ because $b + a \simeq_B a + b$. Therefore rewriting with $A$ modulo $B$ is nonterminating. □

Nevertheless, there is a simple condition that sometimes allows us to infer termination of $A$ modulo $B$ from termination of $A$; we will see later that the same condition also works for the Church-Rosser and local Church-Rosser properties.

**Definition 7.5.2** A $\Sigma$-TRS $A$ and set $B$ of $\Sigma$-equations are said to **commute** iff for any $\Sigma$-term $t_0$, whenever $t_0 \simeq_B t'_0$ and $t'_0 \Rightarrow_A t_1$ there is a $\Sigma$-term $t'_1$ such that $t_0 \Rightarrow_A t'_1$ and $t'_1 \simeq_B t_1$. □

**Lemma 7.5.3** Given a $\Sigma$-MTRS $A$ commuting with $\Sigma$-equations $B$, if $t_0 \Rightarrow_{A/B} t_1 \cdots \Rightarrow_{A/B} t_n$ then there exist $t'_1, \ldots, t'_n$ such that $t_0 \Rightarrow_A t'_1 \cdots \Rightarrow_A t'_n$ and $t'_n \simeq_B t_n$. The same result also holds for weak rewriting modulo $B$.

**Proof:** The first step, $t_0 \Rightarrow_A t'_1$ (with $t'_1 \simeq_B t_1$), is direct from commutativity; next, because $t_1 \Rightarrow_{A/B} t_2$ and $t'_1 \simeq_B t_1$, we get $t'_1 \Rightarrow_A t'_2$ with $t'_2 \simeq_B t_2$; and we can continue in this same way until we get $t'_{n-1} \Rightarrow_A t'_n$ with $t'_n \simeq_B t_n$. The result for weak rewriting modulo $B$ now follows because $\Rightarrow_{A,B}$ is a subrelation of $\Rightarrow_{A/B}$. □

**Proposition 7.5.4** Given a $\Sigma$-TRS $A$ commuting with $\Sigma$-equations $B$, if $A$ is terminating, then $A$ is also terminating modulo $B$; this also holds for weak rewriting modulo $B$.

**Proof:** We prove the contrapositive. Suppose there is an infinite sequence

$$t_0 \Rightarrow_{A/B} t_1 \cdots \Rightarrow_{A/B} t_n \Rightarrow_{A/B} \cdots.$$

Then we can construct an infinite sequence $t_0 \Rightarrow_A t'_1 \cdots \Rightarrow_A t'_n \Rightarrow_A \cdots$ by repeatedly using Lemma 7.5.3. This also holds for weak rewriting because $\Rightarrow_{A,B}$ is a subrelation of $\Rightarrow_{A/B}$. □

We now generalize some results from Chapter 5 to rewriting modulo $B$. For the results that arise from ARS's, we can simply re-apply the ARS result, without doing any new work. For example, instead of generalizing the proof of Proposition 5.5.1 on page 111, we can simply apply its ARS version, which is Proposition 5.8.16 on page 134, to obtain the following, in which we also generalize from $\omega$ to an arbitrary Noetherian poset $P$:

**Proposition 7.5.5** An MTRS $\mathcal{M} = (\Sigma, A, B)$ is ground terminating if there is a function $\rho : T_{\Sigma,B} \to P$, where $P$ is a Noetherian poset, such that for all ground $(\Sigma, B)$-terms $t, t'$, if $t \Rightarrow^1_{A/B} t'$ then $\rho(t) > \rho(t')$. Moreover, the converse holds provided $\mathcal{M}$ is globally finite. □

Note that if we want to define $\rho$ using the initiality of $T_{\Sigma,B}$, then we have to check that the $\Sigma$-structure given to $P$ actually satisfies $B$.

Of course, we also want to generalize Proposition 5.5.4, which gives a termination criterion that is easier to apply in practice, by taking account of the structure of terms. Here we cannot rely on an ARS result, but fortunately the proof generalizes from terms to $B$-classes of terms; again we generalize to an arbitrary Noetherian poset $P$:

**Proposition 7.5.6** Given an MTRS $\mathcal{M} = (\Sigma, A, B)$ and $\rho : T_{\Sigma,B} \to P$ with $P$ Noetherian, if

(1) $\rho(\theta(t)) > \rho(\theta(t'))$ for each $t \to t'$ in $A$ and applicable substitution $\theta : X \to T_{\Sigma,B}$, and

(2) $\rho(t) > \rho(t')$ implies $\rho(t_0(z \leftarrow t)) > \rho(t_0(z \leftarrow t'))$ for each $t, t' \in (T_{\Sigma,B})_s$ and any $t_0 \in T_{\Sigma,B}(\{z\}_s)$ having a single occurrence of $z$,

then $\mathcal{M}$ is ground terminating. □

Note that expressions of the form $\rho(t)$ really mean $\rho([t])$ above. The proof is the same as that of Proposition 5.5.4, with $\overset{1}{\Rightarrow}_{A/B}$ substituted for $\overset{1}{\Rightarrow}_A$. We next generalize Proposition 5.5.6 in the same way, noting that all the concepts in Definition 5.5.3 generalize from $\Sigma$-terms to $B$-classes of $\Sigma$-terms by substituting $\rho([\_])$ for $\rho(\_)$, and that the proof of Proposition 5.5.5 in Appendix B also generalizes, using induction on the $\Sigma$-structure of $\Sigma$-terms that represent $(\Sigma, B)$-terms:

**Proposition 7.5.7** Given an MTRS $\mathcal{M} = (\Sigma, A, B)$ and $\rho : T_{\Sigma,B} \to P$ with $P$ Noetherian, if

(1) each rule in $A$ is strict $\rho$-monotone, and

(2') each $\sigma \in \Sigma$ is strict $\rho$-monotone,

then $\mathcal{M}$ is ground terminating. □

Here the rules must be seen as class rewriting rules, and monotonicity must be interpreted on classes. As before, it is often easiest to define the function $\rho$ by giving its target a $(\Sigma, B)$-algebra structure and then letting initiality define $\rho$, as in the following:

**Example 7.5.8** We can use Proposition 7.5.7 with $P = \omega$ to simplify the ground termination proof of Example 5.5.7 by building in commutativity, i.e., putting it into $B$. The resulting simplified specification is as follows:

```
obj ANDCOMM is sort Bool .
  ops tt ff : -> Bool .
  op _&_ : Bool Bool -> Bool [comm] .
  var X : Bool .
  eq X & tt = X .
  eq X & ff = ff .
endo
```

Letting $\Sigma$ denote the signature of ANDCOMM, we give $\omega$ the structure of a $\Sigma$-algebra by defining $\omega_{\mathrm{tt}} = \omega_{\mathrm{ff}} = 1$ and $\omega_{\&}(m, n) = m + n$. Note that $\omega$ with this structure is a $(\Sigma, B)$-algebra (because addition is commutative), and let $\rho$ denote the resulting unique $\Sigma$-homomorphism $T_{\Sigma,B} \to \omega$. Then by Proposition 7.5.7, we only need to prove

$$\rho(x \,\&\, \mathrm{tt}) > \rho(x)$$
$$\rho(x \,\&\, \mathrm{ff}) > \rho(\mathrm{ff})$$

for all $x \in T_\Sigma$, plus (from condition $(2')$) that

$$\rho(t) > \rho(t') \quad \text{implies} \quad \rho(t_1 \,\&\, t) > \rho(t_1 \,\&\, t')$$

for all $t, t', t_1 \in T_\Sigma$. (Note again that there is an implicit [_] inside each instance of $\rho$ above.) As in Example 5.5.7, the proofs are all trivial – but there are only half as many of them. □

**Example 7.5.9** We can use Proposition 7.5.7 with $P = \omega$ to prove first ground termination and then (non-ground) termination of the basic propositional calculus specification that we have been using as a decision procedure:

```
obj BPROPC is sort Bool .
  ops true false : -> Bool .
  op _and_ : Bool Bool -> Bool [assoc comm prec 2] .
  op _xor_ : Bool Bool -> Bool [assoc comm prec 3] .
  vars P Q R : Bool .
  eq P and false = false .
  eq P and true = P .
  eq P and P = P .
  eq P xor false = P .
  eq P xor P = false .
  eq P and (Q xor R) = (P and Q) xor (P and R) .
endo
```

We let $B$ contain the associative and commutative laws for and and xor, and we define a $B$-algebra structure on $\omega$ as follows, where $p, q$ range over $\omega$:

$$\omega_{\mathrm{true}} = \omega_{\mathrm{false}} = 2$$
$$\omega_{\mathrm{and}}(p, q) = pq$$
$$\omega_{\mathrm{xor}}(p, q) = p + q + 1 .$$

Now observe that $\omega$ with this structure satisfies $B$, because product and addition are both associative and commutative, and that the resulting unique $(\Sigma, B)$-homomorphism $\rho$ satisfies

$$\rho(\mathrm{true}) = \rho(\mathrm{false}) = 2$$
$$\rho(P \text{ and } Q) = \rho(P)\rho(Q)$$
$$\rho(P \text{ xor } Q) = \rho(P) + \rho(Q) + 1 ,$$

and also (by induction) $\rho(P) > 1$ for all $P$. It is now easy to check that all rules and both operations are strict $\rho$-monotone, the least trivial being the distributive law. It then follows that BPROPC is ground terminating. If constants, such as $p, q, r$, are added to the signature, then by defining $\rho$ on them to be 2, the above results still hold, and termination again follows; but since variables are just such constants, we get termination, not just ground termination. □

**Exercise 7.5.1** Use OBJ3 to check all equalities and inequalities needed in Example 7.5.9. □

**Example 7.5.10** We can prove termination of the MTRS DNF of Example 7.3.11 in much the same way. We first consider just the operations and, or and not, with the first fourteen equations, and we let $B$ contain the associative and commutative laws for and and or. Then as in Example 7.5.9 above, we can give a $B$-algebra structure for $\omega$:

$$\begin{align}
\omega_{\text{true}} = \omega_{\text{false}} &= 3 \\
\omega_{\text{and}}(p,q) &= pq \\
\omega_{\text{or}}(p,q) &= p+q+2 \\
\omega_{\text{not}}(p) &= 2^{p+1}.
\end{align}$$

Because $\omega$ with this structure satisfies $B$ (since product and addition are both associative and commutative), the unique $(\Sigma, B)$-homomorphism $\rho$ satisfies

$$\begin{align}
\rho(\text{true}) = \rho(\text{false}) &= 3 \\
\rho(P \text{ and } Q) &= \rho(P)\rho(Q) \\
\rho(P \text{ or } Q) &= \rho(P) + \rho(Q) + 2 \\
\rho(\text{not } P) &= 2^{\rho(P)+1}.
\end{align}$$

and also (by induction) $\rho(P) > 2$ for all $P$. It is easy to check that all fourteen rewrite rules and both operations are strict $\rho$-monotone, the least trivial again being the distributive law. It follows that DNF is ground terminating, and because after adding constants, such as $p, q, r$ to the signature and defining $\rho$ on them to be 3, all of the above results still hold, we get termination.

To extend this to the full MTRS of Example 7.3.11, we define each additional operation on $\omega$ to be one more than $\rho$ of the rightside of its defining equation. For example,

$$\begin{align}
\omega_{\text{xor}}(p,q) &= p2^{q+1} + q2^{p+1} + 2 \\
\omega_{\text{implies}}(p,q) &= 2^{p+1} + q + 2.
\end{align}$$

It is easy to see that these operations and their rules are strict monotone, so the full MTRS is ground terminating by Proposition 7.5.7, and then termination follows using the same trick. □

The above gives a general method for proving ground termination of an extension of an MTRS by derived operations when the MTRS without them has already been proved terminating using Proposition 7.5.7.

**Exercise 7.5.2** Prove termination of the entire PROPC MTRS using the above method for handling derived operations. □

## 7.6 Proving Church-Rosser Modulo Equations

This section generalizes results from Chapter 5 for proving the Church-Rosser property to unconditional rewriting modulo equations. We defer some proofs and even statements of results to Section 7.7.3, which treats the more general case of conditional rewriting modulo equations. We first show that a Church-Rosser TRS is not necessarily Church-Rosser modulo equations:

**Example 7.6.1** Let $\Sigma$ be one-sorted with constants $a, b, 0$, let $A$ have rules $(a + 0) + b \to 0$ and $a + b \to a$, and let $B$ have equations $0 + X = X$ and $X + (Y + Z) = (X + Y) + Z$. Then $(a + 0) + b \overset{1}{\Rightarrow}_{A/B} 0$, and $(a + 0) + b \overset{1}{\Rightarrow}_{A/B} a + b \overset{1}{\Rightarrow}_{A/B} a$, which are both reduced. Therefore $A$ modulo $B$ is not Church-Rosser, although $A$ without $B$ is Church-Rosser. □

The same simple commutativity condition that let us infer termination modulo $B$ from termination without $B$ also works for the Church-Rosser property; the result below is a special case of Proposition 7.7.23, which is proved in Section 7.7.3.

**Proposition 7.6.2** If $A$ is a Church-Rosser (or locally Church-Rosser) $\Sigma$-TRS commuting with $\Sigma$-equations $B$, then $A$ is Church-Rosser (or locally Church-Rosser) modulo $B$. □

The following is a straightforward application of the above plus Proposition 7.5.4:

**Proposition 7.6.3** Any triangular propositional system $T$ is canonical modulo the associative and commutative laws for and and xor.

**Proof:** Associative and commutative laws commute with any rule having a constant as leftside, and Example 5.8.27 showed that any such $T$ is terminating and Church-Rosser (without $B$). □

**Exercise 7.6.1** Give a proof or counterexample for the assertion that if $A$ is Church-Rosser and commutes with $B$ then weak rewriting modulo $B$ is also Church-Rosser. □

Results in Chapter 5 for proving the Church-Rosser property that follow from ARS results immediately extend to MTRS's. One such is the Hindley-Rosen Lemma, which is stated in even greater generality in Section 7.7.3. Here we state another modulo $B$ result, which follows directly from Proposition 5.7.4:

**Proposition 7.6.4** (*Newman Lemma*) A terminating MTRS is Church-Rosser iff it is locally Church-Rosser.   □

Since Example 7.5.9 shows termination of PROPC, it would be very desirable to use Proposition 7.6.4 to prove Hsiang's Theorem (Theorem 7.3.13) by proving the local Church-Rosser property for PROPC. This provides strong motivation for generalizing the Critical Pair Theorem (Theorem 5.6.9) to MTRS's. Unfortunately, this is far from straightforward; some aspects of the problem are treated in Chapter 12. Here we content ourselves with some simple, specialized results, and with showing that the Critical Pair and Orthogonality Theorems do not generalize straightforwardly to the modulo $B$ case. First we generalize the relevant definitions:

**Definition 7.6.5** Two rules of an MTRS with leftsides $\ell, \ell'$, **overlap** iff there exist a subterm $\ell_0$ of $\ell$ not just a variable, and substitutions $\theta, \theta'$ such that $\theta(\ell_0) =_B \theta'(\ell')$. If the two rules are the same, it is required in addition that the corresponding substitution instances of the rightsides are not equivalent modulo $B$, and in this case the rule is called **self-overlapping**. An MTRS is **overlapping** iff it has two rules (possibly the same) that overlap, and then (the $B$-class of) $\theta(\ell_0)$ is called an **overlap** of the rules; otherwise the MTRS is called **non-overlapping**. A **most general overlap** $p$ of $\ell, \ell'$ at $\ell_0$ is an overlap of $\ell, \ell'$ at $\ell_0$ such that any other is equal (modulo $B$) to a substitution instance of $p$, and a **complete overlap set** for $\ell, \ell'$ at $\ell_0$ is a set of overlaps of $\ell, \ell'$ at $\ell_0$ such that any other is equal (modulo $B$) to a substitution instance of some overlap in the set.   □

Note that the subredex need not be proper in the self-overlapping case, as was required in Definition 5.6.2 for ordinary term rewriting. However, it is still true that if the leftsides $\ell, \ell'$ of two rules in $A$ overlap at $\theta(\ell_0)$, then that overlap can be rewritten in two different ways (one for each rule). The following, due to Dr. Monica Marcus, refutes the straightforward modulo $B$ generalization of the TRS orthogonality theorem that replaces all concepts by their modulo $B$ counterparts:

**Example 7.6.6** Let $\Sigma$ be one-sorted with a binary operation $+$ and constants $a, b$, let $A$ have the rules $a + b \to b$ and $b + a \to a$, and let $B$ have the associative law. Then $A$ is non-overlapping modulo $B$, but $a + b + a \stackrel{1}{\Rightarrow}_{A/B} a + a$, and $a + b + a \stackrel{1}{\Rightarrow}_{A/B} b + a \stackrel{1}{\Rightarrow}_{A/B} a$, which are both reduced modulo $B$. Therefore $A$ modulo $B$ is not Church-Rosser. (And it is not hard to see that this MTRS terminates.)   □

The following is proved in Chapter 12, recalling that the associative, commutative, and identity laws are abbreviated A, C, I, respectively:

**Proposition 12.0.2** Given an MTRS $(\Sigma, A, B)$, if $B$ consists of any combination of A, C, I laws for operations in $\Sigma$, except A, I and AI, and if the leftsides $\ell, \ell'$ of two rules in $A$ overlap at a subterm $\ell_0$ of $\ell$, then there is a finite complete overlap set for $\ell, \ell'$ at $\ell_0$. □

Note that any finite complete overlap set contains a minimal such set, in the sense that no subset of it is a complete overlap set; however, there may be more than one such subset.

**Definition 7.6.7** An MTRS $(\Sigma, A, B)$ is said to have **complete overlaps** iff whenever leftsides $\ell, \ell'$ of rules in $A$ overlap at a subterm $\ell_0$ of $\ell$, they have a finite complete overlap set. Each such overlap is called a **superposition** of $\ell, \ell'$, and the pair of rightsides resulting from applying the two rules to the overlap $\theta(\ell_0)$ is called a **critical pair**. If the two terms of a critical pair can be rewritten modulo $B$ to a common term using $A$, then that critical pair is said to **converge** or to **be convergent**. □

The following illustrates the definitions above:

**Example 7.6.8** The first two rules of PROPC (see Example 7.5.9), with leftsides $t = P$ *and false* and $t' = P$ *and true*, have the overlap *true and false* modulo $B$, with $\theta(P) = $ *true* and $\theta'(P) = $ *false*. Then the term $\theta(t) = \theta'(t') = $ *true and false* rewrites to *false* in two different ways. (Only the commutative law for *and* is actually used here.) □

For the Critical Pair Theorem (Theorem 5.6.9) to generalize to the modulo $B$ case, would mean that an MTRS with complete overlaps is locally Church-Rosser if all its critical pairs are convergent. This does *not* hold, and in fact Example 7.6.6 is a counterexample. Chapter 12 discusses some algorithms that generalize unification for computing an analog of critical pairs for MTRS's over certain sets of equations, and thus deciding the local Church-Rosser property.

The following weak modulo $B$ orthogonality result, which follows from the ordinary version, is the best we can do here:

**Proposition 7.6.9** (*Weak Orthogonality Modulo B*) Given an MTRS $\mathcal{M} = (\Sigma, A, B)$, let $R = A \cup B \cup B\breve{}$ (where $B\breve{}$ denotes the converse of $B$). If $R$ is lapse free and orthogonal, and if $B$ is balanced, then $\mathcal{M}$ is Church-Rosser.

**Proof:** $(\Sigma, R)$ is Church-Rosser by Theorem 5.6.4, and since $B$ is balanced and lapse free, any rewrite sequence $t \stackrel{*}{\Rightarrow}_{A/B} t'$ expands to a rewrite sequence $t \stackrel{*}{\Rightarrow}_R t'$, which implies that $\mathcal{M}$ is also Church-Rosser. □

Unfortunately, this is not very useful: the associative law is disqualified because it is self-overlapping; although the commutative law satisfies the assumptions for $B$, it is unlikely to be non-overlapping with interesting rule sets $A$; moreover, $B$ cannot contain identity or idempotent laws, since these (or else their converses) are not lapse free rewrite rules.

**Exercise 7.6.2** Use Propositions 7.6.9 and 7.6.2 for an alternative proof of Proposition 7.6.3. □

### 7.6.1 Adding Constants

The results of Section 5.3 on adding new constants generalize straightforwardly to the modulo $B$ setting, and as before, are important for theorem proving; however, we do not explicitly state them here, but refer the reader to Section 7.7.1, which gives more general results for conditional rules modulo $B$.

## 7.7 Conditional Term Rewriting Modulo Equations

We first develop an ARS version of join conditional term rewriting, which will let us define conditional rewriting modulo equations more easily than by developing it directly; we will also use it again for order-sorted term rewriting in Chapter 10. An unconditional ARS (Definition 5.7.1) consists of a set of "rules," which are really just pairs of elements of the same sort from an indexed set $T$, by convention written $t \to t'$. We generalize this as follows:

**Definition 7.7.1** A **join conditional ARS**, abbreviated **JCARS** or just **CARS**, is a pair $(T, W)$, where $T$ is $S$-sorted and $W$ is a set of **conditional rules** on $T$, which are $(n+1)$-tuples for $n \geq 0$, of pairs of elements of $T$ of the same sort, by convention written in one of the forms

$$t \to t' \text{ if } t_1 = t'_1, \ldots, t_n = t'_n$$
$$\left(\wedge_{i=1}^n t_i = t'_i\right) \Rightarrow t \to t'$$

where the first pair, or **head**, of the tuple is $t \to t'$. Note that unconditional rules are the special case where $n = 0$. Now given a CARS $(T, W)$, define an ordinary ARS on $T$ by

$$\begin{aligned} R_0 &= \{\langle t, t\rangle \mid t \in T\} \\ R_{k+1} &= \{\langle t, t'\rangle \mid (\wedge_{i=1}^n t_i = t'_i) \Rightarrow t \to t' \text{ in } W \text{ and} \\ & \qquad t_i \downarrow t'_i \text{ by } R_k \text{ for } i = 1, \ldots, n\} \cup R_k^\star \end{aligned}$$

for each $k \geq 0$, where $R_k^\star$ denotes the transitive, reflexive closure of $R_k$, and then let

$$R = \cup_{k=0}^\infty R_k \, .$$

We often write $W^\diamond$ for the relation $R$ in the following. Now define an ordinary ARS on $T$ by $t \to t'$ iff there is a rule $(\wedge_{i=1}^n t_i = t'_i) \Rightarrow t \to t'$ in $W$ such that $t_i \downarrow t'_i$ using $W^\diamond$ for $i = 1, \ldots, n$. We call this the **ARS defined by** $W$, and we may write $(T, \to_W)$ or even just $(T, \to)$ for it.

We say that a relation $R$ on $T$ is **join closed under** $W$ iff whenever $t \to t'$ if $t_1 = t'_1, \ldots, t_n = t'_n$ is in $W$ and $t_i \downarrow t'_i$ by $R$ for $i = 1, \ldots, n$ then $\langle t, t' \rangle$ is in $R$. (When $n = 0$, this just means $\langle t, t' \rangle \in R$.) □

**Proposition 7.7.2** Given a CARS $(T, W)$, then $W^\diamond$ (as above) is the least transitive, reflexive relation on $T$ that is join closed under $W$. Moreover, the relation $\xrightarrow{*}_W$ is equal to $W^\diamond$.

**Proof:** We write $R$ for $W^\diamond$. Reflexivity of $R$ follows from the inclusion of $R_0$. To show transitivity, suppose $\langle t_1, t_2 \rangle, \langle t_2, t_3 \rangle \in R$. Then there is some $k$ such that $\langle t_1, t_2 \rangle, \langle t_2, t_3 \rangle \in R_k$, so that $\langle t_1, t_3 \rangle$ is in $R_k^\star$ and hence in $R$. To show join closure under $W$, suppose $t \to t'$ if $t_1 = t'_1, \ldots, t_n = t'_n$ is in $W$ and that $t_i \downarrow t'_i$ by $R$ for $i = 1, \ldots, n$. Then there is some $k$ such that $t_i \downarrow t'_i$ by $R_k$ for $i = 1, \ldots, n$, so that $\langle t, t' \rangle$ is in $R_{k+1}$ and hence is in $R$.

To show minimality, suppose $R'$ is join closed under $W$. Then $R_0 \subseteq R'$, and also $R_k \subseteq R'$ implies $R_{k+1} \subseteq R'$. Therefore $R \subseteq R'$.

For the second assertion, we first show $\to_W \subseteq W^\diamond$, which implies $\xrightarrow{*}_W \subseteq W^\diamond$ since $W^\diamond$ is transitive. So we suppose $t \to_W t'$. Then there exists $k$ such that $t_i \downarrow t'_i$ by $R_k$, which implies that $\langle t, t' \rangle$ is in $R_{k+1}$ and hence in $W^\diamond$. To prove the converse, we show $R_k \subseteq \xrightarrow{*}_W$ for all $k$. $R_0 \subseteq \xrightarrow{*}_W$ since $\xrightarrow{*}$ is reflexive by definition. Next suppose $\langle t, t' \rangle$ is in $R_{k+1}$ but not in $R_k^\star$. Then there is a rule $t \to t'$ if $t_1 = t'_1, \ldots, t_n = t'_n$ in $W$ with $t_i \downarrow t'_i$ by $R_k$ for $i = 1, \ldots, n$. Therefore $t \to_W t'$ since $t_i \downarrow t'_i$ also by $W^\diamond$ since $R_k \subseteq W^\diamond$. □

This result can be used to prove properties of $\xrightarrow{*}_W$. We now apply the CARS machinery to conditional term rewriting modulo equations:

**Definition 7.7.3** A **conditional modulo term rewriting system**, abbreviated **CMTRS**, is $(\Sigma, A, B)$, where $A$ is a set of (possibly) conditional $\Sigma$-rewrite rules, and $B$ is a set of unconditional $\Sigma$-equations. From a given CMTRS $(\Sigma, A, B)$, we define two different CARS's, where $t \to t'$ if $t_1 = t'_1, \ldots, t_n = t'_n$ has sort $s$ and is in $A$, $Y = var(t)$, $\theta : Y \to T_\Sigma$, $u \in T_\Sigma(\{z\}_s)$:

1. For class rewriting, let $W$ be the set of rules of the form $c \to c'$ if $c_1 = c'_1, \ldots, c_n = c'_n$ where $c = [u(z \leftarrow \theta(t))], c' = [u(z \leftarrow \theta(t'))]$, $c_i = [\theta(t_i)]$ and $c'_i = [\theta(t'_i)]$ for $i = 1, \ldots, n$.

2. For term rewriting, let $W$ be the set of rules of the form $v \to v'$ if $v_1 = v'_1, \ldots, v_n = v'_n$ where $v \simeq_B u(z \leftarrow \theta(t)), v' \simeq_B u(z \leftarrow \theta(t'))$, $v_i = \theta(t_i)$ and $v'_i = \theta(t'_i)$ for $i = 1, \ldots, n$.

Definition 7.7.1 now yields an ARS for each of these CARS's. We write $\Rightarrow_{[A/B]}$ for the first, and $\Rightarrow_{A/B}$ for the second, which are conditional class rewriting modulo equations, and conditional term rewriting modulo equations, respectively. The pair $(u, \theta)$ is called a **match**. As before, rewriting extends to terms with variables by extending $\Sigma$ to $\Sigma(X)$, and which case write $c \Rightarrow_{[A/B],X} c'$ and $t \Rightarrow_{A/B,X} t'$, defined on $T_{\Sigma(X),B}$ and $T_{\Sigma(X)}$ respectively. □

All ARS results, e.g., the Newman lemma and the multi-level termination results in Section 5.8.2, apply, because $\Rightarrow_{[A/B]}$ and $\Rightarrow_{A/B}$ are defined by ARS's; Theorem 7.7.7 below shows an equivalence of $\Rightarrow_{[A/B]}$ and $\Rightarrow_{A/B}$. The following is proved similarly to Proposition 7.3.2:

**Proposition 7.7.4** Given $t, t' \in T_\Sigma(Y)$, $Y \subseteq X$ and CMTRS $(\Sigma, A, B)$, then $t \Rightarrow_{A/B,X} t'$ iff $t \Rightarrow_{A/B,Y} t'$, and in both cases $var(t') \subseteq var(t)$. Therefore $t \stackrel{*}{\Rightarrow}_{A/B,X} t'$ iff $t \stackrel{*}{\Rightarrow}_{A/B,Y} t'$, and in both cases $var(t') \subseteq var(t)$. □

Thus both $\Rightarrow_{A/B,X}$ and $\stackrel{*}{\Rightarrow}_{A/B,X}$ restrict and extend reasonably over variables, so we can drop the subscript $X$ and use any $X$ with $var(t) \subseteq X$; also as before, $\stackrel{*}{\Leftrightarrow}_{A/B,X}$ does *not* restrict and extend reasonably, as shown by Example 5.1.15, so we define $t \stackrel{*}{\Leftrightarrow}_A t'$ to mean there exists an $X$ such that $t \stackrel{*}{\Leftrightarrow}_{A,X} t'$. Example 5.1.15 also shows bad behavior for $\simeq_{A/B}^X$ (defined by $t \simeq_{A/B}^X t'$ iff $A \cup B \models (\forall X)\ t = t'$), although again, rule (8) in Chapter 4 (extended to rewriting modulo $B$) implies $\simeq_{A,B}^X$ does behave reasonably when the signature is non-void. Defining $\downarrow_{A/B,X}$ from the ARS, we generalize Proposition 5.1.13, again allowing the subscript $X$ to be dropped:

**Proposition 7.7.5** Given $t, t' \in T_\Sigma(Y)$, $Y \subseteq X$ and CMTRS $(\Sigma, A, B)$, then we have $t_1 \downarrow_{A/B,X} t_2$ if and only if $t_1 \downarrow_{A/B,Y} t_2$, and moreover, these imply $A \cup B \vdash (\forall X)\ t_1 = t_2$. □

We next give a cute proof to illustrate the approach (some additional theory needed for its justification is discussed after Theorem 7.7.10):

**Example 7.7.6** We continue Example 7.3.6 by showing that a ring has no zero divisors (i.e., non-zero elements $a, b$ such that $a * b = 0$) if it satisfies the left cancellation law, that $a * b = a * c$ and $a \neq 0$ imply $b = c$. For this proof, we turn on the "reduce conditions" feature, so that when the conditional rewrite rule for the cancellation law is applied, its condition is automatically checked by reduction; because this rule involves Boolean "and" it is also important that include BOOL be turned on. The result of Example 7.3.6 is used as a lemma.

```
set reduce conditions on .
open RING .   vars-of .
```

```
      ops a b c : -> R .
      eq A * 0 = 0 .    *** the lemma
      [lc] cq B = C if A * B == A * C and A =/= 0 .
      eq a * b = 0 .    *** the assumption
      show rules .
      start b .
      apply .lc with C = 0, A = a at top .
    close
```

Since the result is 0, as expected, the proof is done. Note that both ==
and =/= are used in the conditional equation lc.  □

**Exercise 7.7.1** Prove the converse of the last result of Example 7.3.6, that a ring
satisfies the left cancellation law if it has no zero divisors.  □

We now generalize some basic semantic results on term rewriting
modulo equations to the conditional case, beginning with Theorem 7.3.4,
noting that Proposition 7.2.2 and Theorem 7.2.3 were already stated for
conditional equations. As in the unconditional case, we cannot hope for
completeness due to the semantics of join conditional rewriting.

**Theorem 7.7.7** (*Soundness*) Given a CMTRS $(\Sigma, A, B)$ and $t, t' \in T_\Sigma(X)$, then

$$
\begin{aligned}
t \Rightarrow_{A/B} t' & \quad\text{iff}\quad & [t] \Rightarrow_{[A/B]} [t'] , \\
t \stackrel{*}{\Rightarrow}_{A/B} t' & \quad\text{iff}\quad & [t] \stackrel{*}{\Rightarrow}_{[A/B]} [t'] , \\
[t] \stackrel{*}{\Rightarrow}_{[A/B]} [t'] & \quad\text{implies}\quad & [A] \vdash_B (\forall X)\, [t] = [t'] , \\
t \stackrel{*}{\Leftrightarrow}_{A/B} t' & \quad\text{implies}\quad & A \cup B \vdash (\forall X)\, t = t' .
\end{aligned}
$$

Also $\stackrel{*}{\Leftrightarrow}_{A/B}$ is sound for satisfaction of $A \cup B$, and $\stackrel{*}{\Leftrightarrow}_{[A/B]}$ is sound for
satisfaction of $A$ modulo $B$. Moreover, $\stackrel{*}{\Leftrightarrow}_{A/B} \subseteq \simeq^X_{A \cup B}$ on terms with variables in $X$.

**Proof:** The first assertion follows from the definitions of $\Rightarrow_{A/B}$ and $\Rightarrow_{[A/B]}$ (Definition 7.7.3), and the second follows from the first by induction. The
third follows because $\Rightarrow_{[A/B]}$ rephrases the rule $(+6C_B)$. The fourth follows from the second and third, plus Theorem 7.2.3. The fifth and sixth
follow from the third and fourth plus Theorem 7.2.3.  □

The following generalizes Theorem 7.3.9 to the conditional case:

**Theorem 7.7.8** Given a ground canonical CMTRS $(\Sigma, A, B)$, if $t_1, t_2$ are both normal forms of a ground term $t$ under $\Rightarrow_{A/B}$ then $t_1 \simeq_B t_2$. Moreover, the
$B$-equivalence classes of ground normal forms under $\Rightarrow_{A/B}$ form an initial $(\Sigma, A \cup B)$-algebra, denoted $N_{\Sigma,A/B}$, in the following way, where $[\![t]\!]$
denotes any arbitrary normal form of $t$, and where $[\![t]\!]_B$ denotes the
$B$-equivalence class of $[\![t]\!]$:

(0) interpret $\sigma \in \Sigma_{[],s}$ as $[\![\sigma]\!]_B$ in $N_{\Sigma,A/B,s}$; and

(1) interpret $\sigma \in \Sigma_{s_1...s_n,s}$ with $n > 0$ as the function that sends $(\llbracket t_1 \rrbracket_B, \ldots, \llbracket t_n \rrbracket_B)$ with $t_i \in T_{\Sigma,s_i}$ to $\llbracket \sigma(t_1,\ldots,t_n) \rrbracket_B$ in $N_{\Sigma,A/B,s}$.

Finally, $N_{\Sigma,A/B}$ is $\Sigma$-isomorphic to $T_{\Sigma,A \cup B}$.

**Proof:** For convenience, write $N$ for $N_{\Sigma,A/B}$. The first assertion follows from the ARS result Theorem 5.7.2, using also Theorem 7.7.7. Note that $\sigma_N$ is well defined by (1), by the first assertion, plus the fact that $\simeq_B$ is a $\Sigma$-congruence.

Next, we check that $N$ satisfies $A \cup B$. Satisfaction of $B$ is by definition of $N$ as consisting of $B$-equivalence classes of normal forms. Now let $(\forall X)\ t = t'$ if $C$ be in $A$; we need to prove that $\overline{a}(t) = \overline{a}(t')$ for all $a : X \to N$ that satisfy $C$. The proof follows that of Theorem 7.3.9, except that we must restrict to assignments that satisfy the condition, and that uses of Theorem 7.3.4 must be replaced by uses of Theorem 7.7.7. □

**Definition 7.7.9** A CMTRS $(\Sigma, A, B)$ is **join condition canonical** iff the CMTRS $(\Sigma', A', B)$ is canonical, where: (1) $\Sigma' \subseteq \Sigma$ is least such that if $t_i = t'_i$ is a condition in some rule $r$ in $A$, then $\theta(t_i)$ and $\theta(t'_i)$ are in $T_{\Sigma'}$ for all $\theta : X \to T_\Sigma$ where $X = var(t)$ where $t$ is the leftside of the head of rule $r$; and (2) $A' \subseteq A$ is least such that all conditional rules are in $A'$, and all unconditional rules that can be used in evaluating the conditions of rules in $A$ are also in $A'$. □

It is of course possible that $\Sigma' = \Sigma$, but in many real examples, conditions are relatively simple tests on data types that involve relatively few operations and rules. Consequently, $(\Sigma', A', B)$ is often significantly simpler than $(\Sigma, A, B)$. In OBJ, the operations used in conditions may be built in rather then defined by explicit equations, but in this case, they should be considered defined by a canonical TRS.

**Theorem 7.7.10** (*Completeness*) Given a join condition canonical CMTRS $(\Sigma, A, B)$, the following four conditions are equivalent for any $t, t' \in T_\Sigma(X)$:

$t \overset{*}{\Leftrightarrow}_{A/B} t'$  $\qquad A \cup B \vdash (\forall X)\ t = t'$
$[A] \vdash_B (\forall X)\ t =_B t'$ $\qquad t \simeq_{A \cup B} t'$

Moreover, if $(\Sigma, A, B)$ is Church-Rosser, then $t \downarrow_{A/B} t'$ is also equivalent to the above. Finally, if $(\Sigma, A, B)$ is canonical, then $\llbracket t \rrbracket_A \simeq_B \llbracket t' \rrbracket_A$ is also equivalent.

**Proof:** Equivalence of the last three of the first four assertions has already been proved. The first implies the second by the soundness of rewriting. For the converse, because $(\pm 6C_B^*)$ is complete and is equivalent to bidirectional rewriting, we are done if the conditions of rules can always be evaluated; but this is given by the join condition canonical assumption. Equivalence of the fifth condition assuming Church-Rosser is a general ARS result, and equivalence of the final condition with the fourth is immediate when $\Rightarrow_{A/B}$ is canonical. □

This generalization of Theorem 7.3.10 implies that we can define an operation == that works for a canonical CMTRS $(\Sigma, A, B)$ the same way as for a TRS, CTRS, or MTRS that is canonical: $t==t'$ returns `true` over $(\Sigma, A, B)$ iff $t, t'$ have the same normal form modulo $B$ iff they are provably equal under $A \cup B$ as an equational theory. As before, even when $(\Sigma, A, B)$ is not canonical, if $t == t'$ returns `true` then $t$ and $t'$ are equal modulo $B$. Also as before, if $(\Sigma, A, B)$ is non-canonical, rewriting can be unsound if == occurs in a negative position (such as =/= in a positive position) in a condition; however, it is sound if the system is canonical for the sorts of terms that occur in negative positions, with respect to the subset of rules actually used.

**Exercise 7.7.2** Replace $\downarrow$ in Definition 7.7.1 by $\stackrel{*}{\leftrightarrow}$ to obtain **equality condition abstract rewrite systems**, apply this definition to term rewriting modulo $B$ to obtain **equality condition term rewriting modulo equations**, and then prove the analogs of Proposition 7.7.2 and Theorem 7.7.10, where the latter asserts completeness, not just soundness.    □

The OBJ implementation does not use equality condition rewriting, because it is too inefficient for use in a practical system. On the other hand, most of the literature on conditional rewriting, e.g., [113], uses equality condition rewriting, because it allows stronger theorems to be proved.

**Fact 7.7.11** If a CMTRS has no variables in its rules, then it is terminating iff it is ground terminating, Church-Rosser iff ground Church-Rosser, and hence canonical iff ground canonical.

**Proof:** Let $t$ be a term with variables. Because only subterms (modulo the equations) without variables can be redexes, and because rewriting on these ground subterms of $t$ is terminating (or Church-Rosser), so is rewriting on all of $t$. Therefore the ground properties imply the general properties. The converse is immediate.    □

Applying this result to triangular systems tells us that when variables are treated as constants, ground canonicity is equivalent to general canonicity.

## 7.7.1 Adding Constants

The results of Section 5.3 on adding new constants generalize straightforwardly to conditional rewriting modulo $B$; we state these generalizations explicitly because of their importance for theorem proving, and because they appear to be new in this context.

**Proposition 7.7.12** If a CMTRS $(\Sigma, A, B)$ is terminating, or Church-Rosser, or locally Church-Rosser, then so is $(\Sigma(X), A, B)$, for any suitable countable set $X$ of variable symbols.    □

The ARS proof of Proposition 5.3.1 generalizes, using the ARS $(T_{\Sigma,B}(X_S^\omega), \Rightarrow_{[A/B]})$, where the reader should recall that for a given set $S$ of sorts, $X_S^\omega$ denotes the ground signature with $(X_S^\omega)_s = \{x_s^i \mid i \in \omega\}$ for each $s \in S$, a countable set of new variable symbols distinct from the symbols in $\Sigma$. The proof of Proposition 5.3.4 in Appendix B also generalizes to conditional rewriting modulo $B$, giving the following:

**Proposition 7.7.13** A CMTRS $(\Sigma, A, B)$ is ground terminating if $(\Sigma(X), A, B)$ is ground terminating, where $X$ is a variable set for $\Sigma$; moreover, if $\Sigma$ is non-void, then $(\Sigma, A, B)$ is ground terminating iff $(\Sigma(X), A)$ is ground terminating.  □

**Corollary 7.7.14** If $\Sigma$ is non-void, then a CMTRS $(\Sigma, A, B)$ is ground terminating iff it is terminating.  □

**Exercise 7.7.3** Show that adding any set of constants to DNF gives a terminating MTRS.  □

Proposition 5.3.6 in Section 5.7 can be generalized to the following:

**Proposition 7.7.15** A CMTRS $(\Sigma, A, B)$ is Church-Rosser if and only if $(\Sigma(X_S^\omega), A, B)$ is ground Church-Rosser, and $(\Sigma, A, B)$ is locally Church-Rosser if and only if $(\Sigma(X_S^\omega), A, B)$ is ground locally Church-Rosser.

**Proof:** If we let $T = T_{\Sigma,B}(X)$ and $G = T_{\Sigma,B}(X_S^\omega)$, then the proof given for Proposition 5.3.6 on page 125 goes through as it stands.  □

**Exercise 7.7.4** Use Corollary 7.7.14 and Proposition 7.7.15 to show that PROPC(X) is canonical if PROPC is canonical.  □

### 7.7.2 Proving Termination

Definition 7.5.2 (of commutativity) generalizes to conditional rules, as do Lemma 7.5.3 and Proposition 7.5.4; these results are stated below, and their proofs are exactly the same as for the unconditional case, except that Proposition 7.7.17 uses Lemma 7.7.16.[E31]

**Lemma 7.7.16** Given a $\Sigma$-CMTRS $A$ commuting with $\Sigma$-equations $B$, if $t_0 \Rightarrow_{A/B} t_1 \Rightarrow_{A/B} \cdots \Rightarrow_{A/B} t_n$ then there exist $t_1', \ldots, t_n'$ such that $t_0 \Rightarrow_A t_1' \Rightarrow_A \cdots \Rightarrow_A t_n'$ and $t_n' \simeq_B t_n$. The same result also holds for weak rewriting modulo $B$.  □

**Proposition 7.7.17** If a CTRS $(\Sigma, A)$ commuting with $\Sigma$-equations $B$ is terminating, then the CMTRS $(\Sigma, A, B)$ is also terminating, and the same holds under weak rewriting.  □

We also have the following:

**Proposition 7.7.18** Given a CMTRS $C$, let $C^U$ be the MTRS whose rules are those of $C$ with their conditions (if any) removed. Then $C$ is terminating (or ground terminating) if $C^U$ is.

**Proof:** Any rewrite sequence of $C$ is also a rewrite sequence of $C^U$ and therefore finite. □

The following is a nice application of several results earlier in this chapter:

**Theorem 7.7.19** Given a conditional triangular propositional system, let $T$ be its set of equations, let $A = T \cup P$ where $P$ is the set of equations in PROPC, and let $B$ be the associative and commutative laws for and and xor. Then $T$ and $A$ are terminating modulo $B$.

**Proof:** Let $T', A'$ be the unconditional non-modulo versions of $T, A$, respectively, so that $A' = T' \cup P$. Example 5.8.27 showed that unconditional triangular propositional systems are terminating under rewriting modulo no equations, so $T'$ is terminating. Therefore $T$ is terminating by Proposition 5.8.12, and Proposition 7.7.17 implies that $T$ modulo $B$ is terminating, since $B$ commutes with $T$ because it commutes with any rule with a constant as its leftside.

An argument similar to that of Example 5.8.27 shows that $A'$ is terminating. Let $N = \otimes_{i=1}^{m+1} \omega$ where $m$ is the number of dependent variables in $T$, and note that $N$ is Noetherian. For $t$ a $\Sigma(Z)$-term, let $\psi_i(t)$ be the number of occurrences of $y_i$ in $t$, let $\rho(t)$ be as in Example 7.5.9 with $\rho(z) = 2$ for all $z \in Z$, and let $\tau(t) = (\psi_1(t), \psi_2(t), \ldots, \psi_m(t), \rho(t))$. Then $\tau$ satisfies the hypotheses of Proposition 7.5.7, and thus $A'$ is terminating. Therefore Proposition 5.8.12 implies $A$ is terminating, and so $A$ is also terminating modulo $B$ by Proposition 7.7.17. □

Given a poset $P$, we can define weak and strong $\rho$-monotonicty of conditional rewrite rules modulo $B$, of substitution modulo $B$, and of operations in $\Sigma$, just as in Definition 5.5.3, except that $T_\Sigma$ and $T_{\Sigma(\{z\}_s)}$ are replaced by $T_{\Sigma,B}$ and $T_{\Sigma,B}(\{z\}_s)$, respectively. Note that as before, the inequalities for a rule are only required to hold when all the conditions of the rule converge (modulo $B$). The following is the modulo $B$ generalization of the most powerful termination result (Theorem 5.8.33) for conditional rules in Chapter 5:

**Theorem 7.7.20** Let $(\Sigma, A, B)$ be a CMTRS with $\Sigma$ non-void and with $(\Sigma', A', B)$ a CMTRS with $\Sigma' \subseteq \Sigma$ and $A' \subseteq A$ ground terminating; let $P$ be a poset, and let $N = A - A'$ with $\Sigma'$ the minimal signature. If there is a $\rho : T_{\Sigma,B} \to P$ such that

(1) each rule in $A'$ is weak $\rho$-monotone,

(2) each rule in $N$ is strict $\rho$-monotone,

(3) each operation in $\Sigma$ is strict $\rho$-monotone, and

(4) $P$ is Noetherian, or if not, then for each $t \in T_{\Sigma,s}$ there is some Noetherian poset $P_s^t \subseteq P_s$ such that $t \stackrel{*}{\Rightarrow}_{[A/B]} t'$ implies $\rho(t') \in P_s^t$,

then $(\Sigma, A, B)$ is ground terminating. □

The proof, which is much like that of Theorem 5.8.33, is sketched in Appendix B.

**Exercise 7.7.5** Given a CMTRS $C$, let $C^U$ be the MTRS whose rules are those of $C$ with their conditions (if any) removed. Then $C$ is Church-Rosser (or ground Church-Rosser) if $C^U$ is. □

### 7.7.3 Proving Church-Rosser

As always, ARS results apply directly, including the Newman Lemma and the Hindley-Rosen Lemma (Exercise 5.7.5), so we do not state them here. The results stated here are actually rather weak. Perhaps the most generally useful methods for proving Church-Rosser are based on the Newman Lemma, because it is usually much easier to prove the local Church-Rosser property. As discussed in Section 7.6, and in more detail in Chapter 12, although the Critical Pair Theorem (5.6.9) does not generalize to modulo $B$ rewriting, in many cases the local Church-Rosser property can still be checked by a variant of an algorithm introduced by Knuth and Bendix [117] for the unsorted unconditional non-modulo case. The Hindley-Rosen Lemma applied to conditional rewriting modulo equations gives the following:

**Proposition 7.7.21** Given Church-Rosser CMTRS's $\mathcal{M}_i = (\Sigma, A_i, B)$ for $i \in I$, which **strongly commute** in the sense that if $t_0 \Rightarrow_{A_i/B} t_1$ and $t_0 \Rightarrow_{A_j/B} t_2$ for some $i, j \in I$, then there is some $t_3$ such that $t_1 \stackrel{0,1}{\Rightarrow}_{A_j/B} t_3$ and $t_2 \stackrel{*}{\Rightarrow}_{A_i/B} t_3$, then $\mathcal{M} = (\Sigma, \bigcup_{i \in I} A_i, B)$ is Church-Rosser, where $\stackrel{0,1}{\Rightarrow}$ indicates reflexive closure. □

Proposition 5.2.6 also generalizes, since it follows from the ARS result Proposition 5.7.6:

**Proposition 7.7.22** If $(\Sigma, A, B)$ is a Church-Rosser CMTRS, $A \models (\forall X) \, t =_B t'$ iff $t \downarrow_{A/B} t'$. □

We next prove the following generalization of Proposition 7.6.2:

**Proposition 7.7.23** Given a CTRS $(\Sigma, A)$ commuting with $B$, if $A$ is (locally) Church-Rosser, then the CMTRS $(\Sigma, A, B)$ is also (locally) Church-Rosser.

**Proof:** For the Church-Rosser property, suppose $t_0 \overset{*}{\Rightarrow}_{A/B} t_1$ and $t_0 \overset{*}{\Rightarrow}_{A/B} t_2$. Then by Lemma 7.7.16, we can find $t_1'$ and $t_2'$ such that $t_0 \overset{*}{\Rightarrow}_A t_1'$ and $t_0 \overset{*}{\Rightarrow}_A t_2'$ with $t_1' \simeq_B t_1$ and $t_2' \simeq_B t_2$. Then by Church-Rosser, there is a term $t_3$ such that $t_1' \overset{*}{\Rightarrow}_A t_3$ and $t_2' \overset{*}{\Rightarrow}_A t_3$. Hence $t_1 \overset{*}{\Rightarrow}_{A/B} t_3$ and $t_2 \overset{*}{\Rightarrow}_{A/B} t_3$. So again by Lemma 7.7.16, there exist terms $t_3'$ and $t_3''$ such that $t_1 \overset{*}{\Rightarrow}_A t_3'$ and $t_2 \overset{*}{\Rightarrow}_A t_3''$ with $t_3' \simeq_B t_3$ and $t_3'' \simeq_B t_3$, so that $t_3' \simeq_B t_3''$. Thus $t_1 \overset{*}{\Rightarrow}_{A/B} t_3'$ and $t_2 \overset{*}{\Rightarrow}_{A/B} t_3''$ with $t_3' \simeq_B t_3''$, and we are done. A similar proof works for the local Church-Rosser property. □

The next result uses the above and Proposition 7.7.18:

**Theorem 7.7.24** Given a conditional triangular propositional system $T$ with consistent conditions, let $B$ be the associative and commutative laws for and and xor, and let $A = T \cup P$ where $P$ is the equations in PROPC excluding $B$. Then $T$ and $A$ are ground Church-Rosser modulo $B$, and hence ground canonical.

**Proof:** Proposition 7.7.19 shows termination modulo $B$ for $T$ and $A$. We now show $T$ locally Church-Rosser modulo $B$, using the fact that only the symbols $y_k$ can be rewritten. Suppose $t \overset{1}{\Rightarrow}_{A/B} t_1$ and $t \overset{1}{\Rightarrow}_{A/B} t_2$, with redexes $y_i$ and $y_j$ respectively. If $y_i = y_j$ and they are at the same occurrence in $t$, then $t_1 =_B t_2$ by the consistent conditions hypothesis. Otherwise, we can rewrite $y_j$ in $t_1$ and $y_i$ in $t_2$ to get the same term $t'$. This gives the local Church-Rosser property, so that the Newman lemma implies that $T$ modulo $B$ is Church-Rosser and thus canonical.

Since $P$ modulo $B$ is Church-Rosser, and it is not difficult to check that $\Rightarrow_{P/B}$ and $\Rightarrow_{T/B}$ strongly commute in the sense of Proposition 7.7.21 (the Hindley-Rosen Lemma), it follows that $A$ is Church-Rosser modulo $B$. Therefore $A$ is also canonical. □

## 7.8 Literature

The treatment of equational deduction modulo equations in Section 7.2 may be novel, but it is similar to the treatment of term rewriting modulo equations in Section 7.3, which follows the work of Huet [108] and others, although our exposition is more semantic and algebraic than the standard literature. Basic semantic results on rewriting modulo equations include its equivalence with class rewriting and its relationship with equational deduction (Theorem 7.3.4 and Proposition 7.2.2).

Theorem 7.3.9 is also very fundamental, and although it will not surprise experts, it does not appear in the literature. Hsiang's Theorem (7.3.13) was first proved by Hsiang [106]. The exclusive normal form results such as Proposition 7.3.18 are not as well known as they should be, although they are direct consequences of Hsiang's Theorem. Theorem 7.3.22 says that reduction gives a decision procedure for the predicate calculus; it is of basic importance to this book, and the proof given here, which appears to be new, is a nice example of algebraic techniques in term rewriting theory.

Our algebraic approach to hardware verification was first outlined in [59]. Although influenced by Mike Gordon's clear expositions of hardware verification using higher-order logic, e.g., [93, 94], we disagree with Gordon's claim that higher-order logic is *necessary* for hardware verification. A key insight for this chapter is that equality is already a "bidirectional" (i.e., symmetric) relation, which of course is axiomatized by equational logic. Bidirectionality is needed for many important circuits, and conditional equational logic adds important expressive power. The results about triangular propositional systems seem to be new, and Proposition 7.4.3 and Theorem 7.4.10 justify the use of reduction to prove properties of combinatorial circuits. Proposition 7.4.16 and the subsequent discussion in Section 7.4.3 are very useful. Several of the main techniques from this chapter are illustrated in the proofs of Theorems 7.7.19 and 7.7.24. It is reasonable to conjecture that every triangular system has a most general partial solution, and that general conditional solutions are unique (even though most general unconditional solutions are not unique).

This approach to hardware verification has been applied to many examples, including some that are non-trivial [168, 169, 84], using the 2OBJ theorem prover. An early application of OBJ to hardware specification and testing appears in [159]. Most general solutions may remind some readers of unifiers, and indeed, they *are* a special case of unification understood in a sufficiently broad sense, as for example in [60], though this is not the place to discuss such abstract notions.

Many results from Section 7.5 onward are novel, in that few have been proved for many-sorted rewriting, let alone overloaded many-sorted rewriting, and several are new even for the unsorted case. The results about rules commuting with equations in Proposition 7.6.2 and its generalization Proposition 7.7.23 do not appear to be in the literature. Conditional abstract rewrite systems (Definition 7.7.1) appear to be a new and useful concept. The construction of $W^\diamond$ can be described more abstractly using concepts from Section 8.2: because the relation of the ARS is defined using only Horn clauses, an initiality theorem for Horn clause theories gives the least relation satisfying those sentences. The CARS approach could also have been applied to ordinary CTRS's, simplifying Section 5.8. Theorems 7.7.7, 7.7.8 and 7.7.10 extend the main semantic results of Chapter 5 to conditional term rewriting modulo equations. In particular, Theorem 7.7.8 makes explicit the connection with initial algebra semantics. Proposition 7.7.17 (about commutativity) and the very general Theorem 7.7.20 also appear to be useful new results for proving termination modulo equations.

I thank Prof. Mitsuhiro Okada and Dr. Monica Marcus for many valuable comments on the material in this chapter; the latter has read many parts of the chapter carefully and offered many corrections, though she is of course not responsible for any remaining errors.

---

**A Note to Lecturers:** This chapter contains a great deal of difficult material, much of which would have to be skipped in a one semester or one quarter course. I have found it possible to present the hardware examples with only a pointer to the theory, since intuitions about hardware are strong enough to make the computations convincing. Of course, many proofs can be skipped in lectures and even in readings, especially since this chapter has the structure of a sequence of generalizations, in which similar results appear in gradually increasing generality.

---

# 8 First-Order Logic and Proof Planning

This chapter extends our algebraic approach to full first-order logic. Section 8.2 treats the special case of Horn clause logic, which we show is essentially the same as equational logic. Section 8.3 presents first-order logic syntax and some basics of its model theory; this development is unusual in treating the many-sorted case and in allowing (partially) empty models. Section 8.4 discusses proof planning. Proof rules for existential quantifiers, case analysis, and induction are given in Sections 8.5, 8.6, and 8.7, respectively. The notion of induction is unusually general.

## 8.1 First-Order Signatures, Models and Morphisms

First-order signatures provide symbols for building first-order sentences, but actually defining these sentences and their satisfaction is put off to Section 8.3.

**Definition 8.1.1** Given a set $S$ of **sorts**, an $S$-sorted **first-order signature** $\Phi$ is a pair $(\Sigma, \Pi)$, where $\Sigma$ is an $S$-sorted algebraic signature, i.e., an indexed set of the form $\{\Sigma_{w,s} \mid w \in S^*, s \in S\}$ whose elements are called **function** (or **operation**) **symbols**, and $\Pi$ is an indexed set of the form $\{\Pi_w \mid w \in S^*\}$ whose elements are called **predicate symbols**, where $\pi \in \Pi_w$ is said to have **arity** $w$. □

**Example 8.1.2** Let $\Sigma$ be the algebraic signature $\Sigma^{\mathsf{NATP}}$ of Example 2.3.3, with one sort, Nat, and operations $0, s$; let $\Pi$ have $\Pi_{\mathtt{Nat}} = \{pos\}$, $\Pi_{\mathtt{NatNat}} = \{geq\}$, and $\Pi_w = \varnothing$ otherwise. The signature $\Phi^{\mathsf{NAT}} = (\Sigma, \Pi)$ is adequate for expressing many simple properties of natural numbers. □

Our discussion of semantics for a given signature starts with its models:

**Definition 8.1.3** Given a first-order signature $\Phi = (\Sigma, \Pi)$, a **first-order $\Phi$-model** $M$ consists of a $\Sigma$-algebra, also denoted $M$, together with for each $\pi \in$ $\Pi_w$, a subset $M_\pi \subseteq M_w$, where $M_{s_1\ldots s_n}$ denotes the set $M_{s_1} \times \cdots \times M_{s_n}$. A model $M$ is **nonempty** iff each $M_s$ is nonempty. □

Think of $M_\pi$ as the set of values where $\pi$ is true in $M$. When $w = [\,]$, then $\pi \in \Pi_w$ should represent a relation that is constant, i.e., a truth value. Because $M_{[\,]}$ is a one-point set, say $\{\star\}$, there are only two possible values for $M_\pi \subseteq \{\star\}$, namely $\varnothing$ and $\{\star\}$. We let the first case mean $\pi$ is false, and the second mean it is true.

**Example 8.1.4** If we let $\Phi$ be the signature $\Phi^{\mathsf{NAT}}$ of Example 8.1.2 above, then the standard $\Phi$-model $M$ has $M_{\mathsf{Nat}} = T_\Sigma$ (the natural numbers in Peano notation, with 0 and $s$ interpreted as usual in $T_\Sigma$), with $M_{pos} = \{s(0), s(s(0)), s(s(s(0))),\ldots\}$ and with $M_{geq} = \{\langle m,n\rangle \mid m \geq n\}$ (where $\geq$ has the usual meaning, and $n,m$ are Peano numbers). Of course, there are many other $\Phi$-models, most of which are *not* isomorphic to the natural numbers. For example, there are $\Phi$-models with just one element. □

**Exercise 8.1.1** For $\Phi$ the signature of Example 8.1.2 above, how many $\Phi$-models $M$ are there with $M_{\mathsf{Nat}} = \{0\}$? How many are there with $M_{\mathsf{Nat}} = \{0,1\}$?
□

**Exercise 8.1.2** (a) Give a first-order signature $\Phi$ that is adequate for partially ordered sets, and interpret the natural numbers as a $\Phi$-model.

(b) Give a first-order signature $\Phi$ that is adequate for equivalence relations, and interpret the natural numbers as a $\Phi$-model. □

**Definition 8.1.5** Given a first-order signature $\Phi = (\Sigma, \Pi)$ and given $\Phi$-models $M, M'$, then a $\Phi$-**morphism** $h: M \to M'$ is a $\Sigma$-homomorphism $h: M \to M'$ such that for each $\pi \in \Pi_{s_1\ldots s_n}$

$$(m_1,\ldots,m_n) \in M_\pi \quad \text{implies} \quad (h_{s_1}(m_1),\ldots,h_{s_n}(m_n)) \in M'_\pi$$

for all $m_i \in M_{s_i}$. The **composition** of two $\Phi$-morphisms is their composition as $\Sigma$-homomorphisms. The **identity** $\Phi$-morphism on $M$, denoted $1_M$, is the identity on $M$.

A $\Phi$-morphism $h: M \to M'$ is a $\Phi$-**isomorphism** iff there is a $\Phi$-morphism $g: M' \to M$ such that $h;g = 1_M$ and $g;h = 1_{M'}$. Such a morphism $g$ is called an **inverse** to $h$. □

**Exercise 8.1.3** (a) Show that the composition of two $\Phi$-morphisms is also a $\Phi$-morphism.

(b) Show that $1_M$ satisfies the identity law for composition of $\Phi$-morphisms.

(c) Show that if $h$ is a $\Phi$-isomorphism then it has a unique inverse.

(d) Show that a bijective $\Phi$-morphism is not necessarily a $\Phi$-isomorphism.

□

## 8.2 Horn Clause Logic

Horn clause logic is a sublogic of first-order logic that is essentially the same as conditional equational logic. Although Horn clause notation uses first-order logic symbols, this section will develop its syntax and satisfaction independently.

**Definition 8.2.1** Given a first-order signature $\Phi = (\Sigma, \Pi)$, a $\Phi$-**Horn clause** is an expression of the form

$$(\forall X)\ p_1 \wedge \ldots \wedge p_n \Rightarrow p_0$$

where $X$ is an $S$-sorted set of variable symbols, and the $p_i$, called **atoms**, are each of the form $\pi(t_1, \ldots, t_k)$ such that $\pi \in \Pi_{s_1 \ldots s_k}$ and $t_j \in T_\Sigma(X)_{s_j}$ for $j = 0, \ldots, k$. We may say that $p_0$ is the **head** of the clause, and that $p_1, \ldots, p_n$ is its **body**. As usual, we assume that the components $X_s$ for $X$ are mutually disjoint, and are also disjoint from the symbols in $\Sigma$ and $\Pi$. For $n = 0$, we include Horn clauses of the form $(\forall X)\ p_0$, which are just universally quantified atoms. □

Note that the symbols $\forall$, $\wedge$ and $\Rightarrow$ in a Horn clause do not have any separate meanings, but are parts of one single mixfix symbol, as are the symbols $\forall$ and if in conditional equations.

**Example 8.2.2** For $\Phi$ the signature of Example 8.1.2 above, the following are all Horn clauses:

$(\forall n)\ geq(n, 0)$
$(\forall n, m)\ geq(n, m) \Rightarrow geq(s(n), m)$
$(\forall n, m)\ geq(n, m) \Rightarrow geq(s(n), s(m))$
$(\forall n)\ pos(s(n))\ .$ □

**Exercise 8.2.1** Are all the axioms for partially ordered sets Horn clauses? What about those for equivalence relations? □

Section 3.5 discussed extending an assignment $\theta : X \to M$ where $X$ is a variable set and $M$ is a $\Sigma$-algebra, to a $\Sigma$-homomorphism $\overline{\theta} : T_\Sigma(X) \to M$, where $\overline{\theta}(t)$ is the result of simultaneously substituting $\theta(x)$ for each $x$ into $t \in T_\Sigma(X)$. The following uses this for $\Phi$-models $M$ when $\Sigma$ is the algebraic part of $\Phi$.

**Definition 8.2.3** Given a first-order signature $\Phi = (\Sigma, \Pi)$, a first-order $\Phi$-model $M$, and a $\Phi$-Horn clause $h_c$ of the form $(\forall X)\ p_1 \wedge \cdots \wedge p_n \Rightarrow p_0$, then we say $M$ **satisfies** $h_c$ and write $M \vDash_\Phi h_c$, iff for every assignment $a : X \to M$,

$$\overline{a}(t^i) \in M_{\pi_i}\ \text{for}\ i = 1, \ldots, n\ \text{implies}\ \overline{a}(t^0) \in M_{\pi_0}\ ,$$

where $p_i = \pi_i(t^i)$ with $t^i = (t_1^i, \ldots, t_{k(i)}^i)$, and where $\overline{a}(t^i) = (\overline{a}(t_1^i), \ldots, \overline{a}(t_{k(i)}^i))$, for $i = 1, \ldots, n$.

A **Horn specification** consists of a first-order signature $\Phi = (\Sigma, \Pi)$ and a set $H$ of $\Phi$-Horn clauses; we may write $(\Sigma, \Pi, H)$, or $(\Phi, H)$, or even just $H$. A $\Phi$-model $M$ satisfies $(\Phi, H)$ iff it satisfies each clause in $H$, and then we write $M \vDash_\Phi H$ and call $M$ an $H$-**model**. □

Let *HCL* denote the *institution*[1] of Horn clause logic, consisting of its signatures (first-order signatures), its sentences (Horn clauses), its models and morphisms (first-order models and morphisms), and the notion of satisfaction in Definition 8.2.3 above.

**Exercise 8.2.2** Show that each Horn clause in Example 8.2.2 is satisfied by the model $M_{\text{Nat}}$ of Example 8.1.4, or else that it is not. □

**Example 8.2.4** Letting $\Phi$ be the signature of Example 8.1.2 and $H$ the Horn clauses of Example 8.2.2 gives a Horn specification for the natural numbers with predicates for inequality and positivity. The intended model is the initial model, which exists by Theorem 8.2.6 below. □

### 8.2.1 (⋆) Initial Horn Models

In a precise analogy with the equational case, we have the following definition and theorem:

**Definition 8.2.5** Given a Horn specification $(\Phi, H)$, then an $H$-model $I$ is **initial** iff given any $H$-model $M$, there is a unique $\Phi$-morphism from $I$ to $M$. □

**Theorem 8.2.6** (*Initiality*) Every Horn specification has an initial model.

**Proof:** Given a Horn specification $(\Sigma, \Pi, H)$, we construct an initial model $T_H$ as follows:

1. If $S$ is the sort set of $\Phi = (\Sigma, \Pi)$, let $\overline{S} = S \cup \{\mathsf{B}\}$, where $\mathsf{B} \notin S$ (think of $\mathsf{B}$ as the Booleans).

2. Define an algebraic $\overline{S}$-sorted signature $\overline{\Pi}$ by $\overline{\Pi}_{w,\mathsf{B}} = \Pi_w$ for all $w \in S^+$, and $\overline{\Pi}_{[],\mathsf{B}} = \{true\} \cup \Pi_{[]}$, and $\overline{\Pi}_{w,s} = \emptyset$ otherwise. Let $\overline{\Phi} = (\overline{S}, \Sigma \cup \overline{\Pi})$.

3. Given a $\Phi$-model $M$, let $\overline{M}$ be the $\overline{\Phi}$-model constructed as follows:

    (a) $\overline{M}_s = M_s$ for all $s \in S$;
    
    (b) $\overline{M}_\mathsf{B} = \{\pi(m_1, \ldots, m_n) \mid \pi \in \Pi \text{ and } \langle m_1, \ldots, m_n \rangle \notin M_\pi\} \cup \{true\}$;
    
    (c) $\overline{M}_\sigma = M_\sigma$ for all $\sigma \in \Sigma$;

---

[1]This chapter uses the word "institution" for the signatures, sentences, models, model morphisms, and satisfaction associated with some logical system. This is useful because we work with many different logical systems. Institutions are formalized in [67]; see also the discussion of the satisfaction condition in Section 4.10.

(d) $\overline{M}_\pi(m_1,\ldots,m_n) = true$ if $\langle m_1,\ldots,m_n\rangle \in M_\pi$; and

(e) $\overline{M}_\pi(m_1,\ldots,m_n) = \pi(m_1,\ldots,m_n)$ if $\langle m_1,\ldots,m_n\rangle \notin M_\pi$.

4. Conversely, any $\overline{\Phi}$-algebra $A$ gives a $\Phi$-model $\overline{A}$ by dropping everything involving the sort B, and defining

$$\langle m_1,\ldots,m_n\rangle \in \overline{A}_\pi \text{ iff } A_\pi(m_1,\ldots,m_n) = A_{true}.$$

5. Now let $\overline{H}$ be the set of $\overline{\Phi}$-conditional equations of the form

$$(\forall X)\ p_0 = true \text{ if } \{p_1 = true, \ldots, p_n = true\},$$

one for each Horn clause in $H$ of the form

$$(\forall X)\ p_1 \wedge \cdots \wedge p_n \Rightarrow p_0.$$

6. Finally, define $T_H$ to be $\overline{T_{\overline{H}}}$, where $T_{\overline{H}}$ is the initial $(\overline{\Phi},\overline{H})$-algebra.

We now check that this construction works, i.e., that $T_H$ really is an initial $(\Phi,H)$-model. Let $M$ be any $(\Phi,H)$-model. Then we can use 3 and 5 to check that the algebra $\overline{M}$ satisfies $\overline{H}$. Hence there is a unique $\overline{\Phi}$-homomorphism $h : T_{\overline{H}} \to \overline{M}$, which then gives us a $\Phi$-morphism $\overline{h} : \overline{T_{\overline{H}}} = T_H \to \overline{\overline{M}} = M$. Uniqueness of this morphism follows from the fact that the translations described in 3 and 4 above define a bijective correspondence between $\Phi$-morphisms and $\overline{\Phi}$-homomorphisms. □

This proof translates Horn clauses into conditional equations and Horn models into algebras, and then exploits the existence of initial algebras. It is noteworthy that $\overline{\Pi}$ contains *true* but not *false*, and that the truth values of false atoms are just the atoms themselves. Note that by 3(a) and 6, the carrier $(T_H)_s$ for $s \in S$ is the same as that of $T_\Sigma$, because there are no equations among $\Sigma$-terms.

**Exercise 8.2.3** Give the details of the argument that $\overline{M}$ satisfies $\overline{H}$ for the above proof.   □

**Exercise 8.2.4** Show that if there are no atoms in $H$ then $(T_H)_\pi = \emptyset$ for each $\pi \in \Pi$.   □

Theorem 8.2.6 justifies defining relations "by induction" with Horn clauses: if we define a relation $\pi$ with Horn clauses $H$ and if $M$ is another $H$-model having the same carriers as the initial $H$-model $T_H$, then the unique $\Phi$-morphism $h : T_H \to M$ must be the identity, and so we must have $(T_H)_\pi \subseteq M_\pi$. The same argument works for any initial $H$-model; moreover, for any $H$-model $M$, the image of $(T_H)_\pi$ under the unique homomorphism $h$ to $M$ is contained in $M_\pi$, that is, $h((T_H)_\pi) \subseteq M_\pi$. Thus $(T_H)_\pi$ is the smallest relation satisfying the given formulae.

**Example 8.2.7** We use Theorem 8.2.6 to define the transitive closure of a relation, i.e., the least transitive relation containing the given relation. Here $\Sigma$ has just one sort, Elt, with $\Sigma_{[],\text{Elt}} = X$ for some set $X$, and with all other $\Sigma_{w,s}$ empty; and $\Pi$ has $\Pi_{\text{Elt Elt}} = \{R, R^*\}$, with all other $\Pi_w$ empty. Assuming $R$ is already defined on some set $X$, the following Horn clauses define the transitive closure $R^*$ of $R$, where the variables $x, x', x''$ range over $X$:

$$xRx' \Rightarrow xR^*x'$$
$$xR^*x' \wedge x'R^*x'' \Rightarrow xR^*x'' \ .$$

We can use the construction in the proof of Theorem 8.2.6 to justify an OBJ specification for the transitive closure R* of a relation R. Because this specification works for any R, we give a theory for R; this is preceded by a specification for the auxiliary sort B and its one truth value, denoted tt to distinguish it from OBJ's builtin Boolean value true.

```
obj B is sort B .
  op tt : -> B .
endo

th R is sort Elt .
  ex B .
  op _R_ : Elt Elt -> B .
endth
obj R* is pr R .  ex B .
  op _R*_ : Elt Elt -> B .
  vars X Y Z : Elt .
  cq X R* Y = tt if X R  Y == tt .
  cq X R* Z = tt if X R* Y == tt and Y R* Z == tt .
endo
```

Notice that the module B has initial semantics, R has loose semantics, and R* has again initial semantics. The notation "pr R . ex B ." in R* indicates that the sort Elt is protected but the sort B is only extended; although the single truth value of sort B will not be corrupted, it is likely that other terms of sort B will be added to serve as "false" values for relations. On the other hand, the module R is *protected* in R*, because the equations in R* do not affect the relation R. What all this means is that given any relation R, a unique R* is defined for it; this can be seen by considering the models of R*, and a more formal approach involving second-order quantifiers is also given in Chapter 9. Note that the second equation of the module R* cannot be used for reduction because its condition has variables that are not in its leftside; however OBJ3's apply can be used in equational deduction.

We can give a more idiomatic version of the above specification using builtin truth values; this is valid because if BOOL is protected in the module R, then it is necessarily also protected in R∗, because BOOL is protected in the module R below iff the relation R is correctly defined in whatever model is chosen.

```
th R is sort Elt .
  op _R_ : Elt Elt -> Bool .
endth

obj R* is pr R .
  op _R*_ : Elt Elt -> Bool .
  vars X Y Z : Elt .
  cq X R* Y = true if X R  Y .
  cq X R* Z = true if X R* Y and Y R* Z .
endo
```

(In OBJ, it is more natural to parameterize the object R∗ by the theory R, but since parameterization is not discussed until Chapter 11, we take a simpler approach here.) □

Example 8.2.7 illustrates a general method for replacing relations and Horn clauses by functions and equations. This implies we don't need to add relations and Horn clauses to an implementation of equational logic like OBJ.

**Exercise 8.2.5** (a) Define a relation R on the Peano numbers by

```
N R M = s N == M
```

and use (one of) the above object(s) R∗ and OBJ3's `apply` to show that

```
s 0 R* s s s 0 .
```

Now explain what R∗ is.

(b) Show that any relation is contained in a least equivalence relation, and give corresponding OBJ code defining that relation. Now use OBJ to compute three values of the equivalence closure of some simple (but not wholly trivial) relation. □

**Exercise 8.2.6** Show that X R∗ Y in the second equation of the specification R∗ in Example 8.2.7 can be replaced by X R Y without changing the meaning; say what "without changing the meaning" means in this context.

□

## 8.3 First-Order Logic

This section gives an algebraic treatment of first-order logic syntax, and then defines first-order satisfaction; after that, adding equality predicates and fixing a "data" model are considered; Gödel's completeness and incompleteness theorems are also informally discussed.

### 8.3.1 First-Order Syntax

We define the first-order sentences over a first-order signature $\Phi = (\Sigma, \Pi)$ by first defining $\Sigma$-terms and then defining $\Phi$-formulae; we will have to be careful about variables. Let $S$ be the sort set of $\Phi$, let $\mathcal{X}$ be an $S$-sorted set of variable symbols disjoint from $\Sigma$ and $\Phi$, such that each sort has an infinite number of symbols, and let $X$ be a fixed $S$-indexed subset of $\mathcal{X}$.

A $(\Phi, X)$-**term** is an element of $T_{\Sigma \cup X}$. Recall that the ($S$-indexed) function *Var* is defined on $T_{\Sigma \cup X}$ as follows:

0. $Var_s(\sigma) = \emptyset$ if $\sigma \in \Sigma_{[],s}$ ;
1. $Var_s(x) = \{x\}$ if $x \in X_s$ ;
2. $Var_s(\sigma(t_1, \ldots, t_n)) = \bigcup_{i=1}^{n} Var_{s_i}(t_i)$ for $n > 0$ .

(*Var* is the unique $(\Sigma \cup X)$-homomorphism $T_{\Sigma \cup X} \to \mathcal{P}(X)$ where the $S$-indexed set $\mathcal{P}(X)$ of all subsets of $X$ is given an appropriate $(\Sigma \cup X)$-structure.) We are now ready for the syntax of first-order logic:

**Definition 8.3.1** A (well-formed) $(\Phi, X)$-**formula** is an element of the carrier of the (one-sorted) free algebra $WFF_X(\Phi)$ defined to have the following as its (one-sorted) signature, which we denote $\Omega$ and call the **metasignature**:

0. a constant *true*,
1. a unary prefix operation $\neg$, called **negation**,
2. a binary infix operation $\wedge$, called **conjunction**,
3. a unary prefix operation $(\forall x)$ for each $x$ in $X$, called **universal quantification** over $x$,

plus as its *generators* (i.e., constants not in $\Omega$), the **atomic $(\Phi, X)$-formulae**, which are the elements of

$$G_X = \{\pi(t_1, \ldots, t_n) \mid \pi \in \Pi_{s_1 \ldots s_n} \text{ and } t_i \in (T_{\Sigma \cup X})_{s_i} \text{ for } i = 1, \ldots, n\} .$$

Note that $\Omega$ is infinite if $X$ is, because then there is an infinite number of unary operations $(\forall x)$, but of course all first-order formulae are finite. The symbols in $\Omega$ are called **logical** symbols, whereas those in $\Phi$ are called **non-logical** symbols. Let $WFF(\Phi) = WFF_X(\Phi)$; it contains every $WFF_X(\Phi)$, and its elements are called $\Phi$-**formulae**.

The functions *Var* and *Free*, giving the sets of all **variables**, and of all **free variables**, of $\Phi$-formulae, are defined by the following:

0. $Var(true) = Free(true) = \emptyset$,
1. $Var(\pi(t_1, \ldots, t_n)) = Free(\pi(t_1, \ldots, t_n)) = \bigcup_{i=1}^{n} Var(t_i)$,
2. $Var(\neg P) = Var(P)$, and $Free(\neg P) = Free(P)$,

3. $Var(P \wedge Q) = Var(P) \cup Var(Q)$, and $Free(P \wedge Q) = Free(P) \cup Free(Q)$, and

4. $Var((\forall x)P) = Var(P) \cup \{x\}$, and $Free((\forall x)P) = Free(P) - \{x\}$.

A variable that is not free is called **bound**; let $Bound(P) = Var(P) - Free(P)$. A $\Phi$-**sentence** is a $\Phi$-formula $P$ with no free variables, i.e., with $Free(P) = \emptyset$; $\Phi$-sentences are also called **closed** $\Phi$-**formulae**. A formula that is not closed is called **open**. Let $FoSen(\Phi)$ denote the set of all $\Phi$-sentences. □

**Exercise 8.3.1** Show that the functions *Var* and *Free* are $\Omega$-homomorphisms, by giving $\mathcal{P}(X)$ appropriate $\Omega$-algebra structures. □

We introduce the remaining logical connectives, *false*, $\vee$, $\Rightarrow$, $\Leftrightarrow$, and $(\exists x)$ (the last four are called **disjunction**, **implication**, **equivalence**, and **existential quantification**, respectively), as abbreviations for certain terms over the operations already introduced, as follows:

$$\begin{aligned}
\mathit{false} &= \neg \mathit{true} \\
P \vee Q &= \neg(\neg P \wedge \neg Q) \\
P \Rightarrow Q &= \neg P \vee Q \\
P \Leftrightarrow Q &= (P \Rightarrow Q) \wedge (Q \Rightarrow P) \\
(\exists x) P &= \neg((\forall x) \neg P) \ .
\end{aligned}$$

The symbols $P, Q$ above are variables over formulae, and the five operation symbols on the left sides of the equations extend the metasignature $\Omega$ to a new metasignature $\overline{\Omega}$. Given any $\Omega$-algebra $M$, the five equations above extend $M$ in a unique way to an $\overline{\Omega}$-algebra; moreover, any $\Omega$-homomorphism is automatically an $\overline{\Omega}$-homomorphism between its extended algebras. In particular, the $\Omega$-algebras $WFF_X(\Phi)$ extend to $\overline{\Omega}$-algebras, and the $\Omega$-homomorphisms *Var* and *Free* extend to $\overline{\Omega}$-homomorphisms that correctly handle the new logical symbols. Without these symbols, many of our theorem-proving applications would be much more awkward; this illustrates the conflict between logics for foundations and logics for applications discussed in Section 1.3.

**Exercise 8.3.2** Extend the recursive definitions of *Var* and *Free* so that they directly handle the new symbols in $\overline{\Omega}$. □

## 8.3.2 Scope

It might first appear that formulae like the following are ill-formed or ambiguous:

$(\exists x)(\forall x) \ geq(x, x)$
$(\exists x) \ (pos(x) \wedge (\forall x) \ geq(x, x))$
$(\exists x)(((\forall x) \ pos(x)) \wedge geq(x, x)) \ .$

But because we defined quantifiers as unary operations on expressions, every quantifier has a unique argument, a subformula called its **scope**. Every free instance of the quantifying variable within its scope is said to be **bound** (or **captured**) by that quantifier. Thus, in the first formula above, the two $x$'s in $geq(x,x)$ are bound to the universal quantifier, not to the existential. In the second formula, the first $x$ is bound to the existential quantifier, and the other two are bound to the universal. In the third formula, the first $x$ is bound to the universal and the next two are bound to the existential quantifier. It is poor style to keep reusing the same variable for quantifiers, and the following equivalent formulae would have been clearer:

$$(\exists y)(\forall x)\ geq(x,x)$$
$$(\exists x)\ (pos(x) \wedge (\forall y)\ geq(y,y))$$
$$(\exists y)(((\forall x)\ pos(x)) \wedge geq(y,y))\ .$$

However, the original formulae still have definite structure and meaning, due to our algebraic notion of formula, which does not require any prior definition of scope.

Of course, it is still very possible to write ambiguous formulae, such as

$$(\exists x)\ pos(x) \wedge geq(x,x)\ ,$$

where the argument (scope) of the existential quantifier cannot be determined; however, this is a parsing problem, not a problem in first-order logic as such (see Section 3.7). In fact, it is rather common to write ambiguous formulae when it doesn't matter which parse is taken. For example, in the formula

$$(\exists x)\ pos(x) \wedge (\forall x)\ geq(x,x)\ ,$$

it is not clear whether the existential quantifier acts on the universally quantified subformula, but it does not matter because that subformula is closed (see $E17$ of Exercise 8.3.10). This situation is much the same as in arithmetic when we write $x+y+z$ instead of $x+(y+z)$ or $(x+y)+z$, since we know it doesn't matter because addition is associative. In OBJ, precedence declarations can make quantifiers bind however tightly we wish.

In summary, our algebraic approach to first-order logic syntax has the advantage over the more *ad hoc* approaches usually found in the literature, of a clean separation between structure and parsing, which simply *avoids* complex and confusing definitions of scope.

### 8.3.3 First-Order Semantics

Many different systems of deduction have been given for first-order logic. Kurt Gödel first showed completeness for one of these with respect to a "semantic definition of truth" given by Tarski; this is a notion of satisfaction of first-order sentences by first-order models. All sound and complete systems are equivalent in the sense that they give rise to the same theorems for any theory. This chapter uses (a version of) Tarski's semantics for justifying a set of rules to transform complex proof tasks into Boolean combinations of simpler proof tasks; we call these *proof planning* rules; we do not attempt a completeness proof. More technically, the definitions in this subsection first extend assignments $a : X \to M$ from terms to first-order formulae $P$, and then define the "meaning" or **denotation** $[\![P]\!]$ of a formula $P$ to be the set of all assignments that make $P$ true.

**Definition 8.3.2** Given a first-order signature $\Phi = (\Sigma, \Pi)$, a $\Phi$-model $M$, and an **assignment** (of values in $M$ to variables in $X$), i.e., an $S$-indexed function $a : X \to M$, then we define[E32] $\bar{a} : WFF_X(\Phi) \to B$, where $B = \{true, false\}$ by the following:

0. $\bar{a}(true) = true$.

1. $\bar{a}(\neg P) = \neg \bar{a}(P)$.

2. $\bar{a}(P \wedge Q) = \bar{a}(P) \wedge \bar{a}(Q)$.

3. $\bar{a}((\forall x)P) = true$ iff $\bar{b}(P) = true$ for all $b : X \to M$ such that $b(z) = a(z)$ if $z \neq x$.

4. $\bar{a}(\pi(t_1, \ldots, t_k)) = true$ iff $(\bar{a}(t_1), \ldots, \bar{a}(t_k)) \in M_\pi$.

When $X$ is small, it may be convenient to use the notation $P[x_1 \leftarrow m_1, x_2 \leftarrow m_2, \ldots, x_n \leftarrow m_n]$ instead of $\bar{a}(P)$ with $X = \{x_1, \ldots, x_n\}$ and $a(x_i) = m_i$ for $i = 1, \ldots, n$.

We now define the **denotation** of a $(\Phi, X)$-formula $P$, written $[\![P]\!]_X^M$, or just $[\![P]\!]$, to be

$$\{a : X \to M \mid \bar{a}(P) = true\}.$$

Then $M$ **satisfies** $P \in WFF_X(\Phi)$, written $M \vDash_\Phi P$, iff $[\![P]\!]_X^M = [X \to M]$, i.e., iff all assignments from $X$ to $M$ make $P$ true.[E33] Given a set $A$ of well-formed $\Phi$-formulae, let $M \vDash_\Phi A$ mean $M \vDash_\Phi P$ for each $P \in A$, and let $A \vDash_\Phi P$ mean that $M \vDash_\Phi A$ implies $M \vDash_\Phi P$ for all $\Phi$-models $M$; when the symbol $\vDash$ is used in this way, it may be called **semantic entailment**. Note that $A$ need not be finite. A set $A$ of formulae is **closed** iff all its elements are closed. As usual, we may omit the subscript $\Phi$ on $\vDash_\Phi$ when it is clear from context. Let us write *FOL* for the institution[E34] of first-order logic with this notion of satisfaction. □

Intuitively, 3 in Definition 8.3.2 says $[\![(\forall x)P]\!]$ is the set of assignments that make $P$ true no matter what value they assign to $x$. This suggests that when $P$ has no free variables, $\bar{a}(P)$ should be independent of the values $a(x)$ for *all* $x \in X$. This is made precise in the following:

**Proposition 8.3.3** Given a $(\Phi, X)$-formula $P$ and assignments $a, a' : X \to M$, if $a(x) = a'(x)$ for all $x \in \text{Free}(P)$, then $\overline{a}(P) = \overline{a'}(P)$.

**Proof:** We use induction on the structure of $(\Phi, X)$-formulae. The two base cases are 0 and 4 of Definition 8.3.2. For 0, if $P = \text{true}$, then $\overline{a}(P) = \overline{a'}(P)$ for all $a, a'$. For 4, if $P = \pi(t_1, \ldots, t_k)$, then $\text{Free}(P) = \text{Var}(P) = \bigcup_{i=1}^{k} \text{Var}(t_i)$, and so if $a(z) = a'(z)$ for all $z \in \text{Free}(P)$ then $\overline{a}(t_i) = \overline{a'}(t_i)$ for $i = 1, \ldots, k$, and so $\overline{a}(P) = \overline{a'}(P)$.

There are three "step" cases, of which 1 and 2 are easy, because $\text{Free}(\neg P) = \text{Free}(P)$ and $\text{Free}(P \wedge Q) = \text{Free}(P) \cup \text{Free}(Q)$. For 3, assume $\overline{a}(P) = \overline{a'}(P)$ if $a(z) = a'(z)$ for all $z \in \text{Free}(P)$, and let $Q = (\forall x)P$. Then $\text{Free}(Q) = \text{Free}(P) - \{x\}$, call it $Z$, and so $a(z) = a'(z)$ for all $z \in Z$ implies $\overline{a}(Q) = \overline{a}((\forall x)P) = \text{true}$ iff $\overline{b}(P) = \text{true}$ for all $b$ such that $b(z) = a(z)$ if $z \neq x$. Also $\overline{a'}(Q) = \overline{a'}((\forall x)P) = \text{true}$ iff $\overline{b'}(P) = \text{true}$ for all $b'$ such that $b'(z) = a'(z)$ if $z \neq x$. Now suppose $\overline{a}(Q) = \text{true}$ and let $b' : X \to M$ be such that $b'(z) = a'(z)$ if $z \neq x$. Define $b : X \to M$ by $b(z) = a(z)$ if $z \neq x$ and $b(x) = b'(x)$. Then $b(z) = b'(z)$ for each $z \in \text{Free}(P)$, and so the induction hypothesis gives us $\overline{b'}(P) = \overline{b}(P) = \text{true}$ and also $\overline{a'}(Q) = \text{true}$. Similarly, we can show that $\overline{a'}(Q) = \text{true}$ implies $\overline{a}(Q) = \text{true}$. Therefore $\overline{a}(Q) = \overline{a'}(Q)$ if $a(z) = a'(z)$ for all $z \in Z$. □

**Corollary 8.3.4** If $P$ is a closed $(\Phi, X)$-formula and $M$ is a $\Phi$-model, then either $[\![P]\!]_X^M = \emptyset$ or else $[\![P]\!]_X^M = [X \to M]$.

**Proof:** Since $\text{Free}(P) = \emptyset$, we have $a(z) = a'(z)$ for all $z \in \text{Free}(P)$ for any $a, a'$ at all. Therefore $\overline{a}(P) = \overline{a'}(P)$ for all $a, a'$. Hence $\overline{a}(P) = \text{true}$ for all $a$, or else $\overline{a}(P) = \text{false}$ for all $a$. In the first case, $[\![P]\!] = [X \to M]$ while in the second $[\![P]\!] = \emptyset$. □

That is, any closed formula is either true or else false of any given model.

As usual, $[\![\_]\!]_X^M$ is an $\Omega$-homomorphism to a suitable target algebra, in this case with carrier $\mathcal{A}_X^M = \mathcal{P}([X \to M])$; we usually drop the superscript $M$ and subscript $X$. The $\Omega$-algebra structure for $\mathcal{A}$ is given as follows, for $A, B \subseteq [X \to M]$:

0. $\mathcal{A}_{\text{true}} = [X \to M]$.

1. $\mathcal{A}_\neg(A) = [X \to M] - A$.

2. $\mathcal{A}_\wedge(A, B) = A \cap B$.

3. $\mathcal{A}_{(\forall x)}(A) = \{a : X \to M \mid a'(y) = a(y) \text{ for all } y \neq x \text{ implies } a' \in A\}$.

If we now define $\alpha : \mathcal{G}_X \to \mathcal{A}_X$ by

$$\alpha(\pi(t_1, \ldots, t_n)) = \{a \mid (\overline{a}(t_1), \ldots, \overline{a}(t_n)) \in M_\pi\},$$

then $[\![\_]\!]$ is the unique $\Omega$-homomorphism $\text{WFF}_X(\Phi) \to \mathcal{A}_X$ extending $\alpha$; i.e., $[\![P]\!] = \overline{\alpha}(P)$.

**Definition 8.3.5** First-order $(\Phi, X)$-formulae $P, Q$ are (**semantically**) **equivalent**, written $P \equiv Q$, iff $[\![P]\!]_X^M = [\![Q]\!]_X^M$ for all $M$.  □

Note that $\equiv$ is neither a logical nor a non-logical symbol, but a *metalogical symbol*, used for talking about the satisfaction of formulae. Equivalent formulae are true under exactly the same circumstances and hence can be substituted for each other without changing the truth value of any formula of which they are part. (The ubiquity of this concept reflects the obsession of classical logic with truth.)

**Exercise 8.3.3** Given a first-order signature $\Phi$, a $\Phi$-model $M$, and $(\Phi, X)$-formulae $P, Q$, show the following:

(a) $M \models P \Rightarrow Q$  **iff**  $[\![P]\!]_X^M \subseteq [\![Q]\!]_X^M$ .
(b) $P \equiv Q$  **iff**  $M \models (P \Leftrightarrow Q)$  for all $M$ .
(c) $P \equiv Q$  **implies**  $M \models P$ iff $M \models Q$  for all $M$ .
(d) "**implies**" cannot be replaced by "**iff**" in (c) above.
(e) $[\![(\forall x)\, P]\!]_X^M \subseteq [\![P]\!]_X^M$.

Note that (c) implies that $P \equiv Q$ implies $[\![P]\!] = [X \to M]$ iff $[\![Q]\!] = [X \to M]$.  □

**Exercise 8.3.4** Given a first-order signature $\Phi$, and $\Phi$-formulae $P, Q, R$, show the following:

E1. $P \wedge Q$ $\equiv$ $Q \wedge P$ .
E2. $P \wedge (Q \wedge R)$ $\equiv$ $(P \wedge Q) \wedge R$ .
E3. $P \wedge P$ $\equiv$ $P$ .

Also, for $\Phi$-formulae $A, A', P, P'$, show that

E4.  $A \equiv A'$ and $P \equiv P'$  imply  $A \models P$ iff $A' \models P'$ .  □

E4 is a (weak) version of *Leibniz's principle*, that equal things may be substituted for each other; this supports the importance of $\equiv$, and underlines the somewhat strange view of traditional logic that all true sentences are equal (as are all false sentences). The following builds on E1–E3:

**Notation 8.3.6** In the notation "$A \models \ldots$" where $A$ is a set, we may write $A, P$ or $A \wedge P$ for $A \cup \{P\}$, and write $A, A'$ or $A \wedge A'$ for $A \cup A'$. For example, $A, P, Q \models P, Q$.  □

This notation makes sense because both set notation and conjunction are commutative, associative and idempotent. Any finite set $A$ can be regarded as the conjunction of its sentences, although this does not work if $A$ is infinite.

**Exercise 8.3.5** Let $\Phi$ be the signature of Example 8.1.2 above, let $M$ be the standard $\Phi$-model of Example 8.1.4, and let $X = \{x, y\}$. Now describe $[\![P]\!]$, for $P$ each of the following:

$geq(s(s(s(0))), x)$
$(\forall x) \, geq(s(s(s(0))), x)$
$(\forall x)(\forall y) \, geq(x, y) \wedge pos(x)$
$(\exists y) \, geq(x, y) \wedge pos(x)$
$(\forall y)(\exists x) \, geq(x, y) \wedge pos(x)$
$(\forall x) \, pos(x) \Rightarrow geq(y, x)$
$(geq(x, s(s(0))) \wedge geq(s(s(s(0))), x) \wedge geq(x, y)) \vee$
$\quad (geq(x, y) \wedge geq(y, x))$ □

**Proposition 8.3.7** Given a signature $\Phi$, a $\Phi$-model $M$, and $(\Phi, X)$-formulae $P, Q$, then:

P1.    $M \models P \wedge Q$    **iff**    $M \models P$ **and** $M \models Q$.
P2.    $M \models P \vee Q$    **if**    $M \models P$ **or** $M \models Q$.
P3.    $M \models P \vee Q$    **iff**    $M \models P$ **or** $M \models Q$    **if** $P$ or $Q$ is closed.
P4.    $M \models P \Rightarrow Q$    **iff**    $M \models P$ **implies** $M \models Q$    **if** $P$ is closed.
P5.    $M \models \neg\neg P$    **iff**    $M \models P$.
P5a.    $M \models \neg P$    **iff**    $M \models P$ **is false**    **if** $P$ is closed and $M$ nonempty.[2]
P6.    $M \models (\forall x) P$    **iff**    $M \models P$.
P7.    $M \models P$    **if**    $M \models Q$ **and** $M \models Q \Rightarrow P$.

**Proof:**

P1. Since $[\![P \wedge Q]\!] = [\![P]\!] \cap [\![Q]\!]$, we have $[\![P \wedge Q]\!] = [X \to M]$ iff $[\![P]\!] = [\![Q]\!] = [X \to M]$.

P2. Since $[\![P \vee Q]\!] = [\![P]\!] \cup [\![Q]\!]$, we have $[\![P \vee Q]\!] = [X \to M]$ if $[\![P]\!] = [X \to M]$ or $[\![Q]\!] = [X \to M]$.

P3. If $P$ is closed then $[\![P]\!] = [X \to M]$ or else $[\![P]\!] = \emptyset$ by Corollary 8.3.4. Therefore $[\![P \vee Q]\!] = [X \to M]$ iff $[\![P]\!] = [X \to M]$ or $[\![Q]\!] = [X \to M]$. The argument is the same for the $Q$ closed case.

P4. This follows from Corollary 8.3.4 plus the fact that

$$[\![P \Rightarrow Q]\!] = ([X \to M] - [\![P]\!]) \cup [\![Q]\!].$$

P6. We need

$$[\![(\forall x)P]\!] = [X \to M] \text{ iff } [\![P]\!] = [X \to M],$$

for $x \in X$. By (e) of Exercise 8.3.3, $[\![(\forall x)P]\!] = [X \to M]$ implies $[\![P]\!] = [X \to M]$. Conversely, if $[\![P]\!] = [X \to M]$, then for each $a : X \to M$ and each $b : X \to M$ such that $b(z) = a(z)$ for all $z \neq x$, we have $\overline{b}(P) = $ *true*, so that $a \in [\![(\forall x)P]\!]$. Therefore $[\![(\forall x)P]\!] = [X \to M]$.

---

[2] Recall that this means that all its carriers are non-empty.

$P5, P5a$ and $P7$ are left as exercises. □

$P5$ can be written equivalently as

$$\neg\neg P \equiv P ,$$

and is often called the law of *double negation*. $P6$ says that from the viewpoint of satisfaction, free variables are the same as bound variables that are universally quantified at the outermost level. The inference rule based on $P7$ is traditionally called *modus ponens* and goes back to the ancient Greeks (though the name is Latin); it is closely related to the proof rule $T8$ given later for lemma introduction.

**Example 8.3.8** The "**if**" in $P2$ of Proposition 8.3.7 cannot be replaced by "**iff**." Let $\Phi$ be the signature of Example 8.1.2, let $M$ be its standard model, and let $P, Q$ be the formulae $pos(x)$, $\neg geq(x, s(0))$, respectively. Then $M \models P \vee Q$ holds, but both $M \models P$ and $M \models Q$ are false. □

**Exercise 8.3.6** The following refer to Proposition 8.3.7 above:

(a) Give a signature $\Phi$, a $\Phi$-model $M$ and $\Phi$-formulae $P, Q$ which show that P4 does not hold without the restriction that $P$ is closed.

(b) Show that $M \models P \Rightarrow Q$ implies ($M \models P$ implies $M \models Q$), with neither $P, Q$ required to be closed.

(c) Prove $P5$ and $P5a$.

(d) Prove $P7$. □

The following results for semantic entailment are analoguous to those in Proposition 8.3.7:

**Proposition 8.3.9** Let $\Phi$ be a first-order signature, let $A$ be a set of $(\Phi, X)$-formulae, and let $P, Q$ be $(\Phi, X)$-formulae. Then

| | | | | |
|---|---|---|---|---|
| R0. | $A, P \models P$ . | | | |
| R1. | $A \models P \wedge Q$ | iff | $A \models P$ and $A \models Q$ . | |
| R2. | $A \models P \vee Q$ | if | $A \models P$ or $A \models Q$ . | |
| R3. | $A \models P \Rightarrow Q$ | iff | $A, P \models Q$ | if $P$ is closed . |
| R4. | $A \models \neg P$ | iff | $A, P \models $ *false* | if $P$ is closed . |
| R5a. | $A \models_\Phi (\forall x) P$ | iff | $A \models_\Phi P$ | if $x$ not free in $A$ . |
| R5. | $A \models_\Phi (\forall x) P$ | iff | $A \models_{\Phi(\{x\})} P$ | if $x$ not free in $A$ . |
| R5b. | $A, (\forall x) P \models Q$ | iff | $A, P \models Q$ . | |
| R6. | $A \models P$ | if | $A \models Q$ and $A \models Q \Rightarrow P$ . | |
| R6a. | $A \models P \Rightarrow R$ | if | $A \models P \Rightarrow Q$ and $A \models Q \Rightarrow R$ . | |

**Proof:** $R0, R1, R2$ follow from the definition of semantic entailment, using $P1, P2$ of Proposition 8.3.7. $R3$ follows from the calculation: $A \vDash P \Rightarrow Q$ iff for all models $M$,

$$
\begin{array}{ll}
(M \vDash A) \Rightarrow (M \vDash P \Rightarrow Q)) & \textbf{iff} \\
\neg(M \vDash A) \vee \neg(M \vDash P) \vee (M \vDash Q) & \textbf{iff} \\
\neg(M \vDash A \wedge P) \vee (M \vDash Q) & \textbf{iff} \\
(M \vDash A \wedge P) \Rightarrow (M \vDash Q) \, ,
\end{array}
$$

which is equivalent to $A, P \vDash Q$, where the first $\Rightarrow$ in the first line and all the **iff**s are in the metalanguage, while the second $\Rightarrow$ is in first-order logic, and where the first **iff** uses $P4$ of Proposition 8.3.7.

$R4$ follows from $R3$, by substituting *false* for $Q$ and $\neg P$ for $P$, and using $\neg P \equiv (P \Rightarrow false)$. $R5a, R5b, R6$ follow from $P4, P6, P7$ of Proposition 8.3.7, respectively.

For $R5$, by $R5a$, it suffices to show that

$$A \vDash_\Phi P \text{ iff } A \vDash_{\Phi(\{x\})} P \, ,$$

i.e., to show that the following are equivalent,

$$
\begin{array}{l}
\text{for all models } M, \ (M \vDash_\Phi A \ \Rightarrow \ M \vDash_\Phi P) \\
\text{for all models } M', \ (M' \vDash_{\Phi(\{x\})} A \ \Rightarrow \ M' \vDash_{\Phi(\{x\})} P) \, ,
\end{array}
$$

noting that the $M$ in the first assertion are $\Phi$-models, while the $M'$ in the second are $\Phi(\{x\})$-models. Since $x$ is not free in $A$, it is sufficient to show for all models $M, M \vDash_\Phi P$ iff for all models $M', M' \vDash_{\Phi(\{x\})} P$.

These two expressions respectively equal

$$
\begin{array}{l}
\text{for all models } M, \ [\![P]\!]^M_X \ = \ [X \to M] \\
\text{for all models } M', \ [\![P]\!]^{M'}_{X-\{x\}} = \ [(X - \{x\}) \to M'] \, ,
\end{array}
$$

which are equivalent because an assignment $a : X \to M$ to a $\Phi$-model $M$ is the same thing as an assignment $a' : (X - \{x\}) \to M'$ to a $\Phi(\{x\})$-model $M'$.

$R6a$ is left as an exercise. □

$R3$ is called the *Theorem of Deduction*. $R4$ says that to prove the negation of a closed formula, we can prove that its positive form is inconsistent with our assumptions; this strategy is called *proof by contradiction*. $R5$ is (a version of) the classical first-order logic *Theorem of Constants*, and the heart of its proof is similar to that for the equational case; see also (d) of Exercise 8.3.7 below. Note that in forming $\Phi(\{x\})$, $\Phi$ and $\{x\}$ are disjoint because $\Phi$ and $X$ are. Strictly speaking, we have changed the variable set from $X$ to $X - \{x\}$, so that occurrences of $x$ in $\Phi(\{x\})$-formulae are constants, not variables. $R6$ is the semantic entailment form of *modus ponens*. $R5b$ implies outermost universal quantifiers can be removed from the left of a turnstile; however this should not be done automatically, because it precludes substituting for the variable involved (see Section 8.3.4). $R6a$ expresses the *transitivity* of implication.

**Exercise 8.3.7** The following refer to Proposition 8.3.9:

(a) Give $\Phi, A, P$ showing that "**if**" in $R2$ cannot be replaced by "**iff**".

(b) Prove $R3$ with "**only if**" replacing "**iff**" and without the clause "**if** $P$ is closed.**"

(c) Give $\Phi, A, P$ showing that *neither* direction of the assertion $A \models \neg P$ **iff** $A \not\models P$ is correct.

(d) Generalize $R5$, replacing $x$ by a variable set $X$.

(e) Prove $R6a$.

**Hint:** Very simple choices will work for (a) and (c). □

Mathematics, and especially logic, is often said to deal with *absolute* or *eternal truths*, true under all possible interpretations in all possible models (or "worlds"). Mathematical truths are also said to be *formal truths*, "trivially" true for formal, non-empirical reasons (though of course establishing such truths can be non-trivial); the word "tautological" is also used.

**Definition 8.3.10** A $\Phi$-formula $P$ is a **tautology** iff $M \models P$ for every $\Phi$-model $M$. □

**Exercise 8.3.8** For $\Phi$ the signature of Example 8.1.2, show that each of the following is a tautology, or else show that it is not:

$(\forall x) (geq(x, 0) \Rightarrow pos(x))$
$geq(x, x)$
$geq(x, y) \vee \neg geq(x, y)$
$geq(x, y) \vee geq(y, x)$
$(\forall x) geq(x, x)$
$(\forall x) (geq(x, y) \Rightarrow geq(x, x))$
$(\forall x) (pos(x) \Rightarrow pos(x))$ .

**Hint:** Don't forget that there is no fixed $\Phi$-model here. □

**Exercise 8.3.9** Prove the following equivalences for $P, Q$ first-order formulae:

| | | | |
|---|---|---|---|
| E5. | $true \wedge P$ | $\equiv$ | $P$ . |
| E6. | $false \wedge P$ | $\equiv$ | $false$ . |
| E7. | $true \vee P$ | $\equiv$ | $true$ . |
| E8. | $false \vee P$ | $\equiv$ | $P$ . |
| E9. | $P \wedge \neg P$ | $\equiv$ | $false$ . |
| E10. | $P \vee \neg P$ | $\equiv$ | $true$ . |
| E11. | $\neg \neg P$ | $\equiv$ | $P$ . |
| E12. | $(\forall x)(\forall y) P$ | $\equiv$ | $(\forall y)(\forall x) P$ . |
| E13. | $(\forall x)(\forall x) P$ | $\equiv$ | $(\forall x) P$ . |

□

**Notation 8.3.11** Let $X$ be a variable set with elements $x_1, \ldots, x_n$. Then we may write $(\forall X)P$ for $(\forall x_1) \ldots (\forall x_n)P$. By E12 and E13, ordering and repetition of variables do not matter. Note that $(\forall X)$ does not make sense if $X$ is infinite. We extend existential quantifiers in the same way, to $(\exists X)$ where $X$ is a finite set of variables. □

Results about quantifiers generally extend by induction on the number of quantified variables. Let $\Phi(X)$ denote the first-order signature $(\Sigma(X), \Pi)$ when $\Phi = (\Sigma, \Pi)$.

**Proposition 8.3.12** Let $\Phi$ be a first-order signature, $A$ a set of $\Phi$-sentences, and $P$ a $\Phi$-formula. Then

| | | | |
|---|---|---|---|
| R5aX. | $A \vDash_\Phi (\forall X) P$ | iff | $A \vDash_\Phi P$ . |
| R5X. | $A \vDash_\Phi (\forall X) P$ | iff | $A \vDash_{\Phi(X)} P$ . |

$R5X$ is the classical Theorem of Constants. □

**Exercise 8.3.10** Prove the following equivalences for $P, P', Q, R$ first-order formulae:

| | | | | |
|---|---|---|---|---|
| E14. | $(P \Rightarrow Q) \wedge (P' \Rightarrow Q)$ | $\equiv$ | $(P \vee P') \Rightarrow Q$ . | |
| E15. | $(\forall X) (P \wedge Q)$ | $\equiv$ | $(\forall X)P \wedge (\forall X)Q$ . | |
| E16. | $(\forall X) P$ | $\equiv$ | $P$ | if $P$ is closed . |
| E17. | $(\exists X) (P \vee Q)$ | $\equiv$ | $(\exists X)P \vee (\exists X)Q$ . | |
| E18. | $(\exists X) (P \wedge Q)$ | $\equiv$ | $((\exists X)P) \wedge Q$ | if $Q$ is closed . |
| E19. | $(\exists X) P$ | $\equiv$ | $P$ | if $P$ is closed . |
| E20. | $(\forall X)(\forall Y) P$ | $\equiv$ | $(\forall Y)(\forall X) P$ . | |
| E21. | $(\exists X)(\exists Y) P$ | $\equiv$ | $(\exists Y)(\exists X) P$ . | |
| E22. | $(P \vee Q) \wedge R$ | $\equiv$ | $(P \wedge R) \vee (Q \wedge R)$ . | |

E22, as well as E5–E11, are examples of the very general principle that every equational law of Boolean algebra holds as an equivalence of first-order formulae, i.e., is a tautology; see also E1–E3. □

**Exercise 8.3.11** Let $\Phi$ be a first-order signature and let $P, Q$ be $\Phi$-formulae.

(a) Show that $P$ is a tautology iff $true \vDash_\Phi P$.

(b) Give an example showing that the condition "$Q$ is closed" is necessary in $E18$.

(c) Use Exercise 8.3.10 to show that if $Q$ is closed then $(\forall X)\,(P \vee Q) \equiv ((\forall X)\, P) \vee Q$.  □

We noted earlier that in our syntax for Horn clause logic, the symbols $\forall$, $\wedge$, and $\Rightarrow$ are not themselves logical symbols, but instead together constitute a single mixfix logical symbol. However, this notation does suggest a simple translation into first-order logic, where each symbol is taken as the corresponding logical symbol in first-order logic. This can perhaps be made clearer by adding parentheses, so that the translation of the Horn clause

$$h = (\forall X)\ p_1 \wedge \cdots \wedge p_n \Rightarrow p_0$$

is the first-order formula

$$h' = (\forall X)\,((p_1 \wedge \cdots \wedge p_n) \Rightarrow p_0)\,,$$

where of course the same signature $\Phi$ is used in each case. The following enables us to regard *HCL* as a "subinstitution" of *FOL*:

**Fact 8.3.13** Let $M$ be a $\Phi$-model, let $h_c$ be a Horn clause, and let $h'_c$ be its first order translation. Then

$$M \vDash_\Phi h_c \quad \text{iff} \quad M \vDash_\Phi h'_c\,.$$  □

**Exercise 8.3.12** Prove Fact 8.3.13 from the appropriate definitions of satisfaction.  □

The following demonstrates the very important fact that initial models *do not* always exist for theories over full first-order logic; this implies that it is not (in general) valid to do induction over models defined by sets of first-order sentences, even if they are supposed to be initial.

**Example 8.3.14** Let $\Sigma$ have one sort and two constants, $a, b$, let $\Phi$ have just one relation symbol, $\pi$, and let $A$ consist of the axiom $\pi(a) \vee \pi(b)$. This specification has no initial model: clearly the carrier must be $\{a, b\}$, but there is no way to get a smallest subset for $\pi$; in fact, there are two different equally good (and equally bad) minimal choices, namely $\{a\}$ and $\{b\}$.  □

**Proposition 8.3.15** Given a $(\Phi, X)$-sentence $(\exists x)P$ with $\textit{Free}(P) = \{x\}$ and a $\Phi$-model $M$ with all carriers nonempty,[E35] then $M \vDash (\exists x)P$ iff there is an assignment $a : X \to M$ such that $\overline{a}(P) = \textit{true}$.

**Proof:** By the following computation:

| | | |
|---|---|---|
| $M \models (\exists x)P$ | iff | (by definition of $\exists$) |
| $M \models \neg((\forall x) \neg P)$ | iff | (by *P5a*) |
| not $(M \models (\forall x) \neg P)$ | iff | (by *P6*) |
| not $(M \models \neg P)$ | iff | |
| not $(\llbracket \neg P \rrbracket = [X \to M])$ | iff | |
| not $(\llbracket P \rrbracket = \emptyset)$ | iff | |
| exists $a : X \to M$ with $\overline{a}(P) = true$ . | | □ |

**Exercise 8.3.13** Give examples showing how Proposition 8.3.15 fails if either $M$ has empty carriers, or $P$ has free variables other than $x$. □

### 8.3.4 Substitution

This section extends substitution from terms to first-order formulae, and gives the so-called Substitution Theorem, which will be important for several later developments, including that of quantifiers.

**Definition 8.3.16** Let $\Phi = (\Sigma, \Pi)$ be a first-order signature and let $\theta : X \to T_\Sigma(X)$ be a substitution. Now define $\hat{\theta} : \mathit{WFF}_X(\Phi) \to \mathit{WFF}_X(\Phi)$ recursively as follows:

0. $\hat{\theta}(\pi(t_1, \ldots, t_n)) = \pi(\overline{\theta}(t_1), \ldots, \overline{\theta}(t_n))$ ;
1. $\hat{\theta}(\mathit{true}) = \mathit{true}$ ;
2. $\hat{\theta}(\neg P) = \neg\hat{\theta}(P)$ ;
3. $\hat{\theta}(P \wedge Q) = \hat{\theta}(P) \wedge \hat{\theta}(Q)$ ;
4. $\hat{\theta}((\forall x)P) = (\forall x)\widehat{\theta_x}(P)$ ,

where $\theta_x$ is the substitution that agrees with $\theta$ everywhere on $X$ except $x$, and $\theta_x(x) = x$. We may write $\theta(P)$, or sometimes more elegantly $P\theta$, for $\hat{\theta}(P)$, and call it the result of **applying** $\theta$ to $P$, or of **substituting** $\theta(x)$ in $P$ for each $x \in X$. When $X$ is small, the notation $P[x_1 \leftarrow t_1, \ldots, x_n \leftarrow t_n]$ may be more convenient than $P\theta$. □

The simplicity of this definition, which as usual is recursive over $\Omega$, may come as a pleasant surprise. Notice that $\hat{\theta}$ automatically avoids substituting for bound variables. However, there is a subtle difficulty:

**Example 8.3.17** Let $\Phi$ be the signature of Example 8.1.2, let $X = \{x, y, z\}$, let $\theta(x) = 0$, $\theta(y) = s(x)$, and $\theta(z) = z$. Then

$\theta(\mathit{geq}(y, x))$ $= \mathit{geq}(s(x), 0)$.
$\theta((\forall x)(\mathit{geq}(x, 0) \Rightarrow \mathit{pos}(x)))$ $= (\forall x)(\mathit{geq}(x, 0) \Rightarrow \mathit{pos}(x))$.
$\theta((\forall x)(\mathit{geq}(x, y) \Rightarrow \mathit{geq}(x, x))) = (\forall x)(\mathit{geq}(x, s(x)) \Rightarrow \mathit{geq}(x, x))$.

Note the capture of the $x$ in $s(x)$ by the quantifier in the last formula, although the variable $y$ that it replaced was free in this formula. This phenomenon is called **variable capture**.

Now define a substitution $\tau$ by $\tau(x) = 0$, $\tau(y) = 0$, $\tau(z) = z$. Then $(\theta;\tau)(x) = 0$, $(\theta;\tau)(y) = s(0)$, and $(\theta;\tau)(z) = z$, so that if we let $P$ denote the third formula above, then $P(\theta;\tau) = (\forall x)(geq(x,s(x)) \Rightarrow geq(x,x))$, whereas $(P\theta)\tau = (\forall x)(geq(x,s(0)) \Rightarrow geq(x,x))$. Thus variable capture thwarts the compositionality of substitution. This motivates Definition 8.3.18 below.  □

**Exercise 8.3.14** We can extend the notation $\theta_x$ to $\theta_Z$ for $Z \subseteq X$, by defining $\theta_Z(y) = \theta(y)$ for $y \notin Z$ and $\theta_Z(y) = y$ for $y \in Z$. Show the following for any $P$ in $WFF_X(\Phi)$ and substitution $\theta$:

1. $\theta = \theta_Z$ iff $\theta$ is the identity on (at least) $Z$.

2. $\theta((\forall Z)\ P) = (\forall Z)\ \theta_Z(P)$.

3. $\theta(P) = P$ if $P$ is closed.

4. More generally, $\theta_{Free(P)}(P) = P$.

5. Even more generally, $\theta(P) = \tau(P)$ if $\theta(y) = \tau(y)$ for $y \in Free(P)$.

□

**Definition 8.3.18** Given a $(\Phi, X)$-formula $P$ and a substitution $\theta$, define $\theta$ to be **capture free for** $P$ as follows:

0. $\theta$ is capture free for $P$ if $P$ is atomic;

1. $\theta$ is capture free for *true*;

2. $\theta$ is capture free for $\neg P$ if it is for $P$;

3. $\theta$ is capture free for $P \wedge Q$ if it is for $P$ and for $Q$; and

4. $\theta$ is capture free for $(\forall x)P$ if $\theta_x$ is capture free for $P$ and if $y \neq x$ is a free variable of $P$ then $x$ is not free in $\theta(y)$.

Capture freedom extends from the operations in $\Omega$ to those in $\overline{\Omega}$; for example, $\theta$ is capture free for $(\exists x)P$ under exactly the same conditions as those for $(\forall x)P$.  □

**Proposition 8.3.19** Let $\theta, \tau$ be two substitutions such that $\theta$ is capture free for $P$. Then

1. $(P\theta)\tau = P(\theta;\tau)$, and

2. if $\tau$ is capture free for $P\theta$, then $\theta;\tau$ is capture free for $P$.

**Proof:** We prove 1. by structural induction over $\Omega$. We leave the reader to check the result for $P$ atomic or *true*, and for negation and conjunction. Suppose $P = (\forall x)Q$. Then $(P\theta)\tau = (\forall x)((Q\theta_x)\tau_x)$ and $P(\theta;\tau) = (\forall x)Q(\theta;\tau)_x$; because $\theta$ is capture free for $P$, so is $\theta_x$ for $Q$; thus by the induction hypothesis, $(Q\theta_x)\tau_x = Q(\theta_x;\tau_x)$.

Now we claim $Q(\theta_x;\tau_x) = Q(\theta;\tau)_x$. By 5. of Exercise 8.3.14, it suffices to show $(\theta_x;\tau_x)(y) = (\theta;\tau)_x(y)$ for all $y \in \text{Free}(Q)$. If $x \in \text{Free}(Q)$ then $(\theta_x;\tau_x)(x) = x = (\theta;\tau)_x(x)$; if $y \neq x$ is in $\text{Free}(Q)$, then $(\theta;\tau)_x(y) = (\theta;\tau)(y)$ and also $(\theta_x;\tau_x)(y) = \tau_x(\theta_x(y)) = \tau_x(\theta(y)) = \tau(\theta(y))$, the last equality because $\theta$ is capture free for $P$, since $x$ does not occur in $\theta(y)$.

We also prove 2. by structural induction over $\Omega$. Because capture freedom commutes with negation and conjunction, as does substitution, it suffices to check the induction step for $P = (\forall x)Q$. But $(\theta;\tau)_x$ is capture free for $Q$ because $\theta_x$ is capture free for $Q$ and $\tau_x$ is capture free for $Q\theta_x$, plus the induction hypothesis. Now let $y \neq x$ be a free variable in $Q$; because $x \notin Var(\theta(y))$ and $x \notin Var(\tau(z))$, for any free variable $z \neq x$ of $Q\theta_x$, we have $x \notin Var(\tau(\theta(y)))$. Therefore $\theta;\tau$ is capture free for $P$. □

**Definition 8.3.20** Given a substitution $\theta : X \to T_\Sigma(X)$ and a model $M$, define $[\![\theta]\!]_M : [X \to M] \to [X \to M]$ by $[\![\theta]\!]_M(a) = \theta;\overline{a}$. As usual, we omit the subscript $M$ when context makes it unnecessary. □

The next result does most of the work involved in proving the main result of this subsection; its rather technical proof has been placed in Appendix B to avoid distraction.

**Proposition 8.3.21** If $\theta$ is capture free for $P$, then $[\![\theta(P)]\!]_M = [\![\theta]\!]_M^{-1}([\![P]\!]_M)$ for any model $M$. □

The main result now says that any substitution instance of a valid formula is valid:

**Theorem 8.3.22** (*Substitution*) Let $\Phi = (\Sigma, \Pi)$ be a first-order signature, $A$ a set of $\Phi$-sentences, a $(\Phi, X)$-formula $P$, and $\theta : X \to T_\Sigma(X)$ a substitution that is capture free for $P$. Then

$$A \vDash P \quad \textbf{implies} \quad A \vDash P\theta \;.$$

**Proof:** Fix a model $M$ of $A$. It suffices to show that $[\![P]\!] = [X \to M]$ implies $[\![P\theta]\!] = [X \to M]$, which follows directly from Proposition 8.3.21. □

**Corollary 8.3.23** Let $\Phi = (\Sigma, \Pi)$ be a first-order signature, $A$ a set of $\Phi$-sentences, and $P$ in $WFF_X(\Phi)$. Let $Y \subseteq X$ be a finite variable set and let $\theta : X \to T_\Sigma(X)$ be a substitution such that $\theta_Y = \theta$ (i.e., $\theta$ can only be non-identity outside $Y$). Then

$$A \vDash (\forall Y)\, P \quad \textbf{implies} \quad A \vDash (\forall Y)\, P\theta \;.$$

**Proof:** Let $Q = (\forall Y) P$ and apply$^{E36}$ Theorem 8.3.22 to get

$$A \vDash (\forall Y) P \text{ implies } A \vDash \theta((\forall Y) P) .$$

Now by 2 of Exercise 8.3.14 and because $\theta_Y = \theta$, we get

$$A \vDash (\forall Y) P \text{ implies } A \vDash (\forall Y) P\theta . \qquad \square$$

From this, it further follows that

$$A \vDash (\forall Y) P \text{ implies } A \vDash (\forall Y') P\theta .$$

when $Y' \subseteq Y$ and $\theta = \theta_{Y-Y'}$. This says we can substitute values for variables in $Z = Y - Y'$ and eliminate their quantifiers.

The second result below is needed in Section 8.5.

**Lemma 8.3.24** Let $P$ be a $\Phi$-formula, let $\theta : X \to T_\Sigma(X)$ be a substitution that is capture free for $P$, and let $a : X \to M$ be an interpretation in a $\Phi$-model $M$. Then $\overline{a}(P\theta) = \overline{(\theta;\overline{a})}(P)$.

**Proof:** More precisely, we need to show that $\hat{\theta};\overline{a} = \overline{\theta;\overline{a}}$, which follows by showing that $(\hat{\theta};\overline{a})(x) = (\theta;\overline{a})(x)$ for all $x \in X$, and that $\hat{\theta};\overline{a}$ satisfies the conditions of Definition 8.3.2. $\qquad \square$

**Lemma 8.3.25** Let $P$ be a $\Phi$-formula and let substitutions $\theta, \theta' : X \to T_\Sigma(X)$ be capture free for $P$. Given interpretations $a, a' : X \to M$ in a $\Phi$-model $M$, then $\overline{a}(P\theta) = \overline{a'}(P\theta')$ whenever $\overline{a}(\theta(z)) = \overline{a'}(\theta'(z))$ for all $z \in \text{Free}(P)$.

**Proof:** By Lemma 8.3.24, $\overline{(\theta;\overline{a})}(P) = \overline{a}(P\theta)$ and $\overline{(\theta';\overline{a'})}(P) = \overline{a'}(P\theta')$. Because $(\theta;\overline{a})(x) = (\theta';\overline{a'})(x)$ for all $x \in \text{Free}(P)$, Proposition 8.3.3 gives $\overline{(\theta;\overline{a})}(P) = \overline{(\theta';\overline{a'})}(P)$. Thus $\overline{a}(P\theta) = \overline{a'}(P\theta')$. $\qquad \square$

### 8.3.5 First-Order Logic with Equality

The syntax of **first-order logic with equality** is exactly the same as that of first-order logic, but its signatures are required to have binary (infix) equality predicates, exactly one for each sort $s \in S$, denoted $=_s$; more precisely, we assume that $\{=_s\} \subseteq \Pi_{ss}$ for each $s \in S$. Semantically, first-order logic with equality restricts its models to those where the equality predicates are interpreted as actual identities; that is, for each model $M$ and $s \in S$,

$$M_{=_s} = \{\langle m, m \rangle \mid m \in M_s\} .$$

Satisfaction is as usual. Let us denote this institution *FOLEQ*. It is important to notice that all our definitions and results for *FOL* carry over to *FOLEQ*. This is because the proofs are the same, the only difference being that there are fewer models.

Similarly, **Horn clause logic with equality** is Horn clause logic with the same given equality predicates, interpreted the same way as above; let us denote this institution *HCLEQ*.

The **first-order logic of equality** is the special case of first-order logic with equality where equalities are *the only* predicates. Since $\Phi$ is completely determined by $\Sigma$, we may write $\vDash_\Sigma$ instead of $\vDash_\Phi$ in this context. Again, the definitions, results, and proofs for *FOL* carry over. Let us denote this institution *FOLQ*.

Similarly, the **Horn clause logic of equality** is Horn clause logic with equality where the only predicates are the equalities. In fact, the Horn clause logic of equality is the same as conditional equational logic (see the exercise below); therefore it is another way to view the logic of OBJ. Of course, our algebraic orientation prefers the conditional equational formulation to the Horn clause formulation.

**Exercise 8.3.15** Let $\Phi = (\Sigma, \Pi)$ be a first-order signature with exactly one predicate symbol for each sort, namely the equality. Now define a translation from conditional $\Sigma$-equations $e$ to $\Phi$-Horn clauses $h_e$, and prove that $M \vDash_\Phi e$ iff $M \vDash_\Phi h_e$, for any $\Phi$-model $M$.  □

It now follows that the institution *CEQL* of conditional equational logic is a subinstitution of *FOLEQ*. Hence the rules of deduction for *CEQL* are also valid for *FOLEQ*, of course restricted to the sentences that correspond to conditional equations.

### 8.3.6 Logic over a Fixed Model

Going further along the line of Section 8.3.5, we can give fixed "standard" interpretations not only for equality symbols, but also for any desired sorts and non-logical symbols. For example, if $\Psi$ is the signature $\Phi^{\text{Nat}}$ of Example 8.1.2 and if $\Phi$ is some first-order signature with $\Psi \subseteq \Phi$, then we can fix the interpretation of $\Psi$ to be the standard natural numbers. Define a $\Phi$-**model over** $D$ to be a $\Phi$-model $M$ such that the restriction (reduct) $M|_\Psi$ of $M$ to $\Psi$ is the fixed model $D$. We denote this institution *FOL/D*; then all our definitions and results for *FOL* carry over to *FOL/D*, because the same proofs work on the reduced collection of models. If $A$ is a set of $\Phi$-axioms, then a $(\Phi, A)$-model over $D$ is a $\Phi$-model over $D$ satisfying $A$. Note that for some $A$ there may be no such models, for example, if $A$ implies a $\Psi$-sentence that is false in $D$ (such as $1 = 0$).

Similarly, we can consider the institution *FOLEQ/D* of first-order logic with equality over some fixed $\Psi$-model $D$. If we add a few more arithmetic operations to the signature $\Psi = \Phi^{\text{Nat}}$ of the natural numbers, we get a system to which Gödel's incompleteness theorem applies. This famous result says that any first-order theory rich enough to talk about a certain fragment of arithmetic will always have true sentences that cannot be proved; in other words, no finite set of axioms can be complete for arithmetic. The situations that arise in our applications are often of this kind, since we need to reason about some fixed data types, e.g., natural numbers, integers, lists of natural numbers, etc. In practice, when we stumble over a result that cannot be proved by equational reasoning from the axioms in our theory, we try to prove it using induction. Induction is a second-order axiom, not a first-order axiom, but even so, there is no guarantee that we will find the proof we want by using it.

We can also consider the institution of the first-order logic *of* equality over a fixed model *D*, denoted *FOLQ/D*. The definitions and results for *FOL* again carry over, because the proofs are the same; and the above discussion about incompleteness also applies. *FOLQ/D* is fundamental for this book, because our method is to state proof tasks using formulae in this logic, and then reduce them to a combination of equational proof tasks that can be handled with reduction (see Section 8.4 below). Because *D* is usually defined by initiality with respect to some given equational theory, induction can usually be used to prove additional properties of *D* that are needed (such properties are traditionally called "lemmas").

## 8.4 Proof Planning

For proof planning, we will use the 2-bar semantic entailment turnstile, $\vDash$, in a new way, reading "$A \vDash P$" as indicating the **task** of proving the **goal** *P* from the **assumptions** *A*. With this in mind, we can reformulate the assertions of Proposition 8.3.9 as "proof planning rules," rewrite rules that transform complex proof tasks into combinations of simpler proof tasks. Given a first-order signature $\Phi$, a set *A* of $\Phi$-sentences, and $\Phi$-formulae *P*, *Q*, these rules are as follows:

| | | | | |
|---|---|---|---|---|
| T0. | $A, P \vDash_\Phi P$ | $\rightarrow$ | *true* . | |
| T1. | $A \vDash_\Phi P \wedge Q$ | $\rightarrow$ | $A \vDash_\Phi P$ **and** $A \vDash_\Phi Q$ . | |
| T2. | $A \vDash_\Phi P \vee Q$ | $\rightarrow$ | $A \vDash_\Phi P$ **or** $A \vDash_\Phi Q$ . | |
| T3. | $A \vDash_\Phi P \Rightarrow Q$ | $\rightarrow$ | $A, P \vDash_\Phi Q$ | if *P* is closed . |
| T4. | $A \vDash_\Phi \neg P$ | $\rightarrow$ | $A, P \vDash_\Phi$ *false* | if *P* is closed . |
| T5. | $A \vDash_\Phi (\forall X) P$ | $\rightarrow$ | $A \vDash_{\Phi(X)} P$ . | |

Thus, *T*1 says that to prove $P \wedge Q$ from *A*, we can prove *P* from *A* and *Q* from *A*. These really are rewrite rules rather than equations, because they have a definite left to right orientation. In particular, *T*2 cannot be reversed (although the others can): that is, it "creates truth" rather than "preserves truth," in the sense that if its rightside is true, then its leftside is true, but not *vice versa*, because *R*2 of Proposition 8.3.9 is only valid with "**if**," not "**iff**". Rule *T*4 justifies proofs by contradiction; but since we do not yet have any good way of proving *false*, we must supplement this rule later. Note that the signature subscripts on $\vDash$ are important for $T5$, but not for $T0$–$T4$. We will call the signature that appears as a subscript on the turnstile the **working signature** of the proof task $A \vDash_\Phi P$ in this context.

We say a proof planning rule is **sound** if its rightside is a *sufficient* condition for its leftside; note that this is *opposite* to soundness for rules of deduction, where the rightside is a *necessary* condition for the leftside. For example, if we weaken rule $T2$ to two rules saying that to prove $P \vee Q$ from $A$, it suffices to prove $P$, and it also suffices to prove $Q$, then the right sides are sufficient for their left sides, but far from necessary:

$T2a.$  $A \vDash_\Phi P \vee Q \;\longrightarrow\; A \vDash_\Phi P$ .
$T2b.$  $A \vDash_\Phi P \vee Q \;\longrightarrow\; A \vDash_\Phi Q$ .

We will see that rules like these are adequate for many interesting problems, including the ripple carry adder discussed in Section 8.4.2 below. We will also see that these rules can themselves be expressed and executed in OBJ.

As a first step in making the above intuitions more precise, let us consider the language used for expressing proof tasks. We can see that all the terms in $T0$–$T5$ are Boolean combinations of atoms of the form $A \vDash_\Phi B$, where $A, B$ are $\Phi$-formulae. Since our proof-planning applications involve atomic sentences from the institution *FOLQ/D*, we may write $\vDash_\Sigma$ instead of $\vDash_\Phi$. The terms in $T0$–$T5$ are "metasentences" over *FOLQ/D*: they make assertions about combinations of proof tasks involving sentences in *FOLQ/D*. Of course, most assertions of this form are false.

Our paradigm takes a proof task $A \vDash_\Phi P$ and transforms it into a Boolean combination of proof tasks that can be checked by reduction with OBJ. Proposition 8.3.9 then implies that if we use $T0$–$T5$ for the transformations, and if the OBJ reductions show that all atoms have the expected values, then $A \vDash_\Phi P$ is true, i.e., the proof score consisting of those reductions really does prove $P$ from $A$. If some reductions don't do what we want, then we have failed to prove the result, but in general, this does not mean it isn't true (though there are some cases where failure does imply that the original proof task is false).

The rules in the object META below encode the transformations $T0$–$T5$ in a somewhat abstract way; e.g., they leave signatures undefined, though we have algebraic signatures in mind.[3] Ground terms of sort Meta are metasentences that describe structures of possible proofs; they could also be called "proof terms," because they are possible proofs expressed as terms. This module uses *order-sorted algebra*; for example, the line "subsort BType < Type" means that every BType (for "basic type") is also a `Type`. Order-sorted algebra is developed in Chapter 10, but the OBJ code below should be understandable without a detailed knowledge of Chapter 10.

---
[3] Strictly speaking, the equations here are really rewrite rules, so that the full power of equational logic cannot be used, but only term rewriting.

```
obj META is sorts Meta Sen Sig Type .
  pr QID .
  dfn BType is QID .
  subsort BType < Type .
  subsort Bool  < Sen Meta .

  op _|=[_] _ : Sen Sig Sen -> Meta [prec 11].
  op (_)[_:_] : Sig Id Type -> Sig .
  op _and_ : Meta Meta -> Meta [assoc comm prec 2].
  op _and_ : Sen  Sen  -> Sen  [assoc comm prec 2].
  op _or_  : Meta Meta -> Meta [assoc comm prec 7].
  op _or_  : Sen  Sen  -> Sen  [assoc comm prec 7].
  op _=>_  : Meta Meta -> Meta [prec 9].
  op _=>_  : Sen  Sen  -> Sen  [prec 9].
  op not_  : Meta      -> Meta [prec 1].
  op not_  : Sen       -> Sen  [prec 1].
  op (all_:_ _) : Id Type Sen -> Sen .

  vars A P Q : Sen .  var X : Id .
  var T : Type .  var S : Sig .
  [ass] eq A and P |=[S] P       = true .
  [and] eq A |=[S] (P and Q) =  (A |=[S] P) and (A |=[S] Q) .
  [or]  eq A |=[S] (P or Q)  =  (A |=[S] P) or  (A |=[S] Q) .
  [imp] eq A |=[S] (P => Q)  =  (A and P) |=[S] Q .
  [not] eq A |=[S] (not P)   =  (A and P) |=[S] false .
  [all] eq A |=[S] (all X : T P) =  A |=[S][X : T] P .
endo
```

Note that the Boolean operations `and`, `or`, `=>`, and `not` are triply overloaded, because they are defined for both sentences and metasentences, as well as for OBJ's builtin Booleans. Since we have shown these operations to be associative and commutative, we can include these laws as attributes.

Strictly speaking, the rule `all` should require that the variable X not occur in A, and most texts on first-order logic do give such a "side condition" for this rule. However, it is more natural in our setting to consider this a condition on signatures, since forming $\Sigma(X)$ already requires $X$ to be disjoint from $\Sigma$. This well-formedness condition is easily expressed using so-called "error supersorts" in order-sorted algebra, but because we have not treated that topic, we omit this from the above specification.

Now let's use this machinery to plan some proofs:

**Example 8.4.1** Below is a simple reduction of a compound proof task to a simpler proof task. This computation tells us that if we want to prove a

sentence of the form

$$A \vDash_\Sigma (\forall w_1, w_2 : Bus) \ P_1 \Rightarrow P_2 \ ,$$

then it suffices to prove

$$A, P_1 \vDash_{\Sigma(w_1, w_2 : Bus)} P_2 \ .$$

Here is the OBJ code:

```
open .
ops A1 P1 P2 : -> Sen .
op Sigma : -> Sig .
red A1 |=[Sigma] (all 'w1 : 'Bus (all 'w2 : 'Bus P1 => P2)).
***> should be: A1 and P1 |=[Sigma] ['w1 : 'Bus]
***>                                 ['w2 : 'Bus] P2
close
```

Of course, it works; OBJ does just three rewrites, each an application of a proof rule. This reduction justifies the proof score used in the example of Section 8.4.2. □

**Example 8.4.2** We can use META to plan a proof that the intersection of two transitive relations is transitive. Our proof task has the form

$$(\forall X) \ P_1 \Rightarrow Q_1, \ (\forall X) \ P_2 \Rightarrow Q_2 \ \vDash_\Sigma \ (\forall X) \ P_{12} \Rightarrow Q_{12}$$

where $X$ has variables $x, y, z$ of sort Elt, and where the first clause expresses transitivity of a relation, e.g., $R_1$, with

$$P_1 = (x \ R_1 \ y) \wedge (y \ R_1 \ z)$$
$$Q_1 = x \ R_1 \ z \ ,$$

the second clause expresses transitivity of $R_2$, and the third expresses transitivity of their intersection, which is defined to be $R_1 \wedge R_2$. This definition justifies adding the two lemmas

$$P_{12} = P_1 \wedge P_2$$
$$Q_{12} = Q_1 \wedge Q_2 \ .$$

Now we can write the proof task in OBJ, and reduce it to get a proof plan:

```
open .
op all-X_ : Sen -> Sen .
op Sigma : -> Sig .
vars-of .
eq all-X A = (all 'x : 'Elt (all 'y : 'Elt (all 'z : 'Elt A))).
ops P1 P2 P12 Q1 Q2 Q12 : -> Sen .
eq P12 = P1 and P2 .
eq Q12 = Q1 and Q2 .
red ((all-X (P1 => Q1)) and (all-X (P2 => Q2))) |=[Sigma]
    (all-X (P12 => Q12)).
close
```

OBJ3 does 10 rewrites and produces a rather large term, suggesting that setting up this proof is not completely trivial. □

**Exercise 8.4.1** Execute the reduction in the example above in OBJ3 and interpret the result. Does it make sense? What does it say? Now use OBJ to actually do the proof that has been planned, and interpret the results.
□

**Exercise 8.4.2** In a way similar to the above example and exercise:

(a) Use OBJ to plan and carry out a proof that the union of two symmetric relations (on the same set) is symmetric.

(b) Use OBJ to plan and carry out a proof that the intersection of two equivalence relations (on the same set) is an equivalence relation.
□

The transformation corresponding to *modus ponens* (R6 on page 261) is

T6. $A \vDash_\Phi P \longrightarrow A \vDash_\Phi Q$ **and** $A \vDash_\Phi Q \Rightarrow P$ ,

which is not a rewrite rule, because its rightside contains a variable not in its leftside; hence it cannot be applied automatically by rewriting. But it is still very important for proofs.

We mentioned earlier that to use the rule T4 (proof by contradiction), we need an effective way to prove *false*. Rule T7 below gives one; note that it requires inventing and introducing a new formula $Q$ that can be both proved and disproved. Pure equational logic can never prove disequalities (i.e., negations of equations). But initiality gives us a way forward. If a specification is canonical, then different reduced ground terms necessarily denote distinct elements of its initial model. For example, we know that $0 \neq 1$ and *false* $\neq$ *true* are satisfied in the standard models, so if we can prove $0 = 1$ or *false* = *true*, then we have the desired contradiction.

The second rule below, T8, justifies introducing a "lemma" $Q$ to help prove $P$ from $A$; of course, $Q$ itself must also be valid for $A$. In practice, lemmas are often results about $D$ that require induction, such as the associative and commutative laws for addition. We will see some more substantial lemmas in the proof of the next section.

T7. $A \vDash_\Phi$ *false* $\longrightarrow A \vDash_\Phi Q$ **and** $A \vDash_\Phi \neg Q$ .
T8. $A \vDash_\Phi P \longrightarrow A \vDash_\Phi Q$ **and** $A, Q \vDash_\Phi P$ .

The following justifies the two rules above:

**Proposition 8.4.3** Let $\Phi$ be a first-order signature, let $A$ be a set of $\Phi$-sentences, and let $P, Q$ also be $\Phi$-sentences. Then

R7. $A \vDash_\Phi$ *false* **iff** $A \vDash_\Phi Q$ **and** $A \vDash_\Phi \neg Q$ .
R8. $A \vDash_\Phi P$ **if** $A \vDash_\Phi Q$ **and** $A, Q \vDash_\Phi P$ .

**Proof:**  The first assertion follows from $R1$ of Proposition 8.3.9 and $E9$ of Exercise 8.3.9. The second assertion can be shown directly from the definitions.  □

**Exercise 8.4.3**  (a) Show that "**if**" in R8 above cannot be replaced by "**iff**".

(b) Use $R3$ and $R6$ to show $R8$ if $Q$ is closed.  □

Another useful rule allows us to "strengthen" the axioms (or assumptions) used for a proof. Intuitively, if we can prove something from stronger (i.e., more restrictive) assumptions, then it is also valid under the weaker assumptions:

$R9. \quad A \models P \quad \text{if} \quad A \models A' \text{ and } A' \models P$.

The transformational form of this rule is of course

$T9. \quad A \models P \quad \longrightarrow \quad A \models A' \text{ and } A' \models P$.

This is not a rewrite rule because its rightside contains a variable not in its leftside. This rule may be called the "*wmawlog*" rule, because it justifies the "we may assume without loss of generality" steps that occur at the beginning of many proofs, replacing the original assumptions with others that are stronger or equivalent. (Some proofs that say "we may assume without loss of generality" are actually case analyses, where a relatively easy special case is eliminated; e.g., in showing $n^2 \geq n$, we may assume $n \geq 1$ without loss of generality.)

**Exercise 8.4.4**  Prove soundness of $R9$.  □

The module META2 below expresses $T6$–$T9$ in the same notation as our earlier META module. None of these are rewrite rules, because each has a variable on its rightside that is not on its left. Hence they must be applied "by hand," e.g., with OBJ3's `apply` feature. This makes sense, because creativity is required in choosing suitable $Q$, and this creativity can never be fully automated.

```
obj META2 is pr META .
  var A A' P Q : Sen .   var S : Sig .
  [modp]  eq A |=[S] P      = (A |=[S] Q) and (A |=[S] Q => P) .
  [contd] eq A |=[S] false  = (A |=[S] Q) and (A |=[S] not Q) .
  [lemma] eq A |=[S] P      = (A |=[S] Q) and (A and Q |=[S] P) .
  [astr]  eq A |=[S] P      = (A'|=[S] P) and (A |=[S] A') .
endo
```

Two special cases of $T9$ are especially noteworthy: strengthening the condition of an implication; and substituting values for universally quantified variables. For the first,

$R9a. \quad A, P \Rightarrow Q \models R \quad \text{if} \quad A, P' \Rightarrow Q \models R \text{ and } A \models P' \Rightarrow P$

is the special case of *R9* where $A, P \Rightarrow Q$ is substituted for $A$, where $A, P' \Rightarrow Q$ is substituted for $A'$, where $R$ is substituted for $P$, and then the result is simplified using the rule

$$(\star) \quad A, P \Rightarrow Q \models P' \Rightarrow Q \quad \text{if} \quad A \models P' \Rightarrow P .$$

The resulting transformation rule is

$$T9a. \quad A, P \Rightarrow Q \models R \quad \longrightarrow \quad A, P' \Rightarrow Q \models R \text{ and } A \models P' \Rightarrow P .$$

In case $P'$ is closed, we can use *R3* to put this rule in the somewhat more useful form,

$$T9a. \quad A, P \Rightarrow Q \models R \quad \longrightarrow \quad A, P' \Rightarrow Q \models R \text{ and } A, P' \models P .$$

Note that this rule applies in particular to the condition of a conditional equation.

**Exercise 8.4.5** Prove soundness of $(\star)$. Show how *R9a* justifies strengthening the condition of a conditional equation. □

For our second special case of *T9*, recall from Corollary 8.3.23 that if $Y' \subseteq Y \subseteq X$ are variable sets and $\theta : X \to T_\Sigma(X)$ is a substitution such that $\theta_{X-(Y-Y')} = \theta$ (i.e., $\theta$ is non-identity at most outside $Y - Y'$), then

$$A \models (\forall Y)P \quad \textbf{implies} \quad A \models (\forall Y')\theta P .$$

From this and *T9* we get

$$T9b. \quad A, (\forall Y)P \models Q \quad \longrightarrow \quad A, (\forall Y')\theta P \models Q ,$$

with $Y, Y'$ and $\theta$ as above.

The first nine rules below are similar to the attributes declared for **and** and **or**, simple facts about the extended Boolean connectives; the next three rules help us conclude proofs, and the last rule lets us do *modus ponens* on the leftside of the turnstile. These rules are often useful in simplifying proof plans; it can be shown that applying them never prevents a proof from being found if one exists.

```
obj META3 is pr META2 .
  vars A P Q R : Sen .  var S : Sig .
  eq A and A        = A .
  eq A and true     = A .
  eq A and false    = false .
  eq A and not A    = false .
  eq A or  false    = A .
  eq A or  true     = true .
  eq A => true      = true .
  eq A => false     = not A .
  eq not not A      = A .
  eq P |=[S] P      = true .
```

```
      eq false |=[S] P =  true .
      eq (A and P) |=[S] P  =   true .
      eq A and P and (P => Q) |=[S] R
         = A and P and Q and (P => Q) |=[S] R .
    endo
```

**Example 8.4.4** Here are some proof tasks for which the above rules suffice to show that no proof is needed, because they are already satisfied:

```
    open .
    ops P1 P2  : -> Sen .
    op Phi : -> Sig .
    red P1 |=[Phi]     not P1 => P2 .
    red P1 and P2 |=[Phi]   P1 => P2 .
    red P2 |=[Phi]   P1 => (P1 => P2).
    red true |=[Phi]    false => P1 .
    close
```

(They all reduce to `true`.)                                                      □

We call the following a theorem even though it is easy to prove because of its fundamental importance for our approach to proof planning:

**Theorem 8.4.5** If some task $A \models P$ transforms under the rules $T0$-$T9$ to a Boolean combination of $\models$-atoms having truth values such that the Boolean combination evaluates to `true`, then $A \models P$ is true.

**Proof:** This follows by induction on the length of rule application sequences, using soundness of the individual rules, as stated in assertions $R0$-$R9$.
□

In fact, we can set things up so that the Boolean combination evaluates to true iff each atom is true, by using the rules $T2a$ and $T2b$ instead of $T2$. Then an OBJ proof score based on this proof plan will succeed iff each OBJ evaluation is `true`.

The rule below says that if we can reduce the two sides of an equation to the same thing, then the equation is true:

$$TRW. \quad A \models_\Phi (\forall X)\ t = t' \quad \longrightarrow \quad t \downarrow_{\Phi(X), A^R} t' \ ,$$

where the notation $t \downarrow_{\Phi(X), A^R} t'$ means that the terms $t, t'$ can be rewritten to the same term using the set $A^R$ of rewrite rules of $A$. Although possible, it is not worthwhile expressing this rule in our META framework, because this would require specifying equations, rewriting, etc. in OBJ. Instead, we just note that all atomic clauses should be passed on elsewhere for evaluation, after proof planning is completed. When the institution is *FOLQ/D*, these will all be equations, and rule *TRW* can be used; competition techniques are also possible (see Chapter 12). More interestingly, there are decision procedures for atoms over certain special domains, e.g., Presburger arithmetic.[4]

**Example 8.4.6** We explore some ways that things can go wrong in proofs; failures are unpredictable, irregular, and very common. We first try to plan the easy part of the proof of Exercise 8.2.6, that if a relation R satisfies the equations

```
cq X R* Y = true if X R  Y .
cq X R* Z = true if X R* Y and Y R* Z .
```

then it also satisfies the equation

```
cq X R* Z = true if X R Y and Y R* Z .
```

Our proof task has the form

$$A_1 \wedge A_2 \vDash_\Sigma (\forall X) (P_1 \Rightarrow P_2) ,$$

and we can generate a proof score for it with the following:

```
open META3 .
op all-X_ : Sen -> Sen .
ops A1 A2 P1 P2 : -> Sen .
op Phi : -> Sig .
var A : Sen .
eq all-X A = (all 'x : 'Elt (all 'y : 'Elt (all 'z : 'Elt A))).
red (A1 and A2) |=[Phi] (all-X (P1 => P2)) .
close
```

which yields the proof plan

```
A1 and A2 and P1 |=[Phi] ['x : 'Elt]['y : 'Elt]['z : 'Elt] P2 .
```

But if we translate this into an OBJ proof score, it fails because the second conditional equation (A2) is not a rewrite rule, since the variable Y occurs in its condition but not in its leftside. We could get around this by using OBJ's `apply` feature for A2; but it seems easier to make part of the necessary substitution by hand (the entire substitution would have to be entered by hand to use `apply` anyway), add the resulting rule, and then use reduction. Substituting y for Y in A2 is justified by *T9a*, yielding the first equation in the proof score below:

---

[4]This is a decidable fragment of arithmetic, usually taken to be so-called extended quantifier free Presburger arithmetic for the rationals and integers, with unary minus, addition, subtraction, multiplication by constants, equality, disequality, and the relations $<, \leq, \geq$ and $>$ [164].

```
open R* .
vars-of .
ops x y z : -> Elt .
cq X R* Z = true if X R* y and y R* Z .
eq x R  y = true .
eq y R* z = true .
red x R* z .
close
```

However, this does not work either! This is because OBJ goes into an infinite loop, applying the first equation to itself over and over, with the substitution X = x, Z = y. We can circumvent this by preventing the substitution of y for Z by adding to the condition of the rule. This is justified by *T9b*, and yields the following rule:

```
cq X R* Z = true if Z =/= y and X R* y and y R* Z .
```

However, this still causes an infinite loop, because and does not know that if its first argument is false then the whole conjunction is necessarily false; hence we define and use a more clever conjunction, that uses a partially lazy evaluation (see Section 5.4):

```
open R* .
vars-of .
ops x y z : -> Elt .
var B : Bool .
op _cand_ : Bool Bool -> Bool [strat (1 0)] .
eq false cand B = false .
eq true  cand B = B .
cq X R* Z = true if Z =/= y cand (X R* y and y R* Z) .
eq x R  y = true .
eq y R* z = true .
red x R* z .
close
```

But this does not work either, since OBJ finds a different infinite loop! This one can be prevented by also prohibiting the instantiation of X by y, by adding another conjunct:

```
cq X R* Z = true if (Z =/= y and X =/= y)
                    cand (X R* y and y R* Z) .
```

This (finally!) works, and the proof is done. (However, OBJ3 fails to parse the condition if either pair of the parentheses is omitted; this could be circumvented by declaring a non-default precedence for cand, but it is not worth the trouble.)

It was not so easy to get OBJ3 to execute this simple proof score: four[5] different things went wrong and had to be worked around! These workarounds were: (a) instantiate an equation that was not a rewrite rule to make it one; (b) add conditions to an equation to ensure termination (this was done twice); (c) change the order of evaluation by forcing a rule to fail if one conjunct in its condition fails; and (d) add parentheses to help the parser. All of these are standard "tricks of the trade" for an experienced OBJ user, and I hope this example will help you to use them in the future. In particular, please note how termination was handled: we did not attempt to prove that the rule set was terminating; instead, we discovered experimentally that it was *not* terminating, and then we just strengthened the rules to prevent the particular loop that we found, while preserving correctness. The same approach applies to the Church-Rosser property: when we failed to get the order of evaluation we wanted, we just changed OBJ's evaluation strategy. Our emphasis is on getting a correct proof, rather than on getting a canonical specification.

---

[5]Of course, there were also some typographical errors during the development of this proof; these were caught in the usual way by the OBJ parser, and fixed by the user.

Things are worse for the other half of the proof of Exercise 8.2.6: the proof score that is automatically generated from the proof task is very little help; some entirely new ideas are needed, and initiality must be used. We omit the details, but underline the moral: the proof score automatically generated by our META rules is only adequate for simple proof tasks; for slightly more difficult tasks, small modifications may be sufficient, but in general, some real creativity must be supplied by the user. Nevertheless, close adherence to the transformational approach will guarantee correctness of the proof score, and hence of the proof, if the proof score executes correctly. □

### 8.4.1 Deduction vs. Planning

Our approach to first-order logic has been a bit eccentric: After an algebraic treatment of syntax, we developed a number of properties of semantic entailment, and then applied them to proof planning; we have not considered rules of deduction in the traditional sense at all.

Rules of deduction are used to deduce (infer) something new from something old, such that if the old is true then so is the new. Our purpose in this chapter has been just the opposite: to reduce something we hope is true to some new thing(s), such that if the new are true, then so is the old. Hence, what we call an "elimination rule" corresponds to what is called an "introduction rule" in the standard literature, but applied *backwards*.

To illustrate this, let's consider the traditional rule for conjunction ("and") introduction:

$$\frac{P \quad Q}{P \wedge Q}$$

This says that if we have proved $P$ and $Q$, then we are entitled to say we can prove their conjunction $P \wedge Q$. Since it is awkward to work with tautologies, a more useful formulation is

$$\frac{A \vdash P \quad A \vdash Q}{A \vdash P \wedge Q}$$

where $A$ is some set of axioms, and "$\vdash$" indicates first-order proof;[6] this is very much like what we did for equational logic. But in our present context, where we have the task of proving $P \wedge Q$ (from $A$), the above tells us it is sufficient to prove $P$ and $Q$ separately: that is, we can apply the above rule *backwards* to *eliminate* the conjunction from our goal; this is why we call it "conjunction elimination."

The rule that is usually called conjunction elimination is completely different: it says that if we have proved $P \wedge Q$, then we are entitled to say we have proved $P$. This may be written:

$$\frac{A \vdash P \wedge Q}{A \vdash P}$$

(Of course, there is a similar elimination rule for $Q$.)

The most important property that a rule of deduction can have is **soundness**; an unsound rule cannot guarantee correct proofs. A sequent rule is **sound** iff the result of replacing $\vdash$ by $\vDash$ is a valid implication. For example, soundness of the rule that we call conjunction elimination depends upon the result

$$A \vDash P \text{ and } A \vDash Q \text{ imply } A \vDash P \wedge Q \,.$$

Using the traditional rules in the forward direction gives a *bottom up* proof, starting with what is known, and gradually building up more. By contrast, our proof planning rules build proofs *top down*, starting with what we want, and working down towards what we know. This second kind of proof organization corresponds (roughly) to what is called *natural deduction* in the logic literature. More generally, a rule that transforms what appears on the right of the turnstile is doing *top down* or *backwards* inference, and one that transforms what appears on the left of the turnstile is doing *bottom up* or *forwards* inference.

It is important to notice that "real" proofs (e.g., from textbooks, research papers, lectures, blackboards, etc.) are usually neither top down nor bottom up! In fact, reading a proof written in either of these styles can be pretty tedious. A bottom up version of a complex proof would first present a long list of assumptions and low level results that are completely unrelated to each other; it would then build on top of these a layer of loosely related low level results; and so on upwards; the result would be incomprehensible until the very end (and probably even then). A top down proof would be easier to follow, but would prohibit the use of lemmas, which can make proofs much easier to follow. Thus, natural deduction is not really very natural after all. A brief discussion of the naturalness of proofs appears in Section 8.8.

---

[6] A proof calculus where sentences involve $\vdash$ is sometimes called a **sequent calculus**.

Proof planning rules that are rewrite rules can and generally should be applied automatically, but other kinds of rule require more attention. Therefore only the most routine aspects of proof planning can be completely automated by rewriting; the most interesting rules, such as proof by contradiction, adding lemmas, and induction, require some (often considerable!) ingenuity.

We are now in a position to understand what OBJ "proof scores" really are, and why they work: An OBJ proof score contains the equational reductions that result from applying proof planning rules to an original proof task; such a proof score can be proved valid by appeal to the proof planning rules that produced it.

### 8.4.2 Correctness of a Ripple Carry Adder

This section verifies a ripple carry adder of arbitrary width, i.e., proves that it really does add. The verification makes heavy use of OBJ's abstract data type capabilities. Figure 8.1 shows the structure of this device; it is a cascade connection of $n$ "full adders." In addition, the device has two input buses and one output bus, each $n$ bits wide, plus a final carry bit output. (For those not already familiar with hardware, all of these terms are made precise in the OBJ specifications below.)

The ADT of natural numbers with addition is needed to handle the correctness condition, and the use of multiplication and exponentiation should not be surprising given the nature of binary numbers; but it is interesting to notice how convenient the integers with subtraction really are for this example. $n$-bit wide busses are represented by lists of Booleans of length $n$; this abstract data type is defined using ordersorted algebra, so that inductive proofs have the 1-bit case for their base, and the operation of postpending a bit for their induction step. The result to be verified is that

$$(\forall w_1, w_2) \, (|w_1| = |w_2|) \Rightarrow$$
$$(\#sout^*(w_1, w_2) + 2^{|w_1|} * \#cout(w_1, w_2) \; = \; \#w_1 + \#w_2) \, ,$$

where $w_1$ and $w_2$ range over buses, where $|w|$ and $\#w$ are respectively the length, and the number denoted by, a bus $w$, where $sout^*(w_1, w_2)$ represents the content of the output bus, and where $cout(w_1, w_2)$ is the carry bit. In words, this formula says that given two input buses of the same width, the number on the output bus together with the carry (as highest bit) equals the sum of the numbers on the input buses.

Despite the somewhat complex structure of its terms, this formula is of exactly the form treated in Example 8.4.1. Hence we can prove it by introducing new constants for $w_1$ and $w_2$, then assuming $P_1$, and proving $P_2$ by checking equality of the reduced forms of its two sides. The proofs of the lemmas are by straightforward case analysis and/or induction, and are omitted here. The proof of the main result is by induction on the width of the input buses, starting from width 1.

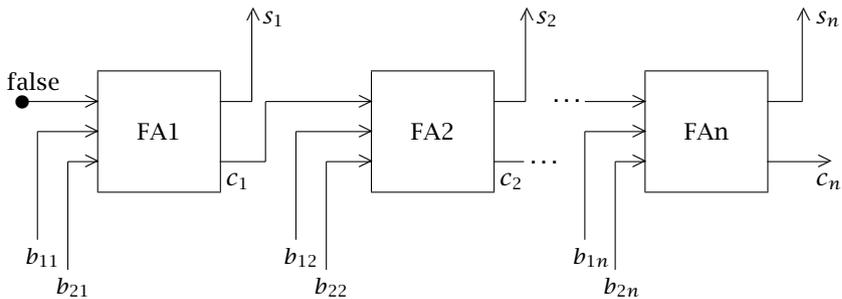

Figure 8.1: A Ripple Carry Adder

The first OBJ module below uses order-sorted algebra to specify the integers; thus "`subsort Nat < Int`" says every natural number is also an integer (see Chapter 10 for details on order-sorted algebra). A number of inductive lemmas are included in this module, such as the distributive law.

```
obj INT is sorts Int Nat .
  subsort Nat < Int .
  ops 0 1 2 : -> Nat .
  op s_ : Nat -> Nat [prec 1] .
  ops (s_)(p_) : Int -> Int [prec 1] .
  op (_+_) : Nat Nat -> Nat [assoc comm prec 3] .
  op (_+_) : Int Int -> Int [assoc comm prec 3] .
  op (_*_) : Nat Nat -> Nat [assoc comm prec 2] .
  op (_*_) : Int Int -> Int [assoc comm prec 2] .
  op (_-_) : Int Int -> Int [prec 4] .
  op -_ : Int -> Int [prec 1] .
  vars I J K : Int .
  eq 1 = s 0 .   eq 2 = s 1 .
  eq s p I = I .
  eq p s I = I .
  eq I + 0 = I .
  eq I + s J = s(I + J) .
  eq I + p J = p(I + J) .
  eq I * 0 = 0 .
  eq I * s J = I * J + I .
  eq I * p J = I * J - I .
  eq I * (J + K) = I * J + I * K .
  eq - 0 = 0 .
  eq - - I = I .
  eq - s I = p - I .
```

```
  eq - p I = s - I .
  eq I - J = I + - J .
  eq I + - I = 0 .
  eq -(I + J) = - I - J .
  eq I * - J = -(I * J) .
  op 2**_ : Nat -> Nat [prec 1] .
  var N : Nat .
  eq 2** 0 = 1 .
  eq 2** s N = 2** N * 2 .
endo

obj BUS is sort Bus .
  extending PROPC + INT .
  subsort Prop < Bus .
  op __ : Prop Bus -> Bus .
  op |_| : Bus -> Nat .
  var B : Prop .    var W : Bus .
  eq | B | = 1 .
  eq | B W | = s | W | .
  op #_ : Bus -> Int [prec 1] .   *** is really -> Nat
  eq # false = 0 .
  eq # true = 1 .
  eq #(B W) = 2** | W | * # B + # W .
endo

***> full adder
obj FADD is extending PROPC .
  ops cout sout : Prop Prop Prop -> Prop .
  vars I1 I2 C : Prop .
  eq sout(I1,I2,C) =  I1 xor I2 xor C .
  eq cout(I1,I2,C) = I1 and I2 xor I1 and C xor I2 and C .
endo

***> n-bit ripple carry adder
obj NADD is protecting FADD + BUS .
  ops cout sout : Bus Bus -> Prop .
  op sout* : Bus Bus -> Bus .
  vars B1 B2 : Prop .    vars W1 W2 : Bus .
  eq cout(B1,B2) = cout(B1,B2,false) .
  eq sout(B1,B2) = sout(B1,B2,false) .
  eq cout(B1 W1,B2 W2) = cout(B1,B2,cout(W1,W2)) .
  eq sout(B1 W1,B2 W2) = sout(B1,B2,cout(W1,W2)) .
  eq sout*(B1,B2) = sout(B1,B2) .
  eq sout*(B1 W1,B2 W2) = sout(B1 W1,B2 W2) sout*(W1,W2) .
endo

obj LEMMAS is protecting NADD .
  vars B1 B2 : Prop .
  eq #(B1 and B2) = # B1 * # B2 .
```

```
    eq # B1 * # B1 = # B1 .
    eq #(B1 xor B2) = # B1 + # B2 - #(B1 and B2)* 2 .
      *** would write up if # : Bus -> Nat .
    vars W1 W2 : Bus .
    ceq | sout*(W1,W2)| = | W1 | if | W1 | == | W2 | .
  endo

  ***> base case
  open LEMMAS .
    ops b1 b2 : -> Prop .
    reduce # sout*(b1,b2)+ # cout(b1,b2)* 2 == # b1 + # b2 .
  close

  ***> induction step
  open LEMMAS .
    ops b1 b2 : -> Prop .
    ops w1 w2 : -> Bus .
    op n : -> Nat .
    eq | w1 | = n .
    eq | w2 | = n .
    eq # sout*(w1,w2) + 2** n * # cout(w1,w2) = # w1 + # w2 .

  red # sout*(b1 w1,b2 w2) + 2** | b1 w1 | * # cout(b1 w1,b2 w2)
      == #(b1 w1) + #(b2 w2).
  close
```

In reducing the above expression to true, OBJ3 did 158 rewrites, many of which were associative-commutative (and many more rewrites were tried but failed), so one would certainly prefer to have this calculation done mechanically, rather than do it oneself by hand! This proof may be about two orders of magnitude easier using induction in OBJ than it would be in a fully manual proof system. In addition, OBJ produced a validated proof score.

**Exercise 8.4.6** Prove the three lemmas in the object LEMMAS above. □

To summarize, we have used OBJ and reduction in two different ways, at two different levels: first at the meta level, to reduce the original proof task to a form that OBJ can directly handle; and then at the "object" level, to do the actual "dirty work" of the proof. Hence, this proof was completely automatic. (Needless to say, this is not always possible.)

### 8.4.3 The Working Signature

After seeing the kind of tricks with signatures that are used to eliminate quantifiers ($R5$ in Section 8.3.3 and $R10$ in the next subsection), a reader may worry that the truth of a goal depends on the signature. The following shows that is not the case.

**Proposition 8.4.7** Let $A$ be a set of $\Phi$-formulae, let $P$ be a $\Phi$-formula, and let $\Phi' \subseteq \Phi$ be such that a sort of $\Phi'$ is void iff it is also void in $\Phi$. If $A, P$ are also $\Phi'$-formulae, then

$$A \vDash_\Phi P \text{ iff } A \vDash_{\Phi'} P .$$

**Proof:** First note that if $M' = M|_{\Phi'}$ for $M$ a $\Phi$-model, then $M \vDash_\Phi A$ iff $M' \vDash_{\Phi'} A$. Also note that by the non-void assumption, any $\Phi'$-model $M'$ extends to a $\Phi$-model $M^*$ such that $M' = M^*|_{\Phi'}$.

Now suppose $A \vDash_\Phi P$ and $M' \vDash_{\Phi'} A$; we will prove that $M' \vDash_{\Phi'} P$. Choose $M^*$ such that $M' = M^*|_{\Phi'}$. Then $M^* \vDash_\Phi A$. Therefore $M^* \vDash_\Phi P$, and hence $M' \vDash_{\Phi'} P$. For the converse, suppose $A \vDash_{\Phi'} P$ and $M \vDash_\Phi A$. Let $M' = M|_{\Phi'}$. Then $M' \vDash_{\Phi'} A$. Therefore $M' \vDash_{\Phi'} P$ and hence $M \vDash_\Phi P$.
□

As long as all formulae parse, and you don't populate an old void sort or depopulate an old non-void sort, the working signature can be as large or as small as you please. This implies that a mechanical theorem prover can effectively ignore the working signature, as our OBJ proof scores in fact do; the above result also helps to justify our frequent practice of dropping the signature subscript from $\vDash$.

## 8.5  Reasoning with Existential Quantifiers

Sentences that involve existential quantifiers can occur either on the assumption or the goal side of the symbol $\vDash$, and must be handled differently in each case. We begin with the assumption case. Because it can be difficult to use assumptions that contain existential quantifiers, it is useful to transform them into a more constructive form. For example, the proof task

$$A, (\exists a, b : Pos)(c = a/b) \vDash_\Phi Q$$

can be transformed to

$$A, c = a/b \vDash_{\Phi(a,b:Pos)} Q .$$

The intuition here is that since we know $a, b$ exist, in trying to prove $Q$ we may as well assume that $a, b$ have been given to us in the signature; in this case $a, b$ are called **Skolem constants**.

More generally, an existential quantifier may lie within the scope of one or more universal quantifiers, as in

$$A, (\forall x : Nat)(\exists y : Nat) \, f(x, y) = 1 \vDash_\Phi Q .$$

In such a case, the choice of $y$ must depend on the value of $x$, so that what is added to the signature must be a *function* of $x$. Hence the result of transforming the above should be

$$A, (\forall x : Nat)\ f(x, y(x)) = 1\ \vDash_{\Phi(y:Nat \to Nat)}\ Q',$$

where $Q'$ denotes the result of substituting $y(x)$ for $y$ in $Q$. Transformations of this kind are justified by Proposition 8.5.1 below.

**Proposition 8.5.1** Given a set $A$ of $\Phi$-formulae plus $\Phi$-formulae[E37] $P, Q$ where $Free(P) = X \cup \{y\}$ with $X = \{x_1 : s_1, \ldots, x_n : s_n\}$ and with $y$ of sort $s$, then

$$R10a.\quad A, P' \vDash_{\Phi(Y)} Q \quad \textbf{implies} \quad A, (\exists y : s)P \vDash_\Phi Q,$$

where $P'$ denotes the result of substituting $y(x_1, \ldots, x_n)$ for $y$ in $P$ and where $Y$ is the declaration $y : s_1 \ldots s_n \to s$. Moreover, under the same assumptions,

$$R10b.\quad A, (\forall X)P' \vDash_{\Phi(Y)} Q \quad \textbf{implies} \quad A, (\forall X)(\exists y:s)P \vDash_\Phi Q.$$

**Proof:** We first prove the implication $R10a$. Let $\theta$ be the substitution that takes $y$ to $y(x_1, \ldots, x_n)$ and is the identity on other variables. Then $P' = P\theta$. Let $M$ be a $\Phi$-model satisfying $(\exists y : s)P$. Then for every $a : X \to M$ there is some $a' : X \cup \{y\} \to M$ such that $a'|_X = a$ and $\overline{a'}(P) = true$. Let $M'$ be a $\Phi(Y)$-model that extends $M$ with a new function $M'_y : M^{s_1 \ldots s_n} \to M_s$ defined by $M'_y(m_1, \ldots, m_n) = a'(y)$ where $a' : X \cup \{y\} \to M$ is an interpretation obtained as above from $a : X \to M$ defined by $a(x_i) = m_i$ for $i = 1, \ldots, n$. Note that $M'|_\Phi = M$ because $M'$ only adds the new operation $M'_y$ to $M$. Also note that $M' \vDash_{\Phi(Y)} P'$: indeed, each interpretation $a : X \to M'$ is actually an interpretation $a : X \to M$ that takes each $x_i$ in $X$ to an $m_i$ in $M_{s_i}$ such that there is an $a' : X \cup \{y\} \to M$ such that $a'|_X = a$, $M'_y(m_1, \ldots, m_n) = a'(y)$ and $\overline{a'}(P) = true$. Because $\overline{a}(y(x_1, \ldots, x_n)) = M'_y(a(x_1), \ldots, a(x_n)) = a'(y)$, then $\overline{a}(\theta(z)) = \overline{a'}(id(z))$ for all $z \in Free(P)$, where $id$ is the identity substitution; now Lemma 8.3.25 implies $\overline{a}(P') = \overline{a'}(P) = true$. Thus $M'$ is a $\Phi(Y)$-model of $P'$. Therefore if $M$ is a $\Phi$-model of both $A$ and $(\exists y:s)P$, then $M'$ is a $\Phi(Y)$-model of both $A$ and $P'$. Hence $M'$ is a $\Phi(Y)$-model of $Q$, and thus $M$ is a $\Phi$-model of $Q$.

$R10b$ now follows from $R10a$ by $n$ applications of $R5b$. □

The new function symbol $y$ is called a **Skolem constant** or a **Skolem function**, depending on whether the quantifier $(\forall X)$ is present. The corresponding transformation rules, called **Skolemization** rules, are

$$T10a.\ A, (\exists y : s)P \vDash_\Phi Q \quad \longrightarrow \quad A, P' \vDash_{\Phi(Y)} Q'.$$
$$T10b.\ A, (\forall X)(\exists y : s)P \vDash_\Sigma Q \quad \longrightarrow \quad A, (\forall X)P' \vDash_{\Phi(Y)} Q'.$$

Note that these rules only apply to formulae on the *left* of the turnstile; a different approach is needed for establishing goals that contain existential quantifiers. Of course, not all existential quantifiers are so polite as to occur only within the scope of universal quantifiers. But since a first-order formula has only a finite number of quantifiers, by patiently applying *T*10 wherever it can be applied (outermost first), all of its existential quantifiers will eventually be eliminated. (Second-order existential quantifiers can be Skolemized by adding further arguments to the quantified function, as discussed in Chapter 9.)

Below is OBJ3 (meta-)code for Skolem constants; Skolem functions can be done in a similar way, but this would require modifying some previous meta code.

```
obj META4 is pr META3 .
  vars A P Q : Sen .   var X : Id .   var T : BType .
  var S : Sig .
  op (exist_:_ _) : Id BType Sen -> Sen .
  eq A and (exist X : T P) |=[S] Q  =  A and P |=[S][X : T] Q .
endo
```

To prove a goal that involves an existential quantifier, it is necessary to show that a suitable value actually exists in all models that satisfy the assumptions. In general, the suitable value will depend upon a choice of other values, because the existential quantifier occurs within the scope of some universal quantifiers in the goal. For example, the sentence

$$(\forall x)(\exists y)\; x + y = 0$$

is satisfied for the integers by choosing $y = -x$. This suggests that if we can find a term expressing the dependency of the existential variable on the universal variables, then we can prove our goal. This proof method is supported by the following:

**Proposition 8.5.2** Given a $\Phi$-sentence $(\forall X)(\exists y : s)P$ with $Free(P) = X \cup \{y\}$, then

$$R11.\quad A \vDash_\Phi (\forall X)P[y \leftarrow t] \quad \textbf{implies} \quad A \vDash_\Phi (\forall X)(\exists y)P\ ,$$

where $t$ is some term over $X$ of sort $s$.

**Proof:** We first assume $A \vDash_\Phi (\forall X)P[y \leftarrow t]$, which by the Theorem of Constants (*R*5), means that $M \vDash_{\Phi(X)} A$ implies $M \vDash_{\Phi(X)} P[y \leftarrow t]$ for all $\Phi(X)$-models $M$. Then we want to show $A \vDash_\Phi (\forall X)(\exists y)P$, i.e., that $M \vDash_{\Phi(X)} A$ implies $M \vDash_{\Phi(X)} (\exists y)P$, for all $\Phi(X)$-models $M$. So we assume $M \vDash_{\Phi(X)} A$, and from this conclude by the assumption that $M \vDash_{\Phi(X)} P[y \leftarrow t]$, and hence[E38] by Proposition 8.3.15, that $M \vDash_{\Phi(X)} (\exists y)P$. □

The corresponding transformation rule is:

$T11.\ A \vDash_\Phi (\forall X)(\exists y)P \quad \longrightarrow \quad A \vDash_\Phi (\forall X)P[y \leftarrow t]\,.$

This rule cannot be expressed in our current meta level formalism, because terms are not specified in it. However, it is easy to express the essence of the rule, by expressing substitutions for variables as new equations, where terms will be handled the usual way in concrete examples.

```
obj META5 is pr META4 .
  vars A P : Sen .  var y : Id .  var T : BType .
  var S : Sig .
  op all-X_ : Sen -> Sen .
  op Eqt : Id -> Sen .
  eq A |=[S] (all-X (exist y : T P)) =
      A and Eqt(y) |=[S][y : T] (all-X P).
  *** where Eqt(y) is the equation y = t
  *** and all-X is one or more universal quantifier
endo
```

Notice that in this formulation of the rule, the status of y is changed from being a variable to being a constant; this is needed so that the new equation will do the substitution.

Proposition 8.5.2 can be applied iteratively to eliminate nested existential quantifiers. For example, a sentence of the form

$(\forall X)(\exists y)(\forall Z)(\exists w)\ P$

can be transformed first to

$(\forall X \cup Z)(\exists w)\ P[y \leftarrow t]$

and then to

$(\forall X \cup Z)\ P[y \leftarrow t][w \leftarrow t']\,,$

provided the restrictions on free variables are satisfied — but of course these are very natural. Below is a simple example.

**Example 8.5.3** Suppose we want to prove

$\mathsf{NAT} \vDash (\forall x, y : \mathsf{Int})(\exists z, w : \mathsf{Int})\ P_1\ \mathrm{and}\ P_2\,,$

where $P_1, P_2$ are linear equations. The goal of this proof task says that these two equations can always be solved for $w, z$ given values for $x, y$. Below we use META5 to plan a proof for this goal; OBJ3's `call-that` feature is used to delay applying the equation that defines the universal quantifiers; without this trick, these quantifiers are turned into constants, and then the quantifier elimination rule in META5 cannot be applied.

```
open META5 .
ops INT P1 P2 : -> Sen .
op Phi : -> Sig .
op Int : -> BType .
var A : Sen .
red INT |=[Phi] (all-X (exist 'z : Int
                           (exist 'w : Int (P1 and P2)))) .
call-that t .
eq all-X A = (all 'x : Int (all 'y : Int A)).
red t .
close
```

The proof plan that results from this is

```
(INT and Eqt('z) and Eqt('w)
  |=[Phi]['z : Int]['w : Int]['x : Int]['y : Int] P2) and
(INT and Eqt('z) and Eqt('w)
  |=[Phi]['z : Int]['w : Int]['x : Int]['y : Int] P1)
```

For a particular instance, suppose that P1 and P2 are respectively the equations

$$\begin{aligned} x - w + 2y &= 3 \\ x + w &= 2z - 5 \,, \end{aligned}$$

for which the solutions are

$$\begin{aligned} w &= x + 2y - 3 \\ z &= x + y + 1 \,. \end{aligned}$$

Then the original goal is proved with the above proof plan by the following, provided the two reductions give `true` (which they do):

```
open INT .
ops x y z w : -> Int .
eq w = x + (2 * y) - 3 .
eq z = x + y + 1 .
red x - w + (2 * y) == 3 .
red x + w          == (2 * z) - 5 .
close                                                          □
```

## 8.6 Case Analysis

Sometimes it is easier to prove a result by breaking the proof (or a part of it) into "cases." For example, in trying to prove a sentence of the form (where $n$ is a natural number variable)

$$(\forall n)\,(n > 0 \Rightarrow Q(n))\,,$$

it might be easier to prove the following two cases separately,

$$(\forall n)\ (n = 1 \Rightarrow Q(n))\,,$$
$$(\forall n)\ (n > 1 \Rightarrow Q(n))\,,$$

than to prove the assertion in its original form. In general, there are many different ways to break a condition like $n > 0$ into cases. Another is

$$(\forall n)\ (n \text{ even } \wedge\ n > 0 \Rightarrow Q(n))\,,$$
$$(\forall n)\ (n \text{ odd} \Rightarrow Q(n))\,,$$

and still another is

$$(\forall n)\ (n = 1 \Rightarrow Q(n))\,,$$
$$(\forall n)\ (n \text{ prime} \Rightarrow Q(n))\,,$$
$$(\forall n)\ (n \text{ composite} \Rightarrow Q(n))\,.$$

Such examples suggest that "cases" are "predicates" (i.e., open formulae) $P_1, \ldots, P_N$ such that $P_1 \vee \cdots \vee P_N$ and $P$ are equivalent, where the sentence to be proved has the form $P \Rightarrow Q$; and they further suggest that a proof by "case analysis" consists of proving $P_i \Rightarrow Q$ for $i = 1, \ldots, N$. We make this more precise as follows:

**Proposition 8.6.1** To prove $A \vDash P \Rightarrow Q$, it suffices to give predicates $P_1, \ldots, P_N$ such that $A \vDash P \Rightarrow (P_1 \vee \cdots \vee P_N)$, and then to prove $A \vDash P_i \Rightarrow Q$ for $i = 1, \ldots, N$.

**Proof:**  Soundness of this proof measure follows from the calculation:

$$A \vDash (P_1 \Rightarrow Q) \wedge \cdots \wedge (P_N \Rightarrow Q) \quad \text{iff}$$
$$A \vDash (P_1 \vee \cdots \vee P_N) \Rightarrow Q \quad \text{implies}$$
$$A \vDash (P \Rightarrow Q)\,.$$

using *E*14 for the first step and *R*6*a* for the second.   □

This justifies the deduction rule

R12. $A \vDash P \Rightarrow Q$ **if** $A \vDash P_i \Rightarrow Q$ for $i = 1, \ldots, N$ **and**
     $A \vDash P \Rightarrow P_1 \vee \cdots \vee P_N\,,$

which in turn justifies the transformation rule

T12. $A \vDash P \Rightarrow Q \longrightarrow A \vDash P_i \Rightarrow Q$ for $i = 1, \ldots, N$ **and**
     $A \vDash P \Rightarrow P_1 \vee \cdots \vee P_N\,.$

**Example 8.6.2** Often a case analysis succeeds because it makes use of new information. For example, we cannot prove

$$A \not\vDash (\forall B)\ \text{not not}\ B = B\,,$$

where $A$ is a ground specification of the Booleans, by proving

$$A \vDash (\forall B : \mathit{Bool})\ \text{not not}\ B = B$$

because it is not true of all models of *A*. However, if we work in the institution *FOLQ/D* where *D* includes the Booleans, then we can prove

$$A \models (\forall B : Bool)\ B = true \lor B = false\ ,$$

using initiality, so that it suffices to prove the two cases,

$$A \models \text{not not } true = true$$
$$A \models \text{not not } false = false\ .$$

In fact, exactly this kind of Boolean case analysis justifies the method of truth tables (as in Example 7.3.16). □

Some typical case analyses are given below, for *x* an integer, *y* a non-zero integer, and *z* a positive integer, respectively:

$$(x = 0) \lor (x > 0) \lor (x < 0)\ ;$$
$$(y > 0) \lor (y < 0)\ ;$$
$$(z = 1) \lor (z > 1)\ .$$

It is also typical that proving the validity of these disjunctions requires induction.

**Exercise 8.6.1** Prove validity of the following case analysis: for all natural numbers *n*, either $n = 0$ or else $(\exists j)\ n = s(j)$. □

**Example 8.6.3** The proof that $\sqrt{2}$ is irrational (discovered by the Pythagoreans in antiquity and kept secret for fear that it would destabilize civilization!) begins by assuming, without loss of generality, that $\sqrt{2} = a/b$ with *a, b* positive relatively prime integers. This step can be justified using case analysis. The initial assumption is $\sqrt{2} = a/b$ with *a, b* positive. Let $gcd(a,b) = g$ (where *gcd* denotes the greatest common divisor). Then either $g = 1$ or $g > 1$. In the first case, *a, b* are already relatively prime, while in the second case we have $a = a'g$ and $b = b'g$ with $a', b'$ positive and relatively prime. Then $\sqrt{2} = a'/b'$. So in either case the initial assumption is justified, and we can proceed with the proof. The rest of the top level proof planning can be done by OBJ:

```
open META4 + NAT .
op NAT : -> Sen .
op NATSIG : -> Sig .
ops 'a 'b : -> NzNat .
op eq : Nat Nat -> Sen .
let P = eq('a * 'a * 2, 'b * 'b).
red NAT |=[NATSIG] not (exist 'a : 'NzNat
                         (exist 'b : 'NzNat P)).
close
```

The result of the reduction is as follows:

```
result Meta: NAT and eq('a * 'a * 2,'b * 'b) |=[(NATSIG]
  ['a : 'NzNat)]['b : 'NzNat] false
```

which says we should assume the negation of the goal, Skolemize 'a and 'b, and then try to derive a contradiction; of course, this leaves out the most difficult parts of the proof, which cannot be automated so easily. □

**Exercise 8.6.2** Give a complete proof plan and OBJ3 proof score for showing $\sqrt{2}$ irrational. □

## 8.7 Algebraic Induction

This section shows how to generalize, justify and use the familiar form of induction that checks base and step cases; this includes so-called "structural induction" but not well-founded induction and similar potentially transfinite methods. Example 8.3.14 showed that first-order specifications in general do not have initial models; therefore induction is not in general valid for proving sentences about structures defined by first-order theories. However, induction *is* valid for proving sentences about structures that are defined (or definable) by initial algebra semantics. Our applications usually have some underlying data values that have been defined in this way, such as the integers or the naturals, and simple inductive results about them are usually needed in all but the simplest proofs. Using the language of Section 8.3.6, this means we are working in the institution $FOLQ/D$, where $D$ is an initial $(\Psi, E)$-algebra. For this institution, satisfaction differs from that of the ordinary first-order logic institution $FOL$, in that for $P$ a first-order $\Psi$-formula, $\emptyset \vDash_\Psi^{FOLQ/D} P$ iff $D \vDash_\Psi^{FOLQ} P$. We can now state the basic justification for inductive reasoning as follows:

$R13.$   $\emptyset \vDash_\Psi^{FOLQ/D} P$   **iff**   $D \vDash_\Psi^{FOLQ} P$   **iff**   $E \models_\Psi P$ ,

where we extend the notation $\models$ of Section 6.4 to first-order sentences, so that $E \models_\Psi P$ means $P$ is satisfied by an initial model of $E$. More generally, we have $A \vDash_\Psi^{FOLQ/D} P$ iff $E \models_\Psi P$, provided $P$ is a $\Psi$-sentence and $A$ is consistent with $D$. We should not neglect to mention a very simple, but still useful further rule, where $P$ is an arbitrary first-order sentence:

$R14.$   $A \models_\Sigma P$   if   $A \vDash_\Sigma P$ .

This rule is sound for $FOLQ/D$ because if $P$ holds for every model of $A$, it certainly holds for an initial model of $A$ (a special case was already mentioned on page 181 in Chapter 6).

In line with $R13$, we have the following result, for which as so often happens, there is a nice semantic proof:

**Proposition 8.7.1** If $M, M'$ are isomorphic $\Psi$-models and if $P$ is a first-order $(\Psi, X)$-formula, then $M \models_\Psi P$ iff $M' \models_\Psi P$.

**Proof:** Let $\psi : M \to M'$ be a $\Psi$-isomorphism with inverse $\rho : M' \to M$, and assume $M \models_\Psi P$. Let $\theta : X \to M'$ be an assignment.[E39] Then $\theta; \rho : X \to M$ is an assignment, so that $P(\theta; \rho) = true$. Therefore $P(\theta; \rho); \psi = (true)\psi = true$. The proof of the converse is similar. □

Formulae that can be proved by induction include sentences of the form $(\forall x : v)P$ where $x$ is free in $P$; in this case, we often write $(\forall x)P(x)$, and call $x$ the **induction variable**. Then $M \models (\forall x)P$ means $M \models \theta_m(P)$ for all $m \in M_v$ where $\theta_m$ is the substitution[E40] with $\theta_m(x) = m$; we may write $P(m)$ for $\theta_m(P)$. If the inductive goal has the form $(\forall x_1)(\forall x_2) \ldots (\forall x_n)P$, then (by E20) we can reorder the quantifiers to put any one of $x_1, \ldots, x_n$ first, say $x_i$, and use it as the induction variable for proving $(\forall X - \{x_i\})P$ where $X = \{x_1, x_2, \ldots, x_n\}$.

Usually more than one induction scheme can be used for a given initial specification; even the natural numbers have many different induction schemes. The most familiar scheme proves $(\forall x)P(x)$ by proving $P(0)$ and then proving that $P(n)$ implies $P(sn)$; let's call this **Peano induction**. But we could also prove $(\forall x)P(x)$ by proving $P(0)$ and $P(s0)$, and then proving that $P(n)$ implies $P(ssn)$; let's call this **even-odd induction**. These two schemes correspond to two different choices of generators: the first uses $0, s\_$, while the second uses $0, s0, ss\_$. These are the first two of an infinite family of induction schemes: for each $n > 0$, the $n$-**jump Peano induction** scheme has $n$ constants $0, \ldots, s^{n-1}0$ and one unary function $s^n\_$.

We insist that before an induction scheme is used, it *should be proved sound*; this requires formalizing the notion of induction scheme for (initial models of) an equational specification $(\Psi, E)$; the formalization will involve a signature $\Gamma$ of generators defined over $(\Psi, E)$ by some new equations that may involve some new auxiliary function symbols.

**Definition 8.7.2** An **inductive goal** for a signature $(V, \Psi)$ is a $V$-sorted family $P$ of first-order $\Psi$-formulae $P_v$, each with exactly one free variable $x_v$ of sort $v \in V$; we may write $P_v(x_v)$ and call $x_v$ the **induction variable** of sort $v$. Usually $P_v(x_v) = true$ for all $v$ except one, say $u \in V$, in which case we identify $P$ with $P_u$ and write $x$ for $x_u$.

Given a specification $T = (\Psi, E)$, an **algebraic induction scheme** for $T$ is a $V$-sorted extension theory $T' = (\Psi', E')$ of $T$ and a subsignature $\Gamma$ of $\Psi'$, written $(\Psi', E', \Gamma)$. Then **inductive reasoning** for an inductive goal $P$ over $T$ using the scheme $(\Psi', E', \Gamma)$ says: first show $P_v(c)$ for each constant $c \in \Gamma_{[],v}$; and then show $P_v(g(t_1, \ldots, t_k))$ for each $g \in \Gamma_{w,v}$ assuming $P_{v_i}(t_i)$ for each $i = 1, \ldots, k$ (where $w = v_1 \ldots v_k$ and each $t_i$ is a ground $\Psi$-term of sort $v_i$) using the equations in $E'$. □

We want to use inductive reasoning to prove $E \models_\Psi (\forall x_v)P_v(x_v)$ for each sort $v \in V$, which we may write for short as $E \models (\forall x)P(x)$. The result below follows from Theorem 6.4.4, that initial algebras have no proper subalgebras:

**Theorem 8.7.3** Given a specification $T = (\Psi, E)$ and an algebraic induction scheme $(\Psi', E', \Gamma)$ over $T$, then inductive reasoning with $(\Psi', E', \Gamma)$ over $T$ is **sound**, in the sense that if the steps of inductive reasoning for $P$ using the scheme are carried out, then $T \models (\forall x) P(x)$, provided $\Gamma$ is **inductive** for $(\Psi', E')$ over $(\Psi, E)$, in the sense that:

(I1) two ground $\Psi$-terms are equal under $E$ iff they are equal under $E'$;

(I2) every ground $\Psi'$-term equals some $\Psi$-term under $E'$; and

(I3) every ground $\Psi$-term equals some $\Gamma$-term under $E'$.

In this case we may also say that $(\Psi', E', \Gamma)$ is **inductive** over $(\Psi, E)$.

**Proof:** We need to show $D \models (\forall x_v) P_v(x_v)$ for each $v \in V$, where $D$ is an initial model for $T$. Because (I1) and (I2) imply that initial models of $T$ and $T'$ are $\Psi$-isomorphic, by Proposition 8.7.1 it suffices to show $D' \models (\forall x_v) P_v(x_v)$ for each $v \in V$, where $D'$ is an initial model for $T'$. To this end, define a $V$-sorted subset $M$ of the initial algebra $D' = T_{\Psi', E'}$ by

$$M_v = \{[t] \mid D' \models P_v(t), t \in T_{\Psi, v}\},$$

for each $v \in V$. Then (I1) and (I2) imply $D' \models (\forall x) P$ iff $M = D'$. To prove $M = D'$, it suffices to show that $M$ is a $\Psi'$-algebra, because $D'$ has no proper $\Psi'$-subalgebras (Theorem 6.4.4). By (I3), it suffices to show that $M$ is a $\Gamma$-algebra. But successfully carrying out the steps of the inductive reasoning shows exactly this. □

Conditions (I1) and (I2) above say that $T'$ is a protecting initial extension of $T$, i.e., that after enriching $T$, there are no new ground terms; and (I3) says that all ground terms are (equal to) $\Gamma$-terms. Theorem 8.7.3 not only justifies the most familiar induction schemes, but also many others, as shown in the examples and exercises below. It is worth noticing that the above theorem applies to any reachable model of $T$, not just to an initial model of $T$ (recall that $D$ is reachable iff the unique $\Psi$-homomorphism $I \to D$ is surjective, where $I$ is an initial model of $T$, i.e., iff $D$ satisfies the "no junk" condition).

**Example 8.7.4** The following take $(\Psi, E)$ to be the usual Peano specification for the natural numbers, with $\Psi$ having just the sort Nat and function symbols $0, s$, and with $E = \emptyset$.

1. Of course, Peano induction takes $\Gamma = \Psi' = \Psi$ with $E' = \emptyset$. In this case, inductivity is trivial.

2. Letting $\Gamma$ contain $0, s0, s^2$, with $\Psi' = \Psi \cup \Gamma$ and $E' = \{s^2(n) = s(s(n))\}$ gives the even-odd induction scheme. Then (I1)–(I3) are easy to prove.  □

**Exercise 8.7.1**  Show that the $n$-jump Peano induction scheme is sound for each $n > 0$.  □

**Example 8.7.5**  A more sophisticated algebraic induction scheme lets $\Gamma$ contain $0, s0$ plus a unary function $\_ \times p$ for each prime $p$, with $(\Psi', E')$ defining the usual binary multiplication and (to help define that) the usual binary addition. The resulting scheme, which we call **prime induction**, says that to prove $P(n)$ for all natural numbers $n$, prove $P(0)$, $P(s0)$, and that $P(n)$ implies $P(n \times p)$ for each prime $p$.

This scheme is inductive, because in this case, (I1) and (I2) just say that after enriching $T$ with addition, multiplication, and primes, the natural numbers are still its ground terms, while (I3) says that every positive number is a product of primes, which is the so-called Fundamental Theorem of Arithmetic, which was first proved by Gauss.  □

**Example 8.7.6**  Prime induction can be used to prove some pretty facts about the so-called *Euler function* $\varphi$, where $\varphi(n)$ is the number of positive integers less than $n$ that are relatively prime to $n$. One of these is the following, which is rather well known as the *Euler formula*,

$$\varphi(n) = n \prod_{p \mid n} \left(1 - \frac{1}{p}\right),$$

where $p$ varies over primes. We can define the Euler function inductively over the prime induction scheme as follows:

$$\begin{aligned}\varphi(1) &= 1 \\ \varphi(np) &= \varphi(n)(p-1) \quad \text{if } p \text{ prime and not } p \mid n \\ \varphi(np) &= \varphi(n)\, p \quad\quad\quad\ \text{if } p \text{ prime and } p \mid n\,.\end{aligned}$$

Or alternatively, we can consider the above as three properties of $\varphi$ that can be proved from its definition as the number of relatively prime numbers less than its argument.

The following is an OBJ proof score for the Euler formula (the specifications NAT and ListOfNat have been omitted):

```
obj PRIME is pr NAT .
  op _|_ : NzNat NzNat -> Bool .
  op prime : NzNat -> Bool [memo].
  op prime : NzNat NzNat -> Bool [memo].
  vars N M : NzNat .
  eq N | M = gcd(N,M) == N .
  eq prime(s 0) = false .
  cq prime(N) = prime(N,p N) if N > s 0 .
```

```
    eq prime(N, s 0) = true .
    cq prime(N,M) = false if M > s 0 and M | N .
    cq prime(N,M) = prime(N,p M) if M > s 0 and not M | N .
  endo

  obj PRIME-DIVISORS is pr PRIME + ListOfNat .
    op pr-div : NzNat -> List .
    op pr-div : NzNat Nat -> List .
    vars N P M : NzNat .  var L : List .
    eq pr-div(s 0)  = nil .
    cq pr-div(P)    = P if prime(P) .
    cq pr-div(N * P) = pr-div(N)      if P | N .
    cq pr-div(N * P) = P pr-div(N)    if prime(P) and not P | N .
    cq pr-div(M)    = pr-div(M,M)     if not prime(M) .
    eq pr-div(N,s 0) = nil .
    cq pr-div(N,P)  = P pr-div(N,p P) if P > s 0 and prime(P)
                                          and P | N .
    cq pr-div(N,P)  = pr-div(N,p P)   if P > s 0 and
                                          not (prime(P) and P | N).
    ops Pi Pip : List -> NzNat .
    eq Pi(nil)  = s 0 .
    eq Pi(N L)  = N * Pi(L) .
    eq Pip(nil) = s 0 .
    eq Pip(N L) = (p N) * Pip(L) .
  endo

  obj EULER is pr PRIME-DIVISORS .
    op phi : NzNat -> NzNat .
    vars N P : NzNat .
    eq phi(s 0) = s 0 .
    cq phi(N * P) = phi(N) * P   if prime(P) and P | N .
    cq phi(N * P) = phi(N) * p P if prime(P) and not P | N .
  endo

  ***> Prove phi(N) * Pi(pr-div(N)) == N * Pip(pr-div(N))
  ***>   for each N:NzNat
  ***> First show the formula for N = 1
  red phi(1) * Pi(pr-div(1)) == 1 * Pip(pr-div(1)) .
  ***> and introduce the basic constants and assumptions
  openr .
  ops n q pq : -> NzNat .
  eq prime(q) = true .
  eq p q = pq .
  close

  ***> Then suppose the property for n and prove it for n * q
  openr .
  eq phi(n) * Pi(pr-div(n)) = n * Pip(pr-div(n)) .
  close
```

```
***> Case where q | n
open .
eq q | n = true .
red phi(n * q) * Pi(pr-div(n * q)) ==
    n * q * Pip(pr-div(n * q)) .
close
***> Case where not q | n
open .
eq q | n = false .
red phi(n * q) * Pi(pr-div(n * q)) ==
    n * q * Pip(pr-div(n * q)) .
close
```

The unary function p is the predecessor function, which is inverse to the successor function s; the function pr-div gives the list of prime divisors of a number; the functions Pi and Pip give the products of a list of numbers, and of the list of predecessors of a list of numbers, respectively. The equation p q = pq tells OBJ that the predecessor of p is positive, an important fact that needs to be proved separately. A total of 148 rewrites were executed in doing this proof. (Special thanks to Grigore Roşu for help with this example.)  □

**Exercise 8.7.2** Prove soundness of the induction scheme for positive natural numbers that shows $P(1)$ and $P(p)$ for each prime $p$, and then shows that $P(m)$ and $P(n)$ imply $P(mn)$ for any positive naturals $m, n$.  □

**Example 8.7.7** We now consider another somewhat sophisticated induction scheme, this one for pairs of natural numbers. The underlying data type is a simple extension of the naturals, with functions for pairing and unpairing,

```
<_,_> : Nat Nat -> 2Nat .
p1,p2 : 2Nat     ->  Nat .
```

subject to the equations

```
p1(< M, N >) = M .
p2(< M, N >) = N .
```

(It is interesting to compare this with the specification for pairs of natural numbers in Example 3.3.10.)

Our induction scheme for this data type uses the functions

```
a,b : Nat  -> 2Nat
f,g : 2Nat -> 2Nat
```

where

```
a(M) = < M, 0 >
b(N) = < 0, N >
f(< M, N >) = < M + N, N >
g(< M, N >) = < M, N + N >
```

where we can think of the $a, b$ as each providing an infinite family of constants.

Then to prove $(\forall p : \text{2Nat})\ P(p)$ for some first-order formula $P$, it suffices to prove the following, where the first two are base cases and the last two are induction steps,

$$(\forall m : \text{Nat}) \quad P(\langle m, 0\rangle)$$
$$(\forall n : \text{Nat}) \quad P(\langle 0, n\rangle)$$
$$(\forall m, n : \text{Nat}) \quad P(\langle m, n\rangle) \Rightarrow P(\langle m+n, n\rangle)$$
$$(\forall m, n : \text{Nat}) \quad P(\langle m, n\rangle) \Rightarrow P(\langle m, n+m\rangle)$$

provided the scheme is inductive, which is Exercise 8.7.3 below. Notice that in the last two steps, we can assume that both $m, n$ are positive without loss of generality (because 0 is covered by the base cases).

To illustrate this induction scheme, we give below an OBJ3 proof score showing that

$$gcd(m, n) = gcd(n, m)$$

by considering $gcd$ as a function $\text{2Nat} \to \text{Nat}$. We first define the inequality relation, and then introduce some lemmas.

```
openr INT .
  op _>_ : Int Int -> Bool .
  vars-of .
  eq s I > I = true .
  eq I > p I = true .
  eq s I > s J = I > J .
  vars M N : Nat .
  eq 0 > M = false .
  eq s M > 0 = true .
  *** some lemmas
  eq I + J + (- J) = I .
  eq I + J + (- I) = J .
  cq I + J > I = true if I > 0 and J > 0 .
  cq I + J > J = true if I > 0 and J > 0 .
  cq I + J > 0 = true if I > 0 and J > 0 .
close

obj 2NAT is sorts 2Nat 2Int .
  pr INT .
  subsort 2Nat < 2Int .
  op <_,_> : Nat Nat -> 2Nat .
  op <_,_> : Int Int -> 2Int .
  ops p1 p2 : 2Nat -> Nat .
```

```
    ops p1 p2 : 2Int -> Int .
    vars M N : Int .
    eq p1(< M, N >) = M .
    eq p2(< M, N >) = N .
  endo

  obj GCD is pr 2NAT .
    op gcd_ : 2Int -> Int .
    vars M N : Int .
    eq gcd < 0, N > = N .
    eq gcd < M, 0 > = M .
    eq gcd < M, M > = M .
    cq gcd < M, N > = gcd(< M - N, N >) if M > N and N > 0 .
    cq gcd < M, N > = gcd(< M, N - M >) if N > M and M > 0 .
  endo

  openr GCD .
  ops m n : -> Nat .
  *** base cases:
  red gcd < m,0 > == gcd < 0,m > .
  red gcd < 0,n > == gcd < n,0 > .
  *** for the induction steps:
  eq m > 0 = true .
  eq n > 0 = true .
  close

  *** induction step computations:
  open .
  eq  gcd < m, n > = gcd < n, m > .   *** induction hypothesis:
  red gcd < m, m + n > == gcd < m + n, m > .
  close

  open .
  eq  gcd < n, m > = gcd < m, n > .   *** induction hypothesis:
  red gcd < m + n, n > == gcd < n, m + n > .
  close
```

These computations require a total of 108 rewrites, many of which check the conditions of equations. □

**Exercise 8.7.3** This involves the material introduced in Example 8.7.7.

(a) Show that the scheme of this example is inductive.

(b) Prove that the equation

< p1(P), p2(P) > = P

holds in any initial model of 2NAT, and use this to conclude that the universal quantification $(\forall P : \text{2Nat})$ is equivalent to $(\forall M, N : \text{Nat})$. □

Notice that in proving that a sentence $P$ holds for some constructor term $t$, we are entitled to assume $P(t')$ for every subterm $t'$ of $t$. This is because when $t = \sigma(t_1, \ldots, t_n)$, we assume $P(t_1), \ldots, P(t_n)$, which in turn were proved assuming that $P$ holds for the top level subterms of $t_1, \ldots, t_n$, etc. Some proofs require these additional assumptions. In the case of the usual Peano induction for the natural numbers, the richer induction principle which includes these additional assumptions is called **strong induction**, or **complete induction**, or **course of values induction**, and it means that in proving $P(n)$, we can assume $P(k)$ for all $k < n$. We will use the same names for the corresponding enrichment of our much more general notion of induction. Below is a simple example for the natural numbers:

**Example 8.7.8** We define a sequence $f$ of natural numbers by $f(0) = 0$, $f(1) = 0$, and

$$f(n) = 3f(div2(n)) + 2 \text{ for } n \geq 2 .$$

The following OBJ proof score shows that $f$ is always even. The module SEQ first defines the auxiliary functions even and div2, which respectively tell if a number is even, and give its quotient when divided by 2. Next four lemmas are stated, a constant $n$ is introduced to eliminate the universal quantifier from the formula to be proved, which is

$$(\forall N) \; even(f(N)) = \text{true} ,$$

and then $n$ is assumed to be at least 2 (we also need to assume it is at least 1, which is easier than introducing another lemma which will deduce that fact). Finally the strong induction hypothesis is stated, the base cases are checked, and then the inductive step is checked, which requires 50 rewrites to get true.

```
obj SEQ is pr NAT .
  op even_ : Nat -> Bool .
  var N : Nat .
  eq even   0 = true .
  eq even s 0 = false .
  eq even s s N = even N .
  op div2_ : Nat -> Nat .
  eq div2 0 = 0 .
  eq div2 s 0 = 0 .
  eq div2 s s N = s div2 N .
  op f_ : Nat -> Nat .
  eq f 0 = 0 .
  eq f s 0 = 0 .
  cq f s N = 3 * f(div2 N) + 2 if N > 0 .
endo
```

```
        openr SEQ .
          vars N M : Nat .
          cq s N > M = true if N > M .
          cq even(N + M) = true if even N and even M .
          cq even(N * M) = true if even M .
          cq N > div2 N = true if N > 0 .
          op n : -> Nat .
          eq n > s 0 = true .
          eq n >   0 = true .
          cq even f N = true if s n > N .
        close

     red even f 0 .
     red even f 1 .
     red even f s n .                                                    □
```

**Exercise 8.7.4** Write a specification for finite sets of naturals having the constructors ∅ and *add* : Nat Set → Set. Now define union, and prove it is associative, commutative, and idempotent. This is a nice example of induction over a specification that is not anarchic.[E41]  □

**Exercise 8.7.5(⋆)** Explore the idea that if specifications $(\Psi, E)$ and $(\Psi', E')$ are equivalent in the (loose) sense of Section 4.10, then they are ground equivalent, in that they have the same initial models (after reducing to a common signature); therefore either one can be used as an induction scheme for the other. Going even further down this road, it might be interesting to consider the two ways of specifying pairs of natural numbers that are given in Examples 3.3.10 and 8.7.3.  □

## 8.8 Literature

Many mathematicians, especially logicians, would say that first-order logic is the most important of all logics, because it is the foundation for set theory, and hence for all of mathematics. It is certainly one of the most intensively studied of all mathematical systems, and from the mid twentieth century has been considered the most classic of all logics. Introductory textbooks on logic include [163, 133] and [48]; there are many more textbooks, and a truly enormous literature of advanced texts and papers.

Most logic texts emphasize the completeness of some set of rules; we have taken a different, some would say eccentric, approach, by developing and using whatever rules we need, subject to proving their soundness based on satisfaction; in this sense, we consider logic to be a kind of open system (see the discussion in Section 1.3). In any case, pure first-order logic is not sufficient for our applications, because of our need for built in data types, induction, and (in the next chapter) second-order logic.

Many mathematicians would also say that the rules of inference of first-order logic are "self evident" logical truths. But this has been challenged by the so-called intuitionism of Brouwer and others, as well as some even more radical approaches, such as Martin-Löf type theory. In particular, the rules of double negation (P5) and proof by contradiction (R4) have been questioned. However, these challenges have less force for reasoning about relatively finitistic applications like numbers, lists, and circuits, which are our main interest in this book.

The material in Section 8.3 builds on the algebraic exposition of first-order logic given in [67]. There are many works on algebraic approaches to logic; some early books are by Paul Halmos [96], Roger Lyndon [125], and Helena Rasiowa [155]; some of the earliest work in this area was done in Poland, and Rasiowa was one of the pioneers. Halmos and Givant [97] give a nice introductory treatment. More advanced approaches involve cylindric algebra, category theory, sheaves, and topoi [100, 44].

Horn clauses are named after Alfred Horn, who for many years was professor of logic at UCLA. Horn clauses are the basis for the syntax of so-called "logic programming" languages, such as Prolog [34]; however, most of the syntax of Prolog does not correspond to Horn clause logic, and the part of its syntax that does correspond has a semantics that differs from the model theory of Horn clause logic.

Our definition of well-formed formulae follows the "initial algebra semantics" advocated by [52] and [89] in using the freeness of certain syntactic algebras. We consider this to be both simpler and clearer than the approaches usually taken in logic.

The semantic definition of truth for first-order logic is originally due to Tarski [175]. This important conceptual advance ushered in the era of so-called "model theory" in mathematical logic, and is the original source for the emphasis on semantics in this book.

The proof of Theorem 8.2.6 is due to Diaconescu, and follows [41]. Those familiar with logic programming may be interested to note that the initial $H$-model $T_H$ is what is usually called a "Herbrand Universe" in that field.

Properties (a) and (b) of Exercise 8.1.3 show that $\Phi$-models with $\Phi$-morphisms form (what is called) a category; then property (c) follows automatically, along with many other useful properties. Category theory also suggests that for some structures bijective morphisms may not be isomorphisms, because bijectivity cannot even be stated for abstract categories (see (d) of Exercise 8.1.3). Such "facts for free" and hints about generalizability motivate the study of category theory, and some basics are given in Chapter 12, along with a few relevant references.

Institutions [67] axiomatize the Tarskian model-theoretic formulation of mathematical logic, by axiomatizing the satisfaction relation between syntax and semantics, indexed by the ambient signature; the main axiom is called the satisfaction condition, a special case of which was treated in Section 4.10. The theory of institutions is developed to some extent in Chapter 14; this theory has been used for many computing science applications, including specification and modularization. The Eqlog system [79, 42] implements the institution *HCLEQ*; because of its use of general equality, Eqlog goes well beyond what standard logic programming languages like Prolog provide, though at some cost in efficiency. The institution *FOLQ/D* is closely related to the hidden algebra institution developed in Chapter 13, which was designed to handle dynamic situations, such as the sequential (i.e., state dependent, time varying) circuits that are discussed in Chapter 9.

Proof planning is a venerable topic in Artificial Intelligence, and there is nothing especially novel about our treatment here, except that we have been very precise about the institutions(s) involved, have put a strong emphasis on equational reasoning, and have accepted the need for human involvement. Inference and proof planning are not equational deduction, because not all rules are reversible; instead they are rewrite rules. If a more logical formulation is really desired, then Meseguer's rewriting logic [134] is applicable.

The formulation of the Substitution Theorem (Theorem 8.3.22) appears to be new. Special thanks to Grigore Roşu for help with its proof, especially Proposition 8.3.21, on which it rests. Skolem constants and functions were introduced by the Norwegian logician Thoralf Skolem in his studies of first-order proof theory in the late 1920s.

The material on algebraic induction in Section 8.7 appears to be new. In general, mathematicians have not seen much value in formalizing exotic variants of ordinary induction, though they have done considerable work in formalizing variants of transfinite induction. On the other hand, computer scientists have been very concerned with inductive techniques for complex data structures; Burstall's work on structural induction pioneered this important area [21]. Some related work has been done in computer science under the name of "cover set" and "test set"; for some recent work, with citations of older literature, see [17, 18]; this work concerns semi-automated inductive proofs of equations for the one-sorted case, largely restricted to anarchic constructors. This material, like much else in this book, is also more general than most of the existing literature in that it treats the overloaded many-sorted case.

Humans have a deep seated desire for the phenomena that they experience to appear coherent, e.g., through some kind of causal explanation. In particular, we want to know *why* a particular step is taken in a proof, not just that it happens to work. Unfortunately, formal proofs usually leave out the motivations for their steps. However, a proof should be more fun (or at least, relatively easier) to read if it is orga-nized to *tell a story*, explaining how various difficulties arise and are overcome. Thus it seems likely that much can be learned about how to effectively present proofs by studying narratives, particularly oral stories as they are actually told, and perhaps also movies. References on narratology include [118, 7, 123, 145, 124], and more information on proofs as narratives can be found in [64]. Semiotics, which is the study of signs, should also be able to help make proofs easier to understand; references on semiotics include [149, 104, 64, 83].

---

**A Note for Lecturers:** Although much of the material in this chapter looks rather technical, there are usually simple underlying intuitions that can be brought out through discussing the examples. (However the proof of Proposition 8.3.21 in Appendix B really is rather technical.)

Some students are troubled by the seemingly circular use of first-order logic to reason about first-order logic. Therefore it should be explained that *informal* first-order (and occasionally higher-order) logic is the language of mathematics, and that the project of this chapter is to *formalize* first-order logic, that is, to make of it a mathematical object, which can then be reasoned about; this formal system has much the same status as the formalizations of the natural numbers that we have been dealing with all along, but of course it is much more complex. The major purpose of this formalization, for this book, is to justify the various ways that we use to mechanize reasoning.

---

# 9 Second-Order Equational Logic

This chapter generalizes ordinary equational logic, which involves only universal quantification over constants, to *second-order* equational logic, which in addition permits universal quantification over operations. We then develop full second-order logic with the same techniques used for first-order logic in Chapter 8, and in particular, we extend first-order quantification to second-order quantification using the techniques of Section 8.3, emphasizing the special case where the only predicates are the equality predicates. This treatment of second-order quantification is a natural extension of our treatment of first-order quantification, and the algebraic techniques used are essentially the same as for the first-order case.

These generalizations are crucial for our approach to verifying *sequential* hardware circuits, the behavior of which changes over time, due to the effects of internal memory and/or external inputs. Although the resulting logic is much simpler than higher-order logic, it is entirely adequate for applications to hardware. These applications extend the approach of Section 7.4 with infinite sequences of Boolean values on wires, instead of single Boolean values, so as to model states, inputs, etc. that vary with (discrete, unbounded) time, as represented by the sequence of natural numbers. The following illustrates some of what can be done in this framework:

**Example 9.0.1** The fact that the union of a relation with its converse is its symmetric closure, i.e., is the smallest symmetric relation containing it, is stated by the second-order formula below, where $R, S$ are relation variables, i.e., they have rank $DD \to $ `Bool` for some sort $D$ (which you should think of as the "domain" of $R$ and $S$), and $x, y$ are ordinary variables of sort $D$:

$$(\forall R, S) \; ( \; [ \; (\forall x, y) \; (S(x,y) = S(y,x)) \land (R(x,y) \Rightarrow S(x,y)) \; ] \Rightarrow$$
$$[ \; (\forall x, y) \; (R(x,y) \lor R(y,x) \Rightarrow S(x,y)) \; ] \; ) \; .$$

After the relevant formal definitions are presented, Example 9.1.9 will give an OBJ proof score that verifies the above formula. □

We begin our formal development with the syntax of equations:

**Definition 9.0.2**  A (second-order) $\Sigma$-**equation** is a signature $X$ of **variable symbols** (disjoint from $\Sigma$) plus two $(\Sigma \cup X)$-terms; we write such equations abstractly in the form

$$(\forall X)\, t = t'$$

and concretely in forms like

$$(\forall x, y, z, f, g)\, f(x, y, z) = g(x, y) + z\,,$$

when $X = \{x, y, z, f, g\}$ and their sorts can be inferred from their uses in the terms. This definition and notation extend to conditional equations in the usual way.  □

To define satisfaction for second-order equations, we extend the notion of an assignment of values in a $\Sigma$-algebra from ground signatures to arbitrary signatures. Such an assignment on $X$ is just an interpretation of $X$ in $M$, that is, an $X$-algebra structure for $M$ in addition to the $\Sigma$-algebra structure it already has. Since $\Sigma$ and $X$ are disjoint, this means that $M$ gets the structure of a $(\Sigma \cup X)$-algebra. Therefore the $\Sigma$-algebra $M$ and the interpretation $a : X \to M$ in $M$ of the variable symbols in $X$ determine a unique $(\Sigma \cup X)$-homomorphism $\overline{a} : T_{\Sigma \cup X} \to M$ by the initiality of $T_{\Sigma \cup X}$. Note that operation symbols in $X$ are interpreted as functions on $M$ in exactly the same way as are operation symbols in $\Sigma$. Now we are ready for

**Definition 9.0.3**  A $\Sigma$-algebra $M$ **satisfies** a second-order $\Sigma$-equation $(\forall X)\, t = t'$ iff for any interpretation $a : X \to M$ we have $\overline{a}(t) = \overline{a}(t')$ in $M$. In this case we may write

$$M \vDash_{\Sigma} (\forall X)\, t = t'$$

omitting some $\Sigma$ subscripts for simplicity. A $\Sigma$-algebra $M$ **satisfies** a set $A$ of (first and second-order) $\Sigma$-equations iff it satisfies each $e \in A$, and in this case we write

$$M \vDash_{\Sigma} A\,.$$

We may also say that $M$ is a $P$-algebra, and write $M \vDash P$ where $P = (\Sigma, A)$. Once again, this extends to conditional equations in the obvious way.  □

It is very pleasing that this is such a straightforward generalization of first-order equations and their satisfaction, obtained by using an *arbitrary* signature instead of a ground signature; in particular, note that initiality plays exactly the same role here as it did in the first-order case.

Perhaps surprisingly, specifications with second-order equations have initial models (see Theorem 9.1.10 below); the proof is nearly the same as for the first-order case (Theorem 6.1.15), but is deferred to the next section because it requires a result given there. The following illustrates the satisfaction of second-order equations, and also shows how easy it is to write second-order equations that identify all elements of any model that satisfies them:

**Example 9.0.4** Many simple second-order equations only have trivial models. For example, given a one-sorted signature, the equation

$$(\forall f, x, y)\ f(x, y) = f(y, x)$$

implies $(\forall x, y)\ x = y$, by taking the function $f(x, y) = x$. Therefore any model satisfying this equation must have just one element. Some other simple equations that have the same effect are $(\forall f, x)\ f(x) = a$ where $a$ is a constant, and $(\forall f, g, x)\ f(x) = g(x)$.   □

## 9.1  Theorem of Constants

The development of first-order equational logic in Chapters 3 and 4 defined variables to be new constants, and used the Theorem of Constants (Theorem 3.3.11) to justify proving equations with variables by regarding the variables as constants, so that ground term deduction could be used; recall that this was necessary for applications such as inductive proofs. We now extend it to our second-order setting, where it will play the same role:

**Theorem 9.1.1** (*Theorem of Constants*) Given disjoint signatures $\Sigma$ and $X$, given a set $A$ of $\Sigma$-equations, and given $t, t' \in T_{\Sigma \cup X}$, then

$$A \vDash_\Sigma (\forall X)\, t = t' \quad \text{iff} \quad A \vDash_{\Sigma \cup X} (\forall \varnothing)\, t = t'\ .$$

**Proof:** Each condition is equivalent to the condition that

$$\overline{a}(t) = \overline{a}(t') \text{ for every } (\Sigma \cup X)\text{--algebra } M \text{ satisfying } A \text{ and}$$
$$\text{every } a \colon X \to M\ ,$$

where $\overline{a} \colon T_{\Sigma \cup X} \to M$ is the unique homomorphism.   □

This is exactly the same proof that we gave for the first-order case; as before, its simplicity arises from using satisfaction and initiality, rather than using some set of rules of deduction. This and the Completeness Theorem for first-order equational logic give the following:

**Corollary 9.1.2** Given disjoint signatures $\Sigma$ and $X$, given a set $A$ of $\Sigma$-equations, and given $t, t' \in T_{\Sigma \cup X}$, then

$$A \vDash_\Sigma (\forall X)\, t = t' \quad \text{iff} \quad A \vdash_{\Sigma \cup X} (\forall \varnothing)\, t = t'\ .$$
   □

Results below on expanding and contracting signatures, which generalize results in Section 4.7, use the following:

**Definition 9.1.3** A signature $\Psi$ is **non-void over** another signature $\Sigma$ iff $\Psi$ is disjoint from $\Sigma$ and $(T_{\Sigma \cup \Psi})_s$ is non-empty for every sort $s$ of $\Psi$. Similarly, a signature $\Psi$ is **non-void relative** to a subsignature $\Phi$ **over** another signature $\Sigma$ if and only if $\Psi$ is disjoint from $\Sigma$, and $(T_{\Sigma \cup \Psi})_s$ is non-empty for every sort $s$ of $\Psi$ iff $(T_{\Sigma \cup \Phi})_s$ is non-empty for every sort $s$ of $\Phi$. □

**Fact 9.1.4** If $\Phi \subseteq \Psi$, if $\Phi$ is non-void over $\Sigma$, and if $\Psi$ is non-void relative to $\Phi$ over $\Sigma$, then $\Psi$ is non-void over $\Sigma$.

**Proof:** By the relative non-void hypothesis, $(T_{\Sigma \cup \Psi})_s \neq \emptyset$ for every sort $s$ of $\Psi$ iff $(T_{\Sigma \cup \Phi})_s \neq \emptyset$ for every sort $s$ of $\Phi$, and the latter is true by the non-voidness of $\Phi$. Therefore the former is also true. □

**Proposition 9.1.5** If $(\forall \Phi) \, t = t'$ is a (second-order) $\Sigma$-equation and if $\Phi \subseteq \Psi$ with $\Psi$ non-void relative to $\Phi$ over $\Sigma$, then $(\forall \Psi) \, t = t'$ is also a $\Sigma$-equation, and for every $\Sigma$-algebra $M$,

$$M \vDash_\Sigma (\forall \Phi) \, t = t' \text{ iff } M \vDash_\Sigma (\forall \Psi) \, t = t' \, .$$

**Proof:** Let $a : \Phi \to M$ be such that $\overline{a}(t) = \overline{a}(t')$. Then $\Phi$ must be non-void over $\Sigma$, and hence so is $\Psi$, by the non-voidness of $\Psi$ relative to $\Phi$. Therefore we can extend $a$ to $b : \Psi \to M$ such that $\overline{b}(t) = \overline{b}(t')$ by choosing interpretations for operations in $\Psi - \Phi$. On the other hand, if there are no such interpretations $a$, then $\Psi$ must be void, so that $\Phi$ is also void, by the non-voidness of $\Psi$ relative to $\Phi$, and hence there are no such interpretations $b$. Therefore the first condition implies the second.

Conversely, if $b : \Psi \to M$ such that $\overline{b}(t) = \overline{b}(t')$, then $\Psi$ is non-void over $\Sigma$, so we can restrict $b$ to $a : \Phi \to M$ such that $\overline{a}(t) = \overline{a}(t')$. On the other hand, if there is no such $b$, then $\Psi$ must be void, and then relative non-voidness implies $\Phi$ is void too, so there is also no such $a$. □

The above result justifies adding extra constant and operation symbols to a proof score under appropriate conditions. Necessity of the non-voidness hypothesis is shown by the specification given in Example 4.3.8, but the following is also interesting:

**Exercise 9.1.1** Give signatures $\Sigma, \Phi, \Psi$ plus a $\Sigma$-model $M$ and a $\Sigma$-sentence $(\forall \Phi) \, t = t'$ such that $\Phi \subseteq \Psi$ where $\Psi$ adds only non-constant operations to $\Phi$, and $M$ satisfies $(\forall \Phi) \, t = t'$, but does not satisfy $(\forall \Psi) \, t = t'$. □

The Theorem of Constants also generalizes in a way similar to that of Proposition 9.1.5:

**Corollary 9.1.6** Given a (second-order) $\Sigma$-equation $(\forall\Phi)\, t = t'$, a signature $\Delta$ disjoint from both $\Phi$ and $\Sigma$, and a set $A$ of $\Sigma$-equations, let $\Psi = \Phi \cup \Delta$. Then

$$A \vDash_\Sigma (\forall\Psi)\, t = t' \text{ iff } A \vDash_{\Sigma \cup \Phi} (\forall\Delta)\, t = t',$$

provided $\Psi$ is non-void relative to $\Phi$ over $\Sigma$.

**Proof:** If $M \vDash_\Sigma A$ then $M \vDash_\Sigma (\forall\Phi)\, t = t'$ iff $M \vDash_\Sigma (\forall\Psi)\, t = t'$ by Proposition 9.1.5, so that $A \vDash_\Sigma (\forall\Phi)\, t = t'$ iff $A \vDash_\Sigma (\forall\Psi)\, t = t'$ iff $A \vDash_{\Sigma \cup \Phi} (\forall\Delta)\, t = t'$, the last step by Theorem 9.1.1.  □

(Theorem 9.1.1 is the special case where $\Delta = \emptyset$.)

Theorem 9.1.1 also justifies a key rule of deduction for second-order equational logic, generalizing the first-order universal quantifier elimination rule (and the transformation T5 of Chapter 8) to the second-order case. Let $\vdash^1$ denote the syntactic derivability relation defined by some complete set of rules $\mathcal{R}^1$ for first-order equations, and let $\vdash^2$ be defined by $\mathcal{R}^1$ plus the following new rule, which we will call **dropping**:

(D) $\quad A \vdash^1_{\Sigma \cup \Phi} (\forall\emptyset)\, t = t' \text{ implies } A \vdash^2_\Sigma (\forall\Phi)\, t = t'$.

**Fact 9.1.7** The rule (D) is sound.

**Proof:** If $A \vdash^1_{\Sigma \cup \Phi} (\forall\emptyset)\, t = t'$, then $A \vDash_\Sigma (\forall\Phi)\, t = t'$ by Corollary 9.1.2, so it is sound to infer $(\forall\Phi)\, t = t'$.  □

We also have the following:

**Theorem 9.1.8** (*Completeness*) If $\mathcal{R}^1$ is a set of rules of deduction for first-order equational logic, defining a relation $\vdash^1$ that is complete, then $\mathcal{R}^2 = \mathcal{R}^1 \cup \{(D)\}$ is complete for unconditional second-order equations, in the sense that it defines a relation $\vdash^2$ such that

$$A \vdash^2_\Sigma (\forall\Phi)\, t = t' \text{ iff } A \vDash_\Sigma (\forall\Phi)\, t = t',$$

for any signature $\Sigma$ and any set $A$ of first-order $\Sigma$-equations.

**Proof:** The direct implication is soundness of (D), which is Fact 9.1.7, plus soundness of $\vdash^1$. For the converse, if $A \vDash_\Sigma (\forall\Phi)\, t = t'$, then Corollary 9.1.2 gives $A \vdash^1_{\Sigma \cup \Phi} (\forall\emptyset)\, t = t'$, and then (D) gives $A \vdash^2_\Sigma (\forall\Phi)\, t = t'$.  □

This result supports deducing a single second-order equation from a set of first-order equations; it does not provide a complete inference system for second-order equational logic, which would instead infer second-order equations from other second-order equations. However, when all sorts involved are non-void, Corollary 9.1.6 allows dropping second-order equations to first-order equations, which could then be used in proofs, although in a limited way, because we cannot substitute for the second-order variables that have been dropped. Nevertheless, the inferences that are supported by the above result are all we need for our applications to the verification of sequential circuits, and many other applications, such as the following:

**Example 9.1.9** We use the machinery developed above to prove the formula of Example 9.0.1. As usual, relations are translated into Boolean-valued functions. The quantifiers for $R$ and $S$ may be considered eliminated by their declarations; note that commutativity of S is given as an attribute in its declaration, instead of an equation. The main implication in the formula is eliminated by applying the rule (D). The quantifiers for $x$ and $y$ are eliminated in the usual way, as are the implications within their two scopes. The case split in the proof is justified by applying the disjunction elimination rule (T2 of Chapter 8) to the formula $R(x, y) \vee R(y, x)$, and then translating the disjuncts to equations.

```
th SETUP is sort D .
  op R : D D -> Bool .
  op S : D D -> Bool [comm] .
  vars X Y : D .
  cq S(X,Y) = true if R(X,Y) .
  ops x y : -> D .
endth
open . *** first case
  eq R(x,y) = true .
  red S(x,y) .
close
open . *** second case
  eq R(y,x) = true .
  red S(x,y) .
close
```

As expected, both reductions give true. Of course, this is a very simple example. □

**Exercise 9.1.2** Extend Example 8.2.7 by proving that for any relation R, its transitive closure R* as defined there, is the least transitive relation containing R. (To do this in OBJ, some ingenuity will be needed in handling the equation for transitivity.) □

**Theorem 9.1.10** (*Initiality*) Given a set $A$ of $\Sigma$-equations, possibly conditional, which may be either first or second-order, let $\equiv$ be the $\Sigma$-congruence on $T_\Sigma$ generated by the relation $R$ having the components

$$R_s \;=\; \{\, \langle t, t' \rangle \mid A \vdash^2 (\forall \emptyset)\; t = t' \text{ where } t, t' \text{ are of sort } s \,\}\,.$$

Then $T_\Sigma/R$, denoted $T_{\Sigma,A}$, is an initial $(\Sigma, A)$-algebra.

**Proof:** [E42] Given any $(\Sigma, A)$-algebra $M$, let $\nu : T_\Sigma \to M$ be the unique homomorphism, and notice that $R \subseteq ker(\nu)$, because $M \models A$ implies $M \models (\forall \emptyset)\ t = t'$ for every $\langle t, t' \rangle \in R$ by Theorem 9.1.8. Now let $v = q; u$ with $u : T_{\Sigma,A} \to M$ be the factorization of $v$ given by Proposition 6.1.12; see Figure 6.2. For uniqueness, if also $u' : T_{\Sigma,A} \to M$, then $q; u' = v$ by the initiality of $T_\Sigma$, and $R \subseteq ker(v)$ because $M \models A$. Therefore $u = u'$ by (2) of Proposition 6.1.12.   □

## 9.2   Second-Order Logic

The algebraic development of first-order logic in Chapter 8 extends straightforwardly to second-order quantification. As before, we assume a fixed first-order signature $\Phi = (\Sigma, \Pi)$ with sort set $S$, but now we let $\mathcal{X}$ be an $(S^* \times S)$-indexed set of variable symbols, disjoint from $\Sigma$ and $\Phi$, such that each sort has an infinite number of symbols. For $X \subseteq \mathcal{X}$, a $(\Phi, X)$-**term** is an element of $T_{\Sigma \cup X}$ and the $(S^* \times S)$-indexed function *Var* on these terms is defined by:

0. $Var_{w,s}(\sigma) = \emptyset$ if $\sigma \in \Sigma_{[],s}$ ;
1. $Var_{w,s}(x) = \emptyset$ if $x \in X_{[],s}$ and $w \neq []$ ;
2. $Var_{[],s}(x) = \{x\}$ if $x \in X_{[],s}$ ;
3. $Var_{w,s}(\sigma(t_1, \ldots, t_n)) = \bigcup_{i=1}^{n} Var_{w,s_i}(t_i)$ if $n > 0$ and $\sigma \notin X_{w,s}$ ;
4. $Var_{w,s}(\sigma(t_1, \ldots, t_n)) = \{\sigma\} \cup \bigcup_{i=1}^{n} Var_{w,s_i}(t_i)$ if $n > 0$ and $\sigma \in X_{w,s}$ .

(As before, *Var* can be seen as a $(\Sigma \cup X)$-homomorphism $T_{\Sigma \cup X} \to \mathcal{P}(X)$.)

Then the well-formed $(\Phi, X)$-**formulae** are defined just as in Definition 8.3.1, elements of the (one-sorted) algebra $WFF_X(\Phi)$, free over the metasignature $\Omega$, except that the universal quantification operations $(\forall x)$ will now include non-constant function symbols, and the atomic $(\Phi, X)$-formula generators should be

$$G_X = \{\pi(t_1, \ldots, t_n) \mid \pi \in \Pi_{s_1 \ldots s_n} \text{ and } t_i \in (T_{\Sigma \cup X})_{s_i} \text{ for } i = 1, \ldots, n\} \ .$$

A $\Phi$-formula is a $(\Phi, X)$-formula for some $X$, and the functions *Var* and *Free*, giving all variables, and all free variables, of $\Phi$-formulae are defined just as in Definition 8.3.1; the notions of bound variable, closed formula, scope, etc. are also the same.

The semantics of first-order formulae in Section 8.3.3 generalizes directly to second-order formulae, along with the results in that section. In particular,[E43] Definition 8.3.2 does not need to be changed at all, except to note that $X$ is an arbitrary signature, not just a ground signature, so that interpretations of $X$ also need to be general; in particular, all the rules given in that section remain sound. The material on substitutions in Section 8.3.4 should be modified a bit, but we will not do so, because we don't need it for this chapter.

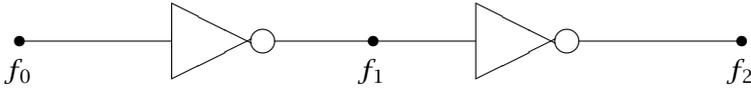

Figure 9.1: Series Connected Inverters

Let us denote the institution of second-order logic that results from the above by *TOL*, and denote the special case where the only predicates are the equality predicates, by *TOLQ*, calling it the **second-order logic of equality**. Also, as in Section 8.3.6, if we fix a signature $\Psi$ and a $\Psi$-model $D$, then we can define the institution *TOLQ/D* to be *TOLQ* with the additional requirement that all its signatures must contain $\Psi$, and that all its models $M$ must be such that their reduct $M|_\Psi$ to $\Psi$ is $D$.

## 9.3 Verification of Sequential Circuits

Whereas combinational circuits can be described by equations that do not involve time, sequential circuits have behaviors that vary with time, and thus require modeling wires that have time-varying values. One common approach is to model such wires as streams of Boolean values. We can still apply the method of Section 7.4 to obtain a system of equations from a circuit diagram, but the variables that model wires now take values that are *functions* from Nat (where the natural numbers represent moments of time) to truth values. As before, we use PROPC to represent the values on wires, rather than just BOOL, so that we can exploit its decision procedure for propositional logic. The following simple example illustrates the approach:

**Example 9.3.1** (*Series Connected Inverters*) We prove that the series connection of two NOT gates (i.e., inverters), each with one unit delay, has the same effect as a two unit delay; see Figure 9.1. The system of equations involved here is

$$f_1(t+1) = \text{not } f_0(t)$$
$$f_2(t+1) = \text{not } f_1(t)$$

each of which is (implicitly) universally quantified over $f_0$, $f_1$, $f_2$ and $t$, where each variable $f_i$ has rank $\langle \text{Nat}, \text{Prop} \rangle$, and where $t$ has sort Nat. We think of $f_0$ as an input variable, $f_2$ as an output variable, and $f_1$ as an internal variable. The behavior that we expect this circuit to have is described by the equation

$$f_2(t+2) = f_0(t) \,,$$

i.e., it functions as a two unit delay. Moreover, its internal behavior (at $f_1$) is described by the first equation of the system,

$$f_1(t + 1) = \text{not } f_0(t) .$$

An OBJ proof score showing that these two terms indeed solve the two inverter system will be given below. The assertion to be verified has the form

$$A \models_\Sigma (\forall \Phi) \, r$$

where $\Sigma$ is the union of the signatures of the objects PROPC and NAT, $A$ is the union of their equations, $\Phi$ is the signature containing three function symbols $f_0, f_1, f_2$ of rank $\langle \text{Nat}, \text{Prop} \rangle$, and $r$ is of the form $(e_1 \wedge e_2) \Rightarrow e_3$, where

$$\begin{aligned} e_1 &= (\forall t) \, f_1(s \, t) = \text{not } f_0(t) \\ e_2 &= (\forall t) \, f_2(s \, t) = \text{not } f_1(t) \\ e_3 &= (\forall t) \, f_2(s \, s \, t) = f_0(t). \end{aligned}$$

We use the transformation rules of Section 8.4 plus R14. By R14 and T6, it suffices to prove that

$$A \models_{\Sigma \cup \Phi} (\forall \emptyset) \, r \, ,$$

which by rule T1 is equivalent to

$$A \cup \{e_1 \wedge e_2\} \models_{\Sigma \cup \Phi} e_3 \, ,$$

which by rule T6 again can be verified by proving

$$A \cup \{e_1 \wedge e_2\} \models_{\Sigma \cup \Phi \cup \{t\}} f_2(s \, s \, t) = f_0(t) \, ,$$

which by rule T3 is equivalent to

$$A \cup \{e_1, e_2\} \models_{\Sigma \cup \Phi \cup \{t\}} f_2(s \, s \, t) = f_0(t) \, ,$$

which is exactly what the proof score below does. This series of deductions at the meta level can be automated using the same techniques as were used in Section 8.4, but we do not give details here.

```
open NAT + PROPC .
ops (f0_)(f1_)(f2_) : Nat -> Prop [prec 9] .
var T : Nat .
eq f1 s T = not f0 T .
eq f2 s T = not f1 T .
op t : -> Nat .
red f2 s s t iff f0 t .
close
```

Notice that the output values of the inverters at time 0 are not determined by the equations given above, and do not enter into the verification. If desired, the following equations could be added

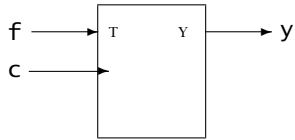

Figure 9.2: Parity of a Bit Stream

```
eq f1 0 = false .
eq f2 0 = false .
```

to give them fixed values, but this is not necessary. □

The sentence proved here is typical of a very large class of sequential hardware verification problems, which have the form

$$A \models_\Sigma (\forall \Phi)\, (C \Rightarrow e)$$

where $A$ defines the abstract data types of the problem, where $C$ is a conjunction of equations defining the circuit, where $e$ is an equation to be proved, and where $\Phi$ may involve second-order quantification.

**Example 9.3.2** We consider a simple circuit to compute the parity of a stream of bits, using just one T (for "toggle") type flip-flop. In Figure 9.2, f is the input stream, c is the clock stream, and y is the output stream; the clock stream just marks cycles to stabilize the flip-flop, and can be ignored for our (logical) purposes. This flip-flop satisfies the following equations,

$$\begin{aligned} y(t+1) &= f(t) + y(t) \\ y(0) &= \textit{false} \end{aligned}$$

from which it follows by induction that

$$(\forall f, y, t)\ y(t+1) = \sum_{i=0}^{t} f(i)\ .$$

The proof is as follows: The base case with $t = 0$ follows because $y(1) = f(0) + y(0) = f(0) + \textit{false} = f(0)$. For the induction step, we assume the above equation, and then prove it with $t+1$ substituted for $t$, by first noting that $y(t+2) = f(t+1) + y(t+1)$ and then applying the above equation. □

**Exercise 9.3.1** Prove correctness of the circuit of Example 9.3.2 using OBJ. □

It is worth remarking that any verification of a combinational circuit "lifts" to a verification of the same circuit viewed as a sequential circuit, by replacing each wire variable, whether an input $i_k$ or a non-input $p_k$, by a function Nat -> Prop, e.g., in the form $f_k(t)$; the reason is that the same proof works for the lifted system, with the same verified property holding at each instant.

## 9.4  Literature and Discussion

The material in this chapter is based on [59], although Proposition 9.1.5 and Theorems 9.1.10 and 9.1.8 are not there, and appear to be new, as does the exposition of second-order logic. The proofs involving universal properties of quotient and freedom seem especially elegant and simple.

It is interesting to compare our method for representing sequential circuits with the more familiar method which represents components using higher-order relations, and represents connections using existential quantification (as in the usual definition of the composition of relations) [93, 94, 25]; see also [167]. By contrast, the representation suggested here uses no relations (except equality in an implicit way), and it represents interconnection by equality of wires. For sequential circuits, both methods represent wires as variables that range over functions, and both methods use second-order quantifiers. However, the results of this chapter show that existential quantification and higher-order relations can be avoided in favor of a simple extension of first-order equational logic by universal quantification over functions, contrary to claims made in [25]. The higher-order logic approach to hardware verification of [25] was claimed to have many benefits, including the following:

1. natural definitions of data types (using Peano style induction principles);

2. the possibility of leaving certain values undefined (such as the initial output of a delay);

3. dealing with bidirectional devices.

But all these benefits can be realized more simply using just second-order equational logic:

1. Chapter 6 showed that initial algebra semantics supports abstract data type definitions in a very natural way, and also supports the use of structural induction principles for such definitions.

2. It is very easy to leave values undefined, such as the values of inverters at time 0; conditional equations can also be used for this purpose.

3. Although it is often advantageous to exploit causality (in the form of an input/output distinction), we are not limited to that case, because equality is bidirectional.

4. Moreover, because equational logic is simpler than higher-order logic, in general its proofs are also simpler.

Of course, our method also has its limits, but fortunately, the problems of greatest interest for hardware verification fall well within its capabilities.

Although the rules and transformations of Chapter 8 were proved for first-order logic, they extend to second-order quantification. We did not use such extended rules in this chapter, because (e.g., in Example 9.3.1) we first applied rule (D) to get rid of the second-order universal quantifiers. This is sufficient for assertions of the form $A \models_\Sigma (\forall \Phi)\, r$ where $r$ contains only first-order quantifiers, but it is not sufficient if $r$ contains second-order quantifiers. Many of the rules in Chapter 8 actually hold for a wide variety of logics, as can be shown using material on institutions in Chapter 14.

It is worth mentioning that the use of a loose extension of a fixed data theory in the institution *TOLQ/D* is closely related to the hidden algebra institution developed in Chapter 13; this should not be surprising, because hidden algebra was designed to handle dynamic situations, of which sequential circuits are a prime example.

---

**A Note for Lecturers:** This short chapter contains some nice examples and some relatively easy theory, which should be included in a course if possible, after the relevant parts of Chapters 7 and 8 have been covered. Proposition 9.1.5 can be skipped, as can the details in Section 9.2.

---

# 10  Order-Sorted Algebra and Term Rewriting

There are many examples where all items of one sort are necessarily also items of some other sort. For example, every natural number is an integer, and every integer is a rational. We can write this symbolically in the form

```
Natural ≤ Integer ≤ Rational ,
```

where `Natural`, `Integer`, and `Rational` are *names* for the sorts of entity involved. If we associate to each such name a *meaning* (i.e., a semantic denotation, also called an *extension*) which is the set of all items of that sort (e.g., the set of all integers), then the subsort relations appear as set-theoretic inclusions of the corresponding extensions. For example, if the usual extensions of `Natural`, `Integer`, and `Rational` are denoted $\mathbb{N}, \mathbb{Z}$, and $\mathbb{Q}$, respectively, then we have

$$\mathbb{N} \subseteq \mathbb{Z} \subseteq \mathbb{Q} .$$

Sort names like `Natural` and `Rational` are *syntactic*, while their extensions are *semantic*. This distinction is formalized below with *order-sorted signatures*, which include a set of sort names with a partial order relation, and a family of operation symbols with sorted arities; an *order-sorted algebra* for a given signature is then an interpretation for these sort and operation names that respects the subsorts and arities. This area of mathematics is called **order-sorted algebra** (hereafter abbreviated **OSA**).

A closely related topic is *overloading*, which allows a single symbol to be used for several different operations. In applying an overloaded operation symbol, we may not even be aware that we are moving among various sorts and operations. For example, we can add $2 + 2$ (two naturals), or $-2/3 + (-2)$ (a rational and an integer), or $2 + 3/29$ (a natural and a rational), or $2/9 + 7/9$ (two rationals). This rich flexibility comes from having both an overloaded operation symbol +, and a subsort relation among naturals, integers, and rationals, in such a way that whichever addition is used, we always get the same result from the same arguments, provided they make sense. We may describe this by saying that + is *subsort polymorphic*.

However, this is only one of several ways that the term "polymorphic" is used. The term was introduced by Christopher Strachey to express the use of the same operation symbol with different meanings in programming languages. He distinguished two main forms of polymorphism, which he called *ad hoc* and *parametric*. In his own words [173]:

> In *ad hoc* polymorphism there is no simple systematic way of determining the type of the result from the type of the arguments. There may be several rules of limited extent which reduce the number of cases, but these are themselves *ad hoc* both in scope and content. All the ordinary arithmetic operations and functions come into this category. It seems, moreover, that the automatic insertion of transfer functions by the compiling system is limited to this class.
>
> Parametric polymorphism is more regular, as illustrated by the following example: Suppose $f$ is a function whose argument is of type $\alpha$ and whose result is of type $\beta$ (so that the type of $f$ might be written $\alpha \to \beta$), and that $L$ is a list whose elements are all of type $\alpha$ (so that the type of $L$ is $\alpha$ **list**). We can imagine a function, say *Map*, which applies $f$ in turn to each member of $L$ and makes a list of the results. Thus $Map[f, L]$ will produce a $\beta$ **list**. We would like *Map* to work on all types of list provided $f$ was a suitable function, so that *Map* would have to be polymorphic. However its polymorphism is of a particularly simple parametric type which could be written $(\alpha \to \beta, \alpha\, \textbf{list}) \to \beta\, \textbf{list}$, where $\alpha$ and $\beta$ stand for any types.

Strachey's distinction is based on the kind of semantic relationship that holds between the different interpretations of an operation symbol, and it suggests a more detailed distinction among the different kinds of polymorphism, in which the less *ad hoc* the relationship is, the easier it is to do type inference, and the closer it is to parametric polymorphism:

- In **strongly ad hoc polymorphism**, an operation symbol has semantically unrelated uses, such as + for both integer addition and Boolean disjunction. (But even in this case, the two instances of + share the associative, commutative, and identity properties.)

- In **multiple representation**, the uses are related semantically, but their representations may be different. For example, in an arithmatic system we may have several representations for the number 2, such as 2.0 and 4/2. Polar and Cartesian coordinate representations of points in the plane are another example.

- **Subsort polymorphism** is where the different instances of an operation symbol are related by subset inclusion such that the result does not depend on the instance used, as with + for natural, integer, and rational numbers. This sense is developed in this chapter.
- **Parametric polymorphism**, as in Strachey's *Map* function, appears in many higher-order functional programming languages, including ML [99, 180], Haskell [107], and Miranda [179].

OBJ supports all four kinds of polymorphism. *Strongly ad hoc* polymorphism is supported by signatures in which the same operation symbol has sorts that are unrelated in the subsort hierarchy. We have already explained that subsort polymorphism is inherent in the nature of OSA. Strachey's implementation of arithmetic involved "transfer functions" (which would now be called "coercions") to change the representation of numbers; but coercions are not needed for subsort polymorphic operations, because subsorts appear as subset inclusions of the data items. Also, for regular signatures (in the sense of Definition 10.2.5 below), expressions involving subsort polymorphism always have a smallest sort. OSA also accommodates coercions and multiple representation polymorphism [138], although we do not treat this topic here. Parametric polymorphism in OBJ is supported by parameterized objects, such as LIST[X], that provide higher-order capabilities in a first-order algebraic setting [61]. So it seems fair to conclude that Strachey was excessively pessimistic about the amount of structure in polymorphism that is not parametric in his sense, since OSA is a simple but rich mathematical theory that is easily implemented and is far from *ad hoc* in the pejorative sense of being arbitrary.

This chapter first generalizes our treatment of many-sorted algebra (MSA) to OSA, with illustrative examples, and then treats retracts, which enable error handling, and order-sorted term rewriting, which provides an operational semantics; details similar to MSA are sometimes omitted.

## 10.1 Signature, Algebra, and Homomorphism

This section introduces and briefly illustrates the three most basic concepts of order-sorted algebra; each is a straightforward extension of the corresponding MSA concept.

**Definition 10.1.1** An **order-sorted signature** $(S, \leq, \Sigma)$ consists of

1. a many-sorted signature $(S, \Sigma)$, with
2. a partial ordering $\leq$ on $S$ such that the following **monotonicity condition** is satisfied,

$$\sigma \in \Sigma_{w_1, s_1} \cap \Sigma_{w_2, s_2} \text{ and } w_1 \leq w_2 \text{ imply } s_1 \leq s_2 \ .$$  □

In OBJ notation, the sort set $S$ and the operation set $\Sigma$ are just the same for OSA signatures as for MSA signatures. The new ingredient is the partial ordering on $S$, which is declared by giving a set of subsort pairs, of the form S1 < S2; these can be strung together in declarations of the form S1 < S2 < $\cdots$ < Sn, which abbreviates S1 < S2, S2 < S3, .... In each case, the declaration must be preceded by the keywords subsort or subsorts, and terminated with a period (preceded by a space, as usual). The partial ordering defined on $S$ is the least such containing the given set of pairs. This syntax is illustrated in the following:

**Example 10.1.2** Below is the signature part of a specification for lists of natural numbers in a Lisp-like syntax:

```
sorts Nat NeList List .
subsorts NeList < List .
op 0 : -> Nat .
op s_ : Nat -> Nat .
op nil : -> List .
op cons : Nat List -> NeList .
op car : NeList -> Nat .
op cdr : NeList -> List .
```

Here cons is a list constructor which adds a new number at the head of a list; car and cdr are the corresponding selectors, which select the *head* and *tail* (also called the *front* and the *rest*) of non-empty lists; and nil is the empty list. However, equations are needed to express these relationships between cons and its selectors, for which see Example 10.2.19 below. □

**Definition 10.1.3** Given an order-sorted signature $(S, \leq, \Sigma)$, an **order-sorted** $(S, \leq, \Sigma)$-**algebra** is a many-sorted $(S, \Sigma)$-algebra $M$ such that

1. $s_1 \leq s_2$ implies $M_{s_1} \subseteq M_{s_2}$

2. $\sigma \in \Sigma_{w_1, s_1} \cap \Sigma_{w_2, s_2}$ and $w_1 \leq w_2$ imply $M_\sigma^{w_1, s_1} = M_\sigma^{w_2, s_2}$ on $M^{w_1}$. □

The second condition says that overloaded operations are consistent under restriction; this expresses subsort polymorphism.

**Example 10.1.4** Letting $\Sigma$ be the signature of Example 10.1.2, define a $\Sigma$-algebra $F$ as follows:

$$F_{\text{Nat}} = \{0\}$$
$$F_{\text{NeList}} = \{0\}$$
$$F_{\text{List}} = \{0, nil\}$$

with $s(0) = 0$, $cons(0, L) = 0$, $car(L) = 0$, $cdr(L) = nil$. Of course, this is not the "intended" standard or initial model, which instead is defined as follows:

$$L_{\text{Nat}} = \{0, s0, ss0, \ldots\}$$
$$L_{\text{NeList}} = L_{\text{Nat}} \cup L_{\text{Nat}}^*$$
$$L_{\text{List}} = L_{\text{NeList}} \cup \{nil\}$$

where $*$ denotes the finite sequence constructor. Finally, the operations of $L$ are defined by

$$cons(N, N_1 \ldots N_n) = NN_1 \ldots N_n$$
$$car(N_1 \ldots N_n) = N_1$$
$$cdr(N_1 \ldots N_n) = N_2 \ldots N_n \text{ if } n > 1$$
$$cdr(N_1 \ldots N_n) = nil \text{ if } n = 1$$

□

**Definition 10.1.5** Given order-sorted $(S, \leq, \Sigma)$-algebras $M, M'$, an **order-sorted $(S, \leq, \Sigma)$-homomorphism** $h : M \to M'$ is a many-sorted $\Sigma$-homomorphism $h : M \to M'$ such that

$$s_1 \leq s_2 \text{ implies } h_{s_1} = h_{s_2} \text{ on } M_{s_1}.$$

(This is also called the **monotonicity condition**.)   □

**Exercise 10.1.1** Adopting the notation of Example 10.1.4, show that there is a unique $\Sigma$-homomorphism $L \to F$. In fact, $F$ is a final $\Sigma$-algebra, and we will see later that $L$ is an initial $(\Sigma, E)$-algebra for the most reasonable and expected equation set $E$.   □

## 10.2 Term and Equation

The main topic of this section is order-sorted equations and their satisfaction. This requires that we first treat OSA terms and substitutions. We will see that there are some slightly subtle points about parsing terms, for which we later introduce the notion of a regular signature.

**Definition 10.2.1** Given an order-sorted signature $\Sigma$, the $S$-indexed set $\mathcal{T}_\Sigma$ of $\Sigma$-**terms** is defined recursively by the following:

1. $\Sigma_{[],s} \subseteq \mathcal{T}_{\Sigma,s}$ for $s \in S$,

2. $s_1 \leq s_2$ implies $\mathcal{T}_{\Sigma,s_1} \subseteq \mathcal{T}_{\Sigma,s_2}$,

3. $\sigma \in \Sigma_{w,s}$ and $t_i \in \mathcal{T}_{\Sigma,s_i}$ for $i = 1, \ldots, n$ imply $\sigma(t_1, \ldots, t_n) \in \mathcal{T}_{\Sigma,s}$ where $w = s_1 \ldots s_n$ and $n > 0$.   □

**Example 10.2.2** Let $\Sigma$ denote the signature of Example 10.1.2 with car and cdr removed. Then the following are terms of sort List:

```
nil
cons(0, nil), cons(s0, nil), ...
cons(0, cons(0, nil)), cons(s0, cons(0, nil)), cons(s0, cons(s0, nil)), ...
...
```

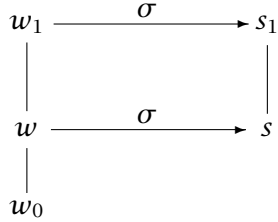

Figure 10.1: Visualizing Regularity

All except `nil` are also of sort `NeList`. □

Notice that conditions 1 and 3 in Definition 10.2.1 are the same as for MSA; condition 2 is needed to satisfy the first condition in Definition 10.1.3. Strictly speaking, we should have used underlined parentheses, $\sigma \underline{(t_1 \ldots t_n)}$, in the above, as we did for MSA terms in Definition 3.7.5. We make this indexed family of $\Sigma$-terms into a $\Sigma$-algebra as follows:

**Definition 10.2.3** Given $\sigma \in \Sigma_{w,s}$ with $w = s_1 \ldots s_n \neq [\,]$, define $(\mathcal{T}_\Sigma)_\sigma : (\mathcal{T}_\Sigma)^w \to (\mathcal{T}_\Sigma)_s$ by $(\mathcal{T}_\Sigma)_\sigma(t_1, \ldots, t_n) = \sigma(t_1, \ldots, t_n)$; and when $w = [\,]$, define $(\mathcal{T}_\Sigma)_\sigma = \sigma$. The resulting $\Sigma$-algebra is called the $\Sigma$-**term** (or sometimes **word**) **algebra**, and denoted $\mathcal{T}_\Sigma$. □

**Example 10.2.4** The terms listed in Example 10.2.2, plus others hinted at there, form the carrier of sort `List` of the term algebra $\mathcal{T}_\Sigma$ for the signature of Example 10.1.2; all but `nil` are also in the carrier of sort `NeList`, while the carrier of sort `Nat` contains the usual Peano numbers. □

**Exercise 10.2.1** Check that $\mathcal{T}_\Sigma$ of Definition 10.2.3 is an order-sorted $\Sigma$-algebra, by checking the two conditions in Definition 10.1.3. □

The next definition is motivated by wanting each $\Sigma$-term to have a unique parse of least sort. Notice that this is precisely what happens in the example arithmetic system that we discussed in the introduction to this chapter: we want each term to have the most specific possible sort; for example, $-1$ should be an `Integer`, but not a `Rational` or a `Natural`. It is natural to achieve this by requiring that the set of ranks that each overloaded operation might have (in a given context) has a least element:

**Definition 10.2.5** An order-sorted signature $(S, \leq, \Sigma)$ is **regular** iff for each $\sigma \in \Sigma_{w_1, s_1}$ and each $w_0 \leq w_1$ there is a unique least element in the set $\{(w, s) \mid \sigma \in \Sigma_{w,s} \text{ and } w \geq w_0\}$. □

This says that the set of possible ranks for $\sigma$ with arity greater than any fixed $w_0$ has a smallest element; see Figure 10.1, in which the vertical lines indicate subsort relations. To explore the consequences of this condition, we consider some examples where it is not satisfied:

**Example 10.2.6** The signature

```
sorts s1 s2 s3 s4 s5 .
subsort s1 < s3 .
subsort s2 < s4 .
op a : -> s1 .
op b : -> s2 .
op f : s1 s4 -> s5 .
op f : s3 s2 -> s5 .
```

is non-regular, because the set of ranks for f with arity at least s1 s2 consists of the two tuples (s1 s4, s5) and (s3 s2, s5), neither of which is less than the other. Therefore the term f(a,b) does not have a least sort, but instead has two incompatible sorts. □

**Exercise 10.2.2** Show that the following is a non-regular signature:

```
sorts Nat NeList List .
subsort Nat < NeList < List .
op 0 : -> Nat .
op s_ : Nat -> Nat .
op nil : -> List .
op cons : NeList List -> NeList .
op cons : List NeList -> NeList .
```

This defines a complex kind of list of natural numbers, in which lists can be elements of other lists, with non-empty lists distinguished from the empty list nil. But its non-regularity shows that this distinction is not sufficiently careful; see Exercise 10.2.4. □

Non-regular signatures can often be made regular by adding some new subsort declarations and/or by changing the ranks of some operations.

**Exercise 10.2.3** Show that adding the subsort declaration s2 < s4 to the signature of Example 10.2.6 gives a regular signature. □

**Exercise 10.2.4** Show how to modify the signature of Exercise 10.2.2 to make it regular. **Hint:** Add a new operation declaration to further overload cons. □

**Proposition 10.2.7** If $\Sigma$ is regular, then for each $t \in \mathcal{T}_\Sigma$ there is a least sort $s \in S$ such that $t \in \mathcal{T}_{\Sigma,s}$. This sort is denoted $LS(t)$.

**Proof:** We proceed by induction on the depth of terms in $\mathcal{T}_\Sigma$. If $t \in \mathcal{T}_\Sigma$ has depth 0, then $t = \sigma$ for some $\sigma \in \Sigma_{[],s1}$ and so by regularity with $w_0 = w_1 = []$, there is a least $s \in S$ such that $\sigma \in \Sigma_{[],s}$; this is the least sort of $\sigma$. Now consider a well-formed term $t = \sigma(t_1 \ldots t_n) \in \mathcal{T}_{\Sigma,s}$ of depth $n+1$. Then each $t_i$ has depth $\leq n$ and therefore by the induction hypothesis, has a least sort, say $s_i$; let $w_0 = s_1 \ldots s_n$. Then $\sigma \in \Sigma_{w',s'}$ for some $w', s'$ with $s' \leq s$ and $w_0 \leq w'$, and so by regularity, there are least $w'$ and $s'$ such that $\sigma \in \Sigma_{w',s'}$ and $w' \geq w_0$; this least $s'$ is the desired least sort of $t$. □

This proof essentially gives an algorithm for finding the least sort parse of any term over a regular signature.

**Theorem 10.2.8** (*Initiality*) If $\Sigma$ is regular and if $M$ is any $\Sigma$-algebra, then there is one and only one $\Sigma$-homomorphism from $\mathcal{T}_\Sigma$ to $M$. □

This result is proved in Appendix B, and as before, its property is called **initiality**. The following example shows that if $\Sigma$ is not regular, then $\mathcal{T}_\Sigma$ is not necessarily initial:

**Exercise 10.2.5** Define an algebra $M$ over the signature of Example 10.2.6 as follows: $M_{s1} = M_{s2} = M_{s3} = M_{s4} = \{1\}, M_{s5} = \{0, 2\}, M_a = M_b = 1, M_f^{s1\,s4,s5}(1,1) = 0, M_f^{s3\,s2,s5}(1,1) = 2$. Now show that if $\mathcal{T}$ is the term algebra for the signature given in Example 10.2.6, then $\mathcal{T}_{s5} = \{\mathsf{f}(\mathsf{a},\mathsf{b})\}$, and conclude from this that $\mathcal{T}$ is not initial. □

However, [68] shows that initial $\Sigma$-algebras do exist even when $\Sigma$ is not regular. Rather than just $\Sigma$-terms, the construction uses terms *annotated* with sort information; the notation $\overline{\mathcal{T}}_\Sigma$ may be used. We omit details, which are very similar to those for the many-sorted case (see Section 3.2). Again as for MSA, we may sometimes write $\mathcal{T}_\Sigma$ when we really mean $\overline{\mathcal{T}}_\Sigma$, and we may ignore the sort annotations, even though they are necessary, because they are implicit in parsing; this is consistent with what is done in implementations of OBJ.

As in the many-sorted case, it is convenient (but not always necessary) for each variable symbol to have just one sort; therefore we assume that any $S$-indexed set $X = \{X_s \mid s \in S\}$ used to provide variables is such that $X_{s_1}$ and $X_{s_2}$ are disjoint whenever $s_1 \neq s_2$, and such that all symbols in $X$ are distinct from those in $\Sigma$; we may use the term **variable set** for such indexed sets. Then as in the many-sorted case, we define the signature $\Sigma(X)$ by $\Sigma(X)_{w,s} = \Sigma_{w,s}$ for $w \neq [\,]$, and $\Sigma(X)_{[\,],s} = \Sigma_{[\,],s} \cup X_s$. We can now form the $\Sigma(X)$-term algebra $\mathcal{T}_{\Sigma(X)}$ and view it as a $\Sigma$-algebra, which is then denoted $\mathcal{T}_\Sigma(X)$. The following is proved in Appendix B:

**Theorem 10.2.9** If $(S, \leq, \Sigma)$ is regular, then $\mathcal{T}_\Sigma(X)$ is a **free** $\Sigma$-algebra on $X$, in the sense that for each $\Sigma$-algebra $M$ and each assignment $a : X \to M$, there is a unique $\Sigma$-homomorphism $\overline{a} : \mathcal{T}_\Sigma(X) \to M$ such that $\overline{a}(x) = a(x)$ for all $x$ in $X$. □

This result also generalizes[E44] to non-regular signatures, using $\overline{\mathcal{T}}_\Sigma(X)$ instead of $\mathcal{T}_\Sigma(X)$, though we omit the details. The same applies to the following, the proof of which is just the same as for Proposition 3.5.1 for MSA:

**Proposition 10.2.10** Given an OSA signature $\Sigma$, a ground signature $X$ disjoint from $\Sigma$, an OSA $\Sigma$-algebra $M$, and a map $a : X \to M$, there is a unique $\Sigma$-homomorphism $\overline{a} : \mathcal{T}_\Sigma(X) \to M$ which extends $a$, in the sense that $\overline{a}_s(x) = a_s(x)$ for each $s \in S$ and $x \in X_s$. □

Substitutions and their composition can now be defined just as in MSA, simply replacing $T_\Sigma(Y)$ by $\mathcal{T}_\Sigma(Y)$:

**Definition 10.2.11** A **substitution** of $\Sigma$-terms with variables in $Y$ for variables in $X$ is an arrow $a : X \to \mathcal{T}_\Sigma(Y)$; the notation $a : X \to Y$ may also be used. The **application** of $a$ to $t \in \mathcal{T}_\Sigma(X)$ is $\overline{a}(t)$. Given substitutions $a : X \to \mathcal{T}_\Sigma(Y)$ and $b : Y \to \mathcal{T}_\Sigma(Z)$, their **composition** $a;b$ (as substitutions) is the $S$-sorted arrow $a;\overline{b} : X \to \mathcal{T}_\Sigma(Z)$. □

We also use Notation 3.5.3 for OSA substitutions: Given $t \in \mathcal{T}_\Sigma(X)$ and $a : X \to \mathcal{T}_\Sigma(Y)$ with $|X| = \{x_1, \ldots, x_n\}$ and $a(x_i) = t_i$ for $i = 1, \ldots, n$, write $\overline{a}(t)$ as $t(x_1 \leftarrow t_1, x_2 \leftarrow t_2, \ldots, x_n \leftarrow t_n)$, omitting $x_i \leftarrow t_i$ when $t_i$ is $x_i$.

**Proposition 10.2.12** OSA substitutions are sort decreasing, in that $LS(\theta(x)) \leq s$ for any $x \in X_s$ and more generally, $LS(\theta(t)) \leq LS(t)$ for any $\Sigma$-term $t$.

**Proof:** The first assertion follows because $\theta(x) \in \mathcal{T}_\Sigma(Y)_s$ and $LS(t) \leq s$ for all $t \in \mathcal{T}_\Sigma(Y)_s$. The second assertion can be proved by an induction similar to that used for Proposition 10.2.7, by applying condition 2 of Definition 10.1.1. □

The composition of substitutions is associative, by exactly the same proof used for the MSA case (Proposition 3.6.5, except using Proposition 10.2.10 instead of 3.5.1):

**Proposition 10.2.13** Given substitutions $a : W \to \mathcal{T}_\Sigma(X)$, $b : X \to \mathcal{T}_\Sigma(Y)$, $c : Y \to \mathcal{T}_\Sigma(Z)$, then $(a;b);c = a;(b;c)$. □

The following consequence of the proof is important for calculations in several proofs:

**Corollary 10.2.14** Given substitutions $a : W \to \mathcal{T}_\Sigma(X)$, $b : X \to \mathcal{T}_\Sigma(Y)$, then $\overline{a};\overline{b} = \overline{(a;\overline{b})}$. □

Definition 10.2.15 below introduces concepts which help define the terms for which our order-sorted equational satisfaction makes sense. Their motivation is that the two terms in an equation must have sorts that are somehow connected. For example, an equation such as

$(\forall n)\ 5n = true$

really cannot make sense; on the other hand, the equation

$(\forall n)\ 5n = i$

where $n$ is a natural and $i$ is the square root of minus one, does make sense, even though it is not satisfied by the standard model of the number system. This is because $5n$ and $i$ are in the same connected component of $S$, while $5n$ and *true* are not, where we say that $s_1$ and $s_2$ are in the same **connected component** of $S$ iff $s_1 \equiv s_2$, where $\equiv$ is the least equivalence relation on $S$ that contains $\leq$. Here is the formalization:

**Definition 10.2.15** A partial ordering $(S, \leq)$ is **filtered** iff for all $s_1, s_2 \in S$, there is some $s \in S$ such that $s_1 \leq s$ and $s_2 \leq s$. A partial ordering $(S, \leq)$ is **locally filtered** iff every connected component of it is filtered. An order-sorted **signature** $(S, \leq, \Sigma)$ is **locally filtered** iff $(S, \leq)$ is locally filtered, and it is a **coherent signature** iff it is both locally filtered and regular. A partial ordering $(S, \leq)$ **has a top** iff it contains a (necessarily unique) maximum element, $u \in S$, such that $s \leq u$ for all $s \in S$. □

Hereafter we assume that *all OSA signatures are coherent* unless otherwise stated. Assuming local filtration is not at all restrictive in practice, because we always add top elements to connected components, or even to the whole partial ordering (see Exercise 10.2.6). Indeed, OBJ3 does just this, with its `Universal` sort. The need for local filtration is shown in Example 10.2.17 below, and is also used in the quotient construction of Definition 10.4.6 and in the application of that construction to order-sorted rewriting modulo equations in Section 10.7.

**Exercise 10.2.6** Show that any filtered partial order is locally filtered, and that any partial order with a top is filtered. Give examples showing that the converse assertions are false. □

**Definition 10.2.16** An order-sorted $\Sigma$-**equation** is a triple $\langle X, t_1, t_2 \rangle$ where $X$ is a variable set, and $t_1, t_2 \in \mathcal{T}_{\Sigma(X)}$ such that $LS(t_1)$ and $LS(t_2)$ are in the same connected component of $(S, \leq)$; we shall of course write $(\forall X)\ t_1 = t_2$, or in concrete cases, things like $(\forall x, y, z)\ t_1 = t_2$. A $\Sigma$-algebra $M$ **satisfies** a $\Sigma$-equation $(\forall X)\ t_1 = t_2$ iff for all assignments $a : X \to M$ we have $\overline{a}(t_1) = \overline{a}(t_2)$.

A **conditional $\Sigma$-equation** is a quadruple $\langle X, t_1, t_2, C \rangle$, where $\langle X, t_1, t_2 \rangle$ is a $\Sigma$-equation and $C$ is a finite set of pairs $\langle u, v \rangle$ such that $\langle X, u, v \rangle$ is a $\Sigma$-equation; we shall write $(\forall X)\ t_1 = t_2$ `if` $C$, or more concretely, things like $(\forall x, y)\ t_1 = t_2$ `if` $u_1 = v_1, u_2 = v_2$. A $\Sigma$-algebra $M$ **satisfies** a conditional $\Sigma$-equation $(\forall X)\ t_1 = t_2$ `if` $C$ iff for all assignments $a : X \to M$ whenever $\overline{a}(u) = \overline{a}(v)$ for all $\langle u, v \rangle \in C$ then $\overline{a}(t_1) = \overline{a}(t_2)$.

Finally, we say that a $\Sigma$-algebra $M$ **satisfies** a set $A$ of $\Sigma$-equations (conditional or not) iff it satisfies each one of them. In this case, we write $M \models A$ or possibly $M \models_\Sigma A$. □

Although satisfaction makes sense when the set of conditions is infinite, we have restricted the definition to finite $C$ because this is needed for both deduction and rewriting. The following gives another reason why local filtration is necessary:

**Example 10.2.17** We show that without local filtration, equational satisfaction is not invariant under isomorphism. Given the specification

```
th NON-LF is sorts A B C .
  subsorts B < A C .
  op a : -> A .
  op b : -> B .
  op c : -> C .
  eq a = c .
endth
```

let $\Sigma$ denote its signature. Then the initial order-sorted $\Sigma$-algebra $T_\Sigma$ has $(T_\Sigma)_A = \{a,b\}$, $(T_\Sigma)_B = \{b\}$, $(T_\Sigma)_C = \{b,c\}$ and it does not satisfy the equation, whereas the $\Sigma$-isomorphic algebra $\mathcal{A}$ with $\mathcal{A}_A = \{b,d\}$, $\mathcal{A}_B = \{b\}$ and $\mathcal{A}_C = \{b,d\}$, does satisfy the equation, where both a and c are interpreted as $d$ in $\mathcal{A}$. (See Exercise 10.7.1 for some further related discussion.) □

The undesirable phenomenon of Example 10.2.17 is impossible for locally filtered signatures:

**Proposition 10.2.18** If $\Sigma$ is a coherent OSA signature and $\mathcal{A}, \mathcal{B}$ are $\Sigma$-isomorphic algebras,[E45] then $\mathcal{A}$ satisfies an equation $(\forall X)\ t = t'$ iff $\mathcal{B}$ does.

**Proof:** By symmetry of the isomorphism relation, it suffices to prove just one direction. So assume $\mathcal{A}$ satisfies the equation, let $f : \mathcal{A} \to \mathcal{B}$ be a $\Sigma$-isomorphism, and let $\beta : X \to \mathcal{B}$ be an assignment. Then $\beta = \alpha; f$ for some assignment $\alpha : X \to \mathcal{A}$. Therefore $\overline{\beta} = \overline{\alpha}; f$, so that if $s \geq LS(t), LS(t')$ then

$$\overline{\beta}_s(t) = f_s(\overline{\alpha}_s(t)) = f_s(\overline{\alpha}_s(t')) = \overline{\beta}_s(t')$$

as desired. □

**Exercise 10.2.7** Generalize Proposition 10.2.18 from unconditional to conditional equations. □

**Example 10.2.19** (*Errors for Lists*) Example 10.1.2 noted that equations are needed to give car and cdr the desired meanings. The following gives those equations in an appropriate object (for convenience, we import the natural numbers instead of defining them from scratch):

```
obj LIST is sorts List NeList .
  pr NAT .
  subsorts NeList < List .
  op nil : -> List .
  op cons : Nat List -> NeList .
  op car : NeList -> Nat .
  op cdr : NeList -> List .
  var L : List .  var N : Nat .
```

```
    eq car(cons(N,L)) = N .
    eq cdr(cons(N,L)) = L .
  endo
```

The initial algebra of this specification is what one would expect, noting that the terms car(nil) and cdr(nil) do not parse, and hence are not in it. However, because these terms, and the many others of which they are subterms, represent errors, what we really want is for them to be proper terms, but of a different sort, so that we can "handle" or "trap" them as errors, without having to invoke any nasty imperative features, as is done in most functional programming languages. The following shows that OSA provides an elegant solution for this problem, and it has even been shown that MSA *cannot* provide a satisfactory solution [138].

```
  obj ELIST is sorts List ErrList ErrNat .
    pr NAT .
    subsort List < ErrList .
    subsort Nat < ErrNat .
    op nil : -> List .
    op cons : Nat List -> List .
    op car : List -> ErrNat .
    op cdr : List -> ErrList .
    var N : Nat .  var L : List .
    eq car(cons(N,L)) = N .
    eq cdr(cons(N,L)) = L .
    op nohead : -> ErrNat .
    eq car(nil) = nohead .
    op notail : -> ErrList .
    eq cdr(nil) = notail .
  endo
```

Now car(nil) is a proper term of sort ErrNat rather then Nat, and similarly for cdr(nil). Therefore we can write equations that have such "error expressions" in their leftsides, as above. Of course more than this is needed to get the right behavior in realistic situations, as further discussed in Section 10.6. (Note that expressions like cons(2, notail) and 1 + car(nil) fail to parse, and hence are not terms for this specification, although if they are executed in OBJ3, some interesting things happen with retracts, as explained in Example 10.6.1.)
□

**Exercise 10.2.8** Define appropriate error supersorts and error messages for operations applied to the empty stack, using the basic specification below as your starting point:

```
  obj STACK is pr NAT .
    sort Stack .
```

```
  op empty : -> Stack .
  op push : Nat Stack -> Stack .
  op top_ : Stack -> Nat .
  op pop_ : Stack -> Stack .
  var X : Nat .
  var S : Stack .
  eq top push(X,S) = X .
  eq pop push(X,S) = S .
endo
```

You should also define and run some test cases for your code.  □

## 10.3 Deduction

Equational deduction also generalizes to order-sorted algebra. The following rules use the same notation as Section 8.4.1, in which the hypotheses of a rule are above a horizontal line, while the conclusion is given below the line.

**Definition 10.3.1 (Order-sorted equational deduction)** Let $A$ be a set of $\Sigma$-equations. If an equation $e$ can be deduced using the rules (1–5C) below, from a given set $A$ of (possibly conditional) equations, then we write $A \vdash e$ or possibly $A \vdash_\Sigma e$.

(1) *Reflexivity:*

$$\frac{}{(\forall X)\ t = t}$$

(2) *Symmetry:*

$$\frac{(\forall X)\ t = t'}{(\forall X)\ t' = t}$$

(3) *Transitivity:*

$$\frac{(\forall X)\ t = t',\ \ (\forall X)\ t' = t''}{(\forall X)\ t = t''}$$

(4) *Congruence:*

$$\frac{(\forall X)\ \theta(y) = \theta'(y)\ \text{ for each } y \in Y}{(\forall X)\ \theta(t) = \theta'(t)}$$

where $\theta, \theta' : Y \to \mathcal{T}_\Sigma(X)$ and where $t \in \mathcal{T}_\Sigma(Y)$.

(5C) *Conditional Instantiation:*

$$\frac{(\forall X)\ \theta(v) = \theta(v')\ \text{ for each } v = v'\ \text{in } C}{(\forall X)\ \theta(t) = \theta(t')}$$

where $\theta : Y \to \mathcal{T}_\Sigma(X)$ and where $(\forall Y)\ t = t'\ \text{if}\ C$ is in $A$.

There is also an unconditional version of (5C):

(5) *Instantiation:*

$$\frac{}{(\forall X)\ \theta(t) = \theta(t')}$$

where $\theta : Y \to \mathcal{T}_\Sigma(X)$ and where $(\forall Y)\ t = t'$ is in $A$.

We can use the same notation for deduction using (1-5) as for using (1-5C) because (5) is a special case of (5C). □

These rules of deduction are essentially the same as for MSA, except for the restriction on substitutions given in Proposition 10.2.12, and the same results as in Chapter 4 for MSA deduction generally carry over. In particular, these rules are sound and complete:

**Theorem 10.3.2** (*Completeness*) Given a coherent order-sorted signature $\Sigma$ then an unconditional equation $e$ can be deduced from a given set $A$ of (possibly conditional) equations iff it is true in every model of $A$, that is, $E \vdash e$ iff $A \vDash e$. □

The proof is given in Appendix B, but it uses results from the next section, because for expository purposes, we have stated this result earlier than required by the logical flow of proof.

**Exercise 10.3.1** Show that rule (0) (*Assumption*) of Chapter 4 (but for OSA) is a special case of rule (5) above. □

It is straightforward to generalize the material on subterm replacement for conditional equations in Section 4.9 to the order-sorted case. The basic rule is as follows:

(6C) *Forward Conditional Subterm Replacement.* Given $t_0 \in \mathcal{T}_\Sigma(X \cup \{z\}_s)$ with $z \notin X$, if
  $(\forall Y)\ t_1 = t_2$ if $C$
is of sort $\leq s$ and is in $A$, and if $\theta : Y \to \mathcal{T}_\Sigma(X)$ is a substitution such that $(\forall X)\ \theta(u) = \theta(v)$ is deducible for each pair $\langle u, v \rangle \in C$, then
  $(\forall X)\ t_0(z \leftarrow \theta(t_1)) = t_0(z \leftarrow \theta(t_2))$
is also deducible.

The substitutions $t_0(z \leftarrow \theta(t_1))$ and $t_0(z \leftarrow \theta(t_2))$ are valid because $LS(\theta(t_i)) \leq LS(t_i)$ by Proposition 10.2.12 and $LS(t_i) \leq s$ by assumption.

Exercises 4.9.1-4.9.4 also generalize, as does the reversed version of (6C):

(-6C) *Backward Conditional Subterm Replacement.* Given $t_0 \in \mathcal{T}_\Sigma(X \cup \{z\}_s)$ with $z \notin X$, if
$$(\forall Y)\; t_2 = t_1 \text{ if } C$$
is of sort $\leq s$ and is in $A$, and if $\theta : Y \to \mathcal{T}_\Sigma(X)$ is a substitution such that $(\forall X)\; \theta(u) = \theta(v)$ is deducible for each pair $\langle u, v \rangle \in C$, then
$$(\forall X)\; t_0(z \leftarrow \theta(t_1)) = t_0(z \leftarrow \theta(t_2))$$
is also deducible.

Soundness of this rule follows as in the MSA case, by applying the symmetry rule, and so we also get:

($\pm 6C$) *Bidirectional Conditional Subterm Replacement.* Given $t_0 \in \mathcal{T}_\Sigma(X \cup \{z\}_s)$ with $z \notin X$, if
$$(\forall Y)\; t_1 = t_2 \text{ if } C \quad \text{or} \quad (\forall Y)\; t_2 = t_1 \text{ if } C$$
is of sort $\leq s$ and is in $A$, and if $\theta : Y \to \mathcal{T}_\Sigma(X)$ is a substitution such that $(\forall X)\; \theta(u) = \theta(v)$ is deducible for each pair $\langle u, v \rangle \in C$, then
$$(\forall X)\; t_0(z \leftarrow \theta(t_1)) = t_0(z \leftarrow \theta(t_2))$$
is also deducible.

We now have the following important completeness result, which is proved in Appendix B:

**Theorem 10.3.3** Given a coherent signature $\Sigma$ and a set $A$ of (possibly conditional) $\Sigma$-equations, then for any unconditional $\Sigma$-equation $e$,
$$A \vdash^C e \quad \text{iff} \quad A \vdash^{(1,3,\pm 6C)} e \,. \qquad \square$$

As in Chapter 4, the rules (+6C), (-6C), and ($\pm 6C$) can each be specialized to the case where $t_0$ has exactly one occurrence of $z$, and these variants are indicated by writing $6_1$ instead of $6$; we do not write them out here (but see Definition 10.7.1 in Section 10.7 below). The following completeness result can now be proved in much the same way as Corollary 4.9.2:

**Corollary 10.3.4** Given a coherent signature $\Sigma$ and a set $A$ of (possibly conditional) $\Sigma$-equations, then for any unconditional $\Sigma$-equation $e$,
$$A \vdash^{(1-5C)} e \quad \text{iff} \quad A \vdash^{(1,3,\pm 6_1 C)} e \quad \text{iff} \quad A \vdash^{(1,2,3,6_1 C)} e \,. \qquad \square$$

**Exercise 10.3.2** Use order-sorted equational deduction to prove the equation $f(a) = f(b)$ for the following specification:

```
th OSRW-EQ is sorts A C .
  subsort A < C .
  ops a b : -> A .
  op c : -> C .
  op f : A -> C .
  eq c = a .
  eq c = b .
endth
```

You may use OBJ's apply feature.  **Hint:** First prove a = b as a lemma.  □

## 10.4  Congruence, Quotient, and Initiality

This section develops some more theoretical topics in order-sorted algebra; in general, they are straightforward extensions of the corresponding MSA topics, the main exception being the treatment of subsorts in quotients. As in MSA, the completeness of deduction (Theorem 10.3.2) can be used to construct initial and free algebras when there are equations, by defining an $S$-sorted relation $\simeq_{A,X}$ on $\mathcal{T}_\Sigma(X)$ for $X$ a variable set, by

$$t_1 \simeq_{A,X} t_2 \text{ iff } A \vdash (\forall X) \; t_1 = t_2$$

using the rules in Definition 10.3.1. Since this relation is an order-sorted congruence in the sense of Definition 10.4.1 immediately below, we can define $\mathcal{T}_{\Sigma,A}(X)$ to be the quotient of $\mathcal{T}_\Sigma(X)$ by $\simeq_{A,X}$ using the quotient construction given in Definition 10.4.6 below. In preparation for the OSA notion of congruence, one should first recall from Definition 6.1.1 that, given a many-sorted signature $(S, \Sigma)$, a **many-sorted $\Sigma$-congruence** $\equiv$ on a many-sorted $\Sigma$-algebra $M$ is an $S$-sorted family $\{\equiv_s | \; s \in S\}$ of equivalence relations, with $\equiv_s$ on $M_s$ such that

(1) given $\sigma \in \Sigma_{w,s}$ with $w = s_1 \ldots s_n$ and $a_i, a'_i \in M_{s_i}$ for $i = 1, \ldots, n$ such that $a_i \equiv_{s_i} a'_i$, then $M_\sigma(a_1, \ldots, a_n) \equiv_s M_\sigma(a'_1, \ldots, a'_n)$ .

**Definition 10.4.1** For $(S, \leq, \Sigma)$ an order-sorted signature and $M$ an order-sorted $\Sigma$-algebra, an **order-sorted $\Sigma$-congruence** $\equiv$ on $M$ is a many-sorted $\Sigma$-congruence $\equiv$ such that

(2) if $s \leq s'$ in $S$ and $a, a' \in M_s$ then $a \equiv_s a'$ iff $a \equiv_{s'} a'$.

An order-sorted $(S, \leq, \Sigma)$-algebra $M'$ is an **order-sorted subalgebra** of another such algebra $M$ iff it is a many-sorted subalgebra such that $M'_s \subseteq M'_{s'}$ whenever $s \leq s'$ in $S$.  □

**Exercise 10.4.1** Show that the intersection of any set of order-sorted $\Sigma$-congruences on an order-sorted $\Sigma$-algebra $M$ is also an order-sorted $\Sigma$-congruence on $M$. Conclude from this that any $S$-sorted family $R$ of binary relations $R_s$ on $M_s$ for $s \in S$ is contained in a least order-sorted $\Sigma$-congruence on $M$. **Hint:** The set of congruences that contain $R$ is non-empty because it contains the relation that identifies everything (for each sort).  □

**Example 10.4.2** Let $M$ be the initial $\Sigma$-algebra $\mathcal{T}_\Sigma$ where $\Sigma$ is the ELIST signature from Example 10.2.19, and let $\equiv$ be the congruence generated by its equations, i.e., the least $\Sigma$-congruence that contains all ground instances of the equations in ELIST. Then `cdr(cons(0,nil))` $\equiv$ `nil` on both sorts List and ErrList, consistent with List < ErrList, while `cdr(cdr(cons(0,nil)))` $\equiv$ `notail` for the sort ErrList.  □

**Fact 10.4.3** $\simeq_{A,X}$ is an order-sorted congruence relation.

**Proof:** It is easy to see that $\simeq_{A,X}$ is reflexive, symmetric, and transitive from rules (1), (2) and (3) of Definition 10.3.1, respectively, and the $\Sigma$-congruence property follows from rule (4). To prove (2) of Definition 10.4.1, suppose that $s \leq s'$ and that $t \simeq_{A,X} t'$ for $t, t' \in \mathcal{T}_\Sigma(X)_s$. Then also $t, t' \in \mathcal{T}_\Sigma(X)_{s'}$ and the same proof that showed $A \vdash (\forall X)\ t = t'$ for sort $s$ also works for sort $s'$, and *vice versa*.  □

Recall from Definition 6.1.5, that given a many-sorted $\Sigma$-homomorphism $f : M \to M'$, the **kernel** of $f$, denoted $ker(f)$, is the $S$-sorted family of equivalence relations $\equiv_f$ defined by $a \equiv_{f,s} a'$ iff $f_s(a) = f_s(a')$, and the **image** of $f$ is the subalgebra $f(M)$ with $f(M)_s = f(M_s)$ for each $s \in S$; it may also be denoted $im(h)$. The following shows that these concepts extend easily from MSA to OSA:

**Proposition 10.4.4** If $f : M \to M'$ is an order-sorted $\Sigma$-homomorphism, then

1. $ker(f)$ is an order-sorted $\Sigma$-congruence on $M$;
2. $f(M)$ is an order-sorted subalgebra of $M'$.

**Proof:** Proposition 6.1.6 showed that each $\equiv_{f,s}$ is an equivalence relation satisfying the congruence property (i.e., (1) above). Property (2) above follows from the fact that $f_s(a) = f_{s'}(a)$ and $f_s(a') = f_{s'}(a')$ for any $s \leq s'$ in $S$ and any $a, a' \in M_s$.

Assertion 2. was proved in Proposition 6.1.6 for MSA, so we need only check (2) of Definition 10.4.1, which is an easy (set-theoretic) consequence of the fact that $f$ is order-sorted.  □

**Example 10.4.5** Let LELIST denote the result of making the specification ELIST of Example 10.2.19 entirely loose, including the imported natural numbers, say with the Peano signature, although we will use ordinary decimal notation for convenience. Let $M$ be the ELIST-algebra with elements of sort List just lists of natural numbers, such as $(2, 3, 5, 7, 11)$; with elements of sort ErrList those of sort List plus notail; and with elements of sort ErrNat the natural numbers plus nohead. Note that expressions like `cons(0, notail)` and `cons(nohead, (0,1,2))` are simply not in this algebra.

Now let $h : M \to M'$ be the $S$-sorted map that sends each natural number to the corresponding list of numbers modulo 3, and that sends notail and nohead to themselves. Then $h$ is a $\Sigma$-homomorphism, and $h(M)$ is the LELIST-algebra $M'$ with $M'_{\text{Nat}} = \{0, 1, 2\}$, with $M'_{\text{List}}$ lists of numbers from $\{0, 1, 2\}$, with $M'_{\mathsf{ErrNat}} = M'_{\mathsf{Nat}} \cup \{\mathsf{nohead}\}$, and $M'_{\mathsf{ErrList}} = M'_{\mathsf{List}} \cup \{\mathsf{notail}\}$.

If we let $R$ denote the kernel of $h$, then $nR_{\mathsf{Nat}}n'$ iff $n - n'$ is divisible by 3, and $\ell R_{\mathsf{List}}\ell'$ iff $\ell$ and $\ell'$ have the same length $N$ and $\ell_i - \ell'_i$ is divisible by 3 for $i = 1, \ldots, N$. Also, $nR_{\mathsf{ErrNat}}n'$ iff $nR_{\mathsf{Nat}}n'$ or $n = n' = \mathsf{nohead}$, and $\ell R_{\mathsf{ErrList}}\ell'$ iff $\ell R_{\mathsf{List}}\ell'$ or $\ell = \ell' = \mathsf{notail}$. □

**Exercise 10.4.2** If $f : M \to M'$ is an OSA $\Sigma$-homomorphism and $M_0 \subseteq M$ is a $\Sigma$-subalgebra, show that $f(M_0)$ is a $\Sigma$-subalgebra of $f(M)$. □

We now define the quotient of an order-sorted algebra by a congruence relation; following [82], the construction exploits local filtration to enable identifications across subsorts:

**Definition 10.4.6** For $(S, \leq, \Sigma)$ a locally filtered order-sorted signature, $M$ an order-sorted $\Sigma$-algebra, and $\equiv$ an order-sorted $\Sigma$-congruence on $M$, the **quotient** of $M$ by $\equiv$ is the order-sorted $\Sigma$-algebra $M/\equiv$ defined as follows: for each connected component $C$, let $M_C = \bigcup_{s \in C} M_s$ and define the congruence relation $\equiv_C$ by $a \equiv_C a'$ iff there is a sort $s \in C$ such that $a \equiv_s a'$. Then $\equiv_C$ is clearly reflexive and symmetric. It is transitive because $a \equiv_s a'$ and $a' \equiv_{s'} a''$ yield $a \equiv_{s''} a''$ for $s'' \geq s, s'$, which exists by local filtration. The inclusion $M_s \subseteq M_C$ induces an injective map $M_s/\equiv_s \to M_C/\equiv_C$ because for $a, a' \in M_s$ we have $a \equiv_s a'$ implies $a \equiv_C a'$ by construction, and conversely $a \equiv_C a'$ implies $a \equiv_{s'} a'$ for some $s' \in C$, and taking $s'' \geq s, s'$ it also implies $a \equiv_{s''} a'$ and therefore it implies $a \equiv_s a'$ by property (2) of the definition of order-sorted congruence. Denoting by $q_C$ the natural projection $q_C : M_C \to M_C/\equiv_C$ of each element $a$ into its $\equiv_C$-equivalence class, we define the carrier $(M/\equiv)_s$ of sort $s$ in the quotient algebra to be the image $q_C(M_s)$. The order-sorted algebra $M/\equiv$ comes equipped with a surjective order-sorted $\Sigma$-homomorphism $q : M \to M/\equiv$ defined to be the restriction of $q_C$ to each of its sorts, and called the **quotient map** associated to the congruence $\equiv$. The operations are defined by $(M/\equiv)_\sigma([a_1], \ldots, [a_n]) = [M_\sigma(a_1, \ldots, a_n)]$, and are well defined because $\equiv$ is an order-sorted $\Sigma$-congruence. □

The following illustrates the above construction:

**Example 10.4.7** For a given $\Sigma$-theory $B$, consider the relation $\equiv$ on $\mathcal{T}_\Sigma$ defined for $LS(t), LS(t') \leq s$ by $t \equiv_s t'$ iff there is a proof that $t = t'$ in which every term used has least sort $\leq s$; it is not difficult to check that $\equiv$ is a $\Sigma$-congruence. Now define $B$ by the following:

```
th OSTH is sorts A C .
  subsort A < C .
  ops a b : -> A .
  op c : -> C .
  eq c = a .
```

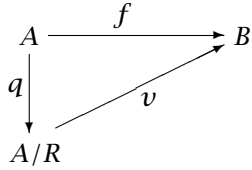

Figure 10.2: Condition (2) of Universal Property of Quotient

```
   eq c = b .
 endth
```

Then under the ordinary quotient construction (as in Appendix C) the ≡-equivalence class $[a]_A$ of a for sort A is $\{a\}$ and also $[b]_A = \{b\}$, whereas $[a]_C = [b]_C = \{a, b, c\}$. However, under the construction of Definition 10.4.6, the equivalence classes of sort $s$ collect all terms of sort $s$ or less that can be proved equal, no matter what other sorts may be involved. So in this case, $[a]_A = \{a, b\}$ and $[a]_B = \{a, b, c\}$. This shows that the construction of Definition 10.4.6 does useful additional work for certain relations, although $\simeq_B$ is not one of these.[E46]  □

**Exercise 10.4.3** Show that the relation ≡ in Example 10.4.7 is a Σ-congruence.  □

The following is straightforward from the definitions:

**Fact 10.4.8** Under the assumptions of Definition 10.4.6, $ker(q) = \equiv$.  □

Exercise 10.4.1 allows us to extend the construction of Definition 10.4.6 to quotients by an arbitrary relation on an algebra.

**Definition 10.4.9** Given an arbitrary $S$-sorted family $R$ of binary relations $R_s$ on $M_s$ for $s \in S$, then the **quotient** of $M$ by $R$, denoted $M/R$, is the quotient of $M$ by the smallest order-sorted Σ-congruence on $M$ containing $R$.  □

**Proposition 10.4.10** (*Universal Property of Quotient*) If Σ is a locally filtered order-sorted signature, if $M$ is an order-sorted Σ-algebra, and if $R$ is an $S$-sorted family of binary relations $R_s$ on $M_s$ for $s \in S$, then the quotient map $q : M \to M/R$ satisfies the following:

(1) $R \subseteq ker(q)$, and

(2) if $f : M \to B$ is any order-sorted Σ-homomorphism such that $R \subseteq ker(f)$, then there is a unique Σ-homomorphism $v : M/R \to B$ such that $q; v = f$ (see Figure 10.2).

**Proof:** (1) follows from Fact 10.4.8 and the definition of ≡ as the smallest congruence that contains $R$.

For (2), let $f : M \to M'$ be an order-sorted $\Sigma$-homomorphism such that $R \subseteq ker(f)$. Then $ker(q) \subseteq ker(f)$ and both are congruences so that for each connected component $C$ we have $ker(q)_C \subseteq ker(f)_C$ and there is a unique function $v_C : (M/R)_C \to M'_C$ such that $v_C \circ q_C = f_C$ for $f_C : M_C \to M'_C$ defined by $f_C(a) = f_s(a)$ if $a \in M_s$ (this is well defined by local filtering). It remains only to check that, restricting $v_C$ to each one of the sorts $s \in C$, the family $\{v_s \mid s \in S\}$ thus obtained is an order-sorted $\Sigma$-homomorphism. Property (2) for order-sorted homomorphisms follows by construction. Let $\sigma \in \Sigma_{w,s}$ with $w = s_1 \ldots s_n$ and let $a_i \in M_{s_i}$ for $i = 1, \ldots, n$. Then (omitting sort qualifications throughout) we have

$$v((M/R)_\sigma([a_1], \ldots, [a_n])) = v([M_\sigma(a_1, \ldots, a_n)]) =$$
$$f(M_\sigma(a_1, \ldots, a_n)) = M'_\sigma(f(a_1), \ldots, f(a_n)) =$$
$$M'_\sigma(v([a_1]), \ldots, v([a_n])) .$$

The case $w = []$ is left for the reader to check. □

The proof of Theorem 10.3.2 in Appendix B shows that the relation $\simeq_{A,X}$ (defined on page 334) is an order-sorted $\Sigma$-congruence. So we now define $\mathcal{T}_{\Sigma,A}(X)$ to be the quotient of $\mathcal{T}_\Sigma(X)$ by $\simeq_{A,X}$. Also, we denote $\mathcal{T}_{\Sigma,A}(\varnothing)$ by $\mathcal{T}_{\Sigma,A}$. The following is also proved in Appendix B:

**Theorem 10.4.11** If $\Sigma$ is coherent and $A$ is a set of (possibly conditional) $\Sigma$-equations, then $\mathcal{T}_{\Sigma,A}$ is an initial $(\Sigma, A)$-algebra, and $\mathcal{T}_{\Sigma,A}(X)$ is a free $(\Sigma, A)$-algebra on $X$, in the sense that for each $\Sigma$-algebra $M$ and each assignment $a : X \to M$, there is a unique $\Sigma$-homomorphism $\overline{a} : \mathcal{T}_{\Sigma,A}(X) \to M$ such that $\overline{a}(x) = a(x)$ for each $x$ in $X$. □

**Example 10.4.12** Theorem 10.4.11 implies that the algebra $M$ of Example 10.4.5 is an initial model for LELIST and hence a model for the original specification ELIST of Example 10.2.19. □

The following theorem generalizes Noether's first isomorphism theorem (Theorem 6.1.7) to OSA:

**Theorem 10.4.13** (*Homomorphism Theorem*) For any $\Sigma$-homomorphism $h : M \to M'$, there is a $\Sigma$-isomorphism $M/ker(h) \cong_\Sigma im(h)$.

**Proof:** Let $f' : A \to f(A)$ denote the corestriction (Appendix C reviews this concept) of $f$ to $f(A)$. Then Proposition 10.4.10 with $R = ker(f') = ker(f)$ gives a (unique) $\Sigma$-homomorphism $v : A/ker(f) \to f(A)$ such that $q; v = f'$, which is surjective because $f'$ is. We will be done if we can show that $v$ is also injective. To this end (and omitting sort subscripts), suppose that $v([a_1]) = v([a_2])$; then $f(a_1) = f(a_2)$, so that $[a_1] = [a_2]$. □

**Exercise 10.4.4** For $M, M', R$ as in Example 10.2.19, check that $M/\equiv$ is $\Sigma$-isomorphic to $M'$. □

It is worthwhile making explicit the following consequence of the proof given in Appendix B of the Completeness Theorem (10.3.2):

**Corollary 10.4.14** Given a coherent order-sorted signature $\Sigma$ and a set $A$ of (possibly conditional) $\Sigma$-equations, then an equation $(\forall X)\, t = t'$ is satisfied by every $\Sigma$-algebra that satisfies $A$ iff it is satisfied by $\mathcal{T}_{\Sigma,A}(X)$.
□

## 10.5 Class Deduction

The theory of class deduction in Section 7.2 easily generalizes to OSA.

**Definition 10.5.1** Given an order-sorted signature $\Sigma$ and a set $B$ of (possibly conditional) $\Sigma$-equations, a (**conditional**) $\Sigma$-**equation modulo** $B$, or $(\Sigma, B)$-**equation**, is a 4-tuple $\langle X, t, t', C \rangle$ where $t, t' \in \mathcal{T}_{\Sigma,B}(X)$ have sorts in the same connected component of $\Sigma$, and $C$ is a finite set of pairs from $\mathcal{T}_{\Sigma,B}(X)$, again with sorts in the same connected components. Usually we write $(\forall X)\, t =_B t'$ if $C$, and may use the same notation with $t, t', C$ all $\Sigma$-terms that represent their $B$-equivalence classes; we may also drop the $B$ subscripts.

Given a $(\Sigma, B)$-algebra $M$, $\Sigma$-**satisfaction modulo** $B$, written $M \vDash_{\Sigma,B} (\forall X)\, t =_B t'$ if $C$, is defined by $\overline{a}(t) = \overline{a}(t')$ whenever $\overline{a}(u) = \overline{a}(v)$ for each $\langle u, v \rangle \in C$, for all $a : X \to M$, where $\overline{a} : \mathcal{T}_{\Sigma,B}(X) \to M$ is the unique $\Sigma$-homomorphism extending $a$. Given a set $A$ of $(\Sigma, B)$-equations, $A \vDash_{\Sigma,B} e$ means $M \vDash_{\Sigma,B} A$ implies $M \vDash_{\Sigma,B} e$ for all $B$-models $M$.
□

As in Section 7.2, class deduction versions of inference rules are obtained just by substituting $\mathcal{T}_\Sigma$ for $T_\Sigma$ and $=_B$ for $=$, assuming that $A$ contains $(\Sigma, B)$-equations; we also use $[A]$ and $[e]$ as in Section 7.2, and the following three results have essentially the same proofs as the corresponding results there. The $B$-class version of rule $(i)$ is denoted $(i_B)$, and $A \vdash_{\Sigma,B} [e]$ denotes **class deduction modulo** $B$ of $[e]$ from $A$, using rules $(1_B - 4_B)$ and $(5C_B)$.

**Proposition 10.5.2** (*Bridge*) Given sets $A, B$ of $\Sigma$-equations and another $\Sigma$-equation $e$ (with $A, B$ and $e$ possibly conditional), then

$$[A] \vdash_B [e] \quad \text{iff} \quad A \cup B \vdash e \,.$$

Furthermore, given any $(\Sigma, B)$-algebra $M$ and a (possibly conditional) $\Sigma$-equation $e$, then

$$M \vDash_{\Sigma,B} [e] \quad \text{iff} \quad M \vDash_\Sigma e \,.$$
□

From this, it is not difficult to prove the following:

**Theorem 10.5.3** (*Completeness*) Given sets $A, B$ of $\Sigma$-equations and another $\Sigma$-equation $e$, all possibly conditional, then the following are equivalent:

(1) $[A] \vdash_B [e]$     (2) $[A] \vDash_B [e]$     (3) $A \cup B \vdash e$     (4) $A \cup B \vDash e$

□

The above result connects OSA class inference and satisfaction with ordinary OSA inference and satisfaction.

**Theorem 10.5.4** (*Completeness*) Given sets $A, B$ of (possibly conditional) $\Sigma$-equations and an unconditional $\Sigma$-equation $e$, then

$$[A] \vdash_B [e] \quad \text{iff} \quad [A] \vdash_B^{(1_B, 3_B, \pm 6C_B)} [e].$$

Moreover, $[A] \vdash_B^{(1_B, 3_B, \pm 6C_B)} [e]$ iff $M \vDash_B [e]$ for all $(\Sigma, A \cup B)$-algebras $M$. □

The above result says that $(6C_B)$, i.e., class rewriting, is complete for class deduction when combined with the reflextive, symmetric ($\pm$), and transitive rules of inference.

## 10.6 Error, Coercion, and Retract

In strongly typed languages, some expressions may not type check, even though they have a meaningful value. For example, given a factorial function defined only on natural numbers, the expression `((- 6)/(- 2))!` is not well-formed, because the parser can only determine that the argument of the factorial is a rational number, possibly negative. It is desirable to give such expressions the "benefit of the doubt," because they could evaluate to a natural (e.g., the above evaluates to 3). **Retract functions** provide this capability, by lowering the sort of a subexpression to the subsort needed for parsing. In this example, the retract function symbol

$r_{\text{Rational,Natural}}$ : `Rational -> Natural`

is automatically inserted by OBJ during rewriting, in a process called **retract rewriting**, to fill the gap between the parsed high sort and the required low sort, yielding the expression `(r`$_{\text{Rational,Natural}}$`((- 6)/(- 2)))!`, which does type check. Then we can use **retract elimination equations**, of the form

$$r_{s,s'}(x) = x$$

where $s' \leq s$ and $x$ is a variable of sort $s'$, to eliminate retracts when their arguments do have the required sorts. When $s' \not\leq s$, the retract remains, providing an error message that pinpoints exactly where the problem occurs and exactly what its sort gap was. For example, 7 + (((- 3)/(- 9))!) evaluates to 7 + ($r_{Rational,Natural}$(1 / 3))!, indicating that the argument to factorial is the rational 1/3. Similar situations arise with the function |_|^2 in Section 10.8 below. And unlike the untyped case, truly nonsensical expressions are detected and rejected at compile time, while any expression that could possibly recover is allowed to be evaluated. By "truly nonsensical" is meant expressions like factorial(false) that contain subexpressions in the wrong connected component (assuming that booleans and natural numbers are in different connected components of the sort poset) and therefore cannot be parsed by inserting retracts. A precise semantics for retracts is given in Section 10.6.1, while the rest of this section is devoted to examples in OBJ.

**Example 10.6.1** (*Lists with Fewer Errors*) As already noted, without retracts, terms like

>    car(cdr(cons(1,cons(2,cons(3,nil)))))

do not parse in the context of the ELIST theory of Example 10.2.19, because the subterm beginning with cdr has sort ErrList, while car requires sort List as its argument. However, the correct answer (which is 2), is obtained by inserting a retract and then reducing the result. The term

>    car(cdr(cdr(cons(1,nil))))

has a somewhat different behavior when retracts are added: it is temporarily accepted as

>    car($r_{ErrList,List}$(cdr($r_{ErrList,List}$(cdr(cons(1,nil))))))

which is then reduced to the form

>    car($r_{ErrList,List}$(nil))

which serves as a very informative error message.                               □

One might think that, since this is a kind of run-time type checking, it is just operational semantics. But our approach requires that the operational semantics agrees with the logical semantics, and retracts have a very nice logical semantics (see Section 10.6.1 below), as well as an operational semantics, which is developed in Section 10.7 below. Moreover, this kind of run-time type checking is relatively inexpensive, and in combination with the polymorphism given by subsorts and by parameterized modules, it provides the syntactic flexibility of untyped languages with all the advantages of strong typing.

The following shows that if deduction is not treated carefully, it can yield unsound results, and that naive attempts to fix this problem can greatly weaken deduction; the discussion following the example above also shows that retracts again provide a nice solution.

**Example 10.6.2** Consider the term f(a) for the following object:

```
obj NON-DED is sorts A B .
  subsorts A < B .
  op a : -> A .
  op b : -> B .
  ops f g : A -> A .
  var X : A .
  eq f(X) = g(X) .
  eq a = b .
endo
```

The first equation can deduce g(a) from f(a), and then the second equation can apparently deduce g(b) from f(a); but g(b) is not a well-formed term! The problem is that, although replacing a by b is sound in itself, it is not sound in the context of g. □

The easiest way to avoid this problem is to prohibit deductions that do not decrease sorts. But this would eliminate many important examples, such as the square norm in Section 10.8 above. A better approach is to prohibit applications yielding terms that don't parse; in fact, Definition 10.3.1 takes this approach, because its rules implicitly assume that every term occurring in them is well-formed. Unfortunately, this prohibits many correct computations, such as that above with factorial, and it also fails to inform the user what went wrong. Retracts allow us to avoid all these difficulties. The result of running OBJ3, which implements retracts, on

```
red f(a) .
```

in the context of NON-DED is the following,

```
reduce in NON-DED : f(a)
rewrites: 2
result A: g(r:B>A(b))
```

which is not only a valid deduction, but also an informative error message.

Example 10.6.2 might raise suspicions that the rules of deduction as stated in Definition 10.3.1 are unsound; but recall that those rules assume that all terms in them are well-formed, so without retracts, they disallow the deduction in Example 10.6.2. Once we formalize retracts as an order-sorted theory, the soundness of deduction using retracts follows from Theorem 10.3.2, because deduction with retracts follows exactly the same rules as deduction without them.

Raising and handling exceptions can also be given a nice semantics using retracts. This is significant because exceptions have both inadequate semantic foundations and insufficient flexibility in many programming and specification languages. Some algebraic specification languages use partial functions, which are simply undefined under exceptional conditions. This can be developed rigorously, e.g., in [111], but it has the disadvantage that neither error messages nor error recovery are possible. OSA with retracts supports both, and is fully implemented in OBJ3 and BOBJ. The following illustrates some capabilities of retracts in this respect:

**Example 10.6.3** (*Lists with Fewer Errors*)  Again in the context of the ELIST theory of Example 10.2.19, in processing large lists, explicit error messages that pinpoint exceptions might be difficult to understand. In this case, we can add some equations to simplify such expressions,

```
var EN : ErrNat .
var EL : ErrList .
eq cons(r:ErrNat>Nat(EN), notail)   = notail .
eq cons(nohead, r:ErrList>List(EL)) = notail .
eq car(r:ErrList>List(EL)) = nohead .
eq cdr(r:ErrList>List(EL)) = notail .
```

in which case

```
red car(cdr(cdr(cdr(cons(1,nil))))) .
```

gives just nohead as its reduced form.                                                    □

The following somewhat open-ended exercise gives a similar but more complex application of retracts:

**Exercise 10.6.1** (⋆) Write a suitable theory for a relational database in which lists represent tuples and so-called "null values," such as

```
op nullNat : -> ErrNat .
```

are treated as exceptions when doing arithmetic, but do not collapse tuples to a single error message. Show that it is also possible to have both kinds of exception in a single theory, by declaring two different error supersorts of Nat.                                                    □

Retracts also support situations where data is represented more than one way, and representations are converted to whatever form is most convenient or efficient for a given context. This kind of **multiple representation** is rather common, but is rarely given semantics. A good example is Cartesian and polar coordinates for points in the plane, as developed in Section 10.9. There are also many cases involving conversion from one sort of data to another in an irreversible way; for example, to apply integer addition to two rational numbers, one might first truncate them; this is called **coercion**. In both multiple representation and coercion, applying functions defined on one representation to data of another is mediated by functions that change the representation, but the conversions between multiple representations are reversible.

A basic theorem about retracts asserts their *consistency*: the theory that results from adding retract function symbols and retract equations to an order-sorted specification is a conservative extension of the original, in the sense that the equational deduction and initial models of the original theory are not disturbed. Thus retracts not only combine the flexibility of untyped languages with the discipline of strong typing, and give satisfactory treatments of exception handling and multiple representation, but the semantics of retracts, both deductive and model theoretic, is just a special case of order-sorted algebra.

**Example 10.6.4** The following is similar to Example 10.6.2, but more subtle.

```
obj MORE-NON-DED is sorts A B C .
  subsorts A < B < C .
  op f : C -> C .
  ops f h : A -> A .
  op g : B -> B .
  op a : -> A .
  var X : B .
  eq f(X) = g(X) .
endo
```

Here `f(X)` has sort C, which looks reasonable because the sort C is greater than the sort B of `g(X)`. But the term `h(f(a))` rewrites to `h(r:B>A(g(a)))`, because the equation obtained from the original by specializing the variable X of sort B to a variable of sort A is not sort decreasing. Therefore, not just the original rules, but also their specializations to rules having variables of smaller sorts should be considered, as discussed in [113]; see also Definition 10.7.10.  □

**Exercise 10.6.2** Use OBJ to verify the assertions about its computations in Examples 10.6.1, 10.6.3, and 10.6.4.  □

### 10.6.1 Semantics of Retracts

We have shown that strong typing is not flexible enough in practice, and have also suggested how OSA can provide the necessary flexibility with retracts. We now develop the formal semantics and prove that retracts are sound under certain mild assumptions. The first step is to extend an order-sorted signature $\Sigma$ to another order-sorted signature $\Sigma^\otimes$ having the same sorts as $\Sigma$, and the same operation symbols as $\Sigma$, plus some new ones called retracts of the form $r_{s',s} : s' \to s$ for each pair $s', s$ with $s' > s$. The semantics of retracts is then given by retract equations

$$(\forall x)\ r_{s',s}(x) = x\ ,$$

for all $s' > s$, where $x$ is a variable of sort $s$.

Given an order-sorted signature $\Sigma$ and a set $A$ of conditional $\Sigma$-equations, extend $\Sigma$ to the signature $\Sigma^\circledast$ by adding the retract operations, and extend $A$ to the set of equations $A^\circledast$ by adding the retract equations. Our requirement for retracts to be well-behaved is that the theory extension $(\Sigma, A) \subseteq (\Sigma^\circledast, A^\circledast)$ should be **conservative** in the sense that for all $t, t' \in \mathcal{T}_\Sigma(X)$,

$$t \simeq_{A(X)} t' \text{ iff } t \simeq_{A^\circledast(X)} t'.$$

This is equivalent in model-theoretic terms to requiring that the unique order-sorted $\Sigma$-homomorphism $\psi_X : \mathcal{T}_{\Sigma,A}(X) \to \mathcal{T}_{\Sigma^\circledast,A^\circledast}(X)$ which leaves the elements of $X$ fixed, is injective. We prove this under the very natural assumption on the algebras $\mathcal{T}_{\Sigma,A}(X)$ that given $X \subseteq X'$, then the unique $\Sigma$-homomorphism $\iota_{X,X'} : \mathcal{T}_{\Sigma,A}(X) \to \mathcal{T}_{\Sigma,A}(X')$ induced by the composite map $X \hookrightarrow X' \to \mathcal{T}_{\Sigma,A}(X')$ (first inclusion, then the natural mapping of each variable to the class of terms equivalent to it) is injective. We will say that a presentation $(\Sigma, A)$ is **faithful** if it satisfies this injectivity condition. Pathological, unfaithful presentations do exist, and for them the extension with retracts is not conservative, as shown by the following example from [78]:

**Example 10.6.5** Let $\Sigma$ have sorts $a, b, u$ with $a, b \leq u$, have an operation $f : a \to b$, have no constants of sort $a$, have constants $0, 1$ of sort $b$, plus $+, \&$ binary infix and $\neg$ unary prefix of sort $b$. Let $A$ have the equations $\neg(f(x)) = f(x)$, $y + y = y$, $y \& y = y$, $y + (\neg y) = 1$, $(\neg y) + y = 1$, $y \& (\neg y) = 0$, $(\neg y) \& y = 0$, $\neg 0 = 1$, $\neg 1 = 0$. Then $(\forall x) 1 = 0$ is deducible from $A$, where $x$ is a variable of sort $a$, although $(\forall \emptyset) 1 = 0$ is *not* deducible from $A$. Thus $(\Sigma, A)$ is not faithful. Note that $\mathcal{T}_{\Sigma,A}$ has $1 \neq 0$ but $\mathcal{T}_{\Sigma^\circledast,A^\circledast}$ has $1 = 0$ because of the first equation and the presence of constants of sort $a$ such as $r_{u,a}(0)$ and $r_{u,a}(1)$. Thus, the extension $(\Sigma, A) \subseteq (\Sigma^\circledast, A^\circledast)$ is not conservative.  □

There are simple conditions on both the signature $\Sigma$ and on the equations $A$ that guarantee faithfulness of a presentation $(\Sigma, A)$. For arbitrary $A$, it is necessary and sufficient[E47] that $\Sigma$ has no **quasi-empty** models, which are algebras $B$ such that $B_s = \emptyset$ for some $s$ but $B_{s'} \neq \emptyset$ for some other sort $s'$ [78]. For arbitrary $\Sigma$, it is sufficient that $A$ is Church-Rosser as a term rewriting system [136]. A proof of the following conservative extension result is given in Appendix B:

**Theorem 10.6.6** If the signature $\Sigma$ is coherent and $(\Sigma, A)$ is faithful, then the extension $(\Sigma, A) \subseteq (\Sigma^\circledast, A^\circledast)$ is conservative.  □

This gives soundness, and Theorem 10.7.18 gives completeness.

## 10.7 Order-Sorted Rewriting

Order-sorted rewriting arises from order-sorted equational logic in much the same way that many-sorted rewriting arises from many-sorted equational logic. We first consider unconditional rewriting, and add the assumption that $B$ contains no conditional equations to our prior assumption that all our order-sorted signatures are coherent (Definition 10.2.15).

**Definition 10.7.1** Given an order-sorted signature $\Sigma$, an unconditional **order-sorted $\Sigma$-rewrite rule** is an order-sorted $\Sigma$-equation $(\forall X)\, t_1 = t_2$ such that $var(t_2) \subseteq var(t_1) = X$; we will write $t_1 \to t_2$. An **order-sorted $\Sigma$-term rewriting system** (or **$\Sigma$-OSTRS**) is a set $A$ of order-sorted $\Sigma$-rewrite rules; we may also write $(\Sigma, A)$. □

Recall that $t_1$ and $t_2$ must lie in the same connected component of the sort set of $\Sigma$ (by Definition 10.2.16), so that, by coherence, there always exists a sort $s$ such that $LS(t_i) \leq s$ for $i = 1, 2$. We begin by restricting the rule of deduction $(+6_1)$, which replaces one subterm in the forward direction, to equations that are rewrite rules:

(RW) *Order-Sorted Rewriting.* Given $t_1 \to t_2$ in $A$ with $LS(t_1), LS(t_2) \leq s$ and $var(t_1) = Y$, and given $t_0 \in \mathcal{T}_\Sigma(X \cup \{z\}_s)$ with exactly one occurrence of $z \notin X$, if $\theta : Y \to \mathcal{T}_\Sigma(X)$ is a substitution, then
$$(\forall X)\, t_0(z \leftarrow \theta(t_1)) = t_0(z \leftarrow \theta(t_2))$$
is deducible.

This rule is sound because it is a restriction of (6), which is already known to be sound. Therefore the following is also sound (in a sense stated formally in Proposition 10.7.9 below):

**Definition 10.7.2** Given a $\Sigma$-OSTRS $A$, **one-step rewriting** is defined for $\Sigma$-terms $t, t'$ by $t \overset{1}{\Rightarrow}_A t'$ iff there exist a rule $t_1 \to t_2$ in $A$ with $var(t_1) = Y$, a term $t_0 \in \mathcal{T}_\Sigma(X \cup \{z\}_s)$ with exactly one occurrence of $z \notin X$, and a substitution $\theta : Y \to \mathcal{T}_\Sigma(X)$, such that $LS(t_1), LS(t_2) \leq s$ and
$$t = t_0(z \leftarrow \theta(t_1)) \text{ and } t' = t_0(z \leftarrow \theta(t_2))\,.$$

The **rewrite relation** is the transitive, reflexive closure of the one-step rewrite relation; we use the notation $t \overset{*}{\Rightarrow}_A t'$ and say $t$ **rewrites to** $t'$ (**under** $A$). □

The words "match," "redex," etc. are used the same way as in the many-sorted case, and the one-step order-sorted rewrite relation gives an abstract rewrite system, so that termination, Church-Rosser, local Church-Rosser, reduced term, and so on all make sense; moreover, the usual proof methods are available for proving these properties. We do not elaborate this, because we will soon generalize order-sorted rewriting to the conditional and modulo equation cases.

**Example 10.7.3** The following illustrates non-trivial overloaded order-sorted rewriting:

```
th OSRW is sorts A B .
  subsort A < B .
  op a : -> A .
  op b : -> B .
  ops f g : B -> B .
  op g : A -> A .
  eq b = a .
  var A : A .
  eq f(A) = A .
endth
red f(g(b)) .
```

The result is `g(a)`, after applying each rule once. □

The relation $\overset{*}{\Leftrightarrow}_A$, the reflexive, symmetric, and transitive closure of $\overset{1}{\Rightarrow}_A$, is order-sorted replacement of equals by equals. An important result for many-sorted rewriting (Theorem 7.3.4) is that $t \overset{*}{\Leftrightarrow}_A t'$ iff $A \vdash (\forall X)\ t = t'$, or otherwise put, that $\overset{*}{\Leftrightarrow}_A$ and $\simeq_{A,X}$ (provability) are equal relations on $T_\Sigma(X)$. By the completeness of many-sorted equational deduction, this also implies that $\overset{*}{\Leftrightarrow}_A$ is complete. Unfortunately, these nice results fail for order-sorted rewriting:

**Exercise 10.7.1** Use the specification OSRW-EQ in Exercise 10.3.2 to show that the relation $\overset{*}{\Leftrightarrow}_A$ is incomplete: $f(a) \overset{*}{\Leftrightarrow}_A f(b)$ does not hold, even though $a \overset{*}{\Leftrightarrow}_A b$ is true, and $f(a)$ and $f(b)$ are provably equal by Exercise 10.3.2. (This observation is due to Gert Smolka.) Examples with this undesirable behavior can be excluded by imposing the conditions in Definition 10.7.10 below, but it is better to use retracts, relying on Theorem 10.7.18; see also Example 10.7.4 below. □

**Exercise 10.7.2** It is also interesting to compare the equivalence classes for $\overset{*}{\Leftrightarrow}_A$ under the ordinary quotient construction with those under the construction of Definition 10.4.6. Use the relation $\equiv$ defined in Example 10.4.7 equals $\overset{*}{\Leftrightarrow}_A$ (though the spec is different from here) to show that the ordinary equivalence classes for $\overset{*}{\Leftrightarrow}_A$ differ from those of Definition 10.4.6. □

**Example 10.7.4** If we write the two equations in the theory OSRW-EQ of Exercise 10.3.2 in the converse order, then OBJ can prove the result with just one reduction, due to of the way that retracts work:

```
th OSRW-EQ-CONV is sorts A C .
  subsort A < C .
```

```
      ops a b : -> A .
      op c : -> C .
      op f : A -> C .
      eq a = c .
      eq b = c .
    endth
    red f(a) == f(b).
    red f(a) .
```

The first reduction gives `true`, even though the normal form of `f(a)` is `f(r:C>A(c))`, as shown by the second reduction, because `f(b)` has exactly the same normal form. In fact, the relation $\overset{*}{\Leftrightarrow}$ is complete for OSA with retracts (by Theorem 10.7.18 below). □

The following is the natural extension of Definition 10.7.1:

**Definition 10.7.5** A **conditional order-sorted rewrite rule** is an order-sorted conditional equation $(\forall X)\ t_1 = t_2$ if $C$ (in the sense of Definition 10.2.16) such that $var(t_1) = X, var(t_2) \subseteq X$, and for each $\langle u, v \rangle \in C$, $var(u) \subseteq X$ and $var(v) \subseteq X$. We use the notation $t_1 \to t_2$ if $C$. □

As with the many-sorted case, it is not straightforward to define the one-step rewrite relation for conditional rules. However, we can use the (Join) Conditional Abstract Rewrite Systems (CARS, Definition 7.7.1), just as we did in Section 7.7 for the many-sorted case; we will include rewriting modulo equations at the same time. For this purpose, we state the order-sorted modulo version of *Forward Conditional Subterm Replacement Modulo Equations*:

($6C_B$) Given $(\forall Y)\ t_1 =_B t_2$ if $C$ in $A$ with $LS(t_1), LS(t_2) \leq s$, and given $t_0 \in \mathcal{T}_\Sigma(X \cup \{z\}_s)$ with $z \notin X$, if $\theta: Y \to \mathcal{T}_\Sigma(X)$ is a substitution such that $(\forall X)\ \theta(u) =_B \theta(v)$ is deducible for each pair $\langle u, v \rangle \in C$, if $t_i' = t_0(z \leftarrow \theta(t_i))$ and $t_i'' \simeq_B t_i'$ for $i = 1, 2$, then $(\forall X)\ t_1'' = t_2''$ is also deducible.

Now the main concepts:

**Definition 10.7.6** An **order-sorted conditional term rewriting system modulo equations** (MCOSTRS) is $(\Sigma, A, B)$ where $A$ is a set of (possibly conditional) $\Sigma$-rewrite rules, and $B$ is a set of (unconditional) $\Sigma$-equations. Given $(\Sigma, A, B)$, define two CARS's as follows, where $t \to t'$ if $t_1 = t_1', \ldots, t_n = t_n'$ is in $A$ with $LS(t), LS(t') \leq s$, $Y = var(t)$, $\theta: Y \to \mathcal{T}_\Sigma$, $u \in \mathcal{T}_\Sigma(\{z\}_s)$ and $z \notin X$:

1. For term rewriting, let $W$ be the set of rules of the form $v \to v'$ if $v_1 = v_1', \ldots, v_n = v_n'$ where $v \simeq_B u(z \leftarrow \theta(t)), v' \simeq_B u(z \leftarrow \theta(t'))$, $v_i = \theta(t_i)$ and $v_i' = \theta(t_i')$ for $i = 1, \ldots, n$.

2. For class rewriting, let $W$ be the set of rules of the form $c \to c'$ if $c_1 = c_1', \ldots, c_n = c_n'$ where $c = [u(z \leftarrow \theta(t))], c' = [u(z \leftarrow \theta(t'))], c_i = [\theta(t_i)]$ and $c_i' = [\theta(t_i')]$ for $i = 1, \ldots, n$.

The classes in the second item are those defined by the quotient construction of Definition 10.4.6 from the provability relation $\simeq_B$. Now Definition 7.7.1 (page 235) yields an ARS $W^\diamond$ for each of these. Write $\Rightarrow_{A/B}$ for the first, which is order-sorted conditional term rewriting modulo $B$, and $\Rightarrow_{[A/B]}$ for the second, which is order-sorted conditional class rewriting modulo $B$. Also, we will use standard terminology ("match," "redex," etc.) in the usual way.  □

As discussed in Section 7.7, OBJ does not use equality semantics for evaluating conditions, even though it is common in the literature (e.g., [113]): it uses join condition semantics for its efficiency and pragmatic adequacy. Because $\Rightarrow_{A/B}$ and $\Rightarrow_{[A/B]}$ are ARS's, all ARS results apply directly, such as the Newman lemma and the multi-level termination results in Section 5.8.2. Also, the following has essentially the same proof as Proposition 7.3.2:

**Proposition 10.7.7** Given $t, t' \in \mathcal{T}_\Sigma(Y)$, $Y \subseteq X$ and MCOSTRS $(\Sigma, A, B)$, then $t \Rightarrow_{A/B,X} t'$ iff $t \Rightarrow_{A/B,Y} t'$, and in both cases $var(t') \subseteq var(t)$. Therefore $t \stackrel{*}{\Rightarrow}_{A/B,X} t'$ iff $t \stackrel{*}{\Rightarrow}_{A/B,Y} t'$, and in both cases $var(t') \subseteq var(t)$.  □

Thus both $\Rightarrow_{A/B,X}$ and $\stackrel{*}{\Rightarrow}_{A/B,X}$ restrict and extend well over variables, so we can drop the subscript $X$ and use any $X$ with $var(t) \subseteq X$; also as before, $\stackrel{*}{\Leftrightarrow}_{A/B,X}$ does *not* restrict and extend well, as shown by Example 5.1.15, so we define $t \stackrel{*}{\Leftrightarrow}_A t'$ to mean there exists an $X$ such that $t \stackrel{*}{\Leftrightarrow}_{A,X} t'$. Example 5.1.15 also shows bad behavior for $\simeq^X_{A/B}$ (defined by $t \simeq^X_{A/B} t'$ iff $A \cup B \vDash (\forall X)\ t = t'$), although again, the concretion rule (8) of Chapter 4 (generalized to order-sorted rewriting modulo $B$) implies $\simeq^X_{A,B}$ does behave reasonably when the signature is non-void. Defining $\downarrow_{A/B,X}$ from the ARS, we generalize Proposition 5.1.13, which again allows the subscript $X$ to be dropped:

**Proposition 10.7.8** Given terms $t, t' \in \mathcal{T}_\Sigma(Y)$, $Y \subseteq X$ and MCOSTRS $(\Sigma, A, B)$, then $t_1 \downarrow_{A/B,X} t_2$ iff $t_1 \downarrow_{A/B,Y} t_2$, and moreover, these imply $A \cup B \vdash (\forall X)\ t_1 = t_2$.  □

Because unconditional order-sorted rewriting modulo no equations is a special case, Exercise 10.7.1 also shows that $\stackrel{*}{\Leftrightarrow}_{A/B}$ is not complete for satisfaction of $A \cup B$. However, we do have:

**Proposition 10.7.9** (*Soundness*) Given an MCOSTRS $(\Sigma, A, B)$ and $t, t' \in \mathcal{T}_\Sigma(X)$, then

$$
\begin{aligned}
t \Rightarrow_{A/B} t' &\quad\text{iff}\quad [t] \Rightarrow_{[A/B]} [t']\ , \\
t \stackrel{*}{\Rightarrow}_{A/B} t' &\quad\text{iff}\quad [t] \stackrel{*}{\Rightarrow}_{[A/B]} [t']\ , \\
[t] \stackrel{*}{\Rightarrow}_{[A/B]} [t'] &\quad\text{implies}\quad [A] \vdash_B (\forall X)\ [t] = [t'] \\
[t] \stackrel{*}{\Leftrightarrow}_{[A/B]} [t'] &\quad\text{implies}\quad A \cup B \vdash (\forall X)\ t = t'\ .
\end{aligned}
$$

Therefore $\stackrel{*}{\Leftrightarrow}_{A/B}$ is sound for satisfaction of $A \cup B$ and $\stackrel{*}{\Leftrightarrow}_{[A/B]}$ is sound for satisfaction of $A$ modulo $B$. Moreover, $\stackrel{*}{\Leftrightarrow}_{A/B} \subseteq \cong^X_{A \cup B}$ on $\mathcal{T}_\Sigma(X)$.

**Proof:** The first assertion follows from the definitions of $\Rightarrow_{A/B}$ and $\Rightarrow_{[A/B]}$ (Definition 10.7.6); then the second follows by induction. The third follows from $\Rightarrow_{[A/B]}$ being a rephrasing of ($6C_B$), and the fourth follows from the third plus Proposition 10.5.2. □

We will consider two ways to render bidirectional term rewriting complete: the first assumes the conditions below, while the second uses retracts (Theorem 10.7.18).

**Definition 10.7.10** An MCOSTRS $(\Sigma, A, B)$ is **sort decreasing** iff $t \stackrel{1}{\Rightarrow}_{A/B} t'$ implies $LS(t) \geq LS(t')$. A $\Sigma$-rule $t \to t'$ if $C$ is **sort decreasing** iff for any substitution $\theta : X \to \mathcal{T}_\Sigma(Y)$ where $X = var(t)$, we have $LS(\theta(t)) \geq LS(\theta(t'))$. An order-sorted $\Sigma$-equation $t = t'$ is **sort preserving** iff for any substitution $\theta : X \to \mathcal{T}_\Sigma(Y)$ where $X = var(t)$, we have $LS(\theta(t)) = LS(\theta(t'))$. □

Notice that these conditions are decidable (provided $\Sigma$ is finite). We will see that they also improve the properties of order-sorted rewriting. The next two results follow [113]. Proposition 10.7.11 is straightforward using induction. Since Theorem 10.7.18 gives completeness with retracts but without the sort decreasing assumption, we do not prove Theorem 10.7.11; however the proofs in [113] carry over to join condition rewriting.

**Theorem 10.7.11** If $(\Sigma, B)$ is sort preserving, then $t \simeq_B t'$ iff $t \stackrel{*}{\Leftrightarrow}_B t'$. Moreover, $(\Sigma, B)$ is sort preserving iff $t \simeq_B t'$ implies $LS(t) = LS(t')$. An MCOSTRS $(\Sigma, A, B)$ is sort decreasing if $A$ is sort decreasing and $B$ is sort preserving. □

**Definition 10.7.12** An MCOSTRS $(\Sigma, A, B)$ is **join condition canonical** if and only if $(\Sigma', A', B)$ is canonical, where: (1) $\Sigma' \subseteq \Sigma$ is least such that if $t_i = t'_i$ is a condition of some rule $r$ in $A$, then $\theta(t_i)$ and $\theta(t'_i)$ are in $T_{\Sigma'}$ for all $\theta : X \to T_\Sigma$ where $X = var(t)$ and $t$ is the leftside of the head of rule $r$; and (2) $A' \subseteq A$ is least such that all conditional rules are in $A'$, and all unconditional rules that can be used in evaluating the conditions of rules in $A$ are in $A'$. □

The following is proved essentially the same way as Theorem 7.7.10:

**Theorem 10.7.13** (*Completeness*) Given a join condition canonical MCOSTRS $(\Sigma, A, B)$, the following four conditions are equivalent for any $t, t' \in \mathcal{T}_\Sigma(X)$:

$t \stackrel{*}{\Leftrightarrow}_{A/B} t'$   $A \cup B \vdash (\forall X)\ t = t'$
$[A] \vdash_B (\forall X)\ t =_B t'$   $t \simeq_{A \cup B} t'$

Moreover, if $(\Sigma, A, B)$ is Church-Rosser, then $t \downarrow_{A/B} t'$ is also equivalent to the above. Finally, if $(\Sigma, A, B)$ is canonical, then $[\![t]\!]_A \simeq_B [\![t']\!]_A$ is also equivalent. □

The next result is proved essentially the same way as Theorem 7.3.9:

**Theorem 10.7.14** Given a ground canonical MCOSTRS $(\Sigma, A, B)$ with $A$ sort decreasing and $B$ sort preserving, if $t_1, t_2$ are two normal forms of a ground term $t$ under $\Rightarrow_{A/B}$ then $t_1 \simeq_B t_2$. Moreover, the $B$-equivalence classes of ground normal forms under $\Rightarrow_{A/B}$ form an initial $(\Sigma, A \cup B)$-algebra, denoted $\mathcal{N}_{\Sigma, A/B}$ or just $\mathcal{N}_{A/B}$, as follows, where $[\![t]\!]$ denotes any arbitrary normal form of $t$, and $[\![t]\!]_B$ denotes the $B$-equivalence class of $[\![t]\!]$:

   (0) interpret $\sigma \in \Sigma_{[],s}$ as $[\![\sigma]\!]_B$ in $\mathcal{N}_{\Sigma, A/B, s}$; and

   (1) interpret $\sigma \in \Sigma_{s_1 \ldots s_n, s}$ with $n > 0$ as the function that sends $([\![t_1]\!]_B, \ldots, [\![t_n]\!]_B)$ with $t_i \in \mathcal{T}_{\Sigma, s_i}$ to $[\![\sigma(t_1, \ldots, t_n)]\!]_B$ in $\mathcal{N}_{\Sigma, A/B, s}$.

Finally, $\mathcal{N}_{\Sigma, A/B}$ is $\Sigma$-isomorphic to $\mathcal{T}_{\Sigma, A \cup B}$. □

As with previous similar results, this justifies the use of rewrite rules and normal forms to represent abstract data types, but now in the very rich setting of conditional order-sorted rewriting modulo equations; Sections 10.8 and 10.9 give examples showing how powerfully expressive this setting can be.

The important Theorem 10.7.18 below shows why OBJ3 works well even without the sort decreasing assumption. But first, we make precise the notion of retract rewriting:

**Definition 10.7.15** Given an MCOSTRS $(\Sigma, A, B)$ with $B$ sort preserving, then the **retract insertion rule** is defined as follows: suppose a term $t$ of least sort $s$ can be rewritten at the top by applying a (possibly conditional) rule $u \to v$ to yield a term $t'$ (i.e., there exists $\theta$ such that $t \simeq_B \theta(u)$ and $t' \simeq_B \theta(v)$), and suppose the least sort $s'$ of $t'$ is not less than or equal to $s$; now let $w(z)$ be a context with variable $z$ of sort $s$ but $s \not\geq s'$; then replace $t$ by $\mathtt{r}\mathtt{:}s''\mathtt{>}s(t')$ where $s'' \geq s, s'$, which exists by local filtration. **Retract rewriting** consists of rewriting with $(\Sigma, A^\circledast, B)$ plus the retract insertion rule. Let $\stackrel{*}{\Rightarrow}_{A^\circledast //B}$ denote retract rewriting. □

The retract insertion rule is sound, because it can be decomposed into two sound deductions: first substitute $t = t'$ into $w(\mathtt{r}\mathtt{:}s''\mathtt{>}s'(z))$ to obtain $w(\mathtt{r}\mathtt{:}s''\mathtt{>}s'(t)) = w(\mathtt{r}\mathtt{:}s''\mathtt{>}s'(t'))$, and then apply retract elimination to obtain $w(t) = w(\mathtt{r}\mathtt{:}s''\mathtt{>}s'(t'))$. Note that, as an optimization, a non-sort decreasing rule can be replaced by the corresponding rule with a retract on its rightside; OBJ3 in fact does this. Also note that a match with $B$ cannot produce terms that require inserting or deleting retracts (although it may manipulate such terms).

Although the relation $\overset{*}{\Leftrightarrow}_{A/B}$ is not complete for $\Sigma$-terms, $t \overset{*}{\Leftrightarrow}_{A^{\circledast}//B} t'$ is complete (and sound) for $\Sigma$-terms. We illustrate this with the specification of Exercise 10.7.1, which was originally used to show the incompleteness of $\overset{*}{\Leftrightarrow}_{A/B}$:

**Example 10.7.16** Example 10.7.4 successfully used the equations of Exercise 10.3.2 backwards, but Exercise 10.7.1 was restricted to forward rewriting. Can we show f(a) = f(b) using $\overset{*}{\Leftrightarrow}_{A/B}$? No, but we can with $\overset{*}{\Leftrightarrow}_{A^{\circledast}//B}$: rewrite the retract term f(r:C>A(c)) in two different ways: first to f(r:C>A(a)) and second to f(r:C>A(b)), using respectively the first and second rules. Then by retract elimination, the first equals f(a) and the second equals f(b). □

**Lemma 10.7.17** If $B$ is sort preserving, then $\simeq_B = \simeq_{B^{\circledast}}$ for $\Sigma$-terms, so that $\simeq_{(A \cup B)^{\circledast}} = \simeq_{A^{\circledast} \cup B}$, again for $\Sigma$-terms.

**Proof:** The first assertion follows because equational reasoning with $B$ cannot insert or delete retracts, since $B$ is sort preserving. The second assertion uses $(A \cup B)^{\circledast} = A^{\circledast} \cup B^{\circledast}$. □

Recall that $t \simeq_E t'$ means $E \models (\forall X)\, t = t'$.

**Theorem 10.7.18** (*Completeness*) If $(\Sigma, A, B)$ is an MCOSTRS with $B$ sort preserving and $A \cup B$ faithful, then $t \overset{*}{\Leftrightarrow}_{A^{\circledast}//B} t'$ iff $A \cup B \vdash_\Sigma (\forall X)\, t = t'$ for any $t, t' \in \mathcal{T}_\Sigma$.

**Proof:** Because order-sorted rewriting is sound, $t \overset{*}{\Leftrightarrow}_{A^{\circledast}//B} t'$ implies $t \simeq_{A^{\circledast} \cup B} t'$ for $t, t' \in \mathcal{T}_\Sigma$, which by Lemma 10.7.17 is equivalent to $t \simeq_{(A \cup B)^{\circledast}} t'$. Theorem 10.6.6 now gives equivalence of this to $t \simeq_{A \cup B} t'$.

For the converse, suppose $t \simeq_{A \cup B} t'$ for $t, t' \in \mathcal{T}_\Sigma$. By the above, this is equivalent to $t \simeq_{A^{\circledast} \cup B} t'$. We will show $t \overset{*}{\Leftrightarrow}_{A^{\circledast}//B} t'$ by simulating the proof for $t \simeq_{A \cup B} t'$ using $\Leftrightarrow_{A^{\circledast}//B}$; the essential difference is that the first can only substitute equals for equals, whereas the second allows first proving and then using lemmas. Using a lemma $u \simeq v$ is the same as applying a rule $u \to v$ (or $v \to u$, which is treated the same way), unless the rule is non-sort decreasing, in which case, when $v' = \theta(v)$ is substituted for $u' = \theta(u)$ in context $w$, an ill-formed $\Sigma$-term $w(v')$ may result. Then retract rewriting will substitute $r:s'>s(v')$ for $u'$, thus obtaining the well-formed $\Sigma^{\circledast}$-term $w(r:s'>s(v'))$. If the proof of $u \simeq v$ involves other lemmas, the same is done recursively, substituting the simulated proof of $u \simeq v$ using $\overset{*}{\Leftrightarrow}_{A^{\circledast}//B}$ for the use of $u \to v$. Doing this recursively for all lemmas finally yields a rewriting sequence for $t \overset{*}{\Leftrightarrow}_{A^{\circledast}//B} t'$. □

The implementation of order-sorted rewriting in OBJ3 achieves almost the effiency of many-sorted rewriting, using clever techniques described in detail in [113]. In addition, retracts are efficiently handled by builtin Lisp code, rather than by interpreting the theory of retracts.

**Exercise 10.7.3** Check whether Example 10.6.4 is sort decreasing according to Definition 10.7.10. Now enrich the theory MORE-NON-DED so that it permits a non-trivial deduction similar to that in Example 10.6.2, involving the intermediate use of retracts, and then show how to accomplish this deduction using the relation $\stackrel{*}{\Leftrightarrow}_{A/B}$. □

**Exercise 10.7.4** Use OBJ's `apply` commands to prove the equation $f(a) = f(b)$ for the specification OSRW-EQ in Exercise 10.3.2. □

## 10.7.1 Adding Constants

The results of Section 7.7.1 on adding new constants generalize straightforwardly to conditional order-sorted rewriting modulo equations; we state these generalizations explicitly because of their importance for theorem proving, and because they appear to be new in this context.

**Proposition 10.7.19** If an MCOSTRS $(\Sigma, A, B)$ is terminating, or Church-Rosser, or locally Church-Rosser, then so is $(\Sigma(X), A, B)$, for any suitable countable variable symbol set $X$. □

**Proposition 10.7.20** An MCOSTRS $(\Sigma, A, B)$ is ground terminating if $(\Sigma(X), A, B)$ is ground terminating, where $X$ is a variable set for $\Sigma$; moreover, if $\Sigma$ is non-void, then $(\Sigma, A, B)$ is ground terminating iff $(\Sigma(X), A, B)$ is ground terminating. □

**Corollary 10.7.21** If $\Sigma$ is non-void, then an MCOSTRS $(\Sigma, A, B)$ is ground terminating iff it is terminating. □

**Proposition 10.7.22** An MCOSTRS $(\Sigma, A, B)$ is Church-Rosser iff $(\Sigma(X_S^\omega), A, B)$ is ground Church-Rosser, is locally Church-Rosser iff $(\Sigma(X_S^\omega), A, B)$ is ground locally Church-Rosser. □

## 10.7.2 Proving Termination

This subsection generalizes termination results for MCTRS's in Section 7.5 to MCOSTRS's.

**Exercise 10.7.5** Generalize Propositions 7.5.6 and 7.5.7 to MCOSTRS's, and give proofs. □

**Exercise 10.7.6** Apply the generalizations of Propositions 7.5.6 and 7.5.7 to MCOSTRS's (that are not MCTRS's) to prove their termination. □

Given a poset $P$, we can define weak and strong $\rho$-monotonicty of order-sorted conditional rewrite rules modulo $B$, of order-sorted substitution modulo $B$, just as in Definition 5.5.3, and of operations in $\Sigma$, except that $T_\Sigma$ and $T_{\Sigma(\{z\}_s)}$ are replaced by $\mathcal{T}_{\Sigma,B}$ and $\mathcal{T}_{\Sigma,B}(\{z\}_s)$, respectively;

note that as before, the inequalities for a rule are only required to hold when all the conditions of the rule converge (modulo $B$). The following generalizes Theorem 7.7.20:

**Theorem 10.7.23** Let $(\Sigma, A, B)$ be a MCOSTRS with $\Sigma$ non-void and $A' \subseteq A$ unconditional and ground terminating; let $P$ be a poset and let $N = A - A'$. If there is $\rho : \mathcal{T}_{\Sigma, B} \to P$ such that

   (1) each rule in $A'$ is weak $\rho$-monotone,

   (2) each rule in $N$ is strict $\rho$-monotone,

   (3) each operation in $\Sigma$ is strict $\rho$-monotone, and

   (4) $P$ is Noetherian, or at least for each $t \in (\mathcal{T}_{\Sigma, B})_s$ there is some Noetherian poset $P_s^t \subseteq P_s$ such that $t \stackrel{*}{\Rightarrow}_{[A/B]} t'$ implies $\rho(t') \in P_s^t$,

then $(\Sigma, A, B)$ is ground terminating. □

**Exercise 10.7.7** Prove Theorem 10.7.23. □

**Exercise 10.7.8** Show that if $C = (\Sigma, A, B)$ is a MCOSTRS and $C^U = (\Sigma, A^U, B)$ where $A^U$ contains the rules in $A$ with their conditions removed, then $C$ is Church-Rosser (or ground Church-Rosser) if $C^U$ is. □

### 10.7.3 Proving Church-Rosser

As always, ARS results apply directly, including the Newman Lemma, the Hindley-Rosen Lemma (Exercise 5.7.5) and Proposition 7.7.22, so we do not state these here; the results that we do state are actually rather weak. Perhaps the most generally useful methods for proving Church-Rosser are based on the Newman Lemma, since it is usually much easier to prove the local Church-Rosser property. As mentioned in Sections 7.6 and 7.7.3, and in more detail in Chapter 12, although the Critical Pair Theorem (5.6.9) does not generalize to modulo $B$ rewriting, the local Church-Rosser property can still in many cases be checked by a variant of the Knuth-Bendix algorithm [117].

**Exercise 10.7.9** Does Proposition 7.6.9 generalize to MCOSTRS's? Give a proof or a counterexample. □

**Exercise 10.7.10** Generalize Proposition 7.7.23 to MCOSTRS's, give a proof, and then apply it to an example (not an MCTRS) to prove the Church-Rosser property. **Hint:** Consider a variant of PROPC where the truth values are a subsort of the propositional expressions. □

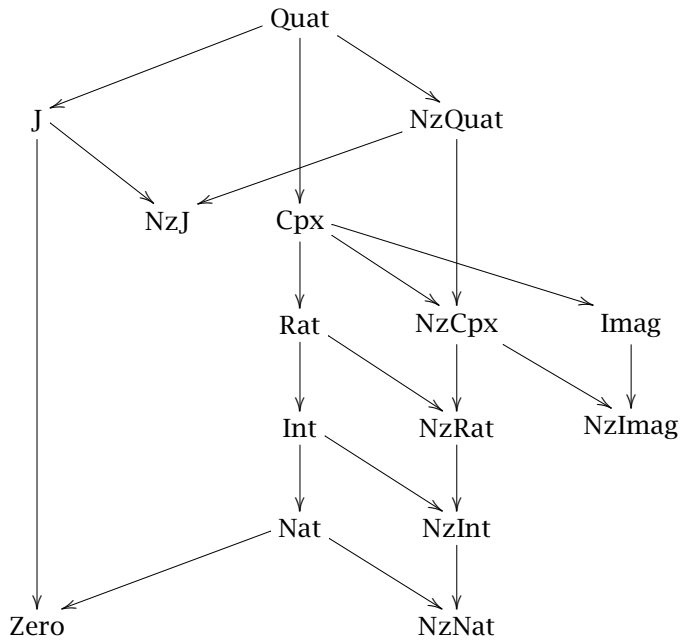

Figure 10.3: Subsort Structure for Number System

## 10.8 A Number System

This section presents an extended example, a rather complete number hierarchy, from the naturals up to the quaternions, including also the integer, rational, and complex numbers, with many of the usual operations upon them; however, it does not include the real numbers, and the complex numbers and quaternions are based on the rationals instead of the reals. A number of test cases are given. This example is from [82], and much of the work on it was done by Prof. José Meseguer and Mr. Tim Winkler. It is interesting to notice that multiplication is not commutative on quaternions, although it is commutative on the subsorts of complexes, rationals, etc., and that this situation is allowed by our notion of overloading, as well as supported by the OBJ implementation. This example is also used in [113], where it is annotated with much information about how its features are efficiently implemented in OBJ3 using techniques that include rule specializations and general variables.

```
obj NAT is sorts Nat NzNat Zero .
  subsorts Zero NzNat < Nat .
  op 0 : -> Zero .
  op s_ : Nat -> NzNat .
```

```
    op p_ : NzNat -> Nat .
    op _+_ : Nat Nat -> Nat [assoc comm] .
    op _*_ : Nat Nat -> Nat .
    op _*_ : NzNat NzNat -> NzNat .
    op _>_ : Nat Nat -> Bool .
    op d : Nat Nat -> Nat [comm] .
    op quot : Nat NzNat -> Nat .
    op gcd : NzNat NzNat -> NzNat [comm] .
    vars N M : Nat .   vars N' M' : NzNat .
    eq p s N = N .
    eq N + 0 = N .
    eq (s N) + (s M) = s s(N + M) .
    eq N * 0 = 0 .
    eq 0 * N = 0 .
    eq (s N) * (s M) = s(N + (M + (N * M))) .
    eq 0 > M = false .
    eq N' > 0 = true .
    eq s N > s M = N > M .
    eq d(0,N) = N .
    eq d(s N, s M) = d(N,M) .
    eq quot(N,M') = if ((N > M')or(N == M')) then
                         s quot(d(N,M'),M') else 0 fi .
    eq gcd(N',M') = if N' == M' then N' else (if N' > M' then
                    gcd(d(N',M'),M') else gcd(N',d(N',M')) fi) fi .
  endo

  obj INT is sorts Int NzInt .
    protecting NAT .
    subsorts NzNat < Nat NzInt < Int .
    op -_ : Int -> Int .
    op -_ : NzInt -> NzInt .
    op _+_ : Int Int -> Int [assoc comm] .
    op _*_ : Int Int -> Int .
    op _*_ : NzInt NzInt -> NzInt .
    op quot : Int NzInt -> Int .
    op gcd : NzInt NzInt -> NzNat [comm] .
    vars I J : Int .  vars I' J' : NzInt .  vars N' M' : NzNat .
    eq - - I = I .
    eq - 0 = 0 .
    eq I + 0 = I .
    eq M' + (- N') = if N' == M' then 0 else
                    (if N' > M' then - d(N',M') else d(N',M') fi) fi .
    eq (- I) + (- J) = -(I + J) .
    eq I * 0 = 0 .
    eq 0 * I = 0 .
    eq I * (- J) = -(I * J) .
    eq (- J) * I = -(I * J) .
    eq quot(0,I') = 0 .
    eq quot(- I',J') = - quot(I',J') .
```

```
    eq quot(I',- J') = - quot(I',J') .
    eq gcd(- I',J') = gcd(I',J') .
  endo

  obj RAT is sorts Rat NzRat .
    protecting INT .
    subsorts NzInt < Int NzRat < Rat .
    op _/_ : Rat NzRat -> Rat .
    op _/_ : NzRat NzRat -> NzRat .
    op -_  : Rat -> Rat .
    op -_  : NzRat -> NzRat .
    op _+_ : Rat Rat -> Rat [assoc comm] .
    op _*_ : Rat Rat -> Rat .
    op _*_ : NzRat NzRat -> NzRat .
    vars I' J' : NzInt .  vars R S : Rat .  vars R' S' : NzRat .
    eq R / (R' / S') = (R * S') / R' .
    eq (R / R') / S' = R / (R' * S') .
    ceq J' / I' = quot(J',gcd(J',I')) / quot(I',gcd(J',I'))
                    if gcd(J',I') =/= s 0 .
    eq R / s 0 = R .
    eq 0 / R' = 0 .
    eq R / (- R') = (- R) / R' .
    eq -(R / R') = (- R) / R' .
    eq R + (S / R') = ((R * R') + S) / R' .
    eq R * (S / R') = (R * S) / R' .
    eq (S / R') * R = (R * S) / R' .
  endo

  obj CPX-RAT is sorts Cpx Imag NzImag NzCpx .
    protecting RAT .
    subsort Rat < Cpx .
    subsort NzRat < NzCpx .
    subsorts NzImag < NzCpx Imag < Cpx .
    subsorts Zero < Imag .
    op _i : Rat -> Imag .
    op _i : NzRat -> NzImag .
    op -_ : Cpx -> Cpx .
    op -_ : NzCpx -> NzCpx .
    op _+_ : Cpx Cpx -> Cpx [assoc comm] .
    op _+_ : NzRat NzImag -> NzCpx [assoc comm] .
    op _*_ : Cpx Cpx -> Cpx .
    op _*_ : NzCpx NzCpx -> NzCpx .
    op _/_ : Cpx NzCpx -> Cpx .
    op _# : Cpx -> Cpx .
    op |_|^2 : Cpx -> Rat .
    op |_|^2 : NzCpx -> NzRat .
    vars R S : Rat . vars R' R" S' S" : NzRat . var A B C : Cpx .
    eq 0 i = 0 .
    eq C + 0 = C .
```

```
    eq (R i) + (S i) = (R + S) i .
    eq -(R' + (S' i)) = (- R') + ((- S') i) .
    eq -(S' i) = (- S') i .
    eq R * (S i) = (R * S) i .
    eq (S i) * R = (R * S) i .
    eq (R i) * (S i) = -(R * S) .
    eq C * (A + B) = (C * A) + (C * B) .
    eq (A + B) * C = (C * A) + (C * B) .
    eq R # = R .
    eq (R' + (S' i)) # = R' + ((- S') i) .
    eq (S' i) # = ((- S') i) .
    eq | C |^2 = C * (C #) .  *** This is an interesting equation.
    eq (S' i) / R" = (S' / R") i .
    eq (R' + (S' i)) / R" = (R' / R") + ((S' / R") i) .
    eq A / (R' i) = A *(((- s 0)/  R') i) .
    eq A / (R" + (R' i)) = A * ((R" / |(R" + (R' i))|^2) +
                                (((- R') / |(R" + (R' i))|^2) i)).
  endo

  obj QUAT-RAT is sorts Quat NzQuat J NzJ .
    protecting CPX-RAT .
    subsorts NzJ Zero < J < Quat .
    subsorts NzCpx < NzQuat Cpx < Quat .
    subsort NzJ < NzQuat .
    op _j : Cpx -> J .
    op _j : NzCpx -> NzJ .
    op -_ : Quat -> Quat .
    op _+_ : Quat Quat -> Quat [assoc comm] .
    op _+_ : Cpx NzJ -> NzQuat [assoc comm] .
    op _*_ : Quat Quat -> Quat .
    op _*_ : NzQuat NzQuat -> NzQuat .
    op _/_ : Quat NzQuat -> Quat .
    op _# : Quat -> Quat .
    op |_|^2 : Quat -> Rat .
    op |_|^2 : NzQuat -> NzRat .
    vars O P Q : Quat .  vars B C : Cpx .  var  C' : NzCpx .
    eq 0 j = 0 .
    eq Q + 0 = Q .
    eq -(C + (B j)) = (- C) + ((- B) j ) .
    eq (C j) + (B j) = (C + B) j .
    eq C * (B j) = (C * B) j .
    eq (B j) * C = (B * (C #)) j .
    eq (C j) * (B j) = -(C * (B #)) .
    eq Q * (O + P) = (Q * O) + (Q * P) .
    eq (O + P) * Q = (O * Q) + (P * Q) .
    eq (P + Q) # = (P #) + (Q #) .
    eq (C j) # = (- C) j .
    eq | Q |^2 = Q * (Q #) .
    eq Q / (C' j) = Q * ((s 0 / (- C')) j) .
```

```
      eq Q / (C + (C' j)) = Q * (((C #) / |(C + (C' j))|^2) +
                                 (((- C') / |(C + (C' j))|^2) j)) .
   endo

      *** now some test cases, preceded by some useful definitions
    open QUAT-RAT .
      ops 1 2 3 4 5 6 7 8 9 : -> NzNat [memo] .
      eq 1 = s 0 .  eq 2 = s 1 .
      eq 3 = s 2 .  eq 4 = s 3 .
      eq 5 = s 4 .  eq 6 = s 5 .
      eq 7 = s 6 .  eq 8 = s 7 .
      eq 9 = s 8 .
      red 3 + 2 .
      red 3 * 2 .
      red p p 3 .
      red 4 > 8 .
      red d(2,8) .
      red quot(7,2).
      red gcd(9,6) .
      red (- 4) + 8 .
      red (- 4) * 2 .
      red 8 / (- 2) .
      red (1 / 3) + (4 / 6) .
      red | 1 + (2 i) |^2 .
      red |(1 + (3 i)) + (1 + ((- 2) i))|^2 .
      red (3 + ((3 i) + ((- 2) i))) / ((2 i) + 2) .
      red (2 + ((3 i) j)) * ((5 i) + (7 j)) .
      red (1 + ((1 i) j)) / (2 j) .
    close
```

The equation that defines the squared norm function $|\_|^2$ is interesting, because given a non-zero rational as input, it should return a non-zero rational, but the rightside of the equation does not parse as a non-zero rational, although it can be *proved* that it always yields one. The attribute [memo] on the constants 1, ...,9 causes OBJ to cache the normal forms of these terms, and then use these cached values instead of recomputing them each time they are needed.

**Exercise 10.8.1** Show that INT, RAT, CPX, and QUAT are rings, where the theory of rings is given by the following theories, where the first defines commutative (also called Abelian) groups with additive notation:

```
  th ABGP is sort Elt .
    op 0 : -> Elt .
    op _+_ : Elt Elt -> Elt [assoc comm id: 0].
    op -_ : Elt -> Elt .
    var X : Elt .
    eq X +(- X) = 0 .
  endth
```

```
th RING is us ABGP .
  op 1 : -> Elt .
  op _*_ : Elt Elt -> Elt [assoc id: 1 prec 30].
  vars X Y Z : Elt .
  eq X * 0 = 0 .  eq 0 * X = 0 .
  eq X *(Y + Z) = (X * Y) + (X * Z) .
  eq (Y + Z)* X = (Y * X) + (Z * X) .
endth
```

Show that INT, RAT, and CPX are commutative rings, where the theory of commutative rings is as above, except that a comm attribute is added for * and the last equation is omitted. Show that QUAT is not a commutative ring. □

**Exercise 10.8.2** Show that INT, RAT, CPX, and QUAT are fields, where the theory of fields is given by the following:

```
th FIELD is us RING .
  sort NzElt .
  subsort NzElt < Elt .
  op _⁻¹ : NzElt -> NzElt [prec 2].
  vars X : NzElt .
  eq X * X⁻¹ = 1 .  eq X⁻¹ * X = 1 .
endth
```

Show that INT, RAT, and CPX are commutative fields, where the theory of commutative fields is as above, except that it imports the theory of commutative rings. Show that QUAT is not a commutative field. □

It is surprising that fields can be specified so simply with order-sorted algebra, but it is not difficult to show that there is an isomorphism between the classes of models of FIELD, and of fields in the ordinary sense, given by the function $\mathcal{U}$, where if $F$ is a model of FIELD, then $\mathcal{U}(F)$ is the partial algebra with $X^{-1}$ undefined when $X = 0$.

**Exercise 10.8.3** Use OBJ to prove in the theory FIELD that if $X, Y$ are non-zero, then so is $X*Y$. Conclude from this that the overload declaration

```
op _*_ : NzElt NzElt -> NzElt [assoc id: 1].
```

can be added to FIELD. □

**Exercise 10.8.4** Use OBJ to prove $(\forall X, Y : \text{Rat})\ (X + i * Y)(X - i * Y) = |X|^2 + |Y|^2$. **Hint:** Do not neglect the cases where $X = 0$ and/or $Y = 0$. □

## 10.9  Multiple Representation and Coercion

This section gives an example based on one in [81, 57], showing how OSA handles multiple representations of a single abstract data type, by providing automatic coercions among the representations. The type here is points in the plane (or vectors from the origin), and the representations are Cartesian and polar coordinates. The specification below uses the module FLOAT, which is OBJ's approximation to the real number field (which cannot be fully implemented on a computer). BOBJ [71, 72], the newest version of OBJ, is needed for its **sort constraints**, which allow users to declare new subsorts of old sorts. The keyword for sort constraints is mb, after the syntax of Maude [30]; the first defines a subsort NNeg of non-negatives for Float, and the second defines a further subsort for angles. Automatic coercions are defined by the three equations with retracts; the first two convert a point in polar coordinates to Cartesian coordinates when the context requires it, and the third does the opposite. Since both the sum and distance functions are only defined for the Cartesian representation, applying them to polar points requires coercion, as illustrated in the three reductions below the specification.

```
obj POINT is pr FLOAT .
  sorts NNeg Point Cart Polar Angle .
  subsorts Angle < NNeg < Float .
  var N : NNeg .   var A : Angle .   var F : Float .
  mb F : NNeg if F >= 0 .
  op _**2 : Float -> NNeg [prec 2].
  eq F **2 = F * F .
  mb F : Angle if 0 <= F and F < 2 * pi .

  subsorts Cart Polar < Point .
  op <_,_> : Float Float -> Cart .
  op _+_ : Cart Cart -> Cart .
  vars F1 F2 F3 F4 : Float .
  eq < F1, F2 > + < F3, F4 > = < F1 + F3, F2 + F4 > .
  op [_,_] : Angle NNeg -> Polar .
  eq r:Point>Cart([A, N]) = <N * cos(A), N * sin(A)> if N > 0 .
  eq r:Point>Cart([A, N]) = <0, 0> if N == 0 .
  eq r:Point>Polar(<F1, F2>) = [atan(F2 / F1), s
                                  qrt(F1 **2 + F2 **2)]
                               if F1 =/= 0 .
  op d : Cart Cart -> NNeg .
  eq d(<F1, F2>, <F3, F4>) = sqrt((F1 - F3)**2 + (F2 - F4)**2).
end

red d(< 1,  1 >, [pi / 3, 3]).
red < 1, 1 > + [pi / 3, 3].
red [pi / 2, 2] + [pi / 3, 3].
```

**Exercise 10.9.1** Add a function to the above code that rotates a point in polar coordinates about the origin, and use it to define a negation function for addition. Now run several test cases on these functions that require coercion. You can download the latest version of BOBJ from
ftp://ftp.cs.ucsd.edu/pub/fac/goguen/bobj/  □

## 10.10  Literature

Order-sorted algebra has evolved through several stages. The older versions OBJT and OBJ1 of OBJ used error algebras [53], which can fail to have initial models [150]. OSA began in 1978 in [54], and was further developed in papers including [82, 113, 152] and [165]. This chapter summarizes many basic results from OSA, mainly following [82]. A rather comprehensive survey up to 1992 is given in [68], and [77] is a less formal introduction with many examples.

Order-sorted rewriting was first treated in depth in [69], and then further developed in [113]. Both papers focus on operational semantics, i.e., on how to efficiently implement order-sorted term rewriting in OBJ: the first translates to many-sorted rewriting, while the second computes a set of rules that work on a term data structure that keeps track of the ranks of operations and the sorts of subterms; this is more efficient, and can be considered a weak form of compilation. Neither of these papers treats join conditional rewriting, so all the results about conditional order-sorted rewriting in Section 10.7 are new (we have argued in previous chapters that join semantics is the most appropriate for OBJ). The semantics of retracts in [78] does not cover rewriting modulo equations or retract rewriting, although this is discussed in [90] under the name "safe rewriting," again without condition rules or modulo equations. Theorem 10.7.18 is an important and perhaps surprising new result. The example in Section 10.9 is also new in this simple form.

**A Note to Lecturers:** This chapter is a culmination of this book, in the sense that we have been gradually building up towards a complete, rigorous treatment of the full operational and algebraic semantics of OBJ3, which is conditional order-sorted rewriting with retracts, modulo equations, with its theories of equational deduction and of algebras as models, together with practical methods for specification and verification, and practical ways to check key properties such as Church-Rosser and termination. This chapter seems to be the only place where such an exposition is available, and many of its results are new, e.g., Theorem 10.7.18. On the other hand, many other results are fairly straightforward generalizations of results in prior chapters. In my opinion, any course on algebraic specification should at least state and illustrate the main theorems of this chapter, making clear the great expressive power that results from combining all these features, although it is not necessary to go over all the counterexamples, auxiliary results, and proofs.

# 11 Generic Modules

> Insert material from [59] here.

With this technique, we can verify the correctness of generic objects in the sense of OBJ, as well as of higher-order functions as used in functional programming.

# 12 Unification

> Use the approach of "What is Unification?" [60], introducing some basic category theory to define unification; see also the discussion at the end of Chapter 8.

The discussion of the Church-Rosser property in Section 5.6 included a claim that it is decidable for terminating TRS's. We now discharge that claim. First, recall from Chapter 5 that, given a TRS $A$, if the left sides $t, t'$ of two rules have an overlap (in the sense of Definition 5.6.2) $\theta(t_0) = \theta'(t')$, where $t_0$ is a subterm of $t$, then $\theta(t)$ can be rewritten in two ways (one for each rule). The following, which is needed in Chapter 5, follows from the existence of most general unifiers, which can be computed as described above:

**Proposition 12.0.1** If terms $t, t'$ overlap at a subterm $t_0$ of $t$, then there is a **most general overlap** $p$, in the sense that any other overlap of $t, t'$ at $t_0$ is a substitution instance of $p$. □

Recall that such a most general overlap is called a **superposition**, and that the pair of terms resulting from applying the two rules to the term $\theta(t)$ is called a **critical pair**. If the two terms in a critical pair can be rewritten to a common term using rules in $A$, then that critical pair is said to **converge** or to **be convergent**. The following was proved in Chapter 5, and is important since it covers non-orthogonal TRS's that cannot be checked with Theorem 5.6.4:

**Theorem 5.6.9** A TRS is locally Church-Rosser if all its critical pairs are convergent. □

With the Newman Lemma (Proposition 5.6.1), this implies:

**Corollary 5.6.10** A terminating TRS is Church-Rosser iff all its critical pairs are convergent, in which case it is also canonical. □

Proofs of canonicity by orthogonality can be mechanized by using an algorithm that checks if each pair of rules is overlapping, and checks left linearity of each rule (which is trivial).

> Section 5.6 promised an algorithm from the above corollary; also the above just repeats material in that section.
>
> Section 5.6 also promises the unification algorithm.
>
> Discuss the Church-Rosser property for the TRS's GROUPC of Example 5.2.8, AND of Example 5.5.7, and MONOID of Exercise 5.2.7. Are these already in Chapter 5?

> Discuss Knuth-Bendix [117], and proof by consistency using completion, which in general is not a very satisfying technique. Do not prove the correctness of Knuth-Bendix completion.
>
> Do MSA case first, then OSA.
>
> Do Example 5.8.37 in algorithmic detail; can copy and edit.
>
> Also discuss completion procedures for matching modulo, as promised in Section 7.3.3.
>
> Check carefully against chapters 5,7,10.

We also need the following:

**Proposition 12.0.2** For $B$ containing any combination of the commutative, associative, and identity laws, if terms $t, t'$ overlap at a subterm $t_0$ of $t$, then there is a **most general overlap** $p$, in the sense that any other overlap of $t, t'$ at $t_0$ is a substitution instance of $p$. □

## 12.1 Blending

The following can now be done using the machinery developed above:

**Exercise 12.1.1** Recalling that Example 7.5.10 shows termination of DNF, use Corollary 5.6.10 to show that the MTRS DNF is locally Church-Rosser, and therefore canonical. □

**Exercise 12.1.2** Show that the normal forms of the MTRS DNF are exactly the disjunctive normal forms of the propositional formulae. □

**Exercise 12.1.3** Recalling that Example 7.5.9 shows termination of PROPC, use Corollary 5.6.10 to complete the proof of Hsiang's Theorem (Theorem 7.3.13) that PROPC is a canonical MTRS. □

# 13 Hidden Algebra

> Not really sure whether to do this; anyway, the following is just a rough sketch of some preliminary ideas for introductory remarks.

Computing applications typically involve states, and it can be awkward, or even impossible, to treat these applications in a purely functional style. Hidden algebra substantially extends ordinary algebra by distinguishing sorts used for data from sorts used for states, calling them respectively visible and hidden sorts. It also changes the notion of satisfaction to *behavioral* (also called *observational*) satisfaction, so that equations need not be literally satisfied, but need only appear to be satisfied under all possible experiments. Hidden algebra is powerful enough to give a semantics for the object paradigm, including inheritance and concurrency.

Whereas initial algebra semantics takes a somewhat static view of data structures and systems, as reflected in the central result that the initial model is unique up to isomorphism, hidden algebra takes a more dynamic view, directly addressing behavior and abstracting away from implementation, with its notion of behavioral satisfaction for equations. In philosophical terms, the evolution from initial algebra to hidden algebra is similar to the evolution from Plato's theory of static eternal ideals, to Aristotle's attempts to confront the kinds of change and development that can be observed especially in the biological realm.

# 14 A General Framework (Institutions)

This book takes the notions of model and satisfaction as basic, and checks the soundness of each proposed rule of deduction with respect to them before accepting it for use in theorem proving. Since we do not assume a single pre-determined logical system, the concepts of model and satisfaction cannot be fixed once and for all. Instead, we adopt a more general approach, influenced by the theory of institutions [67], and gradually enrich our language and models, working upward from simple equational logic.

Assume a concept of signature. For each signature $\Sigma$, assume that there are a class $\mathcal{M}_\Sigma$ of $\Sigma$-models, an algebra $T_\Sigma$ of $\Sigma$-terms, an algebra $\mathcal{F}_\Sigma$ of $\Sigma$-formulae, a concept of $\Sigma$-substitution, and a relation $\models_\Sigma$ of $\Sigma$-satisfaction of formulae by terms. Each concept should be defined recursively, and later concepts can be defined recursively over earlier ones.

Let $\mathcal{R}$ be a language of sentences that can be directly checked by OBJ, e.g., conjunctions of reductions, let $\mathcal{G}$ be a language for goals, and let $\mathcal{L}$ be a "meta" language that includes both $\mathcal{G}$ and $\mathcal{R}$. The **justification** for a proof score is a sequence of applications of proof measures which transforms a goal into an $\mathcal{R}$ sentence, which can then be translated into an OBJ program and run.

Elements of $\mathcal{G}$ express semantic facts which we wish to verify; a typical goal sentence is $E \models e$, where $E$ is a conjunction (perhaps represented as a set) of formulae and $e$ is a single formula. We will call sentences of this form (atomic) **turnstile** sentences. Our aim is then to transform (possibly rather exotic) turnstile sentences into $\mathcal{R}$ sentences that can be checked by OBJ.

(The above can be seen as an attempt to give a more down to earth exposition of the theory of institutions [67].)

---

MORE TO COME HERE.

# A  OBJ3 Syntax and Usage

This appendix gives a formal description of OBJ3 syntax, followed by some practical advice on how to use it. OBJ3 is the latest implementation of OBJ; it is an equational language with a mathematical semantics given by order-sorted equational logic, and a powerful type system featuring subtypes and overloading; these latter features allow us to define and handle errors in a precise way. In addition, OBJ3 has user-definable abstract data types with user-definable mixfix syntax, and a powerful parameterized module facility that includes views and module expressions. OBJ is *declarative* in the sense that its statements assert properties that solutions should have; i.e., they describe the problem. A subset of OBJ is executable using term rewriting, and all of it is provable. See [90] for more details.

OBJ3 syntax is described using the following extended BNF notation: the symbols { and } are used as meta-parentheses; the symbol | is used to separate alternatives; [ ] pairs enclose optional syntax; ... indicates zero or more repetitions of preceding unit; and "$x$" denotes $x$ literally. As an application of this notation, A{,A}... indicates a non-empty list of A's separated by commas. Finally, --- indicates comments in this syntactic description, as opposed to comments in OBJ3 code.

```
--- top-level ---

⟨OBJ-Top⟩ ::= {⟨Object⟩ | ⟨Theory⟩ | ⟨View⟩ | ⟨Make⟩ | ⟨Reduction⟩ |
    in ⟨FileName⟩ | quit | eof |
    start ⟨Term⟩ . |
    open [⟨ModExp⟩] . | openr [⟨ModExp⟩] . | close |
    ⟨Apply⟩ | ⟨OtherTop⟩}...

⟨Make⟩ ::= make ⟨Interface⟩ is ⟨ModExp⟩ endm

⟨Reduction⟩ ::= reduce [in ⟨ModExp⟩ :] ⟨Term⟩ .

⟨Apply⟩ ::=
   apply {reduction | red | print | retr |
     -retr with sort ⟨Sort⟩ |
     ⟨RuleSpec⟩ [with ⟨VarId⟩ = ⟨Term⟩ {, ⟨VarId⟩ = ⟨Term⟩}... ]}
   {within | at}
   ⟨Selector⟩ {of ⟨Selector⟩}...
```

```
⟨RuleSpec⟩ ::= [-][⟨ModId⟩].⟨RuleId⟩
⟨RuleId⟩ ::= ⟨Nat⟩ | ⟨Id⟩

⟨Selector⟩ ::= term | top |
    (⟨Nat⟩...) |
    [ ⟨Nat⟩ [ .. ⟨Nat⟩ ] ] |
    "{" ⟨Nat⟩ {, ⟨Nat⟩}... "}"
  --- note that "()" is a valid selector

⟨OtherTop⟩ ::= ⟨RedLoop⟩ | ⟨Commands⟩ | call-that ⟨Id⟩ . |
    test reduction [in ⟨ModExp⟩ :] ⟨Term⟩ expect: ⟨Term⟩ . | ⟨Misc⟩
  --- "call that ⟨Id⟩ ." is an abbreviation for "let ⟨Id⟩ = ."

⟨RedLoop⟩ ::= rl {. | ⟨ModId⟩} { ⟨Term⟩ .}... .

⟨Commands⟩ ::= cd ⟨Sym⟩ | pwd | ls |
    do ⟨DoOption⟩ . |
    select [⟨ModExp⟩] . |
    set ⟨SetOption⟩ . |
    show [⟨ShowOption⟩] .
  --- in select, can use "open" to refer to the open module

⟨DoOption⟩ ::= clear memo | gc | save ⟨Sym⟩... | restore ⟨Sym⟩... | ?

⟨SetOption⟩ ::= {abbrev quals | all eqns | all rules | blips |
      clear memo | gc show | include BOOL | obj2 | verbose |
      print with parens | reduce conditions | show retracts |
      show var sorts | stats | trace | trace whole} ⟨Polarity⟩ | ?

⟨Polarity⟩ ::= on | off

⟨ShowOption⟩ ::=
    {abbrev | all | eqs | mod | name | ops | params | principal-sort |
      [all] rules | select | sign | sorts | subs | vars}
      [⟨ParamSpec⟩ | ⟨SubmodSpec⟩] [⟨ModExp⟩] |
    [all] modes | modules | pending | op ⟨OpRef⟩ | [all] rule ⟨RuleSpec⟩ |
    sort ⟨SortRef⟩ | term | time | verbose | ⟨ModExp⟩ |
    ⟨ParamSpec⟩ | ⟨SubmodSpec⟩ | ?
  --- can use "open" to refer to the open module

⟨ParamSpec⟩ ::= param ⟨Nat⟩
⟨SubmodSpec⟩ ::= sub ⟨Nat⟩

⟨Misc⟩ ::= eval ⟨Lisp⟩ | eval-quiet ⟨Lisp⟩ | parse ⟨Term⟩ . | ⟨Comment⟩

⟨Comment⟩ ::= *** ⟨Rest-of-line⟩ | ***> ⟨Rest-of-line⟩ |
    *** (⟨Text-with-balanced-parentheses⟩)
⟨Rest-of-line⟩ --- the remaining text of the current line

--- modules ---

⟨Object⟩ ::= obj ⟨Interface⟩ is {⟨ModElt⟩ | ⟨Builtins⟩}... endo

⟨Theory⟩ ::= th ⟨Interface⟩ is ⟨ModElt⟩... endth

⟨Interface⟩ ::= ⟨ModId⟩ [[⟨ModId⟩... :: ⟨ModExp⟩
    {, ⟨ModId⟩... :: ⟨ModExp⟩}... ]]
```

```
⟨ModElt⟩ ::=
    {protecting | extending | including | using} ⟨ModExp⟩ . |
    using ⟨ModExp⟩ with ⟨ModExp⟩ {and ⟨ModExp⟩}... |
    define ⟨SortId⟩ is ⟨ModExp⟩ . |
    principal-sort ⟨Sort⟩ . |
    sort ⟨SortId⟩... . |
    subsort ⟨Sort⟩... { < ⟨Sort⟩... }... . |
    as ⟨Sort⟩ : ⟨Term⟩ if ⟨Term⟩ . |
    op ⟨OpForm⟩ : ⟨Sort⟩... -> ⟨Sort⟩ [⟨Attr⟩] . |
    ops {⟨Sym⟩ | (⟨OpForm⟩)}... : ⟨Sort⟩... -> ⟨Sort⟩ [⟨Attr⟩] . |
    op-as ⟨OpForm⟩ : ⟨Sort⟩... -> ⟨Sort⟩ for ⟨Term⟩ if ⟨Term⟩ [⟨Attr⟩] . |
    [⟨RuleLabel⟩] let ⟨Sym⟩ [: ⟨Sort⟩] = ⟨Term⟩ . |
    var ⟨VarId⟩... : ⟨Sort⟩ . |
    vars-of [⟨ModExp⟩] . |
    [⟨RuleLabel⟩] eq ⟨Term⟩ = ⟨Term⟩ . |
    [⟨RuleLabel⟩] cq ⟨Term⟩ = ⟨Term⟩ if ⟨Term⟩ . |
    ⟨Misc⟩

⟨Attr⟩ ::= [ {assoc | comm | {id: | idr:} ⟨Term⟩ | idem | memo |
    strat (⟨Int⟩... ) | prec ⟨Nat⟩ | gather ({e | E | &}... ) |
    poly ⟨Lisp⟩ | intrinsic}... ]

⟨RuleLabel⟩ ::= ⟨Id⟩... {, ⟨Id⟩... }...

⟨ModId⟩  --- simple identifier, by convention all caps
⟨SortId⟩ --- simple identifier, by convention capitalised
⟨VarId⟩  --- simple identifier, typically capitalised
⟨OpName⟩ ::= ⟨Sym⟩ {"_" | " " | ⟨Sym⟩}...
⟨Sym⟩    --- any operator syntax symbol (blank delimited)
⟨OpForm⟩ ::= ⟨OpName⟩ | (⟨OpName⟩)
⟨Sort⟩   ::= ⟨SortId⟩ | ⟨SortId⟩.⟨SortQual⟩
⟨SortQual⟩ ::= ⟨ModId⟩ | (⟨ModExp⟩)
⟨Lisp⟩   --- a Lisp expression
⟨Nat⟩    --- a natural number
⟨Int⟩    --- an integer

⟨Builtins⟩ ::=
    bsort ⟨SortId⟩ ⟨Lisp⟩ . |
    [⟨RuleLabel⟩] bq ⟨Term⟩ = ⟨Lisp⟩ . |
    [⟨RuleLabel⟩] beq ⟨Term⟩ = ⟨Lisp⟩ . |
    [⟨RuleLabel⟩] cbeq ⟨Term⟩ = ⟨Lisp⟩ if ⟨BoolTerm⟩ . |
    [⟨RuleLabel⟩] cbq ⟨Term⟩ = ⟨Lisp⟩ if ⟨BoolTerm⟩ .

--- views ---

⟨View⟩ ::= view [⟨ModId⟩] from ⟨ModExp⟩ to ⟨ModExp⟩ is ⟨ViewElt⟩... endv |
    view ⟨ModId⟩ of ⟨ModExp⟩ as ⟨ModExp⟩ is ⟨ViewElt⟩... endv

--- terms ---

⟨Term⟩ ::= ⟨Mixfix⟩ | ⟨VarId⟩ | (⟨Term⟩) |
    ⟨OpName⟩ (⟨Term⟩ {, ⟨Term⟩}... ) | (⟨Term⟩).⟨OpQual⟩
  --- precedence and gathering rules used to eliminate ambiguity

⟨OpQual⟩ ::= ⟨Sort⟩ | ⟨ModId⟩ | (⟨ModExp⟩)
⟨Mixfix⟩ --- mixfix operator applied to arguments
```

```
--- module expressions ---

⟨ModExp⟩ ::= ⟨ModId⟩ | ⟨ModId⟩ is ⟨ModExpRenm⟩ |
    ⟨ModExpRenm⟩ + ⟨ModExp⟩ | ⟨ModExpRenm⟩

⟨ModExpRenm⟩ ::= ⟨ModExpInst⟩ * (⟨RenameElt⟩ {, ⟨RenameElt⟩}... ) |
    ⟨ModExpInst⟩

⟨ModExpInst⟩ ::= ⟨ParamModExp⟩[⟨Arg⟩ {,⟨Arg⟩}... ] | (⟨ModExp⟩)

⟨ParamModExp⟩ ::= ⟨ModId⟩ | (⟨ModId⟩ * (⟨RenameElt⟩ {, ⟨RenameElt⟩}... ))

⟨RenameElt⟩ ::= sort ⟨SortRef⟩ to ⟨SortId⟩ | op ⟨OpRef⟩ to ⟨OpForm⟩

⟨Arg⟩ ::= ⟨ViewArg⟩ | ⟨ModExp⟩ | [sort] ⟨SortRef⟩ | [op] ⟨OpRef⟩
--- may need to precede ⟨SortRef⟩ by "sort" and ⟨OpRef⟩ by "op" to
--- distinguish from general case (i.e., from a module name)

⟨ViewArg⟩ ::= view [from ⟨ModExp⟩] to ⟨ModExp⟩ is ⟨ViewElt⟩... endv

⟨ViewElt⟩ ::= sort ⟨SortRef⟩ to ⟨SortRef⟩ . | var ⟨VarId⟩... : ⟨Sort⟩ . |
    op ⟨OpExpr⟩ to ⟨Term⟩ . | op ⟨OpRef⟩ to ⟨OpRef⟩ .
  --- priority given to ⟨OpExpr⟩ case
  --- vars are declared with sorts from source of view (a theory)

⟨SortRef⟩ ::= ⟨Sort⟩ | (⟨Sort⟩)
⟨OpRef⟩ ::= ⟨OpSpec⟩ | (⟨OpSpec⟩) | (⟨OpSpec⟩).⟨OpQual⟩ |
    ((⟨OpSpec⟩).⟨OpQual⟩)
  --- in views if have (op).(M) must be enclosed in (), i.e., ((op).(M))
⟨OpSpec⟩ ::= ⟨OpName⟩ | ⟨OpName⟩ : ⟨SortId⟩... -> ⟨SortId⟩
⟨OpExpr⟩   --- a ⟨Term⟩ consisting of a single operator applied
           --- to variables

--- equivalent forms ---

assoc = associative     comm = commutative
cq = ceq                dfn = define
ev = eval               evq = eval-quiet
jbo = endo              ht = endth
endv = weiv = endview   ex = extending
gather = gathering      id: = identity:
idem = idempotent       idr: = identity-rules:
in = input              inc = including
obj = object            poly = polymorphic
prec = precedence       psort = principal-sort
pr = protecting         q = quit
red = reduce            rl = red-loop
sh = show               sorts = sort
strat = strategy        subsorts = subsort
th = theory             us = using
vars = var              *** = ---
***> = --->
```

```
--- Lexical analysis ---

--- Tokens are sequences of characters delimited by blanks
--- "(", ")", and "," are always treated as single character symbols
--- Tabs and returns are equivalent to blanks (except inside comments)
--- Normally, "[", "]", "_", ",", "{", and "}"
--- are also treated as single character symbols.
```

Although OBJ provides a fully interactive user interface, in practice this is an awkward way to use the system, because users nearly always make mistakes, and mistakes can be very troublesome to correct in an interactive mode. It is much easier to first make a file, then start OBJ, and read the file with the `in` command; then OBJ will report the bugs it finds, based on which you can re-edit and then re-run the file; for complex examples, this cycle can be repeated many times. The author has found it convenient to edit OBJ files in one Emacs buffer, while another Emacs buffer contains a live OBJ; then you can switch between these by switching windows (or buffers); moreover, the results of execution are easily available for consultation and archiving.

OBJ3 can be obtained by ftp from pages linked to the following URL:

> `http://www.cs.ucsd.edu/users/goguen/sys/`

Once OBJ3 is installed, you can invoke it with the command `obj`. A later version of OBJ called BOBJ is also available via the above URL; whereas OBJ3 is implemented in Lisp, BOBJ is implemented in Java, and provides some additional features, including hidden algebra (as in Chapter 13). BOBJ is almost completely upward compatible with OBJ3, except that `apply` commands may need some reorganization, because the internal ordering of rules is different in the two systems.

CafeOBJ [43] is another algebraic specification language that could be used in connection with this text, although syntactical conversion will be needed, since its syntax tends to follow that of C. Information on how to obtain CafeOBJ is also available via the above URL.

---
Also discuss the conversion script once it is available.
---

# B Exiled Proofs

This appendix contains proofs considered too distracting to put in the main body of the text.

## B.1 Many-Sorted Algebra

Most of the proofs for the main results on many-sorted algebra were omitted in Chapter 4, and are also omitted here, because they follow from the more general results on order-sorted algebra that are restated and proved in Section B.5 below. An exception is Theorem 4.9.1, which we prove here as a sort of "warm up" for the proof of Theorem 10.3.3 in Section B.5.

**Theorem 3.2.1** (*Initiality*) Given a signature $\Sigma$ without overloading and a $\Sigma$-algebra $M$, there is a unique $\Sigma$-homomorphism $T_\Sigma \to M$. □

**Theorem 3.2.10** (*Initiality*) Given any signature $\Sigma$ and any $\Sigma$-algebra $M$, there is a unique $\Sigma$-homomorphism $\overline{T}_\Sigma \to M$. □

**Theorem 4.5.4** For any set $A$ of unconditional $\Sigma$-equations and unconditional $\Sigma$-equation $e$,

$$A \vdash e \quad \text{iff} \quad A \vdash^{(1,3,\pm 6)} e \,.$$
□

**Theorem 4.8.3** (*Completeness*) Given a signature $\Sigma$ and a set $A$ of (possibly conditional) $\Sigma$-equations, then for any unconditional $\Sigma$-equation $e$,

$$A \vdash^C e \quad \text{iff} \quad A \models e \,,$$

where $\vdash^C$ denotes deduction using the rules (1,2,3,4,5C). □

Note that Theorem 4.4.2 is the special case of the above where all equations in $A$ are unconditional.

**Theorem 4.9.1** (*Completeness of Subterm Replacement*) For any set $A$ of (possibly conditional) $\Sigma$-equations and any unconditional $\Sigma$-equation $e$,

$$A \vdash^C e \text{ iff } A \vdash^{(1,3,\pm 6C)} e \,.$$

**Proof ($\star$):** Let $X$ be a fixed but arbitrary set of variable symbols over the sort set of $\Sigma$. We will show that for any $e$ quantified by $X$, $A \vdash^C e$ iff $A \vdash^{(1,3,\pm 6C)} e$. For this purpose, we define two binary relations on $T_\Sigma(X)$, for $s \in S$ and $t, t' \in T_\Sigma(X)_s$, by

$$t_1 \equiv_s t_2 \text{ iff } A \vdash^C (\forall X) \, t_1 = t_2 \,,$$

and

$$t_1 \equiv^R_s t_2 \text{ iff } A \vdash^{(1,3,\pm 6C)} (\forall X) \, t_1 = t_2 \,,$$

and then show they are equal. (The superscript "R" comes from "Replacement" in the name of rule ($\pm 6C$).)

Soundness of ($\pm 6C$) and completeness of $\vdash^C$ give us that $A \vdash^{(1,3,\pm 6C)} e$ implies $A \vdash^C e$, which gives us $\equiv^R \subseteq \equiv$.

To show the opposite inclusion, we note that $\equiv$ is the *smallest* $\Sigma$-congruence satisfying a certain property, and then prove that $\equiv^R$ is another $\Sigma$-congruence satisfying that property. The property is *closure under* (5C), in the sense that if $(\forall Y) \, t = t'$ if $C$ is in $A$ and if $\theta : Y \to T_\Sigma(X)$ is a substitution such that $\theta(u) \equiv \theta(v)$ for each pair $(u, v) \in C$, then $\theta(t) \equiv \theta(t')$. That $\equiv$ is the least congruence closed under (5C) follows from its definition.

To facilitate proofs about $\equiv^R$, we define a family of relations on $T_\Sigma(X)$, for $s \in S$, by

$$\equiv^R_{0,s} = \{(t,t) \mid t \in T_\Sigma(X)_s\} \,,$$

and for each $n > 0$,

$$\equiv^R_{n,s} = \{(t_1, t_2) \mid t_1, t_2 \in T_\Sigma(X)_s \text{ and} \\ A \vdash^{(3,\pm 6C)} (\forall X) \, t_1 = t_2 \text{ via a proof of length} \leq n\} \,.$$

Then $\equiv^R = \bigcup_{n \in \omega} \equiv^R_n$.

The relation $\equiv^R$ is reflexive and transitive by definition. To prove its symmetry, we show by induction on $n$ that each relation $\equiv^R_n$ is symmetric. For the induction step, suppose that $t \equiv^R_{n+1} t'$, using symmetry of $\equiv^R_n$. There are just two cases, since the last step in proving $(\forall X) \, t = t'$ must use either the rule (3) or the rule ($\pm 6C$). If the last step used (3), then there exists $t''$ such that $t \equiv^R_n t''$ and $t'' \equiv^R_n t'$. By the induction hypothesis, we have that $t'' \equiv^R_n t$ and $t' \equiv^R_n t''$, which imply that $t' \equiv^R_{n+1} t$. In the second case, where ($\pm 6C$) is used, we again conclude $t' \equiv^R_{n+1} t$, this time by symmetry of ($\pm 6C$). Thus each $\equiv^R_n$ is symmetric, and symmetry of $\equiv^R$ follows from the fact that any union of symmetric relations is symmetric.

To prove $\equiv^R$ is a congruence, we must show that for each operation $\sigma$ in $\Sigma$, $\sigma(t_1,\ldots,t_k) \equiv^R \sigma(t'_1,\ldots,t'_k)$ whenever $t_i \equiv^R t'_i$ for $i = 1,\ldots,k$. For simplicity of presentation (and in fact without loss of generality), we do the proof for $k = 2$, showing by induction on $n$ that if $t_1 \equiv^R_n t'_1$ and $t_2 \equiv^R_n t'_2$ then $\sigma(t_1,t_2) \equiv^R \sigma(t'_1,t'_2)$. For this purpose, we show that $\sigma(t_1,t_2) \equiv^R \sigma(t'_1,t_2)$ and $\sigma(t'_1,t_2) \equiv^R \sigma(t'_1,t'_2)$, and then use transitivity of $\equiv^R$. Since these two subgoals are entirely analogous, we concentrate on the first, $\sigma(t_1,t_2) \equiv^R \sigma(t'_1,t_2)$.

The base case ($n = 0$) is trivial. For the induction step, as in the symmetry proof for $\equiv^R$, there are two cases, where the last step in proving $t_1 \equiv^R t'_1$ is either (3) or else ($\pm$6C).

In the first case, there exists $t_0 \in T_\Sigma(X)$ such that $t_1 \equiv^R_n t_0$ and $t_0 \equiv^R_n t'_1$. Then

$$\sigma(t_1,t_2) \equiv^R \sigma(t_0,t_2) \equiv^R \sigma(t'_1,t_2)$$

by the induction hypothesis, and transitivity of $\equiv^R$ gives the desired result.

For the case where ($\pm$6C) is applied, $A$ contains an equation $e'$,

$$(\forall Y)\ t = t'\ \mathtt{if}\ C\ ,$$

such that there is a substitution $\psi : Y \to T_\Sigma(X)$ such that $\psi(u) \equiv^R_n \psi(v)$ for each pair $(u,v) \in C$, and such that $t_0(z \leftarrow \psi(t)) = t_1$ and $t_0(z \leftarrow \psi(t')) = t'_1$, for some $t_0 \in T_\Sigma(X \cup \{z\})$ with $z \notin X$. By applying ($\pm$6C) with $e'$ to the term $\sigma(t_0,t_2)$ instead of $t_0$, we obtain $\sigma(t_1,t_2) \equiv^R_{n+2} \sigma(t'_1,t_2)$, which concludes our proof that $\equiv^R$ is a congruence.

We still have to show that $\equiv^R$ is closed under (5C). But this follows from the fact that ($\pm$6C) includes (5C), which is Exercise 4.9.4. □

This elegant proof is due to Răzvan Diaconescu. Note that Theorem 4.5.4 is the special case of the above result where all equations in $A$ are unconditional, and that the above result is in turn a special case of Theorem 10.3.3, which covers the order-sorted case.

## B.2  Rewriting

The results in this section concern overloaded many-sorted term rewriting, beginning with the following:

**Proposition 5.3.4**  A TRS $(\Sigma, A)$ is ground terminating if $(\Sigma(X), A)$ is ground terminating, where $X$ is a variable set for $\Sigma$; moreover, if $\Sigma$ is non-void, then $(\Sigma, A)$ is ground terminating iff $(\Sigma(X), A)$ is ground terminating.

**Proof:**  It is clear that ground termination of $(\Sigma(X), A)$ implies that of $(\Sigma, A)$.

For the converse, suppose that $(\Sigma, A)$ is ground terminating and that $\Sigma$ is non-void, so that for each sort $s$ there is some term $a^s \in T_{\Sigma,s}$. Now assume that

$$t_1 \stackrel{1}{\Rightarrow} t_2 \stackrel{1}{\Rightarrow} t_3 \stackrel{1}{\Rightarrow} \cdots$$

is a non-terminating rewrite sequence for $(\Sigma(X), A)$, where the rewrite $t_i \stackrel{1}{\Rightarrow} t_{i+1}$ uses the rule $l^i \to r^i$ in $A$ with $var(l^i) = X^i$ for $i = 1, 2, \ldots$, with $t_0^i$ and $\theta^i : X^i \to T_{\Sigma(X)}$ such that $t_i = t_0^i(z \leftarrow \theta^i(l^i))$ and $t_{i+1} = t_0^i(z \leftarrow \theta^i(r^i))$. Define $g : X \cup \{z\} \to T_{\Sigma(\{z\})}$ by $g(z) = z$ and $g(x) = a^s$ whenever $x \in X$ has sort $s$, and let $\overline{g}$ denote the free extension $T_{\Sigma(X \cup \{z\})} \to T_{\Sigma(\{z\})}$. Then

$$\overline{t}_1 \stackrel{1}{\Rightarrow} \overline{t}_2 \stackrel{1}{\Rightarrow} \overline{t}_3 \stackrel{1}{\Rightarrow} \cdots$$

is a non-terminating rewrite sequence for $(\Sigma, A)$, where each rewrite $\overline{t}_i \stackrel{1}{\Rightarrow} \overline{t}_{i+1}$ uses the rule $l^i \to r^i$ with $\overline{t}_i = \overline{t}_0^i(z \leftarrow \overline{\theta^i}(l^i))$ and $\overline{t}_{i+1} = \overline{t}_0^i(z \leftarrow \overline{\theta^i}(r^i))$, where $\overline{\theta^i} : X^i \to T_\Sigma$ is the composition $\theta^i; \overline{g}$ and where $\overline{t}_0^i = g(t_0^i)$. This works because $\overline{t}_i = \overline{g}(t_i)$ for $i = 1, 2, \ldots$; intuitively, the $\overline{t}_i$ rewrite sequence is the image under $\overline{g}$ of the $t_i$ rewrite sequence. □

The following is used to prove Proposition 5.5.6:

**Proposition 5.5.5** Given a $\Sigma$-TRS $A$ and a function $\rho : T_\Sigma \to \omega$, then $\Sigma$-substitution is strict $\rho$-monotone if every operation symbol in $\Sigma$ is $\rho$-strict monotone; the same holds for weak $\rho$-monotonicity.

**Proof ($\star$):** We use induction on the structure of $\Sigma$-terms. First notice that a term $t_0 \in T_\Sigma(\{z\})$ having a single occurrence of $z$ is either $z$ or else is of the form $\sigma(t_1, \ldots, t_n)$ where each $t_i$ except one is ground, and that one has a single occurrence of $z$. Now define the $z$-**depth** $d$ of a term $t \in T_\Sigma(\{z\})$ having a single occurrence of $z$ as follows:

$d(z) = 0;$

$d(\sigma(t_1, \ldots, t_n)) = 1 + d(t_i)$ where $t_i$ contains the occurrence of $z$.

Notice in particular that $d(\sigma(t_1, \ldots, t_n)) = 1$ iff the unique $t_i$ containing $z$ is $z$.

Now assume that $\rho(t) > \rho(t')$. Then substitution is strict $\rho$-monotone for all terms $t_0$ of $z$-depths 0 and 1; this covers the base cases. For the induction step, assume that strict $\rho$-monotonicity of substitution holds for all terms $t_0$ of $z$-depth less than $m > 0$, and that we are given some $t_0 \in T_\Sigma(\{z\})$ with $z$-depth $m$ and a single occurrence of $z$. Then $t_0$ has the form $\sigma(t_1, \ldots, t_n)$ for some $\sigma \in \Sigma$ and some $n > 0$;

furthermore, if $t_i$ is the term containing $z$ then $d(t_i) = m-1$. Therefore strict $\rho$-monotonicity of substitution holds for $t_i$. We now calculate:

$$t_0(z \leftarrow t) = \sigma(t_1, \ldots, t_i(z \leftarrow t), \ldots, t_n) = \sigma(t_1, \ldots, z, \ldots, t_n)(z \leftarrow \overline{t_i}),$$

where $\overline{t_i} = t_i(z \leftarrow t)$; and similarly, $t_0(z \leftarrow t') = \sigma(t_1, \ldots, z, \ldots, t_n)(z \leftarrow \overline{t_i'})$, where $\overline{t_i'} = t_i(z \leftarrow t')$. Now because strict $\rho$-monotonicity of substitution holds for $t_i$ we have

$$\rho(t_i(z \leftarrow t)) > \rho(t_i(z \leftarrow t')),$$

i.e., $\rho(\overline{t_i}) > \rho(\overline{t_i'})$, and therefore it follows that

$$\rho(t_0(z \leftarrow t)) > \rho(t_0(z \leftarrow t')),$$

as desired. The argument for the second assertion is essentially the same. □

**Theorem 5.6.9** (*Critical Pair Theorem*) A TRS is locally Church-Rosser if and only if all its critical pairs are convergent.

**Sketch of Proof:** The converse is easy. Suppose that all critical pairs converge, and consider a term with two distinct rewrites. Then their redexes are either disjoint or else one of them is a subterm of the other, since if two subterms of a given term are not disjoint, one must be contained in the other. If the redexes are disjoint, then the result of applying both rewrites is the same in either order. If the redexes are not disjoint, then either the rules overlap (in the sense of Definition 5.6.2), or else the subredex results from substituting for a variable in the leftside of the rule producing the larger redex. In the first case, the result terms of the two rewrites rewrite to a common term by hypothesis, since the overlap is a substitution instance of the overlap of some critical pair by Proposition 12.0.1. In the second case, the result of applying both rules is the same in either order, though the subredex may have to be rewritten multiple (or zero) times if the variable involved is non-linear. THIS IS JUST A COPY OF WHAT'S IN SECTION 5.6 ALREADY; SHOULD FILL IN DETAILS HERE OR ELSE OMIT. □

**Proposition 5.8.10** A CTRS $(\Sigma, A)$ is ground terminating if $(\Sigma(X), A)$ is ground terminating, where $X$ is a variable set for $\Sigma$; moreover, if $\Sigma$ is non-void, then $(\Sigma, A)$ is ground terminating iff $(\Sigma(X), A)$ is ground terminating.

**Proof:** We extend the proof of Proposition 5.3.4 above. If a rewrite $t_i \overset{1}{\Rightarrow} t_{i+1}$ in $(\Sigma, A(X))$ uses the rule $l^i \to r^i$ `if` $C^i$ where $var(l^i) = X^i$ and where $C^i$ contains $u_j^i = v_j^i$, then to get the corresponding rewrite $\overline{t_i} \overset{1}{\Rightarrow} \overline{t_{i+1}}$ in $(\Sigma, A)$, we apply $g$ to the conditions as well as to the left and right sides, noting that $\theta^i(u_j^i) \downarrow_A \theta^i(v_j^i)$ on $T_{\Sigma(X)}$ implies $\overline{\theta^i}(u_j^i) \downarrow_A \overline{\theta^i}(v_j^i)$ on $T_\Sigma$. □

### B.2.1 (⋆) Orthogonal Term Rewriting Systems

The proof below that orthogonal term rewriting systems are Church-Rosser was provided by Grigore Roşu, following a suggestion of Joseph Goguen to use the Hindley-Rosen Lemma (Proposition 5.7.5). As in the classic proof of Gerard Huet [108], we use "parallel rewriting" (Definition B.2.1); this should not be confused with the concurrent rewriting of [70, 134], as it is a technical notion especially created for this result. Indeed, the entire proof is rather technical, and suggestions for further simplification would be of interest.

**Definition B.2.1** Given a $\Sigma$-TRS $A$ and $\Sigma$-terms $t, t' \in T_\Sigma(Y)$, the **one-step parallel rewriting** relation, written $t \stackrel{1}{\Rightarrow}_A t'$, holds iff there exists a $\Sigma$-term $t_0 \in T_\Sigma(\{z_1, \ldots, z_n\} \cup Y)$ having exactly one occurrence of each variable $z_i$, and there exist $n$ $\Sigma$-rules $\alpha_i \to \beta_i$ in $A$ and $n$ substitutions $\theta_i : X_i \to T_\Sigma(Y)$ where $X_i = var(\alpha_i)$ for $1 \leq i \leq n$, such that

$$\begin{aligned} t &= t_0[z_1 \leftarrow \theta_1(\alpha_1), \ldots, z_n \leftarrow \theta_n(\alpha_n)] \text{ and} \\ t' &= t_0[z_1 \leftarrow \theta_1(\beta_1), \ldots, z_n \leftarrow \theta_n(\beta_n)] \,. \end{aligned}$$

The **parallel rewriting** relation is the transitive closure of $\Rightarrow_A$, denoted $t \Rightarrow_A^* t'$. □

The relation $\Rightarrow$ is reflexive, as can be seen by taking $n = 0$ in Definition B.2.1. We require that $t_0$ contain exactly one occurrence of each variable $z_i$ only for technical reasons. Note that the $\Sigma$-rules $\alpha_i \to \beta_i$ are not required to be distinct. We may omit the subscript $A$ when it is clear from context, writing $t \Rightarrow^* t'$ instead of $t \Rightarrow_A^* t'$, and also writing $\Rightarrow$ instead of $\stackrel{1}{\Rightarrow}_A$.

**Exercise B.2.1** Given a $\Sigma$-TRS $A$, terms $t_1, t'_1, \ldots, t_n, t'_n \in T_\Sigma(Y)$ such that $t_1 \Rightarrow \Rightarrow_A t'_1, \ldots, t_n \Rightarrow_A t'_n$, and $t_0 \in T_\Sigma(\{z_1, \ldots, z_n\} \cup Y)$, show $t_0[z_1 \leftarrow t_1, \ldots, z_n \leftarrow t_n] \Rightarrow_A t_0[z_1 \leftarrow t'_1, \ldots, z_n \leftarrow t'_n]$. □

Three lemmas precede the main part of the proof. The first justifies using parallel rewriting to prove results about ordinary rewriting.

**Lemma B.2.2** Given a $\Sigma$-TRS $A$, $\Rightarrow_A^* = \Rightarrow_A^*$.

**Proof:** The inclusion $\Rightarrow_A^* \subseteq \Rightarrow_A^*$ follows from the fact that one-step rewriting is the special case of one-step parallel rewriting where $n = 1$ and the single variable is $z$.

Thus it suffices to prove the opposite inclusion, $\Rightarrow_A \subseteq \Rightarrow_A^*$. Suppose that $t \Rightarrow_A t'$ and let $t_0 \in T_\Sigma(\{z_1, \ldots, z_n\} \cup Y)$, as in the definition of one-step parallel rewriting. Let $t_0^i \in T_\Sigma(\{z\} \cup Y)$ denote the term

$$\begin{aligned} &t_0[z_1 \leftarrow \theta_1(\beta_1), \ldots, z_{i-1} \leftarrow \theta_{i-1}(\beta_{i-1}), \\ &\quad z_i \leftarrow z, z_{i+1} \leftarrow \theta_{i+1}(\alpha_{i+1}), \ldots, z_n \leftarrow \theta_n(\alpha_n)] \end{aligned}$$

and let $t^i$ denote the terms $t_0^i[z \leftarrow \theta_i(\beta_i)]$ for $1 \leq i \leq n$.

Because $t = t_0^1[z \leftarrow \theta_1(\alpha_1)]$ and $t^1 = t_0^1[z \leftarrow \theta_1(\beta_1)]$, we get $t \Rightarrow_A t^1$ by the definition of one-step (non-parallel) rewriting. Also because $t^i = t_0^{i+1}[z \leftarrow \theta_{i+1}(\alpha_{i+1})]$ and $t^{i+1} = t_0^{i+1}[z \leftarrow \theta_{i+1}(\beta_{i+1})]$, we get $t^i \Rightarrow_A t^{i+1}$ for $1 \leq i < n$. Finally, since $t^n = t'$, we get the chain of one-step rewrites $t \Rightarrow_A \cdots \Rightarrow_A t^i \Rightarrow_A t^{i+1} \Rightarrow_A \cdots \Rightarrow_A t'$, and therefore $t \Rightarrow_A^* t'$.

□

From now on, we assume $A$ is a fixed $\Sigma$-TRS with $\Sigma$-rules $\alpha_i \to \beta_i$ for $1 \leq i \leq N$. Let $A_i$ denote the $\Sigma$-TRS containing a single $\Sigma$-rule $\alpha_i \to \beta_i$, let $\Rightarrow_i$ denote the relation $\Rightarrow_{A_i}$, let $\Rrightarrow_i$ denote the relation $\Rrightarrow_{A_i}$, and let $X_i = var(\alpha_i)$, the set of variables of $\alpha_i$.

The next lemma is the only place where orthogonality of $A$ is used. In reading its proof, it may help to visualize the various constructions using the picture below.

**Lemma B.2.3** If $A$ is orthogonal and if $\varphi : X_i \to T_\Sigma(Y)$ is a substitution such that $\varphi(\alpha_i) \Rightarrow_j t$ for some $1 \leq i, j \leq N$, then there is some $t' \in T_\Sigma(Y)$ such that $\varphi(\beta_i) \Rrightarrow_j t'$ and $t \Rrightarrow_i t'$.

**Proof:** Because $\varphi(\alpha_i) \Rightarrow_j t$, Definition B.2.1 implies there exist a $\Sigma$-term $t_0 \in T_\Sigma(\{z_1, \ldots, z_n\} \cup Y)$ and substitutions $\theta_k : X_j \to T_\Sigma(Y)$ such that $\varphi(\alpha_i) = t_0[z_1 \leftarrow \theta_1(\alpha_j), \ldots, z_n \leftarrow \theta_n(\alpha_j)]$ and $t = t_0[z_1 \leftarrow \theta_1(\beta_j), \ldots, z_n \leftarrow \theta_n(\beta_j)]$. Therefore $\theta_k(\alpha_j)$ is a subterm of $\varphi(\alpha_i)$ for each $1 \leq k \leq n$. But because $A$ is nonoverlapping, the terms $\alpha_i$ and $\alpha_j$ do not overlap, i.e., there does not exist a non-variable subterm $\alpha_i^k$ of $\alpha_i$ such that $\theta_k(\alpha_j) = \varphi(\alpha_i^k)$. Consequently, the only possibility for $\theta_k(\alpha_j)$ to be a subterm of $\varphi(\alpha_i)$, is to be a subterm (not necessarily proper) of $\varphi(x)$ where $x$ is a variable in $X_i$. Hence for each $1 \leq k \leq n$, there is a variable $x_k$ in $X_i$ such that $\theta_k(\alpha_j)$ is a subterm of $\varphi(x_k)$.

The variables $x_k$ need not be distinct for distinct indices $k$. Because $t_0$ is the term of "positions" of $\theta_k(\alpha_j)$ in $\varphi(\alpha_i)$ for $1 \leq k \leq n$ and because each $\theta_k(\alpha_j)$ is a subterm of $\varphi(x_k)$ and $A$ is left linear, that is, $\alpha_i$ has no more than one occurence of any variable $x$ in $X_i$, we can conclude that for each $x$ in $X_i$ there is a subterm $t_0^x$ of $t_0$ such that $\varphi(x) = t_0^x[z_1 \leftarrow \theta_1(\alpha_j), \ldots, z_n \leftarrow \theta_n(\alpha_j)]$. The term $t_0^x$ is the subterm of $t_0$ that contains the "positions" of each $\theta_k(\alpha_j)$ in $\varphi(x)$ for $1 \leq k \leq n$. It is possible that $\varphi(x)$ does not contain all $\theta_k(\alpha_j)$ as subterms or just does not contain anyone of them, but we still preserve the notation $t_0^x[z_1 \leftarrow \theta_1(\alpha_j), \ldots, z_n \leftarrow \theta_n(\alpha_j)]$ which means that one substitutes only variables $z_k$ that appear in $t_0^x$, that is, variables $z_k$ for those $1 \leq k \leq n$ for which $x_k = x$.

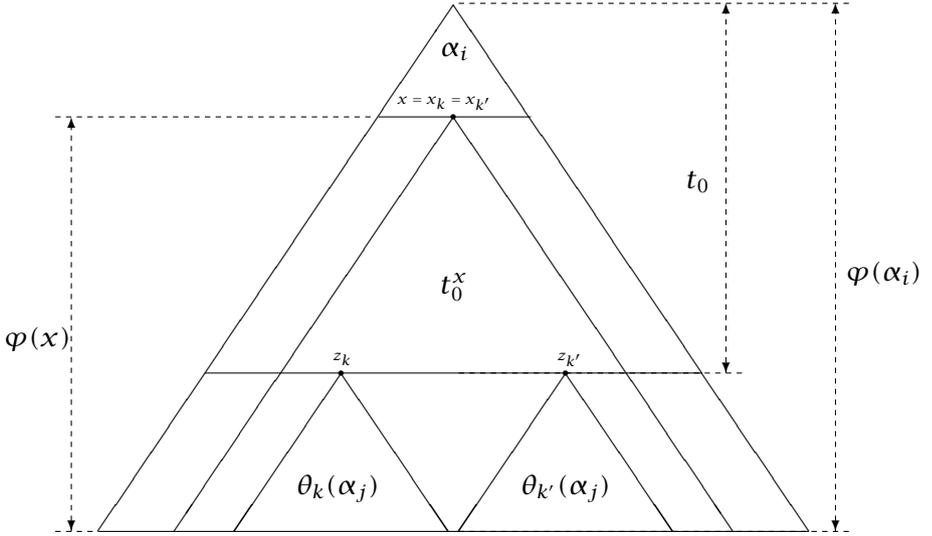

Let $\epsilon : X_i \to T_\Sigma(\{z_1,\ldots,z_n\} \cup Y)$ denote the function for which $\epsilon(x) = t_0^x$. Since $A$ is left linear, no $\alpha_i$ has more than one instance of any variable $x$ in $X_i$, and therefore $\epsilon(\alpha_i) = t_0$.

Now let $\theta^\alpha, \theta^\beta : \{z_1,\ldots,z_n\} \to T_\Sigma(Y)$ be the substitutions such that $\theta^\alpha(z_k) = \theta_k(\alpha_j)$ and $\theta^\beta(z_k) = \theta_k(\beta_j)$ for all $1 \le k \le n$. Then we have $\epsilon; \theta^\alpha = \varphi$, because for each $x$ in $X_i$

$$\begin{aligned}
(\epsilon; \theta^\alpha)(x) &= \\
\theta^\alpha(\epsilon(x)) &= \\
\theta^\alpha(t_0^x) &= \\
t_0^x[z_1 \leftarrow \theta_1(\alpha_j),\ldots,z_n \leftarrow \theta_n(\alpha_j)] &= \\
\varphi(x) \, .
\end{aligned}$$

Let $t'$ be the term $(\epsilon; \theta^\beta)(\beta_i)$, that is $t' = \epsilon(\beta_i)[z_1 \leftarrow \theta_1(\beta_j),\ldots,z_n \leftarrow \theta_n(\beta_j)]$. Then we can show that

$$\begin{aligned}
\varphi(\beta_i) &= \\
(\epsilon; \theta^\alpha)(\beta_i) &= \\
\theta^\alpha(\epsilon(\beta_i)) &= \\
\epsilon(\beta_i)[z_1 \leftarrow \theta_1(\alpha_j),\ldots,z_n \leftarrow \theta_n(\alpha_j)] \, ,
\end{aligned}$$

and by the definition of parallel rewriting, we get $\varphi(\beta_i) \Rightarrow_j t'$. Although $\epsilon(\beta_i)$ may contain multiple occurences of variables $z_1,\ldots,z_n$, this does not modify the one-step parallel rewriting relation (the reader should prove this). On the other hand, because

$$\begin{aligned}
t &= \\
t_0[z_1 \leftarrow \theta_1(\beta_j),\ldots,z_n \leftarrow \theta_n(\beta_j)] &= \\
\epsilon(\alpha_i)[z_1 \leftarrow \theta_1(\beta_j),\ldots,z_n \leftarrow \theta_n(\beta_j)] &= \\
\theta^\beta(\epsilon(\alpha_i)) &= \\
(\epsilon; \theta^\beta)(\alpha_i) \, ,
\end{aligned}$$

it follows that $t \Rightarrow_i t'$, by the definition of one-step ordinary rewriting. □

The above lemma holds even when $n = 0$, that is, when there are no parallel rewrites.

**Lemma B.2.4** If $A$ is orthogonal and if $t, t_1, t_2$ are $\Sigma$-terms such that $t \Rightarrow_i t_1$ and $t \Rightarrow_j t_2$ then there is some $t'$ such that $t_1 \Rightarrow_j t'$ and $t_2 \Rightarrow_i t'$.

**Proof:** There exist $\Sigma$-terms $t_0^i \in T_\Sigma(\{p_1,\ldots,p_m\} \cup Y)$, $t_0^j \in T_\Sigma(\{z_1,\ldots,z_n\} \cup Y)$ and substitutions $\varphi_1,\ldots,\varphi_m : X_i \to T_\Sigma(Y)$ and $\theta_1,\ldots,\theta_n : X_j \to T_\Sigma(Y)$ such that

$$t = t_0^i[p_1 \leftarrow \varphi_1(\alpha_i),\ldots,p_m \leftarrow \varphi_m(\alpha_i)]$$
$$t_1 = t_0^i[p_1 \leftarrow \varphi_1(\beta_i),\ldots,p_m \leftarrow \varphi_m(\beta_i)],$$

and

$$t = t_0^j[z_1 \leftarrow \theta_1(\alpha_j),\ldots,z_n \leftarrow \theta_n(\alpha_j)]$$
$$t_2 = t_0^j[z_1 \leftarrow \theta_1(\beta_j),\ldots,z_n \leftarrow \theta_n(\beta_j)].$$

In the picture below, $t_0^i$ and $t_0^j$ appear as two different "tops" for the term $t$:

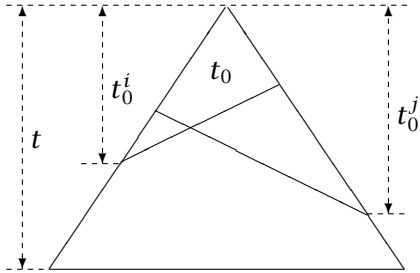

Let $\delta^{\alpha,\alpha} : \{p_1,\ldots,p_m,z_1,\ldots,z_n\} \to T_\Sigma(Y)$ be a substitution such that $\delta^{\alpha,\alpha}(p_l) = \varphi_l(\alpha_i)$ for all $1 \le l \le m$ and $\delta^{\alpha,\alpha}(z_k) = \theta_k(\alpha_j)$ for all $1 \le k \le n$. We get $\delta^{\alpha,\alpha}(t_0^i) = t$ and also $\delta^{\alpha,\alpha}(t_0^j) = t$, that is, $\delta^{\alpha,\alpha}$ is a unifier[1] of the terms $t_0^i$ and $t_0^j$. Because $t_0^i, t_0^j$ are unifiable, they have a most general unifier, say $\psi : \{p_1,\ldots,p_m,z_1,\ldots,z_n\} \to T_\Sigma(\{p_1,\ldots,p_m,z_1,\ldots,z_n\} \cup Y)$. In our case, because $t_0^i$ and $t_0^j$ have exactly one occurrence of the variables $p_1,\ldots,p_m$ and $z_1,\ldots,z_n$ respectively, each $\psi(p_l)$ is either equal to $p_l$ or else is a subterm of $t_0^j$, and each $\psi(z_k)$ is either equal to $z_k$ or else is a subterm of $t_0^i$. The reader can now check that $\psi; \delta^{\alpha,\alpha} = \delta^{\alpha,\alpha}$.

Next we introduce two more substitutions, $\delta^{\alpha,\beta}, \delta^{\beta,\alpha} : \{p_1,\ldots,p_m, z_1,\ldots,z_n\} \to T_\Sigma(Y)$ such that $\delta^{\alpha,\beta}(p_l) = \varphi_l(\alpha_i)$ and $\delta^{\alpha,\beta}(z_k) = \theta_k(\beta_j)$, and $\delta^{\beta,\alpha}(p_l) = \varphi_l(\beta_i)$ and $\delta^{\beta,\alpha}(z_k) = \theta_k(\alpha_j)$, respectively. We now claim that $\varphi_l(\alpha_i) \Rightarrow_j (\psi;\delta^{\alpha,\beta})(p_l)$ for each $1 \le l \le m$. This is because $\varphi_l(\alpha_i) = \delta^{\alpha,\alpha}(p_l) = (\psi;\delta^{\alpha,\alpha})(p_l)$ and because either $\psi(p_l)$ equals $p_l$, in which case $(\psi;\delta^{\alpha,\beta})(p_l) = \varphi_l(\alpha_i)$ and then we use the reflexivity of $\Rightarrow_j$, or else $\psi(p_l)$ contains only distinct variables in $\{z_1,\ldots,z_n\}$ and then

$$\begin{aligned}
\varphi_l(\alpha_i) &= \\
\delta^{\alpha,\alpha}(\psi(p_l)) &= \\
\psi(p_l)[z_1 \leftarrow \theta_1(\alpha_j),\ldots,z_n \leftarrow \theta_n(\alpha_j)] &
\end{aligned}$$

and

$$\begin{aligned}
(\psi;\delta^{\alpha,\beta})(p_l) &= \\
\delta^{\alpha,\beta}(\psi(p_l)) &= \\
\psi(p_l)[z_1 \leftarrow \theta_1(\beta_j),\ldots,z_n \leftarrow \theta_n(\beta_j)].
\end{aligned}$$

---
[1]The notions of unifier and most general unifier are defined in Chapter 12.

Similarly, $\theta_k(\alpha_j) \Rightarrow_i (\psi;\delta^{\beta,\alpha})(z_k)$ for each $1 \le k \le n$.

Suppose that $p_1,\ldots,p_M$ are all the variables of $t_0^i$ such that $p_l \notin var(\psi(z_k))$ for $1 \le k \le n$, and that $z_1,\ldots,z_N$ are the variables of $t_0^j$ such that $z_k \notin var(\psi(p_l))$ for $1 \le l \le m$. Then there is a term $t_0 \in T_\Sigma(\{p_1,\ldots,p_M\} \cup \{z_1,\ldots,z_N\} \cup Y)$ such that $\psi(t_0) = \psi(t_0^i) = \psi(t_0^j)$. In the above picture, $t_0$ is the "intersection" of $t_0^i$ and $t_0^j$. The reader should now check that

$$t = t_0[p_1 \leftarrow \varphi_1(\alpha_i),\ldots,p_M \leftarrow \varphi_M(\alpha_i),\\
z_1 \leftarrow \theta_1(\alpha_j),\ldots,z_N \leftarrow \theta_N(\alpha_j)]$$

$$t_1 = t_0[p_1 \leftarrow \varphi_1(\beta_i),\ldots,p_M \leftarrow \varphi_M(\beta_i),\\
z_1 \leftarrow (\psi;\delta^{\beta,\alpha})(z_1),\ldots,z_N \leftarrow (\psi;\delta^{\beta,\alpha})(z_N)]$$

$$t_2 = t_0[p_1 \leftarrow (\psi;\delta^{\alpha,\beta})(p_1),\ldots,p_M \leftarrow (\psi;\delta^{\alpha,\beta})(p_M),\\
z_1 \leftarrow \theta_1(\beta_j),\ldots,z_N \leftarrow \theta_N(\beta_j)].$$

Because $\varphi_l(\alpha_i) \Rightarrow_j (\psi;\delta^{\alpha,\beta})(p_l)$, we conclude by Lemma B.2.3 that there is some $u_l$ such that $\varphi_l(\beta_i) \Rightarrow_j u_l$ and $(\psi;\delta^{\alpha,\beta})(p_l) \Rightarrow_i u_l$ for all $1 \le l \le M$. Similarly, there is some $v_k$ such that $\theta_k(\beta_j) \Rightarrow_i v_k$ and $(\psi;\delta^{\beta,\alpha})(z_k) \Rightarrow_j v_k$ for all $1 \le k \le N$.

Finally, let $t' = t_0[p_1 \leftarrow u_1,\ldots,p_M \leftarrow u_M, z_1 \leftarrow v_1,\ldots,z_N \leftarrow v_N]$. Then by Exercise B.2.1, we have $t_1 \Rightarrow_j t'$ and $t_2 \Rightarrow_i t'$. □

**Theorem 5.6.4** A $\Sigma$-TRS $A$ is Church-Rosser if it is orthogonal and lapse free.

**Proof:** To prove that $\Rightarrow_i$ and $\Rightarrow_j$ commute for $1 \le i,j \le N$, by Lemma B.2.2 it suffices to prove that $\Rightarrow_i$ and $\Rightarrow_j$ commute.

First, we prove by induction on the length of rewriting with $\Rightarrow_j^*$ that whenever $t \Rightarrow_i t_1$ and $t \Rightarrow_j^* t_2$ there exists a $t'$ such that $t_1 \Rightarrow_j^* t'$ and $t_2 \Rightarrow_i t'$. If the length of the rewrite sequence $t \Rightarrow_j^* t_2$ is zero, then let $t' = t_1$. If it is more than zero, let $t'_2$ be a $\Sigma$-term such that $t \Rightarrow_j^* t'_2 \Rightarrow_j t_2$. By the induction hypothesis, there exists $t''$ such that $t_1 \Rightarrow_j^* t''$ and $t'_2 \Rightarrow_i t''$. Now Lemma B.2.4 gives us $t'$ such that $t'' \Rightarrow_j t'$ and $t_2 \Rightarrow_i t'$. Therefore $t_1 \Rightarrow_j^* t'$ and $t_2 \Rightarrow_i t'$.

Now we prove by induction on the length of rewriting with $\Rightarrow_i^*$ that whenever $t \Rightarrow_i^* t_1$ and $t \Rightarrow_j^* t_2$ there exists $t'$ such that $t_1 \Rightarrow_j^* t'$ and $t_2 \Rightarrow_i^* t'$; that is, $\Rightarrow_i$ and $\Rightarrow_i$ commute. If the length of the rewrite sequence is zero, let $t' = t_2$, and otherwise let $t'_1$ be a $\Sigma$-term such that $t \Rightarrow_i^* t'_1 \Rightarrow_i t_1$. By the induction hypothesis, there exists a term $t''$ such that $t'_1 \Rightarrow_j^* t''$ and $t_2 \Rightarrow_i^* t''$. By the induction above there exists $t'$ such that $t_1 \Rightarrow_j^* t'$ and $t'' \Rightarrow_i t'$.

It now follows that $t_1 \Rightarrow_j^* t'$ and $t_2 \Rightarrow_i^* t'$. Therefore $\Rightarrow_i$ and $\Rightarrow_j$ commute, and so the Hindley-Rosen Lemma (Proposition 5.7.5) gives us that $A$ is Church-Rosser. □

## B.3 Rewriting Modulo Equations

**Theorem 7.7.20** Let $(\Sigma, A, B)$ be a CMTRS with $\Sigma$ non-void and let $(\Sigma, A', B)$ be as ground terminating sub-CMTRS of $\Sigma, A, B$. let $P$ a poset and let $N = A - B$. If there is a $\rho : T_{\Sigma,B} \to P$ such that

(1) each rule in $B$ is weak $\rho$-monotone,

(2) each rule in $N$ is strict $\rho$-monotone,

(3) each operation in $\Sigma$ is strict $\rho$-monotone, and

(4) $P$ is Noetherian, or at least, for each $t \in (T_{\Sigma,B}$ there is some Noetherian poset $P_s^t \subseteq P_s$ such that $t \stackrel{*}{\Rightarrow}_{[A/B]} t'$ implies $\rho(t') \in P_s^t$,

then $(\Sigma, A, B)$ is ground terminating.

**Proof:** ......See proof of Theorem 5.8.20, page 136...... □

## B.4 First-Order Logic

This section restates and proves Proposition 8.3.21, which does most of the work for proving the Subsitution Theorem (Theorem 8.3.22).

**Proposition 8.3.21** If $\theta$ is capture free for $P$, then for any model $M$, $[\![\theta(P)]\!]_M = [\![\theta]\!]_M^{-1}([\![P]\!]_M)$.

**Proof:** We first show by structural induction over $\Omega$ that the required equality holds for every substitution $\tau$ that is capture free for $P$, with $\tau_{Free(P)}$ the identity. The reader is left to check the base cases (where $P$ is a generator in $G_X$ or *true*) and the inductive steps for negation and conjunction. Now suppose $P = (\forall x)Q$. The assertion $a \in [\![\tau((\forall x)Q)]\!]$ is equivalent to $b \in [\![\tau_x(Q)]\!]$ for each $b : X \to M$ with $b(y) = a(y)$ for $y \neq x$, that is, $\tau_x;\overline{b} \in [\![Q]\!]$, because $(\tau_x)_{Free(Q)}$ is the identity, $\tau_x$ is capture free for $Q$, plus the induction hypothesis. Similarly, $a \in [\![\tau]\!]^{-1}([\![(\forall x)Q]\!])$ is equivalent to $\tau;\overline{a} \in [\![(\forall x)Q]\!]$, that is, $b' \in [\![Q]\!]$ for each $b' : X \to M$ with $b'(y) = \overline{a}(\tau(y))$ for $y \neq x$.

Suppose $a \in [\![\tau((\forall x)Q)]\!]$ and let $b' : X \to M$ such that $b'(y) = \overline{a}(\tau(y))$ for $y \neq x$. Define $b : X \to M$ by $b(x) = b'(x)$ and $b(y) = a(y)$ for $y \neq x$; then $\tau_x;\overline{b} \in [\![Q]\!]$. We now claim $b' = \tau_x;\overline{b}$. Indeed, $(\tau_x;\overline{b})(x) = b(x) = b'(x)$, and if $y \neq x$ then $(\tau_x;b)(y) = \overline{b}(\tau_x(y)) = \overline{b}(\tau(y))$. But $x \notin Var(\tau(y))$, because if $y \in Free(P)$ then $x \notin Var(\tau(y))$ because $\tau$ is capture free for $P$, and if $y \notin Free(P)$ then $\tau(y) = y$. Then $\overline{b}(\tau(y)) = a(\tau(y)) = b'(y)$, that is, $b' = \tau_x;\overline{b}$. Therefore $b' \in [\![Q]\!]$, that is, $\tau;\overline{a} \in [\![(\forall x)Q]\!]$.

Conversely, suppose $\tau;\overline{a} \in [\![(\forall x)Q]\!]$ and let $b : X \to M$ such that $b(y) = a(y)$ for $y \neq x$ and let $b'$ be $\tau_x;\overline{b}$. Then $b'(y) = \overline{a}(\tau(y))$ (as above). Therefore $\tau_x;\overline{b} \in [\![Q]\!]$, that is, $a \in [\![\tau((\forall x)Q)]\!]$.

Now let $\theta$ be any substitution and let $\tau$ be the substitution $\theta_{X-Free(P)}$. Then $\tau$ is capture free for $P$, and $\tau_{Free(P)}$ is the identity; therefore $[\![\tau(P)]\!] = [\![\tau]\!]^{-1}([\![P]\!])$. By 5. of Exercise 8.3.14, $\theta(P) = \tau(P)$; therefore it suffices to prove $\theta;\overline{a} \in [\![P]\!]$ iff $\tau;\overline{a} \in [\![P]\!]$ for each $a : X \to M$. By Proposition 8.3.3, it is enough to show that $(\theta;\overline{a})(y) = (\tau;\overline{a})(y)$ for $y \in Free(P)$, which is true because $\theta(y) = \tau(y)$ for $y \in Free(P)$, by construction of $\tau$. $\square$

## B.5 Order-Sorted Algebra

This section provides the omitted proofs for results on order-sorted algebra in Chapter 10.

**Theorem 10.2.8** (*Initiality*) If $\Sigma$ is regular and if $M$ is any $\Sigma$-algebra, then there is one and only one $\Sigma$-homomorphism from $\mathcal{T}_\Sigma$ to $M$.

**Proof:** In this proof we write $\mathcal{T}$ for $\mathcal{T}_\Sigma$. Let $M$ be an arbitrary order-sorted $\Sigma$-algebra; then we must show that there is a unique order-sorted $\Sigma$-homomorphism $h : \mathcal{T} \to M$. We will (1) construct $h$, then (2) show it is an order-sorted $\Sigma$-homomorphism, and finally (3) show it is unique.

(1) We construct $h$ by induction on the depth of terms in $\mathcal{T}$. There are two cases:

(1a) If $t \in \mathcal{T}$ has depth 0, then $t = \sigma$ for some constant $\sigma$ in $\Sigma$. By regularity, $\sigma$ has a least sort $s$. Then for any $s' \geq s$ we define

$$h_{s'}(\sigma) = M_\sigma^{[],s}$$

(1b) If $t = \sigma(t_1 \ldots t_n) \in \mathcal{T}$ has depth $n + 1$, then by regularity there are least $w$ and $s$ with $\sigma \in \Sigma_{w,s}$ where $w = s_1 \ldots s_n \neq []$ and $LS(t_i) \leq s_i$ for $i = 1, \ldots, n$. Then for any $s' \geq s$ we define

$$h_{s'}(t) = M_\sigma^{w,s}(h_{s_1}(t_1), \ldots, h_{s_n}(t_n)),$$

noting that $h_{s_1}(t_1), \ldots, h_{s_n}(t_n)$ are already defined.

(2) We now show that $h$ is an order-sorted $\Sigma$-homomorphism. By construction $h$ satisfies the restriction condition[E48] of Definition 10.1.5. To see that it also satisfies the homomorphism condition of Definition 10.1.5, we again consider two cases:

(2a) $\sigma \in \Sigma_{[],s}$ is a constant. By regularity and monotonicity, $s$ is the least sort of $\sigma$, and we have already defined $h_s(\sigma) = M_\sigma^{[],s}$ as needed.

(2b) We now consider a term $t$ of depth greater than 0, and let $\sigma \in \Sigma_{w',s'}$ with $w' = s_1' \ldots s_n' \neq []$ be such that $t = \sigma(t_1 \ldots t_n) = \mathcal{T}_\sigma^{w',s'}(t_1, \ldots, t_n)$. By regularity and Proposition 10.2.7 there are least $w = s_1 \ldots s_n$ and $s = LS(t)$ such that $t = \sigma(t_1 \ldots t_n) = \mathcal{T}_\sigma^{w,s}(t_1, \ldots, t_n)$. Then $w \leq w'$ and $s \leq s'$ so that (2) of Definition 10.1.3 gives $M_\sigma^{w',s'} = M_\sigma^{w,s}$ on $M^w$. Thus, using the already established fact that $h$ satisfies the restriction condition, we have

$$\begin{aligned} h_{s'}(\sigma(t_1 \ldots t_n)) &= M_\sigma^{w,s}(h_{s_1}(t_1), \ldots, h_{s_n}(t_n)) \\ &= M_\sigma^{w',s'}(h_{s_1'}(t_1), \ldots, h_{s_n'}(t_n)). \end{aligned}$$

(3) Finally, we show the uniqueness of $h$. In fact, we will show that if $h' : \mathcal{T} \to M$ is an order-sorted $\Sigma$-homomorphism, then $h = h'$, by induction on the depth of terms. For depth 0 consider $\sigma \in \Sigma_{[],s}$. Then $s$ is the least sort of $\sigma$, and for any $s \geq s'$, we must have

$$h'_{s'}(\sigma) = h'_s(\sigma) = M_\sigma^{[],s} = h_s(\sigma) = h_{s'}(\sigma),$$

as desired. Now assume the result for depth $\leq n$, and consider a term $t = \sigma(t_1 \ldots t_n) = \mathcal{T}_\sigma^{w',s'}(t_1, \ldots, t_n)$ of depth $n + 1$ with $\sigma \in \Sigma_{w',s'}$ and $w' = s_1' \ldots s_n'$. As in (2b), there are least $w = s_1 \ldots s_n$ and $s = LS(t)$ such that $t = \sigma(t_1, \ldots, t_n) = \mathcal{T}_\sigma^{w,s}(t_1, \ldots, t_n)$ and $M_\sigma^{w',s'} = M_\sigma^{w,s}$ on $M^w$. Then

$$\begin{aligned} h'_{s'}(t) &= M_\sigma^{w',s'}(h'_{s_1'}(t_1), \ldots, h'_{s_n'}(t_n)) \\ &= M_\sigma^{w',s'}(h_{s_1'}(t_1), \ldots, h_{s_n'}(t_n)) \quad \text{(by the induction hypothesis)} \\ &= M_\sigma^{w,s}(h_{s_1}(t_1), \ldots, h_{s_n}(t_n)) \\ &= h_{s'}(t) \end{aligned}$$

as needed. □

**Theorem 10.2.9** (*Freeness*) If $(S, \leq, \Sigma)$ is regular, then $\mathcal{T}_\Sigma(X)$ is a **free** $\Sigma$-algebra on $X$, in the sense that for each $\Sigma$-algebra $M$ and each assignment $a : X \to M$, there is a unique $\Sigma$-homomorphism $\overline{a} : \mathcal{T}_\Sigma(X) \to M$ such that $\overline{a}(x) = a(x)$ for all $x$ in $X$.

**Proof:** The $\Sigma$-algebras $M$ with an assignment $a : X \to M$ are in bijective correspondence with $\Sigma(X)$-algebras $M$. Now the initiality of $\mathcal{T}_\Sigma(X)$ among all $\Sigma(X)$-algebras $A$ (Theorem 10.2.8) gives the desired result. □

**Theorem 10.3.2** (*Completeness*) Given a coherent order-sorted signature $\Sigma$, given $t, t'$ in $\mathcal{T}_\Sigma(X)$, and given a set $A$ of conditional $\Sigma$-equations, then the following assertions are equivalent:

> (C1) $(\forall X) \ t = t'$ is derivable from $A$ using rules (1)–(4) and (5C).
>
> (C2) $(\forall X) \ t = t'$ is satisfied by every order-sorted $\Sigma$-algebra that satisfies $A$.

When all equations in $A$ are unconditional, the same holds replacing rule (5C) by rule (5).

**Proof:** We leave the reader to check *soundness*, i.e., that (C1) implies (C2); this follows as usual by induction from the soundness of each rule of deduction separately. Here we show *completeness*, i.e., that (C2) implies (C1). The structure of this proof is as follows: We are given a $\Sigma$-equation $e = (\forall X) \ t = t'$ that is satisfied by every $\Sigma$-algebra that satisfies $A$, and we wish to show that $e$ is derivable from $A$; to this end, we construct a particular $\Sigma$-algebra $\mathcal{M}$ such that if $\mathcal{M}$ satisfies $e$ then $e$ is derivable from $A$; then we show that $\mathcal{M}$ satisfies $A$.

First, we show that the following property of terms $t, t' \in \mathcal{T}_\Sigma(X)_s$ for some sort $s$, defines an order-sorted $\Sigma$-congruence on $\mathcal{T}_\Sigma(X)$:

> (D) $(\forall X) \ t = t'$ is derivable from $A$ using rules (1–4) plus (5C).

Let us denote this relation $\equiv$. Then rules (1–3) say that $\equiv$ is an equivalence relation on $\mathcal{T}_\Sigma(X)_s$ for each sort $s$. By applying rule (4) to terms $t$ of the form $\sigma(x_1, \ldots, x_n)$ for $\sigma \in \Sigma$, we see that $\equiv$ is a many-sorted $\Sigma$-congruence. Finally, $\equiv$ is also an order-sorted $\Sigma$-congruence, because property (D) does not depend upon $s$.

Now we can form the order-sorted quotient of $\mathcal{T}_\Sigma(X)$ by $\equiv$, which we denote by $\mathcal{T}_{\Sigma,A}(X)$, or within this proof, just $\mathcal{M}$. Then by the construction of $\mathcal{M}$, for each $t, t' \in \mathcal{T}_\Sigma(X)$ we have

> (∗) $[t] = [t']$ in $\mathcal{M}$ iff (D) holds,

where $[t]$ denotes the $\equiv$-equivalence class of $t$.

We next show the key property of $\mathcal{M}$, that

> (∗∗) $(\forall X) \ t = t'$ satisfied in $\mathcal{M}$ implies that (D) holds.

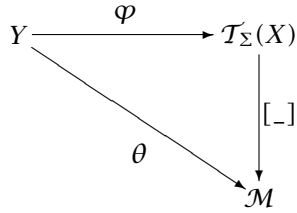

Figure B.1: Factorization of $\theta$

Since the equation $(\forall X)\, t = t'$ is satisfied in $\mathcal{M}$, we can use the inclusion $i_X : X \to \mathcal{M}$ sending $x$ to $[x]$ as an $S$-sorted assignment to see that $[t] = [t']$ in $\mathcal{M}$; then (D) holds by (∗).

We now prove that $\mathcal{M}$ satisfies $A$. Let $(\forall Y)\, t = t'$ if $C$ be a conditional equation in $A$, and let $\theta : Y \to \mathcal{M}$ be an $S$-sorted assignment such that $\theta(u) = \theta(v)$ for each condition $u = v$ in $C$. Then for each $s \in S$ and each $y \in Y_s$, we can choose a representative $t_y \in \mathcal{T}_\Sigma(X)_s$ such that $\theta(y) = [t_y]$ in $\mathcal{M}$. Now let $\varphi : Y \to \mathcal{T}_\Sigma(X)$ be the substitution sending $y$ to $t_y$. Then $\theta(y) = [\varphi(y)]$ for each $y \in Y$, and therefore $\theta(t) = [\varphi(t)]$ in $\mathcal{M}$ for any $t \in \mathcal{T}_\Sigma(Y)$, by the freeness of $\mathcal{T}_\Sigma(Y)$ over $Y$. See Figure B.1.

Therefore, $[\varphi(u)] = [\varphi(v)]$ holds in $\mathcal{M}$, and by the property (∗), the equation $(\forall X)\, \varphi(u) = \varphi(v)$ is derivable from $A$ using (1–4) plus (5C) for each $u = v$ in $C$. Therefore by rule (5C), the equation $(\forall X)\, \varphi(t) = \varphi(t')$ is derivable from $A$, and hence by (∗), $\theta(t) = \theta(t')$ holds in $\mathcal{M}$, and thus the conditional equation $(\forall Y)\, t = t'$ if $C$ holds in $\mathcal{M}$.

Since an unconditional equation is just a conditional equation whose set $C$ of conditions is empty, when every equation in $A$ is unconditional we are reduced to the simplified special case of the above argument where only the rule (5) is needed. □

This result also gives completeness for ordinary MSA, and of course for unsorted algebra, as special cases. Now the initiality and freeness results:

**Theorem 10.4.11** (*Initiality*) If $\Sigma$ is coherent and $A$ is a set of (possibly conditional) $\Sigma$-equations, then $\mathcal{T}_{\Sigma,A}$ is an initial $(\Sigma, A)$-algebra, and $\mathcal{T}_{\Sigma,A}(X)$ is a free $(\Sigma, A)$-algebra on $X$, in the sense that for each $\Sigma$-algebra $M$ and each assignment $a : X \to M$, there is a unique $\Sigma$-homomorphism $\overline{a} : \mathcal{T}_{\Sigma,A}(X) \to M$ such that $\overline{a}(x) = a(x)$ for each $x$ in $X$.

**Proof:** [E49] First notice that the freeness of $\mathcal{T}_{\Sigma,A}(X)$ specializes to the initiality of $\mathcal{T}_{\Sigma,A}$ when $X = \emptyset$, so that it suffices to show the freeness of $\mathcal{T}_{\Sigma,A}(X)$. Let $M$ be an order-sorted algebra satisfying $A$, and let $a : X \to M$ be an assignment for $M$. Then we have to show that there is a unique order-sorted $\Sigma$-homomorphism $a^\& : \mathcal{T}_{\Sigma,A}(X) \to M$ extending $a$, i.e., such that $a^\&(q(x)) = a(x)$ for each $x \in X$, where $q$ denotes the quotient homomorphism $q : \mathcal{T}_\Sigma(X) \to \mathcal{T}_{\Sigma,A}(X)$. The existence of $a^\&$ follows from completeness (Theorem 10.3.2), because the fact that $M$ satisfies $A$ implies that $a^*(t) = a^*(t')$ for every equation $(\forall X)\, t = t'$ that is derivable from $A$ with the rules (1–4) plus (5C), and this implies that $\equiv\, \subseteq ker(a^*)$, and thus by the universal property of quotients (Proposition 10.4.10), there is a unique order-sorted homomorphism $a^\& : \mathcal{T}_{\Sigma,A}(X) \to A$ with $a^* = a^\& \circ q$.

The uniqueness of $a^\&$ now follows by combining the universal property of $\mathcal{T}_\Sigma(X)$ as a free order-sorted algebra on $X$ with the universal property of $q$ as a quotient, as follows: Let $h : \mathcal{T}_{\Sigma,A}(X) \to M$ be another order-sorted homomorphism such that $h(q(x)) = a(x)$ for each $x \in X$. Since $\mathcal{T}_\Sigma(X)$ is a free order-sorted algebra on $X$, we have $a^* = h \circ q$, and by the universal property of $q$ as a quotient we have $h = a^\&$ as desired. □

**Theorem 10.3.3** Given a coherent signature $\Sigma$ and a set $A$ of (possibly conditional) $\Sigma$-equations, then for any unconditional $\Sigma$-equation $e$,

$$A \vdash^C e \quad \text{iff} \quad A \vdash^{(1,3,\pm 6C)} e\,.$$

**Proof:** See page 8 of [82] [OSA1], and Theorem 4.9.1, page 82. □

The model-theoretic proof of Theorem 10.6.6 below uses *naturality* of the family $\psi_X$ of morphisms, which in particular implies commutativity of the following diagram for $X \subseteq X'$, where $\mu_{X,X'}$ is the unique $\Sigma^\otimes$-homomorphism induced by the composite map $X \hookrightarrow X' \to \mathcal{T}_{\Sigma^\otimes,A^\otimes}(X')$:

$$\begin{array}{ccc}
\mathcal{T}_{\Sigma,A}(X) & \xrightarrow{\psi_X} & \mathcal{T}_{\Sigma^\otimes,A^\otimes}(X) \\
\iota_{X,X'} \downarrow & & \downarrow \mu_{X,X'} \\
\mathcal{T}_{\Sigma,A}(X') & \xrightarrow[\psi_{X'}]{} & \mathcal{T}_{\Sigma^\otimes,A^\otimes}(X')
\end{array}$$

**Theorem 10.6.6** If $\Sigma$ is coherent and $(\Sigma, A)$ is faithful, then the extension $(\Sigma, A) \subseteq (\Sigma^\otimes, A^\otimes)$ is conservative.

**Proof:** We have to show that $\psi_X : \mathcal{T}_{\Sigma,A}(X) \to \mathcal{T}_{\Sigma^\otimes,A^\otimes}(X)$ is injective. By the above naturality diagram plus faithfulness, it suffices to show that $\psi_{X'} : \mathcal{T}_{\Sigma,A}(X') \to \mathcal{T}_{\Sigma^\otimes,A^\otimes}(X')$ is injective, where $X' \supseteq X$ is obtained from $X$ by adding a new variable symbol of sort $s$ for each sort $s$ with $X_s = \emptyset$. Now pick an arbitrary variable symbol $x_s^0 \in X_s$ for each $s \in S$. The key step is to make the $(\Sigma, A)$-algebra $\mathcal{T}_{\Sigma,A}(X')$ into a $(\Sigma^\otimes, A^\otimes)$-algebra by defining $r_{s',s} : \mathcal{T}_{\Sigma,A}(X')_{s'} \to \mathcal{T}_{\Sigma,A}(X')_s$ to be the function that sends $[t] \in \mathcal{T}_{\Sigma,A}(X')_s$ to $[t]$, and otherwise sends it to $x_s^0$. It is now easy to see that the retract equations are satisfied. Thus the freeness of $\mathcal{T}_{\Sigma^\otimes,A^\otimes}(X)$ implies that the natural inclusion $X' \to \mathcal{T}_{\Sigma,A}(X')$ induces a unique $\Sigma^\otimes$-homomorphism $q : \mathcal{T}_{\Sigma^\otimes,A^\otimes}(X') \to \mathcal{T}_{\Sigma,A}(X')$ such that $q \circ \psi_{X'}$ is the identity. Therefore $\psi_{X'}$ is injective. □

# C  Some Background on Relations

The first part of this appendix reviews some basic material that is assumed in the body of this text. The approach is oriented towards use in algebra and OBJ, and differs from traditional set-theoretic formalizations like that in Definition C.0.1. Section C.1 implements some of this material in OBJ.

**Definition C.0.1**  A **set-theoretic relation**, from a set $A$ to a set $B$, is a subset $R \subseteq A \times B$; we let $aRb$ mean that $\langle a, b \rangle \in R$. The **image** of $R \subseteq A \times B$ is the set $\{b \mid aRb \text{ for some } a \in A\}$, a subset of $B$, and the **coimage** of $R$ is $\{a \mid aRb \text{ for some } b \in B\}$, a subset of $A$. A **set-theoretic function** from $A$ to $B$ is a set-theoretic relation $f$ from $A$ to $B$ that satisfies the following two properties:

(1) for each $a \in A$ there is some $b \in B$ such that $\langle a, b \rangle \in f$, and

(2) if $\langle a, b \rangle, \langle a, b' \rangle \in f$ then $b = b'$.

When $f$ is a function, we usually write $f(a)$ for the unique $b$ such that $\langle a, b \rangle \in f$.  □

The above is not very satisfactory in some respects. For example, consider the case where $A \subseteq B$ and we want $f$ to be the *inclusion* function from $A$ to $B$. As a set-theoretic relation, this is $\{\langle a, a \rangle \mid a \in A\}$, which is *exactly* the same as the set-theoretic relation for the identity function on $A$ (for a specific example, let $A = \omega^+$ and $B = \omega^*$). But these two functions are *not* the same; although they have the same graph, image, and coimage, they have different *target* sets. Indeed, there are many proofs in this text that use inclusion functions, and that would fail if inclusion and identity functions were the same! Hence, the above formalization of the relation concept is not suitable for our purposes. Instead, we use the following:

**Definition C.0.2**  A **relation from** $A$ **to** $B$ is a triple $\langle A, R, B \rangle$, where $R$ is a set-theoretic relation from $A$ to $B$, called the **graph** of the relation. We write $aRb$ if $\langle a,b \rangle \in R$. If $B = A$, then $R$ is said to be a **relation on** $A$. $A$ and $B$ are called the **source** and **target** of $R$, respectively, or sometimes the **domain** and **codomain** of $R$, respectively, and we may write $R : A \rightarrow B$. We may also use the terms **image** and **coimage**, as defined in Definition C.0.1 for the graph $R$ of the relation. If the graph $R$ satisfies (1) and (2) of Definition C.0.1, then the relation is called a **function**, an **arrow**, or a **map** from $A$ to $B$. □

This differs from Definition C.0.1 in that source and target sets are explicitly given;[1] this allows us to distinguish inclusions from identities, but we may still abbreviate $R : A \rightarrow B$ by just $R$.

**Definition C.0.3** Given a relation $R : A \rightarrow B$ and $A' \subseteq A$, then $\{b \mid \langle a,b \rangle \in R$ for some $a \in A'\}$ is called the **image** of $A'$ **under** $R$, written $R(A')$. Also, given $B' \subseteq B$, then $\{a \mid \langle a,b \rangle \in R$ for some $b \in B'\}$ is called the **inverse image** of $B'$ **under** $R$, written $R^{-1}(B')$. □

The following equivalent formalization of relations is more suitable for mechanization in OBJ (the equivalence is discussed in Chapter 8):

**Definition C.0.4** A **relation from** $A$ **to** $B$ is an arrow $A \times B \rightarrow \{true, false\}$. Its **graph** is the set $\{\langle a,b \rangle \mid R(a,b) = true\}$, and $A, B$ are called its **source** and **target** sets, respectively. □

Here are some further concepts associated with functions that are used in this book:

**Definition C.0.5** A function $f : A \rightarrow B$ is **injective** iff $f(a) = f(a')$ implies $a = a'$ for all $a, a' \in A$, is **surjective** iff for all $b \in B$ there is some $a \in A$ such that $f(a) = b$, and is **bijective** iff it is both injective and surjective. □

**Exercise C.0.1** Given a function $f : A \rightarrow B$, show that $f^{-1}(B) = A$. □

**Exercise C.0.2** Show that a function $f : A \rightarrow B$ is surjective iff its image is $B$. □

**Definition C.0.6** Given a function $f : A \rightarrow B$ and $A' \subseteq A$, then the **restriction** of $f$ to $A'$ is the function $f|_{A'} : A' \rightarrow B$ with graph $\{\langle a,b \rangle \mid a \in A'$ and $\langle a,b \rangle \in f\}$. Also, given $B' \subseteq B$, the **corestriction** of $f$ to $B'$ is the function $f^{-1}(B') \rightarrow B'$ with graph $\{\langle a,b \rangle \mid b \in B'$ and $\langle a,b \rangle \in f\}$. □

We now consider a number of different kinds of relation on a set.

---

[1] Although we can always recover the source of a set-theoretic function because of condition (1).

**Definition C.0.7**  A relation $R$ on a set $A$ is:

- **reflexive** iff $aRa$ for all $a \in A$,
- **symmetric** iff $aRa'$ implies $a'Ra$ for all $a, a' \in A$,
- **anti-reflexive** iff $aRa$ for no $a \in A$,
- **anti-symmetric** iff $aRa'$ and $a'Ra$ imply $a = a'$ for all $a, a' \in A$,
- **transitive** iff $aRa'$ and $a'Ra''$ imply $aRa''$ for all $a, a', a'' \in A$,
- a **partial ordering** iff it is reflexive, anti-symmetric, and transitive,
- a **quasi ordering** iff it is anti-reflexive and transitive, and
- an **equivalence relation** iff it is reflexive, symmetric, and transitive.

It is customary to let $\geq$ and $>$ denote partial and quasi orderings, respectively, and to call a set with a partial ordering a **poset**, in which case the underlying set $A$ may be called the **carrier** of the poset.  □

**Example C.0.8**  Let $A$ be the set of all people. Then the relation "ancestor-of" is transitive and anti-reflexive, and is thus a quasi ordering; but it is not symmetric or reflexive. The "cousin-of" relation is symmetric, but not transitive or reflexive, although the "cousin-of-or-equal" relation is reflexive, symmetric and transitive, and thus is an equivalence relation. The "child-of" relation is anti-reflexive, but has none of the other properties in Definition C.0.7.  □

If $X$ is a set, then $\mathcal{P}(X)$ denotes the set of all subsets of $X$, called the **power set** of $X$.

**Exercise C.0.3**  Let $A = \mathcal{P}(\omega)$. Then what properties does the "subset-of" relation on $A$ have? What about the "proper-subset-of" relation? If $A = \mathcal{P}(X)$, do the answers to these questions vary with $X$? If so, how? If not, why?  □

There is an important bijective correspondence between the partial and the quasi orderings on a set. This expressed precisely in the following:

**Proposition C.0.9**  Given a set $A$ and a relation $R$ on $A$, define $R^Q$ by $aR^Qa'$ iff $aRa'$ and $a \neq a'$, and define $R^P$ by $aR^Pa'$ iff $aRa'$ or $a = a'$. Then $R^Q$ is a quasi order if $R$ is a partial order, and $R^P$ is a partial order if $R$ is a quasi order. Moreover, if $R$ is a quasi order, then $(R^P)^Q = R$, and if $R$ is a partial order then $(R^Q)^P = R$.

**Proof:**  The reader can check the following assertions, which complete the proof: if $R$ is transitive and anti-symmetric, then $R^Q$ is transitive; if $R$ is transitive then so is $R^P$; if $R$ is any relation, then $R^Q$ is anti-reflexive, and $R^P$ is reflexive and anti-symmetric; if $R$ is a partial order, then $aRa'$ iff $aR^Q a'$ or $a = a'$ iff $a(R^Q)^P a'$; and if $R$ is a quasi order, then $aRa'$ iff $aR^P a'$ and $a \neq a'$ iff $a(R^P)^Q a'$. □

We can also have operations and relations on relations. For example:

**Definition C.0.10** If $R$ and $R'$ are relations on $A$, then $R \subseteq R'$ means that $aRa'$ implies $aR'a'$ for all $a, a' \in A$. Also, we define the **union** of $R$ and $R'$, denoted $R \cup R'$, by $a(R \cup R')a'$ iff $aRa'$ or $aR'a'$, and their **intersection**, denoted $R \cap R'$, by $a(R \cap R')a'$ iff $aRa'$ and $aR'a'$. □

**Proposition C.0.11** Every relation $R$ on $A$ is contained in a least transitive relation on $A$, denoted $R^+$ and called the **transitive closure** of $R$.

**Proof:** We define $R^+$ as follows: $aR^+ a'$ iff there exists a finite (possibly empty) list $a_1, a_2, \ldots, a_n$ of elements of $A$ such that $aRa_1$ and $a_1 R a_2$ and $\ldots a_n R a'$. Then $R \subseteq R^+$, and it is not hard to check that $R^+$ is transitive.

Now suppose that $R \subseteq S$ and that $S$ is transitive. If $aR^+ a'$, then there exist $a_1, \ldots, a_n$ such that $aRa_1$ and $a_1 R a_2$ and $\ldots a_n R a'$. Thus $aSa_1$ and $a_1 S a_2$ and $\ldots a_n S a'$ (because $R \subseteq S$), and so transitivity of $S$ gives $aSa'$. Therefore $R^+ \subseteq S$, and hence $R^+$ is minimal. □

**Example C.0.12** If $A$ is the set of all people and $R$ is the "parent-of" relation, then $R^+$ is the "ancestor-of" relation. □

**Proposition C.0.13** Every relation $R$ on a set $A$ is contained in a least transitive and reflexive relation on $A$, denoted $R^*$ and called the **transitive, reflexive closure** of $R$.

**Proof:** Let us define $aR^* a'$ iff $a = a'$ or $aR^+ a'$. Then $R^*$ is transitive, reflexive, and contains $R$. If $S$ is another such relation, then $R^+ \subseteq S$ by Proposition C.0.11 and hence $R^* \subseteq S$ (by reflexivity). □

**Definition C.0.14** Given a relation $R$ on $A$, its **converse** is denoted $R^\smile$, and is defined by $aR^\smile a'$ iff $a'Ra$. □

**Example C.0.15** If $A$ is the set of people and $R$ is the "parent-of" relation, then $R^\smile$ is the "child-of" relation. □

**Proposition C.0.16** Every relation $R$ on $A$ is contained in a least symmetric relation on $A$, namely $R \cup R^\smile$, called the **symmetric closure** of $R$ and denoted $R^\pm$. □

**Proposition C.0.17** Every relation $R$ on $A$ is contained in a least equivalence relation on $A$, namely $(R^\pm)^*$, called the **equivalence relation generated by** $R$, and denoted $R^=$. □

**Exercise C.0.4** Prove Propositions C.0.16 and C.0.17. □

**Exercise C.0.5** Show that a function $f : A \to B$ is bijective iff the converse of its graph is a function $f : B \to A$. □

**Definition C.0.18** If $\equiv$ is an equivalence relation on a set $A$, we let $[a]$ denote the $\equiv$-**equivalence class** of $a \in A$, defined to be $[a] = \{a' \in A \mid a' \equiv a\}$, and we define the **quotient of** $A$ **by** $\equiv$, denoted $A/\equiv$, to be $\{[a] \mid a \in A\}$. □

**Exercise C.0.6** If $\equiv$ is an equivalence relation, and if $[a]$ and $[b]$ are two distinct $\equiv$-equivalence classes, then show that $[a] \cap [b] = \emptyset$. Also, show that $\bigcup \{C \mid C \in A/\equiv\} = A$. □

**Example C.0.19** If $A$ is the set of all people, alive now or in the past, and if $R$ is again the "parent-of" relation, then the hypothesis that all people are descended from Adam and Eve implies that there is just one equivalence class under the relation $R^\equiv$. On the other hand, if there is more than one equivalence class, there may be aliens among us, another species, or some other non-interbreeding population. □

Everything in this section extends to $S$-sorted relations and functions in the style of Section 2.2, using the following set-theoretic representation for $S$-sorted sets: Let $S$ be a set, and let *Set* be some set of sets that includes all sets that are candidates for use in indexed sets within the current context; then an $S$-sorted function $A$ is a function $A : S \to \mathit{Set}$.

## C.1 OBJ Theories for Relations

This section gives OBJ3 code for some of the concepts discussed above. It is intended to be read later than the above material, after the relevant concepts from OBJ have been studied.

To get started, here is OBJ3 code for the theory of relations:

```
th REL is sort Elt .
  op _R_ : Elt Elt -> Bool .
endth
```

Notice that nothing at all is assumed about $R$. We will enrich this theory in various ways in the following development.

**Example C.1.1** The theory of partial ordering relations is given below; a set with a partial ordering is called a "poset."

```
th POSET is sort Elt .
  op _=>_ : Elt Elt -> Bool .
```

```
      vars A B C : Elt .
      eq A => A = true .
      cq A = B if A => B and B => A .
      cq A => C = true if A => B and B => C .
    endth
```

Any initial algebra of the following specification of the natural numbers with their natural ordering will satisfy the above theory

```
    th NAT is sort Nat .
      op 0 : -> Nat .
      op s : Nat -> Nat .
      op _=>_ : Nat Nat -> Bool .
      vars A B : Nat .
      eq A => A = true .
      eq s(A) => 0 = true .
      eq s(A) => s(B) = A => B .
    endth                                                    □
```

**Exercise C.1.1** Write an OBJ3 theory for quasi orders and show how to get a partial order from a quasi order, and *vice versa*. Give three examples of a quasi order. □

**Example C.1.2** (⋆) The transitive closure of a relation R is specified by the following, in which the imported module REL is assumed to define R:

```
    obj TRCL is
      pr REL .
      op _R+_ : Elt Elt -> Bool .
      vars A B C : Elt .
      cq A R+ B = true if A R B .
      cq A R+ C = true if A R B and B R+ C .
    endo
```

However, it is more elegant to define transitive closure as a *parameterized* theory, as follows (Chapter 11 discusses this concept):

```
    obj TRCL[R :: REL] is
      op _R+_ : Elt Elt -> Bool .
      vars A B C : Elt .
      cq A R+ B = true if A R B .
      cq A R+ C = true if A R B and B R+ C .
    endo
```

There are some peculiar points about the OBJ3 specifications above. First, the second conditional equation is not a rewrite rule, because the variable B occurs in the condition but not in the leftside. Therefore OBJ3 will not accept it in an object module, although it will accept it in a theory module. However, initial semantics is needed, because the transitive closure is supposed to be the least transitive relation containing the given one. This means that the above code will not actually run in OBJ3. This is because a variable that is not in the leftside acts as if it were existentially quantified, and OBJ3 cannot handle existential quantifiers. However, the specifications make perfect semantic sense, and in fact would run in Eqlog [79, 42].

Note that under initial semantics, when a R+ b does not equal true, it also does not equal false, but rather equals the term a R+ b, because that term is itself reduced. This means that if we want a version of transitive closure that does equal false when it is not true, then we should replace occurrences of the expression a R+ b by the expression a R+ b == true. □

**Exercise C.1.2** (⋆) Write an OBJ3 parameterized theory specifying the transitive, reflexive closure of a relation. □

**Example C.1.3** Here is an OBJ3 theory for equivalence relations:

```
th EQV is sort Elt .
  op _≡_ : Elt Elt -> Bool .
  vars A B C : Elt .
  eq A ≡ A = true .
  eq A ≡ B = B ≡ A .
  cq A ≡ C = true if A ≡ B and B ≡ C .
endth
```
□

**Exercise C.1.3** (⋆) Write an OBJ3 parameterized theory specifying the transitive, reflexive, symmetric closure of a relation; this is called its **equivalence closure**. □

# D  Social Implications

Practical interest in verification technology is particularly acute for so called "critical systems," which have the property that incorrect operation may result in loss of human life, compromise of national security, massive loss of property, etc. Typical examples are heart pacemakers, flight control systems, automobile brake controllers, encryption systems, nuclear power plants, and electronic fund transfer systems. In this context, it is especially important to understand the limitations of verification, both those limitations that are inherent in the nature of verification, and those that are due to the current state of the art. Unfortunately, there is a temptation to play down, or even cover up, these limitations, due to the lure of fame and fortune. When verification is just an academic exercise, this does little harm; but there is cause for serious concern when manufacturers make advertising claims about the reliability of a critical system (or component thereof) based on its having been verified. As Dr. Avra Cohn [31] says,

> The use of the word 'verified' must *under no circumstances* be allowed to confer a false sense of security.

This is because it could lead to unintentionally and unnecessarily taking severe risks. Indeed, one might well argue that to knowingly make false or misleading claims about the reliability of a critical system should be a criminal offense. It is certainly a grave moral offense.

We should realize that nothing in the real world can have the certainty of mathematical truth. Although we might like to think that '$8 \times 7 = 56$' is always and incontrovertibly true in some mathematical heaven, an actual human being can sometimes misremember the multiplication table, and an actual machine can sometimes drop a bit, break a connector, or burn out a power supply. The most that can truthfully be asserted of a "verified chip" is that certain kinds of design errors have been made far less likely. Of course, this is very much worth pursuing, but there remains a long chain of assumptions that must be satisfied in order that an actual physical instance of a chip whose logic has been verified will operate as intended *in situ*, including the following: the chip must be correctly fabricated; it must be correctly installed; it must be fed correct data and given correct power; the electronic circuits that realize its logic must be correctly designed, and used only within their design limits; the analog circuits that support communication must operate correctly; there must not be excessive electromagnetic radiation around the chip; etc., etc. In addition, human factors are often involved; for example, the user should not override warning signals, and must correctly interpret the output.

However, I wish to concentrate on certain logical issues that are involved. A major point, emphasized by Cohn [31], is that verification is a relationship between two models, that is, between two mathematical abstractions, one of the chip, and the other of the designers' intentions. However, such models can never capture either the totality of any particular chip, or even of the designers intentions for that chip. This is due to a number of factors, including errors made in constructing these necessarily complex abstractions, and deliberately ignoring certain factors (which is of course the very nature of abstraction), such as fluctuations in power levels, aging of physical components, overheating, etc. Moreover, since the languages in which designs and specifications are usually expressed are relatively informal, there must be a translation into some formal language, and errors may be introduced when either the designers or the verifiers misunderstand these informal languages. In addition, the theorem prover must correctly implement some logical system, and the formalism for representing the chip in logic must be correct with respect to that system. Unless a theorem prover is rigorously based on a precise and well-understood logical system, there is little basis for confidence in its "proofs." For example, it is seductive but dangerous to "throw together" several different logical systems, since the combination may fail to have any obvious notions of model and satisfaction, even though the components do have them. Even a theorem prover with a sound logical basis is likely to have some bugs in its code, because it is after all a complex real system itself.

Moreover, it is always possible that the assumptions about how formal sentences represent physical devices are flawed in some subtle ways, for example, relating to signal strength, and it is all too easy to use a theorem prover incorrectly, for example, to give it erroneous input, or to interpret its output incorrectly. Finally, we must note that the current state of the art is not adequate to support the verification of really large or complex systems, although recent advances have been both rapid and significant, and the future looks promising.

# Editors' Notes

## Notes for Chapter 1

E1. [Page 8] The sentence about no new theorems being proved by automated theorem provers is not accurate, because some new theorems in algebra were first proved by automatic first-order provers, like the Robbins conjecture, for example.

E2. [Page 10] The sentence about OBJ semantics can be confusing, because the expression "only because" might suggest that it happens "almost by chance."

E3. [Page 13] The author used to reset counters for each chapter, but the editors have decided to reset them for each section in order to improve readability, since some chapters are quite long.

## Notes for Chapter 3

E4. [Page 41] After defining the concept of satisfaction in Definition 3.3.7, it would be convenient to add a discussion about algebras with empty sorts ($M_s = \emptyset$ for some sort $s \in S$) and the satisfaction of equations for such algebras: $M \models (\forall X)\, t = t'$ if there is no assignment $a: X \to M$ due to $M_s = \emptyset$. See also Section 4.3.2.

E5. [Page 43] The proof of Theorem 3.3.11 is a bit too fast and it is obscured by using the same notation $M$ for the algebra and its underlying $S$-sorted set (see Definition 2.5.1).

E6. [Page 44] After defining satisfaction of conditional equations in Definition 3.4.1, it would be helpful to comment that $M \models (\forall X)\, t = t'$ `if` $C$ when $M$ does not satisfy the conditions $C$.

E7. [Page 52] In the proof of Proposition 3.7.2, the second paragraph proves the contrapositive of first paragraph instead of its converse. Instead, the converse implication does not hold. Consider, for example, the signature $(\{s, s_1, s_2\}, \{f : s \to s_1, f : s \to s_2\})$. There are no overloaded terms because there are no terms, but the signature is not regular.

E8. [Page 53] In Definition 3.7.5 it seems that $M_\Sigma$ denotes the annotated form when there is confusion and the non-annotated form when there is no confusion (as stated at the end, "As with terms in $\overline{T}_\Sigma$, we will usually omit the sort annotation unless it is necessary"). However, it also seems that the author's intention is to denote by $M_\Sigma$ the non-annotated form. Otherwise, Definition 3.7.7 and Exercise 3.7.2 don't make sense. For example, $aaa$ (Exercise 3.7.2) doesn't have five distinct parses because it should be written either $(a(a\, a.A).A).A$ or $(a(a.A\, a).A).A$ or $((a\, a.A).A\, a).A$ or $((a.A\, a).A\, a).A$ or $(a\, a.A\, a).A$.
    **Suggestion:** Similar to $T_\Sigma$ and $\overline{T}_\Sigma$, there could be two notations: $M_\Sigma$ and $\overline{M}_\Sigma$ corresponding to the non-annotated and annotated forms, respectively.

## Notes for Chapter 4

E9. [Page 65] This section on explicit quantification has as title "The Need for Quantifiers." This can be a bit confusing, because what is really necessary is not the quantifiers per se, but the explicit annotation of the set of variables involved in an equation, which instead of being defined as a pair $\langle t, t' \rangle$ is now defined as a triple $\langle X, t, t' \rangle$.

In books like Johnstone's "Notes on Logic and Set Theory" and Lambek & Scott's "Introduction to Higher-Order Categorical Logic", other notations not involving quantifiers are used for this, such as $t =_X t'$.

It is interesting to compare with discussion in page 249 about Horn clause symbols.

E10. [Page 72] In Exercise 4.5.5, it seems it is necessary to add the hypothesis that $A \neq \emptyset$.

E11. [Page 86] In Definition 4.10.6, if $\Sigma'$ contains symbols from $X$, then the translation of $(\forall X) t = t'$ along $\varphi$ is not correctly defined. This is a serious issue which has been considered in Diaconescu's book "Institution-Independent Model Theory": variables are triples of the form $(x, s, \Sigma)$, where $x$ is the name of the variable, $s$ is the sort of the variable, and $\Sigma$ is the signature for which the variable is considered. Then we have $\varphi(x, s, \Sigma) = (x, \varphi(s), \Sigma')$. In this way, variable name clashes are successfully avoided.

## Notes for Chapter 5

E12. [Page 107] The second sentence in the statement of Proposition 5.3.4 is not true: if $(\Sigma, A)$ is ground terminating, then $(\Sigma(X), A)$ may fail to be ground terminating; consider for example the rule $f(Z) \to f(f(Z))$ in Example 5.3.2. The second sentence should be read as "$(\Sigma, A)$ is ground terminating if $(\Sigma(X), A)$ is ground terminating," for the author writes so in the new statement of Proposition 5.3.4 in page 381.

E13. [Page 112] The explaining paragraph after Definition 5.5.3 is confusing because it does not correspond to the definition statement. Instead, the second definition says that $\sigma$ preserves the decrease of the weight, and the third definition says that all operations preserve weight decreasingness.

E14. [Page 118] After Theorem 5.6.4 (Orthogonality), an example showing why the left-linear condition is needed would help the reader (in the same way that Exercise 5.6.2 shows why the lapse free condition is needed). The following $TRS$ is non-overlapping and lapse free but not Church-Rosser because it is not left linear:

$$f(x,x) \to a, f(x, g(x)) \to b, c \to g(c)$$

E15. [Page 126] Proposition 5.3.4 reappears in this page, so note E12 in page 107 also applies here.

E16. [Page 128] In the first paragraph after Definition 5.8.2, it seems unclear whether $R_{k-1}$ should instead be $R_k$.

E17. [Page 131] Proposition 5.8.10 is the conditional generalization of Proposition 5.3.4 (in pages 107 and 126) and the same comment applies: if $(\Sigma, A)$ is ground terminating then $(\Sigma(X), A)$ may fail to be ground terminating. See notes E12 and E15 above.

E18. [Page 147] It would be necessary to check whether this argument is right.

E19. [Page 152] The comment before Definition 5.9.1 about adding "little more information" seems inconsistent because the definition does not take that into account.

E20. [Page 156] This dangling reference probably corresponds to a corollary which may have been removed.

## Notes for Chapter 6

E21. [Page 165] In the proof of the initiality Theorem 6.1.15 there is an important gap, because the author does not prove $T_{\Sigma,A} \models A$, and so we don't know whether $T_{\Sigma,A}$ is a $(\Sigma, A)$-algebra or not. Compare with note E42 for Theorem 9.1.10 in page 312.

E22. [Page 174] Exercise 6.4.1 coincides with Exercise 3.1.1 in Section 3.1.

E23. [Page 177] Example 6.5.7 on commutativity of addition could be used to show how lemmas come up in a proof attempt.

E24. [Page 178] Exercise 6.5.1 coincides with `lemma0` in previous Example 6.5.7.

E25. [Page 180] The `protecting` notion (`pr NAT` in Exercise 6.5.5) has not been explained before. The notion is quickly explained later in Example 7.3.11 (page 197).

## Notes for Chapter 7

E26. [Page 195] In the proof of Theorem 7.3.9, it seems that there is a small gap in the proof of $N \models A \cup B$. Since $[\![\_]\!]_B : T_\Sigma \to N$, we should consider $b : X \to T_\Sigma$ such that $b; [\![\_]\!]_B = a$, because $[\![\_]\!]_B$ is a surjection.

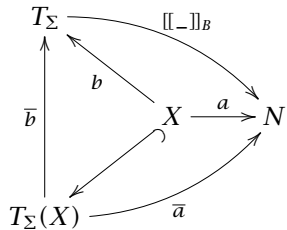

Since $a = b; [\![\_]\!]_B$ and there exists a unique homomorphism $T_\Sigma(X) \to N$ extending $a$, we have $\overline{b}; [\![\_]\!]_B = \overline{a}$. Now applying the given rule to $t$ with substitution $b : X \to T_\Sigma$ gives $\overline{b}(t) \Rightarrow_{A/B} \overline{b}(t')$, so these two terms have the same canonical form, i.e., $[\![\overline{b}(t)]\!]_B = [\![\overline{b}(t')]\!]_B$ and thus $\overline{a}(t) = \overline{a}(t')$.

E27. [Page 201] Here it is stated that the `protecting` notion was defined in Chapter 6, but this is not really the case, as pointed out about Exercise 6.5.5 in page 180 (see note E25 above). The notion is quickly introduced (in Chapter 7) in Example 7.3.11 (page 197).

E28. [Page 203] Corollary 7.3.19 deserves a more detailed proof.

E29. [Page 211] For the statement of Proposition 7.4.3, it needs to be explained how the triangular system $T$ becomes a first-order formula.

E30. [Page 216] As above (see note E29), for the statement of Proposition 7.4.10, it needs to be explained how the conditional triangular system $T$ becomes a first-order formula.

E31. [Page 241] Lemma 7.7.16 and Proposition 7.7.17 refer to weak rewriting modulo, which has not been defined in the conditional case.

## Notes for Chapter 8

E32. [Page 257] In Definition 8.3.2 the author should denote by $\hat{a} : WFF_X(\phi) \to B$ the extension of $a : X \to M$ to well-formed formulae because he denoted by $\overline{a} : T_\Sigma \to M$ the extension of $a$ to terms. As we may notice later, in the definition of substitution (Definition 8.3.16 in page 266), he denotes by $\overline{\theta} : T_\Sigma(X) \to T_\Sigma(X)$ the extension of $\theta : X \to T_\Sigma(X)$ to terms, and by $\hat{\theta} : WFF_X(\phi) \to WFF_X(\phi)$ the extension of $\theta$ to $(\phi, X)$-formulae. So, there seems to be an inconsistency in the notations.

E33. [Page 257] Since formulas and their meaning are defined with respect to a set $X$ of variables, for satisfaction $M \vDash_\Phi P$ to be well defined (in Definition 8.3.2), independence from $X$ must be proved.

However, the definition of satisfaction is clearly ambiguous because it depends on the set of variables $X$. Assume, for example, a $\Phi$-model $M$ such that $M_s = \emptyset$ for some sort $s \in S$, and $[\![P]\!]^M_\emptyset = \emptyset$, i.e., $M \nvDash_\Phi P$, for some sentence (closed formula) $P \in WFF_\emptyset(\Phi)$. Let $X = \{x : s\}$ be the variable set with just one element $x$ of sort $s$. We have $P \in WFF_X(\Phi)$ and $[X \to M] = \emptyset = [\![P]\!]^M_X$. Hence, $M \vDash_\Phi P$, which is a contradiction with our initial assumption $M \nvDash_\Phi P$.

This satisfaction condition should be indexed not only by the signature $\Phi$ but also by the set of variables $X$. For example, the notation $\vDash^X_\Phi$ clears the above ambiguity.

E34. [Page 257] The concept of institution, used at the end of Definition 8.3.2, has only been mentioned previously in the footnote 1 in page 250.

E35. [Page 265] In the statement of Proposition 8.3.15 it could be added that, if $P$ does not contain variables with empty sort, then the nonempty assumption on carriers of $M$ may be omitted.

E36. [Page 269] In the proof of Corollary 8.3.23, Theorem 8.3.22 is applied to $Q = (\forall Y) P$, which requires that $\theta$ is capture free for $Q$.

E37. [Page 288] The hypothesis $Bound(P) \cap X = \emptyset$ must be added to the statement of Proposition 8.5.1; otherwise, it could be possible for $\theta$ not to be capture free for $P$ and then Lemma 8.3.25 could not be applied.

Furthermore, the notation for the function $y : M^{s_1,...,s_n} \to M_s$ should be different from the variable $y : s$. The reason is that $\theta$ has source $X \cup \{y\}$ and target $T_\Sigma(Y)(X \cup \{y\})$. Thus, the target contains $y$ regarded both as variable and as operation symbol.

E38. [Page 289] Proposition 8.3.15 is used in the proof of Proposition 8.5.2 without checking the nonempty carriers condition.

E39. [Page 295] The notation for assignment application to a formula seems different in the proof of Proposition 8.7.1.

E40. [Page 295] In the paragraph after Proposition 8.7.1, $\theta_m$ is not a substitution, but an assignment in general. Then $\theta_m(P)$ is not a sentence.

E41. [Page 303] The concept "anarchic" used in Exercise 8.7.4 has not been defined.

# Notes for Chapter 9

E42. [Page 312] In the proof of Theorem 9.1.10, the author does not prove $T_{\Sigma,A} \vDash A$. Hence, we don't know whether $T_{\Sigma,A}$ is a $(\Sigma, A)$-algebra or not. Compare with note E21 for Theorem 6.1.15 in page 165.

E43. [Page 313] Here the author invokes Definition 8.3.2, meaning that we have to deal here with the same problems as in Chapter 8 (see above notes E32–E33 about this definition).

## Notes for Chapter 10

E44. [Page 326] It seems that the conclusions of Theorem 10.2.9 and Proposition 10.2.10 coincide. However, Theorem 10.2.9 requires signatures to be regular while Proposition 10.2.10 does not. This is rather strange even with the additional explanations.

E45. [Page 329] The calligraphic notation for algebras in Proposition 10.2.18 and the paragraphs above has not been used before.

E46. [Page 336] In Example 10.4.7, a function `op f : A -> A` should be added, in order to point out that `[f(a)] = {f(a),f(b)}` and `f(c) ∉ [f(a)]`.

E47. [Page 345] In the paragraph before Theorem 10.6.6, the statement "for arbitrary $A$, it is necessary and sufficient that $\Sigma$ has no quasi-empty models" needs a proof.

## Notes for Appendix B

E48. [Page 391] In the proof of Theorem 10.2.8, the restriction condition in fact seems to correspond to what is called the monotonicity condition in Definition 10.1.5.

E49. [Page 393] The star notation used in the proof of Theorem 10.4.11 (Initiality) has not been used before. Instead, the overbar notation was preferred.